\documentclass{elsart}
\pdfoutput=1
\usepackage{ifpdf}
\usepackage{natbib}
\usepackage{graphicx}
\usepackage{amsmath}
\usepackage{amssymb}
\newcommand{\mnras}{MNRAS }
\newcommand{\apjl}{ApJ }
\newcommand{\apj}{ApJ }
\newcommand{\apjs}{ApJ }
\newcommand{\aj}{AJ }
\newcommand{\jcap}{JCAP}
\newcommand{\jgr}{JGR}
\newcommand{\nat}{Nature }
\newcommand{\aap}{A\&A }

\newcommand{\memsai}{Mem. Soc. Astron. It.}
\newcommand{\na}{New Astron.}
\newcommand{\nar}{New Astron. Rev.}
\newcommand{\pasj}{Pub. Astro. Soc. of Japan}

\newcommand{\prd}{Phys. Rev. D }

\newcommand{\physrep}{Physics Reports }

\newcommand{\araa}{Annual Review of Astronomy and Astrophysics}

\def\beq{\begin{equation}}
\def\eeq{\end{equation}}

\def\implies{\Longrightarrow}
\long\def\symbolfootnote[#1]#2{\begingroup%
\def\thefootnote{\fnsymbol{footnote}}\footnote[#1]{#2}\endgroup}
\newcommand{\gae}{\lower 2pt \hbox{$\, \buildrel {\scriptstyle >}\over {\scriptstyle
\sim}\,$}}
\newcommand{\lae}{\lower 2pt \hbox{$\, \buildrel {\scriptstyle <}\over {\scriptstyle
\sim}\,$}}
\newcommand{\aprop}{\lower 2pt \hbox{$\, \buildrel {\scriptstyle \propto}\over 
   {\scriptstyle \sim}\,$}}

\begin{document}

% The lineno packages adds line numbers. Start line numbering with
% \begin{linenumbers}, end it with \end{linenumbers}. Or switch it on
% for the whole article with \linenumbers.
% \usepackage{lineno}
% \linenumbers

\begin{frontmatter}

 \title{The Physics of Gamma-Ray Bursts \& Relativistic Jets}
 \author{Pawan Kumar}
 \ead{pk@astro.as.utexas.edu}
 \address{Department of Astronomy, University of Texas at Austin, Austin, TX 78712, USA}
% \author{Pawan Kumar$^1$ \& Bing Zhang$^2$}
 \author{Bing Zhang}
 \ead{zhang@physics.unlv.edu}
 \address{Department of Physics \& Astronomy, University of Nevada Las Vegas, 
Las Vegas, NV 89154, USA}  
% Kavli Institute for Astronomy and Astrophysics and Department of Astronomy,
%Peking University, Beijing 100871, China}

%\begin{keyword}


%\end{keyword}
%\end{frontmatter}

\begin{abstract}
We provide a comprehensive review of major developments in our understanding
of gamma-ray bursts, with particular focus on the discoveries made within the
last fifteen years when their true nature was uncovered. We describe the
observational properties of photons from the radio to multi-GeV bands, both
in the prompt emission and the afterglow phases.
Mechanisms for the generation of these photons 
in GRBs are discussed and confronted with observations to shed light on the
physical properties of these explosions, their progenitor stars and the 
surrounding medium. After presenting observational evidence
that a powerful, collimated, jet moving at close to the speed of light is
produced in these explosions, we describe our current understanding
regarding the generation, acceleration, and dissipation of the jet.
%jet and compare these properties with jets associated with AGNs and pulsars.
We discuss mounting observational evidence that long duration GRBs are 
produced when massive stars die, and that at least some short duration 
bursts are associated with old, roughly solar mass, compact stars.
The question of whether a black-hole or a strongly magnetized,
rapidly rotating neutron star is produced in these explosions is also
discussed. We provide a brief summary of what we have learned about
relativistic collisionless shocks and particle acceleration from GRB afterglow
studies, and discuss the current understanding of radiation mechanism during 
the prompt emission phase. We discuss theoretical predictions of 
possible high-energy neutrino
emission from GRBs and the current observational constraints. Finally, we 
discuss how these explosions may be used to study cosmology, e.g. star formation, metal
enrichment, reionization history, as well as the formation of first stars
and galaxies in the universe.
% and to address fundamental questions in physics, such as Lorentz invariance.
\end{abstract}

\end{frontmatter}

\newpage

\tableofcontents 
%\listoftables

\section{Introduction}
\label{introduction}

This introduction to Gamma-Ray Bursts (GRBs) is meant to provide a
brief summary of their main properties so that someone not interested
in details can obtain a quick overview, in a few pages, of the main 
properties of these explosions from the reading of this introduction.

The serendipitous discovery of Gamma-Ray Bursts (GRBs)
in the late sixties by the Vela satellites\footnote{Vela --- short for velador, 
meaning ``watchman'' in Spanish --- were a group of 12 satellites (including
6 advanced Vela design) that were launched starting from October 17, 1963 
until 1970, and the last satellite was decommissioned in 1984 (even though
they were designed for a nominal life of 6--18 months). Vela satellites
were launched to monitor compliance with the treaty ``banning nuclear weapon 
tests in the atmosphere, in outer space and under water'' signed by the 
governments of the Soviet Union, the United Kingdom and the United States, in
Moscow on August 5, 1963 before being opened for signature by other countries. 
It was ratified by the U.S. Senate on September 24, 1963. The treaty went into 
effect on October 10, 1963.} 
\citep{klebesadel73} puzzled astronomers for several decades: GRBs
are irregular pulses of gamma-ray radiation (typically lasting for less than one
minute), with a non-thermal (broken power-law) spectrum peaking at
$\sim$10--10$^4$ keV,  and are seen a few times a day at random locations
in the sky \citep[e.g.][]{band93,kouveliotou93,meegan92}.
Their spectacular nature, detection at redshift
larger than 9 with current generation of instruments, and their connection 
with supernovae explosions and possibly black-holes formation,
have led to a great deal of time and effort invested to their study
\citep[e.g.][]{fishman95,piran99,meszaros02,zhangmeszaros04,piran04,woosley06,foxmeszaros06,zhangcjaa07,gehrels09}.

The histogram of GRB duration has two distinct peaks. One at 0.3s and the 
other at about 30s, and there is a trough in between the peaks at 2s. Bursts 
with duration less than 2s are classified as short-GRBs and those that last
for more than 2s are called long-GRBs. Based only on the two peaks in 
the duration distribution, and well before anything was known about the
distance or physical origin of GRBs, it was suspected that these peaks
correspond to two physically distinct progenitors. Recent observations
have confirmed that long-GRBs are one possible outcome of the collapses
of massive stars (mass $\gae 15 M_\odot$), and that at least some of 
the short-GRBs arise in the mergers of compact objects in binary systems
(perhaps merger of two neutron stars or a neutron star and a black hole).
The connection between the classifications based on burst duration 
and based on distinct physical origins turns out to be more complicated 
though, and is still not fully understood.

Distances to GRBs were completely uncertain until the launch of 
Compton-Gamma-Ray-Observatory (CGRO), from space shuttle Atlantis, on
5 April 1991 in a low earth orbit at 450 km (in order to avoid the 
Van Allen radiation belt that covers $\sim 10^3 - 6 \times 10^4$ km altitude). 
It carried four instruments that provided a wide energy band coverage of
20 keV --- 30 GeV (at 17 tons, CGRO, was the heaviest astrophysical 
payload flown at that time). CGRO established that these bursts 
are isotropically distributed \citep{meegan92} and their number at the
faint end (but well above the instrument threshold) deviates from the
expected Euclidean count\footnote{The easiest way
to understand this relation is to consider sources of the same intrinsic 
luminosity, $L$, uniformly distributed in an Euclidean space. The observed
flux decreases with distance $R$ as $R^{-2}$, and the total number of sources
within $R$ grows as $R^3$. The observed flux from these sources is 
$>f=L/(4\pi R^2)$. Hence the total number of objects an observer sees
with flux above $f$ scales as $f^{-3/2}$. This argument is easy to 
generalize to consider a more realistic source luminosity function.} 
$N(>f)\propto f^{-3/2}$ 
\citep[e.g.][]{mao92,piran92,fenimore93}.
These two discoveries taken together convinced most astronomers that
GRBs are located at distances much larger than the size of the local
group of galaxies.

The confirmation of the cosmological distance to GRBs was obtained in
1997, when the BeppoSAX satellite, launched on April 30, 1996,  provided 
angular position of bursts to within 4 arc-minutes -- more than a factor 20 
improvement compared with the Compton Gamma-ray Observatory -- which 
enabled optical and radio
astronomers to search for counterparts for these explosions. A rapidly
fading X-ray \& optical emission (the ``afterglow'') accompanying a
GRB was found on February 28, 1997, about a day after the detection of 
a burst, and that led to the determination of redshift for this GRB to be 0.695 
\citep{costa97,frontera98,vanpara97}. This launched a new era in the
study of GRBs which has led to a wealth of new information and a much deeper
understanding of these enigmatic explosions 
\citep[e.g.][]{frail97,kulkarni98,bloom99,zhangmeszaros04,piran04,zhangcjaa07,gehrels09}.

From burst redshift and flux we know that GRBs radiate between
$10^{48}$ and $10^{55}$ ergs, if isotropic. This means that GRBs are
the most energetic explosions in the Universe; the luminosity of the
brightest bursts rivaling that of the entire Universe at all
wavelengths albeit for only a few seconds \citep{kulkarni99}.

Our understanding of GRBs has improved enormously in the last 15 years due to
the observations made by several dedicated $\gamma$-ray/X-ray satellites
(BeppoSAX, KONUS/Wind, HETE-2, Swift, Integral, AGILE, Fermi) and the 
follow-up observations
carried out by numerous ground-based optical, IR, mm and radio observatories.
Much of this progress has been made possible by the monitoring and theoretical
modeling of long-lived afterglow emissions following the burst.

We know from breaks in optical \& X-ray afterglow lightcurves that GRBs are
highly beamed \citep{rhoads99,sari99}, and the true amount of energy release 
in these explosions is $10^{48}-10^{52}$ ergs 
\citep{frail01,panaitescu01,berger03,curran08,liang08,racusin09,cenko10}.

The follow-up of GRBs at longer wavelengths (X-ray, optical, and radio) has
established that afterglow light-curves often decay as a power-law with
time ($F_\nu \aprop t^{-1.0}$) and have a power-law spectrum
($F_\nu \aprop \nu^{-0.9\pm 0.5}$). The synchrotron radiation from the 
external {\em forward-shock} --- which results from the interaction of 
GRB-ejecta with the circumburst medium 
\citep{rees92,paczynski93,meszarosrees93,meszarosrees97} 
--- provides a good fit to the multi-wavelength afterglow data 
for GRBs \citep[e.g.][]{panaitescu02}.

In many cases, the decay of the optical or X-ray afterglow light-curve steepens
to $F_\nu \aprop t^{-2.2}$ at $\sim 1$ day after the burst. The most natural 
explanation for this steepening \citep[foreseen by][]{rhoads99} is that GRB
outflows are not spherical but collimated into narrow jets \citep{sari99}. 
As the ejecta is
decelerated and the strength of the relativistic beaming diminishes, the edge
of the jet becomes visible to the observer. The finite angular extent
of the ejecta leads to an achromatic faster decay of optical \& X-ray
lightcurves. This achromatic transition from a slower to a faster decay of 
lightcurves is called ``jet-break''.

The initial opening angle of the jet and its kinetic energy can be obtained 
by modeling the broadband emission (radio to $X$-ray) of those GRB afterglows
whose light-curve fall-off exhibited a jet-break. 
From these fits it is found that the opening angle of 
GRB jets is in the range of $\sim2$ -- 10 degrees, thus the ejecta collimation 
reduces the required energy budget by a factor $\sim10^2-10^3$ relative to the
isotropic case; the true amount of energy release for most long
duration GRB is found to be $10^{49}\sim10^{52}$ erg 
\citep{rhoads99,sari99,frail01,panaitescu01,berger03,cenko10}. The medium 
within $\sim0.1$ pc of the burst is
found to have a uniform density in many cases, and the density is of the
order of a few protons per ${\rm cm}^3$ \citep{panaitescu02}. This is a
surprising result in the light of the evidence that long
duration GRBs are produced in the collapse of a massive star -- as
suggested by \cite{woosley93,paczynski98,macfadyen99} -- where we
expect the density to decrease with distance from the center as $r^{-2}$ 
due to the wind from the progenitor star 
\citep{dailu98c,chevalier99,chevalier00,ramirezruiz01}.

It was expected from theoretical considerations that GRB outflows are
highly relativistic \citep[e.g.][]{paczynski86,goodman86,fenimore96,piran99}.
A direct observational confirmation of this was provided by measurements 
of radio scintillation for GRB 970508 \citep{goodman97,frail97}, and 
``superluminal'' motion of the radio afterglow of a relatively nearby 
burst GRB 030329 \citep{taylor04} where the blastwaves were found to be 
still mildly relativistic several weeks after the explosion.

The evidence for association of long-duration GRBs (those lasting for more
than 2s) with core collapse SNa comes from two different
kinds of observations: (i) GRBs
are typically found to be in star forming regions of their host galaxies
\citep[e.g.][]{bloom02,fruchter06,christensen04,castroceron06}; 
(ii) For several GRBs, Type Ic supernovae have been detected 
spectroscopically associated with the GRBs.
%\footnote{Many of these
%GRBs with associated supernovae have luminosity significantly lower than 
%the typical long-GRB, and it might be that these are not representative of
%the population, \citep[e.g.][]{bromberg11}. }:
Most of the SNe-associated GRBs have luminosity significantly lower than
typical GRBs\footnote{These low-luminosity GRBs may not be representative
of the main GRB population \cite[e.g.][]{liang07,bromberg11}.}, e.g. 
GRB 980425 \citep{galama98}, 030329 \citep{hjorth03,stanek03}, 
060218 \citep{modjaz06,campana06,pian06}, 100316D \citep{chornock10,starling11}, 
101219B \citep{sparre11}, and 120422A \citep{melandri12}. However, two
nearby high-luminosity GRBs, i.e.
031203 \citep{malesani04} and 130427A \citep{xu13,levan13}, are also
found to be associated with Type Ic SNe.
Additionally, a subset of about a dozen
GRBs show at late-times ($\sim$ 10 days) SNa-like
``bump'' in the optical data and simultaneously a change in color
that is inconsistent with synchrotron emission, and suggests that
optical flux from the underlying supernova is starting to overtake
the GRB afterglow flux \citep{bloom99,woosley06}.

Significant progress toward answering the long standing question regarding 
the nature of short duration GRBs (those lasting for less than 2s) was made
possible by the Swift satellite's more accurate localization of these bursts
(3 arcmin vs. a few degrees for Compton-GRO).
This led to the discovery that a fraction of these bursts are located in
elliptical galaxies, i.e. associated with older stellar population, and were
found to be on average less energetic and at a lower redshift
\citep{gehrels05,fox05,barthelmy05a,berger05,panaitescu06,bloom06,guetta06,nakar07}. These observations are consistent with
the old idea that these bursts originate from neutron star mergers
\citep{eichler89,narayan92}. However, there is no 
conclusive proof for this model as yet.

The Swift satellite, designed for the study of GRBs and launched in November
2004, has X-ray and UV-optical telescopes on board and provides localization 
of bursts to within 3 arcminutes. When Swift's gamma-ray telescope
(Burst and Altert Telescope or BAT) detects a burst, the X-Ray Telescope
(XRT) and the UV-Optical Telescope (UVOT) on board Swift quickly
slew to the GRB position within 60-100 seconds to observe the target,
which provides excellent coverage of the transition from the 
prompt $\gamma$-ray phase to the lower-frequency afterglow emission 
phase\footnote{Prior to the launch of Swift, there was a gap of 
typically about 7-8 hours between the 
detection of a burst in the $\gamma$-ray band and the follow up study of its 
afterglow emissions in the X-ray and lower energy bands.}. 
Swift has provided a wealth of puzzling observations
\citep{tagliaferri05,chincarini05,nousek06}, and revealed that a variety
of physical processes shape the early X-ray afterglow lightcurves
\citep{zhang06}.
Its XRT has found that for about 50\% of GRBs the X-ray
flux decays very rapidly after the burst ($F_x \propto t^{-3}$), followed by a
plateau during which the X-ray afterglow flux decrease is much slower
($F_x \propto t^{-1/2}$) than expected in the standard forward-shock
model. The former feature indicates that the $\gamma$-ray prompt
radiation and afterglows are produced by two different mechanisms or 
arise from different outflows while the latter perhaps suggests that the 
forward shock that powers the afterglow takes a long time (of order several
hours) to become a self-similar blast wave with constant energy
(another possibility is that the observed x-ray radiation is not produced 
in the external shock).

Swift has also discovered episodes of a sharp increase in the X-ray flux
(flares) minutes to hours after the end of the GRB
\citep{burrows05,chincarini07,chincarini10,margutti11}.
The rapid rise time for the X-ray flux, with $\delta t/t\sim 0.1$, rules
out the possibility that flares are produced as a result of inhomogeneity in
the circumstellar medium where the curvature of the relativistic
shock front limits $\delta t\sim R/2c\Gamma^2\sim t$ or $\delta t/t
\sim 1$ \citep{nakarpiran02,ioka05,nakargranot07}. This suggests that the 
central engine in these explosions is active for a time period much longer
than the burst duration\footnote{Well before the discovery of X-ray flares
\cite{katz97} suggested a long lived central engine activity as an 
explanation for the high-energy $\gamma$-rays from GRB 940217 detected 
5000 s after the GRO/BATSE trigger \citep{hurley94}. However, it is 
possible that these high energy photons might have been produced in 
an external shock.} \citep{burrows05,zhang06,fanwei05,lazzati07}.

While the X-ray and optical data for $t\gae10^4$s (time measured from 
$\gamma$-ray trigger) are consistent with external forward 
shock emission, the features seen in the X-ray data prior to $\sim10^4$s 
are not well understood. Similarly the expected achromatic breaks in the
lightcurves (associated with finite jet angle) are seen in some bursts but 
not others \citep{fanpiran06a,panai06b,liang07b,sato07,liang08,curran08,racusin09}.

One of the foremost unanswered questions about GRBs is the physical 
mechanism by which prompt $\gamma$-rays -- the radiation that triggers 
detectors on board GRB satellites -- are produced.
Is the mechanism the popular internal shock model\footnote{According
to the internal shock model, a fraction of jet kinetic energy is converted
to thermal energy when a faster moving segment of the jet collides with a 
slower moving part that was ejected at an earlier time. The thermal 
energy produced is then radiated away as $\gamma$-ray photons via a 
number of different mechanisms such as the synchrotron and inverse-Compton
process.} \citep{rees94},
the external shock model, or something entirely different? Are $\gamma$-ray 
photons generated via the synchrotron process or inverse-Compton process, 
or by a different mechanism? Answers to these questions will help us 
address some of the most important unsolved problems in GRBs -- how is the
explosion powered in these bursts? Does the relativistic jet produced in 
these explosions consist of ordinary baryonic matter, electron-positron
pairs, or is the energy primarily in magnetic fields? 

The Fermi satellite, a multi-purpose high energy satellite launched in
June 2008, has provided useful data extending from $\sim10$keV
to $>$300GeV to help answer some of these questions. It has 
made several important discoveries regarding GRBs 
\citep{abdo09a,abdo09b,ackermann10,ackermann11,zhang11}:
(1) in most cases the high energy photons ($>$10$^2$MeV) are detected with a
delay of a few seconds with respect to the lower energy
emission ($\lae 1$MeV); (2) high energy emission lasts for a time period 
much longer ($\sim10^3$s) than emission below $\sim1$ MeV (which lasts for 
less than 1 minute for most GRBs);
(3) the broad-band prompt $\gamma$-ray spectra are found in most cases to 
  consist
  of one peak and power law functions with different indices at low and high 
  energies with 
  a smooth transition from one to the other over a factor $\lae10$ in 
  frequency (this is the so called ``Band'' spectrum), however in a few cases
 the spectrum has an addition component.

There are several different lines of strong evidence suggesting
that the high energy photons ($>$10$^2$MeV) we observe after the 
prompt phase ($t\gae 10$s)
are produced in the external forward shock via the synchrotron process 
\citep{kumar09,ghisellini10}. On the other hand the origin of prompt
$\gamma$-ray emission, low and high energies, remains a puzzle. 
Some of the proposed models are: synchrotron and inverse-Compton (IC)
radiation processes in internal or external shocks or at sites where
magnetic field in Poynting jet is dissipated \citep[e.g.][]{rees92,
dermermitman99,lyutikov03,zhangyan11}; and photospheric radiation
with contribution from multiple IC scatterings 
\citep[e.g.][]{thompson94,ghisellini99,meszarosrees00,peer06,peer08,giannios07,ioka07,
asano09,lazzati10,beloborodov10,toma11,mizuta11,nagakura11,bromberg11}.

Swift satellite has found GRBs at high redshifts; the highest redshift
GRB discovered to date is at $z=9.4$ when the universe was just
0.52 billion years old or 3.8\% of its current age 
\citep{cucchiara11}. Swift is capable of
detecting bursts of similar intrinsic brightness up to redshift of about 15.
Because of their intrinsically simple spectrum and extremely high luminosity,
GRBs are expected to offer a unique probe of the end of cosmic dark age
 when the first stars and galaxies were forming. 

This review is organized in the order of the best understood GRB 
properties discussed first and the least well understood phenomena described
last. We start with a brief review of radiation physics, and describe the
theory of GRB afterglows which began to be developed even before the
first detection of afterglow radiation.
 We then describe how well the afterglow theory does when confronted
with observations. We first consider the late time
afterglow observations (these observations starting from roughly half a day
after the explosion can last for weeks to months), and what they have taught
us about GRBs and the medium in their vicinity. This is followed by early
afterglow observation --- starting from $\sim$30s (2s) since the burst trigger
for long (short) GRBs and spanning a duration of a few hours --- and our
current understanding of the puzzles they pose. Then the least well
understood of all the data --- properties of the GRB prompt radiation ---
and the strengths and weaknesses of various models proposed to explain 
these observations are reviewed. Next, we take up the properties of the 
central engine, and describe the two leading models: a new-born
hyper-accreting black hole,  and a strongly magnetized, 
rapidly spinning neutron star (magnetar). We then move on to discuss
the possible progenitors of GRBs. We also devote a section to discussion
of possible neutrino emission from GRBs.
%The last part of the review is devoted to application of GRBs to study first-stars, evolution of star formation rate and metal enrichment, high redshift universe, and some aspects of fundamental physics. 

\section{Radiative processes}
\label{rad_process}

We provide in this section a brief overview of a few of the most 
important radiative processes in GRBs which will
be used extensively in this review. There are excellent books
that cover this topic in detail such as the monograph by 
\cite{rybicki79}, and books on high energy astrophysics by
\cite{longair10,krolik99,dermer09,kulsrud05}.
This section is no substitute for the extensive coverage
of this topic provided in these books. The purpose here
is to provide a quick summary of some of the main results
we need in other sections, so as to make this review somewhat
self contained.

We describe synchrotron, inverse-Compton and photo-pion processes 
in this section. A few basic relativity results that are needed 
for understanding of radiative processes are also included here.

\subsection{Photon arrival time from a moving source, Doppler
   shift, Lorentz invariance of power etc.}
\label{relativity}

\begin{figure}
\begin{center}
\includegraphics[width=12cm]{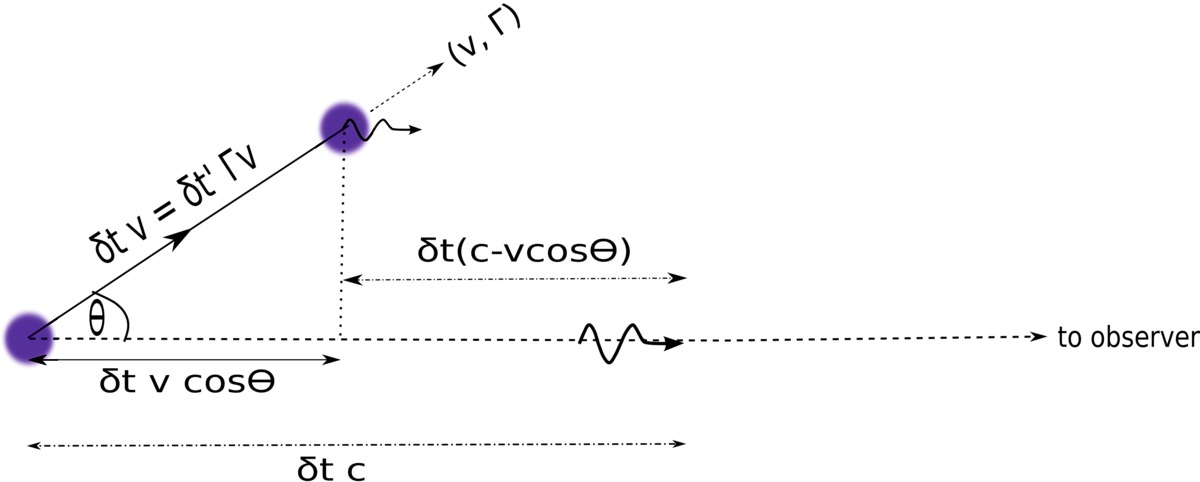}
\caption{The relation between pulse duration in source comoving frame,
$\delta t'$, lab frame ($\delta t$), and the time interval for pulse
received by a distant observer is shown in this figure. The source is
moving with speed $v$ (Lorentz factor $\Gamma$), 
at an angle $\theta$ with respect to observer
line of sight. One photon is emitted when the source was at the location
at the left side of the figure. And a second photon is emitted $\delta t'$ 
later when the photon has already traveled a distance $c \delta t$
toward the observer, and the source is also a distance $v \cos\theta
\delta t$ closer. The difference between these two distances is the
time interval in the observer frame for the arrival of the two 
photons which is given by equation \ref{observer-time}.
}\label{FIG:photon-arrival}
\end{center}
\end{figure}

Consider a source moving with speed $v$, and corresponding Lorentz 
factor $\Gamma$, at an angle $\theta$ with respect to the line of sight 
to the observer located far away from the source. Two photons are
emitted $\delta t'$ apart in the source comoving frame.
In the lab frame (the frame in which
the source is seen to move at speed $v$), the time interval 
of emitting these two photons is $\delta t = \Gamma \delta t'$.
The time difference for the arrival of these photons at the
observer is given by (see fig. \ref{FIG:photon-arrival}):
\begin{eqnarray}
  \delta t_{obs} & = & \delta t + \left[d - v\cos\theta (
     \delta t)\right]/c - d/c =   \delta t (1-v\cos\theta/c) \nonumber \\
& = & \delta t'\Gamma(1 - v\cos\theta/c) = \delta t' {\cal D}^{-1}
%   \approx \delta t'/\Gamma,
   \label{observer-time}
\end{eqnarray}
where $d$ is the distance to the source, 
\begin{equation}
 {\cal D} = [\Gamma (1-v \cos\theta/c)]^{-1}
\end{equation}
is the Doppler factor.
For $\theta \ll 1$ 
and $\Gamma\gg1$. the above expression for $\delta t_{obs}$
can be approximated as
\begin{equation}
    \delta t_{obs} \approx {\delta t'\over\Gamma} \left[ 1 + 
   (\theta\Gamma)^2/2\right] = {\delta t\over\Gamma^2} \left[ 1 +
   (\theta\Gamma)^2/2\right].
   \label{observer-time1}
\end{equation}

The photon frequency in the observer frame, $\nu$, can be
expressed in terms of the comoving frame frequency $\nu'$ 
using the standard Lorentz transformation of photon 4-momentum in
comoving frame --- $\nu'(1, \cos\theta', \sin\theta', 0)$ --- 
to the lab frame 4-momentum $\nu(1, \cos\theta, \sin\theta, 0)$
\begin{equation}
  \nu = \nu'\Gamma\left(1 + v\cos\theta'/c\right)\quad\quad \& \quad\quad
  \nu\cos\theta = \nu'\Gamma(\cos\theta' + v/c)
  \label{photon_LT}
\end{equation}
or
\begin{equation}
  \nu = {\nu'\over \Gamma \left(1 - v\cos\theta/c\right)}\equiv
    \nu' {\cal D},
  \label{doppler}
\end{equation}
which is the standard Doppler shift formula.

\noindent{\bf Relativistic beaming of photons}

The transverse component of momentum does not change under Lorentz 
transformation, i.e. its comoving and lab frame values are the same
\begin{equation}
  \nu\sin\theta = \nu'\sin\theta' \quad\quad {\rm or} \quad\quad
     \sin\theta = \sin\theta'/{\cal D}.
\end{equation}
For large $\Gamma$, $\theta \approx \theta'/\Gamma$, i.e. 
photons are focused in the forward direction such that the 
angular size of the photon beam in the lab frame is smaller than 
it is in the comoving frame by a factor $\sim\Gamma$. The solid angle
for a conical beam of photons in lab frame is smaller than in the 
comoving frame by a factor $\sim\Gamma^2$. A more precise expression for
Lorentz transformation of solid angle is: 
\begin{equation}
   d\Omega = \sin\theta d\theta d\phi = \sin\theta'd\theta'd\phi'/{\cal D}^2
    = d\Omega'/{\cal D}^2.
   \label{LT_solidangle}
\end{equation}

Next we show that the power\footnote{Power is defined as the frequency
integrated total energy radiated per unit time over 4$\pi$ steradians.}
radiated by a particle is Lorentz invariant
when the radiation beam is symmetric under parity transformation in 
particle rest-frame, i.e. the energy radiated per unit solid angle in 
directions ($\theta, \phi$) \& ($\pi-\theta, \pi+\phi$) are equal.
One of the easiest ways to see this 
is to consider the 4-momentum carried away by photons
emitted in time interval $\delta t'$ in the source frame,
which is: $P'\delta t'(1, 0, 0, 0)$; where $P'$ is the
power in source comoving frame, and the space components are
zero because of parity symmetry. The 4-momentum and the
elapsed time in the lab frame are: $\Gamma P' \delta t'(1, v, 0, 
0)$, $\Gamma \delta t'$. Hence the power in the lab frame
is $P = P' \Gamma \delta t'/(\Gamma \delta t') = P'$. 

\noindent{\bf Transformation of specific luminosity and specific intensity}

Another useful result concerns the Lorentz transformation of luminosity.
Let us consider a source that is spherically symmetric and is expanding 
with Lorentz factor (LF) $\Gamma$. The observed specific luminosity, $L_\nu$, 
is the total energy that flows through a surface enclosing the source 
per unit time and frequency. Thus, it follows that
\begin{equation}
   L_\nu = {d E\over d\nu d t_{obs}} = \Gamma {d E'\over d\nu' d t'} = 
    \Gamma L'_{\nu'},
\end{equation}
where we made use of $d\nu d t_{obs}=d\nu' d t'$ (see eqs. \ref{observer-time} 
and \ref{doppler}), and $E = \Gamma E'$ when the 3-momentum vector is zero
as it must for a spherically symmetric radiation source.

The specific intensity is defined as flux per unit frequency and per unit solid 
angle carried by photons traveling within a narrow conical beam with its 
axis perpendicular to surface $dA$, i.e. 
\begin{equation}
   I_\nu \equiv {d E\over d\nu dt_{obs}dA d\Omega}.
\end{equation}
Considering that $E$ transforms as $\nu$, $d\Omega$ transformation is given by 
equation (\ref{LT_solidangle}), and $d\nu dt_{obs}dA$ is Lorentz invariant, 
we find
\begin{equation}
   I_\nu = {\cal D}^3 I'_{\nu'}.
 \label{LT_si}
\end{equation} 

\noindent{\bf Observed lightcurve from a source that is suddenly turned off}

Transient sources such as GRBs can turn off rapidly on a time scale of 
a second or less. Following \cite{fenimore95} and \cite{kumar00} 
we consider a case here where a relativistic, conical, optically thin 
source moving with LF $\Gamma$ turns off abruptly, and 
calculate how the flux declines with time as seen by a far away observer 
in a fixed frequency band. 

We consider the source to be a thin shell, and points in the source 
are specified by ($r, \theta, \phi$) where angle $\theta$ is measured 
with respect to the line of sight to the observer. The source turns 
off suddenly when 
it is at radius $r=R_0$. Photons released at $(r=vt, \theta, \phi)$ arrive at 
the observer with a time delay with respect to a photon emitted at $r=0$ of
\begin{equation}
   t_{obs} = t - r\cos\theta/c = t(1 - v\cos\theta/c) = t/(\Gamma 
   {\cal D}).
   \label{arrival_time}
\end{equation}
We calculate the lightcurve at frequency $\nu$ from the source after
time $t_{0,obs} \approx R_0/(2c\Gamma^2)$ which corresponds to the arrival
of photons from ($R_0, 0, 0$) at the observer. At $t_{obs} > t_{0,obs}$
the observer sees photons which left the source when $r<R_0$
as determined by equation (\ref{arrival_time}). 
The time dependence of the observed flux, when the intrinsic spectrum 
is $I'_{\nu'} = I' \nu'^{-\beta}$, follows from the 
Lorentz transformation of specific intensity. At any given observer time 
$t_{obs}> t_{0,obs}$ we receive radiation from $\theta > \theta_t$; 
where $\theta_t$ is the angle corresponding to time $t_{obs}$ 
such that $t_{obs} = R_0 (1/v - \cos\theta_t/c)$  (see eq. \ref{arrival_time}).
Considering that the observed flux is proportional to the integral of
$I_\nu$ over the solid angle of the source, we find $f_\nu \propto 
\int_{\theta_t} d\theta\, \sin\theta {\cal D}^{-(3+\beta)}$. Or $f_\nu\propto
(1-v\cos\theta_t/c)^{-(2+\beta)}\propto t_{obs}^{-(2+\beta)}$.
A more precise derivation of this result is outlined below.

The specific flux in observer frame from a relativistic source of comoving
specific intensity $I'_{\nu'}$ and spectrum $\propto \nu'^{-\beta}$ is given by
\begin{equation}
   f_\nu(t_{obs}) = \int d\Omega_{obs}\, I_\nu\cos\theta_{obs} = 
    2\pi \int d\theta_{obs}\, {I'_{\nu_0'}{\nu_0'}^\beta\sin2\theta_{obs} [(1+z)
   \Gamma]^{-(3+\beta)}\over 2\nu^{\beta} \left[1 - v\cos(\theta+\theta_{obs})/c
    \right]^{3+\beta} },
\end{equation}
where $\nu_0'$ is a frequency that lies on the powerlaw segment of the
spectrum for $I'_{\nu'}$, and we made use of the Lorentz transformation 
of specific intensity to obtain the last part of the above equation.
The factor $(1+z)^{3+\beta}$ in the above equation takes into account
redshift of frequency for a source at $z$. 

\begin{figure}
\includegraphics[width=13cm]{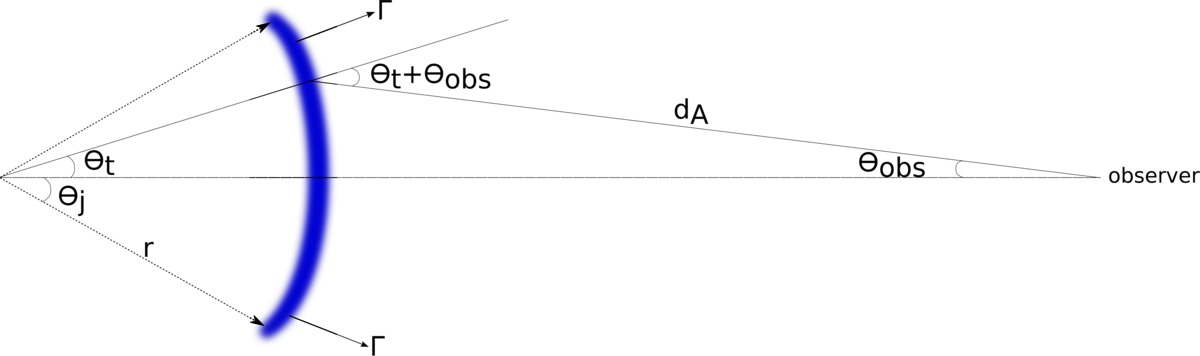}
\caption{A sketch of the various angles and distances for the large angle (or
high latitude) emission when the $\gamma$-ray source turns off suddenly.
}\label{FIG:LAT-sketch}
\end{figure}

Using the law of sine for a triangle (see fig. \ref{FIG:LAT-sketch}), 
$\sin\theta/d_A= \sin\theta_{obs}/r$, the above integral is transformed to
\begin{equation}
f_\nu \approx {2\pi I'{\nu_0'}{\nu_0'}^\beta\nu^{-\beta} \over [(1+z)\Gamma]^{3+\beta}} \left({ R_0
   \over d_A}\right)^2 \int_{\theta_t}^{\pi/2} d\theta\, 
  {\sin\theta\cos\theta\over (1-v\cos\theta/c)^{3+\beta}}.
\end{equation}
We replaced $\theta+\theta_{obs}$ in the denominator with $\theta$
since $\theta_{obs}\ll\theta$. The above integral is straightforward
to carry out and yields
\begin{equation}
   f_\nu(t_{obs}) \propto (1 - v \cos\theta_t/c)^{-(2+\beta)} \nu^{-\beta}
   \propto t_{obs}^{-(2+\beta)}\nu^{-\beta}.
\end{equation}
Thus, the observed radiation does not drop to zero as soon as the source is
turned off, but the flux declines rapidly with time and eventually vanishes 
when $\theta_t$ exceeds the angular size of the source ($\theta_j$).

\subsection{Synchrotron radiation}
\label{synch_rad}

Consider an electron of Lorentz factor $\gamma_e$, and speed $v_e$, moving
perpendicular to the magnetic field of strength $B$. 
The electric field in the electron rest frame is $E = \gamma_e
v_e B/c$, and hence according to the Larmor's formula the power 
radiated due to electron acceleration in this electric field is
\begin{equation}
    P_{syn} = {2 q^4 E^2 \over 3 c^3 m_e^2} = {2 q^4 B^2 \gamma_e^2
  v_e^2\over 3 c^5 m_e^2} = \sigma_T B^2 \gamma_e^2 v_e^2/(4\pi c),
  \label{P_syn}
\end{equation}
where $\sigma_T = 8\pi q^4/(3 m_e^2 c^4)$ is the Thomson 
cross section. Since electric dipole radiation has parity
symmetry, $P_{syn}$ is a Lorentz invariant quantity (see 
\S\ref{relativity}), and
hence the above equation gives the correct synchrotron power from an 
electron as viewed in the lab frame. The average power per electron,
for isotropic pitch angle distribution, is smaller than the
above expression by a factor $3/2$.

\begin{figure}
\begin{center}
\includegraphics[width=12cm]{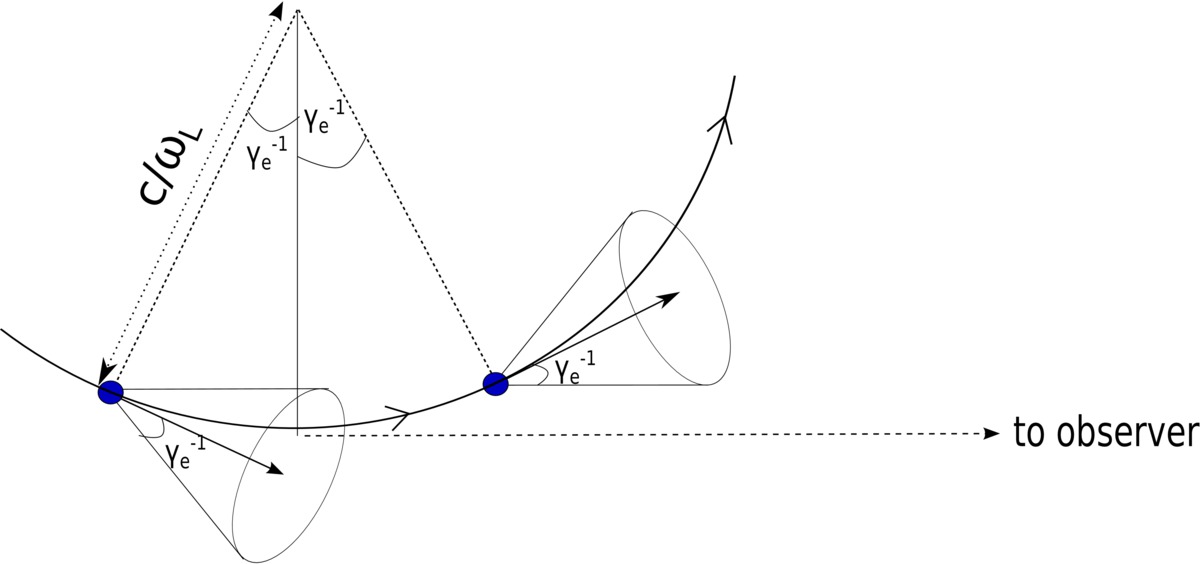}
\caption{The figure shows a segment of electron orbit that
is moving in magnetic field with LF $\gamma_e$. Radiation
from the electron is received by a distant observer only
for a small segment of the orbit when the electron's velocity
vector lies within $\gamma_e^{-1}$ of the observer line of sight
as a result of the beaming of photons in the forward direction
in the lab frame (radiation in electron comoving frame is dipolar
which covers almost $4\pi$ steradians). The observed synchrotron
peak frequency for emission from this electron follows from this
simple property (eq. \ref{syn_omega}).
}\label{FIG:relativistic_beaming}
\end{center}
\end{figure}

\begin{figure}
\begin{center}
\includegraphics[width=11cm]{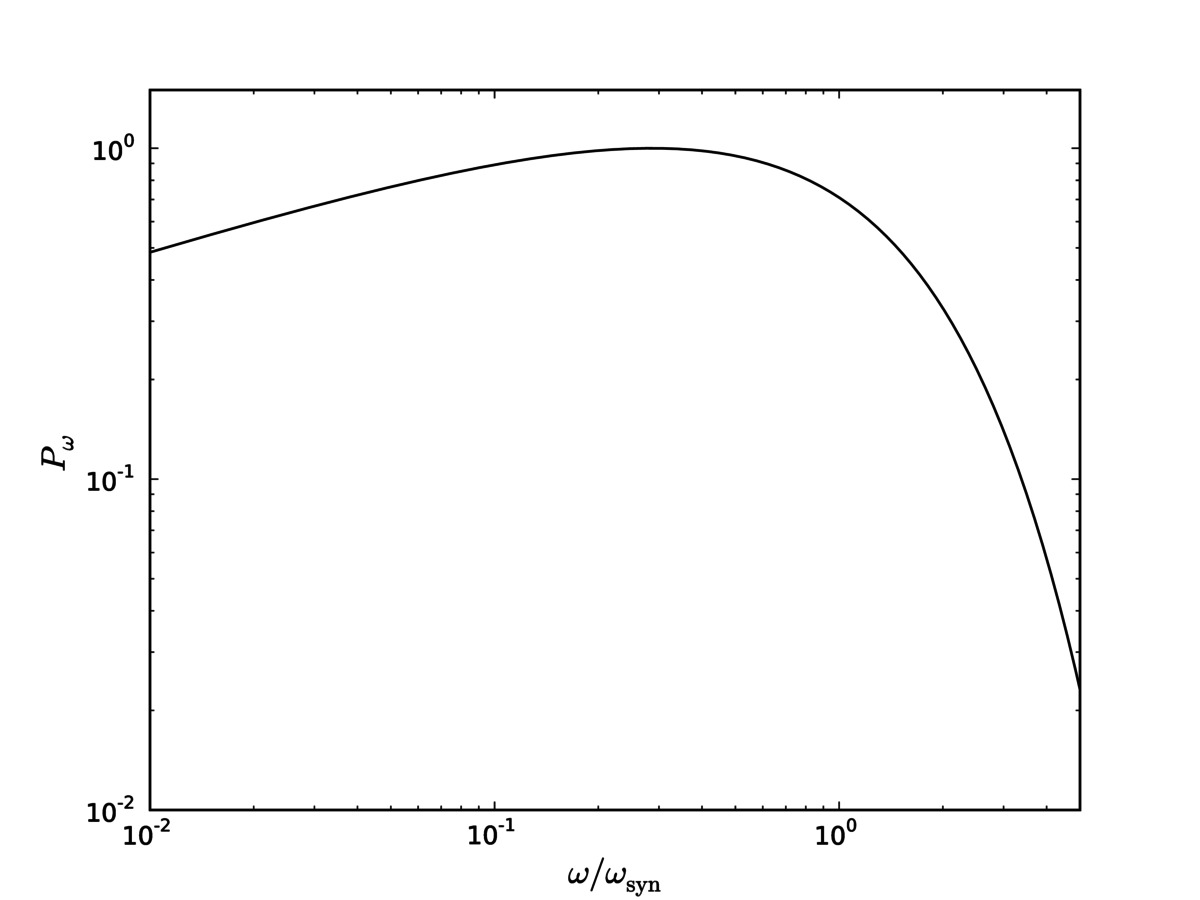}
\caption{This figure shows the synchrotron spectrum for a single
electron; the x-axis is frequency in units of $\omega_{syn}$ (see
eq. \ref{syn_omega}), and the flux is normalized to unity at the peak.
}\label{FIG:syn_spec}
\end{center}
\end{figure}

The Larmor frequency of the electron (or its angular speed) is
\begin{equation}
   \omega_L = {q B \over \gamma_e m_e c}.
\end{equation}
Due to relativistic beaming of photons described in \S\ref{relativity} 
radiation from the electron that a distant observer receives is confined
to the duration when
the electron velocity vector lies within an angle $\gamma_e^{-1}$ 
of the observer line of sight (fig. \ref{FIG:relativistic_beaming}).
The fraction of orbital time when this condition is satisfied is
$\sim 1/\pi\gamma_e$, and therefore the duration of the radiation
pulse received by the observer in each orbit is:
\begin{equation}
   \delta t_{obs} \sim {2\over \gamma_e\omega_L} {1\over 2\gamma_e^2}
        \sim {m_e c \over q B \gamma_e^2},
\end{equation}
where we used equation (\ref{observer-time}) that relates the
comoving time, $\delta t' = \delta t/\gamma_e$, to the observer
frame time duration for photon pulse arrival. The inverse of this
time is the characteristic frequency for synchrotron radiation 
which is given by
\begin{equation}
    \omega_{syn} \sim {q B \gamma_e^2\over m_e c} \quad{\rm and}\quad
     \nu_{syn} = {\omega_{syn} \over 2\pi} \sim {q B \gamma_e^2\over 2\pi 
   m_e c}, 
  \label{syn_omega}
\end{equation}
where $\nu_{syn}$ is the cyclic frequency. A more precise treatment
has an additional factor $(3/2)\sin \alpha$; $\alpha$ is the pitch angle 
between the electron's velocity and the magnetic field.
The synchrotron spectrum peaks at $\sim\nu_{syn}$. The spectrum 
below the peak scales as $P_{syn}(\nu) \propto \nu^{1/3}$ (this behavior is 
determined by the Fourier transform of the synchrotron pulse profile),
and it declines exponentially for $\nu>\nu_{syn}$ (see Fig. 
\ref{FIG:syn_spec}); we
refer to \cite{rybicki79} for the calculation of synchrotron
spectrum. The power per unit frequency $P_{syn}(\nu)$ 
at the peak of the spectrum is
\begin{equation}
   P_{syn}(\nu_{syn}) \sim P_{syn}/\nu_{syn} \sim {\sigma_T B m_e c^2
     \over 2 q}.
    \label{p_syn_peak}
\end{equation}

The synchrotron spectrum for a power-law distribution of electrons, 
$d n_e/d\gamma_e \propto \gamma_e^{-p}$, is $f_\nu\propto\nu^{-(p-1)/2}$.
This follows from adding up contributions to the specific flux at $\nu$
from those electrons with LF larger than
\begin{equation}
   \gamma_\nu \sim \left( {2\pi\nu m_e c \over q B}\right)^{1/2},
  \label{gamma_nu}
\end{equation}
and that leads to
\begin{equation}
   f_\nu = \int_{\gamma_\nu}^\infty d\gamma_e {d n_e\over d\gamma_e}
        P_{syn}(\nu) \propto \nu^{-(p-1)/2},
\end{equation}
where we made use of $P_{syn}(\nu) \propto (\nu/\nu_{syn})^{1/3}$ for
$\nu < \nu_{syn}$, and equation (\ref{syn_omega}) for $\nu_{syn}$.

\subsubsection{Effect of synchrotron cooling on electron distribution}

Another characteristic synchrotron frequency is associated with the
cooling of electrons ($\nu_c$). Let us consider that electrons are
accelerated at some time, and then cool via synchrotron radiation
for time duration $t_0$. Electrons with LF $\gae\gamma_c$ 
(defined below) lose a significant fraction of their energy 
during this time and their LF drops below $\gamma_c$
\begin{equation}
   {d m_e c^2 \gamma_e \over dt} = - {\sigma_T\over 6\pi} B^2 
    \gamma_e^2 c \quad\quad {\rm or} \quad \quad \gamma_c \sim 
    {6\pi m_e c\over \sigma_T B^2 t_0}.
  \label{gam_c}
\end{equation}
The synchrotron frequency corresponding to this LF is defined as
the synchrotron cooling frequency:
\begin{equation}
   \nu_c \equiv {3 q B \gamma_c^2\over 4\pi m_e c} \sim 
       {27 \pi q m_e c\over \sigma_T^2 B^3 t_0^2}.
  \label{nu_c}
\end{equation}
The power-law index of the synchrotron spectrum changes at $\nu_c$
due to the fact that electron distribution function for $\gamma_e
>\gamma_c$ is modified as a result of loss of energy. This can be
seen from the continuity equation for electrons in the energy space:
\begin{equation}
    {\partial \over \partial t}{d n_e\over d\gamma_e} + {\partial 
    \over \partial \gamma_e} \left[ \dot{\gamma_e} {d n_e\over 
   d\gamma_e}\right] = S(\gamma_e),
\end{equation}
where $\dot{\gamma_e} = - \sigma_T B^2 \gamma_e^2/(6\pi m_e c)$ is the
rate of change of $\gamma_e$ due to synchrotron loss, and $S(\gamma_e)$
is the rate at which electrons with LF $\gamma_e$ are injected into
the system. The continuity equation has a steady state solution 
($\partial/\partial t = 0$) for time independent magnetic field which 
is: $d n_e/d\gamma_e\propto \dot{\gamma_e}^{-1} \propto \gamma_e^{-2}$ for 
$\gamma_c < \gamma_e < \gamma_m$;
where $\gamma_m$ is the minimum LF of injected electrons i.e.
$S(\gamma_e) = 0$ for $\gamma_e < \gamma_m$. The synchrotron spectrum
corresponding to this segment of electron distribution function is
$f_\nu \propto \nu^{-1/2}$. For a time dependent magnetic field the
distribution function is not a power law function of $\gamma_e$ with
index 2, and in general its shape evolves with time \citep{uhm14}.

For $\gamma_e > \gamma_c > \gamma_m$, the solution of the continuity equation 
is $d n_e/d\gamma_e \propto \gamma_e^{-p-1}$ in the steady state (for constant
$B$). And the corresponding synchrotron spectrum is $f_\nu \propto 
\nu^{-p/2}$.

\subsubsection{Synchrotron self-absorption frequency}

Yet another characteristic frequency, $\nu_a$, corresponds to the case 
where absorption of photons by the inverse-synchrotron process becomes
important. The easiest way to determine $\nu_a$ is by the 
application of Kirchhoff's law -- the emergent specific flux cannot 
exceed the black-body flux corresponding to the appropriate electron
temperature which is
\begin{equation}
 k_B T \approx {\rm max} (\gamma_a, {\rm min}[\gamma_m,\gamma_c]) m_e c^2/2.7
\end{equation}
where $\gamma_m$, $\gamma_c$ and $\gamma_a$ are electron Lorentz factors
corresponding to synchrotron frequencies $\nu_m$, $\nu_c$ and $\nu_a$,
respectively, and $k_B$ is Boltzmann constant. The synchrotron self-absorption 
frequency ($\nu_a$) is the frequency where the emergent synchrotron flux 
is equal to the black-body flux:
\begin{equation}
   {2 m_e c^2 {\rm max}(\gamma_a,{\rm min}[\gamma_m,\gamma_c]) 
\nu_a^2 \over 2.7 c^2} \approx
         {\sigma_T B m_e c^2 N_>\over 4\pi q}
\label{nu_a}
\end{equation}
where the left side of this equation is the Planck function in the
Rayleigh-Jeans limit, and $N_>$ is the column density
of electrons with LF larger than ${\rm max}(\gamma_a,{\rm min}
[\gamma_m,\gamma_c])$.

\begin{figure}
\begin{center}
\includegraphics[width=13cm]{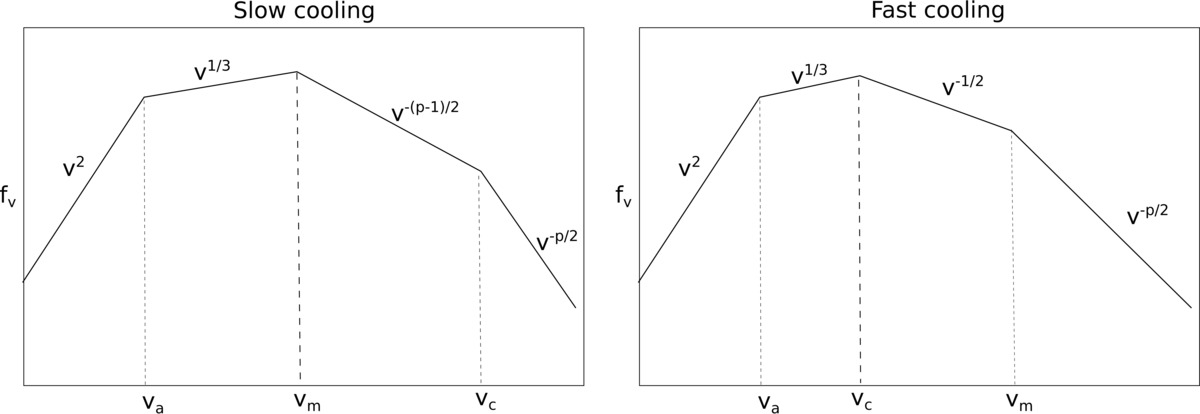}
\caption{Synchrotron spectrum for the case where $\nu_a<\nu_m<\nu_c$ is
shown in the left panel, and for the case $\nu_a<\nu_c<\nu_m$ in the
right panel, e.g. \citep{sari98}.
}\label{FIG:synchro_spectrum2}
\end{center}
\end{figure}

The emergent synchrotron spectrum for a distribution of electrons depends
 on the ordering of these characteristic frequencies. Spectra for 
two particular orderings are shown in fig. \ref{FIG:synchro_spectrum2}.

\subsubsection{Maximum energy of synchrotron photons}
\label{synch_rad_max}

Charged articles are accelerated as they travel back and forth across
a shock front via the first order Fermi process. They gain energy by a
factor $\sim 2$ each time they are scattered from one side to the other 
of a relativistic shock front. The maximum synchrotron frequency 
for an electron in this case turns out to be about $50 \Gamma$ MeV,
and for a proton it is a factor $m_p/m_e$ larger; $\Gamma$ is the
Lorentz factor of shocked plasma with respect to the observer, and
$m_p$ is proton mass.

The minimum time required for acceleration of a charged particle of
mass $m$ while crossing a shock front is of the order of the Larmor time 
$t'_L = m c\gamma/(qB')$; where $\gamma$ is LF of the particle in the 
shock comoving frame, and prime ($'$) refers to quantity measured in the
rest frame of the shocked fluid. 
The particle should not lose more than half its
energy to synchrotron radiation in time $t'_L$, otherwise it will
never get accelerated to LF $\gamma$. This implies that $4 q^4 B'^2
\gamma^2 t'_L/(9 m^2 c^3) < m c^2 \gamma/2$ or $ qB'\gamma^2/(2\pi m
c) < 9mc^3/(16\pi q^2)$. The left side of the last inequality is
the synchrotron frequency for the particle, and the right side
depends on the particle's mass. So we find that the maximum synchrotron
photon energy for an electron (proton) is $\sim$50 MeV (100 GeV) in
shocked fluid comoving frame under the optimistic Bohm diffusion limit.

It is possible to violate this limit, by a factor of a few at least,
when the magnetic field is highly inhomogeneous down stream of the shock
front; synchrotron photons produced when a particle is passing
through a region of much higher-than-average magnetic field can 
have energy larger than the limit described above, e.g. 
\cite{kumarhernandez12}.

\subsection{Inverse-Compton radiation}
\label{ic-radiation} 

When a photon of frequency $\nu$ is scattered by an electron of larger 
energy, the photon gains energy in this process on the average.  
If the electron Lorentz factor is $\gamma_e$, and $h \nu \gamma_e \ll
m_e c^2$, then the average frequency of scattered photon is 
$\nu_s\sim \nu\gamma_e^2$. This is easy to see by viewing the process
from the rest frame of the electron where the angle-averaged 
frequency of the incident photon is $\nu'\sim \nu\gamma_e$ 
(see eq. \ref{doppler} for relativistic Doppler shift). For
$h\nu'\ll m_e c^2$ the scattering is elastic -- the electron
recoil can be neglected -- so that the scattered photon has frequency
$\nu'$ (in electron rest frame) and its angular probability 
distribution is a dipole function.  Transforming the frequency 
of the scattered photon back to the original frame introduces another 
Lorentz boost and that results in $\nu_s \sim \nu\gamma_e^2$.

Consider next an electron moving through a radiation field
where the energy density in photons is $u_\gamma$. The power in 
IC-scattered photons, $P_{ic}$, follows from the energy boost 
by a factor $\gamma_e^2$ for each photon (independent of 
photon energy for the case where $h\nu\gamma_e \ll m_e c^2$
that we are considering here):
\begin{equation}
   P_{ic} \sim \sigma_T \int d\nu {u_\nu c\over h\nu} h\nu\gamma_e^2
     \sim \sigma_T u_\gamma \gamma_e^2 c,
  \label{P_ic1}
\end{equation}
where $u_\nu d\nu$ is energy density in photons of frequency between
$\nu$ and $\nu+\d\nu$; $\int d\nu \, u_\nu = u_\gamma$.
We see from equations (\ref{P_syn}) and (\ref{P_ic1}) that the ratio 
of synchrotron and IC powers is $u_B/u_\gamma$; where $u_B = B^2/8\pi$ 
is the energy density in magnetic field.

A particularly important case of IC radiation is when seed photons
are produced by the synchrotron process, i.e. electrons that produce
seed photons also IC scatter them to typically much larger energies. 
This process --- called synchrotron-self-Compton or SSC --- could be 
important for GRBs and other relativistic sources. 
The relative importance of synchrotron and IC processes for extracting
energy from a population of energetic electrons is specified by the 
Compton Y parameter. Energy density in photons for the synchrotron process is:
\begin{equation}
   u_\gamma = \int dr \int d\gamma_e\, {P_{syn}\over c} {d n_e\over 
     d\gamma_e} = {\sigma_T (\delta R) B^2 \over 6\pi} 
     \int d\gamma_e\, \gamma_e^2
      {d n_e\over d\gamma_e} = {\sigma_T (\delta R) n_e B^2 \over 6\pi} 
      \langle \gamma_e^2 \rangle,
\end{equation}
where $\delta R$ is the radial width of the source, and 
\begin{equation}
   \langle \gamma_e^2 \rangle \equiv {1\over n_e} \int d\gamma_e\, 
   \gamma_e^2 {d n_e\over d\gamma_e}.
\end{equation}
Making use of the expression for $u_\gamma$ for synchrotron radiation
we find the Compton-Y parameter to be
\begin{equation}
    Y \sim P_{ic}/P_{syn} \sim \tau_e \langle \gamma_e^2\rangle, 
   \quad\quad{\rm where} \quad\quad
          \tau_e = \sigma_T (\delta R) n_e
  \label{Y-expression1}
\end{equation}
is the optical depth of the source to Thomson scattering. 

\noindent{\bf IC spectrum}
\smallskip
The spectrum of IC radiation is obtained by convolving electron 
distribution with the seed photon spectrum \citep{rybicki79}:
\begin{equation}
  f_{ic}(\nu_{ic}) \approx {3\sigma_T (\delta R)\over 4} \int d\nu\,
   {\nu_{ic}\over \nu^2} f_{syn}(\nu)\int {d\gamma_e\over \gamma_e^2}
   {d n_e\over d\gamma_e} F\left(\nu_{ic}/4\gamma_e^2\nu\right),
\end{equation}
where 
\begin{equation}
   F(x) \approx {2\over 3} (1-x), \quad\quad  x\equiv\nu_{ic}/(4\gamma_e^2\nu).
\end{equation}

It follows from these equations that the IC spectrum, $f_{ic}(\nu_{ic})$, 
for a $\delta$-function seed photon spectrum (where photons have
frequency $\nu_0$), and a power-law distribution of electrons
with index $p$ which is cutoff at the low energy end at $\gamma_m$,
is proportional to $\nu_{ic}$ for $\nu_{ic} < 4\gamma_m^2\nu_0$.
Therefore, the low energy part of IC spectrum can be significantly 
harder than the hardest possible synchrotron spectrum ($\nu^{1/3}$)
when synchrotron-self-absorption is negligible.
At higher photon energies, $\nu_{ic} > 4\gamma_m^2\nu_0$, the IC 
spectrum has an asymptotic power-law index $\nu_{ic}^{-(p-1)/2}$, 
same as the spectrum for the 
synchrotron process.

\noindent{\bf IC in Klein-Nishina regime}

When photon energy in electron comoving frame approaches (or exceeds)
$m_e c^2$ two effects become important. One of which is that the
electron recoil in the scattering can no longer be ignored. The
other effect is that the cross-section is smaller than $\sigma_T$
and it decreases with increasing photon energy as $\sim \nu^{-1}$.
See \cite{rybicki79} for appropriate equations.
One simple consequence of the recoil effect is that the energy
of the scattered photon is limited to $\sim m_e c^2 \gamma_e/2$ (and is
no longer $\sim\nu_0 \gamma_e^2$) which is obvious from energy 
conservation.

\subsection{Hadronic processes}
\label{hadronic} 

Photo-pion process refers to the production of pions ($\pi^0$, $\pi^+$
and $\pi^-$) in collisions of photons with protons; charge conservation 
requires that $\pi^-$ is produced with at least one $\pi^+$.
The decay of $\pi^+$
produces positrons of very high LF which can then produce high
energy photons via the synchrotron process. A neutral pion $\pi^0$
can directly decay into two photons. The photo-pion process is
likely to be important in those situations where electrons are unable
to be accelerated efficiently to very high LFs whereas protons are. It
also offers a way to beat the well known limit on the maximum
synchrotron photon energy of about $50\Gamma$ MeV for shock accelerated 
electrons(see \S\ref{synch_rad_max}).

The delta resonance for photon-proton interaction,
$p+\gamma\rightarrow\Delta^+$, has the largest cross section, 
$\sigma_{\gamma p} = 5\times10^{-28}$cm$^2$, and 
the lowest energy threshold --- $\sim 200$ MeV for photon in proton rest
frame --- of the photo-proton resonances, and is
therefore the most important photo-pion interaction to consider for many
astrophysical systems. The delta resonance has two main decay channels:
$\Delta^+\rightarrow\pi^++n$ and $\Delta^+\rightarrow \pi^0+p$. The
neutral pions quickly decay in $8.4 \times 10^{-17}$s (in their rest frame)
to two photons, and the outgoing $\gamma$-ray energy is 
at least 67 MeV (in pion rest frame).

The $\pi^+$ decays in $2.6\times 10^{-8}$s to $\mu^+$ \&
$\nu_\mu$, and the anti-muon subsequently decays as \(\mu^+\rightarrow
e^++\bar{\nu}_\mu+\nu_e\). The isospin conservation gives a branching
ratio for $\pi^+:\pi^0$ decay channels of $\Delta^+$ to be $1:2$. However,
when contributions from all the possible resonances as well as the
direct pion production are included the ratio of charged pions to
neutral pions is actually closer to $2:1$ \citep{rachen98}. A detailed
discussion of the photo-pion process in the context of high energy
emission from GRBs is provided in \S\ref{photo_pion}, and its contribution
to neutrinos is described in \S\ref{grb-neutrinos}.

Another relevant process is the Bethe-Heitler pair production process 
$p+\gamma\rightarrow p+e^++e^-$. Its relative importance with respect
to the photon pion process in contributing to radiation from GRBs is described
in \S\ref{bethe_heitler}.

\section{Afterglow theory}
\subsection{Relativistic shocks: basic scalings}
\label{sec:scalings}

%Blandford-McKee self-similar shock solution, synchrotron radiation, and the 
%characteristics of the lightcurve expected from a relativistic shock 
%propagating into a medium with power-law density profile will be discussed.
%The effect on a finite jet-angle on the lightcurve will also be included.

One important piece of the GRB theory is a ``generic'' model that does
not depend on the details of the central engine.
%\citep{rees92,meszarosrees93,meszarosrees97,sari98}. 
This is a relativistic blastwave theory that describes interaction between 
the ``fireball'' --- which moves with Lorentz factor $\Gamma_0$ before 
deceleration \& has total ``isotropic equivalent'' energy $E$ --- and 
the circumburst medium (CBM) described by the density profile, 
$n(R) = (A/m_p) R^{-k}$. Such a fireball--CBM interaction is inevitable 
for any type of energetic explosion. A power-law decaying multi-wavelength 
afterglow was in fact  predicted by \cite{paczynski93,meszarosrees97} 
before the first observational detection of X-ray afterglow in 1997 by 
the BeppoSAX satellite.
A relativistic shock theory was developed by \cite{blandford76}
in the context of AGN jets which turned out to be well suited
for interpreting GRB afterglows in X-ray, optical and radio
bands when they were discovered in 1997
\citep{costa97,vanpara97,frail97}. The self-similar nature of the
blastwave solution naturally explains the power law behavior of
the afterglow lightcurves.

\begin{figure}
\begin{center}
\includegraphics[width=11cm]{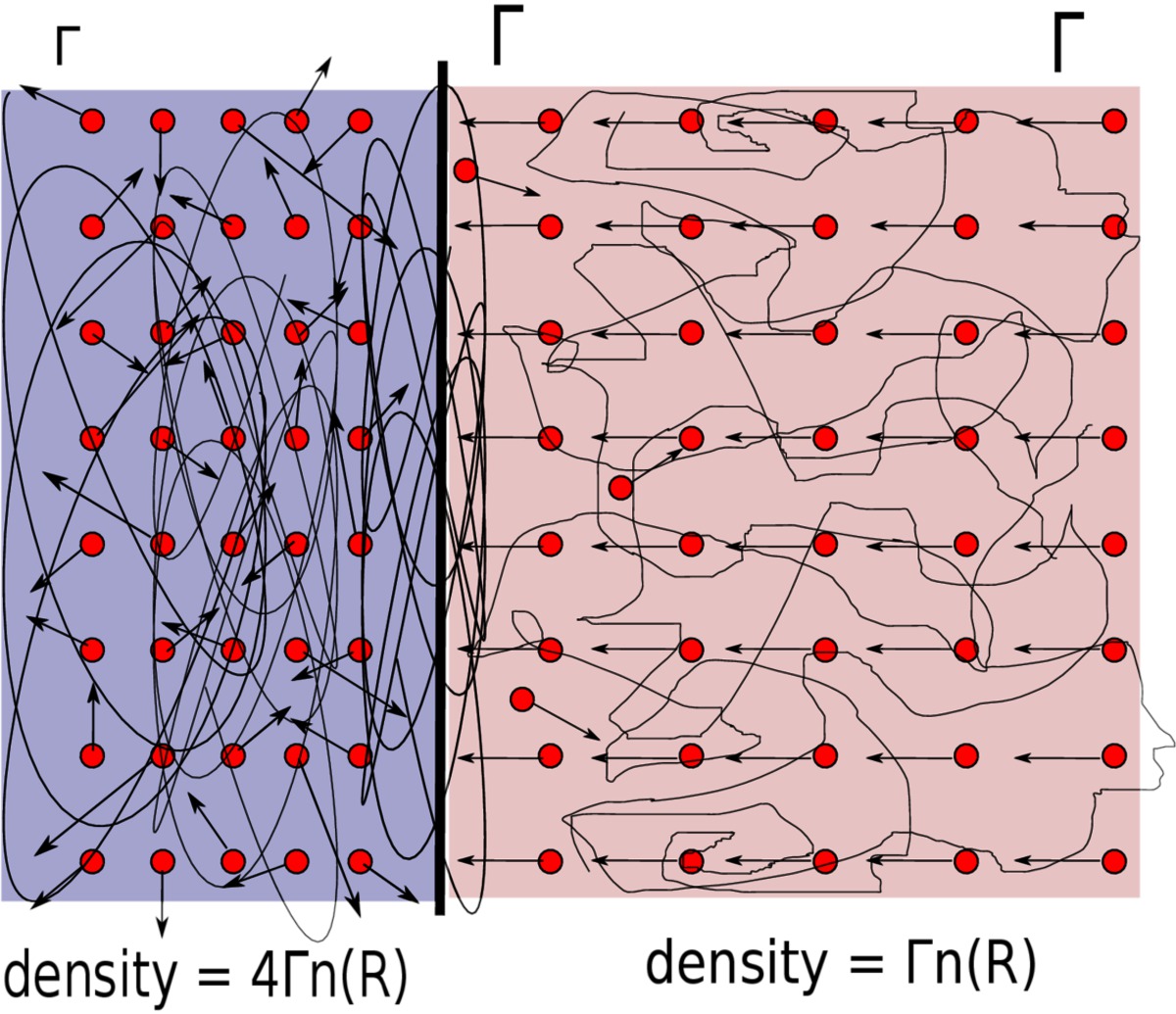}
\caption{A schematic sketch of highly relativistic shock as viewed from 
the mean rest frame of the shocked fluid; lines represent magnetic fields,
and arrows show particle velocity with respect to the shocked plasma. 
The Lorentz factor of the unshocked medium (right hand part of the sketch)
with respect to shocked plasma is $\Gamma$. Cold, upstream, particles stream 
toward the shocked plasma 
with Lorentz factor $\Gamma$ as viewed in this frame, and after crossing 
the front their velocity direction is randomized
but the magnitude of their proper-velocity is nearly unchanged. The shock 
also compresses plasma by a factor 4 (as viewed in the comoving frame of
shocked plasma), and amplifies magnetic fields and accelerates particles.
}\label{FIG:shock_dynamics}
\end{center}
\end{figure}

The basic dynamics of blastwave is easy to understand
using simple physical arguments, and the main results are
sketched in Figure \ref{FIG:shock_dynamics}. 
The emphasis is on trying to provide a physical understanding of the 
key concepts and not on rigorous derivations. For the latter we shall 
provide citations of the relevant literature.

It is best to work in the 
comoving frame of the shocked fluid which is traveling 
with Lorentz factor $\Gamma$ with respect to unshocked fluid.
The density of the unshocked medium in this frame is $\Gamma n$,
and upstream particles are seen to be streaming toward the shocked
fluid with a Lorentz factor $\Gamma$; upstream particles have
thermal energy much smaller than their rest mass. What a shock
does is to randomize the orientation of particle velocity vectors,
without changing their Lorentz factors, when they cross the shock front, 
and therefore 
the mean ``thermal'' energy of protons down-stream of the shock front 
is $\Gamma m_p c^2$ (derivation provided below).
As viewed from the lab frame, the average energy of each down-stream proton 
is $\Gamma^2 m_p c^2$, and hence for a blast wave at radius $R$, the 
total energy in the shocked plasma is 
\begin{equation}
E\approx 4\pi A R^{3-k} c^2 \Gamma^2/(3-k),
\end{equation}
where $A R^{-k}$ is the density of the medium at radius $R$ and $4\pi A R^{3-k}/
  (3-k)$ is the total swept up gas mass. 
 This equation describes the basic dynamics of the
blast wave. For instance, for a constant density CBM, and a non-radiative 
blast wave with constant total energy, the LF $\Gamma\propto R^{-3/2}$.
The deceleration radius -- the distance from the center of explosion (CoE)
 where the blast wave
LF decreases by a factor 2 from its initial value of $\Gamma_0$ 
and the energy imparted to the CBM is E/2 -- is obtained from the above
equation which for a constant density medium is given by
\begin{equation}
   R_d \approx (1.2\times10^{17}{\rm cm}) E_{53}^{1/3} n^{-1/3} 
         \Gamma_{0,2}^{-2/3}.
  \label{R_d_s0}
\end{equation}

Shocks also compress plasma --- for highly relativistic shocks the
compression factor is $4 \Gamma$, i.e. the comoving
frame density of the shocked plasma is $4 \Gamma n$
(quantitative expression is provided in eq. \ref{eq:compression} below)
--- and accelerate
particles to produce a power-law distribution function, and generate 
magnetic fields. These ingredients are all that one needs for calculating
afterglow radiation from the interaction of a relativistic outflow 
with the surrounding medium.

We outline the derivation of the two results mentioned
above, i.e. compression of plasma and entropy produced 
by a blast-wave, and then describe the dynamics for a number of
different situations before taking up the radiation physics of
GRB afterglows.

For a relativistic shock propagating into a cold upstream medium,
the physical condition of the shocked plasma is obtained from
the conservation of baryon number, energy \& momentum fluxes across the 
shock front; the baryon number flux is given by $n'\Gamma c$,
and the momentum and energy fluxes are part of the energy-momentum tensor 
$T^{\mu\nu} = (\rho' c^2 + p') u^\mu u^\nu + p' g^{\mu\nu}$, 
where $\rho' c^2$ \& $p'$ are the total energy density \& pressure
in the plasma rest frame, $u^\mu$ is the 4-velocity, and $g^{\mu\nu}$
is the metric tensor. These conservation equations can be reduced to 
the following three equations \citep{blandford76,rezzolla13}:
\begin{equation}
 \frac{e'_2}{n'_2} = (\gamma_{21}-1) m_p c^2,
\end{equation}
\begin{equation}
 \frac{n'_2}{n'_1} = \frac{\hat\gamma \gamma_{21}+1}{\hat\gamma-1},
\label{eq:compression}
\end{equation}
\begin{equation}
 \gamma_{1s}^2 = \frac{(\gamma_{21}+1)[\hat\gamma(\gamma_{21}-1)+1]^2}
{\hat\gamma(2-\hat\gamma) (\gamma_{21}-1)+2}.
\label{gamma1s}
\end{equation}
Here $m_p$ is proton mass, $c$ is speed of light,
subscript ``2'' and ``1'' denote downstream and upstream,
respectively, $e'$ \& $n'$ are internal energy density \& proton number 
density (in local fluid rest frame), $\gamma_{21}$ is the relative Lorentz 
factor of plasma in region 2 with respect to region 1, $\gamma_{1s}$ is the relative 
Lorentz factor of plasma in region 1 with respect to the shock front, and $\hat\gamma$ 
is the adiabatic index of the fluid. For ultra-relativistic shocks, $\Gamma 
\gg 1$, that describe the afterglow emission from GRB blastwave for 
a few days \citep[e.g.][]{piran99}, one has $\hat\gamma=4/3$, and it follows
from the above conservation equations that 
$e'_2 /n'_2 \simeq \gamma_{21} m_p c^2$ (the average energy of protons
down-stream of the shock front is $\sim\gamma_{21} m_p c^2$), 
$n'_2/n'_1 \simeq 4 \gamma_{21}$ 
(downstream plasma is compressed by a factor of $4\gamma_{21}$, and
$\gamma_{1s} \simeq \sqrt{2} \gamma_{21}$ (the shock front travels 
slightly faster than the downstream fluid).

Once the blastwave enters the self-similar deceleration phase,
some simple scalings can be derived. Let us consider the case of a
constant energy blastwave ($E$) traveling in a constant density CBM ($n$)
as an example. Energy conservation can be written as
\begin{equation}
 E = \frac{4\pi}{3} R^3 n m_p c^2 \cdot \Gamma^2 = {\rm const},
\end{equation}
where $\Gamma = \gamma_{21}$ is the bulk Lorentz factor of the
blastwave (with respect to the unshocked medium), $R$ is the distance of the
shock front from the explosion center, and the factor $\Gamma^2$ takes into 
account average proton thermal energy in the lab frame (proton thermal energy 
in the shocked fluid comoving frame is $m_p c^2\Gamma$).
One therefore finds $\Gamma^2 R^3 =$constant, or
\begin{equation}
 \Gamma \propto R^{-3/2}.
\end{equation}
The time duration for arrival of photons at the observer is smaller than the 
center-of-explosion (CoE) frame time by roughly a 
factor $2\Gamma^2$ due to the blastwave and photons moving in more or
less the same direction and the difference in their speed being $\sim
1/2\Gamma^2$ (see \S\ref{relativity}). Therefore, 
\begin{equation}
 t_{\rm obs} \sim \frac{R}{2\Gamma^2 c} \propto R^4 \propto \Gamma^{-8/3},
\end{equation}
and
\begin{equation}
 \Gamma \propto R^{-3/2} \propto t_{\rm obs}^{-3/8}, ~~~ R \propto 
  t_{\rm obs}^{1/4}.
\label{adiabaticISM}
\end{equation}

More generally, one can consider a power-law stratified density
profile
\begin{equation}
 n = n_0 \left( \frac{R}{R_0}\right)^{-k},
\end{equation}
with $k<3$. The energy conservation equation can be written as 
\begin{equation}
 E = \int n_0 \left(\frac{R}{R_0}\right)^{-k} m_p c^2
\Gamma^2 4 \pi R^2\, d R = {\rm const},
\end{equation}
or $R^{3-k} \Gamma^2 =$constant. Carrying out the same exercise as above, 
one finds the observer frame time
\begin{equation}
 t_{\rm obs} \sim \frac{R}{2\Gamma^2 c} \propto R\Gamma^{-2}
\propto  \left\{
 \begin{array}{l}
   \Gamma^{\frac{2}{k-3}} \cdot \Gamma^{-2} \propto \Gamma^{\frac{8-2k}{k-3}} \\
   R \cdot R^{3-k} \propto R^{4-k}
 \end{array}
\right.
\end{equation}
so that
\begin{equation}
 \Gamma \propto R^{\frac{k-3}{2}} \propto t_{\rm obs}^{\frac{k-3}{8-2k}},
~~ R \propto t_{\rm obs}^{\frac{1}{4-k}}.
\end{equation}
This reduces to (\ref{adiabaticISM}) for $k=0$ (constant density).
For a free wind with constant mass loss rate $\dot M$ and wind 
speed $v_w$, one has 
$\dot M = 4 \pi R^2 n v_w = $constant, or $n\propto R^{-2}$ (or $k=2$).
Plugging in $k=2$, one gets the scaling for a wind medium
\citep{dailu98c,meszaros98,chevalier99,chevalier00}
\begin{equation}
 \Gamma \propto R^{-1/2} \propto t_{\rm obs}^{-1/4}, ~~~ R \propto t_{\rm obs}^{1/2}.
\label{adiabaticWind}
\end{equation}

It is possible that the blastwave energy continuously increases
with time. This is the case for instance when a fireball is fed by a 
long lasting, Poynting-flux dominated, jet (so that the reverse shock,
discussed in \S\ref{RS_radiation}, does not exist or is very weak). 
Then, the dynamics of the blast wave is determined by taking into 
account the additional energy added to it by the outflow from the central 
engine \citep{blandford76,cohen99}. This is particularly relevant when the 
central engine is a
millisecond magnetar \citep{usov92,thompson94,dailu98b,zhangmeszaros01a}.

Consider a central engine with time dependent luminosity --
\begin{equation}
 L(t) = L_0 \left(\frac{t_{obs}}{t_0}\right)^{-q}.
\label{Lt}
\end{equation}
For $q \geq 1$ the injected energy does not grow appreciably with time, 
and the blastwave behavior is essentially same as the constant energy case.
So in the following discussion we will consider the case of $q<1$.

The total energy in the blastwave 
\begin{equation}
 E_{\rm tot} = E_0 + E_{\rm inj}
=E_0 + \int_0^{t_{obs}} L(t)dt = E_0 + \frac{L_0 t_0^q}{1-q}\cdot t_{obs}^{1-q},
\end{equation}
where $E_0$ is the initial energy in the blastwave, and $E_{\rm inj}$
is the injected energy into the blastwave from the long-lasting 
central engine.

The blastwave scaling becomes different when $E_{\rm inj} \gg E_0$ for 
$q<1$. In this case, the total energy 
\begin{equation}
 E_{\rm tot} \sim E_{\rm inj} \propto t_{obs}^{1-q}.
\label{q-equation}
\end{equation}

For the constant density CBM case, one has
\begin{equation}
 \Gamma^2 R^3 \propto t_{obs}^{1-q}.
\end{equation}
Again taking $t_{obs} \propto R/\Gamma^{2}$, one can rewrite 
the above equation as
\begin{equation}
 \Gamma^2 R^3 \propto R^{1-q} \Gamma^{2(q-1)}.
\end{equation}
Regrouping the parameters, one finally has
\begin{equation}
 \Gamma \propto R^{-\frac{2+q}{4-2q}} \propto t_{\rm obs}^{-\frac{2+q}{8}},
~~~ R \propto t_{\rm obs}^{\frac{2-q}{4}}.
\label{injectionISM}
\end{equation}
The limiting case of $q\rightarrow1$ reduces to the constant energy
blastwave dynamics.

For the CBM with density falling off as R$^{-2}$ (wind like medium), one has
\begin{equation}
 \Gamma^2 R \propto t_{obs}^{1-q} \propto R^{1-q} \Gamma^{2q-2}.
\end{equation}
This leads to the following time dependence for blastwave LF and radius
\begin{equation}
 \Gamma \propto R^{\frac{q}{2q-4}} \propto t_{\rm obs}^{-\frac{q}{4}}, ~~~
R \propto t_{\rm obs}^{\frac{2-q}{2}}. 
\label{injectionWind}
\end{equation}
Again this is reduced to the constant energy wind medium when $\rightarrow1$.

An alternative energy injection, or refreshed shock, mechanism is to 
consider a Lorentz factor stratification of the ejecta \citep{rees98}, e.g.
\begin{equation}
M(>\gamma) \propto \gamma^{-s}.
\end{equation}
Mass (and therefore energy) is added to the blastwave when the blastwave
progressively decelerates, so that 
\begin{equation}
 E \propto \gamma^{1-s} \propto \Gamma^{1-s},
\end{equation}
where $\gamma$ is the Lorentz factor of the ejecta, and $\Gamma$ is the
Lorentz factor of the blastwave. Since energy is injected when 
$\Gamma \sim \gamma$, the reverse shock is very weak, one can 
again neglect the reverse shock contribution.

The two energy injection mechanisms can be considered equivalent, as far as 
the blast wave dynamics is considered, and one model can be related to
the other by expressing the injection parameter $s$ in terms of $q$
\citep{zhang06}.
For the constant density CBM one has
\begin{equation}
 \Gamma \propto R^{-3/(1+s)} \propto t_{\rm obs}^{-3/(7+s)}, ~~~ R\propto 
   t_{\rm obs}^{(1+s)/(7+s)}.
  \label{injectionISM_s}
\end{equation}
Therefore, the relation between $s$ and $q$ is obtained by comparing
equations (\ref{injectionISM}) \& (\ref{injectionISM_s}), and requiring
the dynamics for the two forms of energy injections to be the same: 
\begin{equation}
 s = \frac{10-7q}{2+q}, ~~~ q=\frac{10-2s}{7+s}.
\end{equation}
For the wind like CBM one has
\begin{equation}
 \Gamma \propto R^{-1/(1+s)} \propto t_{\rm obs}^{-1/(3+s)}, ~~~ R\propto 
   t_{\rm obs}^{(1+s)/(3+s)},
\end{equation}
so that the equivalency relation between $s$ \& $q$ for a wind-like CBM is
\begin{equation}
 s = \frac{4-3q}{q}, ~~~ q=\frac{4}{3+s}.
\label{s-q}
\end{equation}

\subsection{Afterglow synchrotron spectrum and lightcurve}
\label{FS_radiation}

An instantaneous afterglow spectrum can be characterized by a multi-segment
broken power law \citep{sari98}, separated by three characteristic
frequencies: the typical synchrotron frequency of the accelerated
electrons with the minimum Lorentz factor $\nu_m$, the cooling
frequency $\nu_c$, and the synchrotron self-absorption frequency $\nu_a$.
In the afterglow phase, $\nu_a$ is usually the smallest of these
three frequencies at least for a few months after the explosion 
for a typical CBM density,  and the spectrum
falls into two broad categories depending on the ordering of 
$\nu_m$ and $\nu_c$. The spectrum when $\nu_m < \nu_c$, classified as 
``slow cooling'' case, is (see \S\ref{synch_rad}, Fig. \ref{FIG:syn_spec},
\cite{sari98})
\begin{equation}
f_\nu = \left\{
\begin{array}{ll}
f_{\nu,max} \left(\frac{\nu_a}{\nu_m}\right)^{1/3} \left(\frac{\nu}{\nu_a}\right)^2, 
& \nu<\nu_a \\
f_{\nu,max} \left(\frac{\nu}{\nu_m}\right)^{1/3}, & \nu_a<\nu<\nu_m \\
f_{\nu,max} \left(\frac{\nu}{\nu_m}\right)^{-(p-1)/2}, & \nu_m<\nu<\nu_c \\
f_{\nu,max} \left(\frac{\nu_c}{\nu_m}\right)^{-(p-1)/2} \left(\frac{\nu}{\nu_c}\right)^{-p/2}, 
& \nu>\nu_c
\end{array}
\right.
\end{equation}
and for $\nu_c < \nu_m$, or ``fast cooling'' regime, the emergent spectrum is
\begin{equation}
f_\nu = \left\{
\begin{array}{ll}
f_{\nu,max} \left(\frac{\nu_a}{\nu_c}\right)^{1/3} \left(\frac{\nu}{\nu_a}\right)^2, 
& \nu<\nu_a \\ 
f_{\nu,max} \left(\frac{\nu}{\nu_c}\right)^{1/3}, & \nu_a<\nu<\nu_c \\
f_{\nu,max} \left(\frac{\nu}{\nu_c}\right)^{-1/2}, & \nu_c<\nu<\nu_m \\
f_{\nu,max} \left(\frac{\nu_m}{\nu_c}\right)^{-1/2} \left(\frac{\nu}{\nu_m}
    \right)^{-p/2}. \quad\quad & \nu>\nu_m 
\end{array}
\right.
\end{equation}
Here $f_{\nu,max}$ is the maximum flux density, which is $f_\nu (\nu_m)$
for slow cooling and $f_\nu (\nu_c)$ for fast cooling.
These spectral functions are independent of blast-wave dynamics, although 
the ordering of $\nu_m$, $\nu_c$ and $\nu_a$ and how they evolve with time are 
determined by the dynamics, CBM properties, and micro-physics parameters of 
shocked plasma.
 
The characteristic frequencies $\nu_m$, $\nu_c$ can be calculated from the
synchrotron frequency formula (\S\ref{synch_rad}, eq. \ref{syn_omega})
\begin{equation}
 \nu = \frac{3}{4\pi} \gamma^2 \frac{qB'}{m_e c}
\end{equation}
by replacing $\gamma$ with $\gamma_m$ and $\gamma_c$, where
$\gamma_m$ is the minimum Lorentz factor of electrons in the shock heated
plasma\footnote{The electron distribution function, $d n/d\gamma$,
 peaks at $\gamma_m$. For $\gamma>\gamma_m$, $d n/d\gamma
\propto \gamma^{-p}$, and for $\gamma<\gamma_m$ the distribution function
is uncertain but could be thermal.} which is given by
\begin{equation}
 \gamma_m = g(p) \epsilon_e (\Gamma-1) \frac{m_p}{m_e} \frac{n_p}{n_e},
\end{equation}
$\Gamma$ is the Lorentz factor of the blast wave, $\epsilon_e$ is the
fraction of energy density of shocked fluid given to electrons, $n_p$
\& $n_e$ are the number densities of protons and electrons, respectively, 
and the dimensionless factor $g(p)$ when the maximum LF of electrons
accelerated in the shock is $\gamma_M$ is given by
\begin{equation}
g(p) \simeq \left\{
 \begin{array}{ll}
  \frac{p-2}{p-1}, & p>2 \\
  \ln^{-1} (\gamma_{_M}/\gamma_m), & p=2
 \end{array} 
\right. .
\end{equation}
The above expression for $g$ follows from the requirement that the total 
electron energy --- obtained by integrating the distribution function, 
$dn/d\gamma\propto\gamma^{-p}$, for $\gamma>\gamma_m$ --- is $\epsilon_e$ 
times the energy density on the shocked fluid, 
i.e. $\epsilon_e 4\Gamma(\Gamma-1) n_p m_p c^2$.
The Lorentz factor of electrons that cool on a dynamical time ($t'$) is 
(see \S\ref{synch_rad} for details)
\begin{equation}
 \gamma_c = \frac{6\pi m_e c}{\sigma_{_T} t' {B'}^2 (1+Y)},
\end{equation}
where $Y=u_{\rm syn}/u_{\rm B}$ is the synchrotron self-Compton 
parameter\footnote{The expression for Compton Y parameter is different 
when photon-electron scatterings are in 
 Klein-Nishina limit, i.e. when the energy of a typical photon in the 
rest frame of the electron is larger than $m_e c^2$.},
which is the ratio of the synchrotron photon energy density $u_{\rm syn}$ 
and the magnetic energy density $u_{\rm B} = B^2/8\pi$.

The self-absorption frequency $\nu_a$ can be calculated by equating the
emergent flux at $\nu_a$ to the blackbody flux corresponding to the 
temperature of electrons with synchrotron characteristic 
frequency $\nu_a$ (see \S\ref{synch_rad} for details)
\begin{equation}
 I_\nu^{\rm syn} (\nu_a) = I_\nu^{\rm bb} (\nu_a) \simeq 2 k_B T\cdot
\frac{\nu_a^2}{c^2}, 
\end{equation}
where
\begin{equation}
 k_B T\simeq{\rm max}\left[\gamma_a, \min(\gamma_c, \gamma_m)\right] m_e c^2/3,
\end{equation}
and $\gamma_a$ is the Lorentz factor corresponding to $\nu_a'$, i.e.
$\gamma_a = (4\pi m_e c \nu_a'/3 q B')^{1/2}$.
The comoving magnetic field strength is obtained by taking the energy
density in magnetic field to be $\epsilon_B$ times the energy density
of shocked CBM:
\begin{equation}
 B' \approx [32 \pi m_pc^2 \epsilon_B n_p (\Gamma-1)\Gamma]^{1/2}.
\end{equation}

The time-dependence of spectral break frequencies $\nu_m$, $\nu_c$,
and $\nu_a$ can be calculated from the shock dynamics, or in particular
from the evolution of shock Lorentz factor $\Gamma$. For instance, for 
a constant density CBM, $\Gamma\propto t_{\rm obs}^{-3/8}$ (eq. \ref{adiabaticISM}),
and therefore $B'\aprop \Gamma\propto t_{\rm obs}^{-3/8}$, $\gamma_m\aprop
\Gamma\propto t_{\rm obs}^{-3/8}$, and so $\nu_m\propto B' \gamma_m^2 \Gamma
\aprop \Gamma^4\propto t_{\rm obs}^{-3/2}$; it is easy to show that 
$\nu_m\propto t_{\rm obs}^{-3/2}$ even when the CBM density is not constant
but varies as a power-law function of distance. Similarly it can be
shown that $\nu_c\propto t_{\rm obs}^{-1/2}$ and $\nu_a$ is time independent
for a constant density medium. The full expression for $\nu_m$, $\nu_c$
and $\nu_a$ for a constant density medium is \cite[e.g.][]{granotsari02,gao13b}
\begin{equation}
 \nu_m = 3.3\times 10^{14}~{\rm Hz}~(1+z)^{1/2} \epsilon_{B,-2}^{1/2}
[\epsilon_e g(p)]^2 E_{52}^{1/2} t_{obs,d}^{-3/2},
\end{equation}
\begin{equation}
 \nu_c = 6.3\times 10^{15}~{\rm Hz}~(1+z)^{-1/2} \epsilon_{B,-2}^{-3/2}
E_{52}^{-1/2} n_p^{-1} t_{obs,d}^{-1/2},
\end{equation}
\begin{equation}
 \nu_a = 4.2\times 10^8~{\rm Hz}~(1+z)^{-1} \left[\frac{(p+2)(p-1)}
{(3p+2)}\right]^{0.6} [\epsilon_e g(p)]^{-1} \epsilon_{B,-2}^{1/5}
E_{52}^{1/5} n_p^{3/5},
\end{equation}
whereas for a wind like CBM
\begin{equation}
 \nu_m = 4.0 \times 10^{14}~{\rm Hz}~ (p-0.69) (1+z)^{1/2} \epsilon_{B,-2}^{1/2}
[\epsilon_e g(p)]^2 E_{52}^{1/2} t_{obs,d}^{-3/2};
\end{equation}
\begin{equation}
 \nu_c = 4.4\times 10^{13}~{\rm Hz}~ (3.45-p) \exp(0.45p) (1+z)^{-3/2}
\epsilon_{B,-2}^{-3/2} E_{52}^{1/2} A_*^{-2} t_{obs,d}^{1/2};
\end{equation}
\begin{equation}
 \nu_a = 3.3\times 10^9~{\rm Hz}~ (1+z)^{-2/5} \left(\frac{p-1}{3p+2}\right)^{3/5}
[\epsilon_e g(p)]^{-1} \epsilon_{B,-2}^{1/5} E_{52}^{-2/5} A_*^{6/5} t_{obs,d}^{-3/5};
\end{equation}
where $t_{obs,d}$ is time in observer frame in units of 1 day, and $A_*$ 
is density parameter for a wind like CBM defined as $n(R) = A_* 
(3\times10^{35}) R^{-2}$cm$^{-3}$ with the unit for $R$ in cm; $A_*=1$
corresponds to mass loss rate in the wind of GRB progenitor star of
10$^{-5} M_\odot$yr$^{-1}$ at wind speed of 10$^8$cm/s.

The specific flux at the peak of the spectrum can be written as (see
\S\ref{synch_rad})
\begin{equation}
f_{\nu,max} = {(1+z)L'_{\nu'}\Gamma\over 4\pi D_L^2} \approx (1+z)\frac{N_{tot} 
      P'_{\nu,max}\Gamma}{4\pi D_L^2}~,
  \label{f_peak1}
\end{equation}
where $L'_{\nu'}$ is specific luminosity in jet comoving frame, $N_{tot}$ 
is the total number of electrons that contribute to radiation at frequency 
$\nu$, 
\begin{equation}
P'_{\nu',max} \approx \frac{\sqrt{3} q^3 B'}{m_e c^2},
\end{equation}
is the power radiated per unit frequency for one electron at the peak of the 
spectrum i.e. specific power for an electron with thermal LF 
$\gamma\approx\min(\gamma_c, \gamma_m)$, $z$ is the redshift of the burst, and 
\begin{equation}
D_L = (1+z)\frac{c}{H_0} \int_0^{z} \frac{dz'}{\sqrt{\Omega_m(1+z')^3+\Omega_\Lambda}}
\end{equation}
is the luminosity distance of the burst, $H_0$ is the Hubble's constant, 
$\Omega_m$ and $\Omega_\Lambda$ are density parameters for matter 
and dark energy, respectively.

We see from equation (\ref{f_peak1}) that the peak specific flux, $F_{\nu,max}
\propto N_{tot} B' \Gamma \propto R^3 \Gamma^2$, is time independent for
a constant density medium --- since $R^3 \Gamma^2$ is the total energy 
in the blast wave which is constant for an adiabatic external shock. 
For a wind-like stratified medium $F_{\nu,max}\propto
R \Gamma B' \propto \Gamma^2\propto  t_{\rm obs}^{-1/2}$. The full expression 
for the peak specific flux for these two types of CBM media are
\cite[e.g.][]{granotsari02, yost03}
\begin{eqnarray}
 f_{\nu,max} & = & 1.6 ~{\rm mJy}~(1+z) \epsilon_{B,-2}^{1/2} E_{52}
n_p^{1/2} D_{L,28}^{-2}, ~~~ n_p \propto R^0  \label{F_peak_constant_CBM}\\
 f_{\nu,max} & = & 7.7 ~{\rm mJy}~(p+0.12) (1+z)^{3/2} \epsilon_{B,-2}^{1/2}
E_{52} A_* D_{L,28}^{-2} t_{\rm obs,d}^{-1/2}, ~  n_p\propto R^{-2}. 
      \label{F_peak_wind_CBM}
\end{eqnarray}

Making use of these 
expressions for peak flux, $\nu_m$, and $\nu_c$, it can be shown that the
observed specific flux for $\nu>\max({\nu_m, \nu_c})$ is \citep{kumar00c,freedman01}
\begin{equation}
f_\nu\propto E^{(p+2)/4} \epsilon_e^{p-1} \epsilon_B^{(p-2)/4} 
   t_{obs}^{-(3p-2)/4} \nu^{-p/2}, 
\label{f_nu_nuc}
\end{equation}
which is completely independent of CBM density and its stratification, and very
weakly dependent on $\epsilon_B$, which are the two most uncertain parameters
in afterglow modeling.
This result turns out to be very useful for interpreting high energy GRB
data ($\nu\gae10^2$MeV) obtained by the Fermi/LAT as described in 
\S\ref{high_energy_afterglow}.

\begin{figure}
\begin{center}
\includegraphics[width=13cm]{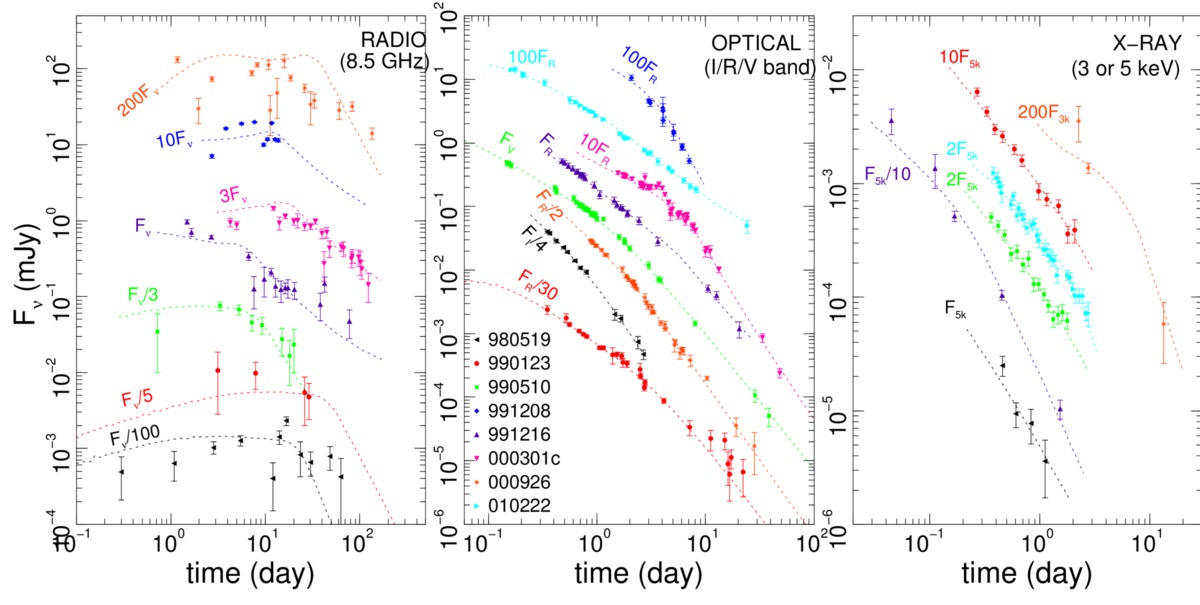}
\caption{Radio, optical, and X-ray model light-curves
   for eight GRB afterglows (legend of middle graph applies to all panels).
   The model light-curves were obtained by $\chi^2$-minimization using radio,
   millimeter, sub-millimeter, near infrared, optical, and $X$-ray data.
   The radio fluctuations are due to scatterings by inhomogeneities in the
   Galactic interstellar medium (Goodman 1997). Fluxes have been multiplied
   by the indicated factors, for clarity (figure from Panaitescu \& Kumar 2001).
}\label{FIG:afterglow_fit}
\end{center}
\end{figure}

The synchrotron radiation mechanism in external shock provides a good 
description of late time ($t\gae10$hr) GRB afterglow radiation from 
radio to X-ray frequencies (see Fig. \ref{FIG:afterglow_fit}), as well
as GeV emission of some GRBs at early times (see \S\ref{high_energy_afterglow}).

\subsection{Reverse shock}
\label{RS_radiation}

\begin{figure}
\begin{center}
\includegraphics[width=8cm]{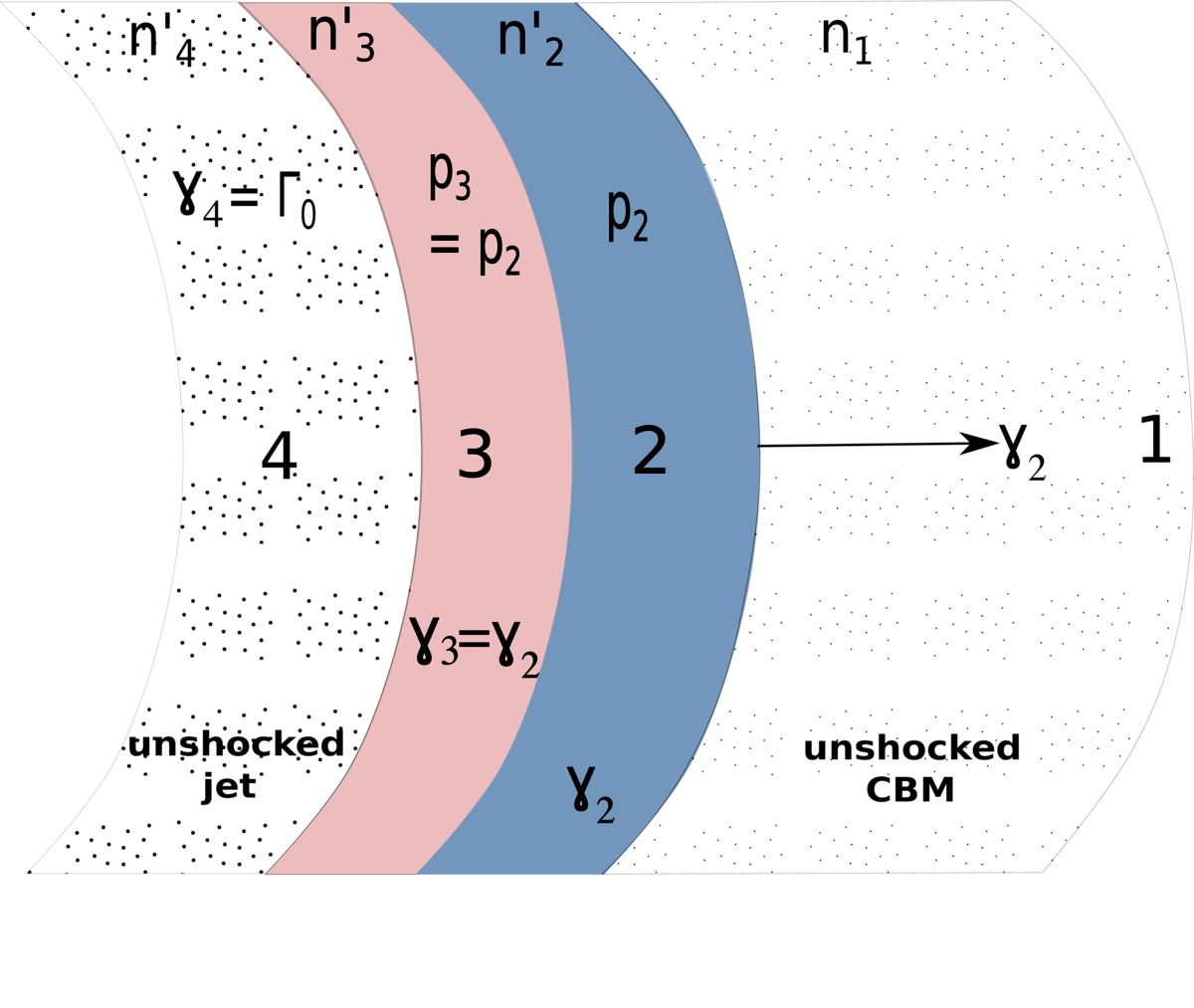}
\caption{This is a schematic sketch of a pair of shocks produced when 
a relativistic jet from a GRB collides with the circum-burst medium (CBM), 
as viewed from the rest frame of unshocked CBM. 
Regions 2 \& 3 represent shocked CBM and GRB jet respectively. They
move together with the same Lorentz factor ($\gamma_2$, as viewed by
a stationary observer in the unshocked CBM), and have
the same pressure but different densities.
}\label{FIG:shock_sketch}
\end{center}
\end{figure}

During the early afterglow phase, a strong reverse shock (RS) propagates across 
the GRB-ejecta to decelerate it if the magnetic field in the ejecta is 
dynamically unimportant, i.e. the magnetization parameter $\sigma\equiv
B'^2/(4\pi n'_p m_p c^2) \ll1$. The RS dynamics is more complicated than the
self-similar solution for forward shock (FS) propagating into CBM. The 
RS--FS system can be separated in four regions (see Fig.\ref{FIG:shock_sketch}):
1. the unshocked medium; 2. shocked medium; 3. shocked ejecta; 4. unshocked
ejecta. These regions are separated by the forward shock front (FS, 
between 1 \& 2), a surface of contact density-discontinuity (between 2 \& 3), 
and the reverse shock front (RS, between 3 \& 4). Radiation from RS-heated GRB 
ejecta was predicted \citep{meszarosrees93b,meszarosrees97,saripiran99b} 
prior to its discovery in 1999 when a very bright optical flash was 
observed from GRB 990123 while $\gamma$-ray burst was still active 
\citep{akerlof99,saripiran99,meszarosrees99}.

A quick derivation of FS--RS system properties, for an unmagnetized 
GRB-jet, follows from the 
requirement of pressure equilibrium across the contact discontinuity
surface (which separates regions 2 \& 3). Let us take the Lorentz factors
of the RS-heated GRB-jet with respect to unshocked jet to be $\gamma_{34}$, and
the shocked CBM with respect to unshocked CBM to be $\gamma_{21}$. We know from
previous discussions regarding relativistic shocks that the pressures
of the shocked fluid in regions 2 \& 3 are $4\gamma_{21}^2 n_1$ \&
$4\gamma_{34}^2 n_4'$, respectively; density in region $i$, in the
local comoving frame, is $n'_i$. The Lorentz factor of 
the unshocked jet (region 4) with respect to the unshocked CBM is 
the jet Lorentz factor $\Gamma_0$ which is equal to $2\gamma_{21}\gamma_{34}$
(this follows from the addition of 4-velocities). Combining this relation 
with pressure equilibrium across the contact discontinuity surface we find: 
\begin{equation}
  \gamma_{34} \approx (n_1/4n'_4)^{1/4}\Gamma_0^{1/2}, \quad\quad
 {\rm and}\quad\quad \gamma_{21} \approx (n'_4/4n_1)^{1/4}\Gamma_0^{1/2}.
\end{equation}
 These equations are only valid
when both RS and FS are relativistic (similar relations are easy to
obtain for a non-relativistic RS). We note that the Lorentz factor
of the shocked jet with respect to CBM ($\gamma_3$) is equal to $\gamma_{21}$ 
since both shocked regions move together at the same speed. With RS/FS 
Lorentz factors in hand it is straightforward to determine various 
thermodynamical variables of the shocked plasma in regions 2 \& 3.
A more detailed derivation of these results can be found in 
\cite{sari95}.

The calculation of radiation from RS is similar to the FS 
emission described in \S\ref{FS_radiation}. For the simplest model,
one assumes a finite radial extent of the GRB-jet (related to the
finite duration of the GRB) and a roughly constant Lorentz factor
of the ejecta\footnote{It is possible that the ejecta has a Lorentz
factor stratification. If so, the RS is long-lived, and may give rise to
interesting features in the RS lightcurve \citep{rees98,sarimeszaros00,uhm07,genet07,uhm12,uhm14c}.}.  In this case the RS lightcurve declines 
rapidly ($\sim t^{-2}$) due to adiabatic cooling of electrons,  
and a decrease of the magnetic field strength, after the RS reaches 
the back end of the jet \citep{saripiran99b}. A useful classification
for RS is based on the dimensionless width of the ejecta defined below
\citep{sari95}
\begin{equation}
 \xi \equiv (l/\Delta)^{1/2} \Gamma_0^{-4/3},
\end{equation}
where $l$ is the Sedov radius (the radius at which the rest mass energy
of the swept up CBM by the blastwave is equal to the initial energy of the GRB), 
$\Delta = c T$ is the 
thickness of the GRB-ejecta in lab frame ($T$ is the burst duration in CoE 
frame), and $\Gamma_0$ is the initial Lorentz factor.
The GRB-ejecta is considered a ``thin shell'' or a ``thick shell'' depending
on whether $\xi>1$ or $\xi<1$, respectively. The FS/RS dynamics for the 
two regimes are
different, and so are the resulting lightcurves. A detailed
study of the RS dynamics and emission signature 
can be found in \cite{kobayashi00}. A joint study of RS/FR emission
signatures can be found in
\citep{kobayashizhang03a,zhang03,kumar03,wu03,kobayashizhang03b,gao13b}.

The shock solution for a jet with arbitrary magnetization $\sigma$, in the 
context of GRBs, can be found in e.g. \cite{zhangkobayashi05,mimica09,narayan11};
\cite{fan04} discuss the case of $\sigma < 1$.
The shock solution of a continuous, relativistic magnetized jet
is described in seminal papers of \cite{kennel84,kennel84b} in context 
of pulsar wind.

The relative importance of the RS and FS emissions depends on the 
ratio of microphysics parameters in these two shock regions, and
on the Lorentz factors of reverse-shock \& forward-shock fronts
\citep{kobayashizhang03a,zhang03,nakarpiran04}. Radiation from the reverse-shock
offers one of the few ways to determine the GRB jet composition, e.g. 
\cite{mcmahon06,nakarpiran04}. However,
this requires separating FS and RS contributions to the afterglow data.
A few different cases of RS/FS signatures are described below
\citep{zhang03,jinfan07}. 

Type I: re-brightening. For standard microphysics parameters, i.e. 
$\epsilon_e=0.1$ and $\epsilon_B=0.01$, for both FS and RS, the optical 
lightcurve usually shows a re-brightening signature. The first peak is 
dominated by the RS emission, while the second peak corresponds to the decline
of $\nu_m$ in the FS below the optical band. Such a pattern has been observed 
in some bursts \citep[e.g.][]{kobayashizhang03a,shao05}, however, 
the color change at the second peak -- associated with the passing of 
$\nu_m$ through the optical band --- has not been confirmed so far.

Type II: flattening. If the the magnetization parameter for the unshocked 
GRB ejecta is not so large as to suppress the RS, and RS has larger magnetic 
field than the FS, then the emission from RS would dominate
as was the case for GRB 990123. The early optical flare is RS dominated, 
and the flare peaks at the time when the reverse shock completes its
passage through the GRB ejecta. The decay of the RS flux in a fixed 
observer frequency band after 
the peak is $\sim t^{-2}$ \citep{meszarosrees99,saripiran99b}. 
This fast decline transitions to a more normal $t^{-1}$ decay when emission from 
the FS takes over. There are quite a few cases of optical afterglows that show 
this behavior \citep{fox03b,li03,zhang03,kumar03,gomboc08}.

Type III: no RS component. In the Swift era, many early optical afterglow lightcurves
have been obtained. To one's surprise, many of these lightcurves show a smooth
hump with the post-decay slope consistent with the FS emission, without the
signature of a RS \citep{molinari07,rykoff09,liang10}. 
This can be due to a Poynting flux dominated GRB-jet that suppresses 
RS \citep{zhangkobayashi05,mimica09}, 
or a very low $\nu_m$ in the RS \citep{jinfan07}.

\subsection{Jet break}
\label{jet_reak}

Evidence suggests that GRB outflows are collimated as anticipated by
\cite{rhoads97}. This is inferred from
an achromatic break seen in many afterglow lightcurves which are 
known as ``jet breaks''.
The steepening of the lightcurve following the ``jet break'' is 
due to two effects \citep[e.g.][]{rhoads99,sari99}. 

The first is the so-called ``edge'' effect 
\citep[e.g.][]{meszarosrees99,panaitescu99b,rhoads99,sari99}. For a jet moving 
relativistically with Lorentz factor $\Gamma$, photons emitted at any point 
on the jet are beamed, as seen in the lab frame, within a $1/\Gamma$ cone.
Thus, for a conical jet with opening angle $\theta_j$, initially when 
$\Gamma > 1/\theta_j$, an observer only sees radiation from a small
fraction of the jet. He then has no knowledge about the finite
collimation angle for the jet, and the jet dynamics resembles that of 
an isotropic fireball; the lightcurve during this phase is the
``pre-jet-break'' lightcurve.
As the jet decelerates, the $1/\Gamma$ cone increases, and the photon
beaming angle becomes comparable to the opening angle of the jet-cone.
Lightcurves display a ``jet break'' when this condition is satisfied.
When $\Gamma < 1/\theta_j$, the observer becomes aware of a deficit of 
flux with respect to an isotropic fireball case, and the lightcurve 
starts to fall off more steeply than the pre-break, isotropic, phase. 
The edge effect involves no change to blastwave dynamics. It is a
geometrical, plus special relativistic, effect, and its effect
on the observed specific flux is to introduce an additional 
factor of $\theta_j^2/(1/\Gamma)^2 \propto \Gamma^2$ to account for the
deficit in the solid angle from which radiation is received compared
with a spherical outflow. For a uniform density CSM case, one has $\Gamma 
\propto t^{-3/8}$, and therefore the post-jet-break lightcurve, for all 
different orderings of synchrotron characteristic frequencies, falls off 
faster than the  isotropic case by a factor $\Gamma^2 \propto t^{-3/4}$; the
temporal behavior of synchrotron characteristic frequencies are unaffected
by the edge effect.

For the case of a CBM with density stratification like that of a wind,
$\Gamma \propto t^{-1/4}$, and post-jet-break lightcurve fall-off
is steeper than the isotropic case by a factor $t^{-1/2}$.
It is found that the jet break in the wind medium is very smooth, 
covering more than 2 orders of magnitude in time, when smearing due 
to integration over equal-arrival-time surface is taken into account
\citep{kumar00b,piran00,granotpiran12}. However, numerical simulations of jet
propagation find that lightcurves make a transition to a steeper 
fall-off, due to jet-break, on a shorter time scale of perhaps  
an order of magnitude \citep[e.g.][]{decolle12}.

The second effect of a finite jet angle on the lightcurve arises due
to its sideways expansion.
\cite{rhoads99} and \cite{sari99} showed that the epoch when the
edge effect kicks in is also the time when sound waves cross the jet in
the transverse direction leading to its sideways expansion. 
The jet opening angle increases as $\theta_j \sim \Gamma^{-1}$ when
the sideways expansion speed in jet rest-frame is of order the sound speed 
which for a relativistic plasma is $c/3^{1/2}$; $\theta_j \sim \Gamma^{-1}$
is a consequence of time dilation plus transverse speed $\sim c$ -- the 
time elapsed in jet comoving frame is a factor $\Gamma$ smaller than the 
lab frame time, and therefore the transverse size of the jet, when it 
expands with speed $\sim c$ in its own rest frame, is approximately $R/\Gamma$.
The transverse speed of the jet, in the lab frame, in this case is 
$v_\theta \sim c/\Gamma$. However, according to the momentum equation for 
a relativistic plasma $\partial (\rho \Gamma^2 v_\theta)/\partial t \sim 
r^{-1}\partial p/\partial\theta$, one has $v_\theta\sim
c/(\Gamma^2\theta_j)$;  for a detailed discussion of this result see e.g. 
\cite{kumargranot03,granotpiran12}. 
The jet evolution for these two different transverse speeds -- $c/\Gamma$ and
$c/(\Gamma^2\theta_j)$ -- are found to be 
not too different \citep{granotpiran12}.

Combining the energy conservation equation for a constant
density CBM --- $E \propto R^3 \Gamma^2 \theta_j^2$ --- with 
$\theta_j \sim \Gamma^{-1}$ after the jet-break, we find that 
the jet radius increase slows down substantially after the jet-break.
Therefore, $\Gamma\propto (R/t_{obs})^{1/2} \aprop t_{obs}^{-1/2}$
after the jet break. A more precise analytic derivation of jet radius and
LF evolution is discussed in e.g. \cite{rhoads99,sari99,piran00,granotpiran12},
 and is given by:
\begin{equation}
 \Gamma \aprop \exp (-R/l_{\rm jet}), ~~~\rightarrow ~~~ \Gamma 
    \propto t_{\rm obs}^{-1/2},
\end{equation}
where
\begin{equation}
 l_{\rm jet} \equiv \left[\frac{E_{\rm jet}}{(4\pi/3) n_p m_p c^2}\right]^{1/3}.
\end{equation}
Therefore, one has
\begin{equation}
 \nu_m \propto \Gamma^4 \propto t_{\rm obs}^{-2};
\end{equation}
\begin{equation}
 \nu_c \propto \Gamma^{-1} t_{\rm obs}^{-2} {B'}^{-3} \propto t_{\rm obs}^0;
\end{equation}
\begin{equation}
 F_{\nu,max} \propto R^3 B' \Gamma \propto R^3 \Gamma^2 \propto t_{\rm obs}^{-1},
\end{equation}
so that the post jet-break afterglow lightcurve, in slow cooling regime,
is given by
\begin{equation}
 f_\nu \propto \left\{
  \begin{array}{ll}
   \nu^{1/3} t_{\rm obs}^{-1/3}, & \nu_a < \nu < \nu_m, \\
   \nu^{-(p-1)/2} t_{\rm obs}^{-p}, & \nu_m < \nu < \nu_c, \\
   \nu^{-p/2} t_{\rm obs}^{-p}, & \nu > \nu_c.
  \end{array}
\right. 
\end{equation}
The flux decay in a band that lies above $\min(\nu_c, \nu_m)$, $\propto 
t^{-p}$, is steeper than when the edge effect alone is considered.

Numerical simulations suggest that sideways expansion of a relativistic jet is
unimportant until $\Gamma$ drops below $\sim2$
\citep{granot01,kumargranot03,cannizzo04,zhangmacfadyen09,decolle12,granotpiran12,vaneerten12a,vaneerten12b}.
Nevertheless, numerical simulations also show that a post-jet-break 
lightcurve is similar to the simple analytical model we have 
described with sideways expansion \citep{zhangmacfadyen09}.
The lightcurve behavior also depends on observer's viewing direction.
Fitting late-time X-ray data with numerical jet models suggests that
the line of sight for most GRBs is mis-aligned from the jet axis.
\citep{zhangbb14b,ryan14}.

The GRB jets are expected to be structured, i.e. the luminosity per unit 
solid angle and Lorentz factor vary with angle across the jet. Several
papers have analyzed jet properties varying with angle as a
 power-law function
\citep{meszaros98,rossi02,zhangmeszaros02b}, or has a Gaussian 
distribution \citep{zhangmeszaros02b,kumargranot03,zhang04}.
For an on-axis observer to a structured jet, the afterglow decay slope 
is steeper than the top hat jet case described above
\citep{meszaros98,daigou01,panaitescu05}. For an off-axis observer,
the viewing angle becomes important for the lightcurve. For a power
law jet, the ``jet break'' time for an off-axis observer is determined
by the viewing angle $\theta_v$ rather than the jet opening angle 
$\theta_j$ as was the case for a ``top hat'' jet model
\citep{zhangmeszaros02b,rossi02,kumargranot03,granotkumar03}.
For a Gaussian jet, the lightcurve is similar to a top hat jet if
the line of sight is inside the Gaussian cone, while it is similar 
to the off-axis power-law case if the line of sight is outside (but not
too much larger than) the
Gaussian cone \citep{kumargranot03,granotkumar03}. 
Structured jets make it possible to understand the GRB phenomenology
within the framework of a  ``quasi-universal''
 \citep{rossi02,zhangmeszaros02b,zhang04} jet, i.e. GRB jets are
similar to each other, and different observed properties are due to
different viewing angles of the observer\footnote{This suggestion
arose from the {\em rough} anti-correlation between $E_{\rm \gamma,iso}$
and $\theta_j$ \citep{frail01,bloom03}, so the original suggestion was 
that GRB jets are {\em quasi}-universal \citep[e.g.][]{zhangmeszaros02b},
rather than strictly universal.}. Such models have
well defined luminosity function \citep{zhangmeszaros02b,rossi02}
and distribution of the observed jet break time \citep{perna03}.
Even though the ``universal'' jet model is challenged by
the data \citep{nakar04}, a ``quasi-universal'' jet, with more free
parameters, is perhaps consistent with various observational constraints 
\citep{lloydronning04,zhang04,daizhang05}.

Another widely discussed jet structure is the two-component jet model.
According to which the GRB outflow is composed of a narrow jet, usually
with higher $L_{\gamma,iso}$ and $\Gamma$, which is surrounded by a 
wider, usually with lower $L_{\gamma,iso}$ and $\Gamma$, jet component.
Depending on the viewing angle, such a two-component jet can account for a
variety of lightcurve features, including an early jet break and late time
re-brightening \citep{huang04,peng05,wu05b}. The model has been applied to
interpret the afterglow data for several busts, such as GRB 030329 
\citep{berger03b} and GRB 080319B \citep{racusin08}. 
The collapsar model of long-duration GRBs offers a natural 
mechanism for generating a two-component jet: 
a narrow, highly relativistic, jet emerging from a star is accompanied
by a wider, less relativistic ``cocoon'' surrounding the jet
\citep{ramirezruiz02,zhangw04}. Alternatively, the narrow jet may be related 
to a magnetically confined proton component, while the wide jet is related
to a neutron component that is not subject to magnetic confinement
\citep{peng05}.

The GRB jets can even be ``patchy'', i.e. the emission comes from many
bright patches or ``mini-jets'' 
within a broad jet cone \citep{kumarpiran00b,yamazaki04}.
Mechanisms to produce patchy jets include non-uniform shells within
the internal shock scenario \citep{kumarpiran00b}, localized Lorentz boosted 
emission regions associated with relativistic outflows in magnetic 
reconnections, or turbulence in a magnetically-dominated jet 
\citep{lyutikov03,narayan09,kumarnarayan09,lazar09,zhangyan11,zhangzhang14}.

An interesting effect associated with relativistic jets of finite opening 
angle is the so-called ``orphan afterglows'', namely, detection of afterglow 
events without the detection of prompt $\gamma$-ray emission itself. 
An observer lying outside the jet cone might not see $\gamma$-rays due to
the strong relativistic beaming of photons in the direction of the jet and 
away from the observer line of sight. 
However, this observer will see the afterglow lightcurve rise initially
as the Doppler beaming factor gradually increases when the $1/\Gamma$ 
cone widens, and the flux will peak when $1/\Gamma$ cone enters 
the line of sight. Subsequently, the lightcurve behaves like a normal 
(post jet-break) afterglow lightcurve \citep{granot02}. An orphan afterglow 
is also possible for a dirty fireball, for which the prompt GRB is not 
detected due to its low Lorentz factor, while the afterglow radiation is
produced when the outflow is decelerated \citep{huang02}. Many authors have 
discussed the detectability of orphan afterglows over the years 
\citep[e.g.][]{totani02,levinson02,nakar02,zou07}.
However, thus far no positive detection has been made\footnote{One possible 
exception was PTF11agg \citep{cenko13}, which is an optical transient with
power law decay but without a $\gamma$-ray trigger. However, \cite{cenko13}
argued that it is unlikely an orphan afterglow seen off-axis.}. This is likely 
due to the combined effect of the faint nature of orphan afterglows
and the difficulty of identifying them. 

\subsection{Other effects}

In this sub-section we describe a number of effects that could modify the
``standard'' afterglow behavior of GRBs ---  which is based on an adiabatic, 
relativistic, blastwave dynamics --- we have discussed thus far. The 
effects of radiative losses on the blastwave dynamics, and afterglow 
lightcurves, have been discussed by a number of authors 
\citep[e.g.][]{rees98,dermer99,meszarosrees99,huang99,bottcher00,nava13},
and we refer to these works for details. In the following subsections 
we describe a few selected effects that can leave a signature on 
afterglow lightcurves. 

\subsubsection{Naked afterglow and high-latitude effect}

When a blast wave encounters a void the observed flux does not 
drop abruptly even though the adiabatic cooling of electrons does 
indeed lead to a very sharp decline of emissivity in the absence of
electron acceleration.
The reason is that photons from parts of the jet lying at an angle
larger than $\Gamma^{-1}$ with respect to the line of sight (high
latitudes) continue to contribute to the observed flux for some 
period of time --- due to the larger path length they have to travel
to get to the observer --- after the jet has run into a void or the
emission is turned off suddenly for some other reason.
A characteristic signature of this ``high latitude'' radiation is 
that the temporal decay index ($\alpha$) is related to the
spectral index ($\beta$) ---  $f_\nu \propto t^{-\alpha} \nu^{-\beta}$ ---
as follows:
\begin{equation}
\alpha = \beta + 2.
\end{equation}
A simple derivation of this result can be found in \S\ref{relativity}, and
for a more complete discussion we refer to \cite{fenimore96,kumar00,dermer04}.
The ``high latitude'' emission contributes to the observed flux 
as the $\gamma$-ray emission winds down, and it probably accounts for the
steeply declining X-ray lightcurve observed by the Swift satellite 
immediately following the prompt $\gamma$-ray phase for some GRBs 
\citep{zhang06}. It also provides a good model for the decay phase of 
X-ray flares when the ``zero time point'' is taken to be close to the
start-time of the flare \citep{liang06}.

\subsubsection{Energy injection}

Energy can be added to a decelerating blastwave, not only in a smooth, 
continuous way (for details of a continuously fed fireball, please see 
more extended discussion in Sec.\ref{sec:scalings}, Eqs.\ref{Lt}--\ref{s-q}), 
but also in discrete steps when fast shells ejected at late times from the 
central engine runs into the hot blastwave. 
This interaction can be described in terms of five \citep{kumarpiran00a}
or six \citep{zhangmeszaros02a} different regions separated by three shocks
and one or two contact continuities, and displays rich afterglow behavior. 
Some abrupt optical rebrightenings detected during the afterglow phase 
\citep[e.g.][]{nardini11} might be related to such
interactions \citep{zhangmeszaros02a}.

\subsubsection{Density bumps}

A blastwave may run into regions of enhanced density in the circum-stellar 
medium. These may lead to bump features in afterglow lightcurves 
\citep{dailu02,lazzati02,daiwu03,peerwijers06}. However, numerical 
calculations \citep{nakar03,nakargranot07,uhm07,uhm14c,geng14}
suggest that this re-brightening feature is expected 
to be very smooth and its amplitude very small in most situations.
The main reason is that due to the relativistic equal-arrival-time surface
effect, the emission received at any observer time comes from different
latitudes and different emission times. This poses some intrinsic constraint
on $\Delta L/L$ with respect to $\Delta t / t$ \citep[e.g.][]{nakar03,ioka05},
making the bumps very smooth. Furthermore, if the observed band (e.g. X-rays)
is above the cooling frequency, then the observed flux is independent of
the ambient density \citep{kumar00c,freedman01}. If there is a long-lasting
reverse shock, the reverse shock lightcurve is more sensitive to the medium
density fluctuations than the forward shock lightcurve \citep{uhm14c}.
A significant afterglow feature due to density fluctuation is expected only
when the long-lasting reverse shock emission outshines the forward shock
emission.

\subsubsection{Synchrotron self-Compton}

The synchrotron self-Compton (SSC) mechanism has two effects on the afterglow 
radiation. First, it introduces an extra cooling to electrons, so that the 
synchrotron cooling frequency is reduced by a factor $(1+Y)^{2}$  
\citep[e.g.][]{weilu98,panaitescu00,sari01}; where $Y=u_{syn}/u_{_B}$ 
is the ratio of synchrotron photon energy density and magnetic field energy
density. Second, the SSC introduces an extra spectral component at high 
energies which could dominate in the GeV band, and might show up in the X-ray 
band at late time if the ambient density is large 
\citep{meszarosrees93b,meszaros94,sari01,zhangmeszaros01b}. 
IC cooling of electrons in the Klein-Nishina 
regime can somewhat flatten the index of synchrotron spectrum in the cooling 
regime \citep[e.g.][]{derishev01,nakar09,daigne11,barniolduran12}, and steepen the 
decay slope of GeV afterglows \citep{wang10}.

The GeV afterglows of most GRBs can be explained as synchrotron radiation
from the forward shock \citep[e.g.][]{kumar09,kumar10}. However, the GeV 
afterglow of GRB 130427A cannot be interpreted as the synchrotron radiation 
only \citep{ackermann14}, and a possible SSC contribution
to the GeV afterglow has been suggested \citep[e.g.][]{fan13a,liu13}.

\subsubsection{Hard electron spectrum}

For $p$ between 1 and 2, the minimum electron Lorentz factor ($\gamma_m$)
depends on the maximum Lorentz factor of shock accelerated electrons ($\gamma_M$) 
and is given by
\begin{equation}
 \gamma_m = \left(\frac{2-p}{p-1} \frac{m_p}{m_e} \epsilon_e \Gamma
\gamma_{_{M}}^{p-2} \right)^{1/(p-1)},
\end{equation}
cf.  \citep{daicheng01,bhattacharya01,resmi08}.
In this case the afterglow decay slopes are systematically shallower than 
when $p>2$, which can be confused with injection of energy to the decelerating 
blastwave especially when spectral information is missing.

\subsubsection{Effect of neutron decay}

The immediate vicinity of the GRB central engine is likely to have high 
temperature for dissociation of nuclei.
A baryonic jet launched from such a site, therefore, is expected to
contain free neutrons along with protons.
Neutrons decouple from protons when the proton-neutron elastic collision
optical depth drops below unity, after which neutrons stream freely
\citep{derishev99,bahcall00,meszarosrees00b,beloborodov03}.
Free neutrons undergo $\beta$-decay
\begin{equation}
 n \rightarrow p^{+}+e^{-}+\bar\nu_e,
\end{equation}
with a mean co-moving life time of just under 15 minutes 
\begin{equation}
 \tau'_n = 881.5 \pm 1.5 {\rm s} .
\end{equation}
The typical radius of neutron decay is
\begin{equation}
 R_\beta = c \tau'_n \Gamma_n \simeq (8\times 10^{15}~{\rm cm})
(\Gamma_n/300),
\end{equation}
which is below the deceleration radius for a uniform density CBM. Since 
neutron decay happens continuously in time (and in distance), neutron decay
are expected to affect both prompt and early afterglow lightcurves. 

The impact of neutron decay on the early afterglow has been studied 
by \cite{beloborodov03b} and \cite{fan05b}, who found that it can lead
to a re-brightening feature in the otherwise
power-law decay lightcurve. The signature is different for the ISM and
wind cases \citep{fan05b}. 

\subsubsection{Radiation front effect}

Gamma-ray photons released during the prompt emission phase move ahead
of the GRB ejecta and interact with the ambient medium before the ejecta 
drives a shock wave into the medium. The CBM profile is modified 
due to this interaction, and as a result the early afterglow emission 
is different from the case of a shock wave moving into an undisturbed
medium \citep{madau00,thompson00,meszaros01,beloborodov02,kumarpanaitescu04}.
The main effect of this interaction between $\gamma$-ray photons and the
CBM is to enrich the medium ahead of the
GRB ejecta with electron-positron pairs which are produced as a result
of $\gamma$-rays scattered by electrons in the medium which then collide
with outward moving $\gamma$-rays associated with the prompt radiation
to produce $e^\pm$; the newly created pairs further scatter $\gamma$-ray
photons thereby leading to another generation of $e^\pm$, and so on.
Thus, the blast-wave propagates through a medium loaded with pairs
which is also moving away from the CoE due to momentum deposited 
to the CBM by outward moving radiation front. These effects are 
particularly important when the CBM density is large.

\subsubsection{Transition to Newtonian phase}

A decelerating, relativistic, blastwave becomes Newtonian when it has
swept up mass from the CBM that is of order the rest mass of the GRB-ejecta
times the initial Lorentz factor, or $E/c^2$; $E$ is the energy of the
blastwave. For a uniform density CBM the radius where the shock becomes
sub-relativistic is $R_N\sim [3E/(4\pi c^2 n_0 m_p)]^{1/3}=1.2\times10^{18}cm 
(E_{52}/n_0)^{1/3}$.
The blastwave dynamics in the Newtonian phase, for a uniform density
CBM, is described by the well known Sedov-van Neumann-Taylor solution:
\begin{equation}
 v \propto R^{-3/2} \propto t_{\rm obs}^{-3/5} \quad{\rm and}\quad 
   R \propto t_{\rm obs}^{2/5}.
\end{equation}
Therefore, $B'\propto t_{\rm obs}^{-3/5}$, $\gamma_m\propto v^2 \propto 
t_{\rm obs}^{-6/5}$,
$\nu_m \propto B' \gamma_m^2 \propto t_{\rm obs}^{-3}$, $\nu_c \propto 
t_{\rm obs}^{-1/5}$, and $F_{\nu,max} \propto t_{\rm obs}^{3/5}$, so that 
\begin{equation}
 f_\nu \propto \left\{ 
  \begin{array}{ll}
   \nu^{-(p-1)/2} t_{\rm obs}^{(21-15p)/10}, & \nu_m < \nu < \nu_c, \\
   \nu^{-p/2} t_{\rm obs}^{(4-3p)/2}, & \nu > \nu_c
  \end{array}
\right.
\end{equation}
For $p=2.3$, the lightcurve decay slopes are $-1.35$ and $-1.45$ for $\nu_m < 
\nu < \nu_c$ and $\nu > \nu_c$, respectively. This decay is steeper than 
the isotropic relativistic case but shallower than the post-jet break phase.
So the lightcurve would show a steepening behavior if relativistic-to-Newtonian
transition happens before the jet break \citep{dailu99,huang99}, while
it would become less steep if the transition happens after the
jet break \citep{livio00}. A generic dynamics model that connects the 
relativistic phase to non-relativistic phase
was developed by \cite{huang99} and improved by \cite{peer12} and \cite{nava13}.
The shock wave evolution in the deep Newtonian regime has been studied by 
\cite{huang03} in the context of GRBs.

Due to host galaxy contamination, observing the Newtonian phase in the optical
band is very difficult. This can be better accomplished in the radio band
for nearby GRBs. For example, late time radio follow-up observations of 
GRB 030329 revealed a brightening of the decaying lightcurve which is 
consistent with the transition to the Newtonian phase when emission from 
the receding counter-jet becomes visible \citep{vanderhorst08,zhangmacfadyen09}.

A complete compilation of characteristic frequencies and light curves of
the analytical synchrotron external shock models in all spectral regimes 
(for different ordering of $\nu_m$, $\nu_c$ and $\nu_a$)
and all temporal phases (including
the forward shock and reverse shock emission during and after the reverse 
shock crossing phase, the pre- and post-jet break self-similar phase,
and Newtonian phase) can be found in an extended review article by
\cite{gao13b}.

\section{Afterglow observations and interpretations}
\label{afterglow}

Broadband GRB afterglows were predicted before their discoveries
\citep{paczynski93,katz94,meszarosrees97}.
Shortly after the publication of the seminal paper by \cite{meszarosrees97}
which provided detailed predictions for the broad-band afterglow based on the
external shock model, the first X-ray and optical afterglows were discovered
for GRB 970228 \citep{costa97,vanpara97}, and the first radio afterglow
was discovered for GRB 970508 \citep{frail97}. Since then, regular 
follow-up observations of GRBs have been carried out, and a large 
amount of broad-band afterglow data have been collected.
Before the launch of the NASA's dedicated GRB mission Swift \citep{gehrels04},
afterglow observations usually started several hours after the burst
trigger. Swift (launched in 2004) has closed this gap, and provides continuous
afterglow data in X-ray, optical and UV bands starting at $\sim 1$ min
after the $\gamma$-ray trigger. This opened a new window to the study
of GRBs. The launch of the high energy mission Fermi has led to the 
discovery of an extended GeV afterglow emission for many bright GRBs. 
We discuss all these topics in this section.

\subsection{Late time afterglow observations and interpretations}

Before launch of the Swift satellite, broad-band, late time 
($t\gae10$ hours) afterglow data had been collected for a moderate sample 
of GRBs. These observations were generally consistent with 
predictions of the external forward shock, synchrotron emission, model. 
The main observational properties of late time afterglow radiation are:

\begin{itemize}
 \item In general the optical afterglow displays a power law decay behavior
$F_\nu \propto t^{-\alpha}$, with a decay index $\alpha \sim 1$
\citep[e.g.][]{wijers97,harrison99}. This is consistent with the prediction
of the standard external shock afterglow model 
\cite[e.g.][]{meszarosrees97,sari98,panaitescu01,panaitescu02,yost03}, see Fig.
  \ref{FIG:afterglow_fit};
\smallskip
 \item A temporal break in the optical afterglow light curve is usually
detected for bright GRBs. The break time
is typically around a day or so, which is followed by a steeper decay
with slope $\alpha \sim 2$ \citep[e.g.][]{harrison99}. This is consistent
with the theoretical prediction of a ``jet break'' \cite[e.g.][]{rhoads99,sari99}.
\smallskip
 \item The radio afterglow light curve initially rises and reaches a peak
around 10 days, after which it starts to decline \cite[e.g.][]{frail00}.
The peak usually corresponds to passage of the synchrotron
injection frequency $\nu_m$, or the synchrotron self-absorption
frequency $\nu_a$, through the radio band.
\smallskip
 \item The broad-band afterglow spectrum can be fit with a broken 
power law, at a fixed observer time \citep{wijers99,harrison99}, 
as one expects for the synchrotron afterglow model.
\smallskip
 \item For bursts with high-quality data (e.g. GRB 021004 and
GRB 030329), richer features in the optical light curves have been 
discovered, which include bumps and wiggles that deviate from the 
simple afterglow model predictions \citep[e.g.][]{holland03,lipkin04}. 
Smooth bumps in afterglow lightcurves with duration  
$\delta t_{obs}\sim t_{obs}$ may be interpreted as due to density
bumps in the external medium \citep{lazzati02,daiwu03,nakargranot07} whereas
sharper features in lightcurves might be due to energy injection from the 
central engine \citep{katz98,kumarpiran00a,zhangmeszaros02a,granot03b}, 
angular fluctuations in energy per unit solid angle 
\citep{kumarpiran00b,yamazaki04b}, or the existence of 
two-component jets \citep{berger03,huang04,racusin08}.
\end{itemize}

\cite{panaitescu01,panaitescu02,yost03} carried out detailed modeling of 
the broad-band afterglow data within the framework of the external shock
model. They found that the late time afterglow data are in line
with the predictions of this model (see Fig. \ref{FIG:afterglow_fit}),
and they were able to derive the micro-physical shock parameters 
($\epsilon_e$, $\epsilon_B$, $p$) using the data which turned out to be
different for different bursts; distribution of $\epsilon_e$ and $\epsilon_B$
are discussed in \S\ref{shock-parameters}. Moreover, the afterglow data 
seem to favor a constant density medium (ISM) for most GRBs rather than the 
stratified density medium expected for a stellar wind. Another interesting 
result is that although the isotropic kinetic energy in the GRB blastwaves 
varies by more than 3 orders of magnitude the jet-corrected afterglow 
energy is clustered within about an order of magnitude.
This, together with the same clustering of the 
jet-corrected $\gamma$-ray energy \citep{frail01,bloom03}, point 
towards a roughly constant energy reservoir for GRBs that were
detected before the {\em Swift} era.

\subsection{Early afterglow observations and interpretations}

The Swift mission carries on board an X-ray telescope \citep[XRT,][]{burrows05b}
and a UV-Optical Telescope \citep[UVOT,][]{roming05} besides the gamma-ray
detector Burst Alert Telescope \citep[BAT,][]{barthelmy05c}. The rapid
slew of XRT and UVOT towards the GRB source allows detections of GRB
early afterglows within less than 100 s after the $\gamma$-ray trigger. As a 
result, Swift has provided a rich trove of early afterglow data which 
revealed many, usually unexpected, interesting features.
The early afterglow data and the ideas to interpret them are summarized below.

\begin{figure}
\begin{center}
\includegraphics[angle=90,width=5.5in]{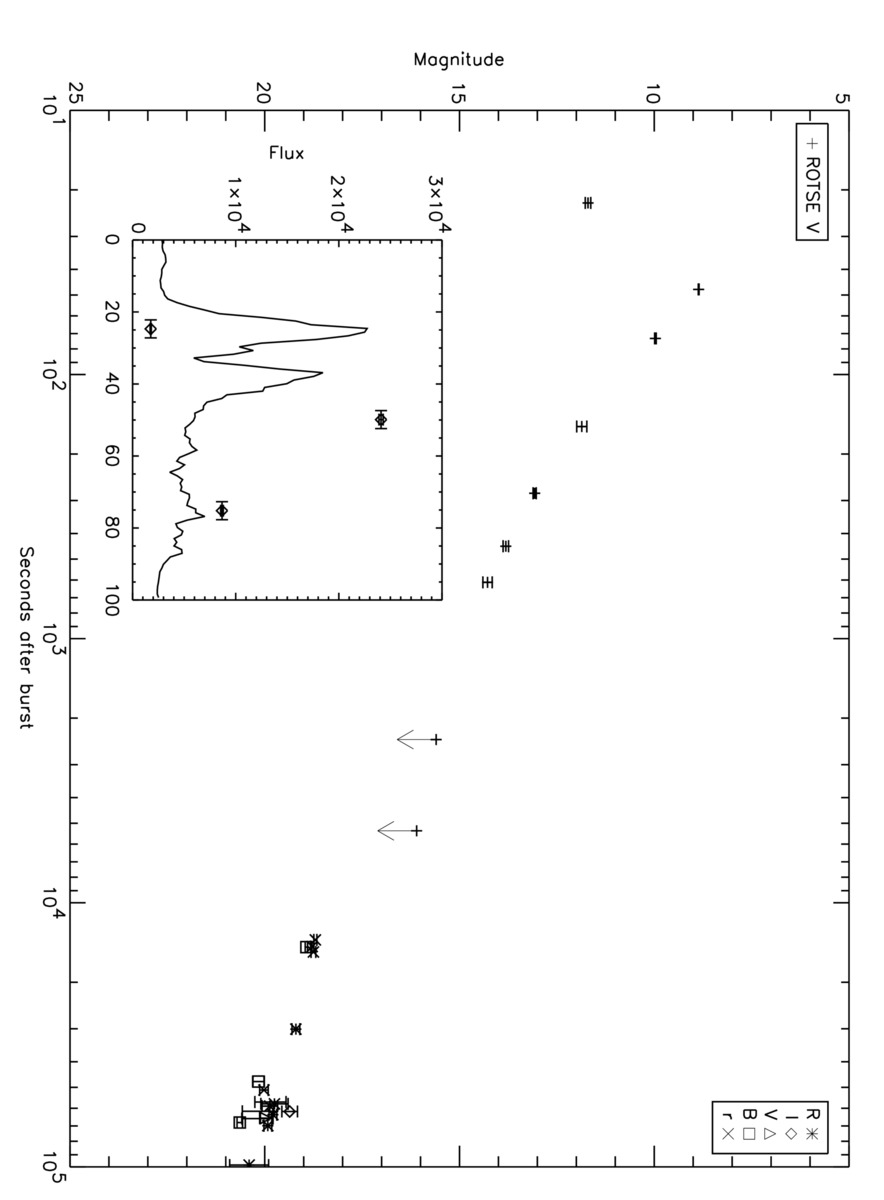}
\caption {The optical flash detected in GRB 990123 by ROTSE \citep{akerlof99}.
The peak flux was about 9th magnitude around 50 seconds after the trigger,
which does not coincide with the $\gamma$-ray peaks.}\label{FIG:990123}
\end{center}
\end{figure}

A bright optical flash was detected during the prompt emission of GRB
990123 which showed a distinct origin from the $\gamma$-ray emission. 
The flash was categorized by a sharp rise and a steep decay 
$F_\nu \propto t^{-2}$ (Fig.\ref{FIG:990123}) \citep{akerlof99}.
This is inconsistent with 
the external forward shock prediction, but is in accord with the theoretical
expectation of emission from the reverse shock 
\citep{meszarosrees97,meszarosrees99,saripiran99,saripiran99b}. 
It was later realized that in order to produce a bright reverse shock 
optical flash such as GRB 990123
\citep[and GRB 021211 and several others,][]{li03,fox03b,gomboc08},
the magnetic field in the reverse shock region should be stronger than
in the external forward shock region \citep{fan02,zhang03,kumar03}, but not
so strong that the magnetization parameter $\gae 0.1$ since in this 
case the magnetic fields would weaken the
reverse shock and the emergent flux would be less than the observed value
\cite[e.g.][]{zhangkobayashi05,mimica09,narayan11}.

Radio flares, possibly associated with optical flashes, were also 
observed for some GRBs such as GRB 990123 and GRB 021004 
\citep{kulkarni99b}. These radio flares peak later (around 1 day), but 
can be interpreted as arising in the reverse shock 
\citep[e.g.][]{saripiran99,kobayashizhang03a}.
 
\begin{figure}
\begin{center}
\includegraphics[angle=-90,width=5.5in]{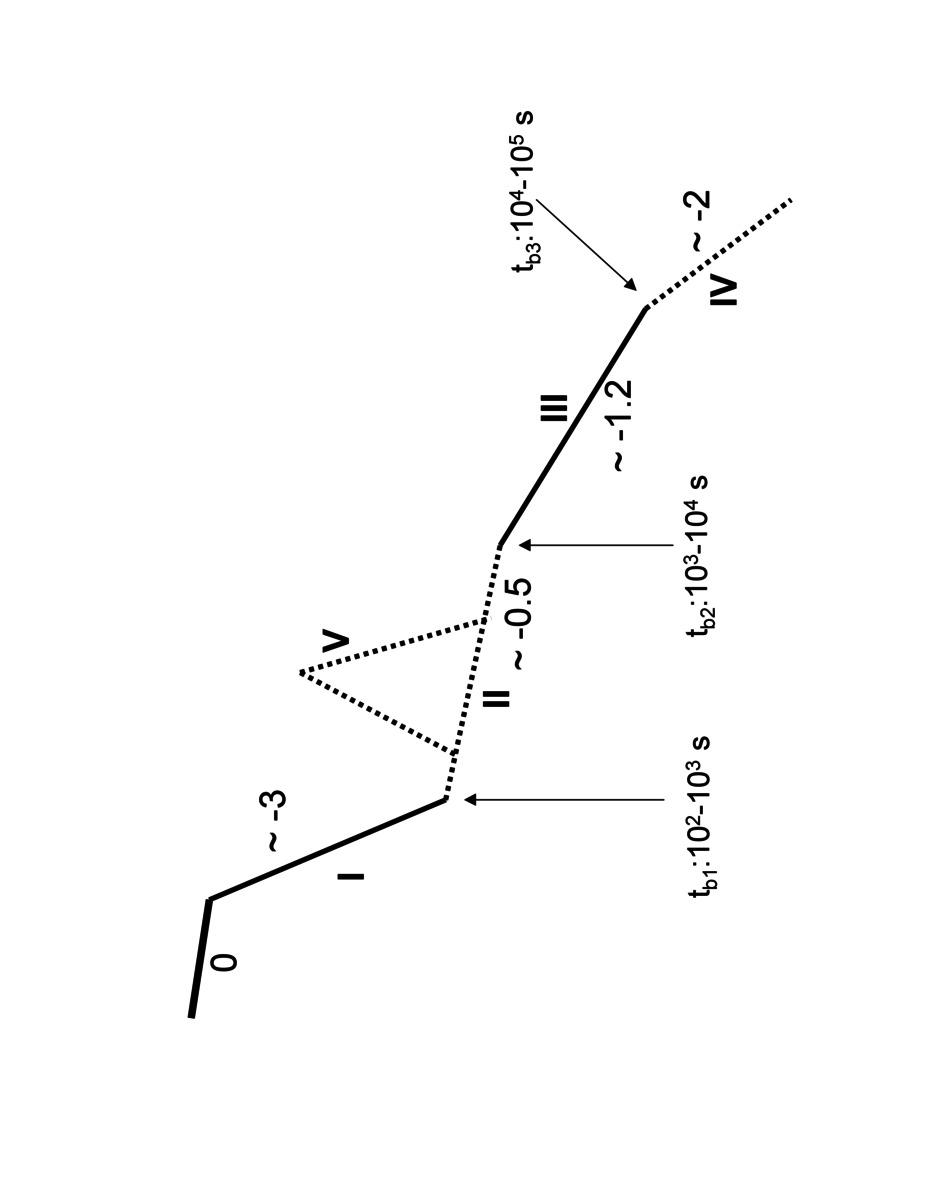}
\caption {The canonical X-ray afterglow light curve, which shows
5 distinct components: I. the steep decay phase which is the tail
of prompt emission; II. the shallow decay
phase (or plateau); III. the normal decay phase; IV. the late
steepening phase; V. X-ray flares. 
The Numerical value provided for each segment of the lightcurve is
the typical decay index for that segment, e.g. the lightcurve decays as
$t^{-3}$ during Phase I. From \cite{zhang06}.
}\label{FIG:canonical}
\end{center}
\end{figure}

%\begin{figure}
%\begin{center}
%\includegraphics[angle=-90,width=4.0in]{XRTdata1.jpg}
%\includegraphics[angle=-90,width=4.0in]{XRTdata2.jpg}
%\caption {Some examples of X-ray afterglow light curve detected by
%Swift XRT. From \cite{nousek06}.
%}
%\label{FIG:XRTdata}
%\end{center}
%\end{figure}

\begin{figure}
\centering
\begin{tabular}{cc}
\begin{minipage}{200pt}
\includegraphics[angle=-90,width=200pt]{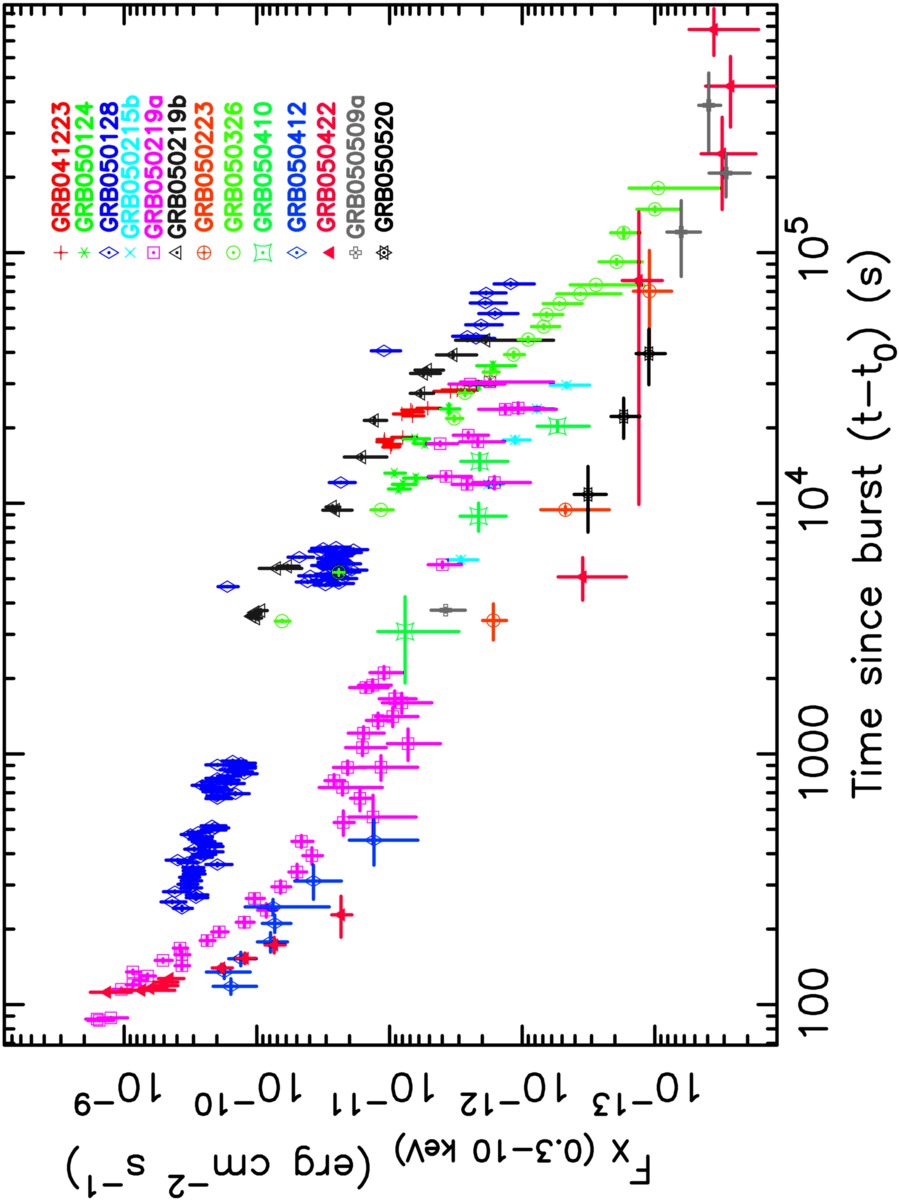}
\end{minipage}
&
\begin{minipage}{200pt}
\includegraphics[angle=-90,width=200pt]{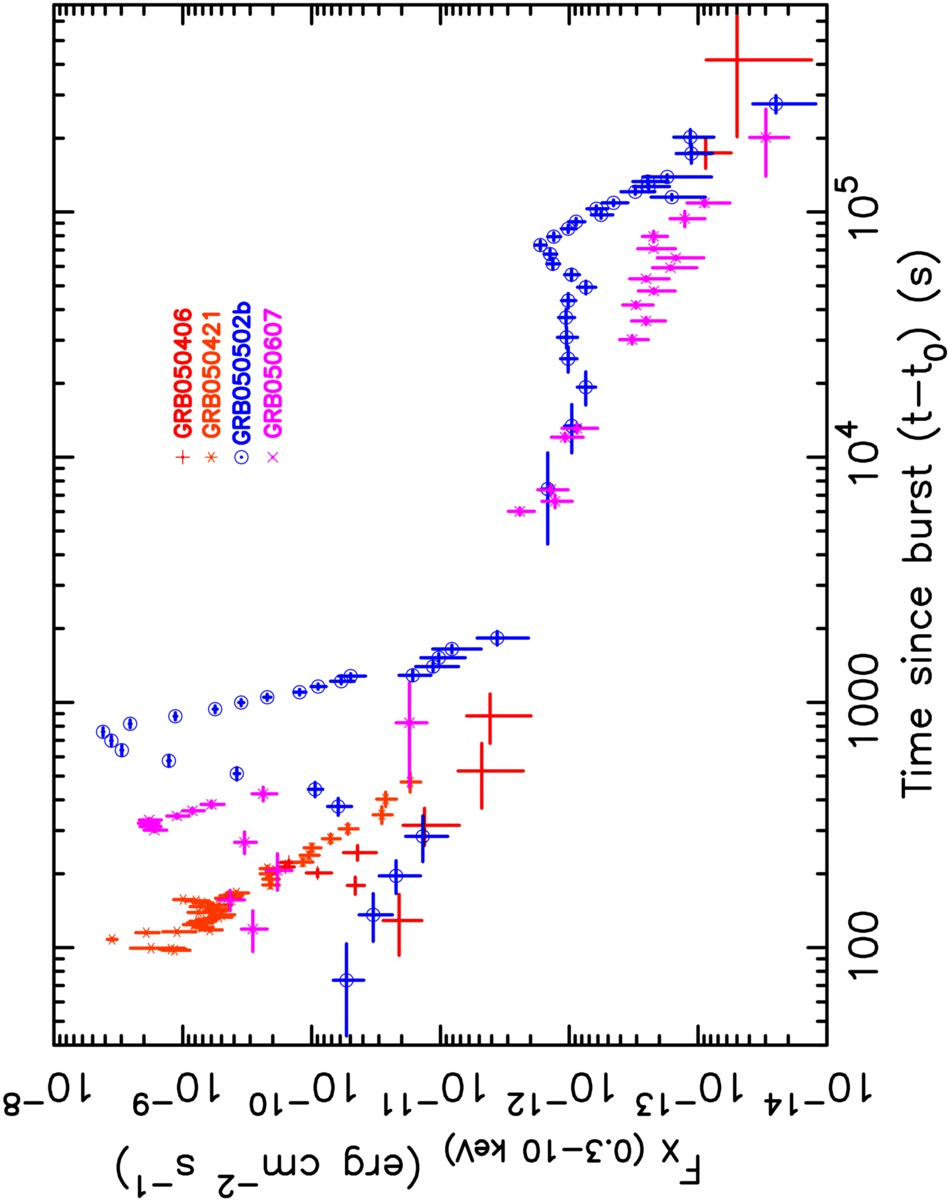}
\end{minipage}
\end{tabular}
\caption{Some examples of X-ray afterglow light curve detected by
Swift XRT. From \cite{nousek06}.
}\label{FIG:XRTdata}
\end{figure}

Swift observations revealed several surprising emission
components in early X-ray afterglow not predicted by the standard
model. The data can be delineated as a canonical lightcurve, which 
generally includes 5 components 
\citep[][see Fig.\ref{FIG:canonical} and Fig.\ref{FIG:XRTdata}]{zhang06,nousek06}.
Not all GRBs show all 5 components. 
The main properties of these 5 components, obtained for a large sample 
of Swift bursts \citep{evans07,evans09}, can be summarized as:
\begin{itemize}
\item I. An early time steep decay phase: it has a temporal decay index 
steeper than -2. When joint XRT/BAT observations were available, it is 
found that this phase is connected to the tail of the prompt emission
\citep{barthelmy05b}. This phase may be simply the high latitude emission 
(described in \S\ref{relativity}) associated with the prompt $\gamma$-ray 
source at $R\gae 10^{15}$cm when the central engine turns off faster 
than the decline of the X-ray lightcurve 
\citep{kumar00,dermer04,zhang06,nousek06,liang06}.
On the other hand if the emission region is at a much smaller radius 
then the rapidly declining X-ray lightcurve reflects the time dependence of 
the central engine activity \citep{fanwei05,barniolduran08}.
\item II. Shallow decay phase (or plateau phase): the temporal decay of flux is
shallow with slope -0.5 or larger, sometimes flat or even slightly 
rising early on. In most GRBs, it is followed by a ``normal'' decay 
with flux decreasing with time as $\sim t^{-1}$. Such data
can be incorporated within the external shock model, with the shallow
decay phase being due to continuous energy injection into the blast
wave \citep{zhang06,nousek06,panai06a}. Occasionally the plateau is 
followed by a very rapid drop \citep[e.g.][]{troja07,liang07b}, which demands 
an ``internal'' origin of the plateau. 
    \item III. Normal decay phase: this is the typical decay 
($\sim t^{-1}$) expected in the standard forward shock model.
    \item IV. Late steep decay phase ($\sim t^{-2}$ or steeper).
Expected in the forward shock model as a jet break.
    \item V. X-ray flares\footnote{One X-ray flare in each of the two
GRBs 011211 and 011121 was detected by BeppoSAX \citep{piro05}, which
was interpreted as onset of external shock afterglow.}: one or more X-ray 
flares can be found in nearly half of GRB X-ray afterglows. These flares 
share many properties with prompt emission pulses. It is widely accepted
that they are powered by late central engine activities
\citep{ioka05,burrows05,fanwei05,zhang06,liang06,lazzati07,chincarini07,maxham09,margutti10}.
   \end{itemize}

 An alternative way to describe the X-ray afterglow is a two-component 
phenomenological model \citep{obrien06,willingale07,ghisellini09}.
According to this method, the X-ray afterglow can be decomposed
into a ``prompt'' component (the prompt emission phase and the subsequent
rapid decay phase), and an ``afterglow'' component (the plateau, normal
decay and the late rapid decay). Although no theoretical model predicts
the specific mathematical form of the two components, this phenomenological
model seems to work well to fit the X-ray afterglow lightcurves of many Swift
GRBs, and to identify X-ray flares or internal plateaus that demand
central engine activities \citep[e.g.][]{lyons10}.

 A puzzling feature seen in a fraction of GRBs is that
the optical and X-ray afterglows are ``chromatic'' 
\citep{panai06b,fan06,liang07,liang08,huang07}\footnote{A recent 
detailed study suggests that about half of GRB afterglows are
consistent with the achromatic hypothesis and the external shock
model, while the others either do not comply with the external shock
closure relations, or show clear ``chromatic'' behavior (X.-G. Wang
et al. 2014, in preparation). 
}. In some cases there is no
temporal break in the optical lightcurve at the epoch when the
X-ray lightcurve makes a transition from Segment II (plateau phase) to
Segment III (normal decay phase) or from Segment III to IV (jet break
phase).  Within the external shock model,
such a chromatic behavior is allowed if there is a significant
spectral change across the temporal break due to, e.g. crossing of a 
spectral break in the X-ray band. The perplexing aspect of the phenomenon 
is that the X-ray spectral index almost never changes across the break. 
This suggests a hydrodynamical or geometrical origin for the break in 
the X-ray lightcurve, but in that case a simultaneous break must also be 
seen in the optical lightcurve. The non-detection of such a break in the 
optical band in some GRBs rules out the one-component forward shock 
model for the broadband afterglow emission observed from these bursts,
and suggests at least two emission sites to account for the optical
and X-ray emissions, respectively.

The unexpected, rich, X-ray lightcurve features detected by Swift 
and the puzzling chromatic behavior of afterglow stimulated
a wave of intense modeling of early afterglow. We provide
a brief summary of various different ideas proposed for explaining
the prominent features in afterglow light-curves.

\subsubsection{Steep decay of early X-ray light-curve}

The standard interpretation of the steep decay (I) phase is that it is 
the tail of prompt emission. The distinct separation between prompt emission
and late afterglow settled down the pre-Swift debate regarding
internal vs. external origin of prompt emission 
\citep[e.g.][]{sari97,dermermitman99}, and established the internal
origin of prompt emission. It is not settled whether
the X-ray flux during the steep decline is simply the high latitude 
emission associated with the rapid cessation of the prompt radiation
\citep{kumar00,zhang06} or emission from a somewhat less rapidly dying
central engine \citep{fanwei05,barniolduran08}.
It is quite common to find a strong spectral softening during the
steep decay phase \citep{zhangbb07}. Such a spectral evolution is not 
expected in the simplest version of the high-latitude
emission models but can be accounted for if the instantaneous
spectrum at the end of prompt emission is characterized by a
power law spectrum with an exponential cutoff \citep{zhangbb09}. 
Detailed analysis of a sample of GRBs suggests that the high-latitude
``curvature effect'' model can explain the steep decay phase of
at least a sample of GRBs \citep{zhangbb09,genet09,mangano11,zhangbb12}.

Other models of the steep decay phase include emission from a rapidly
expanding cocoon \citep{peer06b}\footnote{This model predicts a quasi-thermal
spectrum, which may interpret spectral softening during the X-ray tails.
However, one needs to introduce coincidence to account for the smooth
connection between the prompt emission and the X-ray tails as observed
in many GRBs.}, rapid discharge of hadronic energy 
of the blastwave \citep{dermer07}\footnote{This model requires
both prompt and afterglow emissions to be produced in the external
shock, which is highly unlikely as discussed in \S\ref{prompt_theory}.}, 
high-latitude emission in the
external reverse shock \citep{uhm07,uhm12}\footnote{This model requires
significant suppression of forward shock emission to make the reverse
shock feature to show up, and that is disfavored by the
extensive afterglow data.}, and sweeping of the 
external forward shock synchrotron spectrum with a low maximum 
frequency across the X-ray band \citep{petropoulou11}\footnote{This
model also requires both prompt and afterglow radiation to arise in
the external shock, which is inconsistent with GRB data.}. The latter 
three models have the 
underlying assumption that the prompt emission
itself is also of an external shock origin, since observationally 
the steep decay phase is simply the tail of prompt emission, and
hence are strongly disfavored by the rapid variability seen in $\gamma$-ray
lightcurves.

\subsubsection{Sudden increase in X-ray flux (flares)}
\label{X-ray-flares}

The X-ray flares are usually interpreted as due to re-start of GRB
central engine because of their short rise time of $\delta t_{obs}/t_{obs}\ll1$
\citep{burrows05,zhang06,fanwei05}. Such an interpretation
is directly supported by data analysis. \cite{liang06} assumed that
the decay phase of X-ray flares is dominated by the high-latitude
emission, and searched for the zero point of time ($T_0$) to allow for
the simple prediction $\alpha = 2 + \beta$ \citep{kumar00} to be satisfied. 
They found that the required $T_0$ usually corresponds to the beginning of 
X-ray flare. This is a good evidence for ``re-starting the clock'' when
the central engine comes back to life. Further, more detailed, modeling 
\citep{wu06,lazzati07,maxham09} and data analysis 
\citep{chincarini10,margutti10,margutti11} support this 
interpretation. Other ideas for the origin of X-ray flares include delayed
magnetic dissipation activity as the ejecta decelerates \citep{giannios06}
and anisotropic emission in the blast wave comoving frame 
\citep{beloborodov11b}; however, these models do not account for the 
$T_0$ effect found by \cite{liang06}.

\subsubsection{Plateaus in X-ray light-curves}

The shallow decay (or plateau) phase (II) and the subsequent segments
(III and IV) are more challenging to
interpret. The plausible interpretation for the plateau phase is that 
it arises when energy is injected to the decelerating external shock 
thereby slowing down the decay of the lightcurve, and a transition
to phase III occurs when energy injection is terminated 
\citep{zhang06,granot06b,fan06a,nousek06,panai06a}. 
This model predicts that the shape of lightcurves in the X-ray
\& optical bands should be similar where breaks occur at the same
time in these bands, i.e. an achromatic behavior across the EM spectrum. 
This model indeed works for all those bursts that display the expected
achromatic behavior, e.g. GRB 060614 \citep{mangano07} and GRB 060729 
\citep{grupe07}. 
However, this model cannot explain the data for chromatic afterglows, 
and another mechanism or emission component has to be invoked. 
The most straightforward extension of the external shock model
is to introduce a two-component jet, with the narrow jet dominating
the X-ray band emission while the wide jet dominating the optical
emission \citep[e.g.][]{racusin08}. Some GRBs can be modeled this
way at the price of introducing several additional parameters
that vary significantly from burst to burst \citep{depasquale09}.
A further extension of the external shock model is to include
emission from the reverse shock (RS). \cite{uhm07} and \cite{genet07}
assumed that the external forward shock (FS) does not contribute much to
the observed afterglow radiation, and that a long-lasting RS emission 
is responsible for the chromatic lightcurves observed in X-rays and 
optical bands. Indeed, the RS is
more sensitive than the FS to the ejecta stratification and circumburst
medium density inhomogeneity, and 
is capable of producing a wider variety of lightcurves 
\citep{uhm12,uhm14c}.
One drawback of this proposal is a lack of good reason for suppressing
 the FS emission which is in fact expected to be brighter than the RS 
emission in the X-ray band (for the same microphysics 
parameters in FS \& RS) by at least an order of magnitude, and has been very
successful for interpreting the broad band afterglow data of many GRBs
\citep[e.g.][]{panaitescu01,panaitescu02,yost03}.
A more reasonable possibility might be that the observed lightcurves 
are a superposition of the FS and RS emission, and that sometimes
RS outshines FS emission in certain band. Evolving microphysics 
parameters of the external shock ($\epsilon_e$ \& $\epsilon_B$) has
also been suggested as a possible explanation for the chromatic X-ray 
plateau \citep{ioka06,panaitescu06}.

\cite{shen12} interpreted the shallow decay phase as forward shock
synchrotron radiation during the pre-deceleration, coasting phase
in a wind medium. This model demands a relatively small Lorentz factor
$\Gamma$, which might be at odds with the higher value for $\Gamma$
obtained from the prompt emission data using the pair opacity
argument. Moreover, this model predicts achromatic afterglows, and 
therefore can only explain a sub-sample of GRBs which have
a shallow decay phase in both X-ray and optical bands at the same time. 
\cite{shao07} proposed that 
the X-ray plateau results from the contribution of prompt X-ray
emission scattered by dust in the host galaxy.
However, it predicts strong spectral evolution in X-rays which is not
detected \citep{shen09b}. 
\cite{ioka06} invoked a pre-$\gamma$-ray-trigger outflow to modify the ambient
medium profile in order to account for the shallow decay phase. 
\cite{yamazaki09} assumed a powerful outburst episode that preceded the 
GRB trigger, and suggested that the shallow decay
phase is simply due to a mis-identification of the zero time point.
This scenario predicts an optical flux, which is already ruled out 
by the prompt optical data \citep{birnbaum12}. In general, 
the scenarios of \cite{ioka06} and \cite{yamazaki09} invoked 
a ``prior explosion'' episode, preceding the observed $\gamma$-ray burst by
thousands of seconds, which no known central engine model can account for.

\subsubsection{Steep decay following the plateau in X-ray light-curve}

\begin{figure}
\begin{center}
\includegraphics[angle=-90,width=5.5in]{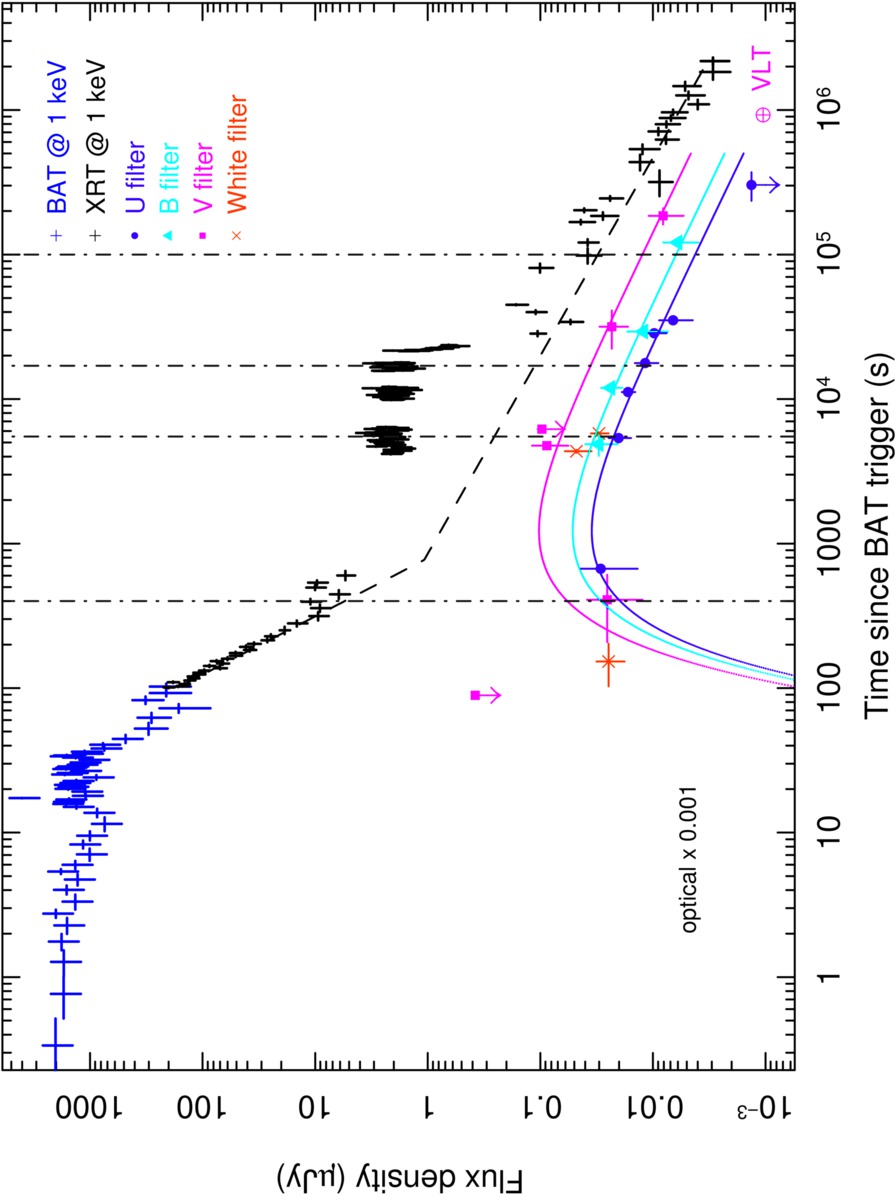}
\caption {The ``internal'' X-ray plateau observed in GRB 070110,
which suggests that the central engine launches a long-lasting
outflow with steady dissipation. From \citep{troja07}.
}\label{FIG:070110}
\end{center}
\end{figure}

Besides X-ray flares, there are a small fraction of GRBs which have plateaus 
in the X-ray lightcurve that are followed by a very steep decay that
is more rapid than a $f_\nu\propto t_{obs}^{-3}$ decline \citep[e.g. 
GRB 070110][]{troja07}, see Fig.\ref{FIG:070110}. These cases of steep 
decline following a plateau are rare \citep{liang07b}, and are not 
included in Fig.\ref{FIG:canonical}. They cannot be explained by the 
external shock model, and can only have an ``internal'' origin
involving direct dissipation of a long-lasting jet.
The existence of flares and these ``internal plateaus''
 \citep{burrows05,chincarini07,falcone07,troja07,lyons10}
suggest that the GRB central engine is long-lived. A more 
extreme opinion is that the entire X-ray emission is powered by a continuous 
jet from a long-lasting central engine, and that the X-ray flux from 
the external shock is buried beneath this emission \citep{ghisellini07}. 
Indeed, the canonical X-ray lightcurve can be matched with the accretion 
history in the collapsar GRB model 
\citep{kumar08b,kumar08a,cannizzo09,lindner10} or with the spindown power 
of a magnetar central engine \citep{yu10,metzger11}.
These models assume that the X-ray luminosity is proportional to the 
accretion power or the spindown power of the central engine. It is
attractive to interpret GRB afterglows that display chromatic behavior
as due to X-ray emission produced via some process internal to 
a continuous jet, and optical flux produced in the external shock. 
GRBs with achromatic lightcurves are cases where the standard forward 
shock emission dominates in both X-ray and optical bands.

Overall, the current data seem to suggest at least three emission sites: 
the erratic component (flares), the broken power-law X-ray component,
and the broken power-law optical component (if chromatic). It is
interesting to note that theoretically, one also naturally has three
emission sites: the FS and RS of the blastwave, and an internal
dissipation site within the relativistic outflow before it encounters
the CBM. In a messy system that invokes late central engine 
activity, the RS is likely long-lived since late ejecta
would continuously pile up onto the blastwave. The ejecta may also
have a wide distribution of Lorentz factor, so that layers with different
Lorentz factors may pile up onto the blastwave at different times.
The internal dissipation site can be either the photosphere of the
outflow, internal shocks, or magnetic dissipation site in a high
$\sigma$ (magnetization parameter) jet. 

Unlike the late time afterglows ($t_{obs}\gae 7$hr) observed before
the Swift mission, which had simple morphology, the afterglow modeling in the 
Swift era is much more complicated due to the complex behavior we see in
the lightcurves for the first few hours following the $\gamma$-ray trigger.
The first step is to disentangle 
various components, and decide which components likely have
an external shock origin and which do not. The traditional modeling
can be only applied to a sample of ``well-behaved'' afterglows
that show clean achromatic behaviors. More detailed studies are
needed to address following questions: What fraction of afterglows
can be interpreted within the standard external shock model? 
Are the differences between the two categories (afterglows that are 
due to FS and those that are not)
due to intrinsic differences in the central engine properties or these
due to external factors such as variations in CBM from one burst
to another? For those bursts that can be interpreted with the standard 
FS model, what are the shock microphysics parameters, and why do they
vary from one burst to another? 

X.-G. Wang et al. (2014, in preparation) 
carried out a detailed study by confronting the joint
X-ray and optical data of a large sample of Swift GRBs with the external
shock models. They found that at least half of the GRBs are consistent 
with the external shock models in both bands and the lightcurves are
achromatic. Only less than 15\% of GRBs in the sample 
are chromatic, which demand two different emission components to account for
the X-ray and optical data, respectively.

It is worth pointing out that short GRBs typically have fainter
afterglows due to their lower energies and probably lower 
circumburst densities \citep{panaitescu01b}. Comparing with the prompt
emission properties, one finds that both long and short GRBs
follow some similar correlations among prompt emission and afterglow
properties \citep{gehrels08,nysewander09,kann11}. This suggests a
similar radiative efficiency and probably also a similar circumburst 
environment for both long and short GRBs \citep{zhang07a,nysewander09}.

In summary, Swift observations have led to the following modified 
understanding of afterglows:
{\em The so-called ``afterglow'', at least for the initial few hours, is 
no longer simply the external forward 
shock emission; instead, it is a superposition of multiple 
components, including emission powered by a long-lasting central engine.}

\subsection{High energy ($>$10$^2$MeV) afterglow radiation}
\label{high_energy_afterglow}

Back in the Compton-Gamma-Ray-Observatory (CGRO) era, one burst detected 
by BATSE, GRB 941017, also triggered the high energy detector EGRET 
\citep{hurley94}. In fact, strong GeV emission was still detectable 
1.5 hours after the trigger when the burst re-emerged from the earth limb. 

We provide a brief summary of the theoretical models that were suggested 
for the delayed, long lasting, high energy photons from GRBs. These include 
internal shocks, e.g.  \cite{bosnjak09}, SSC process operating in the 
external shock \citep{dermer00,zhangmeszaros01b} --- while the reverse shock 
is passing through the GRB jet, two SSC processes (in FS and RS,
respectively) as well as two cross IC processes (FS photons up-scattered
by RS electrons and vice versa) could also contribute to the observed high 
energy flux \citep{wang01,wang01b,granotguetta03,peerwaxman04b,gupta07b,fanpiran08,zou09}. 
Moreover, prompt gamma-rays can be upscattered by electrons in the external FS
or RS and produce high energy emission \citep{meszarosrees94,beloborodov05,fan05c}.
Yet another process for delayed GeV photons from GRBs is up-scattered
CMB photons by high Lorentz factor electron-positron pairs in the
inter-galactic medium \citep{plaga95}; these pairs are produced when
TeV photons from a GRB interact with the cosmic infrared background
radiation. This mechanism can only work when intergalactic magnetic 
field strength is very small, of order $\lae10^{-15}$G, so that electron
deflection angle is small and a collimated GeV front traveling toward
Earth is produced \citep{dailu02,dai02,wang04,murase09}.

The Fermi satellite, with the Large Area Telescope \citep[LAT,][]{atwood09}
and Gamma-ray Burst Monitor \citep[GBM,][]{meegan09} on board, opened
a new window in 2009 to systematically study GRBs above 100 MeV, and to finally
settle the question as to which of the above mentioned mechanisms might
be responsible for producing high energy $\gamma$-ray photons in GRBs.

About 10 GRBs per year are jointly detected by LAT and GBM, allowing
a time-dependent broad-band spectral analysis of these GRBs. This led 
to two interesting observational discoveries, viz. the first photons
of energy $>10^2$MeV typically arrive a few seconds after the GBM 
trigger (or arrival of photons of energy $\lae10 $MeV), and the 
emission in the LAT band ($\gae10^2$MeV) lasts for 
$\sim 10^3$s which is much longer than the typical burst duration
in the GBM band ($\sim$5keV--10MeV) of 10--30s. Moreover, the LAT lightcurve 
usually shows a simple power law decay with time for almost the entire
duration of LAT observation \citep{abdo09c,abdo09a,ghisellini10,zhang11}.

It was realized soon after Fermi discovered these properties of high
energy emission from GRBs that photons of energy $>$10$^2$MeV, after
the prompt phase that lasts for $\sim$30s, are 
produced via the synchrotron process in the external forward-shock
\citep{kumar09,kumar10,ghisellini10}. The reasons for arriving at this 
conclusion are discussed below.

\begin{figure}
\includegraphics[width=13cm]{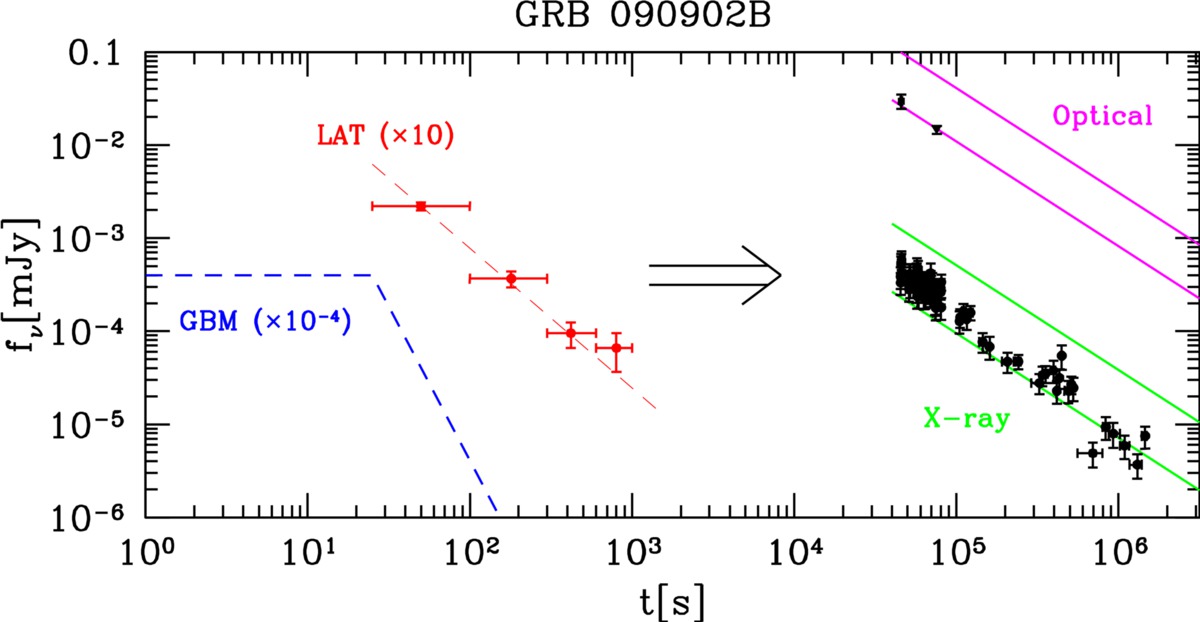}
\caption{
The optical and X-ray fluxes of GRB 090902B predicted at late times
using only the high energy data (photon energy $\gae10^2$MeV) at 50s
(assuming synchrotron emission from external forward shock) are shown
on the right half of this figure (diagonal bands). The predicted flux
are compared with the observed data (discrete points with error bars).
From \cite{kumar10}.
%LAT (X-ray) data red (black) circles, are from \cite{bissaldi09} (\cite{evans09}).
%{\bf Bissaldi et al. 2009 (Evans et al. 2009)}. Optical fluxes are from
%\cite{swenson09} and \cite{guidorzi09} (triangle).
%{\bf Swenson et al. 2009 (square) and Guidorzi et al. 2009 }(triangle). 
%The blue dashed line shows schematically the light curve observed by Fermi/GBM.
%The predicted value for the radio flux at one day has a large range
%(not shown), but consistent with the observed value.
}\label{FIG:predict-090902B}
\end{figure}

It is striking that the spectral index and the decay of the LAT lightcurve 
[$f_\nu(t) \propto \nu^{-1.1} t^{-1.3}$] satisfy the closure relationship
almost prefectly for synchrotron radiation from the shock heated 
circum-burst medium (CBM) 
by the relativistic jet of a GRB\footnote{See \S\ref{sec:scalings} for 
lightcurve scalings, and in particular Eq. \ref{f_nu_nuc}.} when $\nu>\nu_c$ 
(see Fig. \ref{FIG:predict-090902B})\footnote{For a few GRBs, the temporal 
decline of the LAT lightcurve is just slightly steeper (the decay index,
$\alpha$, larger by about 0.1 or 10\%) than what one might
expect from the LAT band spectral index in the regime $\nu>\nu_c$. 
\cite{ghisellini10} suggested that this is due to radiative loses 
affecting the external shock dynamics. 
However, \cite{wang10} showed that the decline of the LAT lightcurve 
is fine for an adiabatic blastwave, and the slightly steeper than
expected decline can be understood as the result of 
IC cooling of high energy electrons 
(those that produce $>$10$^2$MeV photons) which becomes more effective at later 
time; IC cooling of high energy electrons is suppressed at early times
because scatterings are deeper in the Klein-Nishina regime
at earlier times.}.
It was shown by \cite{kumar00c} that when the observation band is
above the synchrotron cooling frequency ($\nu>\nu_c$) then the
specific flux from external forward shock is dependent only on the
blast wave energy and the energy fraction in electrons (see eq. \ref{f_nu_nuc});
the flux is completely independent of the highly uncertain
CBM density, and is insensitive to $\epsilon_B$ ($f_\nu\aprop\,
\epsilon_B^{0.1}$). Therefore, one can confidently predict the flux
in the Fermi/LAT band, to within a factor of a few, from the knowledge
of energy in the prompt $\gamma$-ray radiation for a burst. And
remarkably, it turns out that this predicted flux is consistent with
Fermi/LAT observations for several well studied bursts 
\citep{kumar10} --- Table 1 provides a comparison of
the expected synchrotron flux from external forward-shock at 100 MeV and 
the Fermi/LAT data for five well studied bursts. 

\begin{table*}[H]
%\begin{minipage}[t]{0.8\columnwidth}
\begin{center}
\begin{tabular}{ccccccccc}
\hline
& $z$ & $E_{\gamma}$ & $t_{obs}$ & Exp. flux & Obs. flux & $T_{MeV}$ & $T_{LAT}$ & $T_{LAT}/T_{MeV} $ \\
&     &   (10$^{54}$ erg)    &    (s)    &     (nJy)     &       (nJy)   &      (s)     &     (s)   &                        \\
\hline \hline
080916C  & $4.3$ & $8.8$ & $150$ & $50$ & $67$ & $60$ & $>400$ & $>7$\\
%\hline
090510 & $0.9$ & $0.11$ & $100$ & $9$ & $14$ & $0.3$ & $120$ & $360$ \\
%\hline
090902B & $1.8$ & $3.6$ & $50$ & $300$ & $220$ & $30$ & $700$ & $23$\\
110731A & $2.83$ & $0.6$ & $100$ & $8$ & $\sim 5$ & $7.3$ & $550$ & $75$\\
130427A & $0.34$ & $0.78$ & $600$ & $48$ & $ \sim 40$ & $138$ & $>4300$ & $>30$\\
\hline
%\end{minipage}
\label{TAB:fermi-lat-data}
\end{tabular}
\end{center}
\caption{{\small Comparison of observed flux at 100 MeV and the expected flux 
 from external forward shock. The 4th column is time in observer frame when 
 flux at 100 MeV due to synchrotron radiation in the external forward shock 
 is calculated (which is reported in column 5) and that is compared with the 
 Fermi/LAT measurements (column 6). For the flux calculation we took the energy
 in the blast wave to be $E_{ES} = 3 E_\gamma$, and other parameters were 
 $\epsilon_e=0.2$, $\epsilon_B=10^{-5}$ \& $p=2.4$; the uncertainty in the 
 predicted flux is about a factor 2 due to the uncertainty in 
 $\epsilon_e E_{ES}$ which is the energy carried by electrons in the external 
 shock; we note that the predicted flux is independent of CSM density, and 
 scales as $\epsilon_B^{(p-2)/4} =\epsilon_B^{1/10}$ and hence is almost 
 independent of $\epsilon_B$ as long as the Fermi band lies above the 
 synchrotron cooling frequency. Burst duration in 10 keV---10 MeV band is 
 provided in the column marked $T_{MeV}$, and the time duration that
 $>$100 MeV photons were detected by Fermi/LAT is given in the 2nd last 
 column ($T_{LAT}$). Fermi/LAT lightcurves for all these bursts for 
 $T_{MeV} < t \le T_{LAT}$ show a simple power-law decline.  Considering 
 that $T_{LAT}/T_{MeV} \gg 3$ (the last column) models such as those where 
 prompt MeV photons are IC scattered by $e^\pm$s in the external medium to 
 produce these very long lasting LAT lightcurves \citep[e.g.][]{beloborodov13b}
 are ruled out. }}
\end{table*}

Furthermore, one can determine external shock parameters from early 
time ($t\sim10^2$s) Fermi data and use that to {\em predict} late time 
optical and X-ray flux. Figure \ref{FIG:predict-090902B} shows the result of 
this exercise for Fermi burst GRB 090902B; it shows the comparison 
between the predicted and the observed late time afterglow data, 
which are found to be in good agreement.

\begin{figure}
\begin{center}
\includegraphics[width=13cm]{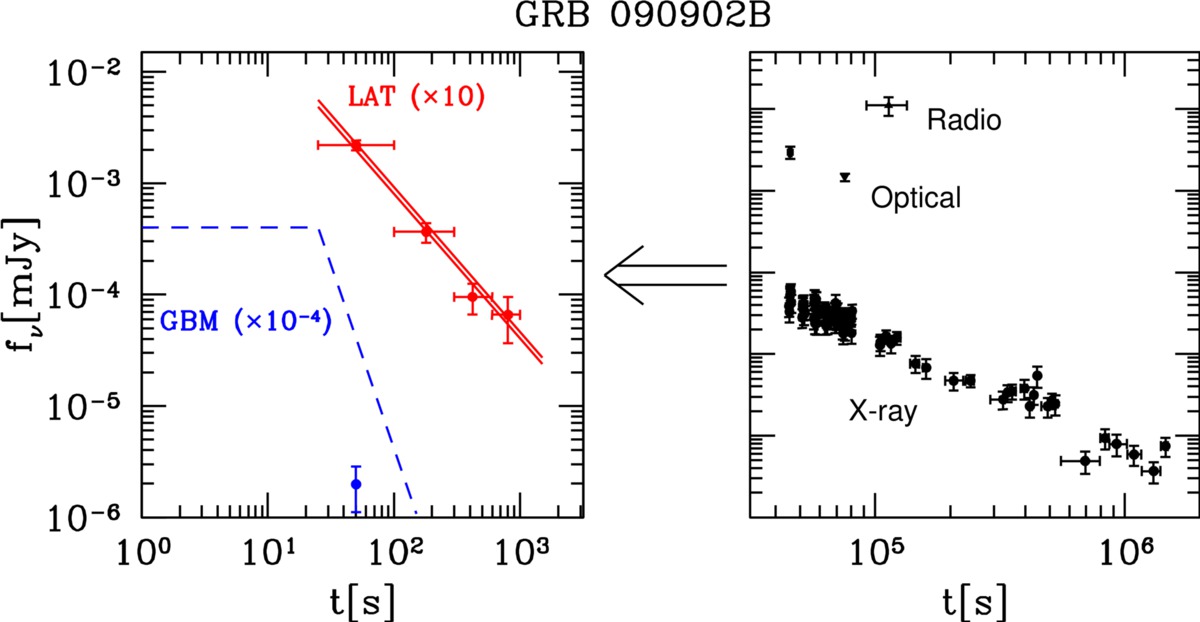}
\caption {Using the X-ray, optical and radio data of GRB 090902B at late
times (right panel) we constrain the external forward shock parameters,
and then use these parameters to predict the 100MeV flux at early times
(left panel). The region between the red lines shows the range for the
predicted flux at 100MeV; note the remarkably narrow range for
the predicted 100 MeV flux and an excellent agreement with the Fermi/LAT
data. The blue point (left panel) indicates the
flux at 100keV and 50s that we expect from the external-shock model; note
that the external-shock flux at 100 keV falls well below
the observed Fermi/GBM flux shown schematically by the dashed line in the
left panel, and that is why the GBM light curve undergoes a rapid decline
with time ($\sim t^{-3}$) at the end of the prompt burst phase.
From \cite{kumar10}.
%The radio flux is taken from van der Host et al. 2009; X-ray data (black
%circles) are from Evans et al. (2009), and optical fluxes are from
%Swenson et al. 2009 (square) and Guidorzi et al. 2009 (triangle).
}\label{FIG:090902B-reverse}
\end{center}
\end{figure}

This exercise can also be carried out in the reverse direction, i.e.
one can determine the external shock parameters from the late time
(t$\gae$0.5 day) X-ray, optical and radio data, and use these parameters
to calculate the flux at 100 MeV at early times ($t\lae10^3$s). We show in
Figure \ref{FIG:090902B-reverse} that this ``predicted flux'' is in 
excellent agreement with the data obtained by Fermi/LAT. These results 
lend strong support to the suggestion that high energy photons from GRBs 
detected by Fermi/LAT, for $t\gae30$s, are produced via the synchrotron 
process in the external shock.

One thing to point out though is that it is not easy to produce
photons with energy more than $\sim 50\Gamma$MeV$\sim$5 GeV via the 
synchrotron process as described in
\S2.2 (but see \cite{kumarhernandez12}, for a way around the maximum 
energy constraint). It is possible that the highest energy photons 
($\gae$5 GeV) detected by Fermi LAT from GRBs are produced via
IC scattering of synchrotron photons.
\cite{zhangmeszaros01b} considered a range of shock micro-physics parameters, 
and identified regimes where synchrotron and SSC dominate
in the GeV--TeV energy range.

\cite{kumar09,kumar10} found that $\epsilon_{_B}$ should be 
small\footnote{$\epsilon_{_B}\sim10^{-6}$ if the CBM particle number 
density is 0.1 cm$^{-3}$, and it is smaller for higher densities.} in the 
highly relativistic external shock for Fermi bursts in order that the flux at 
$\lae 1$ MeV produced in the external shock not exceed the observed value;
Fermi/GBM lightcurves (10 keV -- 10 MeV) fall off very rapidly 
after the prompt phase ($t^{-3}$ or faster), and so the contribution of the
forward shock flux in this band --- which declines with time as 
$\sim t^{-1.2}$ --- at the end of the prompt phase has to be well below 
the observed value in order to make it possible for the GBM lightcurve to 
fall off steeply.

It can be shown that this small magnetic field is sufficient for 
confining high energy electrons of thermal Lorentz factor
$\sim 10^8$ (that produce $\sim 10$ GeV photons), both upstream
and down-stream of the shock front, and for their efficient
acceleration by the first order Fermi mechanism as long as 
these electrons are not exposed to a large flux of a few eV photons 
($\gae10$ mJy in our frame) to cause severe IC losses 
\citep{barniolduran11,piran10}.

Measurements of $\epsilon_B$ for a large sample of GRBs and its
implications are discussed in \S\ref{shock-parameters}.

It is interesting that GeV afterglows almost always follow a 
simple power law ``normal'' decay, while only 5\% of X-ray 
afterglows are a single power law function \citep{liang09,evans09} --
most X-ray lightcurves show a steep-shallow-normal-steep decay behavior. 
Only a few GRBs have jointly triggered both
Swift/LAT and Fermi/LAT. The currently available two cases\footnote{A third 
case of GRB100728A also has simultaneous Swift/XRT and Fermi/LAT observations.
However, for this burst photons
of energy $>$10$^2$MeV were not detected during the prompt $\gamma$-ray phase
but LAT saw emission during the X-ray flares \citep{Abdo11} which perhaps
were due to IC scatterings of X-ray flare photons by electrons in the 
external shock \citep{wang06,he12}.}, i.e. GRB 090510
\citep{depasquale10} and GRB 110731A \citep{ackermann13}, both 
show GeV and X-ray lightcurves to be power law functions of 
time for almost the entire duration of observations starting at 
$\sim 5$s for Fermi/LAT and $\sim10^2$s for Swift/XRT\footnote{The X-ray
data for GRB 090510 shows a jet break at $\sim10^3$s.}. The optical,
X-ray and GeV data for these bursts are consistent with the external 
forward shock model. It would be interesting to find out whether all GRBs 
with GeV afterglows are just those rare cases that display a single power 
law decay X-ray lightcurve\footnote{If future Fermi/LAT observations find 
a burst that has a power-law decay lightcurve, but a complex X-ray afterglow
lightcurve typical for GRBs, then that would constitute yet another 
evidence that the X-ray afterglow emission for 
at least a fraction of bursts is produced not by the FS but by some 
process internal to the relativistic jet.}.

Detailed data analysis \citep{zhang11}
and theoretical modeling \citep{gao09,maxham11,he11,liu11} suggest that the
GeV emission during the prompt phase (when GBM emission is still on)
is likely not dominated by the external shock component, and that the 
external shock emission starts to dominate after the prompt phase.
This is because energy is still being added to the blastwave during
the prompt phase \citep{maxham11}, and observationally, LAT lightcurve
spikes track those in the GBM lightcurves \citep{abdo09a,zhang11} which
is very difficult to produce in external shocks \citep{sari97}.
According to recent observations, some GRBs show a steep to shallow
transition in the GeV lightcurve, which suggests that the radiation
mechanism might be switching from prompt emission to afterglow 
\citep{ackermann13a}.
%(Omodei, 2012, talk presented at ``Fall 2012 GRB Symposium'').
%\citep{omodei12}. 
When the contribution of the
early, steep, phase is subtracted from the Fermi/LAT lightcurve the temporal
slope of the remaining afterglow data is found to be ``normal'' and 
consistent with synchrotron radiation from an adiabatic external 
shock \citep{ackermann13a}
%(Omodei 2012).
%\citep{omodei12}.

\newpage

\section{Collisionless shock properties from GRB afterglow observations}
\label{shock-parameters}

GRB afterglows provide a good laboratory for the study of relativistic
collisionless shocks. In spite of many years of theoretical work several
basic questions regarding collisionless shock remain unanswered. Perhaps
foremost amongst these questions are generation of magnetic fields
down/up stream of the shock front ($\epsilon_B$), particle acceleration 
($p$) and the fraction of energy of shocked plasma that is given to 
electrons ($\epsilon_e$).
The calculation of synchrotron radiation from shocked fluid requires
these three quantities, and hence multi-wavelength GRB afterglow data
can be exploited for their measurement, and that should shed light
on the basic plasma physics of collisionless shocks.

The GRB afterglow flux at any given time is dependent on
at least four parameters when the underlying radiation mechanism is the
synchrotron process --- $E$ (energy in explosion), $n$ (CBM density),
$\epsilon_e$ and $\epsilon_B$ --- even for the simplest, spherical,
 blastwave; there is
a fifth parameter $p$ (electron distribution index) that is readily
determined from the X-ray spectrum, and so can be dropped from the
list of unknown parameters. Therefore, at least four independent 
observations are needed to
determine these four parameters. One might think that observing in four
different energy bands, e.g. radio, mm, infrared and X-ray, would provide
sufficient data to uniquely determine $E$, $n$ etc. However, this is incorrect.
Observations at two different frequencies provide independent pieces
of information only when these frequencies fall on different segments 
of the synchrotron spectrum, such as when one frequency band is below 
the synchrotron peak ($\nu_m$) and the other is above it. Or when one 
frequency band is in the synchrotron-self-absorption regime
whereas the other is not. Another way to emphasize this point is to consider
an example where someone carries out observations of GRB afterglows in
two different frequency bands, say mm and optical, for time periods of
hours and days. This entire observational effort might provide just
one independent piece of information --- equivalent to an observation
carried out at one frequency and at one single snap-shot in time ---  if the
spectrum between mm and optical frequencies for the burst is a single
power-law function for the entire time duration of the observation.
Therefore, measurement of these four different parameters uniquely is 
not possible except for a small number of GRBs that have been followed up 
for a long time in X-ray, optical and radio bands.

The value of $\epsilon_e$ is set by the micro-physics
of relativistic shocks. And if magnetic fields in the shocked fluid is
generated by the Weibel instability \citep{weibel59,medvedev99}, or another
instability based on the local physical condition of the plasma, then
$\epsilon_B$ is also determined by shock micro-physics.
Therefore, based on basic physics considerations, it is expected that
$\epsilon_e$ \& $\epsilon_B$ should be functions of those variables
that characterize a relativistic shock, viz. $E$, $n$ and $\Gamma$ (Lorentz
factor of shock front). 

\begin{figure}
\begin{center}
\includegraphics[width=12cm]{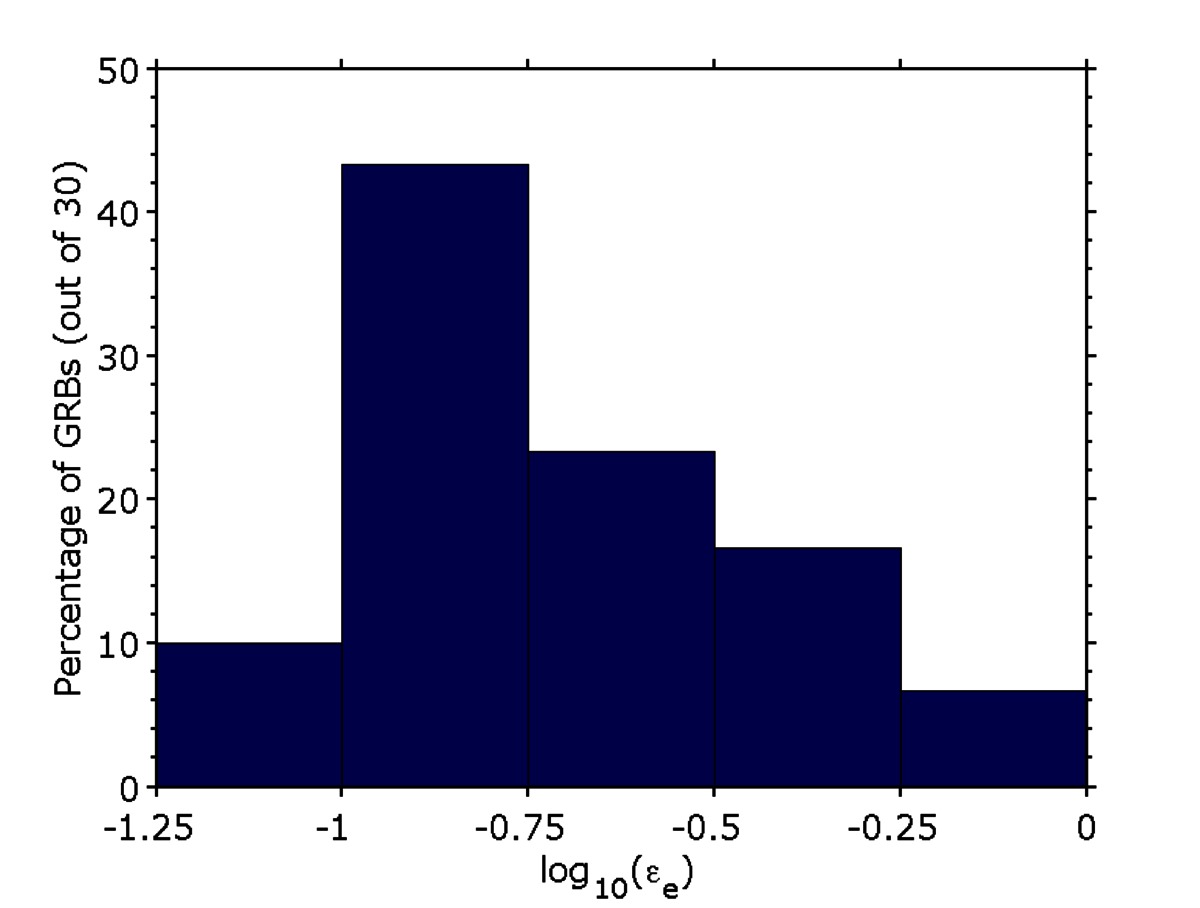}
\caption {Distribution of $\epsilon_e$ for 30 GRBs from published 
literature; \cite{berger03b,berger03a,bjornsson04,cenko10,chandra10,corsi10,
curran07,gao09,gao09a,panaitescu01,panaitescu02,rossi11,soderberg06b,
soderberg07,xu09,yost03}. This figure is taken from  Santana et al. 
(2014), ApJ 785, 29.
}\label{FIG:epsilon_e_histo}
\end{center}
\end{figure}

The afterglow flux at a frequency that lies above $\nu_m$ is proportional 
to $\epsilon_e^{p-1}$, and due to this fairly strong dependence 
$\epsilon_e$ is perhaps one of the most reliably measured parameters.
Figure \ref{FIG:epsilon_e_histo} shows $\epsilon_e$ distribution for a 
sample of 30 GRBs drawn from the published literature. Note that the 
mean value for $\epsilon_e$ for these 30 bursts is 0.2 and the dispersion 
about the mean is a factor 2; $\epsilon_e\sim 0.2$ is consistent with
recent simulations of relativistic collisionless electron-ion shocks
\citep[e.g.][]{sironi11a}\footnote{The simulations by \cite{sironi11a} also find 
the down-stream particle distribution to have a prominent thermal peak 
at electron energy of $\sim m_p\Gamma c^2$ which is not observed
in GRB spectra; where $\Gamma$ is the shock-front Lorentz factor.}. 
These bursts cover a wide range of $E$ 
and $n$. So, to the lowest approximation, $\epsilon_e$ is independent 
of shock strength, and it takes on a nearly universal value that 
varies by a factor $\sim 2$ from one burst to another. 

\begin{figure}
\begin{center}
\includegraphics[width=11.5cm]{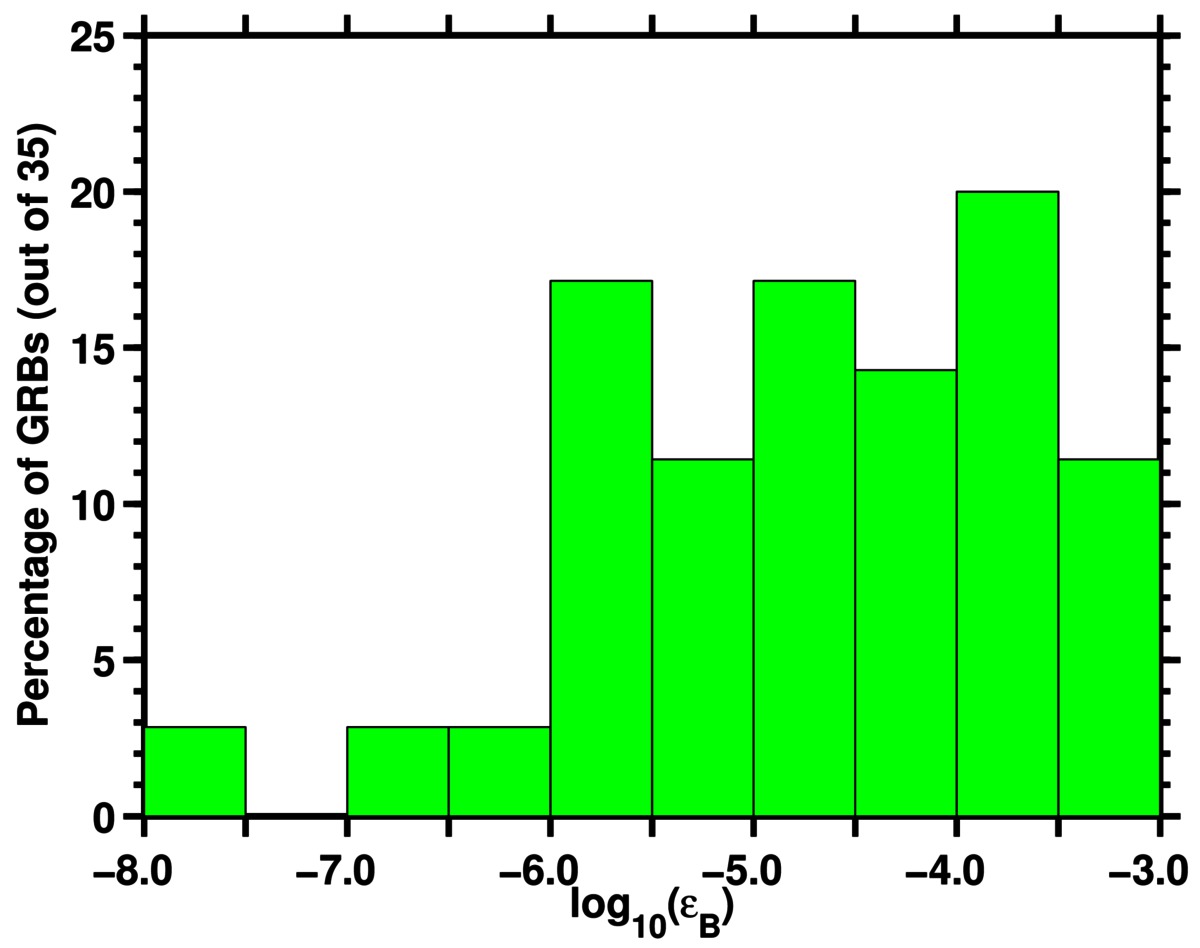}
\caption {Distribution of $\epsilon_B$ for 35 GRBs detected by the
Swift satellite;this figure is taken from  Santana et al. (2014), ApJ 785, 29.
 These $\epsilon_B$ were determined from the
optical afterglow data (see \cite{santana13} for details) 
assuming that $\epsilon_e=0.2$, $n=1$ cm$^{-3}$, and energy
in the external shock is 4 times the energy in prompt $\gamma$-ray
radiation. $p$ is determined from the temporal decay of the lightcurve.
The effect of any error in $n$, $\epsilon_e$ and $E$ on $\epsilon_B$
determination can be estimated from the relation $\epsilon_B \aprop
 E^{-1.6} \epsilon_e^{-1.6} n^{-0.6}$ \citep{santana13}.
}\label{FIG:epsilon_B_histo}
\end{center}
\end{figure}

Assuming that the radiative efficiency for producing prompt $\gamma$-ray 
emission is 20\% for GRBs (so that the energy in blast wave is 4 times 
the energy in prompt $\gamma$-rays), and $\epsilon_e=0.2$ for the external 
shock, one can find $\epsilon_B/n$ from optical afterglow data alone. 
Figure \ref{FIG:epsilon_B_histo} shows a histogram for $\epsilon_B$
for 35 GRBs detected by the Swift satellite \citep{santana13}.
This distribution is
very wide --- the median value of $\epsilon_B$ is about $3 \times 10^{-5}$,
and the distribution spans more than four orders of magnitude.

\begin{figure}
\centering
\begin{tabular}{cc}
\begin{minipage}{200pt}
\includegraphics[width=200pt]{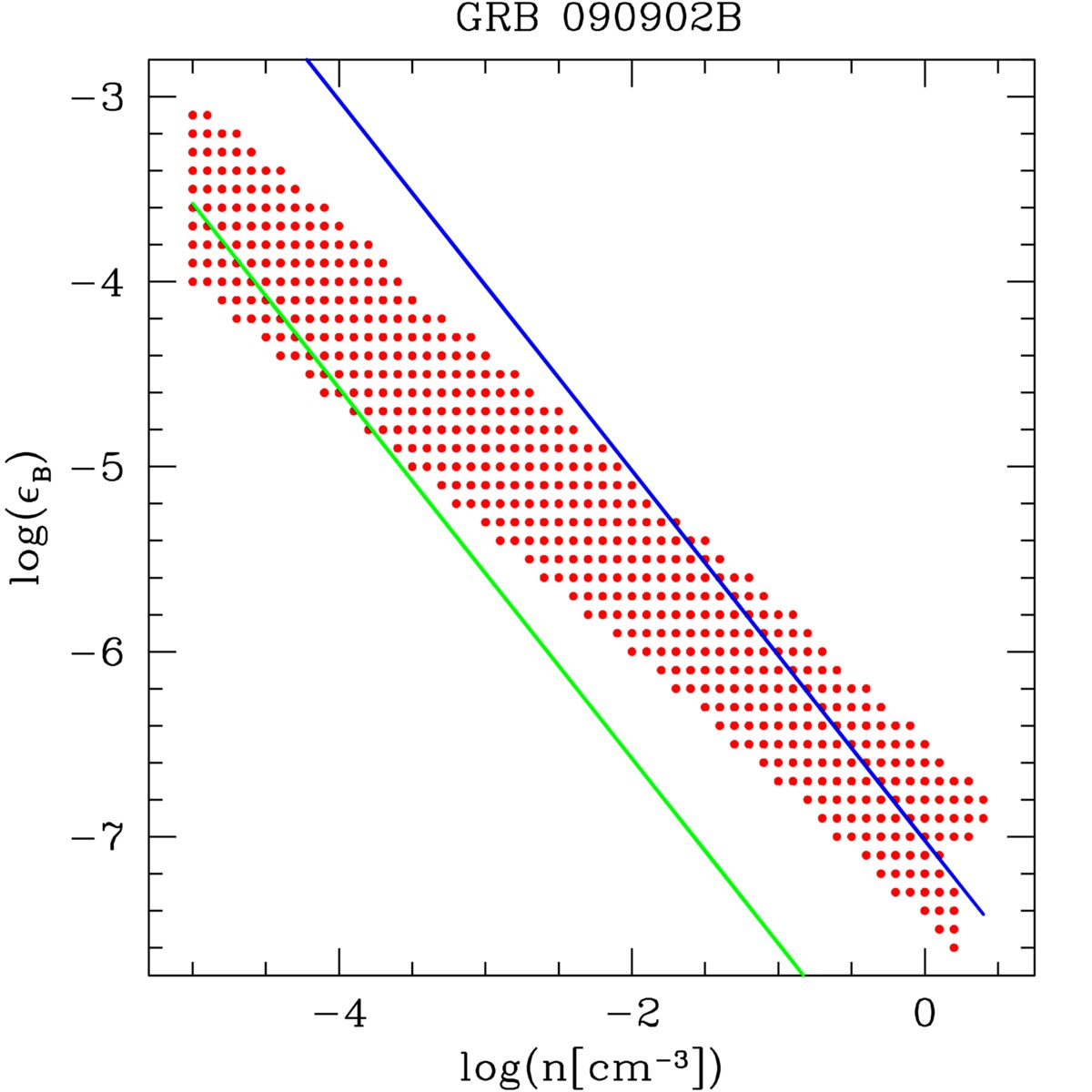}
\end{minipage}
&
\begin{minipage}{200pt}
\includegraphics[width=200pt]{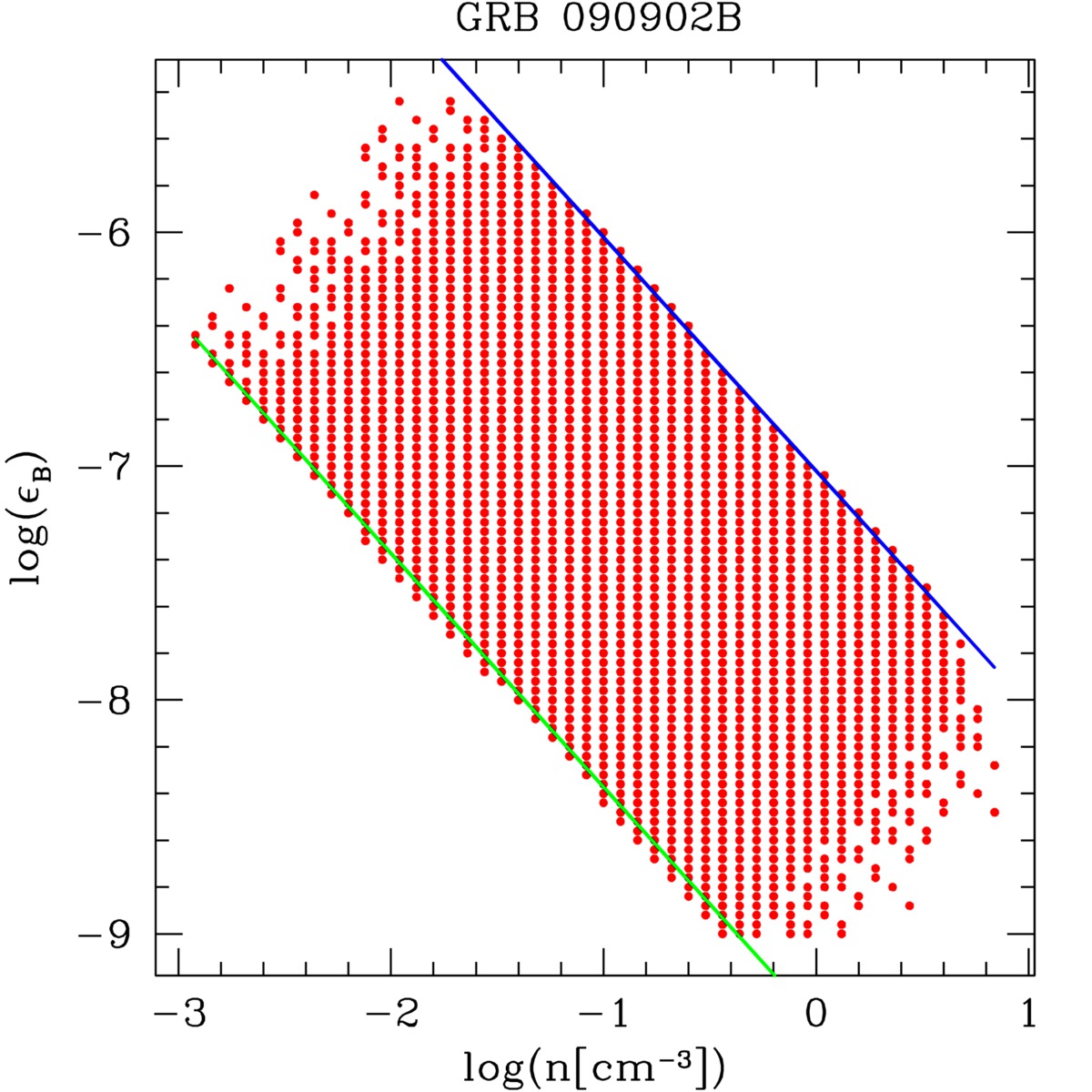}
\end{minipage}
\end{tabular}
\caption{
The {\em left panel} shows $\epsilon_B$--$n$ space (for the forward external 
forward shock going into a uniform density CBM) allowed by the high energy
data for GRB 090902B at t=50s when the shock front Lorentz factor was
$\sim 300$ (see \cite{kumar10}); the discrete points 
reflect the numerical resolution of the calculation. Also shown is the 
expected $\epsilon_B$ for a shock compressed
CBM magnetic field of 5 and 30 $\mu$-Gauss as the green and blue lines
respectively; for a CBM field of strength $B_0$, the value of $\epsilon_B$
downstream of the shock-front resulting from the shock compressed CBM
field is $\approx B_0^2/(2\pi n m_p c^2)$, where $n m_p$ is the CBM mass
density, and $c$ is the speed of light.  The {\em right panel}
shows $\epsilon_B$--$n$ space allowed by the late time ($t>0.5$day) X-ray, 
optical and radio data for GRB 090902B when the shock front Lorentz factor
had dropped to $\sim 10$. Also plotted is the expected
$\epsilon_B$ for a shock compressed CBM magnetic field of
2 and 30 $\mu$-Gauss as the green and blue lines, respectively.
}\label{FIG:epsilon_B_090902B}
\end{figure}

There is no evidence that $\epsilon_B$ depends on shock Lorentz factor.
For a couple of Fermi/LAT bursts one can determine $\epsilon_B$ from 
early time $\gamma$-ray data when the blast wave Lorentz factor was 
larger than $\sim10^2$ (left panel of Figure \ref{FIG:epsilon_B_090902B}), 
and from late time X-ray and
optical data when the Lorentz factor had dropped to $\sim 10$ (result shown 
in Fig. \ref{FIG:epsilon_B_090902B} right panel). And it is found
that the values of $\epsilon_B$ at early and late times are entirely 
consistent with each other. Collisionless shock simulations 
also find no dependence of $\epsilon_B$ on $\Gamma$, e.g. 
\cite{sironi11a}. 

A wide distribution of $\epsilon_B$, which is independent of shock 
Lorentz factor, suggests that magnetic field is unlikely to be 
determined by micro-physics of relativistic collisionless shock alone.

\begin{figure}
\begin{center}
\includegraphics[width=5in,height=4in]{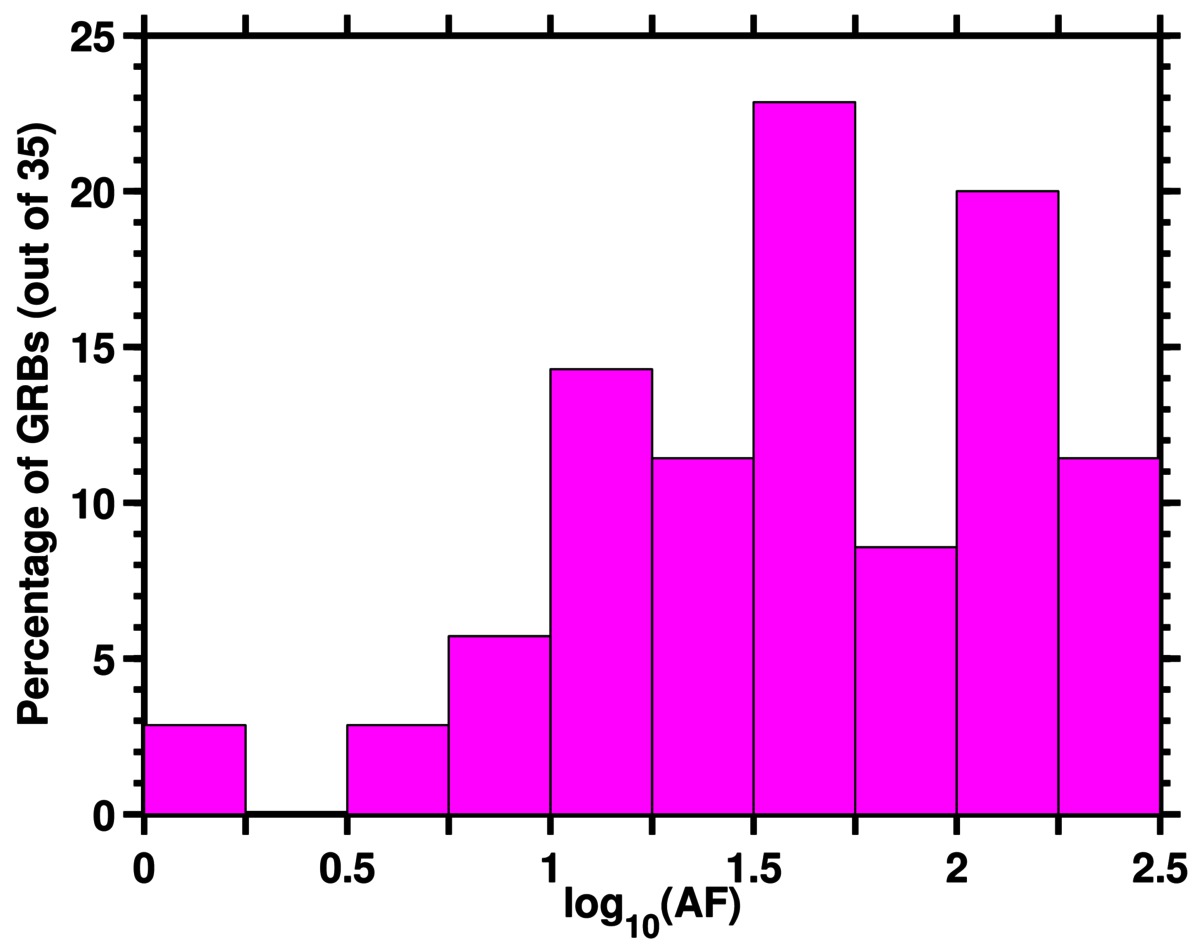}
\caption {
Results for the magnetic field amplification factor (AF) for the optical 
sample shown in Fig. \ref{FIG:epsilon_B_histo}; this figure is taken from 
Santana et al. (2014), ApJ 785, 29. The histogram shows 
results for $p$ calculated from lightcurve decay. A fixed $n = 1$ 
cm$^{−3}$ and $B_0 = 10\mu$G were assumed; $AF\aprop n^{0.2} B_0^{-1}$.
}\label{FIG:amplification_factor_magnetic-field}
\end{center}
\end{figure}

For an upstream magnetic field of strength $B_0$ (in CBM frame), the down 
stream field, due to shock compression alone, is $4B_0\Gamma$ (in shock 
comoving frame). The ratio of energy density in this shock compressed field
 and the energy density of shocked plasma is:  
\begin{equation}
 \epsilon_B^{(sc)} =  {B_0^2\over 2\pi n m_p c^2}.
\end{equation}
The factor by which magnetic field is amplified in GRB external shock
is given by, $AF = \left[\epsilon_B/\epsilon_B^{(sc)}\right]^{1/2}$. 
The amplification factor is very insensitive to the uncertain CBM density, 
$AF \aprop n^{0.2} B_0^{-1}$, since $\epsilon_B\aprop n^{-0.6}$ and 
$\epsilon_B^{(sc)} \propto n^{-1}$. Hence $B_0 \times {\rm AF}$ can be 
determined quite accurately for the
sample of 35 bursts in Fig. \ref{FIG:epsilon_B_histo}, and its distribution
is shown in Fig. \ref{FIG:amplification_factor_magnetic-field}.
The field amplification determined from the afterglow data corresponds to
the average value of magnetic field for the entire volume of the shocked 
plasma that contributes to the observed radiation.
Note that a modest amplification of CBM field, by a factor $\sim 30$, 
down-stream of shock front is all that is required by GRB afterglows;
$AF\sim 10^4$ for equipartition magnetic field.

If magnetic fields were to be generated down-stream by the
Weibel mechanism then we expect $\epsilon_B\sim 0.1$ near the shock-front
\citep{medvedev99}. This field, however, has small coherence length scale
of order the plasma skin depth, and likely decays by a large factor 
over the width of the shocked plasma which is of order $10^8$ skin depth 
for GRB external shocks. This might be the reason for the small average 
$AF$ inferred from afterglow observations; numerical simulations 
\citep{silva03,chang08,sironi11a}, and analysis of GRB afterglow
data \citep{lemoine13} support this general picture of strong field near 
the shock front and their decay down-stream\footnote{An earlier suggestion
was made by \cite{rossi03} that strong magnetic fields only pervade
a few percent of the total thickness of the shocked region. They did
not derive detailed constraints from the data. A similar suggestion was
made by \cite{peerzhang06} for internal shocks.}. 

Another possible mechanism for magnetic field generation is 
shear across the GRB-jet, or density inhomogeneity of the ISM, which
generates turbulence down-stream and leads to a modest field amplification
by about an order of magnitude 
\citep[e.g.][]{milosavljevic06a,sironi07,goodman08,couch08,inoue11}. 
In this case the 
coherence length of the field is large --- of order the shear length 
scale or the size of the system --- and such a field persists throughout 
the down-stream volume. 

\cite{lemoine13} have suggested from the analysis of X-ray and GeV 
afterglow data for four different GRBs that the turbulent magnetic field 
generated in shocks is strong 
near the shock front ($\epsilon_B\sim 10^{-2}$) where GeV photons are 
generated by the synchrotron process, and that the field decays with 
distance ($d'$) from the shock-front so that the value of $\epsilon_B$ 
further down-stream where X-rays are produced is $\sim10^{-6}$;
\cite{lemoine13} find that the X-ray data is consistent with 
$\epsilon_B\aprop d'^{-0.5}$.

The maximum photon energy detected from a burst is $\sim 94$ GeV 
(GRB 130427A), and $>1$GeV photons have been observed
by Fermi/LAT from more than 20 GRBs (\cite{ackermann13a}).
These high energy photons provide a lower limit on the upstream magnetic 
field in the external forward shock. A minimum CBM field strength is
required to ensure that high energy electrons (those that produce GeV photons 
via the synchrotron process down-stream and have LF $\sim 10^8$ in
shock comoving frame) are confined to the shock, and that these
electrons could be turned around on a short time scale while upstream 
before losing a good fraction of their energy to IC scatterings. 
\cite{barniolduran11} showed that a CBM magnetic field of $10 \mu$G 
is sufficient for accelerating electrons to an energy so that they produce 
$\sim 10$ GeV synchrotron photons.

The distribution function for electron energy, just behind the shock 
front, is a power-law function of energy with index $p$. A number of 
calculations suggest that $p$ should be about 2.2 for collisionless 
relativistic shocks independent of shock LF \citep[e.g.][]{bednarz98,
kirk00,achterberg01,lemoine03}.
This expectation is not supported by GRB afterglow spectra which
show that $p$ varies considerably from one burst to another 
\citep{shen06,curran09,curran10}. One possible way out of this discrepancy
is that $p$ calculated from X-ray afterglow spectrum has nothing to do with 
shocks; considering the complexity of X-ray lightcurves it is 
possible that the radiation is not produced in external shocks but rather 
by some other dissipative process internal to the jet 
\citep{ghisellini07,kumar08a}.
The other possibility is that something is missing in theoretical 
calculations of $p$ in relativistic shocks, and in that case the
observed distribution should guide us to the correct model.

\section{Observational properties of GRB prompt radiation} 
\label{prompt-obs}

\subsection{Temporal properties}\label{prompt-temporal}

Observationally, the prompt emission phase of a GRB is conventionally 
defined as the temporal phase during which sub-MeV emission is detected
by the GRB triggering detectors above the background level. 
Quantitatively, the duration of a burst is defined by the so-called
``$T_{90}$'': the time interval between the epochs when $5\%$ and $95\%$
of the total fluence is registered by the detector. Such an observation-based
definition has some limitations: 1. It depends on the energy band of the
detector. A detector with a lower energy bandpass typically gets a longer
$T_{90}$ for the same GRB. 2. It is sensitivity-dependent. A more sensitive
detector (e.g. due to a larger collection area) would detect a longer
duration of a same burst above the background level, and hence, has a
longer $T_{90}$. 3. Some GRBs have clearly separated emission episodes
with long quiescent gaps in between. The parameter $T_{90}$ therefore 
may over-estimate the duration of GRB central engine in these cases.
4. Physically, the emission registered within $T_{90}$ may include
contributions from different sites (e.g. internal dissipation regions
and external shocks). Modelers tend to attribute ``prompt emission''
and ``afterglow'' as emissions from the internal dissipation sites
and the external shock, respectively. Although emission during $T_{90}$
for most GRBs seems to be consistent with an internal origin, the
differentiation between an internal and an external origin of emission
is not straightforward. Throughout this review we stick to the 
observation-defined $T_{90}$ as the duration of ``prompt emission'', 
but limit ourselves to discuss internal dissipation models for 
prompt emission.

The temporal properties of GRBs may be summarized as the following:
\begin{itemize}
\item The duration $T_{90}$ ranges from milliseconds to thousands 
of seconds. The $T_{90}$ distribution includes at least two log-normal
components with a separation line around 2 seconds in the observer
frame in the BATSE energy band (25 - 350 keV) \citep{kouveliotou93}:
a long-duration class with $T_{90}$ peaking at 20-30 s, and a 
short-duration class with $T_{90}$ peaking at 0.2-0.3 s. 
Several papers have suggested that the $T_{90}$ distribution may include
a third, intermediate-duration group 
\cite[e.g.][]{mukherjee98,horvath98,hakkila03,horvath10,veres10}. However,
the recent analysis of \cite{bromberg13} finds little support for a third
class of GRBs.

Statistically, the long-duration group is ``softer'' than the short-duration
group, which means that the ratio between the photon numbers in the detector's 
low-energy and high-energy bands is larger for long GRBs than short GRBs.
  (Fig.\ref{FIG:BATSE}).
The duration distribution is energy-band-dependent and sensitivity-dependent,
so that different detectors give different distributions
\citep{kouveliotou93,sakamoto08,sakamoto11,paciesas12,zhangfw12,qin13}.
\cite{qin13} show that when breaking the Fermi bandpass to different
sub-bandpasses of the previous detectors, similar $T_{90}$ distributions
as previous detectors can be reproduced.
\begin{figure}
\includegraphics[width=13cm]{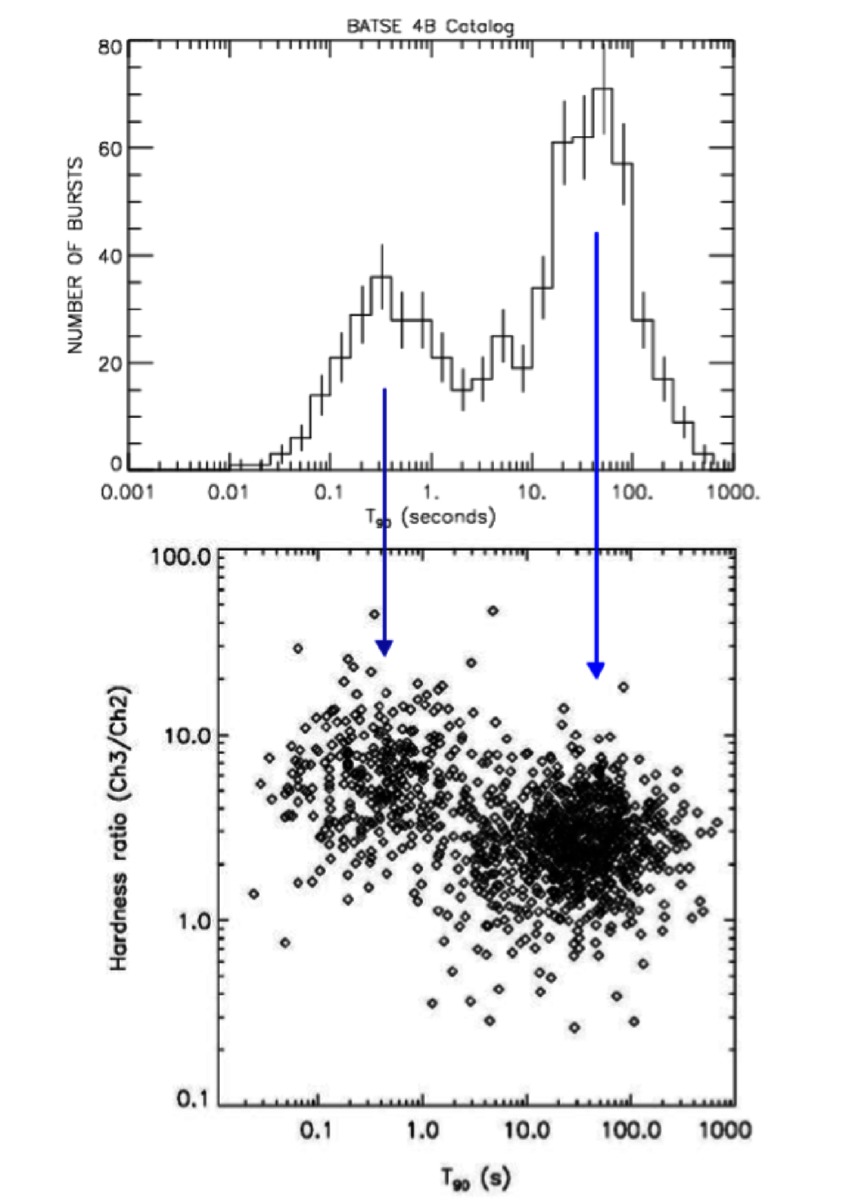}
\caption{Duration and duration/hardness ratio distribution of GRBs detected
by BATSE on board CGRO. Adapted from the BATSE GRB Catalogs 
(http://gammaray.msfc.nasa.gov/batse/grb/catalog/). 
}\label{FIG:BATSE}
\end{figure}
\item The GRB lightcurves are notoriously irregular. Some are 
extremely variable, with detectable minimum variability time scale 
reaching millisecond range, while some others have smooth lightcurves
with relatively simple temporal structures \citep{fishman95}. 
Some GRBs 
have distinct emission episodes separated by long gaps in between.
Some sample lightcurves are presented in Fig.\ref{FIG:lightcurves}.
\begin{figure}
\includegraphics[width=13cm]{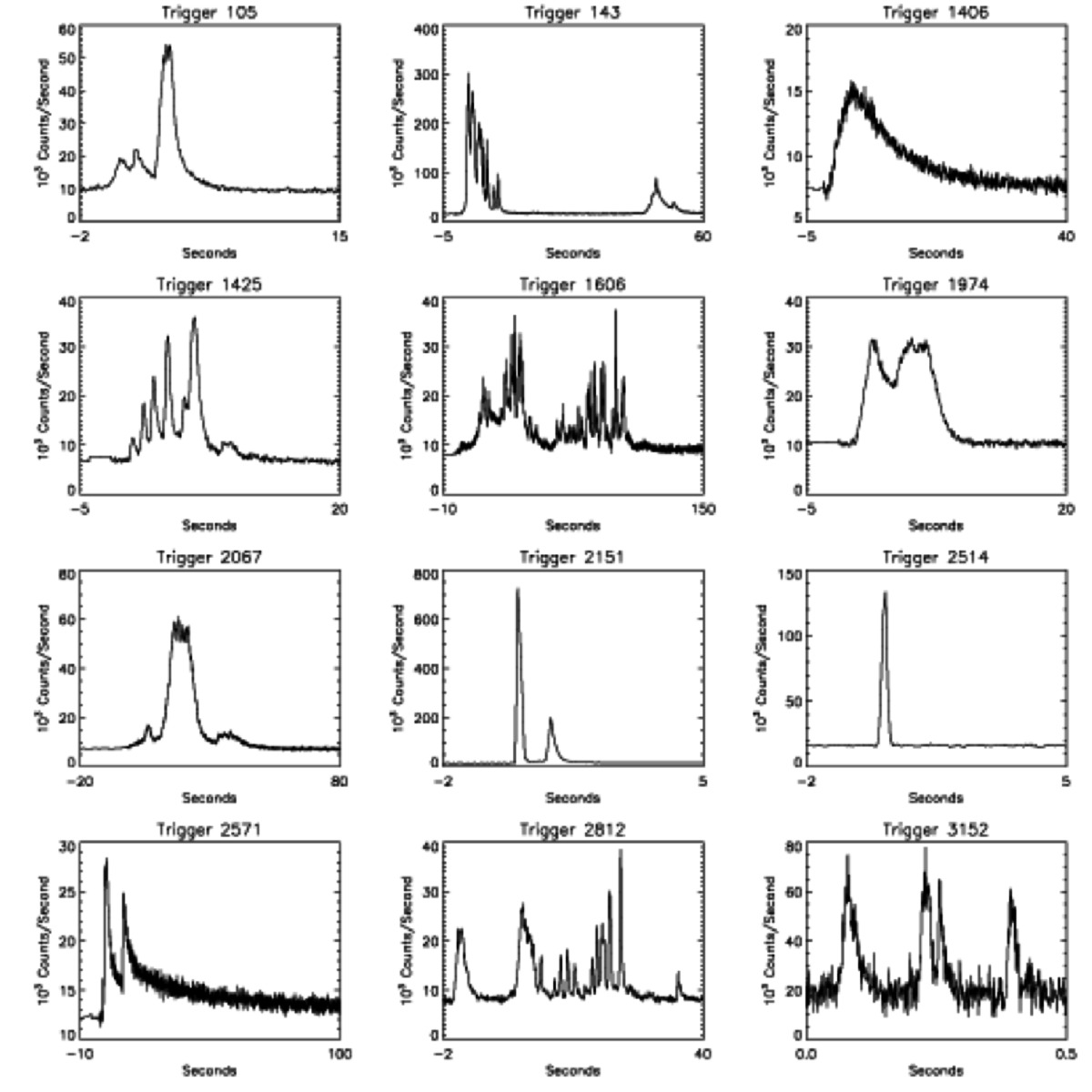}
\caption{Sample lightcurves of GRBs (Fishman \& Meegan, 1995). 
Reproduced, with permission from The Annual Review of Astronomy and
Astrophysics, Volume 33 (c) 1995, pgs 415-458 by Annual Review;
www.annualreviews.org
}\label{FIG:lightcurves}
\end{figure}
\item A fraction of GRBs have a typically
softer and weaker ``precursor'' emission 
well separated from the main burst by 10s to 100s of seconds. 
Subject to definition, the fraction of GRBs with a precursor emission
ranges from 3\% \citep{koshut95} to 12\% \citep{burlon09}. Statistical
studies suggest that the characteristics of the main episode emission
are independent of the existence of the precursor emission, and the
properties of the precursors in some GRBs are similar to those of
the main-episode emission \citep{lazzati05,burlon08,burlon09,hu14}. 
\item Power density spectrum (PDS) analysis of GRB lightcurves reveals 
null periodicity. The PDSs of individual GRBs can be noisy. However,
averaging the PDS of several bright GRBs leads to a power law with
index -5/3 and a sharp break around 1 Hz \citep{beloborodov00}.
\item There is evidence that GRB lightcurves are the superposition of
a slower component and a faster component. This is evidenced by a
gradual depletion of the fast component at low energies \citep{vetere06}, 
and the existence of a distinct low frequency component in a stepwise
low-pass filter correlation analysis \citep{gao12}. 
\item The shape of individual pulses in the lightcurves is typically
asymmetric, with a sharp rising phase and a shallower decay phase. It can 
be fit by a variety of function forms. For some bright, isolated pulses,
the pulse shape is often modeled by a ``FRED'' (fast-rising exponential-decay)
function. 
\item There are quiescent episodes during a burst. The distribution
of the separation times between pulses also satisfies a lognormal
distribution \citep[e.g.][]{mcbreen94,lidermer96,nakarpiran02b}.
\item Lightcurves vary with energy band. Pulses tend to be narrower
in harder bands (e.g. Fig.\ref{FIG:080916C}, Fig.\ref{FIG:090902B}).
The width $w$ of individual pulses is a function of energy $E$:
$w(E) \propto E^{-\alpha}$ with $\alpha \sim 0.3$ to 0.4 
\citep{norris05,liang06}.
\begin{figure}
\includegraphics[width=13cm]{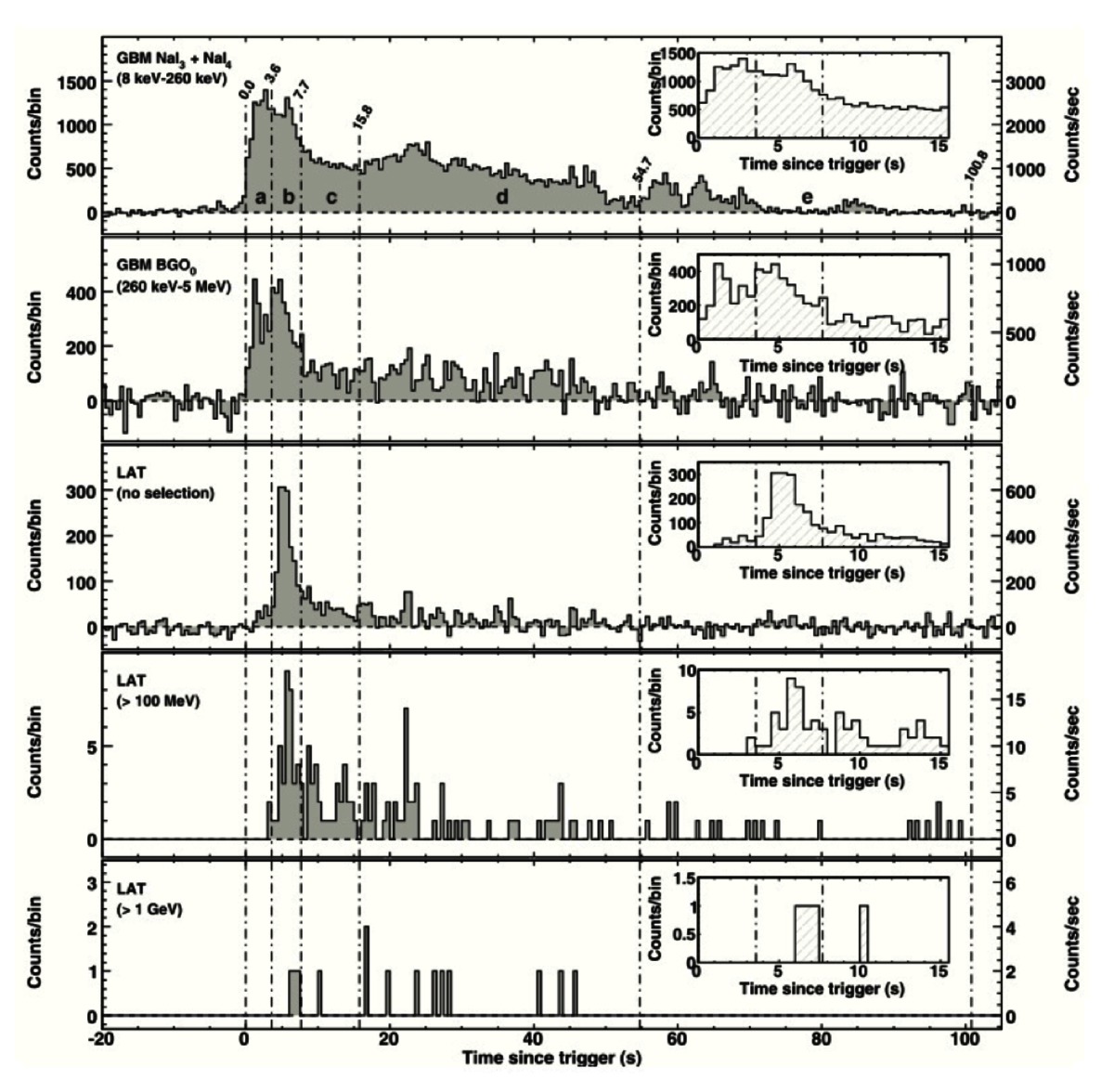}
\caption{Multi-wavelength lightcurves of GRB 080916C as detected by
Fermi. From \cite{abdo09a}.
}\label{FIG:080916C}
\end{figure}

\begin{figure}
\includegraphics[width=13cm]{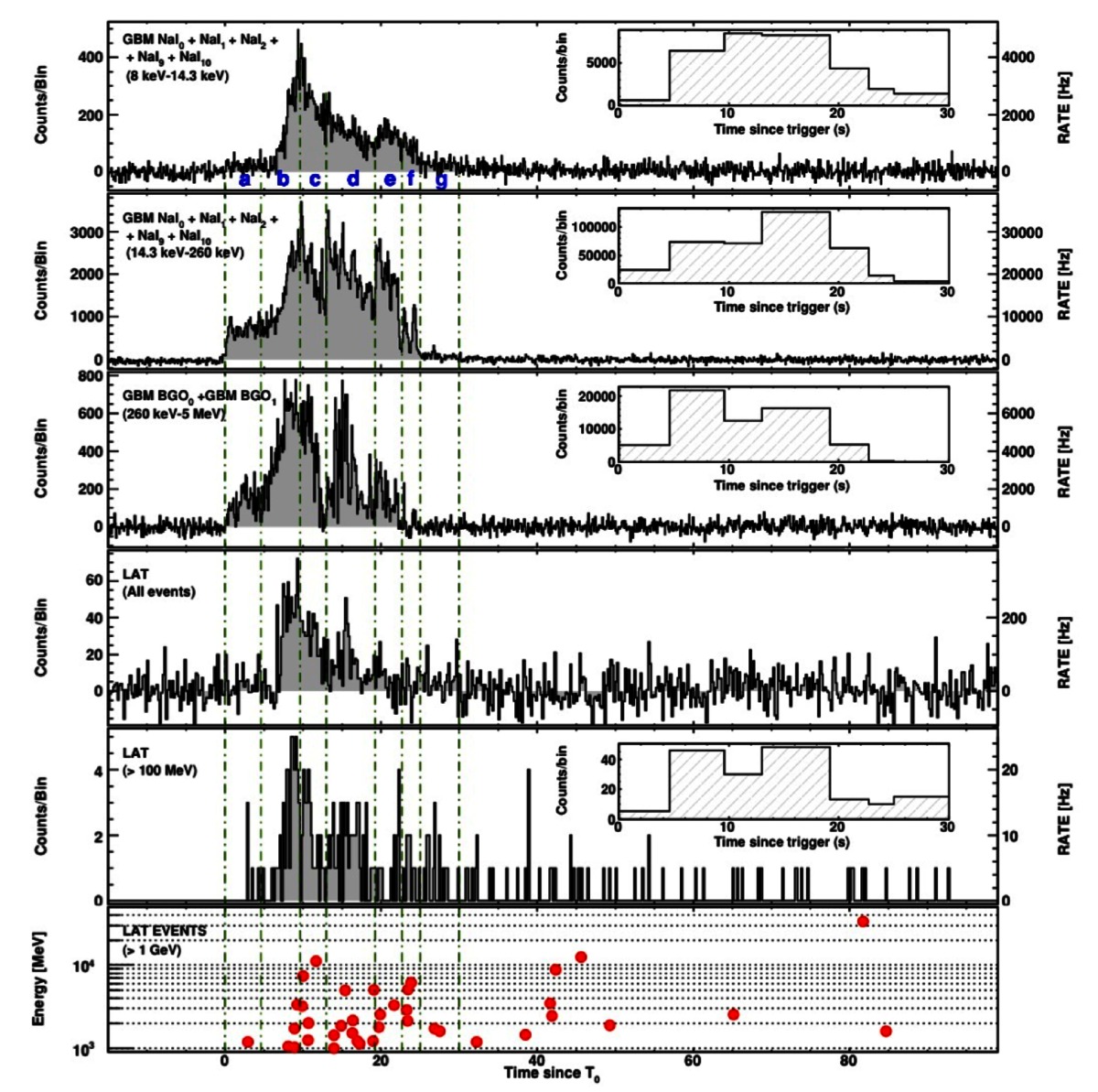}
\caption{Multi-wavelength lightcurves of GRB 090902B as detected by
Fermi. From \cite{abdo09b}.
}\label{FIG:090902B}
\end{figure}
\item A ``spectral lag'', namely, pulses with a lower energy 
being systematically lagged behind those with a high energy, 
is observed in the keV - MeV regime for many long GRBs \citep{norris00,norris02,norris05}. 
Short GRBs do not show significant spectral lags \citep{norris06}. A 
fraction of short GRBs show ``negative'' lags, i.e. high energy pulses
are lagged behind low energy pulses \citep{yi06}. 
\end{itemize}

\subsection{Spectral properties}\label{prompt-spectral}

\subsubsection{Spectral shapes and functions}

The GRB spectra are non-thermal. Spectra are often extracted over the
entire duration of the bursts. This is the time integrated spectrum 
of a GRB. Strong
spectral evolution in some GRBs is observed. Therefore time resolved
spectral information is more essential to understand GRB physics.
Technically, the time bin size cannot be infinitely small, which is 
limited by the requirement that
there are enough photons within each time bin to allow reasonable
spectral fitting to test several plausible spectral models. Therefore,
a time-resolved spectral analysis can be carried out only for bright
GRBs.

\begin{figure}
\includegraphics[width=13cm]{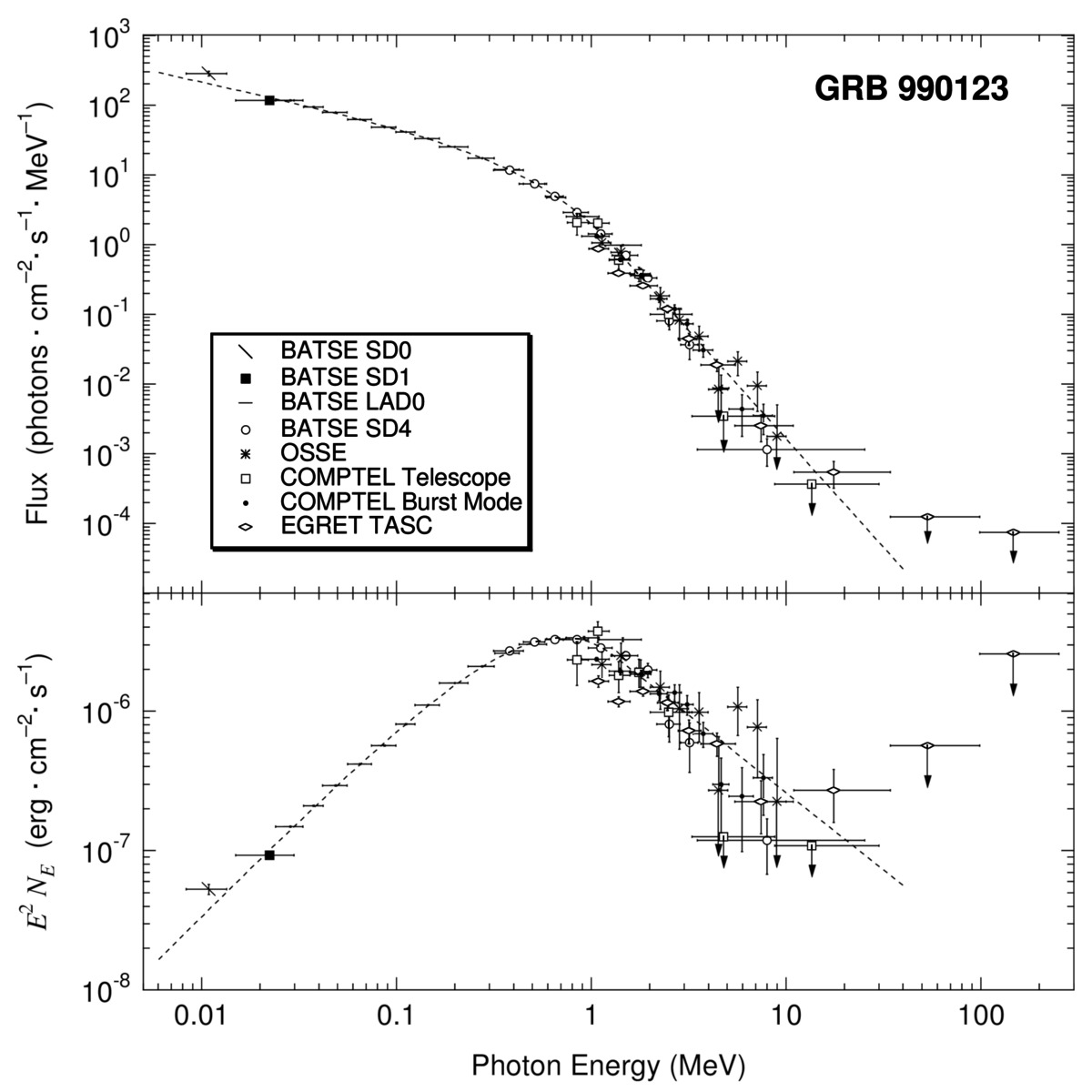}
\caption{A typical Band-function spectrum of GRB 990123. From Briggs
et al. (1999). 
}\label{FIG:Band}
\end{figure}

When the detector's energy band is wide enough, a typical GRB spectrum 
can be fit with a smoothly-joined broken power law known as the 
``Band-function'' \citep{band93}. The photon number spectrum in this
model reads
\begin{equation}
N(E)=\left\{
\begin{array}{l@{\quad \quad\quad}l}
A \left(\frac{E}{100~{\rm keV}}\right)^{\alpha} \exp \left(-\frac{E}{E_0}\right)~,
& E < (\alpha-\beta) E_0~, \\
A \left[\frac{(\alpha-\beta) E_0}{100~{\rm keV}}\right]^{\alpha-\beta} \exp(\beta-\alpha) \left( \frac{E}{100~{\rm keV}}\right)^{\beta}~, & E \geq (\alpha-\beta) E_0~,
\end{array}
\right.
\end{equation}
where $N(E)dE$ is the number of photons in the energy bin $dE$, 
$\alpha$ and $\beta$ (both negative) are the photon spectral 
indices\footnote{Within the GRB afterglow context, the notation $\alpha$
and $\beta$ are also used to define the temporal decay index and flux density
spectral index of the afterglow, with the convention $F_\nu \propto t^{-\alpha}
\nu^{-\beta}$. In this review, we do not differentiate these notations
and keep the convention in the community, but just alert the readers to
pay attention to the possible confusion. The physical meaning of these
notations are usually self-evident within the context of the review.} 
below and above the break energy $E_0$. 
The flux density spectrum ($F_\nu$) usually used in low-energy (optical,
IR, and radio) astronomy 
corresponds to $EN(E)$, and the spectral energy distribution (SED)
corresponds to $E^2N(E)$ or $\nu F_\nu$. The peak of the $E^2 N(E)$ 
spectrum is called the ``E peak'', which is given by
\begin{equation}
E_p = (2+\alpha) E_0~.
\end{equation}
Figure \ref{FIG:Band} gives an example of GRB 990123 whose time integrated
spectrum is well fit by the Band function \citep{briggs99}.

The $E_p$ distribution of GRBs is wide. While bright BATSE GRBs (a sample
of 156 bursts with 5500 spectra) have
$E_p$ clustered around 200-300 keV range \citep{preece00}, lower $E_p$
bursts are found by softer detectors such as {\em HETE-2} and {\em Swift}.
The distribution of $E_p$ seems to form a continuum from several keV to
the MeV range \cite[e.g.][]{bosnjak13}. From hard to soft, bursts are 
sometimes also vaguely
classified as gamma-ray bursts (GRBs, $E_p > 50$ keV), X-ray rich GRBs 
(XRGRBs, $30~{\rm keV} < E_p < 50~{\rm keV}$), and X-ray flashes (XRFs,
$E_p < 30$ keV), with no clear boundaries in between \citep{sakamoto08a}.
For the bright BATSE sample, the two spectral indices have a 
distribution of $\alpha \sim -1 \pm 1$ and $\beta \sim -2^{+1}_{-2}$
\citep{preece00}. Such a distribution is also confirmed for the {\em Fermi}
and INTEGRAL bursts \citep{zhang11,nava11,bosnjak13}.

Spectra for some GRBs can be fitted with a cutoff power-law spectrum, in 
the form
\begin{equation}
N(E)=
A \left(\frac{E}{100~{\rm keV}}\right)^{-\hat\Gamma} \exp 
\left(-\frac{E}{E_{c}}\right)
\end{equation}
This is essentially the first portion of the Band-function, with 
$\alpha$ replaced by $-\hat\Gamma$ ($\hat\Gamma$ is positive). This function 
has been used to fit the prompt spectrum of many {\em HETE-2}, {\em Swift},
and GBM GRBs \citep{sakamoto05,sakamoto08,paciesas12}. 
However, this is mainly due to the narrow
bandpass of the detectors, so that the high energy photon index $\beta$
of the Band-function is not well-constrained. In fact, in most cases when 
a {\em Swift} burst was co-detected by another detector with high-energy
band coverage (e.g. {\em Konus}-Wind, {\em Fermi}-GBM), the 
global spectrum can be still fit by a Band function. 

\begin{figure}
\includegraphics[width=11cm,angle=-90]{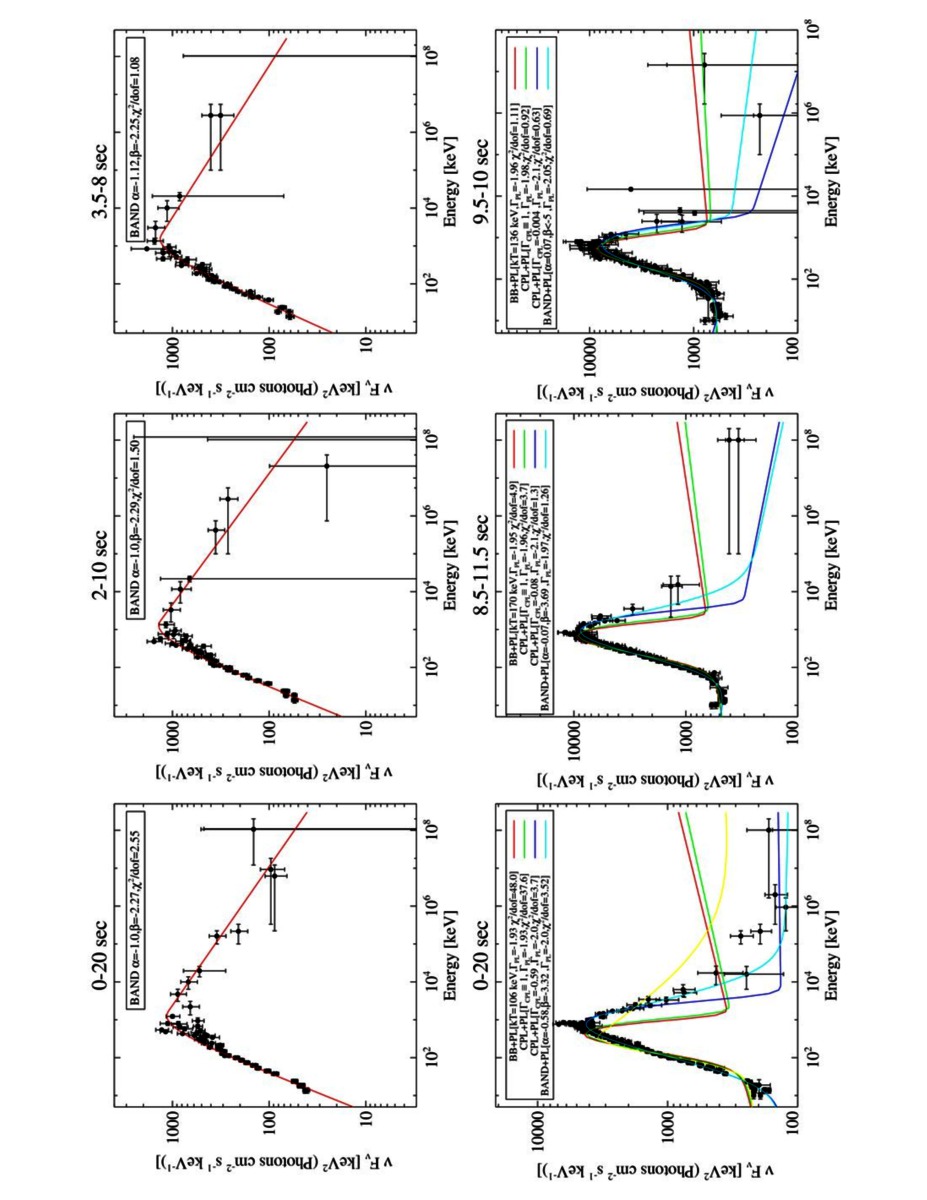}
\caption{Comparison between GRB 080916C that shows no evidence of spectral
narrowing with reducing time bin, and GRB 090902B that shows clear spectral 
narrowing with reducing time bin. From \cite{zhang11}.
}\label{FIG:080916Cvs090902B}
\end{figure}

In the pre-{\em Fermi} era, it was suggested \citep{ryde04,ryde09}
that the observed prompt GRB spectrum is the superposition of a thermal 
(blackbody) component and a non-thermal (power law) component. The traditional
$E_p$ is interpreted as the peak of the thermal component in this model. 
The spectra of some BATSE GRBs could be fit with such a ``hybrid'' model, 
which within the BATSE window may mimic a Band-like spectrum. This model 
however over-predicts the flux in the X-ray range for most GRBs, which 
violates the observational constraints by Beppo-SAX for some BATSE bursts 
\citep{ghirlanda07,frontera13}. A spectral break below the gamma-ray band is needed 
for such a model. {\em Fermi}, with both GBM and LAT on board, significantly
extended the observational spectral window. It is clear now that there are
(at least) two types of prompt emission spectra. The first type, exemplified
by GRB 080916C, has a Band component covering 6-7 orders of magnitude 
\citep{abdo09a}. There is essentially no evidence of superposition between 
a thermal and non-thermal component\footnote{A latest study by S. Guiriec
et al. (2014, in preparation) claims that there is a thermal component
in GRB 080916C. In any case, the amplitude of the blackbody component
is low, which requires significant suppression due to a Poynting flux
dominated flow \citep{zhangpeer09}.}. A second type --- the prototype of which 
is GRB 090902B \citep{abdo09c} --- shows superposition of a thermal-like
spectrum and a non-thermal power law spectrum extending both to low-
and high-energy regimes \citep{ryde10,zhang11}. The difference between
the two types becomes evident when one zooms in the lightcurve and
study the time-resolved spectra \citep{zhang11} (Fig.\ref{FIG:080916Cvs090902B}). 
GRB 080916C shows 
Band-function spectra with essentially no change of spectral indices
as one reduces the time bin. GRB 090902B, on the other hand, shows 
narrowing of the Band component as one goes to smaller time bins, and
eventually can be fit with a quasi-thermal component superposed with
a power law component. 
A systematic analysis of
17 Fermi/LAT GRBs suggest that the first type is very common (14/17),
while the second type is relatively rare (2/17) \citep{zhang11}. 

\begin{figure}
\includegraphics[width=13cm]{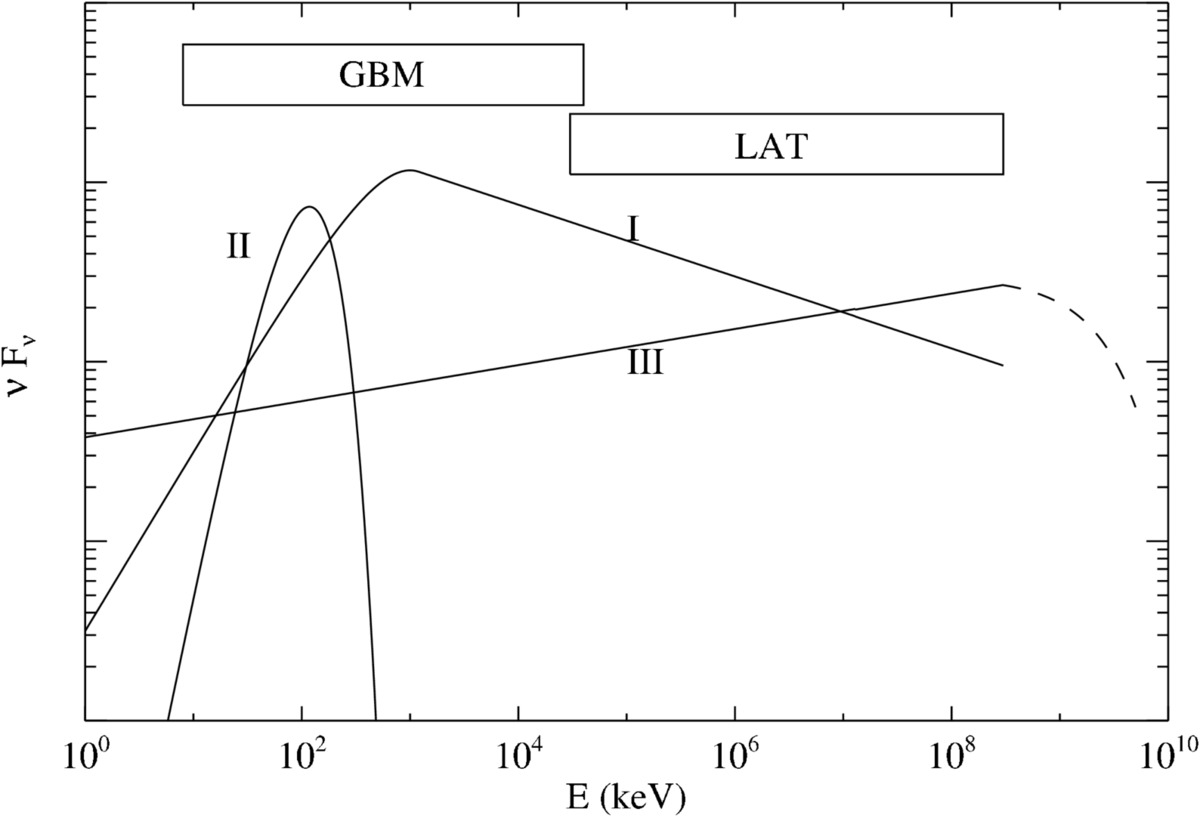}
\caption{The three possible elemental spectrum components that shape the 
observed time-resolved spectra of GRBs. Some components can be suppressed
in some GRBs. Adapted from Zhang et al. (2011).
}\label{FIG:GRBspectrum}
\end{figure}

A synthesized prompt emission spectrum may include three components
\citep{zhang11}:
(I) a non-thermal ``Band'' component; (II) a quasi-thermal component;
and (III) another non-thermal component that can be fit as a power law
extending to high energies (Fig. \ref{FIG:GRBspectrum}). This last component may 
have been detected
in the EGRET burst GRB 941017 \citep{gonzalez03}, and has been clearly 
detected in GRB 090510 and GRB 090902B \citep{abdo09c,ackermann10,zhang11}.
Another {\em Fermi} burst GRB 090926A \citep{ackermann11,zhang11} shows 
late emergence of a high energy component with a potential high
energy cutoff \citep{ackermann11}, which might have the same origin
as the component III. 
The superposition of the first two components (I and II) have
been seen in several GRBs: 100724B \citep{guiriec11}, 110721A 
\citep{axelsson12}, and 120323A \citep{guiriec13}. In all these
cases, the quasi-thermal component is sub-dominant. 
A tentative correlation between the peak energies of the thermal
and non-thermal components was reported \citep{burgess14}.

It is interesting to note that at least some low-luminosity GRBs seem to 
have a somewhat different prompt emission spectrum. An intrinsic cutoff 
power law spectrum is found to correctly 
describe the joint {\em Swift} BAT/XRT prompt emission spectra of 
the low-luminosity GRB 060218
\citep{campana06}.  The $E_p$ of this burst rapidly evolves with time 
from $\sim 80$ keV to 5 keV, with an exponential tail or very steep
power law above $E_p$.  Since GRB 060218 is special in many aspects 
(e.g. nearby, low luminosity, supernova association, extremely long duration,
existence of a thermal component that might be related to shock breakout),
the prompt emission of this burst (and probably also of other nearby 
low-luminosity GRBs) may have a different emission mechanism from the most 
high-luminosity GRBs \citep[e.g.][]{wang07}.

\subsubsection{Spectral evolution}

For bright bursts, time resolved spectral analysis give more clues
about GRB prompt emission. We summarize several interesting features:

\begin{itemize}
\item Regarding the correlation between $E_p$ and flux, it is
found that in general there are two types of behaviors of GRB pulses. 
The first type shows a pattern of ``hard-to-soft'' evolution, 
which means that $E_p$ is decreasing from the very beginning of
the pulse (even during the rising phase of the pulse) 
\citep{norris86}. The second type shows a ``tracking'' behavior:
spectral hardness well tracks intensity ($E_p$ increases during
the rising phase of the pulse) \citep{golenetskii83}. Observationally,
both types of behavior can be seen in a same burst \citep{lu10,lu12},
but see \cite{ghirlanda11}.
Considering a superposition effect, it is suggested that all pulses
are consistent with having a ``hard-to-soft'' evolution'' \citep{hakkila11}.
\item In some bursts, e.g. GRB 080916C \citep{abdo09a,zhang11}, there
exists a trend of ``opening'' of the ``Band'' spectra. Initially, the
spectrum is narrow with a relatively large $\alpha$ and a relatively
small $\beta$. However, this behavior is not representative
in the 17 LAT GRB sample \citep{zhang11}. Most bursts do not show a
clear pattern of evolution trend.
\item A good fraction (but not all) of LAT GRBs show a delayed onset of GeV 
emission with respect to MeV emission as shown in Figures \ref{FIG:080916C}
\& \ref{FIG:090902B} \citep{abdo09a,abdo09c,ackermann10,zhang11}. 
For GRB 080916C, the delayed onset may be related to hardening of 
$\beta$ or the existence of a spectral cutoff early on \citep{zhang11}. 
For GRB 090902B and 090510, it may be related to the delayed onset
of the power law component extending to high energies (component III). 
Several models have been suggested for the delayed onset of GeV emission 
\citep{kumar09,toma09,razzaque10,zhang11,li10,barniolduran11,meszarosrees11,asano12,bosnjak12},
but it is unclear which of these mechanisms operates in GRBs.
\end{itemize}

\subsection{Broad-band prompt emission}

During the prompt phase, it is believed that there should be emission
outside the triggering detectors' bandpass window. Observationally it
is very challenging to obtain a broad-band prompt emission spectrum.
Nonetheless, current observations revealed a sparse picture.

In the high energy regime, Fermi/LAT observations so far suggest that 
most GRBs do not have significant emission beyond 100 MeV
\citep[e.g.][]{beniamini11,ackermann12}. Their
prompt spectra are consistent with the extension of a Band-function
spectrum to the GeV regime \citep{zhang11}, sometimes with a possible
spectral cutoff \citep{ackermann11}. On the other hand, occasionally one
does have bursts with a second component extending to high energies 
\citep[e.g. GRB 090902B and GRB 090510,][]{abdo09b,ackermann10,zhang11}. 
The hard component of these
GRBs have a rising slope in their spectral energy distribution,
suggesting that there could be more energy emitted above the LAT
band. These sources can be ideal targets for ground-based 100 GeV - 
TeV detectors \citep[e.g.][]{kakuwa12,inoue13}.

In the low energy regime, broad-band (optical to sub-MeV gamma-ray)
spectra are available for several GRBs that had a precursor or a
very long duration. Swift XRT and UVOT were able to slew to the
source before the main burst arrives. Examples include GRB 060124
\citep{romano06}, GRB 060218 \citep{campana06}, and GRB
061121 \citep{page07}. For some other bursts, 
early optical observations were carried out by ground-based 
robotic telescopes during the prompt phase, which revealed 
interesting features. Examples include GRB 990123 \citep{akerlof99},
GRB 041219A \citep{blake05,vestrand05}, GRB 050820 \citep{vestrand06},
GRB 080319B \citep{racusin08,beskin10}, and GRB 110205A
\citep{zheng12,cucchiara11b,gendre12}.

So far no burst, with the exception of GRB 130427A \citep{perley13},
 has been simultaneously detected from optical all the way to GeV during 
the prompt phase.

\begin{figure}
\begin{center}
\includegraphics[width=10.7cm,angle=90]{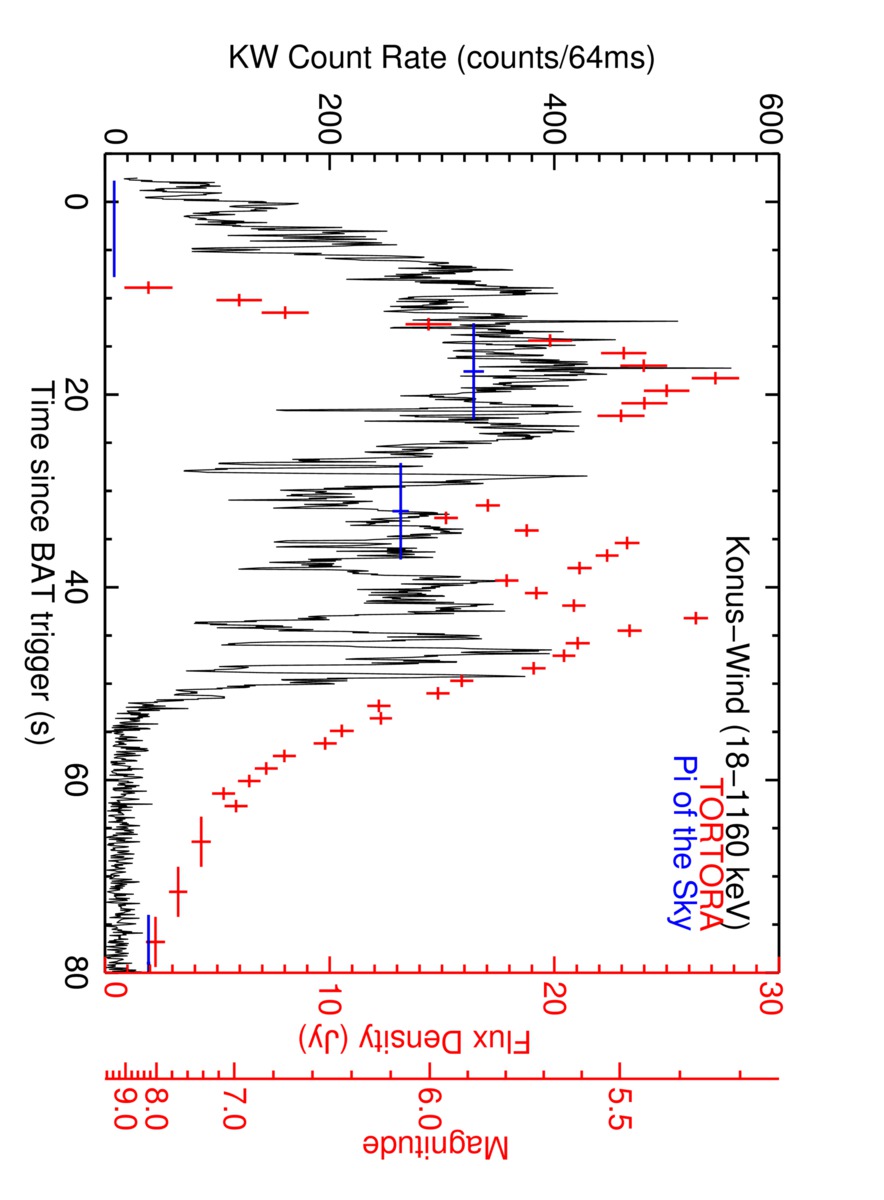}
\includegraphics[width=10.7cm,angle=90]{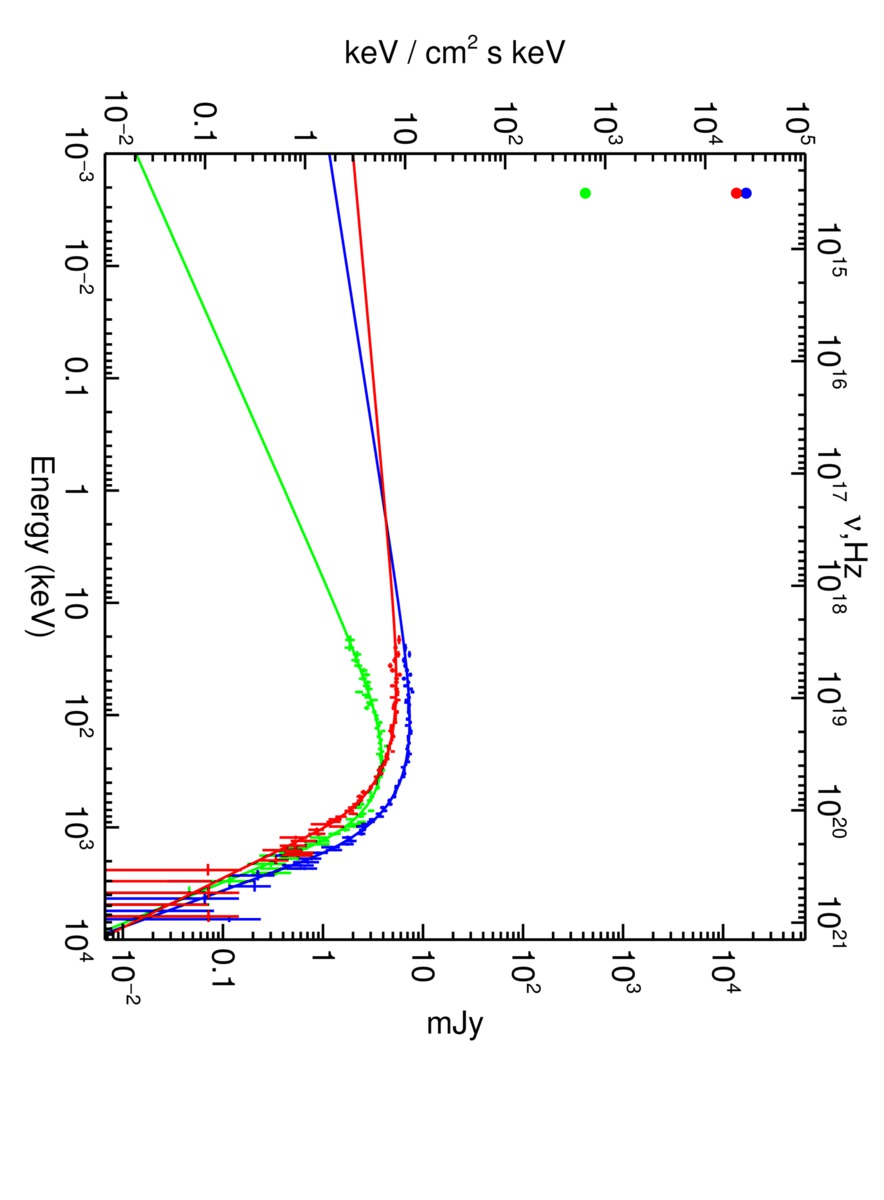}
\caption{Prompt optical and $\gamma$-ray lightcurves and spectra of 
the ``naked-eye'' GRB 080319B. From \cite{racusin08}.
}\label{FIG:080319B}
\end{center}
\end{figure}

Regarding the relation between the prompt optical and gamma-ray
emissions, there are at least three patterns. The first pattern
shows a clear offset between the optical flux peak and gamma-ray
flux peaks. An example is GRB 990123, which showed an optical
peak after all the gamma-ray peaks \citep{akerlof99}. This suggests
different physical origins of the two components. The standard
interpretation is that gamma-rays come from the internal dissipation
region (internal shocks or magnetic dissipation), while optical
comes from the external reverse shock during the early deceleration
of the ejecta by the ambient medium \citep{saripiran99,meszarosrees99}.
The second pattern shows a tracking behavior between the optical
and gamma-ray lightcurves. It was seen in GRB 041219B with sparse
time resolution in the optical data \citep{vestrand05}, and in the 
``naked-eye'' GRB 080319B with high-quality optical and gamma-ray 
data \citep{racusin08,beskin10}  -- see Fig.\ref{FIG:080319B}.
Spectroscopically, although the
optical fluxes are consistent with spectral extension of the 
gamma/X-ray fluxes in GRB 041219B \citep{shen09}, the optical
fluxes in GRB 080319B clearly stand above the spectral extension
of the gamma/X-ray fluxes, suggesting a distinct origin 
\citep{racusin08}. Leading models include attributing optical
and gamma-ray emission to synchrotron and synchrotron self-Compton
emission components, respectively \citep{kumarpanaitescu08,racusin08}, 
invoking two different emission sites \citep{zou09,fan09}, or
two (reverse and forward) shocks in a pair of internal shocks
\citep{meszarosrees99,yu09}. The third pattern shows a mix of both (offset
and tracking) components, as evidenced in GRB 050820 \citep{vestrand06}
and GRB 110205A \citep{zheng12}. Multiple emission
sites have to be invoked to generate these components.

\subsection{Polarization}

Several claims have been made suggesting that the prompt $\gamma$-ray 
emission is linearly polarized with a large degree of polarization.
An analysis of RHESSI data of GRB 021206 suggested a polarization
degree $\Pi = (80\pm 20)\%$ \citep{coburn03}, but the conclusion was
refuted by an independent study \citep{rutledge04}. 
Using the BATSE Albedo Polarimetry System (BAPS) data, 
\cite{willis05} claimed of linear polarization degree 
$\Pi > 35\%$ and $\Pi > 50\%$ for GRB 930131 and GRB 960924,
respectively. Two analysis of the INTEGRAL data of GRB 041219A
led to evidence of linear polarization, but the significance is
only marginal \citep{kalemci07,mcglynn07}. Recently, \cite{yonetoku11}
claimed detection of $\Pi = (27\pm 11)\%$ with $2.9\sigma$ significance
during the prompt emission of GRB 100826A using a GRB polarimeter on 
board a small solar-power-sail demonstrator IKAROS. They also
reported strong polarization for two other bright GRBs \citep{yonetoku12}.

Early polarization measurements were made for a handful of bursts in the 
optical band. Using a ring polarimeter on the robotic Liverpool
Telescope, \cite{mundell07} placed a $2\sigma$ upper limit on $\Pi$ of 
$8\%$ for GRB 060418 at 203 s after trigger. The epoch coincides
with the peak of the forward shock emission. The non-detection is consistent
with the theoretical expectation, since the shocked ambient medium
is not expected to carry significant ordered magnetic fields. 
Observation of another burst GRB 090102 by the same group 
\citep{steele09} revealed a $\Pi=(10\pm 1)\%$ polarization
around 160 s after trigger, and \cite{uehara12} report a polarization
of $10.4\pm2.5$\% for GRB 091208B between 149s \& 706s 
after the burst trigger. These measurements suggest a possibly ordered 
magnetic field configuration. The polarization measurement for GRB 090102
 was during the phase with relatively steep decay ($F \propto t^{-\alpha}$
with $\alpha=1.50 \pm 0.06$) before breaking to a more normal decay phase
($\alpha = 0.97\pm 0.03$) after around 1000 s. 
The steep decay phase of the optical lightcurve is believed to be 
powered by the reverse shock heating of GRB ejecta, which could carry
an ordered magnetic field, and that could account for the polarization
measurement of \cite{steele09}. The polarization measurement for GRB 091208B,
however, was carried out during the phase when the lightcurve decayed with 
$\alpha=0.75 \pm 0.02$ and that suggests that the optical emission was
probably produced in the external forward shock.  Optical polarization for 
GRB 091208B has important implications for the generation of magnetic 
fields in relativistic shocks if FS origin is confirmed, e.g. \cite{uehara12}.
Recently, \cite{mundell13} reported evolution (decrease) of linear polarization
degree in the early optical afterglow of GRB 120308A. This is consistent
with the theoretical expectation of an early RS-dominated (with ordered
magnetic field in the emission region) lightcurve which makes a
transition to a FS-dominated lightcurve (which has a much lower polarization 
degree).  \cite{wiersema14}
reported a circular polarization signature 0.15 days after GRB 121024,
whose physical origin is unclear.

So far, no optical polarization
observation is carried out during the prompt emission phase.

\subsection{Isotropic luminosity function}

The bolometric isotropic $\gamma$-ray energy of GRBs (usually $1-10^4$ keV 
in the rest frame of the GRB), ranges from $\sim 10^{49}$ to
$\sim 10^{55}$ erg. At the peak of the lightcurve, the isotropic
$\gamma$-ray luminosity ranges from $\sim 10^{47}$ to $\sim 10^{54}
~{\rm erg~s^{-1}}$. For high-luminosity long-GRBs (typical ones), the luminosity 
function can be characterized as a broken power law of the following form
\begin{equation}
 \Phi(L) dL = \Phi_0 \left[\left(\frac{L}{L_b}\right)^{\alpha_1}
+\left(\frac{L}{L_b}\right)^{\alpha_2}\right]^{-1} dL,
\end{equation}
where the break luminosity $L_b$ is $\sim 10^{52.2}~{\rm erg~s^{-1}}$.
Several studies agree that the high-luminosity slope is steep: 
$\alpha_2 \sim 2.5$ \citep[e.g.][]{liang07,virgili09,wanderman10}.
The value of $\alpha_1$ depends on whether one introduces a two-component
(low-luminosity vs. high-luminosity) model. If one considers low-luminosity
GRBs (those with luminosity below $\sim 10^{48}-10^{49}~{\rm erg~s^{-1}}$,
typically with long durations, soft spectra, and single-peaked smooth 
lightcurves) as a separate population, which has a distinct bump in the 
luminosity function, then it is found that $\alpha_1\sim 0.5$ for the 
high-luminosity GRB component \citep[e.g.][]{liang07,virgili09}.
On the other hand, if we include low-luminosity bursts to the GRB sample, 
then for the combined luminosity function 
$\alpha_1 \sim 1.2$ \citep{wanderman10}\footnote{Notice that
\cite{wanderman10} defined luminosity function as $\Phi(L) d \log L$
instead of $\Phi(L) dL$. As a result, the two indices reported in that
paper are systematically smaller by 1 than quoted here.}. The
normalization $\Phi_0$ depends on the local rate of GRBs per unit volume, 
which is constrained to be around $1 ~{\rm Gpc^{-3}yr^{-1}}$.

Low-luminosity (LL) GRBs have a higher local event rate 
\citep{soderberg06,liang07} which is inconsistent with a simple
extrapolation of the high-luminosity GRB luminosity function
to low luminosities, and they constitute a distinct class of objects
\citep{liang07,virgili09,bromberg11,bromberg12}. The exact form of the LL-GRB
luminosity function is not well constrained due to the limit
of detectors' sensitivity.

The luminosity function of short-GRBs is also not well constrained
because of the small sample size with redshift measurements.
In order to be able to use short-GRBs with unknown redshifts
to constrain the luminosity function, one needs to
introduce an intrinsic redshift distribution of short GRBs, 
which is unknown. In practice, one can adopt the NS-NS or 
NS-BH merger model to provide an approximate $z$-distribution
which can then be used to constrain the luminosity 
function \citep{guetta09,virgili11}. 
\cite{virgili11} show that compact binary star merger models cannot 
simultaneously reproduce all the observational data of short
GRBs for both the $z$-known (Swift) and $z$-unknown samples.
In these simulations, the short GRB luminosity
function was assumed to take a form similar to the long-GRBs (broken
power law), but the indices are left free parameters to be constrained
by the data.
The sample with known redshift demands a shallow luminosity function 
for short-GRBs with $\alpha_1 < 0.4$. 
This shallow luminosity function in turn usually translates to a shallow 
flux distribution, which is inconsistent with the BATSE data \citep{virgili11}.
There are two possibilities to reconcile the inconsistency between
theory and data. One is that there is a significant contribution 
of massive star bursts to the short-GRB sample. And the other is that 
the delay time scale since star formation for short GRBs to occur has
a typical value of about 2 Gyr. 
Both conclusions are confirmed recently by \cite{wanderman14}, who suggested
that the typical delay time scale is closer to 3 Gyr.

\subsection{Correlations between different observed parameters}

Several observed parameters of the prompt $\gamma$-ray radiation are 
claimed to be correlated. In this section
we summarize these correlations and comment on the ongoing debate of
their validity. 

\subsubsection{$E_{p,z}-E_{\rm \gamma,iso}$ (Amati) and 
$E_{p,z}-L_{\rm \gamma,iso}$ (Yonetoku) relations}

\cite{amati02} discovered that $E_{p,z} \propto {E_{\gamma,iso}}^{1/2}$, 
 where $E_{p,z} = E_p (1+z)$ is photon energy at the peak of the prompt
spectrum in the rest frame of the GRB, and $E_{\gamma,iso}$ is the 
isotropic gamma-ray energy spectrally extrapolated to a standard energy 
band in the GRB rest frame (usually 1 keV - 10,000 keV). 
 Numerically, this relation can be written as
\begin{equation}
\frac{E_{p,z}}{100 ~\rm keV} = C \left(\frac{E_{\gamma,iso}}{10^{52}~{\rm erg}}
\right)^m
\label{amati}
\end{equation}
with $C \sim (0.8-1)$ and $m \sim (0.4 - 0.6)$ \citep{amati06}.
This relation is found for long GRBs with known redshifts 
\citep{amati02,amati06,frontera12}, which covers a wide range of $E_{\gamma,iso}$
and $E_{p,z}$ (from hard GRBs to low luminosity X-ray flashes)
\citep{sakamoto06}.  GRB 980425, a low luminosity GRB with supernova 
association (SN 1998bw), is a significant outlier of this relation.

Several groups have argued that the Amati-relation is an artifact of
observational selection effects 
\citep[e.g.][]{nakarpiran05,band05,butler07,kocevski12b}. 
Counter arguments suggest that selection effects cannot completely
destroy the correlation \citep{ghirlanda08}. In general, a positive
correlation between $E_{p,z}$ and $E_{\gamma,iso}$ seems real, although
the scatter may be wide. 

Similarly, a positive correlation between $E_{p,z}$ and $L_{\gamma,p,iso}$
has been reported \citep{weigao03,yonetoku04}; where $L_{\gamma,p,iso}$ is the
isotropic gamma-ray luminosity of a burst at its peak flux.
Adapted from the original form \citep{yonetoku04}, this relation reads
\begin{equation}
\frac{E_{p,z}}{100~{\rm keV}} \simeq 1.8 \left( \frac{L_{\gamma,p,iso}}{10^{52}
{\rm erg~s^{-1}}}\right)^{0.52}~.
\label{yonetoku}
\end{equation}
This is also a correlation with broad scatter.

A $E_p - L_\gamma$ correlation also exists for the time resolved 
lightcurve of an individual GRB 
\citep{liang04,frontera12,lu12,guiriec13}.
This behavior, at least partially, can be explained by the behavior of
the falling phase of GRB pulses, during which emission clearly softens as
flux decreases \citep[e.g.][]{lu12,preece14}.

Short-GRBs do not fall onto long-GRB Amati-relation. They seem to form
a parallel track above it. In other words, given the same $E_{p,z}$,
short GRBs are systematically less energetic. This can be attributed
to their shorter durations, which hints that luminosity may be more 
intrinsically related to $E_{p,z}$. Indeed, in the $E_{p,z}-L_{\gamma,p,iso}$
space, short and long GRBs are no longer clearly separated, suggesting that 
their radiation processes are similar 
\citep{zhang09,ghirlanda09,guiriec13,tsutsui13}.

\subsubsection{$E_{p,z}-E_{\gamma}$ (Ghirlanda) relation}

Assuming that the afterglow temporal breaks discovered in the pre-Swift era
are jet breaks \citep{rhoads99,sari99}, \cite{ghirlanda04} found a
correlation between $E_{p,z}$ and the geometrically-corrected gamma-ray energy
\begin{equation}
E_\gamma= \frac{E_{\gamma,iso}}{4\pi} 
\int_0^{\theta_j} 2\times 2\pi \sin\theta d\theta
=(1-\cos\theta_j) E_{\gamma,iso} \simeq (\theta_j^2/2) E_{\gamma,iso}~,
\label{Egam}
\end{equation}
where $\theta_j$ is the jet opening
angle inferred from the afterglow temporal break time $t_{obs,j}$. 
The correlation (Ghirlanda relation), in its original form \citep{ghirlanda04}, 
reads
\begin{equation}
\frac{ E_{p,z}}{100~{\rm keV}} \simeq 4.8 \left(\frac{E_\gamma}{10^{51}
{\rm erg}} \right)^{0.7}~,
\end{equation}
which was claimed to be tighter than the Amati relation. In the Swift era, 
interpreting all the afterglow lightcurve breaks as jet breaks has been
questioned. First, the jet-like breaks in X-ray lightcurves
are either systematically earlier than the jet-like breaks in optical
data \citep{liang08,kocevski08}
or no jet-break is detected in XRT observations \citep{sato07,racusin09}. 
Second, achromaticity, a required feature of jet breaks, is not commonly
observed for these late time jet-like breaks \citep{liang08}. Growing
evidence suggests that the X-ray afterglow of a good fraction of GRBs
might not originate from the external shock, and is likely powered by
a long-lasting central engine 
\citep{ghisellini07,kumar08a,kumar08b,cannizzo09,lindner10,metzger11}.
The optical lightcurve may be still related to the external shock.
So jet break time may be obtained for optically-identified breaks only.

\subsubsection{$E_{p,z}-E_{\rm \gamma,iso}-t_{b,z}$ (Liang-Zhang) relation}

Regardless of the interpretation of afterglow temporal breaks, 
\cite{liangzhang05} discovered a fundamental-plane correlation among
$E_{p,z}$, $E_{\gamma,iso}$ and $t_{b,z}$, where $t_{b,z}=t_{obs,b}/(1+z)$ 
is the afterglow lightcurve break time in the rest frame of the burst 
as measured in the {\em optical} band. In its original form, this relation reads
\begin{equation}
\frac{E_{p,z}}{100~{\rm keV}} \simeq 1.09 \left(\frac{E_{\gamma,iso}}{10^{52}
{\rm erg}} \right)^{0.52} \left(\frac{t_{b,z}}{{\rm day}}\right)^{0.64}~.
\end{equation}
Such an empirical correlation is not dependent on interpreting the break 
in the lightcurve to be jet-break.

\subsubsection{$E_{p,z}-L_{\gamma,iso}-T_{0.45}$ (Fermani) relation}

With prompt emission parameters only, \cite{firmani06} discovered
another three-parameter correlation 
\begin{equation}
\frac{ E_{p,z}}{100~{\rm keV}} \simeq 1.37 \left(\frac{L_{\gamma,iso}}{10^{52}
{\rm erg~s^{-1}}} \right)^{0.62} \left(\frac{T_{0.45,z}}{{10 \rm s}}\right)^{-0.30}~.
\end{equation}
Here $T_{0.45,z}=T_{0.45}/(1+z)$, and $T_{0.45}$ 
is the time spanned by the brightest 45\% of the total counts
above the background. Traditionally, the burst duration is defined by $T_{90}$
(or $T_{50}$), the time interval within which 90\% (or 50\%) of the burst
fluence is detected. The main difference between $T_{0.45}$ and $T_{90}$ ($T_{50}$)
is that the former deducts any quiescent period that may exist during the burst,
and therefore better represents the duration of the emission episode of a burst.
The 45\% percentage has no physical significance, which was adopted to achieve
the most significant correlation.

\subsubsection{$E_{\gamma,iso} - \theta_j$ (Frail) relation: constant 
energy reservoir}

\cite{frail01} found that the measured jet opening angle $\theta_j$ of
early GRBs seem to be anti-correlated to $E_{\gamma,iso}$ through
$E_{\gamma,iso} \propto \theta_j^{-2}$. This led to an interesting
conclusion that the jet corrected gamma-ray energy 
$E_\gamma \simeq (\theta_j^2/2) E_{\gamma,iso}$ is roughly constant
for all GRBs. The correlation was confirmed by a later study
\citep{bloom03}, with $E_\gamma$ tightly clustered around
$\sim 10^{51}~{\rm erg}$. Replacing $E_{\gamma,iso}$ by isotropic 
kinetic energy of the afterglow, \cite{panaitescu02} and \cite{berger03} 
found that $E_K\simeq(\theta_j^2/2) E_{\gamma,iso}$ is also roughly 
constant. The implication is that long GRBs have a standard energy reservoir.
Wider jets have low energy concentration, while narrow jets have
high energy concentration. Alternatively, this may be understood
as a universal \citep{zhangmeszaros02b,rossi02} or quasi-universal 
\citep{zhang04} jet for all GRBs, with the measured jet break defined
by observers' viewing angle wrt the jet-axis.

In the Swift era, the Frail relation is found no longer tightly
clustered. Both $E_\gamma$ and $E_K$ are found to have a much
wider distribution than the pre-Swift sample 
\citep{liang08,kocevski08,racusin09}. The Ghirlanda relation
discussed above is in conflict with the Frail relation: instead
of having $E_\gamma$ as a constant, the former relation suggests
a correlation between $E_\gamma$ and $E_{p,z}$.

\subsubsection{Luminosity -- spectral lag ($L-\tau$), Norris relation}

\cite{norris00} discovered an anti-correlation between GRB peak 
luminosity and the delay time (lag), $\tau$, for the arrival of 
low energy photons (25-50 keV) compared with photons of higher energies
(100-300 keV and $>$300 keV) for a sample of BATSE
GRBs.  In its original form, it is written as
\begin{equation}
\frac{L_{\gamma,p,iso}}{10^{53}~{\rm erg~s^{-1}}} 
\simeq 1.3 \left(\frac{\tau}{0.01~{\rm s}}\right)^{-1.14}~,
\end{equation}
where $\tau$ is measured in the observed frame. Several groups 
have later investigated this correlation by considering the
lags in the burst rest frame. One way is to correlate
$L_{\gamma,p,iso}$ with $\tau/(1+z)\times (1+z)^{0.33}
=\tau/(1+z)^{0.67}$ \citep{gehrels06,zhang09}. By doing so,
one has assumed that spectral lag is proportional to the pulse 
width $w$ (which has an energy dependence of $\sim 0.33$ power).
This is valid for individual pulses. For complex bursts with
overlapping pulses, \cite{ukwatta12} argued that it is more
appropriate to investigate a correlation between $L_{\gamma,p,iso}$
and $\tau_z=\tau/(1+z)$. They gave
\begin{equation}
\log \left(\frac{L_{\gamma,p,iso}}{\rm erg~s^{-1}}\right)
=(54.7 \pm 0.4) - (1.2 \pm 0.2) \log \frac{\tau_z}{\rm ms}
\end{equation}
for the lag defined between 100-150 keV and 200-250 keV energy
bands in the rest frame of the GRB source.

There are significant outliers in the luminosity - spectral lag 
correlation. It seems that even though the low-luminosity
GRB 060218 may be moderately accommodated within the correlation
\citep{liang06b}, several other low luminosity GRBs (e.g. GRB 980425,
GRB 031203) and the supernova-less long GRBs (060614 and 060505)
all lie well below the correlation \citep{gehrels06,mcbreen08,zhang09}.
All short-GRBs have negligible lags \citep{yi06}, and do not follow
the correlation.

\subsubsection{Luminosity - variability ($L-V$) relation}

\cite{fenimore00} and \cite{reichart01} proposed a correlation between the
GRB luminosity and the complexity of GRB lightcurves which they parametrize
 as ``variability'' $V$. The definition of variability depends
on how the smoothed background lightcurve is defined, and can be
technically very different among authors. In any case, a positive
correlation $L_{\gamma,p,iso} \propto V^m$ with large scatter 
was found, although the index $m$ ranges from 3.3 \citep{reichart01}
to 1.1 \citep{guidorzi05}. 

\subsubsection{$E_{\gamma,iso}-\Gamma$ and $L_{\gamma,iso}-\Gamma$ relations}
\label{L-Gamma}

A sample of GRBs has high-quality early optical afterglow 
data collected. A good fraction of them show an early hump in
the lightcurve, which is consistent with being due to deceleration
of the blastwave. Assuming such an interpretation, the Lorentz factor
$\Gamma$ of a moderate sample of GRBs was measured. \cite{liang10}
discovered a positive correlation between $\Gamma$ and the 
isotropic gamma-ray energy $\Gamma \propto E_{\gamma,iso}^a$,
with $a \sim 1/4$. The positive correlation was verified by 
\cite{ghirlanda11} and \cite{lv12}. \cite{lv12} further discovered 
a similar correlation between $\Gamma$ and mean isotropic gamma-ray 
luminosity $L_{\gamma,iso}$, i.e. $\Gamma \propto L_{\gamma,iso}^b$,
with $b$ also close to 1/4. \cite{lei13} interpret the correlation within
the framework of both a neutrino-cooling dominated accretion flow model
and a Blandford-Znajek model for
GRB central engine, whereas \cite{fan12} and \cite{lazzati13} suggest that
the photospheric model for prompt $\gamma$-ray radiation can explain this
and other correlations.

\subsection{GRB cosmography}

An exciting prospect of having a tight GRB correlation is to apply
it to measure the geometry of the universe. Since GRBs are typically 
observed at a much higher redshift than the ``standard candle'' SN Ia, 
 they can potentially extend the Hubble diagram to higher redshifts.
This would lead to improvements in the determination of cosmological 
parameters, and in particular help explore the nature of dark energy. 
The difficulty has been to
find a tight enough GRB correlation to conduct such an exercise.

Early efforts in this direction made use of some not-too-tight
correlations (e.g. the $L-\tau$ and $L-V$ correlations by 
\cite{schaefer03} and the Frail correlation by \cite{bloom03})
to construct the GRB Hubble diagram. Since these correlations have 
large scatter, the data cannot 
place meaningful constraints on cosmological parameters. A step forward
was after the tight Ghirlanda-relation \citep{ghirlanda04} was 
discovered. \cite{dai04b}, \cite{ghirlanda04b} and \cite{xu05} show that 
GRBs can serve as a tool to conduct cosmography, and the constrained 
cosmological parameters (even if with large errors) are broadly 
consistent with the $\Lambda$CDM model supported by the combined SN 
Ia and WMAP-CMB data. Another correlation with claimed similar
tightness is the 
Liang-Zhang relation. Applying this relation to the cosmography 
study, \cite{liangzhang05} found constraints on cosmological parameters 
similar to the one obtained using the Ghirlanda relation.
Later, Amati and collaborators suggested that the Amati-relation
can also be used for the purpose of constraining cosmological
parameters \citep[e.g.][]{amati08}; see \cite{amati13} for a review.

We would like to point out two serious limitations we face today when
using GRBs for cosmography. First, due to the {\em intrinsic} dispersion
of the GRB correlations (no clean physics behind the correlations
unlike the Chandrasekhar limit behind the SNe Ia physics), the GRB
candle is much less standard than the SN Ia candle. The efforts 
using GRB data alone have so far led to much poorer constraints on the 
cosmological parameters than other well known methods (e.g. SNe Ia and CMB). 
On the other hand, the advantage of GRBs 
is that they can be detected at much higher redshifts than SNe Ia, so they 
can potentially be used to measure how the dark energy evolves with redshift.
 For example, \cite{schaefer07} applied multiple 
correlations to construct the Hubble diagram of 69 GRBs in the redshift
range from 0.17 to $>6$, and obtained consistency with the concordance
$\Lambda$CDM model without invoking dark energy evolution.
Second, it is not easy to calibrate GRB candles using GRB data alone. A 
robust calibration (e.g. for SNe Ia) requires a low-$z$ sample. However, 
the nearby GRBs tend to have a much lower luminosity than their cosmological
cousins \citep{galama98,campana06}, and likely form a distinct population
\citep{liang07}. Their detected number is also low. One suggested method is 
to consider bursts in a narrow redshift bin to partially calibrate the 
correlation \citep{liangzhang06b,ghirlanda06}. This can well calibrate the 
indices of the correlation, but the normalization parameter still depends on
the adopted cosmological parameters, and can be only ``marginalized''.
A more practical method of calibration is to make use of the SN Ia data 
\citep{liangn08,kodama08}. Taking GRBs in the same redshift range as SN Ia, one
can use the distance moduli of SNe Ia and assign them to GRBs at the same 
redshifts, and then give a cosmology-independent calibration to the GRB 
candles. The derived cosmological parameters using the calibrated
candles are again found consistent with the concordance model
\citep{liangn08}.

\section{Progress toward understanding GRB prompt radiation}
\label{prompt_theory}

The origin of the prompt $\gamma$-ray emission from GRBs is not well understood.
This is due in large part to our lack of knowledge of jet composition, 
energy dissipation and particle acceleration mechanisms. 
A widely used model is the matter-dominated ``fireball'',
which consists of baryons (primarily protons and neutrons), electron \&
positron pairs, and photons. A fireball could be produced in cataclysmic
events such as mergers of binary neutron stars \citep{narayan92} or 
collapses of massive stars \citep{woosley93}. 
The energy in radiation is initially larger than in baryons (including
baryon rest mass) by a factor of about 10$^2$, and as the fireball expands,
baryons get accelerated to a high Lorentz factor. According to this model,
a fraction of the initial thermal energy of the fireball is radiated away
at the photosphere, and at a larger radius internal shocks tap into
the kinetic energy of the jet to accelerate electrons which produce 
non-thermal $\gamma$-rays via the synchrotron and inverse-Compton processes.

Alternatively, the outflow launched from the GRB central engine might be
Poynting-flux-dominated 
\citep[e.g.][]{usov94,katz97a,meszarosrees97b,lyutikov03,zhangyan11}. 
In this case,
jet acceleration, dissipation and particle acceleration are harder to 
calculate, and the model lacks predictive power due to our limited 
understanding of these processes. 

This section provides an overview of GRB prompt emission models. The 
structure of the section is as follows:
\begin{itemize}
\item We begin with a quantitative description of the standard
hot fireball model for GRBs (\S\ref{fireball}). We discuss the
dynamics of fireball evolution as well as the photospheric 
radiation properties. Observational
constraints on the distance of the $\gamma$-ray emission region
from the center of explosion are presented in \S\ref{source_distance}.
\item Next, the internal shock model is discussed in detail. The
topics include conversion of the kinetic energy of the
outflow to thermal energy and the efficiency for producing radiation
(\S\ref{internal_shock}), the difficulty of reproducing the observed
spectrum (\S\ref{shock_synchro} \& \ref{shock_ic}) as well as 
a critical assessment of several recent models that have been proposed to 
explain the nearly flat spectrum ($f_\nu \aprop \nu^0$)  below the
peak.
\item For a GRB jet consisting of protons and neutrons, a fraction of 
the outflow kinetic energy is converted to thermal energy and radiation
when these particles undergo collisions near the photosphere. Whether
this process can explain the GRB prompt radiation is taken up in 
\S\ref{np_collision}.
\item A more general discussion of photospheric radiation, including 
multiple IC scatterings and its application to GRBs is discussed in
  \S\ref{ic_photo}. The hadronic model for the generation of 
prompt $\gamma$-ray
  radiation is analyzed in \S\ref{hadron_model}.
\item Finally, analytical calculations for the acceleration and 
dissipation of Poynting jets are discussed in \S\ref{magnetic_jet}. 
Numerical simulations on magnetic reconnection and particle
acceleration are also reviewed in that section.
\end{itemize}

Our emphasis is on providing physical insights, and not on rigorous 
mathematical derivations, and thus we will make numerous approximations that 
simplify calculations while focusing on the important physical concepts 
underlying various derivations and estimates throughout this section. 

\subsection{Hot fireball model}
\label{fireball}

One of the widely discussed models for GRBs is the so called hot fireball model.
This model was suggested in its currently used form\footnote{The fireball 
model in the context of GRBs and some consequences of high opacity due to 
electron-positron pairs produced by MeV photon collisions were described 
by \cite{cavallo78} well before the work of \cite{paczynski86} and 
\cite{goodman86}.} by \cite{paczynski86} and \cite{goodman86} when 
Paczynski realized that GRBs might be at cosmological distances and 
therefore have luminosity of $\sim 10^{51}$erg s$^{-1}$ produced within
a small volume of radius $\lae 10^7$ cm (from lightcurve variability) and
hence a temperature of $\sim 10^{10}$K so that electron-positron pairs 
coexist with photons in thermal equilibrium.
The energy per proton according to the hot fireball model is of order 
10$^2$ GeV; much of this energy is initially in photons, relativistic $e^\pm$ 
pairs, and neutrinos. The radius where the fireball is produced 
is set by the size of the compact object formed in these 
explosions which is believed to be either a black hole or a 
millisecond magnetar\footnote{A magnetar is a neutron star with magnetic 
field of strength much larger
than a typical pulsar. The surface field of a magnetar is of order $10^{14}$G
or larger.}. 
As the fireball undergoes adiabatic expansion, the energy of photons and 
$e^\pm$s is transferred to protons which are accelerated to a high Lorentz 
factor \citep{shemi90}. The kinetic energy of the outflow is converted back to 
thermal energy and radiated away as $\gamma$-ray photons at some large 
distances from the place where the fireball is produced \citep{rees92,meszarosrees93,rees94}.

The dynamical evolution of a fireball during the acceleration phase 
has been studied analytically \citep{shemi90,meszaros93,piran93,meszarosrees00,meszaros02b}, 
as well as numerically \citep{kobayashi99}. 

\begin{figure}
\begin{center}
\includegraphics[width=13.3cm]{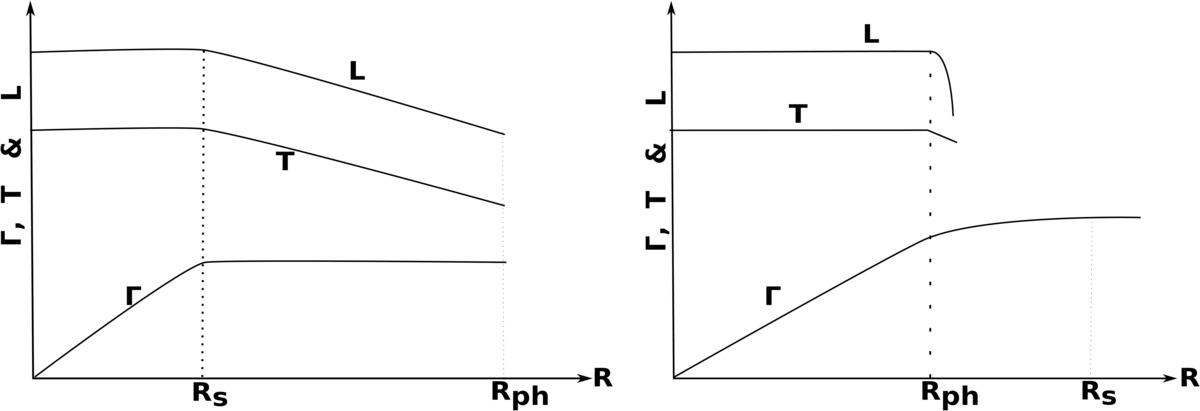}
\caption{Lorentz factor ($\Gamma$), Temperature ($T$) and thermal 
luminosity $L$ (in observer frame) are shown schematically as a function of
fireball radius for the case where $\eta \lae 10^3$ so that the photosphere 
($R_{ph}$) lies above the saturation radius $R_s = \eta R_0$ (left panel). 
$\Gamma$, $T$ \& $L$ for the case where $\eta \gae \eta_* \sim 10^3$, 
so that $R_{ph} < R_s$, is shown in the right panel; $\Gamma$, $T$ \& $L$ 
in this case have the same dependence 
on $R$ for $R<R_{ph}$ as shown in the first segment of the left panel, however, 
the outflow fails to attain the asymptotic LF of $\eta$ in this case (the
slight increase of $\Gamma$ for $R>R_{ph}$ is due to Compton drag on electrons
by photons streaming to larger radius). The time dependence of the observed 
thermal luminosity and temperature, after a brief transient period, mirrors 
temporal fluctuations of the thermal fireball (or outflow) at its base at $R_0$.
}\label{FIG:photospheric_LC1}
\end{center}
\end{figure}

We describe in this subsection the fireball dynamics and the conversion of 
kinetic energy to radiation via collision between fast and slow parts of 
the outflow. The main results for the fireball dynamics are shown in Figure 
\ref{FIG:photospheric_LC1}.

\subsubsection{Dynamics of a Hot Fireball}
\label{fireball_dynamics}

Consider an outflow of luminosity $L$ and initial radius $R_0$ that
is related to the size of the compact object (black hole or
a rapidly rotating neutron star) formed in these explosions. 
The initial temperature of the fireball is
\beq
k_B T_0 \approx k_B\left[ {L\over 4\pi R_0^2 g_0 \sigma_B} \right]^{1/4} = 
     (1.3\,{\rm MeV}) L_{52}^{1/4} R_{0,7}^{-1/2},
 \label{T_R0}
\eeq
where $k_B$ \& $\sigma_B$ are Boltzmann and Stefan-Boltzmann constants, and 
$g_0=2.75$ is half of the effective degrees of freedom for a plasma 
consisting of photons, electrons \& positrons
in thermal equilibrium\footnote{Initially, at radius $R_0$, the temperature is
larger than 1 MeV so that electrons, positrons, and electron neutrinos are 
readily created. However, neutrinos fall out of thermal equilibrium when the
fireball radius is just a little larger than $R_0$ and hence are not counted
towards the effective degree of freedom for particles in 
thermal equilibrium. Each Fermion internal degree of freedom -- spin state -- 
contributes $7/8$ of a Bosonic degree of freedom, so the total for 
a $e^\pm$ and photon plasma comes out to be 5.5. The 2 degree of freedom for
photons is already included in the radiation constant $\sigma_B$, hence 
$g_0=5.5/2=2.75$.};
we are continuing to use the notation $X_n \equiv X/10^n$.

The Lorentz factor of the fireball undergoing adiabatic expansion increases
linearly with radius as long as the energy in radiation per baryon is larger 
than $\sim m_p c^2$, and the fireball is optically thick to Thomson scattering.

\begin{figure}
\begin{center}
\includegraphics[width=13cm]{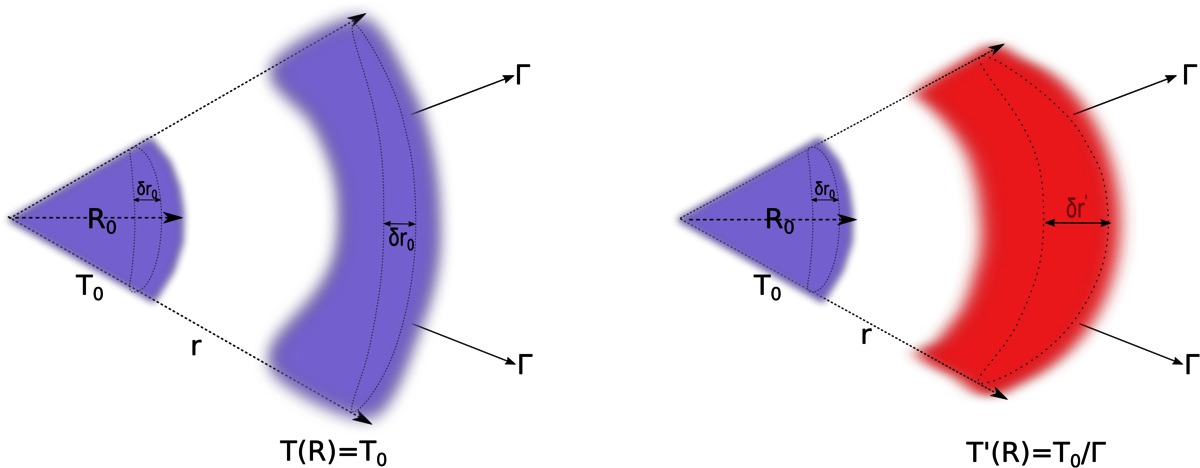}
\caption{Fireball dynamics in lab frame is shown in the left panel and
in the shell comoving frame (right panel). Shown here is a small section 
of the fireball of initial radial width $\delta R_0$ at two different times
when it was at radius $R_0$ and $r$. Its LF when at radius $R_0$ was 
$\sim 1$, and hence its radial widths in the lab and comoving frames were 
the same. At a later time when at radius $r$ its LF increased to $\Gamma$
and its comoving temperature decreased by a factor $\Gamma$.
}\label{FIG:fireball_dyna}
\end{center}
\end{figure}

The fireball dynamics is described by conservation of energy flux and entropy. 
We describe the process as viewed in an inertial frame at rest 
in the GRB host galaxy.
Hereafter we call this rest frame as the ``CoE frame'', sometimes also
called the ``lab frame'' or ``cosmic proper frame''.
Let us consider a spherical shell of radius $r$ and width $\delta r$ in
the CoE frame (the width in the comoving frame is $\delta r'$; see Fig. 
\ref{FIG:fireball_dyna}).
The comoving temperature of the shell is $T'(r)$, and its Lorentz 
factor is $\Gamma(r)$. Its luminosity in the CoE frame does not change
as the shell expands to larger radius, and is given by
\beq
   L = 4\pi r^2 g(r) \sigma_B T'^4(r) \Gamma^2(r).
   \label{jetL1}
\eeq
Moreover, the entropy contained in the shell is
\beq
   s = 4\pi r^2 (\delta r') g(r) T'^3
  \label{entropy1}
\eeq
which is a frame independent, and a conserved quantity for an adiabatically 
expanding shell (ignoring the initial decrease due to neutrino loss).

The width of the shell in the CoE frame does not
change much with $r$ since the front and the back surfaces of the
shell move close to the speed of light and their relative speed
is small. However, the shell-width in the 
comoving frame, which is given by  $\delta r' = \Gamma \delta r$, does 
change with radius as the shell expands and its LF increases.

The equation for the conservation of entropy (eq. \ref{entropy1}) can be
solved for comoving temperature
\beq
  T' = T_0 (R_0/r)^{2/3} (g_0/g)^{1/3} \Gamma^{-1/3},
  \label{TpR1}
\eeq
where $g_0\equiv g(R_0)$. Substituting this into equation (\ref{jetL1}) we find
\beq
   \Gamma(r) = (r/R_0) (g/g_0)^{1/2},
   \label{GamR}
\eeq
and we can solve for $T'$ by eliminating $\Gamma$ from equation (\ref{TpR1}) 
\beq
   T'(r) = T_0 (r/R_0)^{-1} (g_0/g)^{1/2}
  \label{TpR}
\eeq

The Lorentz factor continues to increase as $\Gamma\propto r$ as long as energy 
in photons per baryon in the comoving frame ($3 k_B T' n_\gamma$) is larger 
than $m c^2$ \& the system is optically thick to Thomson scattering so that 
photons and particles are coupled. The terminal value of the Lorentz factor is
\beq
 \Gamma_s = {L\over \dot M c^2} \equiv \eta,
\eeq
which is attained at the radius
\beq
R_s \sim R_0 \Gamma_s
   \label{Rs}
\eeq
provided that the fireball remains optically thick to Thomson scattering 
at $R_s$; $\dot M$ is the baryonic mass flux associated with the outflow.
We will see toward the end of this sub-section that for $\eta \gae 10^3$ 
the fireball becomes optically thin before attaining $\Gamma_s\sim \eta$. 

The optical depth is dominated by $e^\pm$ when $\eta \gae 10^6$, and in this 
case the system becomes transparent to photons when $T'$ drops to about 
20 keV and pairs annihilate.
The number density of electron-positron pairs, in the comoving frame of the 
outflow, at temperature $T'$ is
\beq
   n'_\pm = {2 (2\pi k_B m_e T')^{3/2}\over h^3} \exp\left(-m_e c^2/k_B T'
    \right),
 \label{pairdensity}
\eeq
where $m_e$ is electron mass \& $h$ is Planck's constant. The cross-section
for pair annihilation is
\beq
  \sigma_{e^\pm\rightarrow 2\gamma} = {\sigma_T\over \langle v/c\rangle}
\eeq
where $\langle v\rangle$ is the mean speed of $e^\pm$, and $\sigma_T$ is
Thomson scattering cross-section. Thus, the comoving frame time for a 
positron to annihilate with an electron is
\beq
  t'_{e^\pm\rightarrow 2\gamma} = {2\over \sigma_{e^\pm\rightarrow 2\gamma} 
    n'_\pm \langle v\rangle} \approx {2\over \sigma_T n'_\pm c},
\eeq
where the factor $2$ in the numerator is due to the fact that the 
number density of electrons $=n'_\pm/2$ (ignoring the contribution 
of electrons associated with baryons).
The process of pair annihilation/creation freezes when 
$t_{e^\pm \rightarrow 2\gamma}$ becomes of order the dynamical time $\sim 
r/c\Gamma(r)$. From the above equation we see that the $e^\pm$ freeze-out 
radius is the same as the Thomson-photospheric radius when the baryon loading 
is negligible, i.e. when the electron density is not much larger than 
$n'_\pm/2$ given by equation (\ref{pairdensity}). If the freeze-out were to 
occur during the acceleration phase of the jet then $\Gamma(r)/r \sim 1/R_0$, 
and in that case
\beq
  \sigma_T n'_\pm R_0 \sim 2, \quad\quad {\rm or} \quad\quad n'_\pm \sim 2/(\sigma_T R_0).
\eeq
Substituting for $n'_\pm$ from equation (\ref{pairdensity}) we find
\beq
  T^{'^{3/2}} e^{ -{5.9\times10^9\over T'} } \approx 62 R_{0,7}^{-1}.
\eeq
The solution of this equation is $T'_{freeze}\approx 20.5$ keV. The 
 fireball Lorentz-factor at the freeze-out can be obtained using 
equations (\ref{TpR}) \& (\ref{GamR}) and is
\beq 
  \Gamma_{freeze} \sim T(R_0)/T'_{freeze} \sim 64,
   \label{gam-freeze}
\eeq
independent of the $g$-value, and the radius where the freeze-out occurs is
\beq
  R_{freeze} \sim R_0 \Gamma_{freeze} (g_0/g)^{1/2} \sim 1.7 R_0 \Gamma_{freeze}.
\eeq
For the last part of the above equation we took $g=1$ since at this radius
the entropy is dominated by photons.
The Lorentz factor of the fireball can continue to increase by another
factor of $\sim 2$ due to Compton drag \citep{meszaros93}.

Let us next consider the effect of a non-zero baryon component on the jet 
dynamics, and determine the criterion when the fireball dynamics  
is significantly affected by baryon contamination.

The number density of electrons associated with protons can be obtained 
from the equation for mass outflow, $\dot M$,
\beq
   n_p' = {\dot M\over 4\pi r^2 m_p c\Gamma} = {L \over 4\pi r^2 m_p c^3\eta
   \Gamma} \quad\quad {\rm where} \quad\quad \eta \equiv {L\over \dot M c^2}.
   \label{np1}
\eeq
Therefore, the number density at $R_{freeze}$ is
\beq
   n_p' = {L \over 4\pi R_0^2 m_p c^3 \eta \Gamma_{freeze}^3 }.
   \label{np2}
\eeq

The fireball dynamics beyond $R_{freeze}$ is dominated by 
electrons associated with protons when
\beq
  n'_p(R_{freeze}) > n'_\pm(R_{freeze}) \sim {2\over \sigma_T R_0}.
  \label{np}
\eeq
Or using equations (\ref{np2}), (\ref{np}), (\ref{T_R0}) and (\ref{gam-freeze})
we find
\beq
   \eta < {L \sigma_T \over 8\pi R_0 m_p c^3 \Gamma_{freeze}^3 } \sim
        2\times 10^6 L_{52}^{1/4} R_{0,7}^{1/2}.
\eeq
Whenever this condition is satisfied --- which is likely for most GRBs ---
 the jet continues to accelerate for $r>R_{freeze}$ until $\Gamma(r)\sim
\eta$ or the outflow reaches the Thomson photospheric radius (whichever
comes first).

The Thomson scattering optical depth for a photon at radius $r$ is 
\beq
 \tau_T = \int {dr_1\over c}\, (c - v) \sigma_T n_p  \approx \int {dr_1\over 
   2\Gamma^2}\, \sigma_T n_e \approx \sigma_T n_p' (r/2\Gamma) \approx
    {L\sigma_T \over 8\pi r m_p c^3 \eta\Gamma^2},
\eeq
where we made use of equation (\ref{np1}) for electron density.
Therefore, the photospheric radius, where $\tau_T=1$, is
\beq
   R_{ph} \approx (5.5\times 10^{12} {\rm cm})\, L_{52}\,\eta_2^{-1} 
   \Gamma_2^{-2}.
     \label{rp}
\eeq
The Lorentz factor stops increasing when the outflow reaches the photosphere,
if not before, since at this radius photons decouple from electrons and start 
streaming freely\footnote{Some additional acceleration above the photosphere 
can occur by outward streaming photons dragging electrons along for a while 
\citep{meszaros93}.}. 
Thus, the maximum possible value for Lorentz factor that can be attained
in a hot-fireball is when $R_{ph} \sim \eta R_0$ or
\beq
   \eta_*\equiv\Gamma_{max} \sim 8.5\times 10^2 L_{52}^{1/4} R_{0,7}^{-1/4}
   \label{gammax}
\eeq
The most energetic Fermi bursts from which $>$GeV photons have been
detected approach this theoretical limit on Lorentz factor\footnote{The Fermi
LAT team published lower limits of $\Gamma$ for a few bright LAT GRBs, which
approach 1000 \citep{abdo09a,abdo09b,ackermann10}, and are much larger
than the constrained $\Gamma$ from other GRBs using other methods
\citep{racusin11}. However, these
constraints were based on a simple one-zone model with the emission site
defined at the internal shock radius $R_{\rm IS} \sim \Gamma^2 c \Delta t_{\rm min}$,
which could be an over-estimate if the emission region is not at $R_{\rm IS}$
\citep{gupta08,zhangpeer09}, or when more sophisticated analysis are carried 
out \citep{granot08,zou11,zhao11,hascoet12b}}. For $\eta>\eta_*$
only a fraction of the initial photon energy of the fireball is imparted
to baryons.

\medskip

\subsubsection{Photospheric radiation}
\label{fireball_photosphere}

When $\eta > \eta_* \sim 10^3$ --- the limit given by equation 
(\ref{gammax}) --- the Thomson
photosphere lies within the acceleration zone of the outflow, and the
emergent thermal radiation from the photosphere is 
\beq
   L_{th} = 4\pi R_{ph}^2 \sigma_B T'(R_{ph})^4 \Gamma(R_{ph})^2 \sim 4\pi R_0^2 
    \sigma_B T_0^4 \sim L,
   \label{Lth1}
\eeq
which is of order the central engine luminosity ($L$); in deriving this result 
we made use of the scalings $\Gamma(R_{ph}) \sim R_{ph}/R_0$ (eq. \ref{GamR}) 
and $T'(R_{ph}) \sim T_0 R_0/R_{ph}$ (eq. \ref{TpR}). The observed peak of the 
thermal flash is at a temperature $T'(R_{ph}) \Gamma(R_{ph})/(1+z)\sim T_0/(1+z)
\sim [1.3/(1+z) {\rm MeV}] L_{52}^{1/4} R_{0,7}^{-1/2}$ (see eq. \ref{T_R0}).
This radiation lasts for as long as the central engine is active.

For $\eta \lae 10^3$, the Thomson photosphere lies outside the acceleration 
zone of the outflow, and the emergent thermal luminosity is smaller than 
$L$. The temperature continues to decrease, due to adiabatic cooling, 
beyond the saturation radius as 
\beq
  T(r)\sim T_0 (R_s/r)^{2/3},
  \label{Trp}
\eeq
since the comoving volume of a shell increases with radius as $r^2$ when 
the Lorentz factor of the shell stops increasing with distance; $R_s$
is the saturation radius given by equation (\ref{Rs}). The observed thermal
luminosity while the jet head is between $R_s$ \& $R_{ph}$ varies as 
$r^{-2/3}$ (fig.  \ref{FIG:photospheric_LC1}). Once the jet crosses $R_{ph}$,
the photospheric luminosity is given by \citep[e.g.][]{meszarosrees00,nakar05}
\beq
   L_{th} = 4\pi R_{ph}^2 \sigma_B T'(R_{ph})^4 \Gamma(R_{ph})^2 \sim L 
   (R_s/R_{ph})^{2/3} \sim L R_{0,7}^{2/3} L_{52}^{-2/3} \eta_3^{8/3}.
   \label{Lth2}
\eeq
This equation is valid only when $\eta < \eta_*$;
for $\eta>\eta_*$, $L_{th}\sim L$.
The strong dependence of thermal luminosity on $\eta$ is due to the fact that
the photospheric radius scales as $\eta^{-3}$ and the saturation radius ($R_s$)
increases as $\eta$, and therefore the thermal luminosity --- which scales as
$(R_{ph}/R_s)^{-2/3}$ --- decreases rapidly with decreasing $\eta$.
The peak of the thermal spectrum in the observer frame, when the jet 
becomes optically thin at $r=R_{ph}$, for $\eta<\eta_*$, is
\beq
   h \nu_{th} \sim 3 k_B T'(R_{ph}) \Gamma(R_{ph})/(1+z)\sim \left({4{\rm MeV}\over 
     1+z}\right) L_{52}^{-5/12} R_{0,7}^{1/6} \eta_3^{8/3},
   \label{nuth}
\eeq
which was obtained by making use of $T'(R_s)\Gamma(R_s) \sim T_0$, ~~
$T(R_{ph}) \sim T(R_s) (R_s/R_{ph})^{-2/3}$ and equation (\ref{T_R0}) for $T_0$.
 For $\eta>\eta_*$, the observed peak is at $\sim 3 k_B T_0$.
The observed specific flux at the peak of the spectrum is obtained from
equations (\ref{Lth2}) and (\ref{nuth}), and is given by,
\beq
  f_p^{th} \sim {L_{th} \over 4\pi d_L^2 \nu_{th}}\sim {L (1+z)\over 4\pi d_L^2 
     (3 k_B T_0/h)} \sim [1\, {\rm mJy}] (1+z) L_{52}^{3/4} d_{L28}^{-2} R_{0,7}^{1/2}.
\eeq
The emergent specific flux at the peak of the thermal spectrum is 
approximately 1 mJy, which is independent of the highly uncertain value
of $\eta$ and weakly dependent on jet luminosity $L$, and should be 
detectable by a telescope observing in a band that includes $\nu_{th}$;
we should point out that the observed non-thermal emission between 10 keV 
and $\sim 1$MeV for a typical GRB is also of order 1 mJy.  An observational 
campaign designed to look for this thermal component\footnote{The observed 
spectrum is expected to deviate from the Planck function due
to the fact that photons arriving at any given time at the observer in fact 
originated at different radii and time. Hence, the observed spectrum is a
superposition of Planck functions of different temperatures. In particular,
the observed spectrum below the peak is likely flattened to $f_\nu \aprop
\nu^{1.4}$ from the original $f_\nu\propto \nu^2$ 
shape \citep{beloborodov10,deng14b}.
And if the LF of the jet has an angular dependence such that it peaks at the
jet axis and decreases with angle then the observed spectrum is flattened 
even further and can become $f_\nu \propto \nu^{0}$ \citep{lundman13}.}
between optical and $\gamma$-ray frequencies would help determine the baryonic 
content of GRB jets. For instance, if a thermal component with peak flux of a 
few mJy is not found between 1 eV and 1 MeV, then that would suggest that 
$\eta < 5$.  Since this would contradict a vast amount of data that suggests 
$\Gamma\gae 10^2$ 
\citep{saripiran99b,lithwick01,zhang03,molinari07,abdo09a,abdo09b,liang10} 
--- and therefore $\eta\gae 10^2$ --- the non-detection of a thermal 
component would imply that the GRB jet was launched with a relatively 
small radiation component thereby strengthening
the case for a Poynting outflow \citep[e.g.][]{daigne02,zhangpeer09}. 
On the other hand, a detection of thermal 
signal and measurement of $\nu_{th}$ would help us determine the
amount of baryonic matter in GRB jets; $\nu_{th}$ depends primarily
on $\eta$ (eq. \ref{nuth}) and has a weak dependence on $L$ and $R_0$
as long as photons are not created or processed between $R_0$ and
the photosphere.

\subsection{Distance from the central engine where $\gamma$-rays are produced}
\label{source_distance}

We describe in this sub-section various constraints on the distance from the
center of explosion, $R_\gamma$, where $\gamma$-ray emission is generated. 
Four different ideas have been used to get a handle on 
$R_\gamma$: optical depth to Thomson scattering should be less than $\sim 1$;
detection of high-energy $\gamma$-rays suggests that $e^\pm$ production 
optical depth should also be less than one; detection of a bright optical 
transient during the prompt $\gamma$-ray phase implies that 
the synchrotron-self absorption frequency is below the optical band; 
the rapid decay of X-ray lightcurves at the end of prompt radiation phase 
signals a rapid turn-off of the central engine and this can be used to 
determine $R_\gamma$.
The first three methods provide a lower limit for $R_\gamma$. The last two 
techniques assume a one zone model where the lower frequency photons (optical 
or X-rays) and $\gamma$-rays are produced at the same location --- results 
described below are invalid if this assumption were to turn out to be 
incorrect.

The $\gamma$-ray source distance ($R_\gamma$)
should be smaller than the deceleration radius for the jet, which is of 
order 10$^{17}$cm for a uniform density circum-stellar medium
(see eq. \ref{R_d_s0}), otherwise
much of the energy of the GRB-jet is imparted to the surrounding medium
and not available for $\gamma$-ray radiation\footnote{The $\gamma$-ray
radiation during the prompt phase cannot be produced by the shock heated
GRB circumstellar medium (the so called external shock) since the resulting
lightcurve would vary on timescale of $R_\gamma/(2c\Gamma^2)\sim 10$s 
\citep{sari97}-- instead of 0.1s or less as observed --- and the X-ray 
lightcurve would not show a sharp drop-off at the end of the prompt phase 
as is seen for a good fraction of bursts \citep{tagliaferri05} -- but see 
also \cite{dermermitman99}.}.

The initially highly opaque fireball becomes optically thin at a distance 
$\sim5\times 10^{12} L_{52}\Gamma_2^{-3}$cm (eq. \ref{rp}) --- if baryonic  ---
and so we expect the observed radiation to originate at 
$R_\gamma\gae 10^{12}$cm. 

The detection of high energy $\gamma$-ray photons provides a constraint
on GRB-jet LF ($\Gamma$), which combined with the variability time for GRB 
prompt lightcurve ($\delta t$) gives a rough estimate for the radius where
$\gamma$-rays are produced --- $R_\gamma\sim 2 c (1+z) \delta t \Gamma^2\sim
10^{14}(\delta t)_{-1}$cm \citep[e.g.][]{rees94}; where we used
$\Gamma\sim10^2$ obtained from pair opacity argument described in the 
next two paragraphs. 

Consider that we see a photon of energy $\epsilon_0\gg m_e c^2$,
that lies on a power-law spectrum with photon index $\alpha$, from 
a GRB at redshift $z$. The  isotropic equivalent luminosity for this
burst is $L_\gamma$, the jet LF is $\Gamma$, and consider that $\gamma$-ray
photons are produced at a distance $R_\gamma$ from the center of explosion. 
The photon energy
in the jet comoving frame is $\epsilon_0 (1+z)/\Gamma$, which we assume is
larger than $m_e c^2$, otherwise there is little chance for it to be
converted to a $e^\pm$-pair\footnote{If the comoving energy were to be 
less than $m_e c^2$, then most of photons of energy $\epsilon_0$ can escape 
conversion to $e^\pm$ since they will have to interact with photons of 
larger energy for pair production. And such photons are much smaller in 
number since $\beta\sim 2$ for GRBs.}. This photon can interact with a
photon of jet-comoving frame energy $>(m_e c^2)^2\Gamma/(1+z)\epsilon_0
\equiv \epsilon_1'$ and produce $e^\pm$; the observer frame photon
energy is $\epsilon_1 = \epsilon'_1 \Gamma/(1+z)$.

The number density of photons of energy $\ge\epsilon_1'$ in the jet comoving 
frame is $n'_\gamma \sim L_\gamma (h\nu_p/\epsilon_1)^{\beta-2}/(4 \pi 
R_\gamma^2\Gamma^2 \epsilon'_1 c)$; where $\nu_p$ is frequency at the peak 
of the spectrum in observer frame, and $\beta\sim 2.2$ is the photon index
of the spectrum ($f_\nu\propto \nu^{-\beta+1}$) for $\nu>\nu_p$. The 
probability that the photon of energy 
$\epsilon_0$ will get converted to $e^\pm$ as it tries to escape the jet is
$\sigma_{\gamma\gamma\rightarrow e^\pm} n'_\gamma R_\gamma/\Gamma$;
where the pair production cross-section $\sigma_{\gamma\gamma\rightarrow 
e^\pm}=1.2\times10^{-25}$cm$^2$ at the optimal photon energy.  Therefore,
a lower limit to the LF of jet in order to avoid pair production is
\beq
  \Gamma > \left[ { \sigma_{\gamma\gamma\rightarrow e^\pm} L_\gamma (1+z)
     \epsilon_0\over 4\pi R_\gamma m_e^2 c^5}\right]^{1/2\beta} \left(
    {h\nu_p (1+z)^2\epsilon_0\over m_e^2 c^4}\right)^{(\beta-2)/2\beta}.
\eeq

If we detect a 100 MeV photon from a typical long-duration-GRB at $z=2$,
with $\beta=2$ and $\gamma$-ray luminosity $L_\gamma\sim 10^{52}$erg s$^{-1}$,
 then that suggests $\Gamma\gae 200$. Combining this with the 0.1s 
variability time for the prompt $\gamma$-ray lightcurve implies $R_\gamma \sim 
3\times10^{14}$cm \citep[see, e.g.][]{fenimore93,lithwick01,muraseioka08,gupta08,zhangpeer09}.

Optical photons have been detected during the prompt $\gamma$-ray burst for a
number of GRBs 
\citep{akerlof99,vestrand05,vestrand06,yost07,racusin08,zheng12,kopac13}.
If optical and $\gamma$-ray photons are produced at the same location ---
which is likely the case whenever correlated fluctuations are seen in 
optical and $\gamma$-ray lightcurves or when optical flux declines
rapidly (faster than $t^{-2}$) at the end of the prompt phase --- 
then that suggests that the synchrotron-self-absorption frequency 
is below the optical band. This provides a lower limit on $R_\gamma$ since
for a given $\gamma$-ray flux and spectral peak frequency, the electron
column density increases with decreasing $R_\gamma$, which leads to a larger
self-absorption frequency. This method was used by Shen and 
Zhang (2009) for a sample of 4 GRBs and they found $R_\gamma\gae10^{14}$cm.

One of the major discoveries made by the Swift satellite in regards to
GRBs was the detection of a rapidly declining X-ray lightcurve at the
end of the prompt phase, when the flux declines as $t^{-3}$ or faster
for a duration of a few minutes, and before a slowly declining
``afterglow'' phase sets in (Tagliaferri et al. 2005). This rapid decline
is seen in the majority of GRBs \citep{evans09,liang09} and heralds the winding 
down of central engine activity\footnote{The central engine can continue
to operate, sporadically, as evidenced by sudden increases in X-ray
flux or flaring events, for a period of hours to days e.g. Burrows
et al. (2005), Chincarini et al. (2011).}. Considering that there is 
no discontinuous change in X-ray flux between the prompt phase and the
rapidly declining X-ray phase \citep{barthelmy05b}, the radiation during 
the latter phase should have the same origin as the prompt GRB radiation.
As long as the opening angle of a GRB-jet is larger than $\Gamma^{-1}$
(which seems to be the case from observations of achromatic breaks in 
optical \& X-ray lightcurves several days after the explosion) it is
expected that we will continue to see a tail of prompt radiation 
coming from parts of the jet lying at angles ($\theta$) larger than 
$\Gamma^{-1}$ with respect to the observer line of sight. Radiation from 
larger $\theta$ arrives at a later time due to the larger path length the
light has to travel to get to the observer, and it is also subject to
a much smaller Doppler boosting, thereby leading to a rapidly declining
flux. In fact, this ``large angle radiation'', or the tail of the prompt 
phase, has a well defined, unique signature, that relates the temporal 
decline during the steep phase with the spectral slope at the end of the
prompt phase \citep{fenimore95,kumar00,dermer04,zhangbb09,genet09}:
$f_\nu(t)\propto t^{-2-\beta} \nu^{-\beta}$ (see \S\ref{relativity} 
for a derivation of this result)\footnote{The decay index $\alpha$ depends
on the choice of the zero time. It is usually set to the trigger time,
but can be later for the cases with distinct emission episodes
\citep{zhang06,liang06}.}; so the steeper the flux density
spectral index ($\beta$), the steeper is the temporal
decline during the steep phase. The time when the steep decline 
begins (time measured from the peak of the last pulse in $\gamma$-ray 
lightcurve) is set by the radius where the prompt radiation is produced 
and the jet Lorentz factor: $t_{\rm decline}\sim R_\gamma/(2 c\Gamma^2)$.
And the steep decline lasts for a duration
that is related to the jet opening angle $\theta_j$: 
$t_{\rm tail} \sim (R_\gamma/c) \theta_j^2/2$. Hence, the
timescale for steep decline, and a measurement of $\Gamma$ from 
the onset of ``afterglow'' radiation (or the jet deceleration time) 
enable the determination of $R_\gamma$.  
This idea for the determination of $R_\gamma$ was suggested by 
\cite{lyutikov06}, and \cite{lazzati06}, soon after the discovery of the steep 
decline of X-ray light-curves by the Swift satellite, 
and was recently re-emphasized by \cite{hascoet12}.

A number of GRBs satisfy the ``large angle radiation'' relation, 
$\alpha=2+\beta$, between the temporal decay index ($\alpha$) and spectral 
index ($\beta$) during the steep decline phase of X-ray afterglow lightcurve, 
e.g. \cite{obrien06}, \cite{willingale10}. 
\cite{kumar07} analyzed the data for a sample of 10 of these GRBs that
satisfy this ``closure relation'', and found $R_\gamma\gae 10^{16}$cm 
using the method described above.

\subsection{Internal Shocks: Conversion of outflow kinetic energy to 
   radiation}
\label{internal_shock}

We describe in this section the widely used internal shock model for converting 
the kinetic energy of a baryonic jet to $\gamma$-rays \citep{narayan92,rees94}. 
Consider a relativistic, baryonic outflow, where the LF is time dependent. 
In this case, the faster part of the outflow will catch up with a slower 
moving part ahead of it. The resulting collision produces a pair of shock waves 
that propagate into the fast and slow shells, and a fraction of jet kinetic 
energy is converted to thermal energy. The thermal energy is radiated 
away via synchrotron and the inverse-Compton processes. This model
is called the {\it internal shock model} since shocks are produced within the
jet due to non-zero gradient of velocity.

The main strength of this model lies in its simplicity and its ability
to account for short time scale variability, of order milli-seconds, 
seen in prompt GRB lightcurves. One of its main weaknesses is the 
inability to explain the observed spectrum --- in particular, the 
spectral index below the peak --- and possibly its low 
efficiency. We expand on these points below.

For simplicity, let us consider two shells of masses $m_1$ \& $m_2$ ejected 
from the central engine moving with terminal LFs $\Gamma_1$ and $\Gamma_2$. The 
slower moving shell-1 was launched $\delta t$ time before the other shell. The 
distance from the center of explosion where these shells collide, when
$\Gamma_2 \gae 2\Gamma_1$, is
\beq
   R_{coll} = {v_1 v_2 \delta t\over v_2 - v_1} \approx 2c \Gamma_1^2\delta t.
\eeq
Therefore, the time when the radiation produced in this collision will arrive 
at the observer is given by
\beq
t_{obs} \sim t_0 + R_{coll}/(2c \Gamma_f^2) \sim t_0 + (\delta t)\Gamma_1/\Gamma_2,
 \label{t_internal-shock}
\eeq
where $t_0$ is the time when shell-2 was ejected from the central engine, 
$\Gamma_f$ is the final Lorentz factor of merged shells, which is given by
\beq
    \Gamma_f = {m_1\Gamma_1 + m_2\Gamma_2\over \left(m_1^2 + m_2^2 +
        2 m_1 m_2\Gamma_r\right)^{1/2}},
\eeq
and 
\beq
    \Gamma_r = \Gamma_1\Gamma_2(1 - v_1 v_2/c^2)
\eeq
is the relative LF of the two shells before collision.
We see from equation (\ref{t_internal-shock}) that the variability time of the
GRB lightcurve roughly tracks the engine variability time according to this model
(assuming that $\Gamma_2/\Gamma_1$ does not fluctuate wildly during the course
of engine activity). Therefore, the internal shock model is capable of 
explaining the observed short time scale variability (milli-seconds) 
by linking it to the central engine time-scale, whereas external shocks 
occurring at a much larger radius cannot account for this without 
sacrificing the efficiency for producing $\gamma$-ray emission 
\citep{sari97}; efficiency is somewhat problematic
for internal shocks as well (as discussed below), but it is a much more 
severe problem for external shocks when the requirement of milli-second 
(or even 100 ms) time variability is imposed.

The 4-momenta of the two shells before the collision are 
$\Gamma_i m_i(1, v_i,0,0)$; $i=1, 2$. And the momentum after the collision 
of the merged shells moving together = $\Gamma_f m(1, v_f, 0, 0)$.
It is straightforward to show using the conservation of 4-momentum that 
the thermal energy produced in this collision, where the two shells merge 
and move together, is
\beq
   \Delta E = \Gamma_f(m - m_1 - m_2) = \Gamma_f \left[ \left(m_1^2 + m_2^2 + 
   2 m_1 m_2\Gamma_r\right)^{1/2} - (m_1 + m_2) \right] c^2.
  \label{E_collision}
\eeq
Therefore, the efficiency for producing thermal energy in the collision is given by
\beq
   \epsilon_t = {\Delta E\over (m_1 \Gamma_1 + m_2 \Gamma_2)c^2} = \left[ 1 - 
        {m_1 + m_2 \over \left(m_1^2 + m_2^2 + 2 m_1 m_2\Gamma_r\right)^{1/2}} \right].
    \label{col_efficiency}
\eeq
It is easy to see that for a fixed $\Gamma_r$, the highest efficiency is 
achieved when equal mass shells collide, and in that case the efficiency is 
$\epsilon_{max} = [1 - 2^{1/2}/(1+\Gamma_r)^{1/2}]$. Only a fraction of 
the thermal energy produced in collisions is likely deposited in electrons, 
the rest is taken up by protons and magnetic fields. Since protons are very 
inefficient radiators --- the synchrotron and the IC power
for a proton is smaller than an electron of similar energy by a factor 
$(m_p/m_e)^4$ --- the maximum radiative efficiency one might hope to
get in colliding shells is $\sim \epsilon_{max}/2$ when electrons carry 
50\% of the total thermal energy produced in the collision\footnote{The
radiative efficiency can exceed $\epsilon_{max}/2$ if energy could
be transferred from protons to electrons on a dynamical time. However,
considering that the Coulomb interaction cross-section for relativistic
electrons is smaller than the Thomson cross-section this energy transfer
will have to involve some kind of a collective plasma process.}. 
Therefore, the maximum possible
radiative efficiency when two shells of $\Gamma_2/\Gamma_1 = 20$, or $\Gamma_r=10$,
collide is 28.7\%. However, even in this case of extreme LF contrast between 
colliding shells, and the highly idealized situation of equal shell mass, the 
radiative efficiency in the energy band for a typical GRB instrument 
(10 keV --- 10 MeV) is smaller than the bolometric efficiency calculated above
by a factor of a few (IC scatterings produce much higher energy photons in 
internal shocks that carry away a good fraction of electron energy). More 
detailed calculations find the radiative efficiency for internal shocks in the
observer band --- by considering an ensemble of colliding shells with various 
LF distributions --- to be between 1 and 10 percent 
\citep{kobayashi97,kumar99,panaitescu99,kobayashisari01,maxham09}.  
By contrast, the observed efficiency for prompt $\gamma$-ray emission is 
reported to be much higher --- approaching, or possibly exceeding 
50\% \citep{zhang07a,fanpiran06a}.

\cite{beloborodov00} suggested that internal shocks can be efficient.
However, that is based on the assumption that  
$\Gamma_2/\Gamma_1\gg 10$ for almost all collisions, which is unlikely to be
realistic, and he does not take into account the fact that only a fraction 
of thermal energy is radiated in the observing energy band of 10 keV--10 MeV. 

We next discuss whether synchrotron radiation in internal-shocks can account 
for the observed $\gamma$-ray spectrum. 

\subsection{Viability of Synchrotron radiation mechanism for GRBs for shock heated plasma}
\label{shock_synchro}

We provide in this section some constraints on properties of $\gamma$-ray 
sources --- such as the magnetic field strength, Lorentz factor 
of electrons associated with their random motion in the source 
comoving frame ($\gamma_e$) and the Compton-Y parameter ---
using general theoretical arguments when the radiation
mechanism is synchrotron. We assume
that electrons are accelerated on a time scale short compared with the
dynamical time, and subsequent to the phase of acceleration they
have vanishingly small rate of energy gain, which is the case for 
the Fermi acceleration mechanism in internal and external shocks. 

The $\gamma$-ray source might be highly inhomogeneous where magnetic 
fields might occupy a small fraction of the source volume, and only
a fraction of electrons (and positrons) might be accelerated to high
enough Lorentz factor to radiate in the observed $\gamma$-ray band.
The calculations in this section circumvent these complications by 
focusing only on those electrons that radiate in the observer band, 
and we consider only that part of the
source region where the magnetic field is strong enough for these 
electrons to produce $\gamma$-ray photons. If there are a large number of 
regions with very different values of ($\gamma_e$, $B'$) contributing 
roughly equally to the observed flux, then the simplified
calculation presented here is invalid; however, it would require quite a
coincidence for different pairs of ($\gamma_e, B'$) to have the same
synchrotron frequency.

Let us consider the isotropic equivalent of $\gamma$-ray luminosity for a burst
to be $L_\gamma$, and the frequency at the peak of the $\nu f_\nu$ spectrum to
be $\nu_p$ in the observer frame. 

If we associate the peak of the spectrum, $\nu_p \equiv \nu_{p,5}\times 100$ 
keV, with the synchrotron frequency of electrons with Lorentz factor
$\sim\gamma_e$, then that gives
\begin{equation}
{q B' \gamma_e^2 \Gamma\over 2\pi m_e c (1+z)} = 2.4\times10^{19}\nu_{p,5} \; {\rm Hz}
   \quad \implies \quad B'\gamma_e^2 \Gamma_2 = 8.5\times10^{10} (1+z) 
   \nu_{p,5},
   \label{Bp_syn_peak}
\end{equation}
where $B'$ is the magnetic field in the source comoving frame, and $\Gamma$ 
is the Lorentz factor of the source. 

The radiative cooling time for electrons, in the observer frame, is 
\citep[e.g.][]{sari96}
\begin{equation}
t_c = {3\pi m_e c (1+z)\over \sigma_T B'^2 \gamma_e \Gamma (1+Y)}\sim
    (5\times10^{-16}{\rm s}) (1+z)^{-1} \gamma_e^3 \nu_{p,5}^{-2}\Gamma_2 
    (1+Y)^{-1},
  \label{syn_cool1}
\end{equation}
where $Y$ is the Compton-Y parameter, and the second part of the equation was
obtained by substituting for $B'$ using equation (\ref{Bp_syn_peak}). If $t_c$
is much smaller than the dynamical time, $t_d\sim R(1+z)/(2c\Gamma^2)$, then the
rapid cooling of electrons leads to the spectrum below $\nu_p$ to be 
$f_\nu\propto\nu^{-1/2}$ \citep{sari98}, which is much softer than the 
spectrum for most GRBs; the observed low energy spectrum is often close to 
$\nu^0$ (see \S\ref{prompt-spectral} for detailed information regarding 
observations). The condition that $t_c \gae t_d$ implies
\beq
  \gamma_e \Gamma_2 \gae 1.5\times10^5 R_{15}^{1/3} (1+z)^{2/3} 
   \nu_{p,5}^{2/3} (1+Y)^{1/3}.
  \label{gamma_e_min}
\eeq
Thus, for the synchrotron radiation mechanism to be able to explain the GRB 
peak frequency, and the spectrum below the peak, a very high LF of 
electrons is required\footnote{This condition
on $\gamma_e$ is basically equivalent to the problem discussed in 
\cite{ghisellini00}, where they showed that the synchrotron
cooling time for electrons in internal shock with $\gamma_e\sim 10^2$ is
much smaller than the dynamical time.}: $\gamma_e \gg 10^4$.  This 
large $\gamma_e$ might be a problem for the internal shock model (where the LF 
of the shock front is of order a few) unless just one in $\sim 10^2$ electrons
crossing the shock front are accelerated carrying away $\sim10$\% of
the energy of the shocked fluid as suggested by e.g. \cite{daigne11}. 
Numerical simulations of collisionless 
ion-electron shocks find that these requirement might be
satisfied as long as magnetic fields in the unshocked GRB jet carry less
than $\sim 0.1$\% of the luminosity \citep{sironi11a}; the parameters of
these simulations, however, fall far short of GRB jet conditions, and 
therefore one needs to be careful applying simulation results to GRBs.

A large value for $\gamma_e$ implies a small magnetic field (in order for 
$\nu_p\sim 10^2$keV) and in that case IC losses might dominate. The maximum 
magnetic field strength (corresponding to the minimum value for $\gamma_e 
\Gamma$) can be calculated using equations (\ref{Bp_syn_peak}) \& 
(\ref{gamma_e_min}):
\beq
  B' \sim (4\, {\rm Gauss}) \, R_{15}^{-2/3} (1+z)^{-1/3} \Gamma_2 
   \nu_{p,5}^{-1/3} (1+Y)^{-2/3},
   \label{B_syn}
\eeq
and therefore the energy in magnetic fields is
\beq
  E_B = {\Gamma^2B'^2\over 8\pi}{4\pi R^3\over\Gamma^2}\lae (7\times10^{45} {\rm
   ergs}) R_{15}^{5/3} (1+z)^{-2/3} \Gamma_2^2 \nu_{p,5}^{-2/3} (1+Y)^{-4/3}.
  \label{E_B_syn}
\eeq

The energy in electrons can be calculated from the observed
flux at the peak of the spectrum. The synchrotron flux at 
$\nu_p$ depends on the 
total number of electrons (isotropic equivalent), $N_e$, that radiate
at $\nu_p$, i.e. electrons that have LF $\ge\gamma_e$. The synchrotron 
specific luminosity in the jet comoving frame can be obtained by dividing
the total synchrotron power for $N_e$ electrons by $\nu_p$.
Multiplying this with $\Gamma$ gives the luminosity in observer frame. Thus,
\beq
f_p^{syn} = {N_e (1+z)\over 4\pi d_L^2} {\sqrt{3} q^3 B'\Gamma\over m_e c^2} = 
     {(1.8\times10^{-3}\,{\rm mJy})  N_{e,50} B' \Gamma\over 
    (1+z)^{-1} d_{L,28}^2 }, 
\eeq
or
\beq
   N_{e,50} B' \Gamma_2 \sim 5 f_{p,mJy}^{syn} (1+z)^{-1} d_{L,28}^2,
  \label{fp_syn}
\eeq
where $d_L$ is the luminosity distance, and $f_{p,mJy}^{syn}$ is the observed 
flux at the peak of spectrum in mJy. The total number of electrons that 
contribute to the observed flux at $\nu_p$ is obtained by making use of 
equations (\ref{B_syn}) and (\ref{fp_syn}):
\beq
  N_e \sim 1.2\times10^{50} f_{p,mJy}^{syn} d_{L,28}^2 \nu_{p,5}^{1/3} 
  R_{15}^{2/3} (1+z)^{-2/3} \Gamma_2^{-2} (1+Y)^{2/3}, 
 \label{Ne_syn}
\eeq
and therefore the energy in electrons is
\beq
   E_e \sim \Gamma\,N_e\gamma_e m_e c^2 \gae (1.5\times10^{51} {\rm
   ergs}) f_{p,mJy}^{syn} d_{L,28}^2 R_{15} \Gamma_2^{-2} \nu_{p,5} (1+Y).
  \label{E_e_syn}
\eeq
The second part of the equation was obtained by making use of 
(\ref{gamma_e_min}) \& (\ref{Ne_syn}), and the lower limit to electron energy 
is due to the fact that we only have a lower bound on $\gamma_e$ through
equation (\ref{gamma_e_min}).

The ratio $E_e/E_B\propto \Gamma^{-4}$ is much larger than 1 even when
we consider an extreme value for GRB jet LF of $\sim10^3$, and this suggests
that IC scatterings might carry away a large fraction of electron energy to 
produce very high energy $\gamma$-rays ($\gae$ TeV) thereby significantly
increasing the total energy budget for GRBs. To address this concern we
calculate the Compton-Y parameter. 

The optical depth to Thomson scattering can be calculated using equation
(\ref{Ne_syn}) for $N_e$
\begin{eqnarray}
 \tau_e  =  {\sigma_T N_e \over 4\pi R^2} \sim 6\times10^{-6} f_{p,mJy}^{syn} 
   \nu_{p,5}^{1/3} d_{L,28}^2 (1+z)^{-2/3} R_{15}^{-4/3} \Gamma_2^{-2} 
   (1+Y)^{2/3},
  \label{tau2}
\end{eqnarray}
and with this we can now estimate Compton-Y parameter, which for a typical
GRB with $f_{p,mJy}^{syn}=1$, $\nu_{p,5}=1$, $d_{L,28}=1$ and $z=1$ is given by:
\begin{equation}
  Y \sim {\tau_e \gamma_e^2 \over [h\nu_p(1+z)/m_e c^2] (\gamma_e/\Gamma)} 
  \sim \tau_e \gamma_e \Gamma \sim 10^2 R_{15}^{-1} (1+Y) \Gamma_2^{-2},
  \label{Y_syn}
\end{equation}
 where we have included the Klein-Nishina (K-N) reduction to the photon-electron 
scattering cross-section, since the photon energy in the electron-rest
frame exceeds $m_e c^2$; the factor in the denominator,
  $(h\nu_p[1+z]/m_e c^2) (\gamma_e/\Gamma)$, is the ratio of photon
energy in electron rest frame divided by electron rest-mass energy which
is the factor by which the scattering cross-section in the K-N regime
is smaller than the Thomson cross-section.

Equation (\ref{Y_syn}) has no solution for $Y$ {\em unless} $R_{15} 
\Gamma_2^2 \gae 10^2$. If we take $\Gamma\sim 100$ as is inferred for 
an average GRB jet, then $R\gae 10^{17}$cm, which is equal to or larger than 
the deceleration radius for GRB jets. Thus, there is no synchrotron solution 
for the case where the spectrum below the observed peak ($\nu_p$) is 
$\aprop\nu^0$, unless the source lies at a distance from the central engine 
that is close to the deceleration radius and assuming that electrons are 
not continuously accelerated \citep{kumarmcmahon08,beniamini13}.
The solution corresponding to $R\sim 10^{17}$cm, has $\gamma_e \sim 
7\times10^5$, $Y\lae 1$ (eqs. \ref{tau2} \& \ref{Y_syn}), and the total 
energy in electrons that are 
responsible for producing one pulse in a GRB lightcurve is 
$\sim N_e m_e c^2 \gamma_e\Gamma \gae 10^{53}$erg, and the energy in 
magnetic field is $\sim B^2 R^3/2 \lae 10^{50}$ erg; moreover, the lightcurve
variability time is $\sim R/(2c \Gamma^2) \sim 10^2 R_{17}/\Gamma_2^2$s, 
which is much longer than what observations find for most GRBs.

It is very difficult to get around the low-energy spectrum problem for the
synchrotron model as was pointed out by \cite{ghisellini00} more than a 
decade ago. One possible solution is that electrons are continuously
accelerated (as opposed to acceleration while crossing the shock front 
multiple times but no further energy gain while traveling down-stream of 
the shock front). In this case $B'$ can be larger and $\gamma_e$ smaller ---
so that the radiation can be produced at a smaller $R$ while keeping $Y\lae1$ ---
and the radiative loss of energy for electrons is balanced by energy
gain due to continuous acceleration thereby maintaining the low energy
spectrum to be $f_\nu \aprop \nu^0$ \citep{kumarmcmahon08}; see also
\cite{asano09} and \cite{murase12}, who invoked continued acceleration due 
to MHD turbulence down-stream of the shock front. Alternatively, the
electron cooling problem can be alleviated if magnetic fields were to decay 
rapidly downstream of the shock front \citep{peerzhang06}. However, 
a likely serious problem for this model is the excessive energy requirement,
since the luminosity in the IC component might be much larger than the 
synchrotron emission.

\cite{uhm14} have suggested that the decrease of the magnetic field
with distance from the center of explosion offers another way to 
explain the low energy spectral index for GRB prompt emission. The idea
is that the synchrotron loss rate for $\gamma$-ray emitting electrons decreases
rapidly with time as these electrons move to larger distances where the 
magnetic field is weaker. Therefore, these electrons do not cool 
much to give rise to a $f_\nu \propto \nu^{-1/2}$ spectrum. This mechanism 
works well as long as the magnetic field strength at the radius where electrons
are accelerated is such that the synchrotron cooling time is not much
smaller than the dynamical time. This suggests, using equations 
(\ref{Bp_syn_peak}) and (\ref{syn_cool1}), that $\gamma_e\sim 10^5$
and the magnetic field strength is relatively small (Poynting luminosity 
$\sim 10^{46}R_{15}^2$ erg s$^{-1}$) for this mechanism to 
be effective at explaining the low energy spectral index. While the 
required large emission radius is consistent with constraints of a large
$R_\gamma$ (\S\ref{source_distance}), it is unclear how a small magnetization 
parameter could be achieved in the emission region. 

Yet another solution to the low-energy spectrum problem is for electron cooling
to be dominated by IC scatterings in K-N regime. In this case, the low
energy electron spectrum is $dn_e/d\gamma_e\propto\gamma_e^{-1}$ --- in
the limit of large $\gamma_e$ and IC scatterings in K-N regime as the dominant
energy loss mechanism for electrons \citep{derishev01,nakar09,daigne11} --- and
consequently $f_\nu\aprop \nu^0$. One of the drawbacks with this solution
is the extreme value for $\gamma_e$ required ($\gae10^6$), and even then 
the low energy spectrum is found to be no harder than $\nu^{-0.1}$, which 
fails to account for the observed spectrum for a substantial number of 
GRBs \citep{barniolduran12}.

The bottom line is that the GRB prompt emission can be produced by the
synchrotron process provided that electrons are either continuously
accelerated, or that there is some mechanism that prevents their rapid 
radiative cooling to ensure that the spectrum below the peak is 
consistent with observations.

\subsection{Constraints on Synchrotron-self-Compton mechanism for GRBs}
\label{shock_ic}

The peak frequency and flux at the peak for the SSC case are
$\nu_{syn}\gamma_e^2$ and $f_p^{syn} \tau_e$ respectively; where $\nu_{syn}$ 
and $f_p^{syn}$ are the synchrotron peak frequency and the peak flux, 
$\tau_e$ is the optical depth of the source to 
Thomson scattering and $\gamma_e$ is LF of electrons with characteristic
synchrotron frequency of $\nu_{syn}$.
Equating the IC frequency to the observed peak frequency $\nu_p$, and the 
IC flux to the observed peak flux $f_{p,mJy}$ (in milli-Jansky) provides 
the following constraints:
\begin{eqnarray}
   B'\gamma_e^4 \Gamma_2 & \sim & 8.5\times 10^{10} (1+z)\nu_{p,5}\\
   \tau_e N_{e,55} B' \Gamma_2 & \sim & 5\times 10^{-5}f_{p,mJy} (1+z)^{-1} 
   d_{L,28}^2 \\ \quad {\rm or} \quad \tau_e^2 R_{15}^2 B' \Gamma_2 & \sim &
    2.5\times 10^{-5} f_{p,mJy} (1+z)^{-1} d_{L,28}^2,
\end{eqnarray}
where we made use of equations (\ref{Bp_syn_peak}) \& (\ref{fp_syn}) for 
synchrotron frequency and flux, and substituted for $N_e = 4\pi R^2 
\tau_e/\sigma_T$ in the last part.
Moreover, taking the radiative cooling time for electrons of LF $\gamma_e$
to be of order the dynamical time, for an efficient production of $\gamma$-rays,
and to ensure that the low energy spectrum does not become cooling dominated 
($f_\nu \propto \nu^{-1/2}$) as suggested by data for most GRBs, requires
\begin{eqnarray}
   B'^2 \gamma_e R_{15} (1+Y) \sim 2.3\times10^6 \Gamma_2.
\end{eqnarray}

These equations can be solved for $\gamma_e$, $B'$, $\tau_e$ \& $Y$ to yield:
\begin{eqnarray}
   \gamma_e & \sim & 164 R_{15}^{1/7} \Gamma_2^{-3/7} (1+Y)^{1/7} (1+z)^{2/7}
       \nu_{p,5}^{2/7}, \label{gammae2} \\
   B' & \sim & (120\, {\rm Gauss}) R_{15}^{-4/7} \Gamma_2^{5/7} (1+Y)^{-4/7}
     (1+z)^{-1/7} \nu_{p,5}^{-1/7}, \\
   \tau_e & \sim & 5\times10^{-4} R_{15}^{-5/7} \Gamma_2^{-6/7} (1+Y)^{2/7}
    (1+z)^{-3/7} \nu_{p,5}^{1/14} f_{p,mJy}^{1/2} d_{L,28}, \\
   Y & \sim & \gamma_e^2 \tau_e \sim 300 R_{15}^{-1} \Gamma_2^{-4}
     (1+z)^{1/3} \nu_{p,5}^{3/2} f_{p,mJy}^{7/6} d_{L,28}^{7/3}.  
\end{eqnarray}
Thus, for a solution with $Y\sim 1$ (so that the second 
IC-scattering does not carry away too much energy), we require 
$R \sim 10^{16}$cm (if we take $\Gamma\sim 200$); $\tau_e \sim 5\times
10^{-5}$ at this distance. For most bursts that have optical
data available during the prompt gamma-ray phase, it is found that the specific
flux at the optical band is just a factor 10 or so larger than the
$\gamma$-ray flux, and not a factor $1/\tau_e\sim 10^4$ as
one would expect if $\gamma$-rays are produced via the SSC process.
 For the SSC peak to be at 100 keV,
$\nu_p^{syn} \sim 10^5/\gamma_e^2 \sim (3.7\, eV) R_{15}^{-2/7}\Gamma_2^{6/7}
   (1+Y)^{-2/7}$ --- see equation (\ref{gammae2}) for $\gamma_e$ --- 
and that has a very weak dependence on $R$ and $Y$, and a not particularly
strong dependence on $\Gamma$. So it is unlikely that $\nu_p^{syn}$ 
could be far below or above 2 eV, and therefore it is not possible to suppress 
the optical flux associated with the synchrotron seed field by a factor 
$\sim 10^3$, in order for that to be compatible with the observed optical 
data (or upper limit). Another drawback of the SSC mechanism is that the second 
order SSC would carry more energy, which greatly increases the total energy 
budget of the burst \citep{derishev01}. This point was recently emphasized
by \cite{piran09} who concluded that the SSC mechanism is not viable
for a typical GRB as the energy in seed photons or the second IC component 
is excessive. One other difficulty with the SSC mechanism is that its $E_p$
is very sensitive to the electron injection Lorentz factor ($\propto
\gamma_{\rm inj}^4$), so that it requires fine tuning to obtain the
typical $E_p\sim 10^2$ keV \citep{zhangmeszaros02c}.

The same conclusion can be obtained from the Fermi/LAT data alone. The 
lack of an excess flux at high energies --- there is no evidence 
for departure from a Band function fit for most GRBs e.g. \cite{ackermann13a}
 --- means that the IC 
scattering of photons near the peak ($\nu_p$) into the LAT band should have 
a flux small compared with the Band-function flux.
Let us consider the case where $\gamma_e \lae \Gamma$ (IC scatterings take 
place in the Thomson regime in this case). The lack
of a bump in Fermi/LAT band requires $\tau_e \lae \gamma_e^{2(\beta+1)}/5\sim
   \gamma_e^{-2.5}/5$; where $\beta \sim -2.2$ is high-energy photon index,
and the factor 5 takes into account the fact that any 
departure from a Band-function-fit for most bursts detected by 
Fermi/LAT is less than $\sim 20$\%.
 The implication of this is that $\tau_e \lae 10^{-3}$ for $\gamma_e\gae
 5$, and thus one expects a bright optical flash ($\gae 1$ Jy or 9-mag) 
whenever $\gamma$-rays are produced via the SSC process; we note that
 $\gamma_e^2 \gg 20$
since the spectrum between 10 keV and $\nu_p$ is a flat, single power law,
function. A similar result is obtained for $\gamma_e\gae \Gamma$.

We close this sub-section with a brief discussion of an exceptional burst, 
GRB 080319B (the ``naked-eye'' GRB), which 
was detected to have a very bright optical counterpart during the
prompt phase -- it reached a peak apparent magnitude of 5.8 $\sim30$s
after the GRB trigger -- that roughly tracked the $\gamma$-ray lightcurve
\citep{racusin08}. An attractive possibility for this burst is that 
the optical emission was produced by the synchrotron process, while 
$\gamma$-rays were due to the SSC mechanism \citep{kumarpanaitescu08,racusin08}.
However, the $\gamma$-ray light curve for this GRB varied more rapidly than the 
optical flux, which poses problems for the simplest SSC model 
\citep{resmi12}, but is consistent with the relativistic turbulence 
model \citep{kumarnarayan09}.
People have also invoked a two zone model to interpret $\gamma$-ray and 
optical emissions from this burst \citep{liwaxman08,yu09,fan09,zou09b}.

\subsection{General constraints on electron Lorentz factor ($\gamma_e$)}
\label{electron_lf}

The calculation below, based on very general considerations, 
shows that the Lorentz factor of electrons associated with their 
random motion ($\gamma_e$) in GRB jets, at the site of $\gamma$-ray 
generation, is either less than 2 or larger than $\sim10^2$.

Let us consider the isotropic $\gamma$-ray luminosity in the CoE
frame to be $L_\gamma$. We assume that $\gamma$-rays are produced 
by electrons (and positrons) by some combination of synchrotron 
and IC processes.

We will consider two cases separately. 
\begin{enumerate}
\item Short lived acceleration phase: when electrons in the jet are accelerated 
to a typical LF $\gamma_e$ on a timescale much smaller than the dynamical 
time (for instance, while crossing the shock front back and forth multiple 
times), and subsequently they radiate a part of their energy to produce 
$\gamma$-ray photons of frequency $\nu_p$, but this loss of their energy is 
not compensated by any further acceleration (such is the case for electrons
moving down-stream after crossing the shock-front for the last time); a more
accurate calculation should consider a distribution of electron LF, however
this changes the result by a factor of a few, which is of little concern here.
The energy-luminosity carried by electrons (and positrons) at the end
of the acceleration phase, $L_e$, in this case should be at least as large 
as $L_\gamma$, and so we take
\beq
   L_e = \zeta L_\gamma,
\eeq
where $\zeta\ge1$ is a dimensionless parameter of order no larger than a few
so that the GRB radiative efficiency is roughly of order the observed value.

\item Continuous acceleration: when electrons in the jet are continuously,
or repeatedly, accelerated while they are producing $\gamma$-rays, so 
that the energy they loose to radiation is balanced by the gain from the
acceleration mechanism (the details of this process are unimportant, but such
a scenario could operate in magnetic reconnections inside a current sheet).
The luminosity carried by $e^\pm$s in this case can be much smaller than 
$L_\gamma$.
\end{enumerate}

\subsubsection{Short lived acceleration phase for electrons}
\label{electron_lf1}

The comoving number density of electrons \& positrons, $n_e$, is related 
to the luminosity, $L_e$, carried by e$^\pm$ as follows --
\beq
  L_e = 4\pi R^2 m_e c^3 (\gamma_e -1) \Gamma^2 n_e = \zeta L_\gamma,
\eeq
or
\beq
  n_e = {\zeta L_\gamma \over 4\pi R^2 m_e c^3 (\gamma_e -1) \Gamma^2 },
  \label{ne_Lgam}
\eeq
where R is the distance from the center of explosion, $\Gamma$ is the 
jet LF and $\gamma_e$ is the average LF of electrons (in jet comoving frame)
that emit photons of frequency $\nu_p$ (peak of the $\gamma$-ray spectrum).

The optical depth to Thomson scattering for electrons of LF $\gamma_e$
is given by
\begin{equation}
\tau_e = \sigma_T n_e \min\{c t'_c, R/\Gamma \},
   \label{taue_game}
\end{equation}
where $t'_c$ is the radiative cooling time for electrons of LF $\gamma_e$ in
the jet comoving frame. The upper bound on the cooling time is provided by the
 IC loss of energy, and is given by 
\beq
   t'_{ic} = {4\pi R^2 \Gamma^2 m_e c^2 \over (\gamma_e + 1)\sigma_T L_\gamma }
    = (150 \, s) {R_{15}^2 \Gamma_2^2 \over (\gamma_e + 1) L_{\gamma, 51} }.
   \label{tp_ic_game}
\eeq
The ratio of the cooling and the dynamical time, $t_d' = R/c\Gamma$, is
\beq
{t'_{ic}\over t'_d} = {0.5 R_{15}\Gamma_2^3 \over(\gamma_e + 1)L_{\gamma, 51}}.
   \label{tp_ic_game1}
\eeq
 
Let us assume that the ratio of energy loss rate for an electron of 
LF $\gamma_e$ due to IC scatterings (considered above) and the loss
rate associated with the radiation mechanism that produced photons 
of frequency $\sim \nu_p$ is $Y$. The electron cooling time, $t'_c$, 
can then be written as $t'_c \sim t'_{ic} Y/(Y+1)$.
Making use of this relation, and equations (\ref{ne_Lgam}), (\ref{taue_game}),
and (\ref{tp_ic_game}) we find the optical depth to Thomson scattering 
of electrons responsible for the observed $\gamma$-rays to be
\beq
\tau_e \sim {\zeta \over \gamma_e^2 - 1} {Y \over Y+1},
   \label{taue_game2}
\eeq
as long as $t'_c < t'_d$; the optical depth is smaller than given by the
above equation by a factor $t'_c/t'_d$ when $t'_c>t'_d$. We note that 
$Y$ cannot be much smaller than 1 since that would imply the energy
in magnetic fields (or seed photons that get IC scattered to $\nu_p$) 
to be much larger than the energy in prompt $\gamma$-ray radiation, and
hence a low efficiency for producing prompt radiation, which is 
not supported by the data.

We can now obtain limits on $\gamma_e$ using equation (\ref{taue_game2})
by requiring that IC scatterings of sub-MeV photons should not
produce a bump in the observed spectrum above $\nu_p$ since {\it Fermi} finds
no evidence for such a bump.

Electrons that produce $\gamma$-rays near the peak of the spectrum ($\nu_p$) 
also IC scatter these photons to a frequency $\nu_{ic} \sim \gamma_e^2 \nu_p$. 
The specific flux at $\nu_{ic}$ is $\sim \tau_e f_p$; where $f_p$ is
the flux at $\nu_p$. From equation (\ref{taue_game2}), we see that the IC 
flux exceeds the underlying seed photon flux at $\nu_{ic}$ as long as
the observed $\gamma$-ray spectrum above $\nu_p$ is not shallower than
$\nu^{-1}$ (which is never the case by the definition of the spectral peak),
and $\zeta Y$ is not much less than 1 (which is unlikely due to radiative
efficiency considerations).
Therefore, IC scatterings of sub-MeV photons by electrons would
produce a prominent second peak in the spectrum above $\nu_p$ that could 
lie in the Fermi/GBM or LAT energy band (1 MeV --- 300 GeV) depending on 
the value of $\gamma_e$. 

One way to avoid this second peak (which is not found in GRB spectra) 
is if $\gamma_e$ is less
than $\sim 1.5$ so that the IC and the seed photon peaks merge together
to produce a single peak in the emergent spectrum. Another possibility 
is that $\gamma_e \gae \Gamma\sim 10^2$ so that the IC 
scattering cross-section is reduced due to Klein-Nishina effect, thereby
suppressing the bump in the spectrum above $\sim 10$ MeV.
It could be that the IC scattered photons 
 are converted to electron-positron pairs by 
interacting with lower energy photons as they make their way out of
the source region, and therefore don't contribute to the observed 
flux at high energies \citep{gupta07b}; however, this is not so
likely for $\Gamma\gae 200$ when the pair production optical depth is small
(\S\ref{source_distance}).

The bottom line is that $1.5 \lae \gamma_e \lae 10^2$ can be ruled
out due to the fact that it gives rise to an IC bump above the peak of the 
observed spectrum in the Fermi energy band. Solutions with $\gamma_e 
\lae 1.5$ have $\tau_e\sim 1$, which can be identified as photospheric 
radiation with possibly multiple IC-scatterings accounting for a 
power law spectrum above $\nu_p$.

\medskip
\subsubsection{Continuous/repeated acceleration of electrons}
\label{electron_lf2}

When electrons are continuously, or repeatedly, accelerated while losing 
energy to radiation, their average Lorentz factor in the jet comoving frame 
($\gamma_e$) is such that the energy gain and loss rates are balanced. 
The observed luminosity in this case is:
\begin{equation}
   L_\gamma = 4 \pi R^2 n_e m_e c^3 (\gamma_e-1) \Gamma^2
          {R\over c\Gamma} {1\over t'_c}.
\end{equation}
Therefore, the optical depth to Thomson scattering is
\begin{equation}
   \tau_e = \sigma_T n_e R/\Gamma = \sigma_T {L_\gamma\over 4\pi R^2 \Gamma^2 c}
     {t'_c \over m_e c (\gamma_e -1 ) } = {Y\over (Y+1) (\gamma_e^2 -1) }.
\end{equation}

This optical depth is basically the same as that in equation 
(\ref{taue_game2}), and the constraint on $\gamma_e$ obtained in the previous
subsection holds, i.e. $1.5\lae\gamma_e\lae10^2$ is ruled out.

\medskip
\subsection{Effects of neutrons on jet dynamics and radiation}
\label{np_collision}

GRB jets might be produced by some hydrodynamic
processes in an accretion disk around a black hole or a neutron star. In this
case, the jet composition could include free neutrons 
that are produced by the dissociation of nuclei by $\gamma$-ray photons in 
the inner regions of the disk. These neutrons decouple from protons at 
a radius smaller than the Thomson photosphere 
\citep[e.g.][]{derishev99,bahcall00} ---
due to a smaller cross-section for neutron-proton scattering --- and their
collisions below the photosphere can significantly affect the jet dynamics,
e.g. \cite{rossi06}, and produce positrons (by the decay
of pions) that IC scatter thermal photons and produce $\gamma$-ray radiation 
with peak at $\sim 1$ MeV \citep{beloborodov10}. Neutrons that survive these 
collisions travel to larger distances before decaying and that could affect 
the afterglow radiation from GRBs \citep{beloborodov03b,fan05b}.

Neutrons and protons in the GRB outflow move together as a single fluid as 
long as the timescale for a neutron to collide with a proton is smaller than 
the dynamical time. The collision time, in the jet comoving frame, is
\beq
  t'_{np} = {1\over \sigma_{np} n'_p v}\sim {4\pi R^2 m_p c^2 \eta\Gamma\over
       L\sigma_{nuc} },
   \label{tnp}
\eeq
where $\sigma_{np}=\sigma_{nuc} c/v$ is the cross-section for neutron-proton
scatterings, $\sigma_{nuc}\approx 3\times10^{-26}$ cm$^2$; we made use of 
equation (\ref{np1}) for the proton density ($n'_p$) to arrive at the second 
equality. The scattering is elastic when $v\ll c$ and it becomes inelastic 
that produces pions when $v\sim c$.
The dynamical time in the fluid comoving frame is
\beq
  t'_d\sim R/(c\Gamma).
  \label{td}
\eeq
When $t'_{np} > t'_d$, neutrons and protons decouple, and the radius where this
occurs is
\beq
   R_{np} \sim {\sigma_{nuc} L\over 4\pi m_p c^3 \eta\Gamma^2}.
   \label{R_np}
\eeq
In deriving this equation we have assumed that the luminosity carried by 
neutrons is of the same order as protons, and the LF $\Gamma$ is the smaller 
of proton and neutron LFs.
We note that $R_{np}$ is smaller than Thomson photosphere radius by a factor
$\sim 20$, since $\sigma_T/\sigma_{nuc}\sim 20$. If $R_{np}$ is smaller 
than $R_s$ --- the radius where protons attain their terminal speed --- 
neutrons stop accelerating before protons do,
and the free energy of neutron-proton differential motion is dissipated below
the photosphere, and can be used for producing a non-thermal photon spectrum.
The condition for non-zero differential velocity to arise is
\beq
  R_{np} < R_s \quad\quad {\rm or} \quad\quad \Gamma > \left[ {\sigma_{nuc}L
  \over 4\pi m_p c^3 R_0}\right]^{1/4} = 485 L_{52}^{1/4} R_{0,7}^{-1/4}.
   \label{gamma_np}
\eeq
Thus, a substantial fraction of the kinetic energy of those GRB jets that 
consist of neutrons and protons, and reach terminal LF larger than 
about 500, is dissipated below the Thomson photosphere. A fraction of this
energy goes into producing $e^\pm$s that can scatter photons to possibly 
produce the observed $\gamma$-ray spectrum \citep{beloborodov10}.

Observational evidence does not point to many GRB jets having 
Lorentz factor larger than what is needed for this mechanism to operate 
(eq. \ref{gamma_np}). However, neutrons and protons can 
develop substantial differential velocity even when the condition 
in equation (\ref{gamma_np}) is not satisfied. This can happen, for instance,
in internal shocks where protons are slowed down (and accelerated) in the 
collision, whereas neutrons continue moving at the speed they had before 
the collision \citep{beloborodov10}. However, in this case only a small 
fraction of the energy of differential motion is dissipated, unless shell 
collisions were to take place close to $R_{np}$. If the variability time for 
the central engine is $\delta t$, and the Lorentz-factor 
of the slower part of the outflow is $\Gamma$, then the radius where collisions
take place is $R_{col} \sim 2c\Gamma^2 \delta t= 6\times10^{12}\Gamma_2^2
  (\delta t)_{-2}$ cm; $(\delta t)_{-2}$ is variability time in units of 
10$^{-2}$s. The radius where the probability for n-p collision
drops below one-half is $R_{np} \sim 5\times10^{11} L_{n52} \Gamma_2^{-3}$cm.
Thus, $R_{col}/R_{np}\sim 10 L_{n52}^{-1} \Gamma_2^5 (\delta t)_{-2}$.
For an efficient conversion of outflow kinetic energy to thermal energy
via n--p collisions these radii should be approximately equal, i.e.
$R_{col}\sim R_{np}$, and that requires $50\lae\Gamma<10^2$; considering 
that $R_{col}/R_{np}\propto \Gamma^5$, these limits are quite firm, and 
the allowed range for $\Gamma$ is uncomfortably narrow\footnote{When 
$R_{col}/R_{np}\ll1$, shell collisions affect both protons and neutrons since 
they are well coupled below $R_{np}$. Moreover, any radiation produced in 
such a collision would find itself in a medium of high optical depth and 
hence, the emergent spectrum is nearly thermal, which is inconsistent 
with observed spectra of most GRBs.  For $R_{col}/R_{np}\gg1$, neutrons 
have a small probability for collision and the radiative efficiency is low.}. 
Moreover, given a random distribution of $\delta t$ and $\Gamma$, and assuming
that they are uncorrelated, there should be many collisions where
$R_{col}\gg R_{np}$ or $R_{col}\ll R_{np}$ and the spectra produced in 
these collisions would be very different from those for which 
$R_{col}\sim R_{np}$. 
The problem is that observations do not find big variations of $E_{peak}$ 
and photon indices for time resolved spectra. Nor do they find systematic
 differences in spectra for bursts that show fast variability and those 
that do not (including the extreme case of FRED bursts that have a 
single, smooth, pulse in the lightcurve). 

\begin{figure}
\begin{center}
\includegraphics[width=13cm]{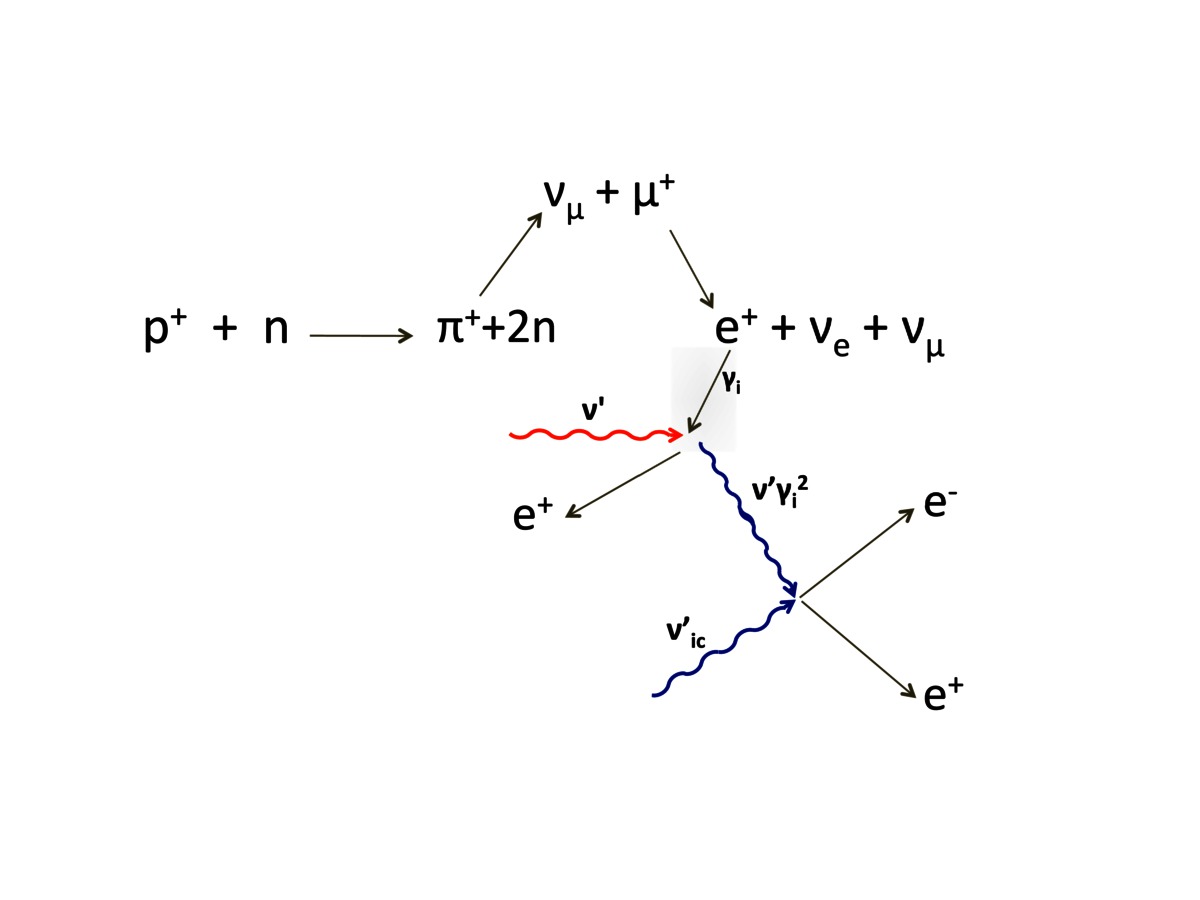}
\caption{This figure provides a quick overview of the $p$--$n$ collision
 process that produces a pion ($\pi^+$), which decays to give a  
positron of LF $\gamma_i\sim 10^2$. This positron inverse Compton 
scatters thermal photons of frequency $\nu'$ to energy $\sim h\nu'
\gamma_i^2> m_e c^2$ (see eq. \ref{epsilon_th_ic}), which in turn collides
with another IC photon to produce $e^\pm$ pair. This cascade to more and 
more pairs continues until $e^\pm$ LF drops below a few.
}\label{FIG:photo-pion}
\end{center}
\end{figure}

Neutron-proton differential velocity could arise for a structured jet where 
neutrons from the outer, slower, part of the jet diffuse toward the middle 
region, where the plasma is moving at a higher speed \citep{meszarosrees00b}.
We now assume that somehow neutron-proton differential velocity gets 
set up in a GRB jet, and look at the sequence of events leading to 
generation of $\gamma$-rays due to neutron-proton collisions for this GRB.
The basic processes are sketched in Fig. \ref{FIG:photo-pion}, and the 
emergent radiation can be understood by using simple physical considerations 
described below; much of this discussion closely 
follows the work of \cite{beloborodov10}.

The inelastic collision of a neutron with a proton produces a pion 
($n+p^+\rightarrow \pi^+ + n + n$; it can also produce a $\pi^-$ or a $\pi^0$). 
The charged pion decays in 26 nano-seconds to a muon, and the muon decays 
in 2.2 $\mu$s to a positron. If the relative LF of a collision between neutron 
and proton is $\Gamma_r \sim 2$, then the LF of the positron produced 
is $\gamma_i\sim \Gamma_r m_\pi/(6 m_e)\sim 100$; the factor $1/6$ accounts
for the fact that $\mu^+$ carries away roughly half of the energy of 
$\pi^+$ when it decays and the decay of a $\mu^+$ imparts roughly equal
energies to $e^+$ and the two neutrinos.

These positrons IC scatter thermal photons produced at the jet launch site,
and carried by the outflow to larger radii, to an energy, in the jet comoving 
frame, that is 
\beq
   \epsilon_{th}'^{ic}\sim 3 k_B T_0 (R_s/R_{np})^{2/3} \gamma_i^2/\Gamma \sim 
   (40 MeV) L_{52}^{-5/12} R_{0,7}^{1/6}\Gamma_3^{5/3}.
   \label{epsilon_th_ic}
\eeq
In deriving this equation, we made use of (\ref{T_R0}), (\ref{Trp}), (\ref{Rs}) 
and (\ref{R_np}), and we assumed that neutrons and protons carry roughly equal
fractions of the jet luminosity, and also that the LF of neutron jet is $\sim4$ 
times smaller than the proton jet; $\Gamma_3\equiv\Gamma/10^3$ is proton jet
 LF\footnote{Beloborodov did not correct for the decrease of thermal photon 
energy --- the factor $(R_s/R_{np})^{2/3}$ --- in his calculations, which has 
an effect on the emergent spectrum.}. The energy of these
scattered photons is larger than $m_e c^2$ and they get converted
to $e^\pm$ pairs since the optical depth for $\gamma+\gamma\rightarrow
e^\pm$ is larger than 1 below the photosphere\footnote{The collision
of thermal photons with IC scattered photons of energy a few MeV cannot 
produce pairs since the thermal photons have too little energy ($\lae 1$ keV)
in the comoving frame. Instead, MeV photons must collide with other MeV,
 non-thermal, photons to produce pairs.}.

The ratio of IC loss time for $e^\pm$ of LF $\gamma_e$ and the dynamical time 
is very small at the photospheric radius ($R_{ph}\sim10^{12}$cm -- eq. \ref{rp})
 even for $\gamma_e\sim 1$ (see eq. \ref{tp_ic_game1}), and thus the LF of 
pairs drops rapidly to order unity. Considering the high optical depth to 
IC scattering for $R\sim R_{np}$, photons are repeatedly scattered by 
$e^\pm$ and a good fraction of energy of the first generation pairs is 
used up in producing more pairs. This process stops when pair LF drops below
$\sim2$, and  IC energy loss rate for $e^\pm$ is balanced by energy gain due 
to interactions 
with protons. The number of pairs created per n-p collision is of order the 
LF of the first $e^\pm$ produced by muon decay. These secondary pairs dominate 
the Thomson scattering optical depth as they outnumber electrons associated
with protons by a factor $\sim 10^2$. The nearly thermal population of $e^\pm$ 
of mildly relativistic temperature ($\gamma_e<2$) scatter thermal photons
multiple times to produce a non-thermal emergent spectrum that has a peak
not far from the peak of the underlying seed-thermal-photons, and the low \&
high energy spectra are $f_\nu \aprop \nu$ \& $\nu^{-1}$ respectively
(see \S\ref{ic_photo} for a more detailed discussion of spectrum produced in 
multiple-IC scatterings). For the peak of the emergent spectrum to be
of order a few $10^2$ keV, n-p collisions should not take place far from 
$R_s\sim 10^{10} R_{0,7}\Gamma_3$ cm, otherwise adiabatic expansion shifts the
thermal peak to an energy below a typical GRB spectral peak. Thus, for a
highly fluctuating central engine, where $\delta t$ has a broad distribution,
many collisions are expected to occur far away from the photosphere and these
should produce $\gamma$-ray spectra with peaks at low energies. However, 
even though variations of $E_p$ has been observed in individual GRBs, 
observations do not find large variations as expected from this model during
the course of a burst.

If a GRB-jet has a non-zero magnetic field --- which is very likely --- 
then there can be significant synchrotron radiation produced by positrons 
from $\pi^+$ decay, and that would modify the IC spectrum.  We note, however, 
that if the synchrotron process were to be the dominant radiation
mechanism below the peak of the spectrum, then the low energy spectral index 
should be $\alpha=-1.5$ -- since the radiative cooling time for $e^\pm$ is very
short at the photospheric radius -- and 
that is too small (soft) for most GRBs. \cite{vurm11} report 
that a combination of thermal and synchrotron radiations results in 
$\alpha\sim -1$, which is roughly where the observed
$\alpha$-distribution peaks. However, it turns out that the addition of a
synchrotron component also steepens the high energy photon index ($\beta<-3$) 
and that is significantly smaller than the average observed value for $\beta$. 
Moreover, it is unclear how a combination of synchrotron and 
thermal spectra can produce a smooth, single peak, Band-function between 
10 keV and ∼ 10$^2$ MeV without some fine tuning of parameters.
Vurm et al. (2011) also find a prominent bump in their spectra at ∼ 300 MeV 
(in the GRB host galaxy rest frame) due to annihilation of pairs of 
LF $\gamma_e\sim 1.05$. Such a feature has never been seen for any GRBs, 
perhaps because the bump is smeared out when data is integrated over a 
finite time interval. This bump 
disappears when the energy fraction in magnetic fields in the jet is 
taken to be larger than $\sim0.5$. However, in this case the observed flux 
above the spectral-peak falls off extremely rapidly \citep{vurm11}, which is 
inconsistent with observations.

\subsection{Prompt $\gamma$-rays from photosphere: processed thermal photons}
\label{ic_photo}

We consider the ``photospheric'' model for the generation of 10 keV--10 MeV 
$\gamma$-rays during the main burst in this sub-section.
According to this model a population of nearly thermal ``seed'' photons 
interact with electrons below the Thomson photosphere to produce GRB 
spectrum.  It is assumed that the 
average seed photon energy is much smaller than the electron's energy. In this 
case, photons typically gain energy by scattering off of electrons,
and the energy continues to increase as they undergo 
multiple scatterings as long as their energy is less than
the average electron energy. There are a number of excellent
articles that discuss multiple-IC scatterings and the emergent spectrum
at great depth \citep{katz76,shapiro76,rybicki79,sunyaev80,ghisellini12}.
Here we provide a simple physical picture of this process and its application
to GRBs.

The average frequency of a photon scattered off of an electron ($\nu_s$) is,
e.g. Rybicki \& Lightman (1979)
\beq
  \nu_s/\nu_i \equiv A_f = 1+ 4 k_BT/m_e c^2 
\eeq
for non-relativistic electron temperature $T$, and
\beq
  \nu_s/\nu_i = 1 + 4\gamma_e^2/3
\eeq
for highly relativistic electrons of LF $\gamma_e$ in Thomson scattering regime; 
where $\nu_i$ is the photon frequency before scattering.

The number of scatterings it takes for the average photon energy to approach that 
of electrons is $N \approx \ln(k_BT/h\nu_i)/\ln A_f$, which for sub-relativistic 
electrons can be rewritten as $N \approx (m_e c^2/4 k_B T) \ln(k_B T/h\nu_i)$.
If these scatterings take place in a medium of Thomson optical depth 
$\tau_T$, then the average number of scatterings suffered by a photon before
it escapes from the surface is $\sim \max(\tau_T, \tau_T^2)$, and in that case
it is useful to define a parameter 
\beq
   Y \equiv \max(\tau_T, \tau_T^2) \max\left[ (4 k_B T/m_e c^2), 4\gamma_e^2/3\right],
\eeq
called the Compton-Y, that captures the information regarding whether photons 
undergo sufficient number of scatterings while traveling through the medium 
to thermalize with electrons or not. 

For $Y\ll 1$, the emergent photon spectrum is not too different from the seed 
photon spectrum except that it can develop a power law tail above the peak, 
which for $\tau_T < 1$ has photon index $\beta = \ln(\tau_T/A_f)/\ln(A_f)$; 
photon index is defined as: $n(\nu)\propto\nu^\alpha$, where $n(\nu)$ is the
number of photons per unit frequency.

For $Y\gg 1$ ---  called saturated Comptonization ---  it follows from the 
discussion above that an average seed photon undergoes sufficient number of 
inverse-Compton scatterings before escaping from the medium so that its energy 
is approximately equal to the average electron energy. If the electrons have
a thermal distribution, then the emergent photon spectrum is Bose-Einstein 
distribution with a non-zero chemical potential, since the number of 
photons is conserved; the spectrum has a Wien shape where the specific flux
below the peak scales as $\nu^3$, instead of $\nu^2$ for a black-body
spectrum. 

For the intermediate case of $Y\sim 1$ \& $\tau_T>1$ --- un-saturated or 
quasi-saturated Comptonization --- the emergent spectrum is more complex 
and is obtained by solving the Kompaneets equation. However, the qualitative 
behavior of the spectrum can be understood using simple arguments
described below\footnote{This simple physical picture closely follows 
a discussion PK had with Lev Titarchuk in Ferrara, Italy, in summer 2011.}.

Consider a slab of optical depth $\tau_T$ (measured from the mid plane of 
the slab to its surface) consisting of hot but non-relativistic electrons.  
Let us assume that photons of energy much smaller than the average electron 
thermal energy are injected at the mid-plane of the slab. These photons
undergo a number of IC scatterings before arriving
at the surface. For $Y\sim 1$ \& $\tau_T>1$, the peak of the photon 
spectrum at the surface of the slab lies at a higher energy than the injected
photons, but the mean number of scatterings is not sufficiently large for the
emergent radiation to have attained thermal equilibrium with electrons.
If the probability of scattering for a photon while crossing the medium
is $p$, then the mean number of scatterings suffered before escape is
$\langle N\rangle \sim \sum_k k p^k (1-p) = p/(1-p)$, or $p \sim \langle 
N\rangle/(1 + \langle N\rangle)$. The peak of the emergent photon spectrum 
in this case lies at frequency  $\nu_p^{ic} \sim A_f^{\langle N\rangle}\nu_p$,
where $\nu_p$ is the
peak of the seed photon spectrum; $\nu_p^{ic} \sim \nu_p \exp(Y)$ is within
a factor of a few of $\nu_p$ since $Y\sim 1$, and hence the photon energy
at the peak of the emergent spectrum is likely to be much smaller than
the mean electron energy. The photon spectral index above the 
peak is $\beta \sim \ln(p/A_f)/\ln(A_f) \sim -1 -1/[\langle N\rangle 
 \ln(A_f)]$. In the limit of a large optical depth, i.e.  $\langle N\rangle 
\sim \tau_T^2\gg1$, and a small gain factor ($k_B T/m_e c^2 \ll 1$)
 $\beta \sim -1 -1/(\tau_T^2 4 k_B T/m_e c^2) = -(1+Y)/Y$.

\begin{figure}
\begin{center}
\includegraphics[width=13cm]{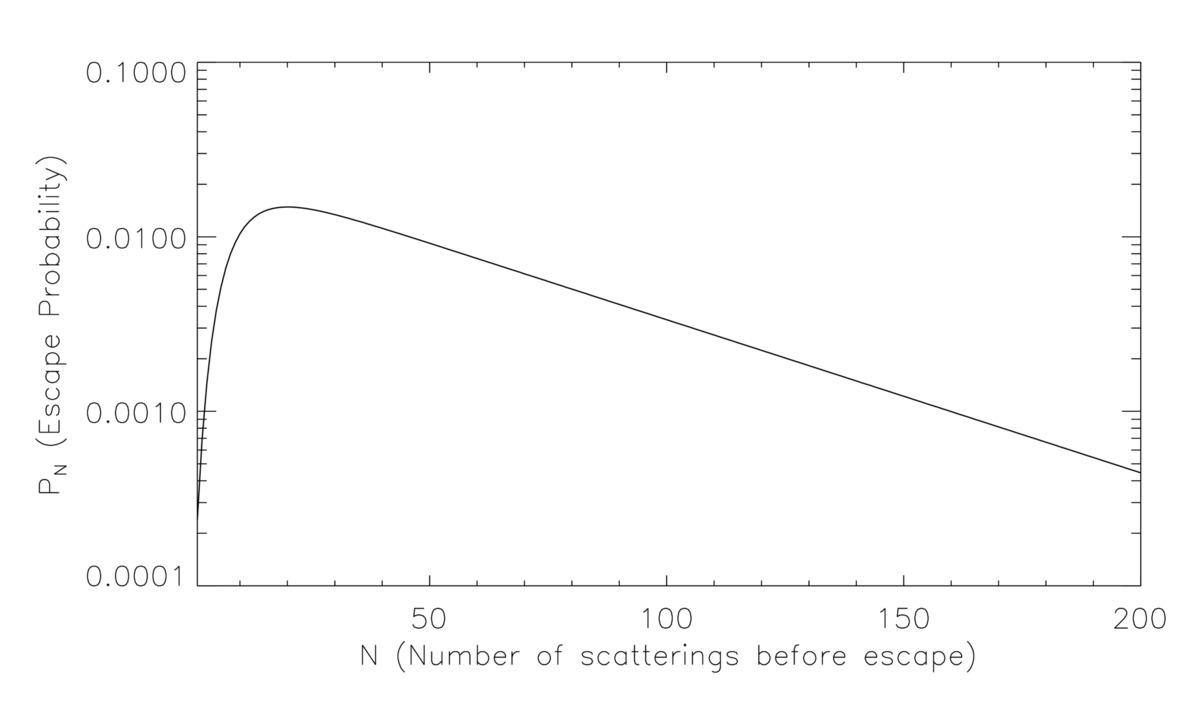}
\caption{The escape probability ($P_N$) for a photon after undergoing
$N$ scatterings within a slab of optical depth $\tau=10$ ($\tau$ is measured 
from the mid-plane of the slab to its surface). The function $P_N$
peaks at $N\sim 20$ and declines exponentially at larger $N$.         
}\label{FIG:rwalk_prob}
\end{center}
\end{figure}

Another derivation for the photon index $\beta$ (for $\nu> \nu_p^{ic}$) 
for the case $\tau_T\gg1$ and $Y\sim 1$ follows from photon escape
probability $P_N$, which is probability that a seed photon released at 
the mid-plane of the slab is scattered $N$ times before escaping at the 
surface. The probability function $P_N$ is shown in Figure 
\ref{FIG:rwalk_prob} for a slab with Thomson scattering optical depth 
of $\tau_T=10$. The escape probability increases for $N \lae 20$ and 
then decreases exponentially for larger $N$; the large $N$ behavior is 
$P_N\propto\exp(-2N/\tau_T^2)$. A photon of initial frequency $\nu$ after
undergoing $N$ scatterings with electrons at temperature $T$ has frequency
$\nu_{ic} \sim \nu\, A_f^N\sim \nu\,\exp(4 k_B T N/m_e c^2)$.
Using the conservation of photon numbers in scatterings, and the
increase to frequency bandwidth after $N$ scatterings by a factor
$A_f^N$, we find $\beta \sim d \ln(P_N/A_f^N)/d\ln(\nu_{ic}) \sim
  -1 - 2/Y$ as long as $\nu_{ic}$ is well below the mean energy of electrons. 
The two expressions we have obtained for $\beta$ differ by $1/Y$ due to 
different approximations made in these two derivations. The exact result, 
obtained from solution of Kompaneets' equation, is $\beta = -1 - 4/(3 Y)$. 
The spectrum for $\nu_{ic} > k_B T/h$ declines exponentially.

The spectrum far below the peak is a rising function of frequency since 
$P_N$ increases for $N\ll \tau_T^2$.
However, the emergent flux is nearly independent of frequency
below the peak $\nu_p^{ic}$, i.e. $\alpha\sim-1$, when $\tau_T\gg1$, 
$Y\sim 1$ and $\nu_p\ll k_B T/h$.
This can be understood using a simple
physical argument provided in \cite{ghisellini12}.
Consider seed photons, and thermal electrons at temperature $T$, that populate 
a region of finite height but infinite length/width uniformly. 
The optical depth of the source to Thomson scattering from mid-plane to
the surface ($\tau_T$) is assumed to be much larger than unity.
Photons escape from a thin surface layer of approximately unit optical depth,
and the number of photons leaving the surface per unit time (integrated over 
frequency), $\dot n_\gamma$, is roughly constant until photons in the medium 
are depleted substantially; the depletion becomes 
severe only when photons from the mid-plane of the medium start arriving 
at the surface. However, the mean energy of emergent photons increases with 
time since later arriving photons come from deeper layers having undergone 
more scatterings. Let us take the spectrum of seed photons to peak at $\nu_0$
and its width to be $\Delta\nu_0$. The emergent instantaneous spectrum 
at time $t$ peaks at $\nu(t)$ and its width is $\Delta\nu(t)$; 
$\Delta\nu/\nu$ is nearly independent of time for roughly the time it
takes for photons to diffuse from the mid-plane to the surface.
The instantaneous specific flux at the peak is $\nu(t)(\dot n_\gamma/\Delta
\nu)$, which is time independent. And therefore 
the emergent specific flux averaged over the diffusion time across the layer is 
nearly independent of frequency 
between $\sim\nu_0$ and $\nu_p\sim \nu_0\exp(Y)$ as $\nu(t)$ sweeps across
this band roughly linearly with time, i.e. $\alpha\approx -1$.

A straightforward prediction of this model is that the spectral-peak 
should shift to larger frequencies with time as photons emerging later
have undergone more number of IC scatterings on average, and thus
have gained more energy. Moreover, the flux should increase with time, at 
first, as the slab radius increases and its optical depth decreases, and 
later on the flux should decline due to the adiabatic cooling of electrons
and photons.

\begin{figure}
\includegraphics[width=13cm]{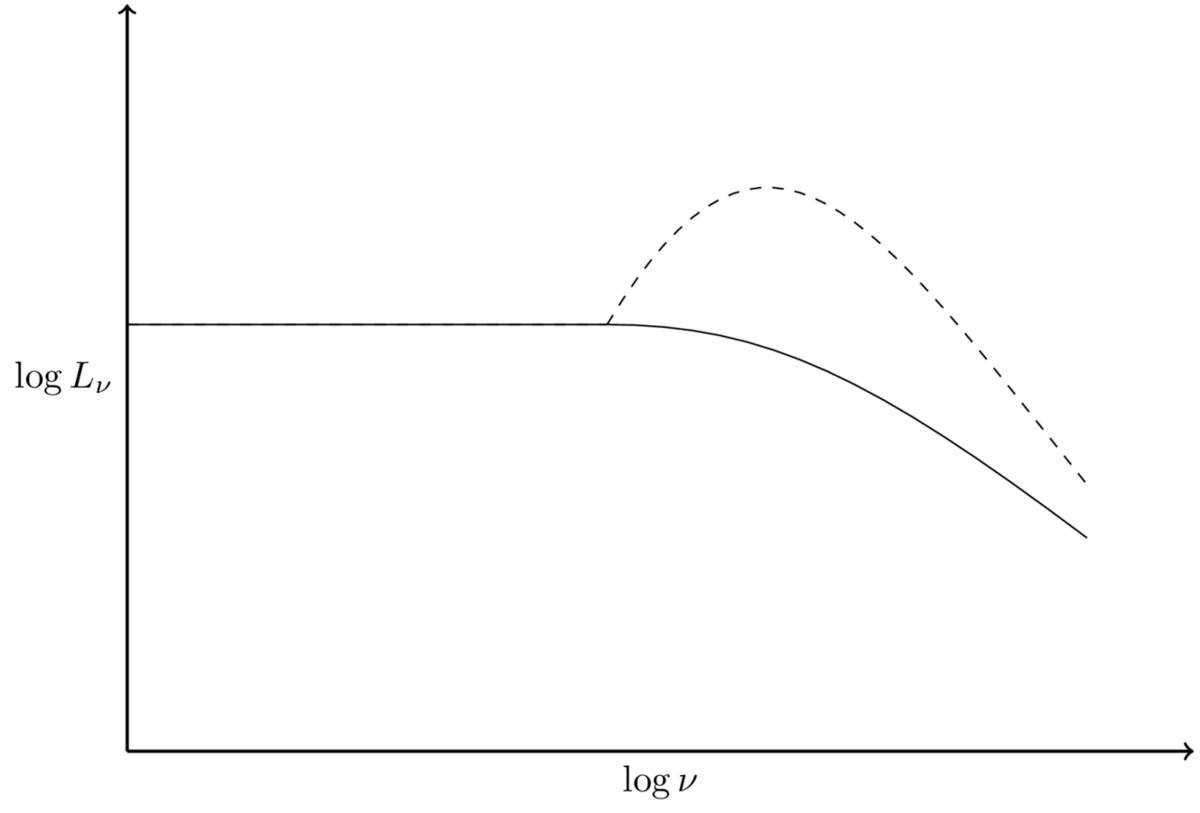}
\caption{Spectrum due to multiple Compton scattering of thermal 
seed photons (dotted line) which shows a sharp rise just below the peak and
a flat shape corresponding to $\alpha=-1$ far below the peak (it is a
schematic drawing). The Band function spectrum (solid line) shows no 
such hump below the peak. 
}\label{FIG:mutpli-IC}
\end{figure}

The peak of the emergent spectrum moves closer to $k_B T/h$ as the 
Compton-$Y$ parameter increases. The specific flux has a sharp rise 
just below the peak when the peak frequency approaches $k_B T/h$.
This rise arises, as shown by e.g. \cite{ghisellini12}, due to accumulation 
of photons in the frequency space as their energies approach $ k_B T$ 
after multiple IC scatterings (see Fig. \ref{FIG:mutpli-IC}).
Such a sharp rise, just below the peak, is never seen in GRB spectra, which
suggests that Compton $Y$ cannot be much greater than 1 (that is to say, if
prompt $\gamma$-rays were to be produced as a result of multiple IC scatterings
at the photosphere).

Energy gained by photons in multiple-IC scatterings at the photosphere
has been suggested as a possible mechanism to explain the prompt $\gamma$-ray 
spectrum, e.g.
\cite{thompson94,ghisellini99,meszarosrees00,eichler00,meszaros02b,rees05,peer06,giannios06,giannios07,ioka07,peer08,asano09,lazzati09,lazzati10,toma11,mizuta11,nagakura11,bromberg11,meszarosrees11,ito13}.

One of the drawbacks of this mechanism is the spectral shape below the peak. 
The observed specific flux for a typical GRB is nearly flat below the peak, 
i.e. $\alpha \approx -1$ over an extended energy band covering more than
an order of magnitude (from $\sim$10 keV to several hundred keV). 
The spectrum produced by multiple IC scatterings, on the other hand, in the 
unsaturated regime with $Y\sim 1$ has $\alpha\sim -1$ over a more
limited bandwidth as described above.

\cite{lundman13} find that photospheric radiation has a
low energy spectral index $\alpha\approx -1$ when the jet LF 
decreases with azimuthal angle $\theta$ from the jet axis while the 
jet luminosity remains constant. This is because in such a configuration,
more photons from larger angles with respect to the observer's line of sight
can contribute to the low-energy flux to flatten the low energy
spectrum. It is unclear whether such a specific jet structure  
applies to the majority of GRBs.

Another suggestion invoked integration over equal-arrival-time hypersurface 
to explain $\gamma$-ray spectra, i.e. photons arriving at any given 
observer time originated at different locations near the photosphere,
with different temperatures, and this superposition makes the observed 
spectrum non-thermal.  \cite{deng14b} investigated
this scenario, and concluded that the low energy photon
spectral index remains essentially intact, i.e. the spectrum below the peak
is much harder than the observed GRB spectra. They also find that it is 
difficult to obtain the commonly observed ``hard-to-soft'' $E_p$ evolution for 
GRB pulses in this model.

One commonly used argument in support of the photospheric origin of the
GRB prompt spectrum is that the observed $E_p$ is narrowly clustered 
around the sub-MeV range \citep{preece00}, which is consistent with the 
photosphere temperature \citep[e.g.][]{beloborodov13}. This corresponds
to the $Y \leq 1$ regime. For such cases, there is a maximum photosphere
temperature (and hence $E_p$) given an observed $\gamma$-ray luminosity
(see eq. \ref{T_R0}).
This defines a ``death line'' of the model in the $E_p$--$L$ plane
\citep{zhang12}. For GRB 110721A, $E_p=15$ MeV early on \citep{axelsson12},
which lies above the death line, and hence the thermal photospheric model is
ruled out. The main spectral component of this burst is well fitted
by a Band function, and that must come from 
 a non-thermal emission process in the optically 
thin region \citep{zhang12,veres12b}. Considering that the Band-function 
parameters of this burst are fairly typical for a GRB, this argument casts 
a doubt on the claim that Band spectra are quasi-thermal emission from 
the photosphere.

There is another issue with the photospheric radiation mechanism for
prompt $\gamma$-ray emission:
electrons need to be heated below the photosphere continuously
while keeping their temperature sub-relativistic ($\ll 1$ MeV). This 
requires some degree of fine tuning for this model as described below.

If the thermal Lorentz factor of electrons is of order unity --
as it is the case for most photospheric models -- then electrons carry a 
tiny fraction of the jet luminosity or the observed $\gamma$-ray luminosity 
of GRBs. Most of the jet energy is in protons (unless the jet has $\gae10^3$
$e^\pm$ pairs per proton) or magnetic fields. Therefore, IC scatterings of
seed photons off of electrons, to produce the observed $\gamma$-ray 
luminosity, requires dissipation of a substantial fraction of jet energy 
below the photosphere and transferring that energy to electrons while 
keeping the electron temperature sub-relativistic. 
The reason that electron temperature should be sub-relativistic (in the 
jet comoving frame) is to prevent IC peaks appearing in
GRB spectra which we have never seen, and also to 
keep Compton-$Y$ from becoming too large otherwise the peak of the emergent
spectrum will appear at $\sim10^2$MeV instead of $\sim10^2$keV.
For a baryonic jet with one electron per proton, $\lae 1$ MeV per electron 
is a tiny fraction (of order 10$^{-3}$) of the jet luminosity. This means 
that electrons need to be heated rapidly and repeatedly, of order 
$\gae 10^3$ times in one dynamical time, at the photosphere, in order 
to transfer a good fraction of jet luminosity
to electrons, which then is passed on to $\gamma$-rays via IC process. 
This requires a certain degree of fine tuning so that electrons
receive a good fraction of jet energy while the temperature 
is kept sub-relativistic -- if the jet energy is transferred to electrons
on a time much shorter than the dynamical time then the temperature would
become relativistic.

\cite{vurm13} provide general constraints on energy dissipation
processes, photon generation mechanisms, and jet LF, for a dissipative
photospheric model to be able to explain the low energy spectral index for 
gamma-ray bursts. They claim that scattering of seed photons by electrons
is not sufficient to be able to account for the observed GRB spectra, and 
that seed photons ought to be produced at a moderate to small Thomson 
optical depths which is a severe requirement for a 
dissipative photosphere model.
\cite{asano13} have derived stringent constraints on the
dissipation radius for the photosphere model to be able to reproduce
the observed GRB spectra. 

\subsection{Hadronic model for prompt $\gamma$-ray radiation}
\label{hadron_model}

Thus far we have considered electrons that are accelerated in shocks or 
otherwise, and these electrons produce $\gamma$-rays via the synchrotron 
and inverse-Compton processes. Protons are also accelerated in shocks and 
attain energy much larger than electrons due to their smaller radiative loss 
rate, and they too could contribute to the observed $\gamma$-ray radiation 
from GRBs as described in a number of papers \citep[e.g.][]{boettcher98,totani98,aharonian00,mucke03,reimer04,gupta07b,asano09b,fanpiran08,razzaque10,asano12,crumley13}. 
This is taken up in subsection \S\ref{proton_synchro} below,
where we show that for the proton-synchrotron process to account for the
observed flux, particularly at photon energies of $\sim$GeV, one requires the 
total energy in GRB explosions to be several orders of magnitude larger 
than the energy we see in $\gamma$-rays. 

High energy protons can contribute to $\gamma$-ray generation in another,
indirect, way. They can produce positrons of very large Lorentz factor by 
the photo-pion
and Bethe-Heitler processes. Both of these processes involve collisions
between energetic protons and photons to produce $e^\pm$ directly
(Bethe-Heitler process) or via generation of pions (photo-pion process)
which decay to positrons and neutrinos. 
Photo-pion and Bethe-Heitler processes, although inefficient for producing
high energy electrons compared with the Fermi acceleration process 
operating in shocks, can be important in those situations where we
need electrons of energy larger than the maximum that a shock can
deliver. These processes and the radiation they produce are described 
in the next two sub-sections. 

\subsubsection{Photo-pion process for producing high energy photons}
\label{photo_pion}

The basics of photo-pion process is described in \S\ref{hadronic}.
In this section, we provide an estimate of the energy required in protons
for producing a certain observed flux in photons of $>$100MeV via
the photo-pion process. We consider photons at the peak of the prompt 
GRB spectrum ($\nu_p$) colliding with high energy protons to produce pions,
and positrons produced by the decay of these pions emitting
high energy photons by the synchrotron process; a more precise numerical
calculation \citep[e.g.][]{asano09b,asano12} that takes into account photon 
and proton spectra gives results for the energy requirement that is within 
an order of magnitude of the estimate provided below.

The photon energy at the peak of the spectrum, in the jet comoving frame,
is $\nu'_p = \nu_p (1+z)/\Gamma$, if the observed peak is at $\nu_p$
for a burst located at redshift $z$, and the jet is moving with LF
$\Gamma$.  The threshold photon energy, in the proton rest frame, for
photo-pion production is approximately 200 MeV. Therefore, the Lorentz
factor of a proton for pion production, when interacting with photons
of energy $\nu_p'$, must satisfy
\begin{equation}
\gamma_p'\gae 2\times10^4 \nu_{p,6}^{-1} (1+z)^{-1}\Gamma_2,
  \label{gamp_pion}
\end{equation} 
where $\nu_{p,6}$ is the observed spectral peak in MeV.

At the threshold energy, the pion LF in the jet frame is also equal to 
$\gamma_p'$, since it is more or less at rest in the proton-rest frame.
The decay of a $\pi^+$ (half life 26 ns) produces $\mu^+$ 
and $\nu_\mu$, and the muon decays to $e^+$ and neutrinos in 2.2 $\mu$s 
on average. The positron carries roughly 1/4 the energy of the pion,
and therefore, the Lorentz factor of the $e^+$ in the jet rest frame is
\begin{equation}
\gamma_e'\sim 50 \gamma_p'\sim 10^6 \Gamma_2 \nu_{p,6}^{-1} (1+z)^{-1}.
  \label{game_pion}
\end{equation}

For a GRB of observed isotropic luminosity $L_\gamma$ (integrated over Band 
function spectrum), the number density of photons in the comoving frame of 
the jet is
\begin{equation}
n_\gamma'\sim \frac{L_\gamma (1+z)^{-1}} {4\pi R^2 \Gamma c h\nu_p}\approx
2\times10^{14}L_{\gamma,52}R_{15}^{-2}\Gamma_{2}^{-1}\nu_{p,6}^{-1}(1+z)^{-1}\ \mathrm{cm}^{-3},
\label{nphoton'}
\end{equation}
where $R$ is the distance from the center of the explosion in centimeters.

Given the cross-section for the delta resonance, 
$\sigma_{\gamma p}=5\times10^{-28}\ \mathrm{cm^2}$, the optical depth for pion
production for a photon of frequency $\sim\nu_p'$ interacting with a proton of
LF given by equation (\ref{gamp_pion}) is 
\begin{equation}
\tau_{\gamma p}\approx\sigma_{\gamma p} n_\gamma' \frac{R}{\Gamma}\approx 0.8
L_{\gamma,52}R_{15}^{-1}\Gamma_{2}^{-2}\nu_{p,6}'^{-1}(1+z)^{-1}.
\end{equation}

The magnetic field in the source comoving frame can be constrained by the 
requirement that the synchrotron frequency for positrons produced by the 
photo-pion process has some desired frequency $\nu$. Using equation 
(\ref{game_pion}), this condition leads to:
\begin{equation}
   \frac{qB'\gamma_e'^2\Gamma}{2\pi m_ec(1+z)}\sim 1.6\times
      10^{-4}\nu_{8}\quad \implies \quad B'\sim (10^2\,{\rm G})\,\nu_{8}
    \nu^2_{p,6} (1+z)^3\Gamma_{2}^{-3}\ ,
\label{MagField}
\end{equation}
where $\nu_8$ is frequency in unit of 10$^8$eV. 

We now use this magnetic field to determine the total energy in protons 
so that the photo-pion process results in a desired level of flux at $\nu_8$.
The observed synchrotron flux at $\nu$ is related to the number
of positrons, $N_{e^+}$, that radiate at that frequency --
\begin{equation}
   f_\nu=1.2\ \mu \mathrm{Jy}\ N_{e,50} B'\Gamma d_{L,28}^{-2}(1+z).
\end{equation}
Thus, the number of $e^+$ needed to produce the observed flux at $\nu$ is
\begin{equation}
  N_{e^+}\approx8\times10^{47}\frac{f_{\nu,\mu Jy}d^{2}_{L,28}}{B'\Gamma_{2}(1+z)},
\label{ElectronEst}
\end{equation}
where $f_{\nu,\mu Jy}$ is observed specific flux in $\mu$Jy.

The number of protons with energy above the pion production threshold
required to produce the necessary number of positrons (eq. \ref{ElectronEst}) 
is given by
\begin{equation}
N_p\approx \frac{N_e}{\tau_{\gamma p}}\approx
10^{48}f_{\nu,\mu Jy}d^{2}_{L,28}\Gamma_{2}R_{15}\nu_{p,6}B'^{-1}
    L_{\gamma,52}^{-1}, 
\end{equation}
and the energy in these protons is
\begin{equation}
E_p\approx N_p(\gamma_p' m_p c^2)\Gamma\approx 3.0\times 10^{51}
\ \frac{\Gamma_{2}^3f_{\nu,\mu Jy}d_{L,28}^2R_{15}}{B'
  L_{\gamma,52}(1+z)} \ \mathrm{erg}.
  \label{photo-pion-Ep}
\end{equation}

It is more useful to consider the luminosity carried by these protons ($L_p$)
for determining the efficiency of the photo-pion process for high energy 
$\gamma$-ray production. The proton-luminosity is related to $E_p$ via
\beq
   L_p = E_p \Gamma\times \max\{ t_{dyn}'^{-1}, t_{cool}'^{-1} \},
   \label{photo-pion-Lp}
\eeq
where $t'_{dyn}$ is the dynamical time in the jet comoving frame
\begin{equation}
   t'_{dyn}=\frac{R}{2c\Gamma}\approx (170\, {\rm s}) R_{15}\Gamma_{2}^{-1},
   \label{photo-pion-tdyn}
\end{equation}
and $t'_{cool}$ is the synchrotron cooling time\footnote{To be more
  precise, $t'_{cool}$ is the radiative cooling time which includes
  inverse-Compton and synchrotron contributions. However, for
  positrons of LF $\gae 10^6$ considered here, the IC scattering lies
  in the Klein-Nishina regime, and so for a large part of the GRB
  parameter space, synchrotron losses dominate.} for a positron with
LF $\gamma_e'$ (eq. \ref{game_pion}) that is moving in a magnetic field
given by equation (\ref{MagField}),
\begin{equation}
  t'_{cool}=\frac{6\pi m_ec}{\sigma_T B'^2 \gamma_e'}\approx (8\times10^{-2}
   {\rm s}) \frac{\Gamma_{2}^5}{\nu^2_{8}\nu_{p,6}^3(1+z)^5}.
   \label{photo-pion-tcool}
\end{equation}
Substituting, equations (\ref{MagField}), (\ref{photo-pion-Ep}),
(\ref{photo-pion-tdyn}) \& (\ref{photo-pion-tcool}) into (\ref{photo-pion-Lp}),
we find the proton luminosity to be \citep{crumley13}
\begin{equation}
L_{p} =  \left\{\begin{array}{ll}
\hskip -4pt 2\times10^{49}\ \Gamma_{2}^{8}f_{\nu,\mu Jy}d_{L,28}^2 L^{-1}_{\gamma,52} \nu^{-1}_{8} \nu^{-2}_{p,6}(1+z)^{-4}\ \mathrm{erg\,s}^{-1} \quad
&  t'_{cool} > t'_{dyn}\\  \\
\hskip -4pt 4\times10^{52}\
\Gamma_{2}^{2}f_{\nu,\mu Jy}d_{L,28}^2\nu_8\nu_{p,6}L^{-1}_{\gamma, 52}(1+z)R_{15}\ 
   \mathrm{erg\,s}^{-1} &  t'_{cool} < t'_{dyn}.
\end{array}
\right. \label{E_jet_photo-pion}
\end{equation}
This proton luminosity is based on taking the magnetic field strength 
as given in equation (\ref{MagField}). It might be tempting to think 
that a larger magnetic field might reduce $L_p$. However, that turns 
out not to be the case because even though a larger $B'$ means that 
$e^+$ LF ($\gamma_e'$) needed for producing a photon of a desired synchrotron 
frequency is smaller ($\gamma_e'\propto B'^{-1/2}$), it also means that 
a smaller fraction of particles produced by the
photo-pion process can radiate at this frequency at any given time since the
synchrotron cooling time decreases as $B'^{-2} \gamma_e'^{-1}\propto 
B'^{-3/2}$; the latter effect overwhelms the net gain of the former
as can be readily seen by the dependence of $L_p$ on $\nu_8$ in
equation (\ref{E_jet_photo-pion})\footnote{If the synchrotron cooling time for 
$e^\pm$ radiating at $\nu_8$ is larger than the dynamical time, then a 
magnetic field of strength higher than that given by eq. \ref{MagField} 
does reduce the energy requirement for protons; the dependence on $B$ is 
weak though. However, for the vast majority of allowed GRB parameter space, 
the cooling time is smaller than the dynamical time.}.

We can assess the viability of the photo-pion process for producing
$>$10$^2$MeV photons detected by the Fermi satellite from a number of
highly luminous GRBs. Let us consider the data for a particular
burst, GRB 080916C, as an example. This burst was at a redshift of
4.3, $d_{L,28}=12$, the peak of the observed spectrum was at 400 keV,
and the flux at 100 MeV during the burst was $f_{\nu}\sim 3
\mu$Jy. The $\gamma$-ray isotropic luminosity for GRB 080916C was
$L_{\gamma,52}\sim 20$, and the jet LF is estimated to be
$\Gamma_{2}\sim 9$ \citep[e.g.][]{abdo09a,greiner09b}.
 For these parameters we find $t'_{cool} <
t'_{dyn}$ as long as $R>10^{15}$cm, and in that case the required
luminosity in protons of LF $\gae10^5$ is $L_p \sim 1.5\times10^{56}
R_{15}$ erg/s. This is larger than the $\gamma$-ray luminosity by a 
factor $\sim 700$, if the radiation is produced at $R=10^{15}\ 
{\rm cm}$. For $R<10^{15}$cm, $t'_{\rm dyn}<t'_{\rm cool}$ and the 
proton luminosity is independent of $R$.
The total luminosity carried by protons, if their
distribution function extends down to LF $\sim 10$ with $p=2.4$, is
another factor of $\sim 40$ larger, and that makes the photo-pion
process unacceptably inefficient for this burst. Moreover, the high
proton luminosity required for the photo-pion process is inconsistent with
the upper limit on high-energy neutrino flux from GRBs provided by the
IceCube observations \citep{icecube12}.

\subsubsection{Bethe-Heitler process}
\label{bethe_heitler}

We assess the viability of the Bethe-Heitler process in this sub-section
for producing high energy $\gamma$-rays detected by Fermi/LAT from a
number of bursts.

The cross-section for Bethe-Heitler pair production process --- 
$p+\gamma\rightarrow p+e^++e^-$ --- has a strong dependence on the 
angle between the outgoing electron and the incident photon in the 
nuclear rest frame. Assuming that the protons and photons are isotropic 
in the jet's rest frame, and using the head on approximation, 
\textit{i.e.} the angle between the photon and proton is zero in 
the nuclear rest frame, \(\epsilon''=2\gamma_p'\epsilon'\) 
[where $\epsilon' = h\nu'/(m_e c^2)$], the equation for the rate of 
production of secondary electrons is:
\begin{equation}
\frac{d\dot{N}_e}{d\gamma_e'}=2c\int_0^\infty{d\epsilon'\ 
    n'(\epsilon')\int_1^\infty{d\gamma_p'\ N_p(\gamma_p')
      \frac{d\sigma_{BH}(\epsilon',\gamma_p')}{d\gamma_e'}}},
   \label{pairpro}
\end{equation}
where $N_p$ is the number of protons in the shell with LF $\gamma_p'$, and
$n'(\epsilon')$ is the number density of photons in the jet comoving frame
with dimensionless energy $\epsilon'$ defined above.
The differential cross section in the Born approximation
integrated over angles, in the highly relativistic regime, was derived by 
Bethe \& Maximon (1953) (see \cite{rachen96}, for a recent review)
\begin{equation}
\frac{d\sigma_{BH}}{d\gamma''_+}=\frac{3\alpha_f\sigma_T}{2\pi
{\epsilon''}^3}
\left({\gamma''_+}^2+{\gamma''_-}^2
+\frac{2}{3}\gamma''_+\gamma''_-\right)
\left(\log{\frac{2\gamma''_+\gamma''_-}{\epsilon''}}-\frac{1}{2}.
  \right)
\end{equation}
Where $\gamma''_+,\gamma''_-$ are the Lorentz factors of the positron 
and electron, respectively, in the nuclear rest frame, as are all other
variables in the above equation, and $\alpha_f\approx 1/137$ is the 
fine-structure constant.
The differential cross-section peaks sharply when the angle between 
the incoming photon and the outgoing $e^\pm$ in the nuclear rest frame 
($\theta''_\pm$) is $\sim1/\gamma''_\pm$.
So, for $\gamma_p'\gg\gamma''_\pm$, the Lorentz factor of $e^\pm$ in the jet
rest frame is
\begin{equation}
\gamma_{\pm}'=\gamma_p'\gamma_\pm''
\left(1-\beta_p'\beta''_\pm\cos{\theta''_\pm}\right)\approx
\frac{\gamma_p'\gamma_\pm''}{2}\left(\gamma_p'^{-2}
+{\gamma_\pm''}^{-2}+{\theta''_\pm}^{-2}\right)
\approx \frac{\gamma_p'}{\gamma''_{\pm}}.
   \label{gamma_epm_Bethe}
\end{equation}
Therefore, pairs produced via the Bethe-Heitler process have LF (in jet comoving
frame) that is smaller than the proton LF most of the time. 

If $\epsilon''\ll m_p/m_e\sim 10^3$, the proton recoil can be neglected 
and the following equality holds
\begin{equation}
   \gamma''_+ + \gamma''_- = \epsilon''.
\end{equation}
For large $\epsilon''$, the differential cross-section decreases
extremely rapidly when $\gamma_\pm''<2$. Therefore, we restrict
ourselves to $\gamma_\pm''\geq 2$. In this regime, the differential
cross section simplifies as follows
\begin{equation}\label{diff-cross-NRF}
\frac{d\sigma_{BH}}{d\gamma''_+}\approx\frac{\alpha_f\sigma_T}{\epsilon''},
\quad \mbox{if}\quad 2\le\gamma_+''\le\epsilon''-2.
\end{equation}
Or writing equation (\ref{diff-cross-NRF}) in terms of
quantities in the jet comoving frame, using $\epsilon''\approx
2\gamma_p'\epsilon'$, we find:
\begin{equation}
  \frac{d\sigma_{BH}}{d\gamma_+'}\approx\frac{\alpha_f\sigma_T}{2\epsilon'
     \gamma_+'^2}, \quad \mbox{if}\quad \frac{1}{2\epsilon'}\leq\gamma_+'\leq 
     \frac{\gamma_p'}{2}.
   \label{diff-cross-jet}
\end{equation}

The integral in equation (\ref{pairpro}) can be simplified when we
exclude the part of the $\gamma_p'$---$\epsilon'$ plane where the
cross-section, $\sigma_{BH}$, is small, i.e. $\epsilon'\lae \gamma_e'^{-1}$ 
and $\gamma_p'\lae 2\gamma_e'$. The cross-section in the remainder
of the plane is given by equation (\ref{diff-cross-jet}). With these
approximations, and taking the photon spectrum to be Band function
with indices $\alpha$ \& $\beta$, the integral in equation (\ref{pairpro}) is
straightforward to calculate and the result is \citep{crumley13}
\begin{equation}
\frac{d\dot{N}_e}{d\gamma_e'}\approx
 \left\{  \begin{array}{ll}
   \hskip -4pt \frac{2c\alpha_f\sigma_T}{\beta(p+1)\gamma_e'}n'({\epsilon'_p}
   N_p(\gamma'_i) \left(\frac{\gamma_e' \epsilon_p'}{5}\right)^{-\beta}
   \left(\frac{2\gamma_e'}{\gamma_i'}\right)^{-p} & \mbox{for }
      \frac{\gamma_i'}{2}\leq\gamma_e'\leq 5/\epsilon_p'\\  \\
  \hskip -4pt \frac{2c\alpha_f\sigma_T\epsilon_p'}{5\beta(p+1)}n'({\epsilon'_p})
    N_p(\gamma'_i)\left(\frac{10}{\epsilon_p'\gamma_i'}\right)^{-p}
    \left(\frac{\epsilon_p'\gamma_e'}{5}\right)^{-\alpha-p-1} & \mbox{for }
    5/\epsilon_p' \leq\gamma_e'\leq 5/\epsilon'_{min}\\
\end{array}   \right.
\label{pairpro_general}
\end{equation}
where $\epsilon'_p = h\nu_p(1+z)/(\Gamma m_e c^2)$ is the dimensionless photon 
energy at the peak of the spectrum in the jet comoving frame (which is of 
order 10$^{-2}$ for a typical long-GRB), $\gamma'_i$ is the minimum LF
of protons in the jet comoving frame, and $\epsilon'_{min}$ is the
dimensionless photon energy (jet comoving frame) below which the source
becomes opaque due to synchrotron absorption and the 
spectrum declines rapidly; the value of $\epsilon'_{min}$ is poorly constrained
by GRB observations, but theoretical calculations suggest it is likely of 
order 10$^{-7}$, which corresponds to a synchrotron self-absorption frequency
of a few eV in the observer frame.

The peak cross-section for the Bethe-Heitler process is roughly 10 times
larger than the cross section for the photo-pion $\Delta$-resonance.
For any given proton LF, the photon energy required for the former 
process is roughly 50 times smaller than the photo-pion process.
Moreover, protons of LF $\gamma_p'$, produce $e^\pm$ with an average
Lorentz factor $\sim\gamma_p'/4$ via the Bethe-Heitler process (eqs. 
\ref{gamma_epm_Bethe} \& \ref{diff-cross-NRF}), whereas 
$\gamma_e'\sim 50\gamma'_p$ for the $\Delta$-resonance of the photo-pion 
process (eq. \ref{game_pion}). 

Therefore, the ratio of the rate of generation of $e^\pm$ with LF
$\gae\gamma_e'$ by the Bethe-Heitler and photo-pion processes is $\sim
10\times(10^{4})^{-\alpha-1}\times (200)^{-p+1}$; where the first
factor is the ratio of the cross-sections for the two processes, the
second factor accounts for the larger number of photons that
participate in the Bethe-Heitler process --- the dimensionless photon
threshold energy for producing $e^\pm$ of LF $\gamma_e'$ by the B-H
process is $\epsilon'\sim \gamma_e'^{-1}$ and for the photo-pion it is
$10^4 \gamma_e'^{-1}$ --- and the third factor is due to the fewer
number of protons that are capable of producing positrons of LF
$\gae\gamma'_e$ via the Bethe-Heitler process. For $\gamma_e' \gae
10^6$, the threshold photon energy for both processes lies below the
peak of the spectrum, and in that case $\alpha\sim -1$. Thus, for these
high energy positrons, the Bethe-Heitler is less efficient than the
photo-pion process by a factor $\sim10^2$. However, for $\gamma_e'\lae
10^3$, the threshold photon energy lies above the peak of the spectrum,
and the Bethe-Heitler process is a lot more efficient than the
photo-pion process \citep{crumley13}. Whether the Bethe-Heitler or photo-pion 
process are more important for the intermediate regime, 
$10^3\lae\gamma_e'\lae10^6$, depends on the spectral indices, proton 
distribution index $p$, and $\epsilon_p'$.

Relativistic shocks are believed to accelerate electrons to $\gamma_e'\gg10^3$ 
efficiently via the Fermi mechanism, and that might suggest that the 
Bethe-Heitler process can't compete with it and play an important role 
for GRBs. However, Bethe-Heitler  might be important for those GRBs where 
the number of $e^\pm$s produced by this process is larger than the number of 
electrons that came with protons in GRB jets. We quantify this 
condition below.

 Let us consider the isotropic luminosity carried by protons in a GRB 
jet to be $L_p$, which is a factor $\eta_p$ larger than the $\gamma$-ray
luminosity: $L_P = \eta_p L_\gamma$. The co-moving number density of 
electrons associated with protons is
\beq
n_e'=n_p'\approx 2 \times 10^9 \eta_p L_{\gamma,52}
\Gamma_2^{-2} R_{15}^{-2}.
\eeq
The number density of $e^\pm$ produced by the Bethe-Heitler
process is \citep{crumley13}
\beq 
  n_{BH}' \approx \alpha_f\sigma_Tn_\gamma' n_p' R/\Gamma \Rightarrow
  \frac{n_{BH}'}{n_e'} < \alpha_f\sigma_Tn_\gamma' R/\Gamma, 
\eeq
where $\alpha_f\approx 1/137$ is the fine-structure constant, and the 
inequality is due to the fact that only a fraction of protons have sufficient
energy for pair production.
Since the optical depth to Thomson scattering associated with proton-electrons
is \( \tau_T = \sigma_T n_e' R/\Gamma\), we find
\beq
\frac{n_{BH}'}{n_e'}<\alpha_f\frac{n_\gamma'\tau_T}{n_e'}\sim
10^3\tau_T\eta_p^{-1}\Gamma_2\nu_{p,6}^{-1}(1+z)^{-1} 
\eeq 
where we used equation (\ref{nphoton'}) for $n_\gamma'$. The Bethe-Heitler 
process is likely important whenever $n_{BH}'/n_e' > 1$.

\subsubsection{Proton synchrotron model for producing $>$10$^2$MeV photons}
\label{proton_synchro}

Protons are easier to accelerate in shocks due to their lower rate of 
radiative losses, and the Lorentz factor that protons can attain
is much larger than the maximum LF electrons can be accelerated to. It
is easy to show that the maximum synchrotron photon energy for protons
(accelerated in shocks) is a factor $m_p/m_e$ larger than that for
electrons, i.e. instead of a maximum energy of $\sim50$ MeV for
electron-synchrotron photons (see \S\ref{synch_rad_max}), the proton 
synchrotron process can produce photons of energy $10^2$GeV (in the 
jet comoving frame).
 For this reason, whenever photons of energy larger than $\sim 10^2\Gamma$ MeV
are detected from a source, proton synchrotron process is suggested as a
possible radiation mechanism for the generation of these photons
\citep[e.g.][]{boettcher98,totani98,aharonian00,mucke03,reimer04,razzaque10,crumley13}.

However, due to the low radiative efficiency of the proton-synchrotron
process, the energy requirement to produce $\gae$GeV photon flux at 
the level observed by the Fermi satellite, for a number of GRBs, is
found to be highly excessive \citep{crumley13}.

Let us consider protons of LF $\gamma_i$ in the GRB-jet comoving frame.
The synchrotron frequency for these protons is
\begin{equation} \label{pro_i}
\nu_i=\frac{qB'\Gamma\gamma_i^2}{2\pi m_pc(1+z)}\approx
6.3\times10^{-10} B'\Gamma_2\gamma_i^2(1+z)^{-1}\ \mathrm{eV}.
\end{equation}
A reasonable upper limit for $B'$ is obtained by requiring that the
luminosity carried by magnetic fields is not much larger than the 
$\gamma$-ray luminosity, $L_\gamma$, in order to avoid
low radiative efficiency of GRBs ($\lae10$\%) which is not supported 
by observations. The luminosity carried by magnetic fields is 
 $R^2 B'^2 c\Gamma^2$. Therefore, $B'\lae 2\times10^3 L_{51}^{1/2}
R_{15}^{-1}\Gamma_2^{-1}$ Gauss. Substituting this into equation
(\ref{pro_i}), we find $\gamma_i\gae 2\times10^7 L_{51}^{-1/4} R_{15}^{1/2}$ 
in order to produce photons of energy $\sim 1$GeV.

The typical proton Lorentz factor associated with the random component of 
velocity, in a relativistic shock, is approximately equal to the LF of the 
shock front with respect to unshocked fluid if every proton crossing the 
shock front is accelerated; proton LF is proportionally
larger if only a small fraction of protons are accelerated and the 
remaining ones are ``cold'' downstream of the shock front. Considering 
that the LF for GRB internal shocks is of order a few 
to perhaps a few tens, the typical proton LF should be $\sim10$--10$^3$
(the larger value corresponds to when only 1 in $\sim10^2$ protons are 
accelerated as suggested by some simulations, e.g. \cite{sironi11a}). 
Considering that $\gamma_i \gg 10^3$, the proton synchrotron 
spectrum should extend down to photon energies of $\sim 10$ eV, and the
spectrum between 10 eV and 1 GeV should be $f_\nu\propto\nu^{-(p-1)/2}$;
where $p\gae 2.2$ is the power law index for the proton distribution function. 
Thus, if the proton synchrotron flux at 1 GeV matches the observed value, 
then this process would overproduce the flux below MeV\footnote{The observed 
spectra are often $f_\nu\aprop \nu^0$ below the peak of the spectrum which 
lies at a few hundred keV. The proton-synchrotron spectrum, as we have discussed,
is $\nu^{-0.6}$ or steeper between $\sim10$ eV and GeV, and therefore
it would dominate below $\sim 1$ MeV, in clear conflict with data.}. 
Another problem with this process is the excessive energy requirement 
described below.

The synchrotron flux $f_{\nu_i}$ at $\nu_i$ is
\begin{equation}\label{pro_flux}
f_{\nu_i}\approx 7 B' N_{52}\Gamma_2(1+z)d^{-2}_{L,28}\ \mu\mathrm{Jy},
\end{equation}
where $N$ is the total number of protons (assuming an isotropic source),
in a region of comoving radial width $\delta R'=R/\Gamma$, from which 
radiation at $\nu_i$ is received at a fixed observer time.
In order to account for $\sim0.1\mu$Jy flux at 1 GeV observed for 
several GRBs at $z\approx 2$ \& $d_{L,28}\approx 4.5$ 
\citep[e.g.][]{abdo09a,abdo09c,ackermann10,ackermann11},
it is required that $N\gae 5\times10^{47} 
R_{15} L_{51}^{-1/2}$. Therefore, the energy in protons in the shell 
of thickness $\delta R'$, 
responsible for the GeV emission, is:
\beq
  E_{proton} = m_p c^2 N\Gamma\gamma_i\gae 10^{54} R_{15}^{3/2} 
   L_{51}^{-3/4}\Gamma_2\; {\rm erg},
\eeq
and the luminosity carried by these very high LF protons is \citep{crumley13}: 
\beq
L_{proton}\sim {E_{proton} c \over R/\Gamma^2} \gae
 10^{54} R_{15}^{1/2} L_{51}^{-3/4} \Gamma_2^3\; {\rm erg\, s}^{-1}.
\eeq

GRBs from which GeV photons are detected have $200\lae\Gamma\lae10^3$ 
\citep{abdo09a,abdo09c,zou11,hascoet12b}.
For these bursts, the requirement on luminosity carried by protons of 
LF $\gae 10^7$ is $L_p \gae 10^{55}$erg s$^{-1}$, and the total proton 
luminosity --- most of which is in protons of LF $\ll 10^7$ ---
is at least $10^{56}$erg s$^{-1}$. This makes the energy
requirement for the proton-synchrotron process a factor $\sim10^3$
larger than the energy in $\gamma$-rays, and therefore this process
is too inefficient to account for the observed GeV emission from GRBs.

\subsection{Magnetic jet model}
\label{magnetic_jet}

Magnetic outflows in the astrophysical context have been extensively 
investigated for decades in order to understand properties of jets 
associated with active galactic nuclei (AGNs), micro-quasars, pulsars 
and relatively more recently GRBs.

We consider in this section an outflow where magnetic fields carry a 
substantial fraction of the luminosity at the base of the 
jet where it is launched. We describe how such a Poynting jet can be 
accelerated by converting the magnetic field energy to bulk kinetic energy 
of the jet, and how radiation might be produced.

A class of magnetic jet models has been developed that is based on the force-free
electrodynamics approximation (or ``magnetodynamics'') in which the plasma 
inertia is ignored e.g. \cite{komissarov02}. This approach has limited 
application because the neglect of the inertia term means that these
models cannot account for the transformation of magnetic energy to
jet kinetic energy, which is an important process of interest to
many astrophysical systems.

The acceleration of a magnetic jet can proceed either by dissipation of field
(if the magnetic field has the right geometry and scale, such as the stripped 
configuration of a pulsar wind) or by adiabatic expansion of the outflow.

Analytical and semi-analytical solutions have been found for a limited
class of configurations that are characterized by steady-state, axisymmetric,
dissipation-less flows. For instance, \cite{li92} described a 
self-similar solution for a cold magnetic outflow. They find that the jet
acceleration takes place over a very extended range of distance, well past 
the fast magnetosonic surface --- the surface where the magnetosonic wave
speed is equal to the flow speed --- until the magnetization parameter 
($\sigma$) drops to order unity (for comparison, for a radial wind, the
outflow LF saturates at $\sim\sigma_0^{1/3}$ and $\sigma$ does not decrease
below $\sim\sigma_0^{2/3}$); $\sigma \equiv B'^2/[4\pi (\rho' c^2+p')]$, 
and $\sigma_0$ is the initial magnetization parameter of the outflow.  
\cite{vlahakis03}, \cite{vlahakis03b} and \cite{beskin06} extended this work 
and found an exact self-similar solution for an initially hot, axisymmetric,
 magnetic jet\footnote{A non-axisymmetric jet is typically subject to 
instabilities \citep{lyubarsky10,heinz00}, and that can substantially 
increase the efficiency of jet acceleration.}. 

Recent advances in numerical solutions for relativistic MHD have led to 
significant progress in our understanding of the magnetic jet launching 
mechanism,
propagation and acceleration \citep[e.g.][]{komissarov01,komissarov04,mckinney06,komissarov07a,komissarov07b,mckinney07,tchekhovskoy08,komissarov09,mckinney09,tchekhovskoy09,tchekhovskoy10}.

The plan for this sub-section is that we first discuss a steady state, 
axisymmetric outflow, and show that the asymptotic Lorentz factor is
limited to $\sim\sigma_0^{1/3}$ for spherically symmetric systems, 
as pointed out by \cite{goldreich70}. For an outflow of a finite
opening angle ($\theta_j$), the asymptotic value of LF is larger by a factor
$\theta_j^{-2/3}$ provided that it is collimated by the pressure of an
external medium and causal contact across the jet in the transverse direction
is maintained.

Next, we drop the assumption of steady state and describe the acceleration
of an impulsive outflow of finite radial extent due to adiabatic
expansion. Jet acceleration when ideal-MHD approximation breaks down, 
magnetic field is dissipated, and its energy is converted to the bulk 
kinetic energy of plasma is taken up last. 

\subsubsection{Adiabatic expansion and acceleration of a Poynting jet}
\label{magnetic_jet_adiabatic}

We consider in this sub-section an axisymmetric, highly magnetized, time
independent outflow. The magnetization parameter of the outflow, $\sigma$, 
is defined as the ratio of Poynting flux and energy flux carried by particles,
\beq
  \sigma = {B^2 \over 4\pi (\rho' c^2 + p') \Gamma^2} = {{B'}^2 \over 4\pi 
    (\rho' c^2 + p')},
   \label{sigma_def}
\eeq
where $B'$, $\rho'$ and $p'$ are magnetic field strength, internal
plus rest mass energy density, and pressure as measured in the local plasma 
comoving frame; $B$ and $\Gamma$ are magnetic field strength and 
outflow LF as measured in the CoE frame. The base of the outflow is at
$R=R_0$, where the magnetization parameter is $\sigma_0 \equiv 
\sigma(R_0)$ and the Lorentz factor is $\Gamma_0$; $\sigma_0\gg1$.

The conservation of energy flux for a cold magnetized outflows governed 
by the non-dissipative ideal MHD equations is
\beq
  R^2 \left[ \pi\rho' c^2 \Gamma^2 v + B'^2\Gamma^2 v/4 \right] \theta_j(R)^2
    = L,
\eeq
where $\theta_j(R)$ is half-angular-size of the jet at radius $R$, $v$ is the
proper velocity of the jet corresponding to $\Gamma$, and the second term 
is the Poynting luminosity (electric field in the outflow comoving frame 
vanishes). The equation for the conservation of mass flux is
\beq
  \pi R^2 \theta_j(R)^2 \rho' \Gamma v = \dot{M}.
\eeq
These two equations can be combined to give
\beq
   \Gamma(1 + \sigma) = L/\dot{M}c^2 = \Gamma_0(1+\sigma_0).
   \label{gamma_sigma}
\eeq
As the outflow moves to larger distances, $\sigma$ decreases and $\Gamma$ 
increases and their product remains constant. According to these 
conservation laws, it is allowed for the magnetic energy to be entirely 
converted to outflow kinetic energy, and in that case the outflow LF 
attains a value of $(1+\sigma_0)\Gamma_0\approx \sigma_0$. For 
a steady, spherical, outflow, however, the LF stops increasing 
when $\Gamma\approx\sigma_0^{1/3}$ \citep{goldreich70}. The reason for 
this is that when 
$\Gamma \gae \sigma_0^{1/3}$, causal contact is only maintained in a
narrow region of the outflow and magnetic pressure gradients can no longer
accelerate the flow. To see this, let us consider a signal propagating
at a speed $c_s'$ and at an angle $\theta'$ with respect to the radial 
direction in the comoving frame. The signal speed and direction in the 
CoE frame are $c_s$ and $\theta$. The 4-velocity in the outflow
frame is $\Gamma'_s(1, c_s'\cos\theta', c_s'\sin\theta', 0)$, and
in the CoE frame $\Gamma_s(1, c_s\cos\theta, c_s\sin\theta, 0)$.
Taking the outflow velocity and LF to be $v$ and $\Gamma$, and 
Lorentz transforming the CoE frame 4-velocity for signal propagation to 
the comoving frame, we find
\beq
  \Gamma'_s = \Gamma\Gamma_s \left( 1 - v c_s \cos\theta/c^2 \right),
\eeq
which can be solved to determine the signal speed, $c_s$, in CoE frame
\beq
   {c_s \over c} = { v\cos\theta/c + \left[ v^2\cos^2\theta/c^2 + 
   (\Gamma'_s/\Gamma)^2 - 1 \right]^{1/2} (\Gamma'_s/\Gamma) \over 
     v^2\cos^2\theta/c^2 + {\Gamma'_s}^2/\Gamma^2 }.
\eeq
The signal propagation in the CoE frame is confined to a narrow cone of 
half-opening angle, $\theta_s$, with axis along the direction of 
outflow velocity, when $\Gamma'_s/\Gamma <1$. This angle can be 
obtained by setting the discriminant to zero in the above equation. 
We thus find,
\beq
   \sin \theta_s = {\Gamma'_s c'_s\over \Gamma v},
\eeq
which is a relativistic generalization of the familiar expression for ``Mach 
cone'' opening angle when an object moves at a speed faster than the signal 
speed in the medium. Points outside the ``Mach cone'' are not in causal 
contact with the apex of the cone.

\begin{figure}
\begin{center}
\includegraphics[width=8.5cm]{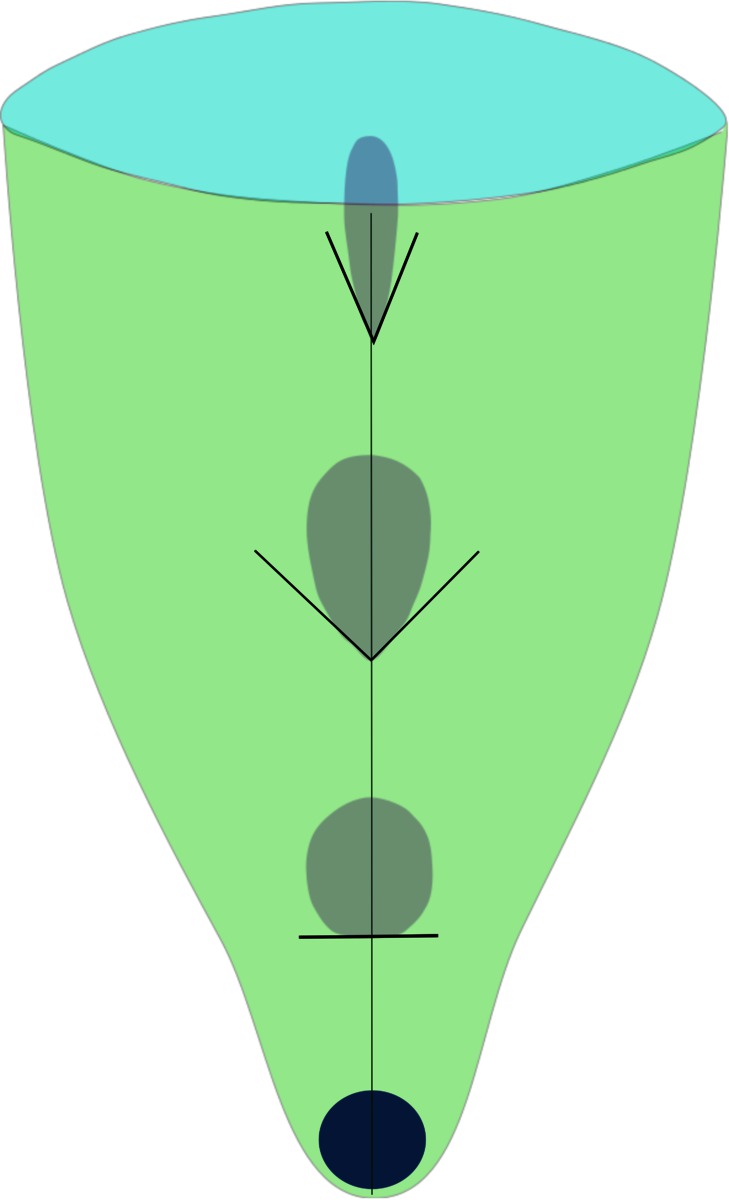}
\caption{The darker shaded regions show the causally connected part of the jet 
at different distances from the jet launching site as seen by a lab
frame observer. Fast magnetosonic signal propagation is 
confined to cones of decreasing opening angle as the jet accelerates 
with distance. When the opening angle of the ``causal cone'' becomes 
smaller than the jet opening angle then the jet is no longer in causal 
contact with the external medium in the direction normal to the jet axis 
and its acceleration can not be influenced by the pressure stratification
of the GRB progenitor star.
}\label{FIG:mach_cone_magnetic_jet}
\end{center}
\end{figure}

The fast-magnetosonic wave proper-velocity in the jet comoving frame is 
\beq
   \Gamma'_s c'_s = (B'^2/4\pi\rho')^{1/2} = c\, \sigma^{1/2} 
   \approx c (\sigma_0/\Gamma)^{1/2}.
  \label{magnetosonic_speed}
\eeq
We used equations (\ref{sigma_def}) \& (\ref{gamma_sigma})
for deriving the last two relations. Thus, $\theta_s < \pi/2$ when 
$\Gamma > \sigma_0^{1/3}$, and only part of the outflow is causally
connected in the lateral direction (see Fig. \ref{FIG:mach_cone_magnetic_jet}). 
For a collimated outflow with opening angle $\theta_j$,
lateral causal contact can be maintained as long as $\Gamma \lae
\sigma_0^{1/3} \theta_j^{-2/3}$. During this phase, the acceleration
of the magnetic jet is governed by pressure stratification of the
surrounding medium. 

Far away from the CoE, magnetic fields are predominantly
toroidal (transverse to the radial direction), and the field
falls off as $1/R$ if the jet diameter
increases linearly with $R$. In this case, $\sigma \propto B^2/\rho$
has no explicit dependence on $R$, and therefore $\Gamma$ does not increase 
with distance (see eq. \ref{gamma_sigma}).
In order for $\Gamma$ to increase with $R$,
the separation between neighboring magnetic field lines must increase 
faster than $R^1$.  This can only be done if 
different parts of the jet are in causal contact so that as the pressure
of the ambient medium (eg. GRB progenitor star) decreases with radius, a
signal can propagate from the outer to the inner part of the jet and
field lines can fan outward in response. 

If the pressure of the ambient medium decreases as $p\propto R^{-a}$, then 
the transverse size of the jet and the jet Lorentz factor both increase as
$R^{a/4}$ for $a \le 2$ \citep{komissarov09}\footnote{The result $\Gamma\propto
R^{a/4}$ is easy to understand. To maintain pressure equilibrium, the 
magnetic field in the jet comoving frame falls off as $R^{-a/2}$ since the 
pressure on the sideways surface of the jet, which is perpendicular to the 
jet velocity, is same in the jet comoving frame and the star's rest frame.
 Let us take the jet transverse
radius to increase with $R$ as $R_\perp(R)$. The transverse and the
radial components of the magnetic field, in the rest frame of the
star, scale as $B_\phi\propto R_\perp^{-1}$ and $B_r\propto R_\perp^{-2}$ 
respectively. These field components in the jet rest frame vary as 
$R_\perp^{-1}/\Gamma(R)$ and $R_\perp^{-2}$. $B_\phi'$ (jet comoving frame 
transverse field) should not be much stronger than $B_r'$, otherwise the
jet becomes unstable and constricted. And it is difficult to have $B_r'>
B_\phi'$ over an extended interval of $R$ since that requires the jet
to continue to flare up rapidly in the transverse direction, which is 
prevented by the pressure of the external medium. Taking $B_\phi'\sim B_r'$ 
(for a jet in lateral pressure balance with the external medium) we find 
$\Gamma\aprop R_\perp$, and therefore the magnetic pressure falls off
as $\Gamma^{-4}$. Equating the magnetic pressure with the external pressure,
we finally obtain $\Gamma\propto R^{a/4}$.
 We are grateful to Jonathan Granot, for pointing out the simple, physical, 
arguments in this footnote.}.  
Thus, the radius where the jet Lorentz factor is equal to the 
fast-magnetosonic wave Lorentz factor is given by $R_{ms}\sim R_0 
\sigma_0^{4/3a}$; where $R_0$ is the radius where the jet is launched. 
For $a\gae2$, the central region of the jet ceases to be
in causal contact with the external medium at radius $R_{ncc}\sim R_0 
\sigma_0^{4/3a} \theta_j(R)^{-8/3a}$, and consequently the jet acceleration
is more or less terminated at $R\sim R_{ncc}$.

A steady state collimated outflow, with a small opening angle $\theta_j(R)$, 
confined by the pressure of an external medium, can accelerate to a terminal 
LF  $\Gamma \sim \min\left\{\sigma_0, \sigma_0^{1/3} \theta_j^{-2/3}\right\}$
while causal contact with the external medium is maintained. This 
suggests that for an efficient 
acceleration of jet to LF $\Gamma\sim\sigma_0$, $\theta_j\sigma_0$ should be
less than 1, whereas GRB afterglow observations suggest $\theta_j\Gamma
\sim 10$ \citep{panaitescu02}. MHD simulations of high magnetization 
jets carried out by \cite{komissarov10a} and \cite{tchekhovskoy10} 
find that magnetic field lines fan outward
rapidly when the jet emerges from the surface of the progenitor star
into the surrounding vacuum. This leads to a sudden increase to the jet 
LF by a factor of a few to $\sim10$ for long duration GRBs while their
 jet opening angle remains essentially unchanged. The rapid acceleration phase 
ceases when the rarefaction wave crosses the jet in the transverse direction. 
This short lived phase of sudden acceleration could be responsible for 
$\theta_j \Gamma \sim 10$ as GRB observations suggest. However, jets 
produced by short duration GRBs, which are not collimated by the envelope
of a star, are unlikely to undergo this sudden acceleration phase, 
and yet for these bursts $\theta_j \Gamma \gae 10$. This poses an interesting
puzzle regarding jet acceleration mechanism for a Poynting flux dominated jet.

We now drop the steady state assumption, and consider the acceleration of
a magnetic outflow of a finite, short, duration. An outflow of a short 
spatial extent can undergo efficient acceleration while traveling 
in vacuum as a result of adiabatic expansion, e.g. \cite{contopoulos95}
who considered adiabatic expansion and acceleration of a Newtonian jet,
and \cite{granot11} showed that a relativistic outflow can attain 
the limiting LF of $\sigma_0$ while traveling in vacuum and 
its $\sigma$ can decrease well below unity as a result of continued
adiabatic expansion. The adiabatic expansion and acceleration of a 
spherical, relativistic, outflow of short spatial extent with $\sigma_0\gg1$ is 
described next.

Consider a thin shell of magnetized plasma undergoing adiabatic expansion
in vacuum driven by magnetic pressure. For simplicity, we
will consider the magnetic field in the shell to be uniform and of 
strength $B(R)$ in the CoE frame when the shell is at radius $R$. The field
orientation is transverse to the radial vector. Consider
two spherical surfaces within this magnetized shell, one of which lies
close to the front end of the shell and the other somewhere in the middle.
These surfaces are frozen into the shell plasma and move with them as the
shell expands. The separation between the surfaces is $\xi(R)$ 
in the CoE frame when the shell is at a distance $R$ from the center. 
The difference in speed between these surfaces, in the CoE frame, is
$\sim c/[2\Gamma(R)^2]$; the front end of the shell is moving faster. The 
plasma in the shell is in causal contact in the radial direction (the 
causal contact in the transverse direction extends only to distance 
$\sim R/\Gamma$).

The separation between the surfaces increases with $R$ as
\beq
  \xi(R) = \xi(R_i) + \int_{R_i}^R {dr\over 2\Gamma^2(r)}.
\eeq
The magnetic field strength can be obtained by flux conservation across
a planar annulus of width $\xi$ that is perpendicular to the two spherical 
surfaces
\beq
   B(R) = B_i {\xi_i R_i\over \xi(R) R},
\eeq
where $B_i \equiv B(R_i)$ \& $\xi_i \equiv \xi(R_i)$.

The total electro-magnetic energy contained in between the two spherical 
surfaces, in the CoE frame, is 
\beq
  E_B = B^2 R^2 \xi(R) = B^2_i R_i^2 \xi^2_i/\xi(R) \sim
    {L\xi_i\over c}\left[ 1 - \int_{R_i}^R {dr\over 2\Gamma^2(r) \xi_i} 
   \right].
\eeq
The last step in the above equation was 
obtained by assuming that $[\xi(R)/\xi_i -1]\ll 1$, which is a fine
approximation to describe the dynamics of the shell because 
when $\xi(R)\sim 2\xi_i$, the electro-magnetic energy drops by a
factor 2 and $\Gamma\sim\sigma_i/2$, i.e. the shell LF is close to 
attaining its terminal value.

The rate of increase of the kinetic energy of the plasma contained between
the two spherical surfaces should be equal to the rate of decrease of 
electro-magnetic energy, i.e.
\beq
  {d\over dR} \left[4\pi R^2 \xi \rho c^2 \Gamma\right] = -{d E_B\over dR}
   \sim {L\over 2c\Gamma^2},
\eeq
or
\beq
  {d (M c^2\Gamma)\over dR} \sim {L\over 2c\Gamma^2},
\eeq
where $M=4\pi R^2\xi\rho$ is the plasma mass contained within the 
shell of thickness $\xi$, which does not change with time for a cold
outflow undergoing adiabatic expansion. The solution of the above equation 
is straight forward to obtain and is given by
\beq
   \Gamma(R) \sim \Gamma_i\left[ 1 + {3\sigma_i\over 2\Gamma_i^2} {(R-R_i)\over
       \xi_i}\right]^{1/3},
   \label{gamma_adia_mag}
\eeq
where $\Gamma_i=\Gamma(R_i)$ and $\sigma_i=\sigma(R_i)$. Note that 
$\Gamma$ attains a value $\sim\Gamma_i\sigma_i^{1/3}$ when the jet
has traveled a distance $\sim \xi_i \Gamma_i^2$ (Granot et al.  2011);
for an outflow of a finite opening angle, 
$\Gamma_i\sim\sigma_0^{1/3}\theta_j^{-2/3}$ is the
LF at the time when the central part of the jet loses causal contact 
with the surrounding medium, so that any further acceleration of the jet
results from its radial expansion.

The LF increases with radius as $R^{1/3}$ until it approaches 
$\Gamma_i\sigma_i\sim\sigma_0$ at $R_s \sim \xi_i\sigma_0^2$ (eq. 
\ref{gamma_adia_mag} is not valid beyond this radius). The overall momentum 
conservation of the outflow is maintained by the back-end of the shell
slowing down to a LF order unity, while the outer part of the shell, which
contains most of the energy and momentum, accelerates to $\Gamma\sim\sigma_0$.

For $R\gg R_s$, the LF is approximately constant, and therefore $\xi\aprop R$
(the radial width of the jet increases linearly with $R$),
$B\propto R^{-2}$ and $\sigma\propto R^{-1}$, e.g. \cite{granot11}. So the 
shell magnetization
can drop to well below unity at large distances, and shell collisions 
can then in principle convert the jet kinetic energy to internal energy 
efficiently.

\subsubsection{Magnetic dissipation and jet acceleration}
\label{magnetic_jet_dissipation}

If the magnetic field geometry in a high $\sigma_0$ outflow is such as 
to promote reconnection and dissipation, then a fraction of the magnetic energy 
can be converted to jet kinetic energy (eq. \ref{gamma_sigma}) and a 
 fraction goes into accelerating electrons \& protons by the electric 
field in the current sheet. Magnetic field reversing
directions on short length scales, such as the stripped wind from a fast 
rotating pulsar, or scrambled field lines that result from repeated internal 
collisions of ordered, magnetized, outflow \citep{zhangyan11}, are examples 
where reconnection is expected to take place. In the former case, the 
dissipation of magnetic energy and jet acceleration are gradual processes that 
take place over an extended range of radius, and is discussed below. In the
latter case, the dissipation can be sudden, which is triggered by an
instability when the magnetic geometry becomes sufficiently tangled.

The dissipation of magnetic energy, and jet acceleration, for a
reversing magnetic field geometry, or any similar 
configuration that is conducive to reconnection, is described 
using a simplified picture that should capture the basic physics of a 
rather complex process.

Two effects control the acceleration of a jet when magnetic field is dissipated.
One of which is a drop in thermal plus magnetic pressure when magnetic field 
is dissipated which can speed up magnetic reconnections (conversion of 
magnetic energy to thermal energy of particles and photons, even in the 
absence of any radiative loss, leads to a drop in the total pressure 
because the magnetic pressure is equal to the energy density, 
whereas the pressure for a relativistic fluid is one-third its 
energy density). The other process is the conversion of thermal energy
(from magnetic dissipation) to jet kinetic energy as a result of adiabatic 
expansion. For the simplified calculations presented below, we will ignore 
radiative losses and assume that a good fraction of the magnetic energy
dissipated is converted to bulk kinetic energy of the jet.

Let us consider an outflow which has a uniform magnetic field of strength $B$
whose direction reverses on a length scale $\ell_0$ (these quantities are 
in the CoE frame). The magnetic field is toroidal, and thus $B(R)\propto 
R^{-1}$. We assume that there is no differential velocity across stripes of 
radial width $\sim\ell_0$, and therefore the length scale over which the
magnetic field reverses direction ($\ell_0$) does not change with $R$.

Let us consider a highly simplified model for reconnection where we assume,
following \cite{drenkhahn02} and \cite{drenkhahn02b}, that the reconnection
speed in the comoving frame of the jet is a fraction of the Alfv\'en speed, i.e.
the speed at which plasma from outside the current sheet flows into it 
is $v'_{in} = \epsilon V'_A$ (see Figs. \ref{FIG:parker_current_sheet}
and \ref{FIG:petschek_current_sheet} for schematic sketches of possible
reconnection scenarios). For a high $\sigma$ plasma, the LF corresponding
to the Alfv\'en speed is $\sigma^{1/2}\gg1$, and thus we take $v'_{in} = 
\epsilon c$.

The radial width of the region where the magnetic field has been
dissipated when the jet has traveled a distance $R$ from the center of
explosion is
\beq
   w(R) \sim v'_{in} \left[{ R\over c \Gamma(R)}\right] {1\over \Gamma(R)} =
      {\epsilon R\over \Gamma^2(R)},
\eeq
where the factor $R/(c\Gamma)$ is time elapsed in the jet comoving frame,
and $\Gamma^{-1}$ transforms the comoving length to the CoE frame.

The total energy -- magnetic, thermal and jet kinetic energies -- contained 
within a segment of jet of radial width $\ell_0$ should not change as the 
jet propagates to larger radii since the net energy flux across the segment 
is zero for a uniform system. Thus, any loss of magnetic energy should 
show up as an increase to the kinetic energy of the jet when the thermal 
energy share can be ignored.  Therefore,
\beq
   4\pi R^2 \rho \Gamma c^2 \ell_0 \sim w r^2 B^2 \sim w L/c,
\eeq
where $L$ is the total luminosity carried by the jet (which is a conserved 
quantity in absence of radiative losses). Since the mass flux associated with
the jet is $\dot{M} = 4\pi R^2 \rho c$, we can rewrite the above equation as
\beq
   \dot{M} \Gamma c^2 \sim {L w\over \ell_0}, \quad\quad{\rm or} \quad\quad
      \Gamma \sim {L\over \dot{M} c^2} {\epsilon R \over \Gamma^2 \ell_0}.
\eeq
 From equation (\ref{gamma_sigma}) $L/(\dot{M} c^2)=\sigma_0$, and thus
we arrive at the desired scaling for jet LF with $R$
\beq
   \Gamma(R) \sim (\epsilon \sigma_0)^{1/3} (R/\ell_0)^{1/3}.
 \label{gamma_R_B_dissipation}
\eeq
It should be noted that the increase of $\Gamma$ with distance for the 
magnetic dissipation model (eq. \ref{gamma_R_B_dissipation}) is same as that 
for the adiabatic expansion case given by equation (\ref{gamma_adia_mag}), 
even though the underlying physical processes are very different. The 
reason for the identical scaling is that the increase of $\Gamma$ 
with $R$ is ultimately set by the speed of rarefaction waves for the
adiabatic expansion model and the reconnection speed (for magnetic dissipation) 
which are both of order $c$.

The jet LF attains its terminal value, $\sim \sigma_0$, when $w(R_s)\sim 
\ell_0$, i.e. when the magnetic field in the entire slab of width 
$\ell_0$ has undergone reconnection. The radius where this occurs, $R_s$, 
is estimated from the above expression for $w$ and is given by
\beq
   R_s \sim \ell_0 \sigma_0^2/\epsilon.
\eeq
If the magnetic field in the outflow reverses direction on the length scale of
light-cylinder radius for a milli-second pulsar, or the Schwarzschild
radius for a $\sim 10 M_\odot$ black hole, believed to be likely central 
engines for long-GRBs, then $\ell_0 \sim 10^7$cm. Numerical simulations 
for relativistic reconnection find $\epsilon\sim 10^{-2}$
\citep[e.g.][]{takahashi11}.  Therefore,
$R_s\sim 10^{15} \sigma_{0,3}^2$cm, which is much larger than the LF 
saturation radius for a thermal fireball ($10^{10}$cm).

\cite{mckinney12} studied the reconnection process of GRBs 
in a striped wind in detail.
They found a transition from collisional reconnection to collisionless
reconnection at a radius around $10^{13}-10^{14}$ cm, and they identify this
as the GRB prompt emission site.

\medskip
\subsubsection{Basic reconnection physics} 
\label{reconnection}

A general discussion of magnetic field dissipation and jet acceleration 
when the direction reverses on a short distance scale is described in the 
previous subsection. In this subsection, we describe some aspects of the 
physics of magnetic dissipation in a reconnection layer, or current sheet,
for an electron--proton plasma.

\begin{figure}
\includegraphics[width=13cm]{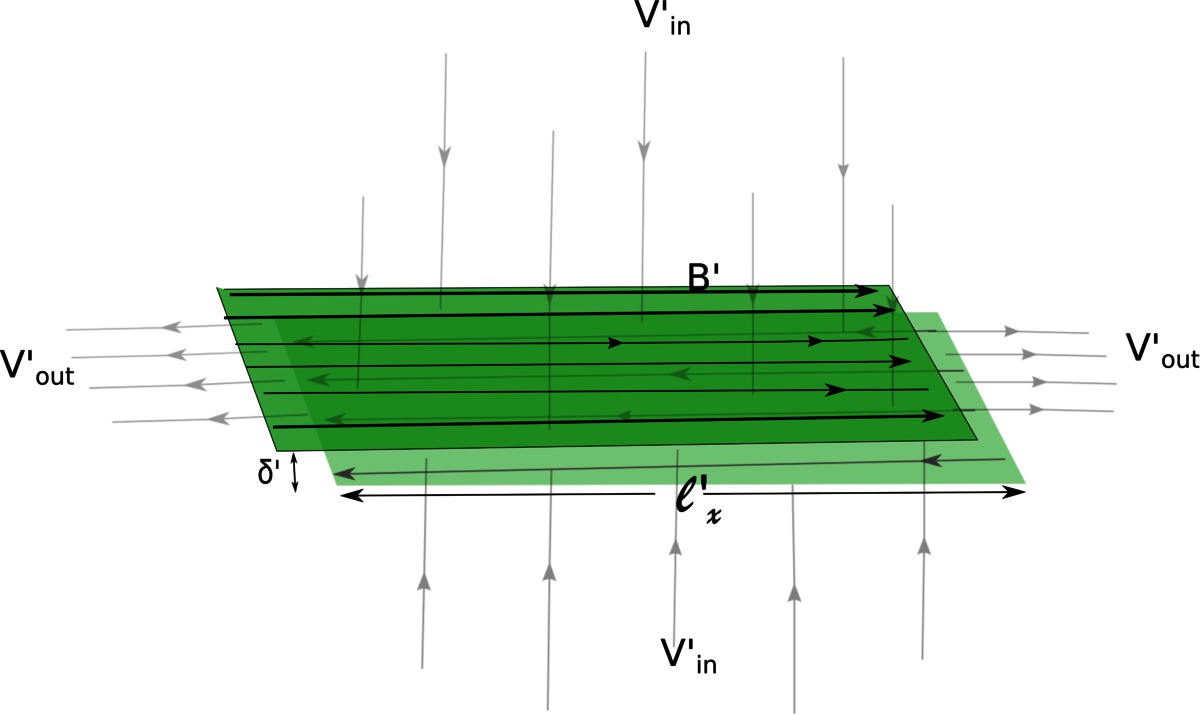}
\caption{A schematic sketch of current-sheet, and plasma from outside 
the sheet flowing toward it, for Sweet-Parker magnetic reconnection. The
sketch only shows the region in the immediate vicinity of the current sheet.
Magnetic field lines outside of the sheet are curved away from the center of 
the current sheet.
}\label{FIG:parker_current_sheet}
\end{figure}

According to \cite{sweet58} and \cite{parker57}, plasma consisting of 
oppositely oriented magnetic field can undergo forced reconnection
where the magnetic field is dissipated on a time scale much shorter than 
the diffusion time. The basic configuration is a thin current sheet of
width $\delta'$, and length, $\ell'_x$, where magnetic field is dissipated due 
to its large gradient across this region. Plasma carrying magnetic fields 
of strength $B'$ flows into this region at speed $v'_{in}$, and is squirted 
out of the thin current sheet at proper-velocity $v'_{out}\gamma'_{out}$ 
(see Fig. \ref{FIG:parker_current_sheet}).
The basic features of Sweet-Parker reconnection for a 
relativistic plasma -- with magnetization parameter $\sigma\gg1$ --- can be 
obtained from the conservation of mass and energy flux at the surface of 
the current sheet, and the pressure balance. The mass and energy flux 
conservation equations are
\begin{equation}
   n_1'' \ell'_x v'_{in} = n_2'' \delta' v'_{out}\gamma'_{out},
 \label{mass_flux_reconnect}
\end{equation}
\begin{equation}
   (B'^2/4\pi) \ell_x' v'_{in} = n_2'' \delta' m_p c^2 v'_{out}\gamma_{out}'^2
    \gamma_t',
  \label{energy_flux_reconnect}
\end{equation}
where $n_1''$ \& $n_2''$ are plasma densities outside and inside the current 
sheet, respectively, as measured in the local plasma rest frame, and 
$\gamma_t'$ is the Lorentz factor associated with 
the random velocity component of protons, in the mean rest frame of plasma, 
inside the current sheet. The ratio of these equations give
\begin{equation}
  {B'^2 \over 4\pi n_1'' m_p c^2 } \equiv \sigma = \gamma_A'^2 = 
   \gamma'_{out}\gamma'_t, 
 \label{gamma_t_reconnet}
\end{equation}
where $V'_A\gamma_A'=B'/(4\pi n''_1 m_p)^{1/2}$ is the Alf\'ven wave 
proper-velocity outside the current sheet. If $\gamma'_t$ were to be of order 
unity, then the Lorentz factor of the plasma leaving the current-sheet is 
$\sim \gamma_A'^2$ \citep{lyutikov03c}. \cite{lyubarsky05} has suggested
that $\gamma'_{out} \sim 1$, however, his argument is based on making 
ad hoc assumptions regarding the length scale over which $\gamma'_{out}$
changes and the strength of the magnetic field component perpendicular
to the current sheet, which might not apply to GRB jets.

The magnetic pressure outside the current sheet should roughly equal 
the pressure inside the sheet provided by the transverse ``thermal'' 
motion of protons. This yields
\begin{equation}
   {B'^2 \over 8\pi}\sim n_2'' \gamma_t' m_p c^2 \quad\quad {\rm or} \quad\quad
     {n_2''\over n_1''} \sim {\sigma\over \gamma_t'} \sim \gamma'_{out},
   \label{n2_n1}
\end{equation}
where we used equation (\ref{gamma_t_reconnet}) to obtain the last equality. 

The plasma inflow velocity toward the current sheet ($v'_{in}$), which
is the speed at which forced reconnection can proceed, is regulated
by the requirement that the rate at which magnetic flux 
flows into the current sheet should not exceed the rate at which 
magnetic field is dissipated inside the sheet (otherwise magnetic field will 
build up and prevent plasma from entering the sheet).
Let us assume that the diffusion coefficient for magnetic
field dissipation is $\eta$, which could be microscopic or turbulent in
origin. The time scale for magnetic field dissipation in the current sheet
is
\begin{equation}
   t_{B,dissi}' \approx {\delta'^2\over \eta}.
\end{equation}
Therefore, the effective speed at which magnetic field dissipation
proceeds inside the current sheet is given by
\begin{equation}
   {\delta'\over t_{B,dissi}'} \sim {\eta \over \delta'}.
\end{equation}

The speed for plasma flowing into the current sheet, $v'_{in}$, should be
roughly this speed, i.e. $v'_{in} \sim \eta/\delta'$.
Using equations (\ref{mass_flux_reconnect}) \& (\ref{n2_n1}) we find
\begin{equation}
   v'_{in} \sim (V'_A v'_{out})^{1/2} \gamma_{out}' s^{-1/2},
   \label{inflow_speed}
\end{equation}
where $s$ is the Lundquist number, defined as
\begin{equation}
   s \equiv {\ell_x' V_A'\over \eta}.
\end{equation}

\begin{figure}
\begin{center}
\includegraphics[width=13cm]{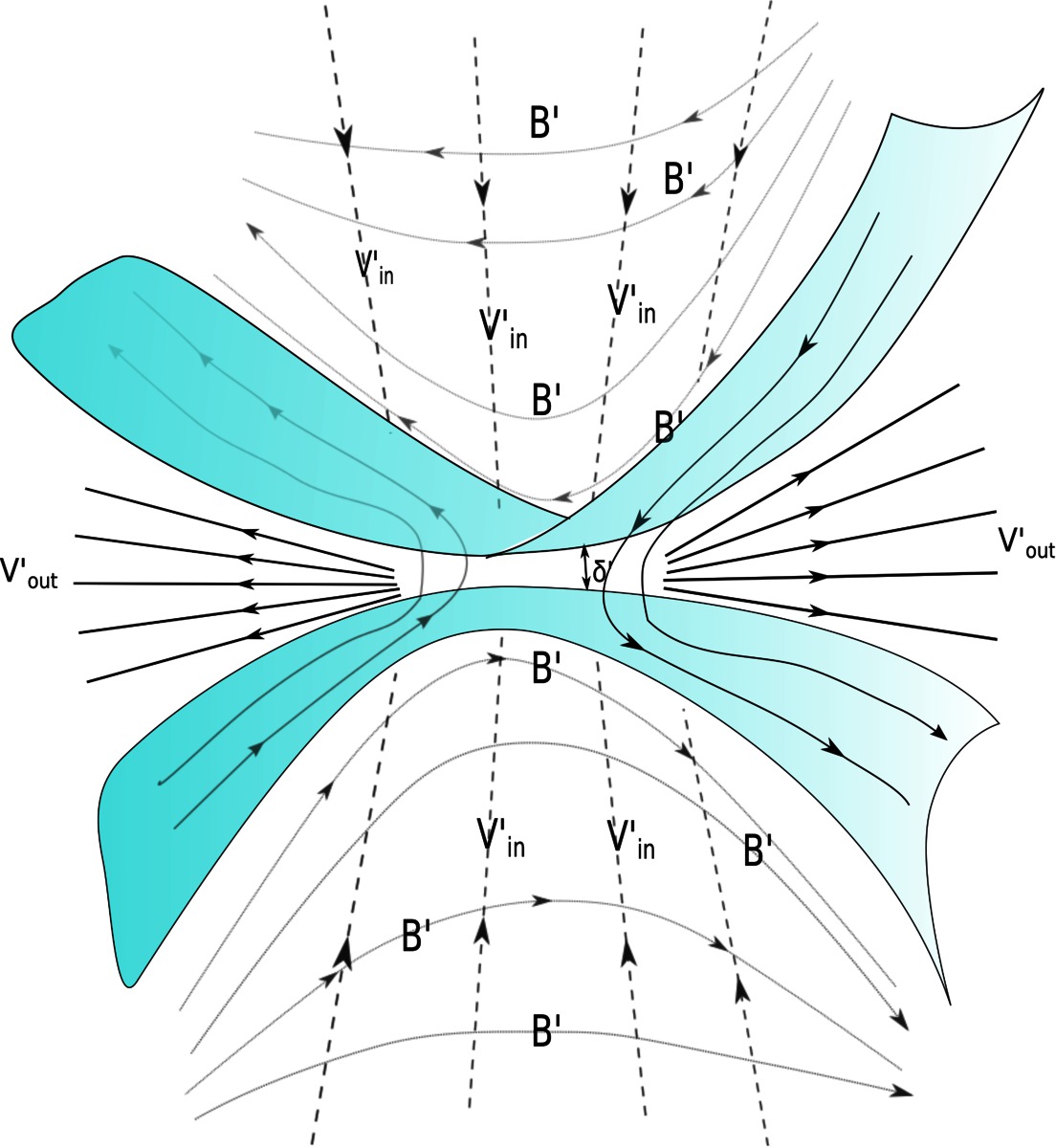}
\caption{A schematic sketch of plasma inflow, and current-sheet, for 
Petschek magnetic reconnection. Much of the plasma flowing toward the current
sheet does not pass through it, but instead is redirected by standing shock 
waves; the inflow and outflow regions are separated by stationary slow 
mode shocks.
}\label{FIG:petschek_current_sheet}
\end{center}
\end{figure}

For a typical GRB jet with $\Gamma\sim10^2$, $\ell_x'\sim R/\Gamma$
(size of causally connected region in the jet comoving frame), 
isotropic jet luminosity of 10$^{52}$erg s$^{-1}$ carried by magnetic fields 
($B'\sim 10^4$G $R_{15}^{-1}$), \& $V'_A\sim c$, we find $s\sim 10^{11}$
in the Bohm diffusion limit, i.e. when $\eta= c R_L'$  ($R_L'$ is the proton
Larmor radius), and hence the speed at which reconnection is expected to
proceed according to Sweet-Parker mechanism is $\sim 10^{-5}$c, which is 
much too slow to be of practical interest. A fast steady-state reconnection 
scenario was proposed by \cite{petschek64}, which invokes a much shorter 
length for the resistive layer, thereby significantly increasing the speed
at which reconnection can proceed (see. fig. \ref{FIG:petschek_current_sheet}). 
However, according to resistive MHD simulations the Petschek model is
unstable, unless the magnetic diffusivity increases near the 
X-point \citep[e.g.][]{uzdensky00}.
Simulations also find that the Alfv\'enic tearing instability of Sweet-Parker
current sheet \citep[e.g.][]{loureiro07,samtaney09} could increase
the reconnection rate significantly.

\cite{lazarian99} proposed that reconnection in magnetic fields 
with stochastic geometry can proceed rapidly, thanks to the turbulent nature
 of the magnetized fluid that both broadens the reconnection zone and allows 
many independent reconnection events to occur simultaneously.
Moreover, once reconnection gets started in one localized region, 
it can trigger many other reconnection events as a result of the plasma 
squirting out of the current sheet at speed $V'_A$ and stirring up magnetic 
fields in neighboring regions. Three-dimensional numerical simulations of 
reconnection carried out by \cite{kowal09} provide support for this 
turbulence model. 

In the presence of turbulence, magnetic fields reconnect on the length scale 
$\lambda_\parallel$ for magnetic field fluctuation, rather than the much 
larger global scale $\ell_x'$. Accordingly, it is the effective Lundquist 
number
\beq
s = {\lambda_\parallel V_A'\over \eta},
\eeq
that determines the speed at which reconnection proceeds.

In turbulent reconnection, the global reconnection rate is
larger by a factor $\sim \ell_x'/\lambda'_\parallel$, since 
there are $\sim \ell_x'/\lambda_\parallel'$ reconnection sites along
any random direction cutting across the outflow \citep{zhangyan11}. 
As a result, the small local reconnection speed $V'_{in}\sim V'_A/s^{1/2}$
is adequate to power a GRB as long as $(\ell_x'/\lambda'_\parallel)s^{1/2} 
\sim 1$, or
\beq
   \lambda'_\parallel \sim \ell_x'^{2/3} (\eta/c)^{1/3}.
\eeq
We note that the effective speed at which magnetic field needs to be 
dissipated should be of order the speed of light (comoving frame) in 
order to obtain the large luminosity of GRBs at a reasonable efficiency 
of $\gae 10$\% \citep{zhangyan11}.

\medskip
\subsubsection{Forced reconnection of magnetic fields for 
   a high-$\sigma$ GRB jet and prompt $\gamma$-ray radiation} 
\label{forced-reconnection}

The magnetic field geometry for a black hole central engine is 
likely to have a non-alternating toroidal configuration,
which is less amenable to dissipation via reconnection. 
\cite{zhangyan11} have suggested the possibility that if the Lorentz 
factor of the outflow varies with time, then multiple internal collisions 
can scramble magnetic fields and ignite reconnections, thereby converting 
a fraction of magnetic energy to thermal energy and radiation, and named 
it the ICMART model
(Internal Collision-induced MAgnetic Reconnection and Turbulence).
Since magnetic fields are stretched in the transverse direction, thereby
straightening out field lines in between episodes of collisions when the 
shell undergoes adiabatic expansion, frequent collisions with large
relative LF are required in order to sufficiently tangle up the magnetic
fields for efficient reconnection. 

According to the ICMART model, $\gamma$-rays are produced in highly 
localized reconnection regions via the synchrotron process. The Lorentz 
factor of the outflow in these regions could be relativistic, say,
of the order of the Alfv\'en wave 
LF;  $\gamma_t'\sim\gamma_A' = (1+\sigma)^{1/2}$.
 Therefore, the observed radiation is dominated by those
regions where the outflow velocity is pointing in our direction,
and the observed duration of a $\gamma$-ray pulse from one of these regions
is at most $\sim R/(c\Gamma^2 \sigma)$; where $R$ is the distance from the 
center of explosion where the magnetic energy is being dissipated, and $\Gamma$
is the LF of the jet associated with its mean bulk velocity. A $\gamma$-ray 
pulse can be of even shorter duration if the size of the reconnection 
region is much smaller than the transverse size of the jet. This idea of 
producing rapid variability of $\gamma$-ray lightcurves even when radiation 
is produced at a large distance (R$\gae 10^{15}$cm) --- as suggested by a 
number of observations discussed in \S\ref{source_distance} ---  is a 
generic feature of relativistic turbulence models described by 
\cite{lyutikov03,narayan09,kumarnarayan09,lazar09}, and the ICMART model 
of \cite{zhangyan11}. 

One of the positive features of the relativistic turbulence model is its 
high radiative efficiency\footnote{The internal shock model seems incapable of 
explaining the observed GRB spectrum and has low radiative efficiency 
in the MeV band as discussed in \S\ref{shock_synchro}.}, and unlike the 
internal-shock model, it is capable of explaining the observed GRB spectra
\citep{kumarnarayan09}. Since the model invokes synchrotron emission of
particles in an ordered magnetic field (which is being rapidly distorted),
the observed emission is expected to be highly polarized, with the
polarization degree decreasing with time during the course of a broad pulse
\citep{zhangyan11}. According to the model, only a small amount of energy 
should come out in the IC component, which is consistent with Fermi 
observations \citep{kumarnarayan09}.

\cite{lazar09} criticized the relativistic turbulence model by suggesting
that it tends to
produce too spiky light curves, with each pulse being symmetric.
The ICMART model invokes an exponential
growth of the number of mini-jets due to the reconnection-turbulence
``avalanche'', which abruptly discharges the magnetic field energy.
As a result, at any instant, an observer would receive emission
from many mini-jets that beam in random directions. While those beaming
towards the observer make rapid spikes, the other off-beam jets contribute
to the broad component \citep{zhangzhang14}. The rising wing of a broad
pulse is defined by this exponential growth of the number of mini-jets,
while the decaying wing is controlled by the high-latitude curvature
effect. As a result, this model produces an asymmetric broad pulse for each
ICMART event. A GRB is composed of multiple ICMART events, and the 
simulated light curves are roughly in
line with observed lightcurves \citep[e.g.][]{gao12}. 
%However, it
%is unclear how relativistic turbulence models can produce highly 
%asymmetric pulse profiles that are found in prompt $\gamma$-ray 
%lightcurves, and several other weaknesses of the model are discussed
%in \cite{lazar09}. 
However, full numerical simulations invoking a high-$\sigma$, high-$\Gamma$
outflow with strong (cascade) magnetic dissipation are not available.
Recent numerical simulations of relativistic 
magnetohydrodynamical turbulence, e.g. \cite{zrake12}, shed some
light on the power spectrum of velocity field, but we are far from 
being able to simulate anything close to the parameters expected for GRB
jets.

It is expected that a Poynting jet would suffer rapid dissipation of magnetic  
energy at the deceleration radius --- the distance from the CoE where 
roughly half of the jet energy is transferred to the circum-stellar 
medium --- if it were to be able to travel to this radius with its
high magnetization parameter intact. This is because a collision with the 
circum-stellar medium sends a strong megneto-sonic wave 
into the outflow. This can lead to development of a large gradient in the
magnetic field, and trigger current driven instabilities that dissipate
magnetic fields \citep{lyutikov03}. The ensuing acceleration 
of particles then produces $\gamma$-rays via the synchrotron mechanism. A 
signature of this mechanism is that $\gamma$-rays are generated 
close to the deceleration radius.

\medskip
\subsubsection{Particle acceleration}
\label{magnetic_acceleration}

\begin{figure}
\includegraphics[width=13cm]{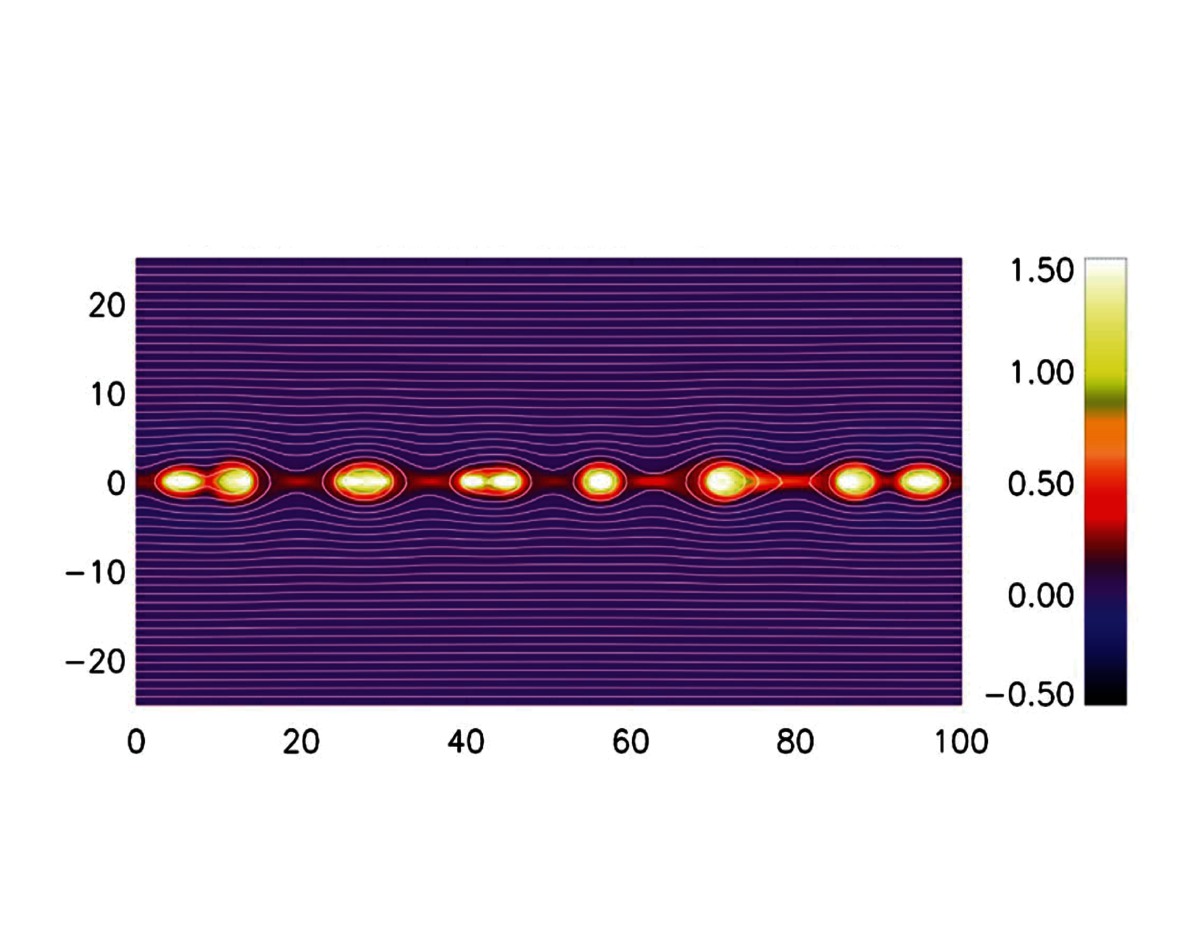}
\caption{Formation of magnetic islands due to tearing instability is
shown in this numerical simulation result taken from Hesse \& Zenitani
(2007). Plotted are magnetic field lines and the component of current density 
perpendicular to the figure plane with color coded strength (color bar 
to the right).
}\label{FIG:magnetic-island}
\end{figure}

Particles are accelerated in current sheets, where magnetic field
dissipation takes place,  via a number of different processes. These 
sheets have regions where the electric field is larger than the local magnetic 
field and where particles can be accelerated to relativistic speeds
by the electric field. Tearing instability of the current
sheet, in the non-linear phase, produces a number
of magnetic islands (plasmoids) moving close to the Alfv\'en speed
(see Fig. \ref{FIG:magnetic-island}). Particles
are also accelerated via the Fermi mechanism by scattering off of these
plasmoids. Moreover, converging inflow of plasma toward the current
sheet provides another venue for particle acceleration via the first-order 
Fermi process \citep[e.g.][]{giannios10}.
These processes together produce a hard
spectrum for accelerated particles that cuts-off steeply at some
LF to ensure overall energy conservation.

The minimum electric field inside the current sheet is $E' = {\bf v}_{in}' 
\times {\bf B}'/c$ (where $v_{in}'$ is the speed with which plasma flows 
into the sheet). This
follows from the induction equation for time independent reconnection, i.e. 
$\nabla\times {\bf E}' = 0$, according to which the electric field
parallel to the current sheet inside the sheet has the same magnitude 
as the outside field.
 The electric field inside the sheet can be significantly 
larger than this due to particle inertia and non-zero divergence of
anisotropic pressure tensor (terms in the generalized Ohm's law equation), 
\cite[e.g.][]{hesse07}.

This field can rapidly accelerate particles to high LFs, provided that
the particle trajectory passes through the region where the electric field
is larger than the magnetic field. A number of people have calculated
particle acceleration and their distribution function by following
particle trajectory in the combined electric and magnetic fields inside
current sheets \citep[e.g.][]{giannios10,uzdensky11,bessho12}.
A number of groups have carried out numerical Particle-In-Cell (PIC) 
studies of reconnection and particle acceleration 
\citep[e.g.][]{zenitani01,zenitani07,zenitani08,jaroschek04,bessho07,bessho12,petri07,liu11b,sironi11a,sironi12,kagan13}  --- see \cite{zweibel09} for a more
complete discussion of the extensive literature. According to several 
of these simulations, much of the particle acceleration takes place
near X-points, which are located in between magnetic islands, due to 
the reconnection electric field \citep[e.g.][]{zenitani01,jaroschek04,sironi12},
 and some acceleration 
occurs due to first-order Fermi process as particles are reflected back 
and forth between converging islands \citep[e.g.][]{drake06,sironi12}.
However, little acceleration takes place while 
particles are trapped to an island.
Presence of a non-zero guide field does not change the acceleration 
process significantly unless its strength becomes of order the 
reversing magnetic field (the field undergoing reconnection)
in which case fewer particle pass through X-points and hence fewer 
particles are accelerated by the reconnection electric field 
and the mean thermal LF of accelerated particles is lower
\citep{zenitani08,sironi12}.

\begin{figure}
\includegraphics[width=13cm]{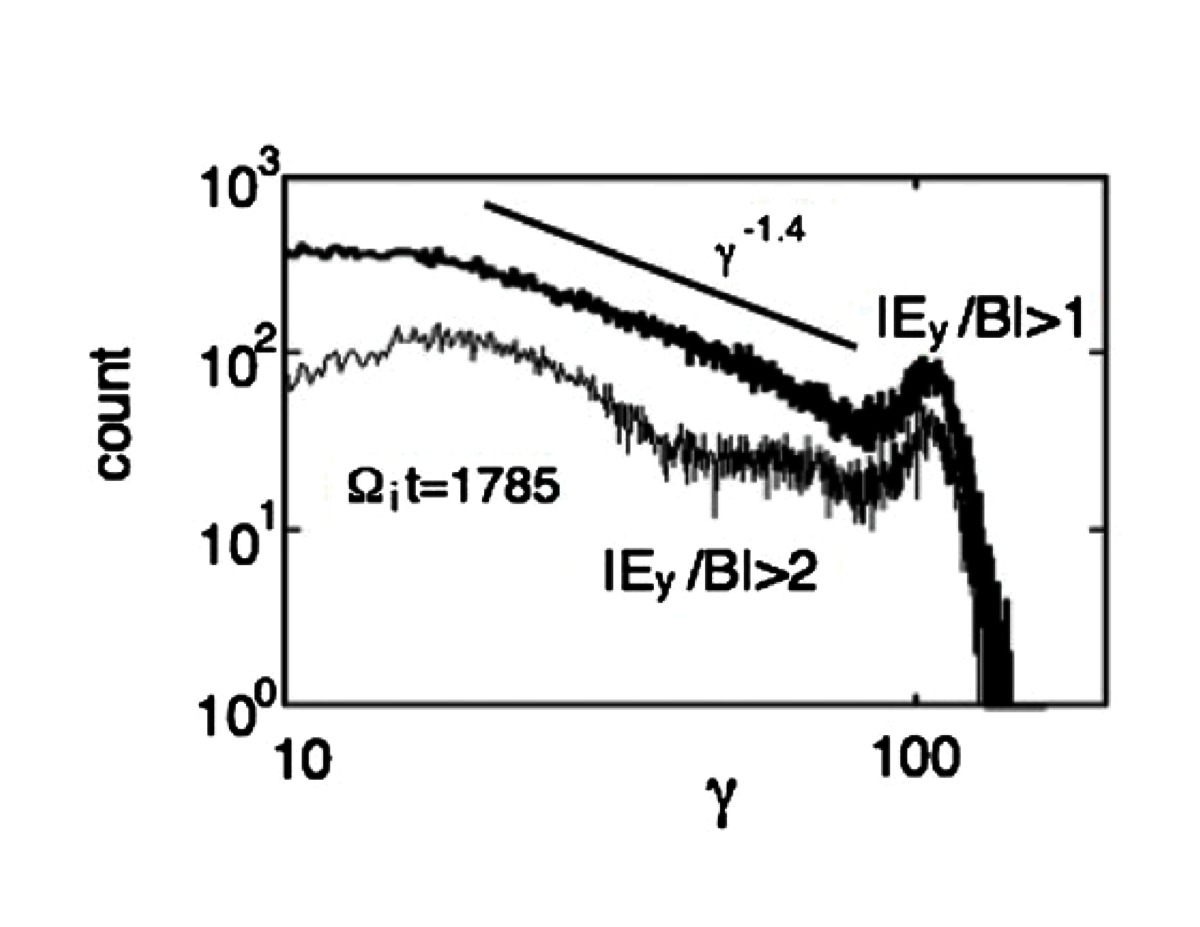}
\caption{Spectrum of particles accelerated in a current sheet according to
recent numerical simulations of \cite{bessho12}.
}\label{FIG:besso-particle-spectrum}
\end{figure}

The power-law index for the non-thermal electron distribution in 
magnetic dissipation, $p\equiv -d\ln n/d\ln\gamma$, is reported to be 
about $1$ \citep{romanova92,zenitani01,larrabee03}.
The distribution function must steepen at some LF
in order to keep the total energy finite. In fact, several papers claim that
the distribution function in reconnection layer falls off 
exponentially at high LF (\cite{bessho12}
 find the fall off to be proportional to $\exp(-\gamma^{1/2})$ --- see 
Fig. \ref{FIG:besso-particle-spectrum} --- and  \cite{kagan13}
find the particle spectrum to be a superposition of two
thermal peaks); in contrast, $p<3$ for particles accelerated
in shocks almost up to the LF where particles are no longer confined 
to the system. Although numerical simulations don't offer a precise
answer as to the dependence of the particle terminal LF -- where the 
distribution function begins its steep decline -- on parameters such 
as $\sigma$, guide field strength, and relevant length scale of the system, 
energy conservation suggests that the average thermal LF of particles should 
be of order $\sigma$  as long as most of the particles
flowing into the current sheet undergo acceleration (which is expected,
since the reconnection electric field is fairly wide-spread in the sheet).
Results of the recent PIC simulations of \cite{sironi12}
for a relativistic striped wind are consistent with this expectation.

\cite{cerutti12,cerutti14} report strong, energy dependent, 
anisotropy in the distribution of accelerated particles such that higher
energy particles are more concentrated along the electric field in the
current sheet, which is perpendicular to the plasma inflow direction and
the outside alternating-magnetic field, i.e. more high energy particles 
are found along the guide field; according to their simulations the 
anisotropy increases with increasing energy, and similar beaming 
effect is also found in simulations of \cite{kagan13}. This result, 
if correct, has 
important implications for lightcurve variability for 
relativistic outflows associated with neutron stars and black holes.

\medskip
\subsection{Some off-the-beaten-track ideas for GRB prompt radiation 
mechanism}
\label{other_prompt_models}

The models we have described thus far as to how $\gamma$-ray prompt radiation 
is produced in GRBs are ideas widely discussed in the published 
literature and are popular amongst active researchers in this field.  Since 
our understanding of the $\gamma$-ray generation mechanism remains elusive, 
there is a possibility that none of these models survive a closer scrutiny
and more detailed future investigations. 
Therefore, we briefly discuss several other proposals put forward for the 
origin of GRB prompt radiation which have received somewhat limited attention 
but might contribute toward our eventual understanding of these enigmatic 
bursts. All of these models have strengths and weaknesses, but we are going 
to have to let the reader decide this for herself by reading the original 
papers since this section is already very long.

\cite{lazzati00} proposed that photons from the GRB progenitor
star, or from the cocoon produced by the passage of the jet through the
star,  undergo IC scattering by cold electrons in the highly relativistic
jet to produce the prompt $\gamma$-ray radiation. Many variations of the 
general idea of converting the kinetic energy of a relativistic jet to 
$\gamma$-ray radiation by inverse-Compton scatterings have been published.
For instance, \cite{broderick05}
considered IC scattering of supernova light
by a relativistic jet which is produced by accretion of supernova ejecta 
onto a neutron star companion; the system just prior to the supernova
was a helium star-neutron star binary system.

Dar, de R\'ujula and Dado have spent a huge amount of effort in developing 
a model for GRBs they refer to as the cannon-ball model.
They have written numerous papers analyzing different aspects of this model,
and carried out detailed comparisons with GRB data. We refer the interested 
person to their review article that describes the cannon-ball model and its
application to GRB observations \citep{dar04}.
The basic idea is that small packets of plasma, or ``cannon-balls'', are 
shot out from the GRB central engine moving at close to the speed of 
light (LF of these ``cannon balls'' is $\sim10^3$), and cold electrons 
in these objects IC scatter ambient radiation --- which was either 
produced by the supernova that preceded the launch of these bullets by 
several hours, or it resulted from episodes of intense stellar activity 
just prior to the death of the progenitor star, and then scattered by the 
dense wind of the star, resulting in a nearly isotropic radiation field
--- to produce the prompt $\gamma$-ray lightcurve.
The absence of diffractive scintillation in the 
VLBI data of a relatively nearby burst, GRB 030329 at $z=0.168$, 
is at odds with the expectation of the cannonball model, and so 
is the smaller than expected proper motion \citep{taylor04};
but see \cite{dado04} for an interpretation of the Taylor et al. 
observations according to the cannonball model.

Another model, developed extensively by Ruffini and his collaborators, is the 
so called ``fireshell'' model for GRBs, which suggests a unified picture  
for long- and short- GRBs \citep[][and references therein]{ruffini01a,bernardini07}.
According to this model, a spherically symmetric $e^\pm$ shell is ejected when 
a charged black-hole is formed in a catastrophic collapse \citep{ruffini01b}.
This model is characterized by the energy in this shell 
(related to the black-hole charge and mass) and the amount of baryonic matter: 
fireshells of low baryonic loading (fractional rest-mass energy in baryons 
less than 10$^{-5}$) produce short GRBs, whereas a larger baryon loading 
leads to long-GRBs. Ruffini et al. suggest that a radiation pulse is produced
when the fireshell becomes transparent (they call it P-GRB), which could
be either the precursor to a GRB or the first pulse of the
main burst. The fireshell interacting with blobs in the circum-stellar
medium produces more pulses of radiation in the prompt
$\gamma$-ray lightcurve \citep{bianco05}.
The basic physical processes at work in the generation of $e^\pm$ shell are 
described in an extensive review article \citep{ruffini10}.
The question of astrophysical processes responsible for the formation of a 
charged black-hole --- a central element of the fireshell model --- however, 
is unclear to us. It is also unclear how a spherical fireshell gives rise
to a jet break, and how to explain the association of long GRBs with 
supernovae but not short GRBs, and the different distribution for the location
of these two classes of GRBs in their host galaxies.

A number of people have invoked a precessing jet to explain the complex
$\gamma$-ray lightcurve of GRBs, e.g. 
\cite{lei07,romero10,liu10,fargion12}. However, it is difficult 
to avoid high baryon loading for a
precessing jet, which instead of moving straight through an evacuated polar 
cavity, is likely to collide with the cavity wall and entrain a lot
of baryonic matter, and therefore might not attain the high LF inferred
for GRBs.

\cite{ioka11}
 suggested a model where a jet of very low baryonic content and low 
magnetization undergoes internal shocks while still radiation dominated; 
it is suggested that the jet is confined by the external pressure of the 
progenitor 
star so that its cross-section increases more slowly than a conical jet, and
therefore it continues to be radiation dominated out to a much larger radius. 
In such radiation dominated shocks, thermal photons cross the 
shock front multiple times and their energy increases as a result of 
the first order Fermi process, i.e. energy is transferred from the bulk kinetic
energy of the jet to photons, when the photons are scattered by the
converging outflow associated with the shock. The emergent photon 
spectrum in this case has a non-thermal power-law shape above the peak.

\cite{titarchuk12} have proposed a two step process for the
generation of $\gamma$-rays by inverse-Compton scatterings.
The first step involves Compton scattering of low energy photons from the
GRB progenitor star, in the high Compton-Y parameter regime, by electrons of 
energy $\sim 10^2$ keV in the radiation dominated sub-relativistic
outflow produced in the explosion. The outflow subsequently expands and
becomes relativistic, and relatively cold electrons in the jet 
inverse-Compton scatter photons produced in step 1 that results in a 
non-thermal, Band, Spectrum. \cite{titarchuk12} have provided a detailed fit 
to the observed GRB data using this model.

\cite{kazanas02} proposed a ``supercritical pile'' model for GRBs. The
idea is that as a relativistic GRB blastwave propagates in the interstellar
medium, the Bethe-Heitler process ($p\gamma \rightarrow pe^+e^-$) may
reach resonance, namely, the typical energy of synchrotron radiation
of the pairs is just the one to ensure Bethe-Heitler kinetic condition,
and the column density of the photons also satisfy runaway production of
$e^+e^-$ pairs. Similar to the external shock model, this model invokes
an ``external'' site to discharge the kinetic energy of the blastwave,
which has difficulties to account for the GRB variability data.

\section{GRB central engine}

Although the progenitors of long- and short- GRBs might be very different
objects, the basic nature of the central engine --- the mechanism by which 
highly luminous relativistic jet is produced --- is expected to be similar 
for these bursts. The details of the process can be somewhat
different. The discussion below mostly focuses on long duration GRBs,
but short GRBs will be discussed whenever noticeable differences with long
GRBs exist.
%with long GRBs  and the discussion in this section applies mostly to 
%long-GRBs (information regarding binary star mergers and 
For a more detailed discussions of short GRB central
engine, please see e.g. 
\cite{ruffert99,rosswog03,aloy05,lee07,nakar07,baiotti08,dessart08,gehrels09,rezzolla11} 
and references therein.

The central engine of GRBs has not been positively identified, however, 
observations have narrowed down candidates to a small number of possibilities. 
Any successful 
GRB central engine model should be able to satisfy the following 
requirements: (1) Ability to launch an extremely energetic and luminous
jet whose luminosity greatly exceeds the Eddington luminosity;
(2) The jet must be clean, i.e. energy per baryon $\gg m_p c^2$,
 so that the outflow can reach ultra-relativistic speed with Lorentz factor 
greater than $\sim10^2$;
(3) The engine should be intermittent as suggested by the erratic
rapidly variable light curves\footnote{It is possible that the variability
observed in prompt $\gamma$-ray lightcurves is due to relativistic turbulence
at the location where $\gamma$-rays or produced, and in that case the 
jet luminosity from the GRB central engine might be a smooth function of time.
It is also possible that variability is introduced by the interaction of jet
with the stellar envelope. See \S\ref{forced-reconnection} for detailed 
discussion.}; (4) The central engine should be able to re-activate 
at later times to power softer flares\footnote{Some ideas to interpret
X-ray flares without invoking a re-activation of  central engine have been
proposed, but these proposals are not well developed to interpret the
entire X-ray flare phenomenology. See \S\ref{X-ray-flares} for a detailed 
discussion.}. 

Two types of widely discussed central engines satisfy 
these requirements: a 
hyper-accreting stellar-mass black hole 
\citep{woosley93,popham99,lee00,narayan01,dimatteo02,wangdx02,mckinney05,uzdensky06,chen07,lei09,lei13,yuan12,nagataki09,nagataki11},
and a rapidly spinning, highly magnetized, neutron star or ``fast magnetar''
\citep{usov92,thompson94,dailu98b,kluzniak98,wheeler00,zhangmeszaros01a,dai06,bucciantini08,bucciantini09,metzger11}. 
We describe these models in the following subsections.

\subsection{Hyper-accreting black holes}

If a GRB is powered by accretion onto a stellar mass black hole, 
a very high accretion rate is required. In general, one can write
\begin{equation}
 L_{\rm GRB} = \zeta \dot M c^2=1.8 \times 10^{51}~{\rm erg~s^{-1}}
~\zeta_{-3} \left(\frac{\dot M}{1~{\rm M_\odot~s^{-1}}}\right),
\end{equation}
where $\zeta$ is a dimensionless number that represents the efficiency of 
converting accretion power to radiation power. For reasonable values, the 
accretion rate for a typical cosmological GRB is 
$(0.01 - {\rm several})~{\rm M_\odot~s^{-1}}$. 

At these high accretion rates the plasma is extremely hot and forms a 
thick disk or torus around the central black hole/neutron-star.
Photons are trapped in the accretion flow, and neutrino cooling might be
effective only for a fraction of the burst duration close to the central
engine, so that the accretion flow is advection dominated (ADAF) or convective (CDAF) 
throughout much of the volume.  Close to the inner disk radius, the temperature is 
so high that neutrino cooling does become effective for at least some time, and 
in that case the disk temperature drops, density increases, and the geometrical 
shape of the flow is that of a thin disk; this is called 
 neutrino-dominated accretion flow (NDAF).

The accreting BH likely carries large angular momentum. 
This is naturally formed in a rapidly rotating
core. Due to the large
accretion rate, the BH can spin up further rather quickly. If a strong
magnetic field threads the spinning BH and is connected with
an external astrophysical load, BH spin energy can be tapped
via the Blandford-Znajek mechanism \citep{blandford77}.

In general, a GRB jet can be launched from a hyper-accreting BH via
three possible mechanisms:

\begin{itemize}
 \item Neutrino annihilation along the spin axis of a NDAF can drive
a hot jet with properties similar to what is conjectured in the
hot fireball model;
 \item Blandford-Znajek mechanism can launch a Poynting-flux-dominated
outflow from the central engine;
 \item The accretion disk can be also highly magnetized. A plausible,
but less well studied, mechanism is that differential
accretion would lead to accumulation of vorticity and energy within
the accretion disk, leading to eruption of magnetic blobs.
\end{itemize}

We discuss these mechanisms in the following subsections.

\subsubsection{Neutrino-dominated accretion flow (NDAF) and
advection-dominated accretion flow (ADAF)}

The structure of the GRB accretion disk depends on the mass of the
black hole $M$, accretion rate $\dot M$, radius $r$ from the central engine, 
and the poorly known viscosity (usually parametrized by a dimensionless
parameter $\alpha$). At the high accretion rate required for GRBs, the
disk temperature is very high. Above a critical accretion rate, the disk
is cooled by significant neutrino emission and is in the NDAF regime. 
Below the critical rate, neutrino cooling
is not important. The disk becomes much thicker, significant thermal
energy is ``advected'' into the black hole, and the disk is in the ADAF
regime.  For a given GRB accretion
disk, there is a characteristic radius below and above which the disk is 
approximately in the NDAF and ADAF regimes, respectively
\citep[e.g.][]{chen07}.

To derive the structure of a GRB accretion disk, one needs to solve a
set of equations \citep{popham99}, including mass conservation equation, 
energy equation, radial momentum equation, angular momentum conservation
equation, equation of state, and cooling and heating of plasma.
 In general, numerical calculations are needed to precisely
solve the GRB accretion disk problems. An approximate solution 
to the disk structure in the NDAF and ADAF regimes can be written in 
the following forms using results of numerical calculations 
\citep{popham99,narayan01,kohri02,kohri05,chen07,yuan12}:

NDAF:
\begin{eqnarray}
 \rho & = & 1.2\times 10^{14}~{\rm g~cm^{-3}}~ 
\alpha_{-2}^{-1.3} \dot M_{-1} 
\left(\frac{M} {3 M_\odot}\right)^{-1.7} \left(\frac{r}{r_g}\right)^{-2.55} \\
%\nonumber \\
 T & = & 3\times 10^{10}~{\rm K}~ \alpha_{-2}^{0.2} 
\left(\frac{M} {3 M_\odot}\right)^{-0.2} \left(\frac{r}{r_g}\right)^{-0.3} \\
%\nonumber \\
 V_r & = & 2\times 10^6 ~{\rm cm~s^{-1}}~ \alpha_{-2}^{1.2}
\left(\frac{M} {3 M_\odot}\right)^{-0.2} \left(\frac{r}{r_g}\right)^{0.2} 
\end{eqnarray}

ADAF:
\begin{eqnarray}
 \rho & = & 6\times 10^{11}~{\rm g~cm^{-3}}~ 
\alpha_{-2}^{-1} \dot M_{-1} 
\left(\frac{M} {3 M_\odot}\right)^{-2} \left(\frac{r}{r_g}\right)^{-1.5} \\
%\nonumber \\
 T & = & 3\times 10^{11}~{\rm K}~ \alpha_{-2}^{-1/4} 
\left(\frac{M} {3 M_\odot}\right)^{-0.5} \left(\frac{r}{r_g}\right)^{-5/8} \\
%\nonumber \\
 V_r & = & 10^8 ~{\rm cm~s^{-1}}~ \alpha_{-2}
 \left(\frac{r}{r_g}\right)^{-0.5} .
\end{eqnarray}
Here $\rho$, $T$, and $V_r$ are the density, temperature, and radial
velocity of the accretion flow, $\alpha$ is the viscosity parameter, 
$M$ is the black hole mass, $\dot M_{-1} = \dot M/0.1 M_\odot~{\rm 
s^{-1}}$ is accretion rate, and $r_g = 2 GM/c^2$ is the Schwarzschild 
radius of the black hole. It is possible that for a given accretion rate, 
different parts of an accretion disk (different $r$ ranges) belong to 
different regimes, i.e.  ADAF or CDAF, \citep[e.g.][]{chen07}.

Neutrino and anti-neutrino emission from an NDAF with power 
$\dot E_\nu$ would lead to $\nu \bar\nu$ annihilation, and generate a 
hot photon and electron-positron gas, which expands under its thermal
pressure as a fireball. The annihilation power \citep{zalamea11,lei13}
\begin{equation}
 \dot E_{\nu\bar\nu} \simeq 1.1\times 10^{52}~{\rm erg~s^{-1}}~
\left(\frac{M}{M_\odot}\right)^{-3/2} \left(\frac{\dot M}
{\rm M_\odot/s}\right)^{9/4}
\end{equation}
launches an outflow with luminosity of order given by the 
above equation. 

Neutrinos can also interact with protons through weak interaction and
transfer momentum to protons. This gives rise to a neutrino-driven
baryon wind. The baryon-loading rate is \citep{qian96,lei13}
\begin{equation}
 \dot M_\nu = 10^{-6}~{\rm M_\odot~s^{-1}}~ \dot E_{\nu,52}^{5/3}
\left< \left(\frac{\epsilon_\nu}{10~{\rm MeV}}\right)^2\right>^{5/3}
r_6^{5/3} \left(\frac{M}{3 M_\odot}\right)^{-2} \left(\frac{h}{r}\right)^{-1}.
\end{equation}
One can then calculate the amount of baryon loading in a $\nu\bar\nu$
annihilation jet \cite{lei13}:
\begin{equation}
 \eta = \frac{\dot E_{\nu\bar\nu}}{\dot M_\nu c^2},
\end{equation}
where $\eta$ is the ``dimensionless entropy'' of the fireball, which 
is essentially the terminal Lorentz factor of the baryon loaded fireball
at the end of its acceleration phase. Given a range of black hole mass, 
accretion rate, and spin rate, one can simulate the distribution of $\eta$ and
$\dot E_{\nu\bar\nu}$. 

\subsubsection{Blandford-Znajek mechanism}

The rotational energy of a BH with angular momentum $J$ can be written as:
\begin{equation}
E_{\rm rot}= 1.8 \times 10^{54} f_{\rm rot}(a_*) \frac{M}{M_\odot} {\rm erg},
\end{equation}
where
\begin{equation}
f_{\rm rot}(a_*)=1-\sqrt{(1+q)/2 },
\end{equation}
$q=\sqrt{1-a_*^2}$, and $a_*=Jc/GM^2$ is the BH spin parameter. 
For a maximally rotating BH ($a_*=1$), one has $f(1)=0.29$. 

Then the total power of Poynting flux from the BZ process can be estimated
as \citep{lee00,li00b,wangdx02,mckinney05,lei13}
\begin{equation}
\dot{E}_{\rm BZ}=1.7 \times 10^{50} ~{\rm erg \ s^{-1} }~ 
a_*^2 \left(\frac{M}{M_{\odot}}
\right)^2 B_{15}^2
%\left(\frac{B_{\rm BH}}{10^{15} {\rm G}}\right)^2 
F(a_*).
\end{equation}
The spin-dependent function $F(a_*)$ needs full general relativity to
solve \citep{blandford77}. An analytical approximation gives
\citep{lee00,wangdx02}
\begin{equation}
F(a_*)=\left[\frac{1+h^2}{h^2}\right]
\left[\left(h+\frac{1}{h}\right) \arctan h-1\right],
\label{Fa}
\end{equation}
where
\begin{equation}
h =\frac{a_*}{1+q},
\end{equation}
and so $F(0)=2/3$, and $F(1)=\pi -2$.
\cite{tchekhovskoy10,tchekhovskoy12} investigated this function
numerically and obtained an analytical fit to the numerical model.
The results are similar to Equation (\ref{Fa}) at most $a_*$ values
and only slightly deviates from (becomes lower than) Equation (\ref{Fa})
when $a_*$ is close to 1.

A major uncertainty in estimating the BZ power is the strength of 
magnetic fields. Depending on how $B$ is estimated (e.g.
magnetic pressure vs. ram pressure balance or equipartition with
the gas pressure), the BZ power is different. Numerically, 
\cite{tchekhovskoy12} estimated the BZ power by feeding a spinning
black hole with high magnetic flux. The BZ efficiency 
(defined as $ \eta_{\rm BZ} \equiv {\left<\dot E_{\rm BZ}\right>} /
{\left<\dot M\right> c^2} \times 100\%$) was found to exceed 100\% under certain
conditions. This suggests that the jet power indeed comes from
the BH spin, not from accretion. Evidence for BZ mechanism at work in long-GRB
central engines is also found in 2-D general-relativistic MHD simulations 
\citep[e.g.][]{nagataki09,nagataki11}.

In any case, to maintain a high BZ power, accretion rate should be
still very high. Neutrino emission/annihilation, and neutrino-driven
wind still occur from the disk. This has two implications
\citep{lei13}: First, the total jet power should be the sum of the
BZ power and the neutrino-annihilation power, so that the jet launched
from the base has both a ``hot'' (neutrino annihilation) and ``cold'' 
(Poynting flux) component. Second, due to the magnetic barrier, 
protons cannot drift into the central magnetically dominated jet.
Baryon loading, however, may proceed through neutron drift \citep{levinson03}.
 This results in much cleaner jets.

\subsubsection{Magnetic jets launched from the accretion disk}

A less well studied, yet plausible, mechanism invokes magnetic blobs
launched from the accretion disk.
\cite{uzdensky06} applied the ``magnetic tower''
mechanism (self-collimated toroidal magnetic jet structure produced by
a differentially-rotating central disk or a magnetar)
suggested by \cite{lynden-bell96} for producing AGN jets 
to the collapsar model of GRBs. The magnetic fields in the disk tend to
twist, wind up, and erupt, forming episodic magnetic bubbles.
This gives rise to an intrinsically episodic magnetically
launched jet even if the accretion rate is not episodic.
Baryon loading in such a model is however not easy to estimate.
Besides the neutrino-driven baryon load, corona materials
can be trapped in magnetic blobs, the amount of which is difficult
to estimate.

\subsubsection{Effects of the stellar envelope in long GRBs}

For a long GRB formed from the collapse of a massive star, the jet has to 
propagate through the stellar envelope. For a matter-dominated jet,
Kelvin-Helmholtz instability develops at the lateral-surface of the jet 
where there is substantial differential motion wrt the star. This 
induces variability in a jet even when the central engine has little 
fluctuation \citep{morsony07}. The envelope also collimates the
jet so that it has an opening angle of a few degrees when it emerges at the
stellar surface. The propagation of the jet through the envelope of the star, 
outside the iron core, produces a hot cocoon which can be very
effective in collimating the jet 
\citep[e.g.][]{aloy00,meszarosrees01,ramirezruiz02,matzner03,bromberg11b}. 
When the cocoon erupts at the stellar surface it makes 
 a wider, weaker, and less relativistic jet surrounding
the central narrow, stronger, and highly relativistic jet
\citep{ramirezruiz02,matzner03,zhangw03,zhangw04}.

For a highly-magnetized jet, the strong magnetic pressure prevents
ambient material from entering the jet. The hot cocoon surrounding the jet
also helps in its collimation and acceleration, and magnetic jets require
less expenditure of energy to punch through the star than baryonic jets
of same luminosity and cross-section at the launching site 
\citep{tchekhovskoy09,tchekhovskoy10,levinson13,bromberg14}. 

\subsubsection{Black hole engine in short GRBs}

For NS-NS or NS-BH mergers, the material in the accretion torus
has high densityand total mass of order 0.1 M$_\odot$. The duration of 
accretion is short, which is 
suitable for producing short GRBs \citep[e.g.][]{rosswog03,aloy05}. Lacking a 
heavy envelope, the jet is expected to be less collimated. In any case, the 
black hole vicinity is permeated by tidal debris launched during the merger 
and baryons launched from a neutrino wind from the hot accretion flow. These 
materials can collimate the GRB jet to $\sim 10-20$ degrees 
\citep[e.g.][]{rezzolla11,nagakura14}.

\subsection{Millisecond magnetars}

An alternative possibility for the GRB central engine is a 
rapidly spinning (period $P \sim$ 1 ms), highly magnetized
(surface magnetic field $B_s \sim 10^{15}$ G) neutron star
known as a millisecond magnetar. Such a magnetar, when spinning
down, has the right parameters to power a GRB \citep{usov92,wheeler00}.
The total spin energy
(which is the main power source of a millisecond magnetar) for a 
magnetar with initial spin period $P_0 \sim 1$ ms is
\begin{equation}
 E_{\rm rot} \simeq \frac{1}{2}I\Omega^2 \simeq 2 \times 10^{52}
~{\rm erg}~ \frac{M}{1.4 M_\odot} R_6^2 P_{0,-3}^{-2}.
\label{Emax}
\end{equation}
This equation gives an upper limit to GRB energy when the central engine is 
a magnetar. If a GRB violates this constraint, then the magnetar model is 
ruled out for that GRB. A systematic study of GRB prompt emission and
afterglow data suggests that all the magnetar-candidate GRBs 
appear to have collimation-corrected energy
in electromagnetic radiation that is smaller than this limit,
while some other GRBs (presumably having a black hole central engine)
do violate such a limit \citep{luzhang14}.

Making the simplest assumption of dipolar spindown, the total luminosity 
for a magnetar is given by
\begin{equation}
 L(t) = L_0 \frac{1}{(1+{t}/{t_0})^2} \simeq \left\{
 \begin{array}{ll}
  L_0, & t \ll t_0, \\
  L_0 (t/t_0)^{-2}, & t \gg t_0,  
 \end{array}
\right.
\end{equation}
where 
\begin{equation}
 t_0 = \frac{3 c^3 I}{B_p^2 R^6 \Omega_0^2} \simeq 20.5
~{\rm s}~ (I_{45} B_{p,16}^{-2} P_{0,-3}^2 R_6^{-6})
\label{tsd}
\end{equation}
is the characteristic spindown time scale, and 
\begin{equation}
 L_0 = \frac{I \Omega_0^2}{2 t_0} = \frac{B_p^2 R^6 \Omega^4}
{6c^3} \simeq 1.0 \times 10^{51}~{\rm erg~s^{-1}} (B_{p,16}^2
P_{0,-3}^{-4} R_6^6)
\end{equation} 
is the typical spindown luminosity. For $P \sim 1$ ms, and
$B_p \gae 10^{16}$ G, the typical spindown luminosity and time scale
coincide with the typical luminosity and duration of a GRB \citep{usov92}.

The mechanism by which a new-born magnetar might power a GRB has been
studied in detail in recent years \citep{bucciantini08,bucciantini09,
metzger11,kiuchi12,siegel14}. 
During the early phase of evolution, the simple dipole spindown formula
is not adequate to describe the relevant physics.
The evolution of a magnetar-powered GRB is well summarized by
\cite{metzger11}: a new born neutron star is initially very hot which leads to a
heavy baryon loading of the wind from magnetar due to neutrino driven mass loss
from the surface, and such an outflow has too small a terminal Lorentz factor
to power a GRB. After $\sim 10$ s or so, the neutron star cools down, the
neutrino driven baryonic wind diminishes, and a jet with $\sigma > 100$ is 
produced. This phase lasts for about half a minute when $\sigma$ 
increases rapidly due
to an abrupt drop in neutrino wind and that according to \cite{metzger11}
terminates the prompt GRB phase. During the prompt phase, erratic
lightcurves can be powered by magnetic dissipation instabilities
\citep{kluzniak98,ruderman00}. The energy budget during the prompt emission
phase is from the differential
rotation of the neutron star, which generates magnetic energy via a dynamo
mechanism. After this phase, the magnetar 
continuously spins down and injects energy as a Poynting flux.
Late magnetic activities arising from the residual differential rotation
of the neutron star can power X-ray flares after the prompt phase
\citep{dai06}.

The spin down power of a magnetar can leave an interesting signature
in the GRB afterglow lightcurve \citep{zhangmeszaros01a}; see also
\cite{dailu98b} for the case of a millisecond neutron star central engine 
with a normal ($\sim 10^{12}$ G) magnetic field. The basic feature is that
Poynting flux from the neutron star spin down can be directly injected into 
the blastwave. If the injected energy exceeds the energy 
deposited during the prompt phase, the external forward shock 
afterglow lightcurve would show a shallow decay phase.
The shallow decay phase of GRB afterglow lightcurves requires such an energy 
injection mechanism, and a magnetar central engine could perhaps offer a 
plausible explanation for this behavior \citep{dai04,zhang06,metzger11,dallosso11}. On the 
other hand, another model that invokes stratification of ejecta Lorentz 
factor \citep{rees98,granot06,uhm12} could also explain these plateaus. 

The discovery of an ``internal plateau'' for GRB 070110 \citep{troja07}
which was followed by a very rapid decay ($t^{-9}$) of the X-ray flux
rules out an external shock origin for the X-ray emission, and requires 
 internal dissipation of a long-lived jet to account for this steep decline.
A smooth lightcurve during the plateau is easier to understand when the
power source is the spin down of a neutron star.
From the observed luminosity and duration of the plateau, one can infer 
the parameters of the neutron star which turns out to be consistent with a
fast magnetar: $P_0 \sim 1$ ms, and $B_p \gae 3 \times 10^{14}$G when
the jet opening angle is assumed to be a large angle ($\sim 18^{\rm o}$) 
\citep{lyons10}. 

Such a magnetar signature also shows up in several short 
GRBs \citep[e.g.][]{rowlinson10,rowlinson13}. This suggests that
NS-NS mergers may also give rise to a supra-massive, likely highly
magnetized, millisecond magnetar \citep[e.g.][]{dai06,
gaofan06,fanxu06,metzger08,liu08,anderson08,giacomazzo11}.
Since the magnetar spindown time scale is typically 20s or more (Eq. \ref{tsd}),
one challenge of this model is to produce a short duration prompt emission.
Mechanisms discussed in the literature
include a brief accretion phase \citep{metzger08}, a brief differential
rotation phase \citep[e.g.][]{fan13b}, and phase transition \citep{chengdai96,chen13}.

Since the millisecond magnetar wind is essentially isotropic (data are
consistent with such a hypothesis \citep{luzhang14}), a post-merger
supra-massive millisecond magnetar is expected to emit bright electromagnetic
emission in the off-jet directions. \cite{zhang13} proposed that
NS-NS merger-induced gravitational wave bursts 
can have a bright early X-ray afterglow powered by a supra-massive
magnetar even if they are not associated with short GRBs (jet misses
earth). Such a magnetar also powers a bright multi-wavelength 
afterglow \citep{gao13a} and a bright ``merger'' nova \citep{yu13,metzger14}.
The collapse of the supra-massive neutron star into a black hole would
give distinct observational signatures, such as a sharp decline in
the X-ray lightcurve \citep{rowlinson10,rowlinson13}. 

\subsection{Models of late central engine activities}

Shortly after the observations of the first afterglow 
\cite{Katz98a} and \cite{katz98} suggested the 
possibility that some of the afterglow flux arises due to 
long lasting central engine activity, stressing that a strong 
variability in afterglow light curves cannot be produced via an 
external shocks.
Swift observations indeed find that GRB central engine activity lasts for
much longer than the duration of prompt emission.  There are two types of 
extended engine activity. One is erratic, manifested as late X-ray 
flares \citep{burrows05,chincarini07,falcone07}; the other is where 
the power output is steady for an extended period which we see 
as ``internal X-ray plateaus'' \citep{troja07,liang07,lyons10}.
A successful central engine model should be able to interpret these
diverse properties.

\subsubsection{X-ray flares}

A number of different models have been suggested for X-ray flares:
\begin{itemize}
 \item \cite{king05} proposed that a collapsing massive star may fragment 
into many blobs, which are accreted onto the central compact object at 
different times; blobs accreted at late times give rise to X-ray 
flares. Since short GRBs also have flares \citep[e.g. GRB 050724][]{barthelmy05b},
and a number of them are found in elliptical galaxies with very low star
formation rates e.g. the host galaxy of GRB 050724, this suggests that
the X-ray flare mechanism should also apply to progenitors that are not
massive stars.
 \item \cite{perna06} argued that the outer part of accretion disk, for 
long and short GRBS, is susceptible to gravitational instability and could 
fragment into clumps, and the accretion of these clumps produces X-ray flares;
short GRBs require some extreme conditions for this mechanism to work.
 \item \cite{proga06} argued that accumulation of magnetic flux near the
black hole during accretion can temporarily build up a ``magnetic barrier'',
which shuts down accretion for some time. When accretion resumes after 
accumulating enough material, an X-ray flare is produced. Such a process
may repeat itself to power multiple flares. A similar scenario was
proposed by \cite{cao14} to interpret extended emission of short GRBs.
 \item \cite{dai06} invoked a post-merger differentially-rotating millisecond
neutron star to power X-ray flares following short GRBs within the framework
of the NS-NS merger progenitor.
 \item \cite{lee09} suggested that post merger accretion disk may undergo
``phase transition'' triggered by He-synthesis, which would temporarily
launch a powerful wind to shut down accretion. Accretion resumes later to
power an X-ray flare.
\end{itemize}

\subsubsection{X-ray ``plateaus''}

The ``internal plateau'' observed in X-ray afterglow lightcurves 
requires energy dissipation within the jet of a long-lasting
central engine \citep{troja07}. A plausible scenario is that it is 
due to continuous energy injection from a magnetar wind 
\citep{metzger11,siegel14}, and the abrupt decay of flux at the end of 
plateau may be related to collapse of the magnetar to a black hole after 
it has lost enough angular momentum that it can no longer support itself
against the force of gravity \citep{troja07,zhang14}.

\begin{figure}
\begin{center}
\includegraphics[width=12cm]{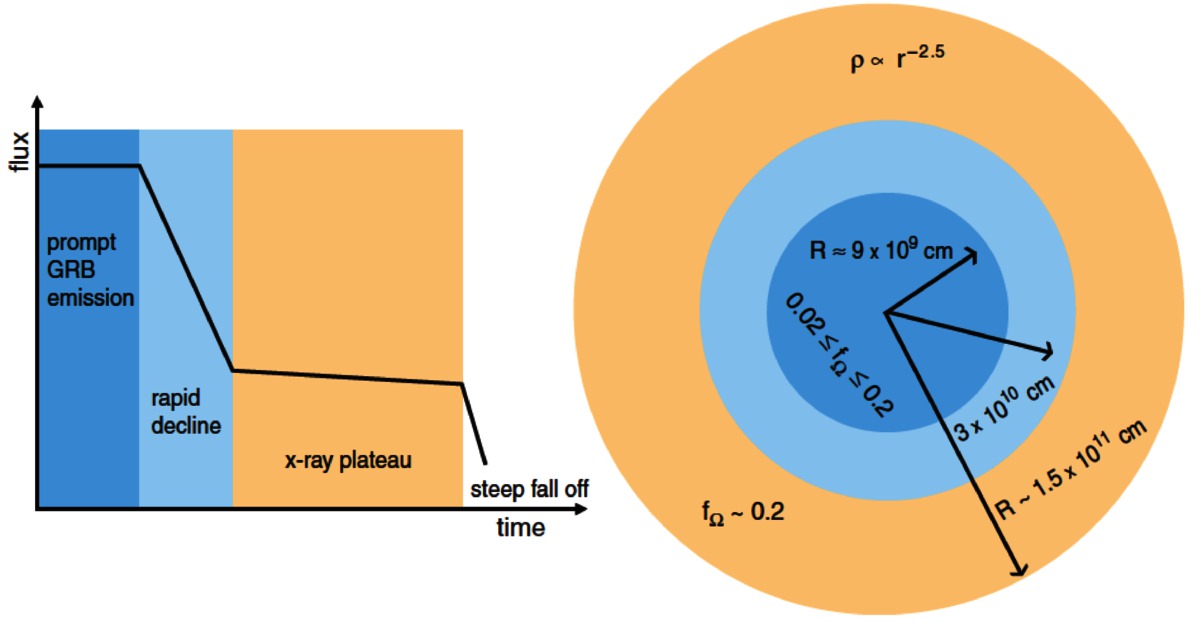}
\caption{The panel on the left shows a schematic X-ray lightcurve
with the following four segments: a prompt emission phase, a steep decline
phase, a plateau phase, and a post-plateau phase. The panel on the right shows 
how the different segments in the LC are related to the accretion of 
different parts of the progenitor star. The radii ($r$) and spin parameters 
($f_\Omega\equiv \Omega/\Omega_k$) of the various zones can be estimated from 
the X-ray data \citep{kumar08b}; $\omega_k(r)$ is Keplerian rotation rate at $r$.
}\label{FIG:x-ray-LC-invert}
\end{center}
\end{figure}

Regular X-ray plateaus (those followed by a normal decay $\propto t^{-1}$)
may be interpreted as due to energy being added for a period of plateau 
duration to a decelerating external shock \citep{zhang06,nousek06}. Some 
afterglows show achromatic behavior in both X-ray and optical bands, which 
can be easily explained by this model.
However, some others show a chromatic behavior, which requires that the
X-ray emission is powered by a different source from the optical emission.
One possibility is that the entire observed X-ray afterglow of these GRBs
is powered by a long-lasting central engine model. It can be from a 
millisecond magnetar without collapsing into a black hole 
\citep[e.g.][]{yu10} or from a hyper-accreting black hole.
Assuming a hyper-accreting black hole model for GRBs, 
\cite{kumar08b,kumar08a} showed that the morphology of a canonical X-ray light 
curve --- the steep decline of flux at the end of the prompt phase and a plateau
following that --- is similar to the time dependence of accretion rate onto 
the central object. 
The time dependence of the rate at which stellar material is added to 
the accretion disk, and the rate at which mass falls onto the central object, is 
a function of the density profile of the progenitor star \citep{kumar08a}. 
The duration of the steeply declining early
X-ray lightcurve -- or the beginning of the plateau -- is 
set by the dynamical timescale of the stellar core i.e.
$(R_c^3/G M_c)^{1/2}$; where $R_c$ and $M_c$ are the radius and mass of the
progenitor star's core. The X-ray flux \& its rate of decline during 
the plateau is determined by the mass, radius, and the rotation
rate of the stellar envelope, and therefore, the X-ray data can be 
{\it inverted} to obtain the GRB progenitor star structure 
as outlined in \cite{kumar08b}, and the result is shown in 
Figure \ref{FIG:x-ray-LC-invert}.

\subsection{Difference between the two types of engines}

If we ever detect a milli-second pulsation in X-ray
lightcurve of GRB prompt or afterglow radiation then that will clinch the case 
for the magnetar model. However, in the absence of this signature we have to
look at other possible ways of determining whether the GRB  central engine is
an accreting black-hole or a milli-second magnetar.
 We describe a few of the main properties of the GRB prompt and afterglow
lightcurves and how these could shed light on the nature of the
central engine. 

\begin{enumerate}

\item As already mentioned, a magnetar based model for GRBs cannot have 
   total energy in the burst exceeding $2 \times 10^{52}~{\rm erg}$ (eq.
\ref{Emax}) which is the rotational energy of a neutron star that is spinning
at close to the breakup speed; this total energy is the sum of the energy 
emitted during the prompt phase and the afterglow. On the other hand, a BH 
based central engine can in principle produce bursts with energy much larger 
than 2x10$^{52}$erg.

A major complication in determining the total energy output of a burst is the 
unknown collimation angle. Jet opening angle is measured for several long GRBs 
from the achromatic break in their multi-wavelength afterglow curves. 
Assuming that the opening angle for the jet during the prompt phase is same
as the angle determined from afterglow data, the total collimation-corrected
energy is found to be typically smaller than the upper limit of 
$2 \times 10^{52}~{\rm erg}$ \citep{frail01,panaitescu02,bloom03,berger03}. 
However, some bursts have energy close to or above this limit and that poses a 
challenge for the magnetar model \citep{liang08,cenko10,luzhang14}. 

Based on energy considerations alone, all that one can say is that 
 magnetar model could produce less energetic bursts, and the BH model is needed
for the most powerful explosions. 

\item The specific angular momentum for a milli-second magnetar, $j\sim R_{ns}^2
\Omega$, is $6 \times 10^{15}$ cm$^2$ s$^{-1}$, whereas for the innermost stable 
circular orbit of a maximally rotating Kerr black hole it is $2GM/3^{1/2}c
\sim 5\times10^{16} M_1$ cm$^2$ s$^{-1}$; where $M_1$ is BH mass in units of 
10$M_\odot$. Thus, the requirement on the rotation rate of the GRB progenitor
star is more severe for a BH model. 

\item The steep decline of X-ray lightcurve at the end of the prompt GRB phase 
 indicates that the central engine turns off very rapidly. A sharp decline 
   of accretion rate onto the newly formed BH soon after the core 
   collapse is found analytically and numerically for the collapsar model 
   e.g. \cite{kumar08a},\cite{lindner10}. The steep decline occurs when 
  the accretion 
  flow makes a transition from a NDAF to an ADAF, and at roughly the same time
  an accretion shock forms and pushes back the in-falling gas that further 
  reduces the accretion rate.

  It is much more difficult for a magnetar to be turned off as
rapidly as observations suggest. The luminosity of a magnetar wind, according to 
the dipole radiation model, fall off as $t^{-2}$ which is far too slow. However,
it is also the case that pulsar braking index $n$ -- defined as 
$d\Omega/dt \propto \Omega^n$ --  according to the dipole model is $3$ whereas
the index for 6 well studied pulsar is found to be between 1.4 and 2.9 
\citep{lorimer05}. If the braking
index for a newly born magnetar where to be like these {\it much older} pulsars
then that would suggest a faster fall off of the luminosity than $t^{-2}$
and that might explain the observed steep decline of X-ray lightcurves.

Another possibility for the steep decline of lightcurve for the magnetar model 
is suggested by \cite{metzger11}. According to these authors the steep decline 
is associated with a sharp increase of the magnetization parameter ($\sigma_0$) 
from $10^4$ to $10^9$ as the neutron star becomes transparent to neutrinos, 
and the neutrino-driven mass-loss drops rapidly. They invoke inefficient 
acceleration, and radiation, for very high $\sigma$ wind as the reason for the
steeply declining X-ray lightcurve. It is not clear that a high $\sigma$ wind
is necessarily difficult to accelerate in a typical long-GRB setting. Adiabatic
expansion of a wind of short spatial extent can lead to rapid acceleration 
e.g. \cite{granot11}; see \S\ref{magnetic_jet_adiabatic} for details.
Moreover, for $\sigma\gae10^6$ the jet becomes transparent to photons
in the transverse direction while still inside the star and that causes
a severe inverse-Compton drag on electrons bringing them quickly to almost 
standstill, and the resulting low current results in the dissipation of 
magnetic energy. If the Poynting jet with $\sigma\gae 10^6$ were to 
somehow escape this fate, it will become charge starved before reaching 
the deceleration radius, and consequently magnetic energy will be dissipated 
quickly. So it seems unlikely that a transition to high $\sigma$ outflow 
necessarily leads to rapidly falling lightcurve, and hence a rapid turn-off 
of the central engine poses a challenge for the Magnetar model.  

\item A plateau in jet luminosity can arise, according to the black-hole central
engine, when the outer part of the star with $\rho \aprop R^{-2.5}$ is accreted 
onto the black hole \citep{kumar08a,lindner10}.

According to the magnetar model the X-ray plateau is associated with its 
spindown time scale. Therefore, this model predicts that the average 
luminosity during the plateau should be inversely proportional to the 
duration of the plateau. This is something that observers should be able 
to verify and thus help determine the correct model for the GRB central 
engine; recent analysis of X-ray plateau data seems consistent with 
this expectation, e.g. \cite{bernardini12,xuhuang12,dainotti13,luzhang14}.

Another point to note is that it is more natural for a magnetar model
to produce a smooth lightcurve during the plateau, as is in fact observed, 
than for an accreting black hole model. The sharp decline of X-ray flux
at the end of the plateau for a few GRBs \citep[e.g.][]{troja07} might 
seem inconsistent with the magnetar model. However, a sharp decline
could perhaps arise when a supra-massive milli-second magnetar's rotation speed 
falls below a threshold value so that the centrifugal force is no longer 
able to prevent its collapse to a black hole.
A black hole central engine may be also abruptly stopped if the accretion
disk is suddenly blown away by a disk wind. However, the
combination of a flat X-ray lightcurve (which would require a constant
accretion rate) and a subsequent very rapid drop ($\propto t^{-9}$) 
is difficult to arrange for the black hole central engine model.

\item It is much easier to understand X-ray flares when the central engine is 
a magnetar than when it is a black hole \citep[e.g.][]{kluzniak98,dai06}. 
X-ray flares, for a magnetar model, are analogous to Soft Gamma-ray Repeaters 
(SGRs). However, the energy of flares should not exceed $\sim B^2 R_{ns}^3/5
 \sim 10^{48} B_{15}^2$ erg if produced by dissipation of magnetic fields
in a magnetar; bright X-ray flares in GRBs almost certainly violate 
this limit modulo the uncertainty regarding the beaming angle.
Note that the magnetic field strength, $B$, cannot be much larger than 
$\sim 10^{15}$G, otherwise the duration of the plateau would be much 
smaller than the observed value of $10^3 - 10^4$s.          

According to \cite{chincarini10} and \cite{margutti11} the average X-ray 
flare luminosity decreases with time (measured from $\gamma$-ray trigger) 
as $\sim t^{-2.7}$,
and the energy in flares scales as $\sim t^{-1.8}$. Accretion onto a BH can 
give this steep decline in the CDAF regime. However, the sharp rise and fall off
of X-ray flare lightcurve is puzzling to understand in this model.
It is unclear why a magnetar model --- where X-ray flare is produced by the
dissipation of some fraction of the energy of the neutron star's magnetic
field --- should have flare luminosity falling off as $\sim t^{-2.7}$.

\end{enumerate}

\section{Progenitors of GRBs}
\label{progenitor}

\subsection{Two physically distinct types of GRBs}

Gamma-ray observations led to identification of two
phenomenological classes of GRBs in the duration-hardness
($T_{90} - {\rm HR}$) plane: long/soft vs. short/hard
\citep{kouveliotou93}. The boundary between the two classes
is vague. The duration separation line is around 2 seconds 
in the BATSE band (30 keV - 2 MeV). Long and short GRBs roughly 
comprise 3/4 and 1/4 of the total population of the BATSE sample, 
but the short GRB fraction is smaller for other detectors
\citep{sakamoto08,sakamoto11,paciesas12,zhangfw12,qin13}.
This is because the duration $T_{90}$ of a GRB is energy-dependent
and detector-sensitivity-dependent \citep{qin13}. It is possible
that a short GRB detected by BATSE would appear as ``long'' by
a detector with a softer bandpass (e.g. Swift). Indeed, in the
Swift era, about 2\% of GRBs have a short/hard spike typically 
shorter than or around 2s, but with an extended emission (EE) 
lasting 10's to $\sim 100$ seconds \citep{norris06}. 
So the unfortunate consequence of the $T_{90}$ classification 
is that the membership to a certain category of the {\em same} GRB 
could change when the detector is changed. 
Nonetheless, the confusion in $T_{90}$ classification 
mostly arises in the ``grey'' area between the two classes.

Follow-up afterglow and host galaxy observations of GRBs led to
the identification of at least two broad categories of progenitor.
Observations led by BeppoSAX, HETE2, and Swift suggest that 
at least some long GRBs are associated with supernova Type Ic 
\citep[e.g.][]{galama98,hjorth03,stanek03,campana06,pian06}.
Most long GRB host galaxies are found to be dwarf star-forming 
galaxies \citep{fruchter06}.
These facts establish the connection between 
long GRBs and deaths of massive stars \citep{woosley93}.
The breakthrough led by Swift unveiled that some nearby
short GRBs (or short GRBs with EE) have host galaxies that are
elliptical or early-type, with little star formation
\citep{gehrels05,barthelmy05a,berger05}. Some others occur 
in star-forming galaxies, but the GRB location has a large
offset from the host galaxy where the local star formation rate is low
\citep{fox05,fong10}. All these point towards another type of
progenitor that does not involve massive stars, but is likely
related to compact stars, such as NS-NS or NS-BH mergers 
\citep[e.g.][]{eichler89,paczynski91,narayan92}.

The cozy picture that long GRBs are all physically related to 
massive star core collapses while short GRBs all physically
related to compact star mergers was soon destroyed by
several observations. GRB 060614 and GRB 060605 are both
nearby long-duration GRBs, but deep searches show no association
of a supernova accompanying the GRB 
\citep{gehrels06,galyam06,fynbo06,dellavalle06}, unlike other
nearby long GRBs. Moreover, the gamma-ray properties of GRB 060614
share many properties with short GRBs \citep{gehrels06}, and
it would resemble GRB 050724 (a smoking gun ``short'' GRB that
has a definite non-massive star origin) if it were somewhat less
luminous \citep{zhang07b}. Although theoretically some massive star
core collapses can have faint supernova signals 
\citep[e.g.][]{nomoto06}, the available data for GRB 060614 do not demand
such a scenario, since except the long duration all other properties
are similar to those of other nearby short GRBs. Rather, it suggests
that some GRBs that are not related to massive stars can have a
long duration. Later, it was noticed that the three GRBs with the
highest redshifts as of end of 2012, i.e. GRB 080913 at $z = 6.7$ 
\citep{greiner09},  GRB 090423 at $z=8.2$ \citep{tanvir09,salvaterra09},
and GRB 090429B at $z=9.4$ \citep{cucchiara11} all have a ``rest-frame
duration'' $T_{90}/(1+z)$ shorter than 2 seconds\footnote{Simulations
suggest \citep[e.g.][]{lv14,littlejohns13} that when a long GRB is
progressively moved to high redshifts, the observer-frame duration
may not increase noticeably due to the fact that some signals are buried
below the background. As a result, $T_{90}/(1+z) < 2$s may not carry
a direct clue about the progenitor, and rather could be due to a
``tip-of-iceberg'' selection effect.}. Yet, various
arguments suggest that they still originate from deaths of massive 
stars \citep{zhang09}. Later, an observer-frame short GRB
090426 at $z=2.609$ was discovered, which shared many properties of long 
GRBs with a massive star origin \citep{levesque10,antonelli09,xin11,thone11}.
Independent arguments suggest that at least some short GRBs,
especially those at high redshifts with high luminosities, are
probably not related to compact star mergers
\citep{zhang09,virgili11,cui12,bromberg12}.

\cite{bromberg12} found that there exists a plateau in the  $dN/dT_{90}$
duration distribution of GRBs (for all samples with different
detectors) and argued that this is an evidence for a massive
star origin; the plateau, according to them, is due to the finite time
it takes for GRB jets to clear a cavity and make their way out of the 
star. \cite{bromberg13}
suggest that 40\% of the short-GRBs detected by the Swift satellite could 
arise from collapse of a massive star, and that the
distinction between long and short bursts is detector dependent.
This is consistent with previous works \citep[e.g.][]{zhang09,virgili11,cui12} 
that arrived at this conclusion using very different arguments.
The host galaxy data, on the other hand, suggest that contamination
of massive star GRBs in the short GRB sample may not be large
\citep{fong10,berger14}. Based on an analysis invoking an
``amplitude'' parameter $f$ (ratio between the peak flux and
background flux) of short GRBs, \cite{lv14} claimed that the massive
star contamination becomes progressively important at low $f$
values due to the ``tip-of-iceberg'' effect.

In view of the confusions in classification, suggestions have been
made to distinguish the phenomenological classes (long vs. short) 
and the physically-motivated classes (massive star or Type II
GRBs vs. compact star or Type I GRBs)
\citep{zhangnature06,zhang07b,zhang09}.
The challenge is how to identify the physical class based on
data. \cite{zhang09} summarized a list of
multi-wavelength observational criteria that could be connected
to the physical nature of a GRB, and suggested to apply them
to identify the physical class of a GRB. In particular, the
observational criteria that are mostly related to the physical
nature of GRBs include supernova association, host galaxy
properties, as well as the location within the host galaxy.
A flowchart to diagnose the physical category of a GRB based
on multiple observational criteria was proposed \citep{zhang09}, 
which was applied to study long and short GRBs observed in the
Swift era \citep{kann10,kann11}. 

The multi-wavelength data cannot be immediately obtained when
a GRB is detected. So it is important to find a way of determining 
the physical class of a GRB from prompt $\gamma$-ray data alone.
Several attempts have been made toward this goal. For example, 
\cite{lv10} showed that for GRBs with known redshifts, the 
parameter $\varepsilon \equiv E_{\gamma,iso,52}/E_{p,z,2}^{5/3}$ 
has a more pronounced bimodal distribution; where $E_{\gamma,iso,52}$ is
the GRB isotropic energy in units of $10^{52}$ erg, and
$E_{p,z,2}$ is the peak of the $\gamma$-ray spectrum in 
units of 100 keV in GRB host galaxy rest frame 
at redshift $z$. The high-$\varepsilon$ vs.
low-$\varepsilon$ categories are found to be more closely related 
to massive star GRBs vs. compact star GRBs, respectively.
\cite{lv14} suggested the ``amplitude'' of an observed 
lightcurve should be taken into account to classify GRBs
based on duration. In particular, a low-amplitude short GRB
can be the ``tip-of-iceberg'' of a long GRB, whose longer
emission episode is buried beneath the background level.
The rest-frame-short nature of high-$z$ GRBs can be naturally
accounted for with this effect \citep[see also][]{kocevski13}. 

Sub-classes likely exist within these two broadly defined 
progenitor classes. For example, within the massive star GRBs,
the low-luminosity bursts typically have smooth lightcurves and
long durations, are more abundant, and probably form a distinct
population in the luminosity function \citep{liang07,virgili09}.
Physically, they may mark unsuccessful jets, and their emission
is from a trans-relativistic shock breaking out from the star
\citep[e.g.][]{bromberg11,nakar12}. Regular high-luminosity
GRBs, in contrast, have successful jets, as manifested by the
erratic variability in the lightcurves. The separation between the
two populations is not so clear cut though, as several low luminosity
GRBs with successful jets driven by a central engine have been
observed (e.g. GRB 120422A, \cite{zhangbb12}).

Another potential sub-category of massive star GRBs was proposed
to interpret several ``ultra-long'' GRBs \citep[e.g.][]{gendre13,levan14}.
The ultra-long durations (of order $10^3$--$10^4$ seconds) of these 
events have led to the suggestion that they might be associated with 
a blue supergiant progenitor, in contrast to the standard Wolf-Rayet 
stars.  The afterglow properties of these GRBs are not very different 
from the normal ones \citep[e.g.][]{virgili13}. Considering
that long-lasting X-ray flares exist in a good fraction of GRBs,
the possibility that these are the long-duration tail of the
normal long GRBs is not ruled out \citep{zhangbb14a}.

\subsection{Massive star GRBs}

The progenitor of GRBs is hard to identify, since the progenitor
is already destroyed when the GRB occurs. We do not know much 
regarding the GRB progenitors other than that there are two
physically distinct types with one type related to deaths of 
massive stars and the other type not related to massive stars.

We know better the progenitor of GRBs that are associated with 
massive stars. This is because considerable amount of information 
is available about the properties of the supernovae that are associated
with these GRBs and about their host galaxies.

\subsubsection{Properties of supernovae associated with GRBs}

A handful of long GRBs have ironclad associations with 
spectroscopically-identified SNe. The
list includes GRB 980425/SN 1998bw at $z=0.0085$ \citep{galama98}, 
GRB 030329/SN 2003dh at $z=0.168$ \citep{stanek03,hjorth03}, 
GRB 031203/SN 2003lw at $z=0.105$ \citep{malesani04},
GRB 060218/SN 2006aj at $z=0.033$ \citep{pian06,campana06},
GRB 100316D/SN 2010bh at $z=0.059$ \citep{starling11},
GRB 101219B/SN 2010ma at $z=0.55$ \citep{sparre11}, 
GRB 120422A/SN 2012bz at $z=0.283$ \citep{melandri12}, and 
GRB 130427A/SN 2013cq  at $z=0.34$ \citep{xu13,levan13}.
There are a lot more cases of association with various degrees of
confidence level, some with a light curve bump as well as some 
spectroscopic evidence of the SN, some others with a clear light
curve bump that is consistent with other GRB-SN associations 
\citep[e.g.][]{hjorth11}. 
An optimistic statement that the data are consistent with the 
hypothesis that all long GRBs are associated with an underlying
SN was made \citep{zeh04,woosley06}, until the null search results
for GRB 060614 and 060505 were reported in 2006\footnote{Observational 
properties (e.g. relatively short hard spike, short
spectral lag, low specific star formation rate) of GRB 060614 make
it more consistent with belonging to the compact star GRB category
\citep{gehrels06,galyam06,zhang07b}. Indeed, \cite{zhang07b} shows
that it would look rather similar to the smoking-gun compact star
GRB 050724 if it were somewhat less energetic. The case of GRB 
060505 is more controversial, but it is by no means a typical long 
GRB.} \citep{galyam06,fynbo06,dellavalle06}.

The GRBs with firmly established associations with SNe are 
typically nearby events. As of July 2014, all bursts with
SNe association, with the exception of GRBs 030329 and 130427A, are
low luminosity GRBs or X-ray flashes. This is likely a selection 
effect, since the faint SN signal (especially the spectral features) 
can be only detected at low redshifts, and when the SNa flux is not too faint 
compared with the afterglow emission in the optical band. High luminosity 
GRBs have optical afterglows that are brighter than SNe, and they 
are typically 
observed at $z > 1$. For these events, it is very difficult
to detect their associated SNe. In any case, the identifications
of SN 2003dh associated with a garden variety GRB 030329 at
$z=0.168$, and SN 2013cq associated with the nearby, high-luminosity
GRB 130427A at $z=0.34$ suggest that high luminosity GRBs are also 
associated with SNe Ic, whose properties are similar to those associated with
low luminosity GRBs.

The spectroscopically identified SNe associated with GRBs are 
of the Type Ic. These SNe are produced by core collapses of
massive stars whose hydrogen and helium envelopes have been
striped before the explosions, so the progenitors were most likely
Wolf-Rayet stars \citep{woosley06}. 

Not all Type Ic SNe have GRB associations. A systematic radio survey
of Type Ibc SNe suggests that less than 3\% are associated with
GRBs \citep{soderberg07}.  The GRB-associated SNe are consistent
with being  broad-lined Type Ic, suggesting
a large kinetic energy. They have diverse peak brightness,
rise time, light curve width, and spectral broadness.  Compared
with regular Type Ic SNe, the few GRB-associated SNe appear to 
represent the brighter end of the Type Ic population. However,
when non-detections and upper limits on SN light are taken
into account, the GRB-associated Type Ic SNe may not be all that 
different compared with normal Type Ic SNe \citep{woosley06}.

\subsubsection{Host galaxy properties of long GRBs}

The majority of long-GRB host galaxies are irregular, star-forming 
galaxies, and a few are spiral galaxies with active star formation
\citep{fruchter06}. Occasionally, one can have a GRB located in a
galactic halo, \cite[e.g. GRB 070125,][]{cenko08}. 
A systematic study suggests that long-GRBs occur in the brightest region
of the host galaxy, suggesting a very high specific star formation
rate at the burst site \citep{fruchter06}. All these properties 
are consistent with the massive star origin of long GRBs.
Nonetheless, cases of long GRBs with relatively low local specific 
star formation rate have been also discovered \citep[e.g.][]{levesque12}.

One controversial aspect is whether long-GRBs prefer low metallicity
environments. Claims that long GRB-hosts are relatively metal poor
have been made \citep[e.g.][]{fynbo03,prochaska04,fruchter06}. It was
noted that GRB-SNe occur in environments that have a systematically
lower metallicity than broad-lined Type Ic SNe \citep{modjaz08}. 
Counter-arguments suggest that this apparent metal poor property
of long GRB hosts is not intrinsic, but is rather a consequence
of anti-correlation between star-formation and metallicity seen in
general galaxy population \citep{savaglio09}.
Recently, \cite{graham13} compared the metallicity of the hosts
of long GRBs, broad-lined Type Ic SNe, and Type II SNe to each other,
and to the metallicity distribution of local star-forming galaxies 
in the SDSS sample, and concluded that such an anti-correlation is
not enough to explain the data, and that long GRBs indeed favor a
low metallicity environment. This is consistent with the expectation
of collapsar model of GRBs \citep{macfadyen99}, as well as 
numerical simulations of GRB host galaxy luminosity function
\citep{niino11}. Nonetheless, some dark GRBs are found to 
be located in relatively metal-rich host galaxies 
\citep[e.g.][]{holland10,perley13b}, suggesting that low metallicity 
may not be the critical condition to produce a GRB.

\subsubsection{Progenitor of long GRBs}

With the above observational constraints, the progenitor of long
GRBs can be narrowed down to massive stars with rapid spin
(as required to launch a jet), relatively low metallicity, and stripped
of their hydrogen and helium envelope. However, the
explicit type of star is not identified. Theoretical arguments
favor a Wolf-Rayet star with mass larger than
10 $M_\odot$ \citep[but not too large,][]{woosley11}. The
leading candidate is a massive star directly collapsing
to a black hole -- the collapsar model \citep{woosley93,macfadyen99}.
But models invoking binary stars \citep[e.g.][]{fryer99}
or a magnetar central engine \citep[e.g.][]{wheeler00,bucciantini09}
are also viable candidates.

\subsection{Compact star GRBs}

A detailed review of observational evidences that many short GRBs are
related to compact star mergers is presented in \cite{berger14}.

\subsubsection{Non-detection of SN light}

Deep searches of SNe associated with short-GRBs have been carried out for all
nearby bursts. The upper limits on SNa luminosity vary from case to case
\citep[e.g.][]{kann11,berger14}, but so far no positive detection has been 
made. This is consistent with a compact star origin (rather than massive
star origin) of these GRBs.

A weaker than supernova optical/IR signal, dubbed ``macronova'', 
``kilonova'', ``r-process nova'', or ``mergernova'' by various groups, 
has been predicted to be associated with NS-NS
or NS-BH mergers \citep{lipaczynski98,kulkarni05,metzger10,barnes13}.
Recently, a bright near-IR emission component was detected from
the short-GRB 130603B with HST \citep{tanvir13,berger13}, which is 
consistent with the recent prediction of a ``kilonova'' -- luminosity
being $\sim 10^3$ times the luminosity of a typical nova \citep{barnes13}. 
If a supra-massive magnetar is born during the merger, such a
merger-nova could be brighter due to the additional energy injection
into the ejecta from the magnetar \citep{yu13,metzger14}. 
It is speculated that most spin energy
of the magnetar is carried away by gravitational waves, but 
extreme conditions are required to excite such a strong gravitational
wave radiation in order to fit the data.
More observations are needed to establish whether all short-GRBs are
accompanied by a merger-nova, and to determine whether the central engine 
in these explosions is a black hole or a magnetar.

\subsubsection{Host galaxy properties of short GRBs}

\cite{fong10} systematically analyzed the host galaxy properties
of 10 nearby short GRBs and compared them with the hosts of long
GRBs and Type II SNe. They found that short-GRB host galaxies have
exponential disk profiles, but with a medium size twice as large
as long-GRB hosts. More importantly, the GRB site has large
offsets from the central star-forming regions. The accumulative
fraction as a function of fractional flux is very different
from long-GRBs, which show strong concentration to the brightest
region in the host galaxy \citep{fruchter06}, 
and is also very different from that of the core-collapse 
supernovae. The short-GRBs appear to under-represent their host galaxy 
light in contrast to long-GRBs. This is consistent with the compact
star merger scenarios, since compact stars born
in asymmetric supernovae most likely received a ``kick'',
so that the binary system drifted away from the star forming
regions when mergers occur \citep{bloom02b}.

There is a population of short GRBs that are ``hostless''.
They may be ``kicked'' away from their host, or reside in distant
faint host galaxies \citep{berger11}.

\subsubsection{Progenitor of short GRBs}

Observations suggest that the progenitor of short GRBs is different
from that of long GRBs. However, the explicit progenitor type is
not identified. Leading candidates include mergers (or for a small
fraction, collisions) of NS-NS and
NS-BH systems \citep{eichler89,paczynski91,rosswog13}. An alternative
candidate is accretion induced collapse of a NS to BH 
\citep{qin98,macfadyen05}. A small fraction of short GRBs can be
the giant flares of soft gamma-ray repeaters in nearby galaxies
\citep{palmer05}.

One should be cautious and not jump to conclusion that all short/hard GRBs
in the BATSE sample are due to the compact star merger origin.
Even though the NS-NS merger model is claimed to be able to 
reproduce the Swift short GRB data \citep{nakar06} and the BATSE
short GRB data \citep{guetta05b}, the model cannot
simultaneously explain the $z$-known Swift sample, and the
$z$-unknown BATSE sample \citep{virgili11}. In particular, 
the $z$-known sample demands a shallow luminosity function
in order not to over-produce nearby low-luminosity short GRBs.  
For a reasonable redshift distribution of NS-NS mergers, such a
shallow luminosity function is always translated to a shallow
flux distribution, which is inconsistent with the BATSE data.
A possible way out of this problem might be that some high-redshift, 
high-luminosity short-GRBs are related to massive stars 
\citep{zhang09,bromberg12}.

Since giant flares of Galactic soft gamma-ray repeaters (SGRs)
also generate a short, hard, emission episode, it has been speculated
that some SGR giant flares in the nearby galaxies can give rise to
apparent short hard GRBs \citep{palmer05}. Searchs in nearby
galaxies for well localized short GRBs suggest that such SGR
contamination to the observed short-GRB population is low, below
15\% \citep{tanvir05,nakar06b}.

\subsection{Gravitational wave diagnosis of GRB progenitor}

Probably the most definite diagnosis of GRB progenitor can be
made when gravitational waves are jointly detected with GRBs.
Different progenitors have distinct gravitational wave signatures
\citep[e.g.][]{bartos13}. In particular, compact star mergers 
have a characteristic in-spiral chirp signal \citep{flanagan98}, 
detection of which would give definite identification of the
short GRB progenitor \citep[e.g.][]{kochanek93}. For NS-NS mergers, the 
post-merger product can be either a black hole or a supra-massive neutron 
star, which would give different gravitational wave signatures: a black hole 
engine would show a ``ring-down'' signal after the merger phase 
\citep[e.g.][]{flanagan98,piran02,kobayashimeszaros03,baiotti08}, while 
a supra-massive neutron star would give extended gravitational wave signals due
to a secular bar-mode instability \citep[e.g.][]{baiotti08,corsi09}. 
%%These two references do not fit here. They have been cited elsewhere.
%or X-ray
%flares and plateau for a finite duration of time until the neutron star 
%slows down and collapses to a blackhole \citep{gaofan06,fanxu06}.

The gravitational wave signal due to a massive star 
core collapse is subject to large uncertainties. If collapse
is asymmetric, bar-mode instability may develop in the accretion
disk, so that strong gravitational waves can be released from
the central engine of long GRBs \citep{piran02,kobayashimeszaros03,ott12}. 

Detecting gravitational waves from astrophysical objects is
challenging. The upcoming advanced
gravitational wave detectors such as Advanced LIGO \citep{ligo} and
Advanced VIRGO \citep{virgo} are expected to expand the detection
horizon to a few hundred Mpc as early as 2015.
Detecting the electromagnetic counterparts of gravitational wave
sources would increase the signal-to-noise ratio of the gravitational
wave signal and confirm its astrophysical origin. If the final
product of a compact star merger is a black hole, the electromagnetic 
signals associated with the gravitational wave burst 
include a short GRB, an optical ``macronova'' 
\citep{lipaczynski98,kulkarni05,metzger10}, and a long-lasting radio afterglow 
due to the interaction of the ejecta with the surrounding matter
\citep{rezzolla10,nakar11,shibata11,metzger12,piran13,kyutoku13a,kyutoku13b}.
These signals are either beamed (short GRB) or very faint. On the
other hand, if a NS-NS merger leaves behind a supra-massive millisecond
magnetar which is possible based on uncertainties of our understanding
of equation-of-state of nuclear matter \citep{dai06,giacomazzo13}, then
very bright electromagnetic counterparts can be detected with gravitational
wave bursts without a short GRB association. The signals include a bright
early X-ray afterglow due to
internal dissipation of a proto-magnetar wind \citep{zhang13}, bright
broad-band afterglow of a magnetar-powered ejecta \citep{gao13a}, as 
well as a merger-nova brighter than the ``kilo-nova'' predicted for
a black hole central engine \citep{yu13,metzger14}.
The planned multi-messenger observations of GRBs would greatly
enrich our understanding of GRB physics.

\section{High energy neutrinos from GRBs}
\label{grb-neutrinos}

As energetic, non-thermal photon emitters, GRBs are believed to be
efficient cosmic ray accelerators as well. The standard scenario
invokes first order Fermi acceleration mechanism in relativistic shocks,
both in internal shocks and the external (forward and reverse)
shocks. Alternatively, magnetic reconnection can also accelerate
cosmic rays to high energies. 

The maximum proton energy can reach
the ultra-high energy (UHE) range \citep{waxman95,vietri95,milgrom95}.
The maximum energy of the shock accelerated protons can be estimated
by the condition $t'_{\rm acc} = {\rm min} (t'_{\rm dyn}, t'_c)$,
where $t'_{\rm acc} = \xi (\gamma_p m_p c/eB')$, $t'_{\rm dyn}$, and 
$t'_c$ are the acceleration, dynamical, and cooling time scales in the 
co-moving frame. For example, within the internal shock framework,
when we ignore proton cooling via the photo-pion process (which can
be important for UHE protons), the maximum proton energy is
\begin{equation}
E_{\rm p,max} \simeq 4\times 10^{20}~{\rm eV}~ \xi^{-1} \left(
\frac{\epsilon_{B,-1} L_{\gamma,52}}{\epsilon_{e,-1}}\right)^{1/2}
\Gamma_{2.5}^{-1}~,
\end{equation}
which is in the UHE range. Protons with energies below this maximum
value can produce neutrinos of different energies.

A GRB has multiple emission sites that can accelerate protons. These
same sites usually are also permeated with photons. 
If protons in a GRB jet can be accelerated to
an energy $E_p$ so that the condition
\begin{equation}
E_p E_\gamma \gae \frac{m_\Delta^2 - m_p^2}{2} \left(\frac{\Gamma}
{1+z}\right)^2 = 0.147 ~{\rm GeV}^2 \left(\frac{\Gamma}{1+z}\right)^2
\label{pgamma}
\end{equation}
is satisfied, significant neutrino emission is possible via the $p\gamma$ 
mechanism at the $\Delta$-resonance
\begin{equation}
p\gamma \rightarrow (\Delta^+ \rightarrow)  \left\{
 \begin{array}{ll}
  n \pi^+ \rightarrow n \mu^+ \nu_\mu \rightarrow n e^+ \nu_e\bar\nu_\mu \nu_\mu, & {\rm fraction}~ 1/3 \\
  p \pi^0 \rightarrow p\gamma\gamma, & {\rm fraction}~ 2/3.
 \end{array}
\right.
\end{equation}
Here $\Gamma$ is the bulk Lorentz 
factor, $E_\gamma$ is photon energy in observer frame, $m_\Delta=1.232$ GeV 
and $m_p = 0.938$ GeV are the rest masses of $\Delta^+$ and proton, 
respectively. At $\Delta$-resonance, about 20\% of the proton energy goes to
$\pi^+$ ($\epsilon_{\pi^+} \sim 0.2 \epsilon_p$), whose energy 
is evenly distributed to 4 leptons ($\epsilon_\nu \sim 0.25 
\epsilon_{\pi^+}$). So overall
\begin{equation}
 E_\nu \sim 0.05 E_p.
\label{EnuEp}
\end{equation}

Due to the high
compactness of the ejecta, the $p\gamma$ interaction can have
high optical depth, so that $\pi^+$ are copiously generated.
$\pi^+$ decay and subsequent $\mu^+$ decay generate neutrinos
($\nu_\mu$ and $\nu_e$) and anti-neutrinos ($\bar\nu_\mu$).

Another important neutrino production mechanism is hadronic
collisions, including $pp$ and $pn$ processes, e.g.
\begin{eqnarray}
 pp & \rightarrow & pn \pi^{+}/K^{+} \rightarrow pn\mu^+ \nu_\mu
\rightarrow pne^+ \nu_e \bar\nu_{\mu} \nu_\mu \nonumber \\
 pn & \rightarrow & pp\pi^{-}/K^{-} \rightarrow pp\mu^{-} \bar\nu_\mu
\rightarrow ppe^{-}\bar\nu_e\nu_\mu\bar\nu_\mu \nonumber \\
 pn & \rightarrow & nn\pi^{+}/K^{+} \rightarrow nn\mu^{+} \nu_\mu
\rightarrow nne^{+}\nu_e\bar\nu_\mu\nu_\mu. 
% pp & \rightarrow & pp\pi^0 \rightarrow pp\gamma\bar\gamma, \nonumber \\
% pn & \rightarrow & pn\pi^0 \rightarrow pn\gamma\bar\gamma.
\end{eqnarray}
%\begin{equation}
% pn\rightarrow pp\pi^{-} \rightarrow pp\mu^{-} \bar\nu_\mu
%\rightarrow ppe^{-}\bar\nu_e\nu_\mu\bar\nu_\mu,
%\end{equation}
%\begin{equation}
% pn\rightarrow nn\pi^{+}É \rightarrow nn\mu^{+} \nu_\mu ...
%\rightarrow nne^{+}\nu_e\bar\nu_\mu\nu_\mu ...,
%\end{equation}
%\begin{equation}
% pn\rightarrow pn\pi^0 \rightarrow pn\gamma\bar\gamma.
%\end{equation}
Free neutrons will subsequently decay: $n\rightarrow pe^{-}\bar\nu_e$.
These processes are important in a dense environment, such as
inside the progenitor star.

\subsection{PeV neutrinos}

For GRBs, a guaranteed target photon source for $p\gamma$
interaction is the burst itself. For the 
typical peak photon energy $E_\gamma \sim$ several hundred keV, the 
corresponding neutrino energy 
is in the sub-PeV regime \citep{waxman97}.
The standard model invokes internal shocks as the site of
both gamma-ray photon emission and proton acceleration
\citep{rees94,waxman97}. Alternatively, photons can be generated
at the photosphere 
\citep{rees05,peer06,thompson07,peer08,giannios08,beloborodov10,lazzati10,ioka10} 
or from magnetic field dissipation beyond the internal shock
radii \citep{lyutikov03}. Protons can be accelerated in the same
site or a different site from the gamma-ray emission region. Over the
years, PeV neutrino flux from GRBs has been calculated both analytically
and numerically \citep{waxman97,razzaque03,razzaque03b,guetta04,murase06b,murase08b,wangdai09,gao12}.  We describe here a general formalism for calculating 
the strength of the neutrino signal that can be applied to any of the above 
mentioned models for GRB emission \citep{zhangkumar13}:

For an observed ``Band''-function photon flux spectrum 
\begin{eqnarray}
 F_\gamma(E_\gamma) &  = & \frac{dN(E_\gamma)}{d E_\gamma} \nonumber \\
 & = & f_\gamma  \left\{
 \begin{array}{ll}
  \left(\frac{\epsilon_\gamma}{\rm MeV}\right)^{\alpha_\gamma}
  \left(\frac{E_\gamma}{\rm MeV}\right)^{-\alpha_\gamma}, & 
%{\rm for}~
E_\gamma < \epsilon_\gamma \nonumber \\
  \left(\frac{\epsilon_\gamma}{\rm MeV}\right)^{\beta_\gamma}
  \left(\frac{E_\gamma}{\rm MeV}\right)^{-\beta_\gamma}, & 
%{\rm for}~
E_\gamma \geq \epsilon_\gamma
 \end{array},
\right.
\end{eqnarray}
the observed neutrino number spectrum can be expressed as \citep{waxman97,icecube10}
\begin{eqnarray}
& F_\nu(E_\nu) = \frac{dN(E_\nu)}{d E_\nu} 
%= f_\nu 
\nonumber \\
=f_\nu & \left\{
 \begin{array}{ll}
  \left(\frac{\epsilon_{\nu,1}}{\rm GeV}\right)^{\alpha_\nu}
  \left(\frac{E_\nu}{\rm GeV}\right)^{-\alpha_\nu}, & 
%{\rm for}~
E_\nu < \epsilon_{\nu,1} \nonumber \\
  \left(\frac{\epsilon_{\nu,1}}{\rm GeV}\right)^{\beta_\nu}
  \left(\frac{E_\nu}{\rm GeV}\right)^{-\beta_\nu}, & 
%{\rm for}~
\epsilon_{\nu,1} \leq E_\nu < \epsilon_{\nu,2} \nonumber \\ 
  \left(\frac{\epsilon_{\nu,1}}{\rm GeV}\right)^{\beta_\nu}
  \left(\frac{\epsilon_{\nu,2}}{\rm GeV}\right)^{\gamma_\nu-\beta_\nu}
  \left(\frac{E_\nu}{\rm GeV}\right)^{-\gamma_\nu}, & 
%{\rm for}~
 E_\nu \geq \epsilon_{\nu,2}
 \end{array},
\right.
\end{eqnarray}
where
\begin{equation}
 \alpha_\nu = p+1-\beta_\gamma, ~\beta_\nu = p+1-\alpha_\gamma,
~\gamma_\nu = \beta_\nu + 2,
\end{equation}
and $p$ is the proton spectral index defined by $N(E_p)dE_p
\propto E_{p}^{-p} dE_p$. 
The indices $\alpha_\nu$ and $\beta_\nu$ are derived by assuming
that the neutrino flux is proportional to the $p\gamma$ optical depth
$\tau_{p\gamma}$. This is valid when the fraction of proton energy
that goes to pion production, i.e.
$f\equiv 1-(1-<\chi_{p\rightarrow \pi}>)^{\tau_{p\gamma}}$, 
is proportional to $\tau_{p\gamma}$
($<\chi_{p\rightarrow\pi}> \simeq 0.2$ is the average fraction of
energy transferred from protons to pions), which is roughly valid 
when $\tau_{p\gamma} < 3$. 
In general, one can write \citep{zhangkumar13}
\begin{equation}
 \epsilon_{\nu,1} = \epsilon_{\nu,1}^0 {\rm min}
(1,(\tau_{p\gamma}^p/3)^{1-\beta_\gamma}),
\label{eps_nu_1}
\end{equation}
where
\begin{equation}
 \epsilon_{\nu,1} = \epsilon_{\nu,1}^0 =
7.3\times 10^5 ~{\rm GeV}~(1+z)^{-2}~
\Gamma_{2.5}^2 \epsilon_{\rm \gamma,MeV}^{-1},
\end{equation}
\begin{equation}
 \epsilon_{\nu,2} = 3.4\times 10^8~{\rm GeV}~ (1+z)^{-1}~
\epsilon_{_B}^{-1/2} L_{w,52}^{-1/2} \Gamma_{2.5}^2 R_{14},
\label{eps2}
\end{equation}
and
\begin{equation}
 \tau_{p\gamma}^p \equiv \tau_{p\gamma}(E_{p}^p)
\simeq \frac{\Delta R'}{\lambda'_{p\gamma}(E_{p}^p)}=0.8 L_{\gamma,52}
\Gamma_{2.5}^{-2} R_{14}^{-1} \epsilon_{\rm \gamma,MeV}^{-1},
\label{fpi}
\end{equation}
$\lambda'_{p\gamma}(E_{p}^p)$ is the comoving proton mean free path 
for $p\gamma$ interaction at $E_{p}^p$ ($E_p^p$ is the energy of protons that
interact with peak energy photons at $\Delta$-resonance), $\Delta R'$ is the 
comoving width of the jet, $R$ denotes the distance of proton acceleration
site (rather than the photon emission site if the two sites are
different) from the central engine, $\epsilon_{_B}$ is the fraction
of dissipated jet energy in magnetic fields, and $L_w$ is the luminosity 
of the dissipated wind. We further define
\begin{equation}
 f_{\gamma/p}\equiv \frac{L_\gamma}{L_p}, 
\label{fgamma/p}
\end{equation}
and
\begin{equation}
 f_{p} \equiv \frac{\int_{E_{p,1}}^{E_{p,2}} dE_p E_p^2 dN(E_p)/dE_p}
{\int_{E_{p,min}}^{E_{p,max}}dE_p E_p^2 dN(E_p)/dE_p}
 \simeq  \frac{\ln (\epsilon_{\nu,2}/\epsilon_{\nu,1})}
{\ln (E_{p,max}/E_{p,min})}~({\rm for}~p=2),
\label{fp}
\end{equation}
where $E_{p,1}$ \& $E_{p,2}$ 
are proton energies corresponding to $\epsilon_{\nu,1}$ and 
$\epsilon_{\nu,2}$, respectively (Eq.\ref{EnuEp}),
and $E_{p,max}$ and $E_{p,min}$ are the maximum and minimum proton
energy. One can then normalize the neutrino spectrum with the total 
photon fluence \citep{icecube10}
\begin{equation}
 \int_0^\infty dE_\nu E_\nu F_\nu(E_\nu) =
 \frac{1}{8} \frac{f_p}{f_{\gamma/p}} 
[1-(1-<\chi_{p\rightarrow \pi}>)^{\tau_{p\gamma}^p}] 
 \int_{\rm 1~keV}^{\rm 10 MeV}
dE_\gamma E_\gamma F_\gamma(E_\gamma).
\end{equation}
The coefficient 1/8 is the product of 1/4 (4 leptons share the energy of
one $\pi^+$) and 1/2 (on average roughly half of $p\gamma$ interactions
go to the $\pi^+$ channel when all the $\pi^+$ processes besides
$\Delta^+$ resonance, e.g. direct-pion production, and multiple pion
production, are taken into account).

Over the years, the IceCube Collaboration have been searching
for high energy neutrino signals coincident with GRBs in time and
direction, and progressively deeper non-detection upper limits have
been placed \citep{icecube10,icecube11,icecube12}. The current IceCube 
upper limit was claimed to be at least a factor of 3.7 smaller 
than the theoretical predictions 
for neutrino flux from GRBs according to the internal-shock model, 
which has raised further doubt regarding the viability
of GRBs as sources of UHECRs \citep{icecube12}.
More detailed, follow-up, calculations \citep{lizhuo12,hummer12,he12} 
suggest that the current limit is still not deep enough to provide 
significant constraints on the validity of the internal shock model.
However, the model would be severely challenged if the upper 
limit continues to go down in the next few years\footnote{As of
July 2014, the IceCube upper limit on neutrino flux for GRBs
goes down roughly by another factor of 3 since the upper limit
reported in \cite{icecube12} (2014, A. Karle, I. Taboada, 
private communications), placing even tighter constraints
on the internal shock and photosphere models.}. 

The internal shock model fails to explain prompt $\gamma$-ray spectra.
Alternative prompt emission models (e.g.
dissipative photosphere models and large-radius magnetic dissipation
models) have been widely discussed in the literature. These 
different models have different predictions for the neutrino flux.
\cite{zhangkumar13} compared the predictions of different models
and concluded that the current upper limit already constrains the
photosphere model unless $f_{\gamma/p} > 0.1$ or protons are not 
accelerated to the desired energy to satisfy the $\Delta$-resonance
condition. The internal shock model is barely constrained by the
current data \citep{he12}. On the other hand, magnetic dissipation
models that invoke a large emission radius \citep[e.g. the ICMART
model;][]{zhangyan11} predict a much lower neutrino flux, which is
consistent with the current null result. If in the next few years
the neutrino flux limit continues to go down, it would favor the magnetic
dissipation models and further constrain the parameter space of
the matter-dominated models.

The recent nearby, very bright GRB 130427A did not show a
positive PeV neutrino signal. This non-detection makes even
tighter constraints on the internal shock model and the 
photosphere model of GRBs \citep{gaoshan13}.

Low-luminosity GRBs are more common and form a distinct population
in the GRB luminosity function \citep{soderberg06,liang07,virgili09}.
If these GRBs produce successful jets, 
since they are softer and have lower Lorentz factors, the characteristic
neutrino energy in the traditional internal shocks 
is higher than that of the high-luminosity GRBs. These GRBs
give a neutrino background in the sub EeV range with a flux level 
comparable to that by
high luminosity GRBs \citep{murase06,gupta07a}.

\subsection{Other neutrino emission components from GRBs}

GRBs also have other sites that generate high energy neutrinos. Since
the seed photon energies and Lorentz factor can be different at different
sites, the characteristic energies of neutrinos are also different.

At the deceleration radius, 
The typical target photon energy may be $\sim 1$ eV and $\sim 1$ keV
for the reverse and forward shock, respectively. Given $\Gamma \sim 100$,
the $\Delta$-resonance condition gives the corresponding 
neutrino energy $\epsilon_\nu \sim 5\times 10^{19}$ eV and
$\sim 5\times 10^{16}$ eV for the forward and reverse shock, 
respectively \citep{waxman00,dailu01b,dermer02}. This is broadly in the
EeV regime. 

 Due to a smaller Lorentz factor when the jet has not completed the acceleration
phase while inside the star, internal shocks and proton acceleration can occur 
before the jet reaches the stellar surface.  Neutrinos can be generated via 
both $p\gamma$ and $pp/pn$ collision mechanisms (if the envelope is large 
enough) \citep{meszaroswaxman01,razzaque03,razzaque03b,murase13}. Taking
$\Gamma \sim 10$, and $E_\gamma \sim 5$ keV (X-ray
photons trapped in the jet), one can estimate
$E_p \sim 2\times 10^{13}$ eV, and the typical neutrino
energy $\epsilon_\nu \sim 10^{12}$ eV (or TeV). Since this
mechanism applies to both successful and failed GRBs, 
detecting this neutrino emission can probe 
failed jets in core-collapsing massive stars.
Recent studies suggest that a relativistic photon-mediated shock
is inefficient in accelerating protons 
\citep[e.g.][]{levinson08}. This would suppress high-energy neutrinos
in successful GRBs, but low-luminosity GRBs remain good candidates
to generate the high-energy neutrino background observed by IceCube
\citep{murase13}. 

For a neutron-rich ejecta, protons and neutrons can decouple
and move with different Lorentz factors (see \S\ref{np_collision}).
If the relative speed between the two components is larger than
about 0.5C, then pions, muons, and neutrinos are produced in inelastic
collisions between protons and neutrons \citep{bahcall00,meszarosrees00b};
pion mass is about 140 MeV and proton mass 940 MeV.
The neutrinos produced by this process have energies $\sim$10--10$^2$ GeV. 
In this energy
range, the atmospheric neutrino background is very strong, therefore,
detecting these neutrinos from GRBs is very difficult 
with ground-based detectors. However, time- and space-coincidence
with GRBs can help significantly reduce the background problem and
improve the chances of detecting these quasi-thermal neutrinos 
with 10 year observations with IceCube \citep[e.g.][]{murase13b}.

Finally, the GRB central engine is expected to produce copious 
MeV neutrinos \citep{kumar99}. These MeV neutrinos are generic feature of
all core collapse events. Positive detections have been made for SN 1987A. 
In order to detect MeV neutrinos from a GRB, the GRB
has to be very close to earth. The event rate of such nearby
GRBs whose MeV neutrinos are detectable is extremely low.

\section{GRBs from the first stars (pop III stars) and their use for 
    investigating the high redshift universe}

The universe was essentially devoid of stars until the redshift of $\sim 15$--20,
when the first stars were born, and the strong UV radiation from them 
contributed to the reionization of the universe, and bringing to an end
the cosmic dark age \citep[e.g.][]{tumlinson00,schaerer02,venkatesan03,bromm04}. 
A fraction of these stars likely ended their life as GRBs. In this section,
we describe how GRBs can be used to study the end of the cosmic dark ages.

According to the cold dark matter (CDM) paradigm of hierarchical structure
formation, the first generation of stars, or pop III stars, are expected
to have formed in dark matter halos of mass $\sim 10^6 M_\odot$ which
decoupled from general expansion, and collapsed at about a redshift of 20
\citep[e.g.][]{tegmark97,yoshida03}.

The first stars, free of metals, were born with mass larger than their metal
rich descendants. Ten years ago it was thought that the typical mass of these 
stars might be more than $10^2 M_\odot$ \citep[e.g.][]{bromm99b,abel00,nakamura01}.
The formation of metal free stars, in
the absence of magnetic fields, is easier to understand than the much 
more complex physics behind the formation of later generation of stars 
where one needs to consider effects of magnetic fields and a complex network 
of radiative processes involving a rich variety of atoms.
 The characteristic mass scale of
pop III stars is set by the Jeans mass for a primordial cloud that is 
cooled to a temperature of order 200 K by the rotational-vibrational transitions
of molecular hydrogen; an upper limit to the density of the primordial 
clouds ($n\sim 10^4$ cm$^{-3}$) from which stars are born is obtained 
by the requirement that
the time scale for collisional de-excitation of $H_2$ should be longer than
the time it takes for radiative transition to lower energy state (otherwise the cloud
would be unable to cool and form stars). 
An additional complication one needs to deal with is the fragmentation of 
clouds while it is undergoing collapse. Earlier simulations had underestimated
this effect, and newer, higher resolution, simulations find that the typical
mass of pop III stars is close to $\sim 40 M_\odot$ \citep[e.g.][]{stacy10}.

Population III stars should have had a weak wind 
--- massive stars have radiation driven winds which are launched by photons 
scattering off of metals, the most efficient of which for this purpose are 
the iron group of elements --- and hence they retain their angular momentum and 
are rapidly spinning at the time of their death. These conditions -- high mass 
and rapid rotation rate -- are conducive to formation of an accretion disk when 
the star undergoes collapse at the end of its nuclear burning life cycle,
 and could produce a relativistic jet and a GRB. It is, therefore, speculated 
that a fraction of population III stars should produce GRBs when they die.

The recent discoveries of GRBs at redshifts of 8.26 \citep{tanvir09,salvaterra09}
and 9.4 \citep{cucchiara11} have established
that GRBs did indeed occur when the universe was young ($z=9.4$ 
corresponds to 525 million years after the big bang). 
These bursts were fairly typical of long-GRBs in terms of their luminosity
and spectral properties, and bursts like these can be detected by the
Swift satellite up to redshift of $\sim 15$. Thus, if the first stars to 
form in the universe, at the end of the cosmic dark age, were massive, 
and rapidly rotating, as suggested by theoretical calculations, then 
they should produce GRBs \citep{woosley11,suwa11,nagakura12} and these 
can be detected by Swift and future GRB missions.

\begin{figure}
\begin{center}
\includegraphics[width=12cm]{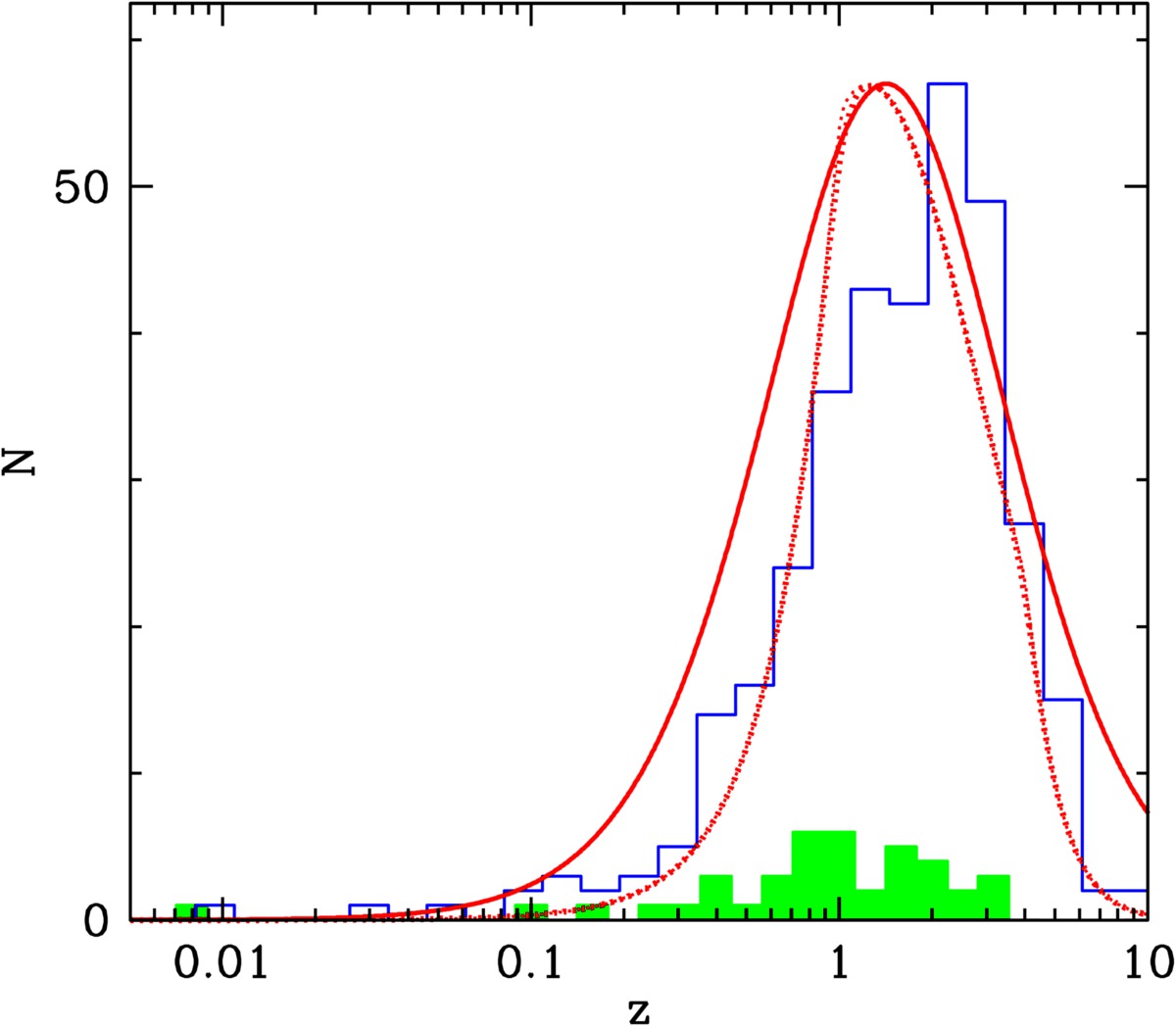}
\caption{This figure, showing redshift distribution of GRBs, is adapted from
\cite{gehrels09}; the z-distributions of Swift GRBs and pre-Swift 
GRBs are in blue and green histograms, respectively 
(the higher redshift of Swift bursts, $\langle 
z\rangle \sim 2.5$, compared to $\langle z\rangle\sim 1.2$ for pre-swift 
bursts is due to the higher sensitivity of Swift/BAT). The thick solid red 
curve is the evolution of a comoving volume element of the Universe for 
$\Lambda$-CDM model; the thin dotted red curve is a convolution of the 
comoving volume with a model for the star-formation rate as calculated 
by \cite{yuksel08}. Credit: John Cannizzo.
}\label{FIG:grb-z-dist}
\end{center}
\end{figure}

The observed redshift distribution of bursts is shown in Figure 
\ref{FIG:grb-z-dist}, along with the star formation rate. It is clear 
that GRB rate is falling off less rapidly than the star formation rate 
at $z>2$, which might be related to the claimed lower metallicity of 
GRB host galaxies.  Correcting for the detector's sensitivity selection 
effect against detection of high-$z$ GRBs, this high-$z$ excess effect 
is even more significant.  Detailed studies of this high-$z$ excess 
effect have been carried out in the last several years
\citep[e.g.][]{kistler08,campisi10,qin10,virgili11,robertson12,trenti13}.
The general conclusion from these studies is that the excess requires 
a low-metallicity preference for GRB progenitors, a possible evolution of
GRB luminosity function, or even both \citep{daigne06,virgili11}. Current 
observations of high-$z$ GRBs suggest that their rate per unit star
formation is increasing with $z$, but the rate is not as large
as previously estimated \citep[e.g.][]{bromm02,bromm06,desouza11}.

GRBs have some advantages over other astronomical objects for exploring the
high redshift universe including the fact that the GRB afterglow is a factor 
$\sim 10^4$ brighter than the brightest quasars 
and its spectrum is a featureless powerlaw function. Due to their
extreme luminosity \citep{lamb00}, a favorable negative $k$-correction
and time-dilation effect of optical/IR/radio afterglow \citep{ciardi00,gou04},
GRBs and their afterglows can be detected to a redshift $z \sim 20$. If 
a IR camera can observe the early afterglow phase and take a spectrum,
one would be able to identify them as a high-$z$ GRB through the 
Gunn-Peterson ``trough'' \citep{barkana07}. Since the intrinsic GRB 
afterglow spectra are featureless, all the lines that are observed are due to 
atomic/molecular absorption in the burst-host-galaxy and by gas in the 
intergalactic medium. Thus, one can learn about gas density and 
composition at high redshifts from afterglow observations. 
By studying the damped Lyman $\alpha$ systems of the high-$z$ GRB hosts
\citep{nagamine08,pontzen10}, one can gain insights on structure formation
in the early universe.
High redshift GRBs could also serve as bright background sources for
21cm absorption by neutral hydrogen \citep[e.g.][]{ioka05b,toma11b,ciardi13}
that would facilitate investigation of gas distribution at the dawn of 
galaxy formation, and for determining the reionization history of the 
universe \citep{kawai06,totani06}.  Although we 
are far from there, properties of high-$z$ GRB progenitor stars can, 
in principle, be obtained from the prompt and early X-ray lightcurves. 

Determining the atomic/molecular abundance of ISM at high redshifts would 
provide insight into the history of star formation and supernovae that
enriched the ISM with metals cooked inside the first generation of stars.
Infrared spectra of GRB afterglows would provide this information 
at distances of $\gae10$ pc from the site of explosion. The medium within 
about 10pc of the GRB is, however, completely ionized by the extreme 
luminosity of GRBs and their X-ray and UV afterglows. Thus, little 
information regarding the medium in the immediate vicinity of GRBs --- which 
contains information about the GRB progenitor star and its mass loss 
history ---  can be obtained from afterglow spectra. We have to
resort to other methods in order to obtain information regarding the 
ISM density within a few parsec of the GRB progenitor star.

\section{Concluding thoughts and future prospects}

It has been a long wild ride for people working on gamma-ray bursts to figure out the
true nature and origin of these cosmic explosions. We provide a brief summary of things 
we have learned, and the questions still unanswered, regarding these powerful transient 
events.

\smallskip
\noindent{\bf What are the things we know with confidence?}
\smallskip

The distance to these events is well established from afterglow observations, and hence,
we know the isotropic equivalent of energy release in $\gamma$-rays. The mean redshift
for long duration bursts detected by the Swift satellite is $z\sim2.5$, and for short
bursts it is $\sim 0.3$ \citep{gehrels09}. The median isotropic energy in long (short) bursts 
is $\sim10^{52}$ erg ($\sim10^{50}$ erg).

These explosions have an outflow speed (whatever is its composition) that is 
close to that of light, with a Lorentz factor of a few hundred. There are 
numerous lines of evidence for the high outflow speed. The most direct ones 
are from radio observations that determine the angular size of the ejecta 
with time. The time dependence of ejecta size has been determined either 
directly using VLBI maps for a relatively nearby burst 
\citep{taylor04}, or by using erratic variations of flux by a factor 
$\sim2$ (scintillation) for a period of a few weeks --- which is produced 
when the source size is small --- followed by a smooth decline when the 
source angular size becomes larger than the electron fluctuation scale in 
the inter-stellar medium \citep{goodman97,frail97}. These angular size
measurements show that the LF of the outflow a few days after the
explosion was $\sim10$ which when extrapolated back to 
about 1 minute -- when the GRB blast wave started decelerating due to
interaction with the interstellar medium -- yields a LF of $\sim 10^2$.

It is also certain that at least a part of the afterglow radiation in 
GeV, X-ray, optical, 
radio bands, is produced via the synchrotron process when the relativistic ejecta from the 
explosion drives a strong shock into the surrounding medium.

The relativistic outflow produced in GRBs are highly collimated. This is 
required by consideration of total energy --- most energetic explosions 
release $10^{55}$ergs, if isotropic, and exceed the energy one can 
realistically expect from a stellar mass object --- and also confirmed 
at least in a few cases where achromatic jet break is seen 
in X-ray and optical afterglow lightcurves.

There are at least two physically distinct types of GRBs.
Most long duration bursts ($t>2$s) occur in star forming galaxies,  
and several of these have spectroscopically identified 
supernova associated with them, giving a direct confirmation of 
their origin in the collapse of massive stars. On the other hand, several 
short duration GRBs have been found in low star forming galaxies, or low
star forming regions in star forming galaxies, 
which implies that these were not produced in the collapse of massive, 
short lived, stars. A likely (but not proven) possibility is that
they are produced from binary compact star mergers.

The black hole or neutron star produced in these explosions remains 
{\it active} and continues to produce relativistic jets for a long period 
of time, hours to days, as 
evidenced by flares seen in the X-ray band and occasionally in the optical. 

\smallskip
\noindent{\bf What we wish to know and how to get there?}
\smallskip

Some of the foremost unanswered questions regarding GRBs are: 
\begin{itemize}
\item what is the composition of jet/ejecta (baryonic, $e^\pm$ or 
   magnetic outflow)? 

\item how are $\gamma$-rays, particularly of energy less than $\sim 10$MeV, 
  produced?

\item is a black hole or a rapidly rotating, highly magnetized, neutron star 
  (magnetar) produced in GRBs?

\item what is the mechanism by which relativistic jets are launched? And 
  how is their energy dissipated when they reach a large distance 
  ($\gae10^{12}$cm) from their launching site?

\item what are the properties of long and short duration GRB progenitor stars?

\end{itemize}

Since GRBs are short lived transient events, where the flux changes on time
scale of seconds or less, it has been difficult to coordinate
observing campaigns to cover the very broad segment of electromagnetic spectrum
over which these bursts have significant radiation and capture their temporal
behavior.  
An ideal observational campaign would be to obtain broad band data, from
radio to GeV, for a period of several days starting from $\gamma$-ray trigger 
time. This demands
wide-field, high-sensitivity detectors in a wide bandpass. Such detectors are not
available, but the Swift observatory \citep{gehrels04} and ground-based robotic 
optical telescopes with rapid slew capability have made possible monitoring of 
these events in optical and X-ray bands with excellent time coverage.
Joint triggers by Swift and Fermi LAT \citep{atwood09}, 
even though rare, can provide prompt data from $\sim 10$ keV to $10^2$GeV. 
Future missions such as SVOM \citep{paul12} and UFFO \citep{grossan12,park13} 
would continue to allow rapid follow up observations of GRBs, and move the 
field forward.

Light on jet composition can be shed by measuring optical flux and spectrum
during the prompt and the very early afterglow phase for a large sample of GRBs,
e.g. with UFFO. Early observations in even longer
wavelengths with telescopes such as The Expanded Very Large Array 
(EVLA, http://www.aoc.nrao.edu/evla/), 
Atacama Large Millimeter/submillimeter Array 
(ALMA, http://www.almaobservatory.org), and The LOw Frequency ARray 
(LOFAR, http://www.lofar.org) would also be very useful.
These observations will help us determine whether
the jet is heated by a reverse shock -- which is strong for a baryonic jet
and weak for jets of high magnetization ($\sigma$) -- and 
when the jet starts interacting with the external medium. Firm determination of 
a thermal component to prompt $\gamma$-ray radiation is also an important, 
although not unique, signature of baryonic outflows. Detection or 
an order of magnitude improvement to the currently available upper limit
for neutrino flux from GRBs would also be very important 
for answering the question of jet composition.

Measurement of polarization of $\gamma$-ray radiation will both help 
determine the jet composition and shed light on the radiation mechanism. 
Polarization measurement of early optical radiation is also very useful 
for this purpose. 

Detection of gravitational waves from GRBs would rapidly advance our 
understanding of the GRB progenitor and the central engine. 

We need a concerted theoretical effort to model multi-wavelength prompt and 
afterglow data to be able to extract information regarding the birth of a 
black hole/neutron star and the GRB progenitor star properties. 
Numerical simulations designed to study fundamental physical processes in 
extreme conditions, such as particle acceleration in relativistic shocks and 
in magnetic reconnection regions, and acceleration and dissipation of a 
Poynting jet, are also needed to attack this complex problem from ``bottom-up''.
Collecting more data, although very satisfying, is only 
truly rewarding when at least equal effort is invested to try to understand 
what the data is telling us about the underlying physics and astrophysics of 
the object.  Otherwise we just keep collecting more data and don't advance 
any real understanding, which is like collecting fossil specimen, or 
cataloging plants/animals, without trying to figure out the underlying 
cause for the origin and diversity of life.

\noindent{\bf Acknowledgments}

PK thanks Rodolfo Barniol Duran, Jonathan Granot, 
Milos Milosavljevi{\'c, Ramesh Narayan, Tsvi Piran, Craig Wheeler for 
numerous discussions. He is especially indebted to Alin Panaitescu 
for nearly a decade long collaboration, and countless illuminating 
discussions.

We are grateful to Tsvi Piran and his group at the Hebrew university for 
carefully reading a draft of this review and discussing it over a period 
of about two months, and for providing extensive feedback that helped 
improve this work. We would like a thank a number of colleagues
for providing comments, in particular Lorenzo Amati,
\v Zeljka Bo\v snjak, Patrick Crumley, Rodolfo Barniol Duran, Zigao Dai, 
Yizhong Fan, Filippo Frontera, Jonathan Granot, Bruce Grossan, Kunihito Ioka, 
Peter M{\'e}sz{\'a}ros, Kohta Murase, Shigehiro Nagataki, Asaf Pe'er, 
Rodolfo Santana, Kenji Toma, Ryo Yamazaki, and Xiangyu Wang. 
We are indebted to the referee for taking the time
to read this long review article and for providing extensive, constructive,
comments and criticisms.
Any remaining error or omission is obviously the responsibility 
of the authors. 
We also thank John Beacom, John Cannizzo, Patrick Crumley, Neil Gehrels,
Savannah Kumar, Rodolfo Santana, and Bin-Bin Zhang 
for help with several figures. 

PK is grateful to Dan Jaffe, Chairman of the Astronomy department 
at UT, for a semester off teaching and other administrative duties 
which were big help in completing this review, and he is indebted for 
the hospitality extended to him by KIAA, Beijing, and Jeremy Heyl at UBC 
where he worked uninterrupted for several weeks in very serene settings.
BZ acknowledges UNLV faculty senate sabbatical committee for approving
one-year on leave from UNLV, and the hospitality of KIAA and 
Department of Astronomy, Peking University.
This work was funded in part by NSF grants AST-0909110 (PK),
AST-0908362 (BZ), and NASA grants NNX10AP53G and NNX14AF85G (BZ).

%----------


\begin{thebibliography}{889}
\expandafter\ifx\csname natexlab\endcsname\relax\def\natexlab#1{#1}\fi
\expandafter\ifx\csname url\endcsname\relax
  \def\url#1{\texttt{#1}}\fi
\expandafter\ifx\csname urlprefix\endcsname\relax\def\urlprefix{URL }\fi

\bibitem[{{Abbasi} et~al.(2012){Abbasi}, {Abdou}, {Abu-Zayyad}, {Ackermann},
  {Adams}, {Aguilar}, {Ahlers}, {Altmann}, {Andeen}, {Auffenberg}, and
  et~al.}]{icecube12}
{Abbasi}, R., {Abdou}, Y., {Abu-Zayyad}, T., {Ackermann}, M., {Adams}, J.,
  {Aguilar}, J.~A., {Ahlers}, M., {Altmann}, D., {Andeen}, K., {Auffenberg},
  J., et~al., Apr. 2012. {An absence of neutrinos associated with cosmic-ray
  acceleration in {$\gamma$}-ray bursts}. \nat 484, 351--354.

\bibitem[{{Abbasi} et~al.(2010){Abbasi}, {Abdou}, {Abu-Zayyad}, {Adams},
  {Aguilar}, {Ahlers}, {Andeen}, {Auffenberg}, {Bai}, {Baker}, and
  et~al.}]{icecube10}
{Abbasi}, R., {Abdou}, Y., {Abu-Zayyad}, T., {Adams}, J., {Aguilar}, J.~A.,
  {Ahlers}, M., {Andeen}, K., {Auffenberg}, J., {Bai}, X., {Baker}, M., et~al.,
  Feb. 2010. {Search for Muon Neutrinos from Gamma-ray Bursts with the IceCube
  Neutrino Telescope}. \apj 710, 346--359.

\bibitem[{{Abbasi} et~al.(2011){Abbasi}, {Abdou}, {Abu-Zayyad}, {Adams},
  {Aguilar}, {Ahlers}, {Andeen}, {Auffenberg}, {Bai}, {Baker}, and
  et~al.}]{icecube11}
{Abbasi}, R., {Abdou}, Y., {Abu-Zayyad}, T., {Adams}, J., {Aguilar}, J.~A.,
  {Ahlers}, M., {Andeen}, K., {Auffenberg}, J., {Bai}, X., {Baker}, M., et~al.,
  Apr. 2011. {Limits on Neutrino Emission from Gamma-Ray Bursts with the 40
  String IceCube Detector}. Physical Review Letters 106~(14), 141101.

\bibitem[{{Abbott} et~al.(2009){Abbott}, {Abbott}, {Adhikari}, {Ajith},
  {Allen}, {Allen}, {Amin}, {Anderson}, {Anderson}, {Arain}, and et~al.}]{ligo}
{Abbott}, B.~P., {Abbott}, R., {Adhikari}, R., {Ajith}, P., {Allen}, B.,
  {Allen}, G., {Amin}, R.~S., {Anderson}, S.~B., {Anderson}, W.~G., {Arain},
  M.~A., et~al., Jul. 2009. {LIGO: the Laser Interferometer Gravitational-Wave
  Observatory}. Reports on Progress in Physics 72~(7), 076901.

\bibitem[{{Abdo} et~al.(2009{\natexlab{a}}){Abdo}, {Ackermann}, {Ajello},
  {Asano}, {Atwood}, {Axelsson}, {Baldini}, {Ballet}, {Barbiellini}, {Baring},
  and et~al.}]{abdo09b}
{Abdo}, A.~A., {Ackermann}, M., {Ajello}, M., {Asano}, K., {Atwood}, W.~B.,
  {Axelsson}, M., {Baldini}, L., {Ballet}, J., {Barbiellini}, G., {Baring},
  M.~G., et~al., Nov. 2009{\natexlab{a}}. {A limit on the variation of the
  speed of light arising from quantum gravity effects}. \nat 462, 331--334.

\bibitem[{{Abdo} et~al.(2009{\natexlab{b}}){Abdo}, {Ackermann}, {Ajello},
  {Asano}, {Atwood}, {Axelsson}, {Baldini}, {Ballet}, {Barbiellini}, {Baring},
  and et~al.}]{abdo09c}
{Abdo}, A.~A., {Ackermann}, M., {Ajello}, M., {Asano}, K., {Atwood}, W.~B.,
  {Axelsson}, M., {Baldini}, L., {Ballet}, J., {Barbiellini}, G., {Baring},
  M.~G., et~al., Nov. 2009{\natexlab{b}}. {Fermi Observations of GRB 090902B: A
  Distinct Spectral Component in the Prompt and Delayed Emission}. \apjl 706,
  L138--L144.

\bibitem[{{Abdo} et~al.(2011){Abdo}, {Ackermann}, {Ajello}, {Baldini},
  {Ballet}, {Barbiellini}, {Baring}, {Bastieri}, {Bechtol}, and
  {Bellazzini}}]{Abdo11}
{Abdo}, A.~A., {Ackermann}, M., {Ajello}, M., {Baldini}, L., {Ballet}, J.,
  {Barbiellini}, G., {Baring}, M.~G., {Bastieri}, D., {Bechtol}, K.,
  {Bellazzini}, R. e.~a., Jun. 2011. {Detection of High-energy Gamma-Ray
  Emission During the X-Ray Flaring Activity in GRB 100728A}. \apjl 734, L27.

\bibitem[{{Abdo} et~al.(2009{\natexlab{c}}){Abdo}, {Ackermann}, {Arimoto},
  {Asano}, {Atwood}, {Axelsson}, {Baldini}, {Ballet}, {Band}, {Barbiellini},
  and et~al.}]{abdo09a}
{Abdo}, A.~A., {Ackermann}, M., {Arimoto}, M., {Asano}, K., {Atwood}, W.~B.,
  {Axelsson}, M., {Baldini}, L., {Ballet}, J., {Band}, D.~L., {Barbiellini},
  G., et~al., Mar. 2009{\natexlab{c}}. {Fermi Observations of High-Energy
  Gamma-Ray Emission from GRB 080916C}. Science 323, 1688--.

\bibitem[{{Abel} et~al.(2000){Abel}, {Bryan}, and {Norman}}]{abel00}
{Abel}, T., {Bryan}, G.~L., {Norman}, M.~L., Sep. 2000. {The Formation and
  Fragmentation of Primordial Molecular Clouds}. \apj 540, 39--44.

\bibitem[{{Acernese} et~al.(2008){Acernese}, {Alshourbagy}, {Amico},
  {Antonucci}, {Aoudia}, {Astone}, {Avino}, {Baggio}, {Ballardin}, and
  {Barone}}]{virgo}
{Acernese}, F., {Alshourbagy}, M., {Amico}, P., {Antonucci}, F., {Aoudia}, S.,
  {Astone}, P., {Avino}, S., {Baggio}, L., {Ballardin}, G., {Barone}, F. e.~a.,
  Jun. 2008. {Status of Virgo}. Classical and Quantum Gravity 25~(11), 114045.

\bibitem[{{Achterberg} et~al.(2001){Achterberg}, {Gallant}, {Kirk}, and
  {Guthmann}}]{achterberg01}
{Achterberg}, A., {Gallant}, Y.~A., {Kirk}, J.~G., {Guthmann}, A.~W., Dec.
  2001. {Particle acceleration by ultrarelativistic shocks: theory and
  simulations}. \mnras 328, 393--408.

\bibitem[{{Ackermann} et~al.(2014){Ackermann}, {Ajello}, {Asano}, {Atwood},
  {Axelsson}, {Baldini}, {Ballet}, {Barbiellini}, {Baring}, and
  {Bastieri}}]{ackermann14}
{Ackermann}, M., {Ajello}, M., {Asano}, K., {Atwood}, W.~B., {Axelsson}, M.,
  {Baldini}, L., {Ballet}, J., {Barbiellini}, G., {Baring}, M.~G., {Bastieri},
  D. e.~a., Jan. 2014. {Fermi-LAT Observations of the Gamma-Ray Burst GRB
  130427A}. Science 343, 42--47.

\bibitem[{{Ackermann} et~al.(2011){Ackermann}, {Ajello}, {Asano}, {Axelsson},
  {Baldini}, {Ballet}, {Barbiellini}, {Baring}, {Bastieri}, {Bechtol}, and
  et~al.}]{ackermann11}
{Ackermann}, M., {Ajello}, M., {Asano}, K., {Axelsson}, M., {Baldini}, L.,
  {Ballet}, J., {Barbiellini}, G., {Baring}, M.~G., {Bastieri}, D., {Bechtol},
  K., et~al., Mar. 2011. {Detection of a Spectral Break in the Extra Hard
  Component of GRB 090926A}. \apj 729, 114.

\bibitem[{{Ackermann} et~al.(2013{\natexlab{a}}){Ackermann}, {Ajello}, {Asano},
  {Axelsson}, {Baldini}, {Ballet}, {Barbiellini}, {Bastieri}, {Bechtol},
  {Bellazzini}, and et~al.}]{ackermann13a}
{Ackermann}, M., {Ajello}, M., {Asano}, K., {Axelsson}, M., {Baldini}, L.,
  {Ballet}, J., {Barbiellini}, G., {Bastieri}, D., {Bechtol}, K., {Bellazzini},
  et~al., Nov. 2013{\natexlab{a}}. {The First Fermi-LAT Gamma-Ray Burst
  Catalog}. \apjs 209, 11.

\bibitem[{{Ackermann} et~al.(2013{\natexlab{b}}){Ackermann}, {Ajello}, {Asano},
  {Baldini}, {Barbiellini}, {Baring}, {Bastieri}, {Bellazzini}, {Blandford},
  {Bonamente}, and et~al.}]{ackermann13}
{Ackermann}, M., {Ajello}, M., {Asano}, K., {Baldini}, L., {Barbiellini}, G.,
  {Baring}, M.~G., {Bastieri}, D., {Bellazzini}, R., {Blandford}, R.~D.,
  {Bonamente}, E., et~al., Feb. 2013{\natexlab{b}}. {Multiwavelength
  Observations of GRB 110731A: GeV Emission from Onset to Afterglow}. \apj 763,
  71.

\bibitem[{{Ackermann} et~al.(2012){Ackermann}, {Ajello}, {Baldini},
  {Barbiellini}, {Baring}, {Bechtol}, {Bellazzini}, {Blandford}, {Bloom}, and
  {Bonamente}}]{ackermann12}
{Ackermann}, M., {Ajello}, M., {Baldini}, L., {Barbiellini}, G., {Baring},
  M.~G., {Bechtol}, K., {Bellazzini}, R., {Blandford}, R.~D., {Bloom}, E.~D.,
  {Bonamente}, E. e.~a., Aug. 2012. {Constraining the High-energy Emission from
  Gamma-Ray Bursts with Fermi}. \apj 754, 121.

\bibitem[{{Ackermann} et~al.(2010){Ackermann}, {Asano}, {Atwood}, {Axelsson},
  {Baldini}, {Ballet}, {Barbiellini}, {Baring}, {Bastieri}, {Bechtol}, and
  et~al.}]{ackermann10}
{Ackermann}, M., {Asano}, K., {Atwood}, W.~B., {Axelsson}, M., {Baldini}, L.,
  {Ballet}, J., {Barbiellini}, G., {Baring}, M.~G., {Bastieri}, D., {Bechtol},
  K., et~al., Jun. 2010. {Fermi Observations of GRB 090510: A Short-Hard
  Gamma-ray Burst with an Additional, Hard Power-law Component from 10 keV TO
  GeV Energies}. \apj 716, 1178--1190.

\bibitem[{{Aharonian}(2000)}]{aharonian00}
{Aharonian}, F.~A., Nov. 2000. {TeV gamma rays from BL Lac objects due to
  synchrotron radiation of extremely high energy protons}. \na 5, 377--395.

\bibitem[{{Akerlof} et~al.(1999){Akerlof}, {Balsano}, {Barthelmy}, {Bloch},
  {Butterworth}, {Casperson}, {Cline}, {Fletcher}, {Frontera}, {Gisler},
  {Heise}, {Hills}, {Kehoe}, {Lee}, {Marshall}, {McKay}, {Miller}, {Piro},
  {Priedhorsky}, {Szymanski}, and {Wren}}]{akerlof99}
{Akerlof}, C., {Balsano}, R., {Barthelmy}, S., {Bloch}, J., {Butterworth}, P.,
  {Casperson}, D., {Cline}, T., {Fletcher}, S., {Frontera}, F., {Gisler}, G.,
  {Heise}, J., {Hills}, J., {Kehoe}, R., {Lee}, B., {Marshall}, S., {McKay},
  T., {Miller}, R., {Piro}, L., {Priedhorsky}, W., {Szymanski}, J., {Wren}, J.,
  Apr. 1999. {Observation of contemporaneous optical radiation from a
  {$\gamma$}-ray burst}. \nat 398, 400--402.

\bibitem[{{Aloy} et~al.(2005){Aloy}, {Janka}, and {M{\"u}ller}}]{aloy05}
{Aloy}, M.~A., {Janka}, H.-T., {M{\"u}ller}, E., Jun. 2005. {Relativistic
  outflows from remnants of compact object mergers and their viability for
  short gamma-ray bursts}. \aap 436, 273--311.

\bibitem[{{Aloy} et~al.(2000){Aloy}, {M{\"u}ller}, {Ib{\'a}{\~n}ez},
  {Mart{\'{\i}}}, and {MacFadyen}}]{aloy00}
{Aloy}, M.~A., {M{\"u}ller}, E., {Ib{\'a}{\~n}ez}, J.~M., {Mart{\'{\i}}},
  J.~M., {MacFadyen}, A., Mar. 2000. {Relativistic Jets from Collapsars}. \apjl
  531, L119--L122.

\bibitem[{{Amati}(2006)}]{amati06}
{Amati}, L., Oct. 2006. {The $E_{p,i}-E_{iso}$ correlation in gamma-ray bursts:
  updated observational status, re-analysis and main implications}. \mnras 372,
  233--245.

\bibitem[{{Amati} et~al.(2002){Amati}, {Frontera}, {Tavani}, {in't Zand},
  {Antonelli}, {Costa}, {Feroci}, {Guidorzi}, {Heise}, {Masetti}, {Montanari},
  {Nicastro}, {Palazzi}, {Pian}, {Piro}, and {Soffitta}}]{amati02}
{Amati}, L., {Frontera}, F., {Tavani}, M., {in't Zand}, J.~J.~M., {Antonelli},
  A., {Costa}, E., {Feroci}, M., {Guidorzi}, C., {Heise}, J., {Masetti}, N.,
  {Montanari}, E., {Nicastro}, L., {Palazzi}, E., {Pian}, E., {Piro}, L.,
  {Soffitta}, P., Jul. 2002. {Intrinsic spectra and energetics of BeppoSAX
  Gamma-Ray Bursts with known redshifts}. \aap 390, 81--89.

\bibitem[{{Amati} et~al.(2008){Amati}, {Guidorzi}, {Frontera}, {Della Valle},
  {Finelli}, {Landi}, and {Montanari}}]{amati08}
{Amati}, L., {Guidorzi}, C., {Frontera}, F., {Della Valle}, M., {Finelli}, F.,
  {Landi}, R., {Montanari}, E., Dec. 2008. {Measuring the cosmological
  parameters with the E$_{p,i}$-E$_{iso}$ correlation of gamma-ray bursts}.
  \mnras 391, 577--584.

\bibitem[{{Amati} and {Valle}(2013)}]{amati13}
{Amati}, L., {Valle}, M.~D., Dec. 2013. {Measuring Cosmological Parameters with
  Gamma Ray Bursts}. International Journal of Modern Physics D 22, 30028.

\bibitem[{{Anderson} et~al.(2008){Anderson}, {Hirschmann}, {Lehner},
  {Liebling}, {Motl}, {Neilsen}, {Palenzuela}, and {Tohline}}]{anderson08}
{Anderson}, M., {Hirschmann}, E.~W., {Lehner}, L., {Liebling}, S.~L., {Motl},
  P.~M., {Neilsen}, D., {Palenzuela}, C., {Tohline}, J.~E., May 2008.
  {Magnetized Neutron-Star Mergers and Gravitational-Wave Signals}. Physical
  Review Letters 100~(19), 191101.

\bibitem[{{Antonelli} et~al.(2009){Antonelli}, {D'Avanzo}, {Perna}, {Amati},
  {Covino}, {Cutini}, {D'Elia}, {Gallozzi}, {Grazian}, {Palazzi},
  {Piranomonte}, {Rossi}, {Spiro}, {Stella}, {Testa}, {Chincarini}, {di Paola},
  {Fiore}, {Fugazza}, {Giallongo}, {Maiorano}, {Masetti}, {Pedichini},
  {Salvaterra}, {Tagliaferri}, and {Vergani}}]{antonelli09}
{Antonelli}, L.~A., {D'Avanzo}, P., {Perna}, R., {Amati}, L., {Covino}, S.,
  {Cutini}, S., {D'Elia}, V., {Gallozzi}, S., {Grazian}, A., {Palazzi}, E.,
  {Piranomonte}, S., {Rossi}, A., {Spiro}, S., {Stella}, L., {Testa}, V.,
  {Chincarini}, G., {di Paola}, A., {Fiore}, F., {Fugazza}, D., {Giallongo},
  E., {Maiorano}, E., {Masetti}, N., {Pedichini}, F., {Salvaterra}, R.,
  {Tagliaferri}, G., {Vergani}, S., Dec. 2009. {GRB 090426: the farthest short
  gamma-ray burst?} \aap 507, L45--L48.

\bibitem[{{Asano} et~al.(2009){Asano}, {Inoue}, and
  {M{\'e}sz{\'a}ros}}]{asano09b}
{Asano}, K., {Inoue}, S., {M{\'e}sz{\'a}ros}, P., Jul. 2009. {Prompt
  High-Energy Emission from Proton-Dominated Gamma-Ray Bursts}. \apj 699,
  953--957.

\bibitem[{{Asano} and {M{\'e}sz{\'a}ros}(2012)}]{asano12}
{Asano}, K., {M{\'e}sz{\'a}ros}, P., Oct. 2012. {Delayed Onset of High-energy
  Emissions in Leptonic and Hadronic Models of Gamma-Ray Bursts}. \apj 757,
  115.

\bibitem[{{Asano} and {M{\'e}sz{\'a}ros}(2013)}]{asano13}
{Asano}, K., {M{\'e}sz{\'a}ros}, P., Sep. 2013. {Photon and neutrino spectra of
  time-dependent photospheric models of gamma-ray bursts}. \jcap 9, 8.

\bibitem[{{Asano} and {Terasawa}(2009)}]{asano09}
{Asano}, K., {Terasawa}, T., Nov. 2009. {Slow Heating Model of Gamma-ray Burst:
  Photon Spectrum and Delayed Emission}. \apj 705, 1714--1720.

\bibitem[{{Atwood} et~al.(2009){Atwood}, {Abdo}, {Ackermann}, {Althouse},
  {Anderson}, {Axelsson}, {Baldini}, {Ballet}, {Band}, {Barbiellini}, and
  et~al.}]{atwood09}
{Atwood}, W.~B., {Abdo}, A.~A., {Ackermann}, M., {Althouse}, W., {Anderson},
  B., {Axelsson}, M., {Baldini}, L., {Ballet}, J., {Band}, D.~L.,
  {Barbiellini}, G., et~al., Jun. 2009. {The Large Area Telescope on the Fermi
  Gamma-Ray Space Telescope Mission}. \apj 697, 1071--1102.

\bibitem[{{Axelsson} et~al.(2012){Axelsson}, {Baldini}, {Barbiellini},
  {Baring}, {Bellazzini}, {Bregeon}, {Brigida}, {Bruel}, {Buehler},
  {Caliandro}, and et~al.}]{axelsson12}
{Axelsson}, M., {Baldini}, L., {Barbiellini}, G., {Baring}, M.~G.,
  {Bellazzini}, R., {Bregeon}, J., {Brigida}, M., {Bruel}, P., {Buehler}, R.,
  {Caliandro}, G.~A., et~al., Oct. 2012. {GRB110721A: An Extreme Peak Energy
  and Signatures of the Photosphere}. \apjl 757, L31.

\bibitem[{{Bahcall} and {M{\'e}sz{\'a}ros}(2000)}]{bahcall00}
{Bahcall}, J.~N., {M{\'e}sz{\'a}ros}, P., Aug. 2000. {5-10 GeV Neutrinos from
  Gamma-Ray Burst Fireballs}. Physical Review Letters 85, 1362--1365.

\bibitem[{{Baiotti} et~al.(2008){Baiotti}, {Giacomazzo}, and
  {Rezzolla}}]{baiotti08}
{Baiotti}, L., {Giacomazzo}, B., {Rezzolla}, L., Oct. 2008. {Accurate
  evolutions of inspiralling neutron-star binaries: Prompt and delayed collapse
  to a black hole}. \prd 78~(8), 084033.

\bibitem[{{Band} et~al.(1993){Band}, {Matteson}, {Ford}, {Schaefer}, {Palmer},
  {Teegarden}, {Cline}, {Briggs}, {Paciesas}, {Pendleton}, {Fishman},
  {Kouveliotou}, {Meegan}, {Wilson}, and {Lestrade}}]{band93}
{Band}, D., {Matteson}, J., {Ford}, L., {Schaefer}, B., {Palmer}, D.,
  {Teegarden}, B., {Cline}, T., {Briggs}, M., {Paciesas}, W., {Pendleton}, G.,
  {Fishman}, G., {Kouveliotou}, C., {Meegan}, C., {Wilson}, R., {Lestrade}, P.,
  Aug. 1993. {BATSE observations of gamma-ray burst spectra. I - Spectral
  diversity}. \apj 413, 281--292.

\bibitem[{{Band} and {Preece}(2005)}]{band05}
{Band}, D.~L., {Preece}, R.~D., Jul. 2005. {Testing the Gamma-Ray Burst Energy
  Relationships}. \apj 627, 319--323.

\bibitem[{{Barkana} and {Loeb}(2007)}]{barkana07}
{Barkana}, R., {Loeb}, A., Apr. 2007. {The physics and early history of the
  intergalactic medium}. Reports on Progress in Physics 70, 627--657.

\bibitem[{{Barnes} and {Kasen}(2013)}]{barnes13}
{Barnes}, J., {Kasen}, D., Sep. 2013. {Effect of a High Opacity on the Light
  Curves of Radioactively Powered Transients from Compact Object Mergers}. \apj
  775, 18.

\bibitem[{{Barniol Duran} et~al.(2012){Barniol Duran}, {Bo{\v s}njak}, and
  {Kumar}}]{barniolduran12}
{Barniol Duran}, R., {Bo{\v s}njak}, {\v Z}., {Kumar}, P., Aug. 2012.
  {Inverse-Compton cooling in Klein-Nishina regime and gamma-ray burst prompt
  spectrum}. \mnras 424, 3192--3200.

\bibitem[{{Barniol Duran} and {Kumar}(2009)}]{barniolduran08}
{Barniol Duran}, R., {Kumar}, P., May 2009. {Adiabatic expansion, early x-ray
  data and the central engine in GRBs}. \mnras 395, 955--961.

\bibitem[{{Barniol Duran} and {Kumar}(2011)}]{barniolduran11}
{Barniol Duran}, R., {Kumar}, P., Mar. 2011. {Implications of electron
  acceleration for high-energy radiation from gamma-ray bursts}. \mnras 412,
  522--528.

\bibitem[{{Barthelmy} et~al.(2005{\natexlab{a}}){Barthelmy}, {Barbier},
  {Cummings}, {Fenimore}, {Gehrels}, {Hullinger}, {Krimm}, {Markwardt},
  {Palmer}, {Parsons}, {Sato}, {Suzuki}, {Takahashi}, {Tashiro}, and
  {Tueller}}]{barthelmy05c}
{Barthelmy}, S.~D., {Barbier}, L.~M., {Cummings}, J.~R., {Fenimore}, E.~E.,
  {Gehrels}, N., {Hullinger}, D., {Krimm}, H.~A., {Markwardt}, C.~B., {Palmer},
  D.~M., {Parsons}, A., {Sato}, G., {Suzuki}, M., {Takahashi}, T., {Tashiro},
  M., {Tueller}, J., Oct. 2005{\natexlab{a}}. {The Burst Alert Telescope (BAT)
  on the SWIFT Midex Mission}. Space Science Reviews 120, 143--164.

\bibitem[{{Barthelmy} et~al.(2005{\natexlab{b}}){Barthelmy}, {Cannizzo},
  {Gehrels}, {Cusumano}, {Mangano}, {O'Brien}, {Vaughan}, {Zhang}, {Burrows},
  {Campana}, {Chincarini}, {Goad}, {Kouveliotou}, {Kumar}, {M{\'e}sz{\'a}ros},
  {Nousek}, {Osborne}, {Panaitescu}, {Reeves}, {Sakamoto}, {Tagliaferri}, and
  {Wijers}}]{barthelmy05b}
{Barthelmy}, S.~D., {Cannizzo}, J.~K., {Gehrels}, N., {Cusumano}, G.,
  {Mangano}, V., {O'Brien}, P.~T., {Vaughan}, S., {Zhang}, B., {Burrows},
  D.~N., {Campana}, S., {Chincarini}, G., {Goad}, M.~R., {Kouveliotou}, C.,
  {Kumar}, P., {M{\'e}sz{\'a}ros}, P., {Nousek}, J.~A., {Osborne}, J.~P.,
  {Panaitescu}, A., {Reeves}, J.~N., {Sakamoto}, T., {Tagliaferri}, G.,
  {Wijers}, R.~A.~M.~J., Dec. 2005{\natexlab{b}}. {Discovery of an Afterglow
  Extension of the Prompt Phase of Two Gamma-Ray Bursts Observed by Swift}.
  \apjl 635, L133--L136.

\bibitem[{{Barthelmy} et~al.(2005{\natexlab{c}}){Barthelmy}, {Chincarini},
  {Burrows}, {Gehrels}, {Covino}, {Moretti}, {Romano}, {O'Brien}, {Sarazin},
  {Kouveliotou}, {Goad}, {Vaughan}, {Tagliaferri}, {Zhang}, {Antonelli},
  {Campana}, {Cummings}, {D'Avanzo}, {Davies}, {Giommi}, {Grupe}, {Kaneko},
  {Kennea}, {King}, {Kobayashi}, {Melandri}, {M\'esz\'aros}, {Nousek}, {Patel},
  {Sakamoto}, and {Wijers}}]{barthelmy05a}
{Barthelmy}, S.~D., {Chincarini}, G., {Burrows}, D.~N., {Gehrels}, N.,
  {Covino}, S., {Moretti}, A., {Romano}, P., {O'Brien}, P.~T., {Sarazin},
  C.~L., {Kouveliotou}, C., {Goad}, M., {Vaughan}, S., {Tagliaferri}, G.,
  {Zhang}, B., {Antonelli}, L.~A., {Campana}, S., {Cummings}, J.~R.,
  {D'Avanzo}, P., {Davies}, M.~B., {Giommi}, P., {Grupe}, D., {Kaneko}, Y.,
  {Kennea}, J.~A., {King}, A., {Kobayashi}, S., {Melandri}, A., {M\'esz\'aros},
  P., {Nousek}, J.~A., {Patel}, S., {Sakamoto}, T., {Wijers}, R.~A.~M.~J., Dec.
  2005{\natexlab{c}}. {An origin for short {$\gamma$}-ray bursts unassociated
  with current star formation}. \nat 438, 994--996.

\bibitem[{{Bartos} et~al.(2013){Bartos}, {Brady}, and {M{\'a}rka}}]{bartos13}
{Bartos}, I., {Brady}, P., {M{\'a}rka}, S., Jun. 2013. {How gravitational-wave
  observations can shape the gamma-ray burst paradigm}. Classical and Quantum
  Gravity 30~(12), 123001.

\bibitem[{{Bednarz} and {Ostrowski}(1998)}]{bednarz98}
{Bednarz}, J., {Ostrowski}, M., May 1998. {Energy Spectra of Cosmic Rays
  Accelerated at Ultrarelativistic Shock Waves}. Physical Review Letters 80,
  3911--3914.

\bibitem[{{Beloborodov}(2000)}]{beloborodov00}
{Beloborodov}, A.~M., Aug. 2000. {On the Efficiency of Internal Shocks in
  Gamma-Ray Bursts}. \apjl 539, L25--L28.

\bibitem[{{Beloborodov}(2002)}]{beloborodov02}
{Beloborodov}, A.~M., Feb. 2002. {Radiation Front Sweeping the Ambient Medium
  of Gamma-Ray Bursts}. \apj 565, 808--828.

\bibitem[{{Beloborodov}(2003{\natexlab{a}})}]{beloborodov03b}
{Beloborodov}, A.~M., Mar. 2003{\natexlab{a}}. {Neutron-fed Afterglows of
  Gamma-Ray Bursts}. \apjl 585, L19--L22.

\bibitem[{{Beloborodov}(2003{\natexlab{b}})}]{beloborodov03}
{Beloborodov}, A.~M., May 2003{\natexlab{b}}. {Nuclear Composition of Gamma-Ray
  Burst Fireballs}. \apj 588, 931--944.

\bibitem[{{Beloborodov}(2005)}]{beloborodov05}
{Beloborodov}, A.~M., Jan. 2005. {Optical and GeV-TeV Flashes from Gamma-Ray
  Bursts}. \apjl 618, L13--L16.

\bibitem[{{Beloborodov}(2010)}]{beloborodov10}
{Beloborodov}, A.~M., Sep. 2010. {Collisional mechanism for gamma-ray burst
  emission}. \mnras 407, 1033--1047.

\bibitem[{{Beloborodov}(2013)}]{beloborodov13}
{Beloborodov}, A.~M., Feb. 2013. {Regulation of the Spectral Peak in Gamma-Ray
  Bursts}. \apj 764, 157.

\bibitem[{{Beloborodov} et~al.(2011){Beloborodov}, {Daigne}, {Mochkovitch}, and
  {Uhm}}]{beloborodov11b}
{Beloborodov}, A.~M., {Daigne}, F., {Mochkovitch}, R., {Uhm}, Z.~L., Feb. 2011.
  {Is gamma-ray burst afterglow emission intrinsically anisotropic?} \mnras
  410, 2422--2427.

\bibitem[{{Beloborodov} et~al.(2013){Beloborodov}, {Hascoet}, and
  {Vurm}}]{beloborodov13b}
{Beloborodov}, A.~M., {Hascoet}, R., {Vurm}, I., Jul. 2013. {On the origin of
  GeV emission in gamma-ray bursts}. ArXiv e-prints.

\bibitem[{{Beniamini} et~al.(2011){Beniamini}, {Guetta}, {Nakar}, and
  {Piran}}]{beniamini11}
{Beniamini}, P., {Guetta}, D., {Nakar}, E., {Piran}, T., Oct. 2011. {Limits on
  the GeV emission from gamma-ray bursts}. \mnras 416, 3089--3097.

\bibitem[{{Beniamini} and {Piran}(2013)}]{beniamini13}
{Beniamini}, P., {Piran}, T., May 2013. {Constraints on the Synchrotron
  Emission Mechanism in Gamma-Ray Bursts}. \apj 769, 69.

\bibitem[{{Berger}(2011)}]{berger11}
{Berger}, E., Jan. 2011. {The environments of short-duration gamma-ray bursts
  and implications for their progenitors}. \nar 55, 1--22.

\bibitem[{{Berger}(2014)}]{berger14}
{Berger}, E., Aug. 2014. {Short-Duration Gamma-Ray Bursts}. \araa 52, 43--105.

\bibitem[{{Berger} et~al.(2013){Berger}, {Fong}, and {Chornock}}]{berger13}
{Berger}, E., {Fong}, W., {Chornock}, R., Sep. 2013. {An r-process Kilonova
  Associated with the Short-hard GRB 130603B}. \apjl 774, L23.

\bibitem[{{Berger} et~al.(2005){Berger}, {Kulkarni}, {Fox}, {Soderberg},
  {Harrison}, {Nakar}, {Kelson}, {Gladders}, {Mulchaey}, {Oemler}, {Dressler},
  {Cenko}, {Price}, {Schmidt}, {Frail}, {Morrell}, {Gonzalez}, {Krzeminski},
  {Sari}, {Gal-Yam}, {Moon}, {Penprase}, {Jayawardhana}, {Scholz}, {Rich},
  {Peterson}, {Anderson}, {McNaught}, {Minezaki}, {Yoshii}, {Cowie}, and
  {Pimbblet}}]{berger05}
{Berger}, E., {Kulkarni}, S.~R., {Fox}, D.~B., {Soderberg}, A.~M., {Harrison},
  F.~A., {Nakar}, E., {Kelson}, D.~D., {Gladders}, M.~D., {Mulchaey}, J.~S.,
  {Oemler}, A., {Dressler}, A., {Cenko}, S.~B., {Price}, P.~A., {Schmidt},
  B.~P., {Frail}, D.~A., {Morrell}, N., {Gonzalez}, S., {Krzeminski}, W.,
  {Sari}, R., {Gal-Yam}, A., {Moon}, D.-S., {Penprase}, B.~E., {Jayawardhana},
  R., {Scholz}, A., {Rich}, J., {Peterson}, B.~A., {Anderson}, G., {McNaught},
  R., {Minezaki}, T., {Yoshii}, Y., {Cowie}, L.~L., {Pimbblet}, K., Nov. 2005.
  {Afterglows, Redshifts, and Properties of Swift Gamma-Ray Bursts}. \apj 634,
  501--508.

\bibitem[{{Berger} et~al.(2003{\natexlab{a}}){Berger}, {Kulkarni}, and
  {Frail}}]{berger03}
{Berger}, E., {Kulkarni}, S.~R., {Frail}, D.~A., Jun. 2003{\natexlab{a}}. {A
  Standard Kinetic Energy Reservoir in Gamma-Ray Burst Afterglows}. \apj 590,
  379--385.

\bibitem[{{Berger} et~al.(2003{\natexlab{b}}){Berger}, {Kulkarni}, {Pooley},
  {Frail}, {McIntyre}, {Wark}, {Sari}, {Soderberg}, {Fox}, {Yost}, and
  {Price}}]{berger03b}
{Berger}, E., {Kulkarni}, S.~R., {Pooley}, G., {Frail}, D.~A., {McIntyre}, V.,
  {Wark}, R.~M., {Sari}, R., {Soderberg}, A.~M., {Fox}, D.~W., {Yost}, S.,
  {Price}, P.~A., Nov. 2003{\natexlab{b}}. {A common origin for cosmic
  explosions inferred from calorimetry of GRB030329}. \nat 426, 154--157.

\bibitem[{{Berger} et~al.(2003{\natexlab{c}}){Berger}, {Soderberg}, {Frail},
  and {Kulkarni}}]{berger03a}
{Berger}, E., {Soderberg}, A.~M., {Frail}, D.~A., {Kulkarni}, S.~R., Apr.
  2003{\natexlab{c}}. {A Radio Flare from GRB 020405: Evidence for a Uniform
  Medium around a Massive Stellar Progenitor}. \apjl 587, L5--L8.

\bibitem[{{Bernardini} et~al.(2007){Bernardini}, {Bianco}, {Caito}, {Dainotti},
  {Guida}, and {Ruffini}}]{bernardini07}
{Bernardini}, M.~G., {Bianco}, C.~L., {Caito}, L., {Dainotti}, M.~G., {Guida},
  R., {Ruffini}, R., Oct. 2007. {GRB 970228 and a class of GRBs with an initial
  spikelike emission}. \aap 474, L13--L16.

\bibitem[{{Bernardini} et~al.(2012){Bernardini}, {Margutti}, {Mao}, {Zaninoni},
  and {Chincarini}}]{bernardini12}
{Bernardini}, M.~G., {Margutti}, R., {Mao}, J., {Zaninoni}, E., {Chincarini},
  G., Mar. 2012. {The X-ray light curve of gamma-ray bursts: clues to the
  central engine}. \aap 539, A3.

\bibitem[{{Beskin} et~al.(2010){Beskin}, {Karpov}, {Bondar}, {Greco},
  {Guarnieri}, {Bartolini}, and {Piccioni}}]{beskin10}
{Beskin}, G., {Karpov}, S., {Bondar}, S., {Greco}, G., {Guarnieri}, A.,
  {Bartolini}, C., {Piccioni}, A., Aug. 2010. {Fast Optical Variability of a
  Naked-eye Burst Manifestation of the Periodic Activity of an Internal
  Engine}. \apjl 719, L10--L14.

\bibitem[{{Beskin} and {Nokhrina}(2006)}]{beskin06}
{Beskin}, V.~S., {Nokhrina}, E.~E., Mar. 2006. {The effective acceleration of
  plasma outflow in the paraboloidal magnetic field}. \mnras 367, 375--386.

\bibitem[{{Bessho} and {Bhattacharjee}(2007)}]{bessho07}
{Bessho}, N., {Bhattacharjee}, A., May 2007. {Fast collisionless reconnection
  in electron-positron plasmas}. Physics of Plasmas 14~(5), 056503.

\bibitem[{{Bessho} and {Bhattacharjee}(2012)}]{bessho12}
{Bessho}, N., {Bhattacharjee}, A., May 2012. {Fast Magnetic Reconnection and
  Particle Acceleration in Relativistic Low-density Electron-Positron Plasmas
  without Guide Field}. \apj 750, 129.

\bibitem[{{Bhattacharya}(2001)}]{bhattacharya01}
{Bhattacharya}, D., Jun. 2001. {Flat Spectrum Gamma Ray Burst Afterglows}.
  Bulletin of the Astronomical Society of India 29, 107--114.

\bibitem[{{Bianco} and {Ruffini}(2005)}]{bianco05}
{Bianco}, C.~L., {Ruffini}, R., Nov. 2005. {Exact versus Approximate Solutions
  in Gamma-Ray Burst Afterglows}. \apjl 633, L13--L16.

\bibitem[{{Birnbaum} et~al.(2012){Birnbaum}, {Zhang}, {Zhang}, and
  {Liang}}]{birnbaum12}
{Birnbaum}, T., {Zhang}, B., {Zhang}, B.-B., {Liang}, E.-W., May 2012.
  {Observational constraints on the external shock prior emission hypothesis of
  gamma-ray bursts}. \mnras 422, 393--400.

\bibitem[{{Bj{\"o}rnsson} et~al.(2004){Bj{\"o}rnsson}, {Gudmundsson}, and
  {J{\'o}hannesson}}]{bjornsson04}
{Bj{\"o}rnsson}, G., {Gudmundsson}, E.~H., {J{\'o}hannesson}, G., Nov. 2004.
  {Energy Injection Episodes in Gamma-Ray Bursts: The Light Curves and
  Polarization Properties of GRB 021004}. \apjl 615, L77--L80.

\bibitem[{{Blake} et~al.(2005){Blake}, {Bloom}, {Starr}, {Falco}, {Skrutskie},
  {Fenimore}, {Duch{\^e}ne}, {Szentgyorgyi}, {Hornstein}, {Prochaska},
  {McCabe}, {Ghez}, {Konopacky}, {Stapelfeldt}, {Hurley}, {Campbell}, {Kassis},
  {Chaffee}, {Gehrels}, {Barthelmy}, {Cummings}, {Hullinger}, {Krimm},
  {Markwardt}, {Palmer}, {Parsons}, {McLean}, and {Tueller}}]{blake05}
{Blake}, C.~H., {Bloom}, J.~S., {Starr}, D.~L., {Falco}, E.~E., {Skrutskie},
  M., {Fenimore}, E.~E., {Duch{\^e}ne}, G., {Szentgyorgyi}, A., {Hornstein},
  S., {Prochaska}, J.~X., {McCabe}, C., {Ghez}, A., {Konopacky}, Q.,
  {Stapelfeldt}, K., {Hurley}, K., {Campbell}, R., {Kassis}, M., {Chaffee}, F.,
  {Gehrels}, N., {Barthelmy}, S., {Cummings}, J.~R., {Hullinger}, D., {Krimm},
  H.~A., {Markwardt}, C.~B., {Palmer}, D., {Parsons}, A., {McLean}, K.,
  {Tueller}, J., May 2005. {An infrared flash contemporaneous with the
  {$\gamma$}-rays of GRB 041219a}. \nat 435, 181--184.

\bibitem[{{Blandford} and {McKee}(1976)}]{blandford76}
{Blandford}, R.~D., {McKee}, C.~F., Aug. 1976. {Fluid dynamics of relativistic
  blast waves}. Physics of Fluids 19, 1130--1138.

\bibitem[{{Blandford} and {Znajek}(1977)}]{blandford77}
{Blandford}, R.~D., {Znajek}, R.~L., May 1977. {Electromagnetic extraction of
  energy from Kerr black holes}. \mnras 179, 433--456.

\bibitem[{{Bloom} et~al.(2003){Bloom}, {Frail}, and {Kulkarni}}]{bloom03}
{Bloom}, J.~S., {Frail}, D.~A., {Kulkarni}, S.~R., Sep. 2003. {Gamma-Ray Burst
  Energetics and the Gamma-Ray Burst Hubble Diagram: Promises and Limitations}.
  \apj 594, 674--683.

\bibitem[{{Bloom} et~al.(2002{\natexlab{a}}){Bloom}, {Kulkarni}, and
  {Djorgovski}}]{bloom02b}
{Bloom}, J.~S., {Kulkarni}, S.~R., {Djorgovski}, S.~G., Mar.
  2002{\natexlab{a}}. {The Observed Offset Distribution of Gamma-Ray Bursts
  from Their Host Galaxies: A Robust Clue to the Nature of the Progenitors}.
  \aj 123, 1111--1148.

\bibitem[{{Bloom} et~al.(1999){Bloom}, {Kulkarni}, {Djorgovski},
  {Eichelberger}, {C{\^o}t{\'e}}, {Blakeslee}, {Odewahn}, {Harrison}, {Frail},
  {Filippenko}, {Leonard}, {Riess}, {Spinrad}, {Stern}, {Bunker}, {Dey},
  {Grossan}, {Perlmutter}, {Knop}, {Hook}, and {Feroci}}]{bloom99}
{Bloom}, J.~S., {Kulkarni}, S.~R., {Djorgovski}, S.~G., {Eichelberger}, A.~C.,
  {C{\^o}t{\'e}}, P., {Blakeslee}, J.~P., {Odewahn}, S.~C., {Harrison}, F.~A.,
  {Frail}, D.~A., {Filippenko}, A.~V., {Leonard}, D.~C., {Riess}, A.~G.,
  {Spinrad}, H., {Stern}, D., {Bunker}, A., {Dey}, A., {Grossan}, B.,
  {Perlmutter}, S., {Knop}, R.~A., {Hook}, I.~M., {Feroci}, M., Sep. 1999. {The
  unusual afterglow of the {$\gamma$}-ray burst of 26 March 1998 as evidence
  for a supernova connection}. \nat 401, 453--456.

\bibitem[{{Bloom} et~al.(2002{\natexlab{b}}){Bloom}, {Kulkarni}, {Price},
  {Reichart}, {Galama}, {Schmidt}, {Frail}, {Berger}, {McCarthy}, {Chevalier},
  {Wheeler}, {Halpern}, {Fox}, {Djorgovski}, {Harrison}, {Sari}, {Axelrod},
  {Kimble}, {Holtzman}, {Hurley}, {Frontera}, {Piro}, and {Costa}}]{bloom02}
{Bloom}, J.~S., {Kulkarni}, S.~R., {Price}, P.~A., {Reichart}, D., {Galama},
  T.~J., {Schmidt}, B.~P., {Frail}, D.~A., {Berger}, E., {McCarthy}, P.~J.,
  {Chevalier}, R.~A., {Wheeler}, J.~C., {Halpern}, J.~P., {Fox}, D.~W.,
  {Djorgovski}, S.~G., {Harrison}, F.~A., {Sari}, R., {Axelrod}, T.~S.,
  {Kimble}, R.~A., {Holtzman}, J., {Hurley}, K., {Frontera}, F., {Piro}, L.,
  {Costa}, E., Jun. 2002{\natexlab{b}}. {Detection of a Supernova Signature
  Associated with GRB 011121}. \apjl 572, L45--L49.

\bibitem[{{Bloom} et~al.(2006){Bloom}, {Prochaska}, {Pooley}, {Blake}, {Foley},
  {Jha}, {Ramirez-Ruiz}, {Granot}, {Filippenko}, {Sigurdsson}, {Barth}, {Chen},
  {Cooper}, {Falco}, {Gal}, {Gerke}, {Gladders}, {Greene}, {Hennanwi}, {Ho},
  {Hurley}, {Koester}, {Li}, {Lubin}, {Newman}, {Perley}, {Squires}, and
  {Wood-Vasey}}]{bloom06}
{Bloom}, J.~S., {Prochaska}, J.~X., {Pooley}, D., {Blake}, C.~H., {Foley},
  R.~J., {Jha}, S., {Ramirez-Ruiz}, E., {Granot}, J., {Filippenko}, A.~V.,
  {Sigurdsson}, S., {Barth}, A.~J., {Chen}, H.-W., {Cooper}, M.~C., {Falco},
  E.~E., {Gal}, R.~R., {Gerke}, B.~F., {Gladders}, M.~D., {Greene}, J.~E.,
  {Hennanwi}, J., {Ho}, L.~C., {Hurley}, K., {Koester}, B.~P., {Li}, W.,
  {Lubin}, L., {Newman}, J., {Perley}, D.~A., {Squires}, G.~K., {Wood-Vasey},
  W.~M., Feb. 2006. {Closing in on a Short-Hard Burst Progenitor: Constraints
  from Early-Time Optical Imaging and Spectroscopy of a Possible Host Galaxy of
  GRB 050509b}. \apj 638, 354--368.

\bibitem[{{Bosnjak} et~al.(2013){Bosnjak}, {Gotz}, {Bouchet}, {Schanne}, and
  {Cordier}}]{bosnjak13}
{Bosnjak}, Z., {Gotz}, D., {Bouchet}, L., {Schanne}, S., {Cordier}, B., Sep.
  2013. {The spectral catalogue of INTEGRAL gamma-ray bursts: results of the
  joint IBIS/SPI spectral analysis}. ArXiv e-prints.

\bibitem[{{B\"ottcher} and {Dermer}(1998)}]{boettcher98}
{B\"ottcher}, M., {Dermer}, C.~D., Jun. 1998. {High-energy Gamma Rays from
  Ultra-high-energy Cosmic-Ray Protons in Gamma-Ray Bursts}. \apjl 499, L131+.

\bibitem[{{B{\"o}ttcher} and {Dermer}(2000)}]{bottcher00}
{B{\"o}ttcher}, M., {Dermer}, C.~D., Mar. 2000. {Early Gamma-Ray Burst
  Afterglows from Relativistic Blast Waves in General Radiative Regimes}. \apj
  532, 281--285.

\bibitem[{{Bo{\v s}njak} et~al.(2009){Bo{\v s}njak}, {Daigne}, and
  {Dubus}}]{bosnjak09}
{Bo{\v s}njak}, {\v Z}., {Daigne}, F., {Dubus}, G., May 2009. {Prompt
  high-energy emission from gamma-ray bursts in the internal shock model}. \aap
  498, 677--703.

\bibitem[{{Bo{\v s}njak} and {Kumar}(2012)}]{bosnjak12}
{Bo{\v s}njak}, {\v Z}., {Kumar}, P., Mar. 2012. {Magnetic jet model for GRBs
  and the delayed arrival of $>$100 MeV photons}. \mnras 421, L39--L43.

\bibitem[{{Briggs} et~al.(1999){Briggs}, {Band}, {Kippen}, {Preece},
  {Kouveliotou}, {van Paradijs}, {Share}, {Murphy}, {Matz}, {Connors},
  {Winkler}, {McConnell}, {Ryan}, {Williams}, {Young}, {Dingus}, {Catelli}, and
  {Wijers}}]{briggs99}
{Briggs}, M.~S., {Band}, D.~L., {Kippen}, R.~M., {Preece}, R.~D.,
  {Kouveliotou}, C., {van Paradijs}, J., {Share}, G.~H., {Murphy}, R.~J.,
  {Matz}, S.~M., {Connors}, A., {Winkler}, C., {McConnell}, M.~L., {Ryan},
  J.~M., {Williams}, O.~R., {Young}, C.~A., {Dingus}, B., {Catelli}, J.~R.,
  {Wijers}, R.~A.~M.~J., Oct. 1999. {Observations of GRB 990123 by the Compton
  Gamma Ray Observatory}. \apj 524, 82--91.

\bibitem[{{Broderick}(2005)}]{broderick05}
{Broderick}, A.~E., Aug. 2005. {Supernovae in helium star-compact object
  binaries: a possible {$\gamma$}-ray burst mechanism}. \mnras 361, 955--964.

\bibitem[{{Bromberg} et~al.(2014){Bromberg}, {Granot}, {Lyubarsky}, and
  {Piran}}]{bromberg14}
{Bromberg}, O., {Granot}, J., {Lyubarsky}, Y., {Piran}, T., Feb. 2014. {The
  dynamics of a highly magnetized jet propagating inside a star}. ArXiv
  e-prints.

\bibitem[{{Bromberg} et~al.(2011{\natexlab{a}}){Bromberg}, {Nakar}, and
  {Piran}}]{bromberg11}
{Bromberg}, O., {Nakar}, E., {Piran}, T., Oct. 2011{\natexlab{a}}. {Are
  Low-luminosity Gamma-Ray Bursts Generated by Relativistic Jets?} \apjl 739,
  L55.

\bibitem[{{Bromberg} et~al.(2011{\natexlab{b}}){Bromberg}, {Nakar}, {Piran},
  and {Sari}}]{bromberg11b}
{Bromberg}, O., {Nakar}, E., {Piran}, T., {Sari}, R., Oct. 2011{\natexlab{b}}.
  {The Propagation of Relativistic Jets in External Media}. \apj 740, 100.

\bibitem[{{Bromberg} et~al.(2012){Bromberg}, {Nakar}, {Piran}, and
  {Sari}}]{bromberg12}
{Bromberg}, O., {Nakar}, E., {Piran}, T., {Sari}, R., Apr. 2012. {An
  Observational Imprint of the Collapsar Model of Long Gamma-Ray Bursts}. \apj
  749, 110.

\bibitem[{{Bromberg} et~al.(2013){Bromberg}, {Nakar}, {Piran}, and
  {Sari}}]{bromberg13}
{Bromberg}, O., {Nakar}, E., {Piran}, T., {Sari}, R., Feb. 2013. {Short versus
  Long and Collapsars versus Non-collapsars: A Quantitative Classification of
  Gamma-Ray Bursts}. \apj 764, 179.

\bibitem[{{Bromm} et~al.(1999){Bromm}, {Coppi}, and {Larson}}]{bromm99b}
{Bromm}, V., {Coppi}, P.~S., {Larson}, R.~B., Dec. 1999. {Forming the First
  Stars in the Universe: The Fragmentation of Primordial Gas}. \apjl 527,
  L5--L8.

\bibitem[{{Bromm} and {Larson}(2004)}]{bromm04}
{Bromm}, V., {Larson}, R.~B., Sep. 2004. {The First Stars}. \araa 42, 79--118.

\bibitem[{{Bromm} and {Loeb}(2002)}]{bromm02}
{Bromm}, V., {Loeb}, A., Aug. 2002. {The Expected Redshift Distribution of
  Gamma-Ray Bursts}. \apj 575, 111--116.

\bibitem[{{Bromm} and {Loeb}(2006)}]{bromm06}
{Bromm}, V., {Loeb}, A., May 2006. {High-Redshift Gamma-Ray Bursts from
  Population III Progenitors}. \apj 642, 382--388.

\bibitem[{{Bucciantini} et~al.(2008){Bucciantini}, {Quataert}, {Arons},
  {Metzger}, and {Thompson}}]{bucciantini08}
{Bucciantini}, N., {Quataert}, E., {Arons}, J., {Metzger}, B.~D., {Thompson},
  T.~A., Jan. 2008. {Relativistic jets and long-duration gamma-ray bursts from
  the birth of magnetars}. \mnras 383, L25--L29.

\bibitem[{{Bucciantini} et~al.(2009){Bucciantini}, {Quataert}, {Metzger},
  {Thompson}, {Arons}, and {Del Zanna}}]{bucciantini09}
{Bucciantini}, N., {Quataert}, E., {Metzger}, B.~D., {Thompson}, T.~A.,
  {Arons}, J., {Del Zanna}, L., Jul. 2009. {Magnetized relativistic jets and
  long-duration GRBs from magnetar spin-down during core-collapse supernovae}.
  \mnras 396, 2038--2050.

\bibitem[{{Burgess} et~al.(2014){Burgess}, {Preece}, {Ryde}, {Veres},
  {M{\'e}sz{\'a}ros}, {Connaughton}, {Briggs}, {Pe'er}, {Iyyani}, {Goldstein},
  {Axelsson}, {Baring}, {Bhat}, {Byrne}, {Fitzpatrick}, {Foley}, {Kocevski},
  {Omodei}, {Paciesas}, {Pelassa}, {Kouveliotou}, {Xiong}, {Yu}, {Zhang}, and
  {Zhu}}]{burgess14}
{Burgess}, J.~M., {Preece}, R.~D., {Ryde}, F., {Veres}, P., {M{\'e}sz{\'a}ros},
  P., {Connaughton}, V., {Briggs}, M., {Pe'er}, A., {Iyyani}, S., {Goldstein},
  A., {Axelsson}, M., {Baring}, M.~G., {Bhat}, P.~N., {Byrne}, D.,
  {Fitzpatrick}, G., {Foley}, S., {Kocevski}, D., {Omodei}, N., {Paciesas},
  W.~S., {Pelassa}, V., {Kouveliotou}, C., {Xiong}, S., {Yu}, H.-F., {Zhang},
  B., {Zhu}, S., Apr. 2014. {An Observed Correlation between Thermal and
  Non-thermal Emission in Gamma-Ray Bursts}. \apjl 784, L43.

\bibitem[{{Burlon} et~al.(2009){Burlon}, {Ghirlanda}, {Ghisellini}, {Greiner},
  and {Celotti}}]{burlon09}
{Burlon}, D., {Ghirlanda}, G., {Ghisellini}, G., {Greiner}, J., {Celotti}, A.,
  Oct. 2009. {Time resolved spectral behavior of bright BATSE precursors}. \aap
  505, 569--575.

\bibitem[{{Burlon} et~al.(2008){Burlon}, {Ghirlanda}, {Ghisellini}, {Lazzati},
  {Nava}, {Nardini}, and {Celotti}}]{burlon08}
{Burlon}, D., {Ghirlanda}, G., {Ghisellini}, G., {Lazzati}, D., {Nava}, L.,
  {Nardini}, M., {Celotti}, A., Sep. 2008. {Precursors in Swift Gamma Ray
  Bursts with Redshift}. \apjl 685, L19--L22.

\bibitem[{{Burrows} et~al.(2005{\natexlab{a}}){Burrows}, {Hill}, {Nousek},
  {Kennea}, {Wells}, {Osborne}, {Abbey}, {Beardmore}, {Mukerjee}, {Short},
  {Chincarini}, {Campana}, {Citterio}, {Moretti}, {Pagani}, {Tagliaferri},
  {Giommi}, {Capalbi}, {Tamburelli}, {Angelini}, {Cusumano}, {Br{\"a}uninger},
  {Burkert}, and {Hartner}}]{burrows05b}
{Burrows}, D.~N., {Hill}, J.~E., {Nousek}, J.~A., {Kennea}, J.~A., {Wells}, A.,
  {Osborne}, J.~P., {Abbey}, A.~F., {Beardmore}, A., {Mukerjee}, K., {Short},
  A.~D.~T., {Chincarini}, G., {Campana}, S., {Citterio}, O., {Moretti}, A.,
  {Pagani}, C., {Tagliaferri}, G., {Giommi}, P., {Capalbi}, M., {Tamburelli},
  F., {Angelini}, L., {Cusumano}, G., {Br{\"a}uninger}, H.~W., {Burkert}, W.,
  {Hartner}, G.~D., Oct. 2005{\natexlab{a}}. {The Swift X-Ray Telescope}. Space
  Science Reviews 120, 165--195.

\bibitem[{{Burrows} et~al.(2005{\natexlab{b}}){Burrows}, {Romano}, {Falcone},
  {Kobayashi}, {Zhang}, {Moretti}, {O'Brien}, {Goad}, {Campana}, {Page},
  {Angelini}, {Barthelmy}, {Beardmore}, {Capalbi}, {Chincarini}, {Cummings},
  {Cusumano}, {Fox}, {Giommi}, {Hill}, {Kennea}, {Krimm}, {Mangano},
  {Marshall}, {M{\'e}sz{\'a}ros}, {Morris}, {Nousek}, {Osborne}, {Pagani},
  {Perri}, {Tagliaferri}, {Wells}, {Woosley}, and {Gehrels}}]{burrows05}
{Burrows}, D.~N., {Romano}, P., {Falcone}, A., {Kobayashi}, S., {Zhang}, B.,
  {Moretti}, A., {O'Brien}, P.~T., {Goad}, M.~R., {Campana}, S., {Page}, K.~L.,
  {Angelini}, L., {Barthelmy}, S., {Beardmore}, A.~P., {Capalbi}, M.,
  {Chincarini}, G., {Cummings}, J., {Cusumano}, G., {Fox}, D., {Giommi}, P.,
  {Hill}, J.~E., {Kennea}, J.~A., {Krimm}, H., {Mangano}, V., {Marshall}, F.,
  {M{\'e}sz{\'a}ros}, P., {Morris}, D.~C., {Nousek}, J.~A., {Osborne}, J.~P.,
  {Pagani}, C., {Perri}, M., {Tagliaferri}, G., {Wells}, A.~A., {Woosley}, S.,
  {Gehrels}, N., Sep. 2005{\natexlab{b}}. {Bright X-ray Flares in Gamma-Ray
  Burst Afterglows}. Science 309, 1833--1835.

\bibitem[{{Butler} et~al.(2007){Butler}, {Kocevski}, {Bloom}, and
  {Curtis}}]{butler07}
{Butler}, N.~R., {Kocevski}, D., {Bloom}, J.~S., {Curtis}, J.~L., Dec. 2007. {A
  Complete Catalog of Swift Gamma-Ray Burst Spectra and Durations: Demise of a
  Physical Origin for Pre-Swift High-Energy Correlations}. \apj 671, 656--677.

\bibitem[{{Campana} et~al.(2006){Campana}, {Mangano}, {Blustin}, {Brown},
  {Burrows}, {Chincarini}, {Cummings}, {Cusumano}, {Della Valle}, {Malesani},
  {M{\'e}sz{\'a}ros}, {Nousek}, {Page}, {Sakamoto}, {Waxman}, {Zhang}, {Dai},
  {Gehrels}, {Immler}, {Marshall}, {Mason}, {Moretti}, {O'Brien}, {Osborne},
  {Page}, {Romano}, {Roming}, {Tagliaferri}, {Cominsky}, {Giommi}, {Godet},
  {Kennea}, {Krimm}, {Angelini}, {Barthelmy}, {Boyd}, {Palmer}, {Wells}, and
  {White}}]{campana06}
{Campana}, S., {Mangano}, V., {Blustin}, A.~J., {Brown}, P., {Burrows}, D.~N.,
  {Chincarini}, G., {Cummings}, J.~R., {Cusumano}, G., {Della Valle}, M.,
  {Malesani}, D., {M{\'e}sz{\'a}ros}, P., {Nousek}, J.~A., {Page}, M.,
  {Sakamoto}, T., {Waxman}, E., {Zhang}, B., {Dai}, Z.~G., {Gehrels}, N.,
  {Immler}, S., {Marshall}, F.~E., {Mason}, K.~O., {Moretti}, A., {O'Brien},
  P.~T., {Osborne}, J.~P., {Page}, K.~L., {Romano}, P., {Roming}, P.~W.~A.,
  {Tagliaferri}, G., {Cominsky}, L.~R., {Giommi}, P., {Godet}, O., {Kennea},
  J.~A., {Krimm}, H., {Angelini}, L., {Barthelmy}, S.~D., {Boyd}, P.~T.,
  {Palmer}, D.~M., {Wells}, A.~A., {White}, N.~E., Aug. 2006. {The association
  of GRB 060218 with a supernova and the evolution of the shock wave}. \nat
  442, 1008--1010.

\bibitem[{{Campisi} et~al.(2010){Campisi}, {Li}, and {Jakobsson}}]{campisi10}
{Campisi}, M.~A., {Li}, L.-X., {Jakobsson}, P., Sep. 2010. {Redshift
  distribution and luminosity function of long gamma-ray bursts from
  cosmological simulations}. \mnras 407, 1972--1980.

\bibitem[{{Cannizzo} and {Gehrels}(2009)}]{cannizzo09}
{Cannizzo}, J.~K., {Gehrels}, N., Aug. 2009. {A New Paradigm for Gamma-ray
  Bursts: Long-term Accretion Rate Modulation by an External Accretion Disk}.
  \apj 700, 1047--1058.

\bibitem[{{Cannizzo} et~al.(2004){Cannizzo}, {Gehrels}, and
  {Vishniac}}]{cannizzo04}
{Cannizzo}, J.~K., {Gehrels}, N., {Vishniac}, E.~T., Jan. 2004. {A Numerical
  Gamma-Ray Burst Simulation Using Three-Dimensional Relativistic
  Hydrodynamics: The Transition from Spherical to Jetlike Expansion}. \apj 601,
  380--390.

\bibitem[{{Cao} et~al.(2014){Cao}, {Liang}, and {Yuan}}]{cao14}
{Cao}, X., {Liang}, E.-W., {Yuan}, Y.-F., Jul. 2014. {Episodic Jet Power
  Extracted from a Spinning Black Hole Surrounded by a Neutrino-dominated
  Accretion Flow in Gamma-Ray Bursts}. \apj 789, 129.

\bibitem[{{Castro Cer{\'o}n} et~al.(2006){Castro Cer{\'o}n}, {Micha{\l}owski},
  {Hjorth}, {Watson}, {Fynbo}, and {Gorosabel}}]{castroceron06}
{Castro Cer{\'o}n}, J.~M., {Micha{\l}owski}, M.~J., {Hjorth}, J., {Watson}, D.,
  {Fynbo}, J.~P.~U., {Gorosabel}, J., Dec. 2006. {Star Formation Rates and
  Stellar Masses in z \~{} 1 Gamma-Ray Burst Hosts}. \apjl 653, L85--L88.

\bibitem[{{Cavallo} and {Rees}(1978)}]{cavallo78}
{Cavallo}, G., {Rees}, M.~J., May 1978. {A qualitative study of cosmic
  fireballs and gamma-ray bursts}. \mnras 183, 359--365.

\bibitem[{{Cenko} et~al.(2008){Cenko}, {Fox}, {Penprase}, {Cucchiara}, {Price},
  {Berger}, {Kulkarni}, {Harrison}, {Gal-Yam}, {Ofek}, {Rau}, {Chandra},
  {Frail}, {Kasliwal}, {Schmidt}, {Soderberg}, {Cameron}, and {Roth}}]{cenko08}
{Cenko}, S.~B., {Fox}, D.~B., {Penprase}, B.~E., {Cucchiara}, A., {Price},
  P.~A., {Berger}, E., {Kulkarni}, S.~R., {Harrison}, F.~A., {Gal-Yam}, A.,
  {Ofek}, E.~O., {Rau}, A., {Chandra}, P., {Frail}, D.~A., {Kasliwal}, M.~M.,
  {Schmidt}, B.~P., {Soderberg}, A.~M., {Cameron}, P.~B., {Roth}, K.~C., Apr.
  2008. {GRB 070125: The First Long-Duration Gamma-Ray Burst in a Halo
  Environment}. \apj 677, 441--447.

\bibitem[{{Cenko} et~al.(2010){Cenko}, {Frail}, {Harrison}, {Kulkarni},
  {Nakar}, {Chandra}, {Butler}, {Fox}, {Gal-Yam}, {Kasliwal}, {Kelemen},
  {Moon}, {Ofek}, {Price}, {Rau}, {Soderberg}, {Teplitz}, {Werner}, {Bock},
  {Bloom}, {Starr}, {Filippenko}, {Chevalier}, {Gehrels}, {Nousek}, and
  {Piran}}]{cenko10}
{Cenko}, S.~B., {Frail}, D.~A., {Harrison}, F.~A., {Kulkarni}, S.~R., {Nakar},
  E., {Chandra}, P.~C., {Butler}, N.~R., {Fox}, D.~B., {Gal-Yam}, A.,
  {Kasliwal}, M.~M., {Kelemen}, J., {Moon}, D.-S., {Ofek}, E.~O., {Price},
  P.~A., {Rau}, A., {Soderberg}, A.~M., {Teplitz}, H.~I., {Werner}, M.~W.,
  {Bock}, D.~C.-J., {Bloom}, J.~S., {Starr}, D.~A., {Filippenko}, A.~V.,
  {Chevalier}, R.~A., {Gehrels}, N., {Nousek}, J.~N., {Piran}, T., Mar. 2010.
  {The Collimation and Energetics of the Brightest Swift Gamma-ray Bursts}.
  \apj 711, 641--654.

\bibitem[{{Cenko} et~al.(2013){Cenko}, {Kulkarni}, {Horesh}, {Corsi}, {Fox},
  {Carpenter}, {Frail}, {Nugent}, {Perley}, {Gruber}, {Gal-Yam}, {Groot},
  {Hallinan}, {Ofek}, {Rau}, {MacLeod}, {Miller}, {Bloom}, {Filippenko},
  {Kasliwal}, {Law}, {Morgan}, {Polishook}, {Poznanski}, {Quimby}, {Sesar},
  {Shen}, {Silverman}, and {Sternberg}}]{cenko13}
{Cenko}, S.~B., {Kulkarni}, S.~R., {Horesh}, A., {Corsi}, A., {Fox}, D.~B.,
  {Carpenter}, J., {Frail}, D.~A., {Nugent}, P.~E., {Perley}, D.~A., {Gruber},
  D., {Gal-Yam}, A., {Groot}, P.~J., {Hallinan}, G., {Ofek}, E.~O., {Rau}, A.,
  {MacLeod}, C.~L., {Miller}, A.~A., {Bloom}, J.~S., {Filippenko}, A.~V.,
  {Kasliwal}, M.~M., {Law}, N.~M., {Morgan}, A.~N., {Polishook}, D.,
  {Poznanski}, D., {Quimby}, R.~M., {Sesar}, B., {Shen}, K.~J., {Silverman},
  J.~M., {Sternberg}, A., Jun. 2013. {Discovery of a Cosmological, Relativistic
  Outburst via its Rapidly Fading Optical Emission}. \apj 769, 130.

\bibitem[{{Cerutti} et~al.(2012){Cerutti}, {Werner}, {Uzdensky}, and
  {Begelman}}]{cerutti12}
{Cerutti}, B., {Werner}, G.~R., {Uzdensky}, D.~A., {Begelman}, M.~C., Aug.
  2012. {Beaming and Rapid Variability of High-energy Radiation from
  Relativistic Pair Plasma Reconnection}. \apjl 754, L33.

\bibitem[{{Cerutti} et~al.(2014){Cerutti}, {Werner}, {Uzdensky}, and
  {Begelman}}]{cerutti14}
{Cerutti}, B., {Werner}, G.~R., {Uzdensky}, D.~A., {Begelman}, M.~C., Feb.
  2014. {Three-dimensional Relativistic Pair Plasma Reconnection with Radiative
  Feedback in the Crab Nebula}. \apj 782, 104.

\bibitem[{{Chandra} et~al.(2010){Chandra}, {Frail}, {Fox}, {Kulkarni},
  {Berger}, {Cenko}, {Bock}, {Harrison}, and {Kasliwal}}]{chandra10}
{Chandra}, P., {Frail}, D.~A., {Fox}, D., {Kulkarni}, S., {Berger}, E.,
  {Cenko}, S.~B., {Bock}, D.~C.-J., {Harrison}, F., {Kasliwal}, M., Mar. 2010.
  {Discovery of Radio Afterglow from the Most Distant Cosmic Explosion}. \apjl
  712, L31--L35.

\bibitem[{{Chang} et~al.(2008){Chang}, {Spitkovsky}, and {Arons}}]{chang08}
{Chang}, P., {Spitkovsky}, A., {Arons}, J., Feb. 2008. {Long-Term Evolution of
  Magnetic Turbulence in Relativistic Collisionless Shocks: Electron-Positron
  Plasmas}. \apj 674, 378--387.

\bibitem[{{Chen} and {Labun}(2013)}]{chen13}
{Chen}, P., {Labun}, L., Oct. 2013. {Electromagnetic signal of the QCD phase
  transition in neutron star mergers}. \prd 88~(8), 083006.

\bibitem[{{Chen} and {Beloborodov}(2007)}]{chen07}
{Chen}, W.-X., {Beloborodov}, A.~M., Mar. 2007. {Neutrino-cooled Accretion
  Disks around Spinning Black Holes}. \apj 657, 383--399.

\bibitem[{{Cheng} and {Dai}(1996)}]{chengdai96}
{Cheng}, K.~S., {Dai}, Z.~G., Aug. 1996. {Conversion of Neutron Stars to
  Strange Stars as a Possible Origin of {$\gamma$}-Ray Bursts}. Physical Review
  Letters 77, 1210--1213.

\bibitem[{{Chevalier} and {Li}(1999)}]{chevalier99}
{Chevalier}, R.~A., {Li}, Z.-Y., Jul. 1999. {Gamma-Ray Burst Environments and
  Progenitors}. \apjl 520, L29--L32.

\bibitem[{{Chevalier} and {Li}(2000)}]{chevalier00}
{Chevalier}, R.~A., {Li}, Z.-Y., Jun. 2000. {Wind Interaction Models for
  Gamma-Ray Burst Afterglows: The Case for Two Types of Progenitors}. \apj 536,
  195--212.

\bibitem[{{Chincarini} et~al.(2010){Chincarini}, {Mao}, {Margutti},
  {Bernardini}, {Guidorzi}, {Pasotti}, {Giannios}, {Della Valle}, {Moretti},
  {Romano}, {D'Avanzo}, {Cusumano}, and {Giommi}}]{chincarini10}
{Chincarini}, G., {Mao}, J., {Margutti}, R., {Bernardini}, M.~G., {Guidorzi},
  C., {Pasotti}, F., {Giannios}, D., {Della Valle}, M., {Moretti}, A.,
  {Romano}, P., {D'Avanzo}, P., {Cusumano}, G., {Giommi}, P., Aug. 2010.
  {Unveiling the origin of X-ray flares in gamma-ray bursts}. \mnras 406,
  2113--2148.

\bibitem[{{Chincarini} et~al.(2005){Chincarini}, {Moretti}, {Romano}, {Covino},
  {Tagliaferri}, {Campana}, {Goad}, {Kobayashi}, {Zhang}, {Angelini}, {Banat},
  {Barthelmy}, {Beardmore}, {Boyd}, {Breeveld}, {Burrows}, {Capalbi},
  {Chester}, {Cusumano}, {Fenimore}, {Gehrels}, {Giommi}, {Hill}, {Hinshaw},
  {Holland}, {Kennea}, {Krimm}, {La Parola}, {Mangano}, {Marshall}, {Mason},
  {Nousek}, {O'Brien}, {Osborne}, {Perri}, {M\'esz\'aros}, {Roming},
  {Sakamoto}, {Schady}, {Still}, and {Wells}}]{chincarini05}
{Chincarini}, G., {Moretti}, A., {Romano}, P., {Covino}, S., {Tagliaferri}, G.,
  {Campana}, S., {Goad}, M., {Kobayashi}, S., {Zhang}, B., {Angelini}, L.,
  {Banat}, P., {Barthelmy}, S., {Beardmore}, A.~P., {Boyd}, P.~T., {Breeveld},
  A., {Burrows}, D.~N., {Capalbi}, M., {Chester}, M.~M., {Cusumano}, G.,
  {Fenimore}, E.~E., {Gehrels}, N., {Giommi}, P., {Hill}, J.~E., {Hinshaw}, D.,
  {Holland}, S.~T., {Kennea}, J.~A., {Krimm}, H.~A., {La Parola}, V.,
  {Mangano}, V., {Marshall}, F.~E., {Mason}, K.~O., {Nousek}, J.~A., {O'Brien},
  P.~T., {Osborne}, J.~P., {Perri}, M., {M\'esz\'aros}, P., {Roming}, P.~W.~A.,
  {Sakamoto}, T., {Schady}, P., {Still}, M., {Wells}, A.~A., Jun. 2005. {Prompt
  and afterglow early X-ray phases in the comoving frame. Evidence for
  Universal properties?} ArXiv Astrophysics e-prints.

\bibitem[{{Chincarini} et~al.(2007){Chincarini}, {Moretti}, {Romano},
  {Falcone}, {Morris}, {Racusin}, {Campana}, {Covino}, {Guidorzi},
  {Tagliaferri}, {Burrows}, {Pagani}, {Stroh}, {Grupe}, {Capalbi}, {Cusumano},
  {Gehrels}, {Giommi}, {La Parola}, {Mangano}, {Mineo}, {Nousek}, {O'Brien},
  {Page}, {Perri}, {Troja}, {Willingale}, and {Zhang}}]{chincarini07}
{Chincarini}, G., {Moretti}, A., {Romano}, P., {Falcone}, A.~D., {Morris}, D.,
  {Racusin}, J., {Campana}, S., {Covino}, S., {Guidorzi}, C., {Tagliaferri},
  G., {Burrows}, D.~N., {Pagani}, C., {Stroh}, M., {Grupe}, D., {Capalbi}, M.,
  {Cusumano}, G., {Gehrels}, N., {Giommi}, P., {La Parola}, V., {Mangano}, V.,
  {Mineo}, T., {Nousek}, J.~A., {O'Brien}, P.~T., {Page}, K.~L., {Perri}, M.,
  {Troja}, E., {Willingale}, R., {Zhang}, B., Dec. 2007. {The First Survey of
  X-Ray Flares from Gamma-Ray Bursts Observed by Swift: Temporal Properties and
  Morphology}. \apj 671, 1903--1920.

\bibitem[{{Chornock} et~al.(2010){Chornock}, {Berger}, {Levesque}, {Soderberg},
  {Foley}, {Fox}, {Frebel}, {Simon}, {Bochanski}, {Challis}, {Kirshner},
  {Podsiadlowski}, {Roth}, {Rutledge}, {Schmidt}, {Sheppard}, and
  {Simcoe}}]{chornock10}
{Chornock}, R., {Berger}, E., {Levesque}, E.~M., {Soderberg}, A.~M., {Foley},
  R.~J., {Fox}, D.~B., {Frebel}, A., {Simon}, J.~D., {Bochanski}, J.~J.,
  {Challis}, P.~J., {Kirshner}, R.~P., {Podsiadlowski}, P., {Roth}, K.,
  {Rutledge}, R.~E., {Schmidt}, B.~P., {Sheppard}, S.~S., {Simcoe}, R.~A., Apr.
  2010. {Spectroscopic Discovery of the Broad-Lined Type Ic Supernova 2010bh
  Associated with the Low-Redshift GRB 100316D}. ArXiv e-prints.

\bibitem[{{Christensen} et~al.(2004){Christensen}, {Hjorth}, and
  {Gorosabel}}]{christensen04}
{Christensen}, L., {Hjorth}, J., {Gorosabel}, J., Oct. 2004. {UV star-formation
  rates of GRB host galaxies}. \aap 425, 913--926.

\bibitem[{{Ciardi} et~al.(2013){Ciardi}, {Labropoulos}, {Maselli}, {Thomas},
  {Zaroubi}, {Graziani}, {Bolton}, {Bernardi}, {Brentjens}, {de Bruyn},
  {Daiboo}, {Harker}, {Jelic}, {Kazemi}, {Koopmans}, {Martinez}, {Mellema},
  {Offringa}, {Pandey}, {Schaye}, {Veligatla}, {Vedantham}, and
  {Yatawatta}}]{ciardi13}
{Ciardi}, B., {Labropoulos}, P., {Maselli}, A., {Thomas}, R., {Zaroubi}, S.,
  {Graziani}, L., {Bolton}, J.~S., {Bernardi}, G., {Brentjens}, M., {de Bruyn},
  A.~G., {Daiboo}, S., {Harker}, G.~J.~A., {Jelic}, V., {Kazemi}, S.,
  {Koopmans}, L.~V.~E., {Martinez}, O., {Mellema}, G., {Offringa}, A.~R.,
  {Pandey}, V.~N., {Schaye}, J., {Veligatla}, V., {Vedantham}, H., {Yatawatta},
  S., Jan. 2013. {Prospects for detecting the 21 cm forest from the diffuse
  intergalactic medium with LOFAR}. \mnras 428, 1755--1765.

\bibitem[{{Ciardi} and {Loeb}(2000)}]{ciardi00}
{Ciardi}, B., {Loeb}, A., Sep. 2000. {Expected Number and Flux Distribution of
  Gamma-Ray Burst Afterglows with High Redshifts}. \apj 540, 687--696.

\bibitem[{{Coburn} and {Boggs}(2003)}]{coburn03}
{Coburn}, W., {Boggs}, S.~E., May 2003. {Polarization of the prompt
  {$\gamma$}-ray emission from the {$\gamma$}-ray burst of 6 December 2002}.
  \nat 423, 415--417.

\bibitem[{{Cohen} and {Piran}(1999)}]{cohen99}
{Cohen}, E., {Piran}, T., Jun. 1999. {Radiative Efficiencies of Continuously
  Powered Blast Waves}. \apj 518, 346--355.

\bibitem[{{Contopoulos}(1995)}]{contopoulos95}
{Contopoulos}, J., Sep. 1995. {A Simple Type of Magnetically Driven Jets: an
  Astrophysical Plasma Gun}. \apj 450, 616.

\bibitem[{{Corsi} et~al.(2010){Corsi}, {Guetta}, and {Piro}}]{corsi10}
{Corsi}, A., {Guetta}, D., {Piro}, L., Dec. 2010. {GeV emission from short
  gamma-ray bursts: the case of GRB 081024B}. \aap 524, A92.

\bibitem[{{Corsi} and {M{\'e}sz{\'a}ros}(2009)}]{corsi09}
{Corsi}, A., {M{\'e}sz{\'a}ros}, P., Sep. 2009. {Gamma-ray Burst Afterglow
  Plateaus and Gravitational Waves: Multi-messenger Signature of a Millisecond
  Magnetar?} \apj 702, 1171--1178.

\bibitem[{{Costa} et~al.(1997){Costa}, {Frontera}, {Heise}, {Feroci}, {in't
  Zand}, {Fiore}, {Cinti}, {Dal Fiume}, {Nicastro}, {Orlandini}, {Palazzi},
  {Rapisarda\#}, {Zavattini}, {Jager}, {Parmar}, {Owens}, {Molendi},
  {Cusumano}, {Maccarone}, {Giarrusso}, {Coletta}, {Antonelli}, {Giommi},
  {Muller}, {Piro}, and {Butler}}]{costa97}
{Costa}, E., {Frontera}, F., {Heise}, J., {Feroci}, M., {in't Zand}, J.,
  {Fiore}, F., {Cinti}, M.~N., {Dal Fiume}, D., {Nicastro}, L., {Orlandini},
  M., {Palazzi}, E., {Rapisarda\#}, M., {Zavattini}, G., {Jager}, R., {Parmar},
  A., {Owens}, A., {Molendi}, S., {Cusumano}, G., {Maccarone}, M.~C.,
  {Giarrusso}, S., {Coletta}, A., {Antonelli}, L.~A., {Giommi}, P., {Muller},
  J.~M., {Piro}, L., {Butler}, R.~C., Jun. 1997. {Discovery of an X-ray
  afterglow associated with the {$\gamma$}-ray burst of 28 February 1997}. \nat
  387, 783--785.

\bibitem[{{Couch} et~al.(2008){Couch}, {Milosavljevi{\'c}}, and
  {Nakar}}]{couch08}
{Couch}, S.~M., {Milosavljevi{\'c}}, M., {Nakar}, E., Nov. 2008. {Shock
  Vorticity Generation from Accelerated Ion Streaming in the Precursor of
  Ultrarelativistic Gamma-Ray Burst External Shocks}. \apj 688, 462--469.

\bibitem[{{Crumley} and {Kumar}(2013)}]{crumley13}
{Crumley}, P., {Kumar}, P., Mar. 2013. {Hadronic models for Large Area
  Telescope prompt emission observed in Fermi gamma-ray bursts}. \mnras 429,
  3238--3251.

\bibitem[{{Cucchiara} et~al.(2011{\natexlab{a}}){Cucchiara}, {Cenko}, {Bloom},
  {Melandri}, {Morgan}, {Kobayashi}, {Smith}, {Perley}, {Li}, {Hora}, {da
  Silva}, {Prochaska}, {Milne}, {Butler}, {Cobb}, {Worseck}, {Mundell},
  {Steele}, {Filippenko}, {Fumagalli}, {Klein}, {Stephens}, {Bluck}, and
  {Mason}}]{cucchiara11b}
{Cucchiara}, A., {Cenko}, S.~B., {Bloom}, J.~S., {Melandri}, A., {Morgan}, A.,
  {Kobayashi}, S., {Smith}, R.~J., {Perley}, D.~A., {Li}, W., {Hora}, J.~L.,
  {da Silva}, R.~L., {Prochaska}, J.~X., {Milne}, P.~A., {Butler}, N.~R.,
  {Cobb}, B., {Worseck}, G., {Mundell}, C.~G., {Steele}, I.~A., {Filippenko},
  A.~V., {Fumagalli}, M., {Klein}, C.~R., {Stephens}, A., {Bluck}, A., {Mason},
  R., Dec. 2011{\natexlab{a}}. {Constraining Gamma-Ray Burst Emission Physics
  with Extensive Early-time, Multiband Follow-up}. \apj 743, 154.

\bibitem[{{Cucchiara} et~al.(2011{\natexlab{b}}){Cucchiara}, {Levan}, {Fox},
  {Tanvir}, {Ukwatta}, {Berger}, {Kr{\"u}hler}, {K{\"u}pc{\"u} Yolda{\c s}},
  {Wu}, {Toma}, {Greiner}, {Olivares}, {Rowlinson}, {Amati}, {Sakamoto},
  {Roth}, {Stephens}, {Fritz}, {Fynbo}, {Hjorth}, {Malesani}, {Jakobsson},
  {Wiersema}, {O'Brien}, {Soderberg}, {Foley}, {Fruchter}, {Rhoads},
  {Rutledge}, {Schmidt}, {Dopita}, {Podsiadlowski}, {Willingale}, {Wolf},
  {Kulkarni}, and {D'Avanzo}}]{cucchiara11}
{Cucchiara}, A., {Levan}, A.~J., {Fox}, D.~B., {Tanvir}, N.~R., {Ukwatta},
  T.~N., {Berger}, E., {Kr{\"u}hler}, T., {K{\"u}pc{\"u} Yolda{\c s}}, A.,
  {Wu}, X.~F., {Toma}, K., {Greiner}, J., {Olivares}, F.~E., {Rowlinson}, A.,
  {Amati}, L., {Sakamoto}, T., {Roth}, K., {Stephens}, A., {Fritz}, A.,
  {Fynbo}, J.~P.~U., {Hjorth}, J., {Malesani}, D., {Jakobsson}, P., {Wiersema},
  K., {O'Brien}, P.~T., {Soderberg}, A.~M., {Foley}, R.~J., {Fruchter}, A.~S.,
  {Rhoads}, J., {Rutledge}, R.~E., {Schmidt}, B.~P., {Dopita}, M.~A.,
  {Podsiadlowski}, P., {Willingale}, R., {Wolf}, C., {Kulkarni}, S.~R.,
  {D'Avanzo}, P., Jul. 2011{\natexlab{b}}. {A Photometric Redshift of z \~{}
  9.4 for GRB 090429B}. \apj 736, 7.

\bibitem[{{Cui} et~al.(2012){Cui}, {Nagataki}, {Aoi}, and {Xu}}]{cui12}
{Cui}, X.-H., {Nagataki}, S., {Aoi}, J., {Xu}, R.-X., Sep. 2012. {Origins of
  short gamma-ray bursts deduced from offsets in their host galaxies
  revisited}. Research in Astronomy and Astrophysics 12, 1255--1268.

\bibitem[{{Curran} et~al.(2010){Curran}, {Evans}, {de Pasquale}, {Page}, and
  {van der Horst}}]{curran10}
{Curran}, P.~A., {Evans}, P.~A., {de Pasquale}, M., {Page}, M.~J., {van der
  Horst}, A.~J., Jun. 2010. {On the Electron Energy Distribution Index of Swift
  Gamma-ray Burst Afterglows}. \apjl 716, L135--L139.

\bibitem[{{Curran} et~al.(2009){Curran}, {Starling}, {van der Horst}, and
  {Wijers}}]{curran09}
{Curran}, P.~A., {Starling}, R.~L.~C., {van der Horst}, A.~J., {Wijers},
  R.~A.~M.~J., May 2009. {Testing the blast wave model with Swift GRBs}. \mnras
  395, 580--592.

\bibitem[{{Curran} et~al.(2008){Curran}, {van der Horst}, and
  {Wijers}}]{curran08}
{Curran}, P.~A., {van der Horst}, A.~J., {Wijers}, R.~A.~M.~J., May 2008. {Are
  the missing X-ray breaks in gamma-ray burst afterglow light curves merely
  hidden?} \mnras 386, 859--863.

\bibitem[{{Curran} et~al.(2007){Curran}, {van der Horst}, {Wijers}, {Starling},
  {Castro-Tirado}, {Fynbo}, {Gorosabel}, {J{\"a}rvinen}, {Malesani}, {Rol},
  {Tanvir}, {Wiersema}, {Burleigh}, {Casewell}, {Dobbie}, {Guziy}, {Jakobsson},
  {Jel{\'{\i}}nek}, {Laursen}, {Levan}, {Mundell}, {N{\"a}r{\"a}nen}, and
  {Piranomonte}}]{curran07}
{Curran}, P.~A., {van der Horst}, A.~J., {Wijers}, R.~A.~M.~J., {Starling},
  R.~L.~C., {Castro-Tirado}, A.~J., {Fynbo}, J.~P.~U., {Gorosabel}, J.,
  {J{\"a}rvinen}, A.~S., {Malesani}, D., {Rol}, E., {Tanvir}, N.~R.,
  {Wiersema}, K., {Burleigh}, M.~R., {Casewell}, S.~L., {Dobbie}, P.~D.,
  {Guziy}, S., {Jakobsson}, P., {Jel{\'{\i}}nek}, M., {Laursen}, P., {Levan},
  A.~J., {Mundell}, C.~G., {N{\"a}r{\"a}nen}, J., {Piranomonte}, S., Oct. 2007.
  {GRB060206 and the quandary of achromatic breaks in afterglow light curves}.
  \mnras 381, L65--L69.

\bibitem[{{Dado} et~al.(2004){Dado}, {Dar}, and {De Rujula}}]{dado04}
{Dado}, S., {Dar}, A., {De Rujula}, A., Jun. 2004. {The Discovery of a
  Hyperluminal Source in the Radio Afterglow of GRB 030329}. ArXiv Astrophysics
  e-prints.

\bibitem[{{Dai} and {Zhang}(2005)}]{daizhang05}
{Dai}, X., {Zhang}, B., Mar. 2005. {A Global Test of a Quasi-universal
  Gamma-Ray Burst Jet Model through Monte Carlo Simulations}. \apj 621,
  875--883.

\bibitem[{{Dai}(2004)}]{dai04}
{Dai}, Z.~G., May 2004. {Relativistic Wind Bubbles and Afterglow Signatures}.
  \apj 606, 1000--1005.

\bibitem[{{Dai} and {Cheng}(2001)}]{daicheng01}
{Dai}, Z.~G., {Cheng}, K.~S., Sep. 2001. {Afterglow Emission from Highly
  Collimated Jets with Flat Electron Spectra: Application to the GRB 010222
  Case?} \apjl 558, L109--L112.

\bibitem[{{Dai} and {Gou}(2001)}]{daigou01}
{Dai}, Z.~G., {Gou}, L.~J., May 2001. {Gamma-Ray Burst Afterglows from
  Anisotropic Jets}. \apj 552, 72--80.

\bibitem[{{Dai} et~al.(2004){Dai}, {Liang}, and {Xu}}]{dai04b}
{Dai}, Z.~G., {Liang}, E.~W., {Xu}, D., Sep. 2004. {Constraining ${\Omega}_{M}$
  and Dark Energy with Gamma-Ray Bursts}. \apjl 612, L101--L104.

\bibitem[{{Dai} and {Lu}(1998{\natexlab{a}})}]{dailu98b}
{Dai}, Z.~G., {Lu}, T., May 1998{\natexlab{a}}. {Gamma-ray burst afterglows and
  evolution of postburst fireballs with energy injection from strongly magnetic
  millisecond pulsars}. \aap 333, L87--L90.

\bibitem[{{Dai} and {Lu}(1998{\natexlab{b}})}]{dailu98c}
{Dai}, Z.~G., {Lu}, T., Jul. 1998{\natexlab{b}}. {Gamma-ray burst afterglows:
  effects of radiative corrections and non-uniformity of the surrounding
  medium}. \mnras 298, 87--92.

\bibitem[{{Dai} and {Lu}(1999)}]{dailu99}
{Dai}, Z.~G., {Lu}, T., Jul. 1999. {The Afterglow of GRB 990123 and a Dense
  Medium}. \apjl 519, L155--L158.

\bibitem[{{Dai} and {Lu}(2001)}]{dailu01b}
{Dai}, Z.~G., {Lu}, T., Apr. 2001. {Neutrino Afterglows and Progenitors of
  Gamma-Ray Bursts}. \apj 551, 249--253.

\bibitem[{{Dai} and {Lu}(2002)}]{dailu02}
{Dai}, Z.~G., {Lu}, T., Dec. 2002. {Spectrum and Duration of Delayed MeV-GeV
  Emission of Gamma-Ray Bursts in Cosmic Background Radiation Fields}. \apj
  580, 1013--1016.

\bibitem[{{Dai} et~al.(2006){Dai}, {Wang}, {Wu}, and {Zhang}}]{dai06}
{Dai}, Z.~G., {Wang}, X.~Y., {Wu}, X.~F., {Zhang}, B., Feb. 2006. {X-ray Flares
  from Postmerger Millisecond Pulsars}. Science 311, 1127--1129.

\bibitem[{{Dai} and {Wu}(2003)}]{daiwu03}
{Dai}, Z.~G., {Wu}, X.~F., Jul. 2003. {GRB 030226 in a Density-Jump Medium}.
  \apjl 591, L21--L24.

\bibitem[{{Dai} et~al.(2002){Dai}, {Zhang}, {Gou}, {M{\'e}sz{\'a}ros}, and
  {Waxman}}]{dai02}
{Dai}, Z.~G., {Zhang}, B., {Gou}, L.~J., {M{\'e}sz{\'a}ros}, P., {Waxman}, E.,
  Nov. 2002. {GeV Emission from TeV Blazars and Intergalactic Magnetic Fields}.
  \apjl 580, L7--L10.

\bibitem[{{Daigne} et~al.(2011){Daigne}, {Bo{\v s}njak}, and
  {Dubus}}]{daigne11}
{Daigne}, F., {Bo{\v s}njak}, {\v Z}., {Dubus}, G., Feb. 2011. {Reconciling
  observed gamma-ray burst prompt spectra with synchrotron radiation?} \aap
  526, A110.

\bibitem[{{Daigne} and {Mochkovitch}(2002)}]{daigne02}
{Daigne}, F., {Mochkovitch}, R., Nov. 2002. {The expected thermal precursors of
  gamma-ray bursts in the internal shock model}. \mnras 336, 1271--1280.

\bibitem[{{Daigne} et~al.(2006){Daigne}, {Rossi}, and {Mochkovitch}}]{daigne06}
{Daigne}, F., {Rossi}, E.~M., {Mochkovitch}, R., Nov. 2006. {The redshift
  distribution of Swift gamma-ray bursts: evidence for evolution}. \mnras 372,
  1034--1042.

\bibitem[{{Dainotti} et~al.(2013){Dainotti}, {Petrosian}, {Singal}, and
  {Ostrowski}}]{dainotti13}
{Dainotti}, M.~G., {Petrosian}, V., {Singal}, J., {Ostrowski}, M., Jul. 2013.
  {Determination of the intrinsic Luminosity Time Correlation in the X-ray
  Afterglows of GRBs}. ArXiv e-prints.

\bibitem[{{Dall'Osso} et~al.(2011){Dall'Osso}, {Stratta}, {Guetta}, {Covino},
  {De Cesare}, and {Stella}}]{dallosso11}
{Dall'Osso}, S., {Stratta}, G., {Guetta}, D., {Covino}, S., {De Cesare}, G.,
  {Stella}, L., Feb. 2011. {Gamma-ray bursts afterglows with energy injection
  from a spinning down neutron star}. \aap 526, A121.

\bibitem[{{Dar} and {de R{\'u}jula}(2004)}]{dar04}
{Dar}, A., {de R{\'u}jula}, A., Dec. 2004. {Towards a complete theory of
  gamma-ray bursts}. \physrep 405, 203--278.

\bibitem[{{De Colle} et~al.(2012){De Colle}, {Ramirez-Ruiz}, {Granot}, and
  {Lopez-Camara}}]{decolle12}
{De Colle}, F., {Ramirez-Ruiz}, E., {Granot}, J., {Lopez-Camara}, D., May 2012.
  {Simulations of Gamma-Ray Burst Jets in a Stratified External Medium:
  Dynamics, Afterglow Light Curves, Jet Breaks, and Radio Calorimetry}. \apj
  751, 57.

\bibitem[{{de Pasquale} et~al.(2009){de Pasquale}, {Evans}, {Oates}, {Page},
  {Zane}, {Schady}, {Breeveld}, {Holland}, {Kuin}, {Still}, {Roming}, and
  {Ward}}]{depasquale09}
{de Pasquale}, M., {Evans}, P., {Oates}, S., {Page}, M., {Zane}, S., {Schady},
  P., {Breeveld}, A., {Holland}, S., {Kuin}, P., {Still}, M., {Roming}, P.,
  {Ward}, P., Jan. 2009. {Jet breaks at the end of the slow decline phase of
  Swift GRB light curves}. \mnras 392, 153--169.

\bibitem[{{De Pasquale} et~al.(2010){De Pasquale}, {Schady}, {Kuin}, {Page},
  {Curran}, {Zane}, {Oates}, {Holland}, {Breeveld}, and
  {Hoversten}}]{depasquale10}
{De Pasquale}, M., {Schady}, P., {Kuin}, N.~P.~M., {Page}, M.~J., {Curran},
  P.~A., {Zane}, S., {Oates}, S.~R., {Holland}, S.~T., {Breeveld}, A.~A.,
  {Hoversten}, E.~A. e.~a., Feb. 2010. {Swift and Fermi Observations of the
  Early Afterglow of the Short Gamma-Ray Burst 090510}. \apjl 709, L146--L151.

\bibitem[{{de Souza} et~al.(2011){de Souza}, {Yoshida}, and {Ioka}}]{desouza11}
{de Souza}, R.~S., {Yoshida}, N., {Ioka}, K., Sep. 2011. {Populations III.1 and
  III.2 gamma-ray bursts: constraints on the event rate for future radio and
  X-ray surveys}. \aap 533, A32.

\bibitem[{{Della Valle} et~al.(2006){Della Valle}, {Chincarini}, {Panagia},
  {Tagliaferri}, {Malesani}, {Testa}, {Fugazza}, {Campana}, {Covino},
  {Mangano}, {Antonelli}, {D'Avanzo}, {Hurley}, {Mirabel}, {Pellizza},
  {Piranomonte}, and {Stella}}]{dellavalle06}
{Della Valle}, M., {Chincarini}, G., {Panagia}, N., {Tagliaferri}, G.,
  {Malesani}, D., {Testa}, V., {Fugazza}, D., {Campana}, S., {Covino}, S.,
  {Mangano}, V., {Antonelli}, L.~A., {D'Avanzo}, P., {Hurley}, K., {Mirabel},
  I.~F., {Pellizza}, L.~J., {Piranomonte}, S., {Stella}, L., Dec. 2006. {An
  enigmatic long-lasting {$\gamma$}-ray burst not accompanied by a bright
  supernova}. \nat 444, 1050--1052.

\bibitem[{{Deng} and {Zhang}(2014)}]{deng14b}
{Deng}, W., {Zhang}, B., Apr. 2014. {Low Energy Spectral Index and E$_{p}$
  Evolution of Quasi-thermal Photosphere Emission of Gamma-Ray Bursts}. \apj
  785, 112.

\bibitem[{{Derishev} et~al.(1999){Derishev}, {Kocharovsky}, and
  {Kocharovsky}}]{derishev99}
{Derishev}, E.~V., {Kocharovsky}, V.~V., {Kocharovsky}, V.~V., Aug. 1999. {The
  Neutron Component in Fireballs of Gamma-Ray Bursts: Dynamics and Observable
  Imprints}. \apj 521, 640--649.

\bibitem[{{Derishev} et~al.(2001){Derishev}, {Kocharovsky}, and
  {Kocharovsky}}]{derishev01}
{Derishev}, E.~V., {Kocharovsky}, V.~V., {Kocharovsky}, V.~V., Jun. 2001.
  {Physical parameters and emission mechanism in gamma-ray bursts}. \aap 372,
  1071--1077.

\bibitem[{{Dermer}(2002)}]{dermer02}
{Dermer}, C.~D., Jul. 2002. {Neutrino, Neutron, and Cosmic-Ray Production in
  the External Shock Model of Gamma-Ray Bursts}. \apj 574, 65--87.

\bibitem[{{Dermer}(2004)}]{dermer04}
{Dermer}, C.~D., Oct. 2004. {Curvature Effects in Gamma-Ray Burst Colliding
  Shells}. \apj 614, 284--292.

\bibitem[{{Dermer}(2007)}]{dermer07}
{Dermer}, C.~D., Jul. 2007. {Rapid X-Ray Declines and Plateaus in Swift GRB
  Light Curves Explained by a Highly Radiative Blast Wave}. \apj 664, 384--396.

\bibitem[{{Dermer} et~al.(1999){Dermer}, {Chiang}, and
  {B{\"o}ttcher}}]{dermer99}
{Dermer}, C.~D., {Chiang}, J., {B{\"o}ttcher}, M., Mar. 1999. {Fireball Loading
  and the Blast-Wave Model of Gamma-Ray Bursts}. \apj 513, 656--668.

\bibitem[{{Dermer} et~al.(2000){Dermer}, {Chiang}, and {Mitman}}]{dermer00}
{Dermer}, C.~D., {Chiang}, J., {Mitman}, K.~E., Jul. 2000. {Beaming, Baryon
  Loading, and the Synchrotron Self-Compton Component in Gamma-Ray Bursts}.
  \apj 537, 785--795.

\bibitem[{{Dermer} and {Menon}(2009)}]{dermer09}
{Dermer}, C.~D., {Menon}, G., 2009. {High Energy Radiation from Black Holes:
  Gamma Rays, Cosmic Rays, and Neutrinos}. Princeton Univerisity Press, 2009.

\bibitem[{{Dermer} and {Mitman}(1999)}]{dermermitman99}
{Dermer}, C.~D., {Mitman}, K.~E., Mar. 1999. {Short-Timescale Variability in
  the External Shock Model of Gamma-Ray Bursts}. \apjl 513, L5--L8.

\bibitem[{{Dessart} et~al.(2008){Dessart}, {Ott}, {Burrows}, {Rosswog}, and
  {Livne}}]{dessart08}
{Dessart}, L., {Ott}, C., {Burrows}, A., {Rosswog}, S., {Livne}, E., Jun. 2008.
  {Neutrino signatures and the neutrino-driven wind in Binary Neutron Star
  Mergers}. ArXiv e-prints 806.

\bibitem[{{Di Matteo} et~al.(2002){Di Matteo}, {Perna}, and
  {Narayan}}]{dimatteo02}
{Di Matteo}, T., {Perna}, R., {Narayan}, R., Nov. 2002. {Neutrino Trapping and
  Accretion Models for Gamma-Ray Bursts}. \apj 579, 706--715.

\bibitem[{{Drake} et~al.(2006){Drake}, {Swisdak}, {Che}, and {Shay}}]{drake06}
{Drake}, J.~F., {Swisdak}, M., {Che}, H., {Shay}, M.~A., Oct. 2006. {Electron
  acceleration from contracting magnetic islands during reconnection}. \nat
  443, 553--556.

\bibitem[{{Drenkhahn}(2002)}]{drenkhahn02b}
{Drenkhahn}, G., May 2002. {Acceleration of GRB outflows by Poynting flux
  dissipation}. \aap 387, 714--724.

\bibitem[{{Drenkhahn} and {Spruit}(2002)}]{drenkhahn02}
{Drenkhahn}, G., {Spruit}, H.~C., Sep. 2002. {Efficient acceleration and
  radiation in Poynting flux powered GRB outflows}. \aap 391, 1141--1153.

\bibitem[{{Eichler} and {Levinson}(2000)}]{eichler00}
{Eichler}, D., {Levinson}, A., Jan. 2000. {A Compact Fireball Model of
  Gamma-Ray Bursts}. \apj 529, 146--150.

\bibitem[{{Eichler} et~al.(1989){Eichler}, {Livio}, {Piran}, and
  {Schramm}}]{eichler89}
{Eichler}, D., {Livio}, M., {Piran}, T., {Schramm}, D.~N., Jul. 1989.
  {Nucleosynthesis, neutrino bursts and gamma-rays from coalescing neutron
  stars}. \nat 340, 126--128.

\bibitem[{{Evans} et~al.(2009){Evans}, {Beardmore}, {Page}, {Osborne},
  {O'Brien}, {Willingale}, {Starling}, {Burrows}, {Godet}, {Vetere}, {Racusin},
  {Goad}, {Wiersema}, {Angelini}, {Capalbi}, {Chincarini}, {Gehrels}, {Kennea},
  {Margutti}, {Morris}, {Mountford}, {Pagani}, {Perri}, {Romano}, and
  {Tanvir}}]{evans09}
{Evans}, P.~A., {Beardmore}, A.~P., {Page}, K.~L., {Osborne}, J.~P., {O'Brien},
  P.~T., {Willingale}, R., {Starling}, R.~L.~C., {Burrows}, D.~N., {Godet}, O.,
  {Vetere}, L., {Racusin}, J., {Goad}, M.~R., {Wiersema}, K., {Angelini}, L.,
  {Capalbi}, M., {Chincarini}, G., {Gehrels}, N., {Kennea}, J.~A., {Margutti},
  R., {Morris}, D.~C., {Mountford}, C.~J., {Pagani}, C., {Perri}, M., {Romano},
  P., {Tanvir}, N., Aug. 2009. {Methods and results of an automatic analysis of
  a complete sample of Swift-XRT observations of GRBs}. \mnras 397, 1177--1201.

\bibitem[{{Evans} et~al.(2007){Evans}, {Beardmore}, {Page}, {Tyler}, {Osborne},
  {Goad}, {O'Brien}, {Vetere}, {Racusin}, {Morris}, {Burrows}, {Capalbi},
  {Perri}, {Gehrels}, and {Romano}}]{evans07}
{Evans}, P.~A., {Beardmore}, A.~P., {Page}, K.~L., {Tyler}, L.~G., {Osborne},
  J.~P., {Goad}, M.~R., {O'Brien}, P.~T., {Vetere}, L., {Racusin}, J.,
  {Morris}, D., {Burrows}, D.~N., {Capalbi}, M., {Perri}, M., {Gehrels}, N.,
  {Romano}, P., Jul. 2007. {An online repository of Swift/XRT light curves of
  {$\gamma$}-ray bursts}. \aap 469, 379--385.

\bibitem[{{Falcone} et~al.(2007){Falcone}, {Morris}, {Racusin}, {Chincarini},
  {Moretti}, {Romano}, {Burrows}, {Pagani}, {Stroh}, {Grupe}, {Campana},
  {Covino}, {Tagliaferri}, {Willingale}, and {Gehrels}}]{falcone07}
{Falcone}, A.~D., {Morris}, D., {Racusin}, J., {Chincarini}, G., {Moretti}, A.,
  {Romano}, P., {Burrows}, D.~N., {Pagani}, C., {Stroh}, M., {Grupe}, D.,
  {Campana}, S., {Covino}, S., {Tagliaferri}, G., {Willingale}, R., {Gehrels},
  N., Dec. 2007. {The First Survey of X-Ray Flares from Gamma-Ray Bursts
  Observed by Swift: Spectral Properties and Energetics}. \apj 671, 1921--1938.

\bibitem[{{Fan} and {Piran}(2006{\natexlab{a}})}]{fanpiran06a}
{Fan}, Y., {Piran}, T., Jun. 2006{\natexlab{a}}. {Gamma-ray burst efficiency
  and possible physical processes shaping the early afterglow}. \mnras 369,
  197--206.

\bibitem[{{Fan} and {Piran}(2006{\natexlab{b}})}]{fan06a}
{Fan}, Y., {Piran}, T., Jun. 2006{\natexlab{b}}. {Gamma-ray burst efficiency
  and possible physical processes shaping the early afterglow}. \mnras 369,
  197--206.

\bibitem[{{Fan} et~al.(2009){Fan}, {Zhang}, and {Wei}}]{fan09}
{Fan}, Y., {Zhang}, B., {Wei}, D., Jan. 2009. {Naked-eye optical flash from
  gamma-ray burst 080319B: Tracing the decaying neutrons in the outflow}. \prd
  79~(2), 021301--+.

\bibitem[{{Fan} et~al.(2002){Fan}, {Dai}, {Huang}, and {Lu}}]{fan02}
{Fan}, Y.-Z., {Dai}, Z.-G., {Huang}, Y.-F., {Lu}, T., Oct. 2002. {Optical Flash
  of GRB 990123: Constraints on the Physical Parameters of the Reverse Shock}.
  Chinese Journal of Astronomy and Astrophysics 2, 449--453.

\bibitem[{{Fan} and {Piran}(2008)}]{fanpiran08}
{Fan}, Y.-Z., {Piran}, T., Sep. 2008. {High-energy {$\gamma$}-ray emission from
  gamma-ray bursts before GLAST}. Frontiers of Physics in China 3, 306--330.

\bibitem[{{Fan} et~al.(2006){Fan}, {Piran}, and {Xu}}]{fan06}
{Fan}, Y.-Z., {Piran}, T., {Xu}, D., Sep. 2006. {The interpretation and
  implication of the afterglow of GRB 060218}. Journal of Cosmology and
  Astro-Particle Physics 9, 13--+.

\bibitem[{{Fan} et~al.(2013{\natexlab{a}}){Fan}, {Tam}, {Zhang}, {Liang}, {He},
  {Zhou}, {Yang}, {Jin}, and {Wei}}]{fan13a}
{Fan}, Y.-Z., {Tam}, P.~H.~T., {Zhang}, F.-W., {Liang}, Y.-F., {He}, H.-N.,
  {Zhou}, B., {Yang}, R.-Z., {Jin}, Z.-P., {Wei}, D.-M., Oct.
  2013{\natexlab{a}}. {High-energy Emission of GRB 130427A: Evidence for
  Inverse Compton Radiation}. \apj 776, 95.

\bibitem[{{Fan} and {Wei}(2005)}]{fanwei05}
{Fan}, Y.~Z., {Wei}, D.~M., Nov. 2005. {Late internal-shock model for bright
  X-ray flares in gamma-ray burst afterglows and GRB 011121}. \mnras 364,
  L42--L46.

\bibitem[{{Fan} et~al.(2004){Fan}, {Wei}, and {Zhang}}]{fan04}
{Fan}, Y.~Z., {Wei}, D.~M., {Zhang}, B., Nov. 2004. {{$\gamma$}-ray burst
  internal shocks with magnetization}. \mnras 354, 1031--1039.

\bibitem[{{Fan} et~al.(2012){Fan}, {Wei}, {Zhang}, and {Zhang}}]{fan12}
{Fan}, Y.-Z., {Wei}, D.-M., {Zhang}, F.-W., {Zhang}, B.-B., Aug. 2012. {The
  Photospheric Radiation Model for the Prompt Emission of Gamma-Ray Bursts:
  Interpreting Four Observed Correlations}. \apjl 755, L6.

\bibitem[{{Fan} et~al.(2013{\natexlab{b}}){Fan}, {Wu}, and {Wei}}]{fan13b}
{Fan}, Y.-Z., {Wu}, X.-F., {Wei}, D.-M., Sep. 2013{\natexlab{b}}. {Signature of
  gravitational wave radiation in afterglows of short gamma-ray bursts?} \prd
  88~(6), 067304.

\bibitem[{{Fan} and {Xu}(2006)}]{fanxu06}
{Fan}, Y.-Z., {Xu}, D., Oct. 2006. {The X-ray afterglow flat segment in short
  GRB 051221A: Energy injection from a millisecond magnetar?} \mnras 372,
  L19--L22.

\bibitem[{{Fan} et~al.(2005{\natexlab{a}}){Fan}, {Zhang}, and {Wei}}]{fan05b}
{Fan}, Y.~Z., {Zhang}, B., {Wei}, D.~M., Jul. 2005{\natexlab{a}}. {Early
  Optical Afterglow Light Curves of Neutron-fed Gamma-Ray Bursts}. \apj 628,
  298--314.

\bibitem[{{Fan} et~al.(2005{\natexlab{b}}){Fan}, {Zhang}, and {Wei}}]{fan05c}
{Fan}, Y.~Z., {Zhang}, B., {Wei}, D.~M., Aug. 2005{\natexlab{b}}. {Early
  Photon-Shock Interaction in a Stellar Wind: A Sub-GeV Photon Flash and
  High-Energy Neutrino Emission from Long Gamma-Ray Bursts}. \apj 629,
  334--340.

\bibitem[{{Fargion}(2012)}]{fargion12}
{Fargion}, D., 2012. {GRBs by thin persistent precessing lepton Jets: the long
  life GRB110328 and the Neutrino signal }. \memsai 83, 312.

\bibitem[{{Fenimore} et~al.(1993){Fenimore}, {Epstein}, {Ho}, {Klebesadel},
  {Lacey}, {Laros}, {Meier}, {Strohmayer}, {Pendleton}, {Fishman},
  {Kouveliotou}, and {Meegan}}]{fenimore93}
{Fenimore}, E.~E., {Epstein}, R.~I., {Ho}, C., {Klebesadel}, R.~W., {Lacey},
  C., {Laros}, J.~G., {Meier}, M., {Strohmayer}, T., {Pendleton}, G.,
  {Fishman}, G., {Kouveliotou}, C., {Meegan}, C., Nov. 1993. {The intrinsic
  luminosity of {$\gamma$}-ray bursts and their host galaxies}. \nat 366,
  40--42.

\bibitem[{{Fenimore} et~al.(1995){Fenimore}, {in 't Zand}, {Norris}, {Bonnell},
  and {Nemiroff}}]{fenimore95}
{Fenimore}, E.~E., {in 't Zand}, J.~J.~M., {Norris}, J.~P., {Bonnell}, J.~T.,
  {Nemiroff}, R.~J., Aug. 1995. {Gamma-Ray Burst Peak Duration as a Function of
  Energy}. \apjl 448, L101+.

\bibitem[{{Fenimore} et~al.(1996){Fenimore}, {Madras}, and
  {Nayakshin}}]{fenimore96}
{Fenimore}, E.~E., {Madras}, C.~D., {Nayakshin}, S., Dec. 1996. {Expanding
  Relativistic Shells and Gamma-Ray Burst Temporal Structure}. \apj 473,
  998--+.

\bibitem[{{Fenimore} and {Ramirez-Ruiz}(2000)}]{fenimore00}
{Fenimore}, E.~E., {Ramirez-Ruiz}, E., Apr. 2000. {Redshifts For 220 BATSE
  Gamma-Ray Bursts Determined by Variability and the Cosmological
  Consequences}. ArXiv Astrophysics e-prints.

\bibitem[{{Firmani} et~al.(2006){Firmani}, {Ghisellini}, {Avila-Reese}, and
  {Ghirlanda}}]{firmani06}
{Firmani}, C., {Ghisellini}, G., {Avila-Reese}, V., {Ghirlanda}, G., Jul. 2006.
  {Discovery of a tight correlation among the prompt emission properties of
  long gamma-ray bursts}. \mnras 370, 185--197.

\bibitem[{{Fishman} and {Meegan}(1995)}]{fishman95}
{Fishman}, G.~J., {Meegan}, C.~A., 1995. {Gamma-Ray Bursts}. \araa 33,
  415--458.

\bibitem[{{Flanagan} and {Hughes}(1998)}]{flanagan98}
{Flanagan}, {\'E}.~{\'E}., {Hughes}, S.~A., Apr. 1998. {Measuring gravitational
  waves from binary black hole coalescences. I. Signal to noise for inspiral,
  merger, and ringdown}. \prd 57, 4535--4565.

\bibitem[{{Fong} et~al.(2010){Fong}, {Berger}, and {Fox}}]{fong10}
{Fong}, W., {Berger}, E., {Fox}, D.~B., Jan. 2010. {Hubble Space Telescope
  Observations of Short Gamma-Ray Burst Host Galaxies: Morphologies, Offsets,
  and Local Environments}. \apj 708, 9--25.

\bibitem[{{Fox} et~al.(2005){Fox}, {Frail}, {Price}, {Kulkarni}, {Berger},
  {Piran}, {Soderberg}, {Cenko}, {Cameron}, {Gal-Yam}, {Kasliwal}, {Moon},
  {Harrison}, {Nakar}, {Schmidt}, {Penprase}, {Chevalier}, {Kumar}, {Roth},
  {Watson}, {Lee}, {Shectman}, {Phillips}, {Roth}, {McCarthy}, {Rauch},
  {Cowie}, {Peterson}, {Rich}, {Kawai}, {Aoki}, {Kosugi}, {Totani}, {Park},
  {MacFadyen}, and {Hurley}}]{fox05}
{Fox}, D.~B., {Frail}, D.~A., {Price}, P.~A., {Kulkarni}, S.~R., {Berger}, E.,
  {Piran}, T., {Soderberg}, A.~M., {Cenko}, S.~B., {Cameron}, P.~B., {Gal-Yam},
  A., {Kasliwal}, M.~M., {Moon}, D.-S., {Harrison}, F.~A., {Nakar}, E.,
  {Schmidt}, B.~P., {Penprase}, B., {Chevalier}, R.~A., {Kumar}, P., {Roth},
  K., {Watson}, D., {Lee}, B.~L., {Shectman}, S., {Phillips}, M.~M., {Roth},
  M., {McCarthy}, P.~J., {Rauch}, M., {Cowie}, L., {Peterson}, B.~A., {Rich},
  J., {Kawai}, N., {Aoki}, K., {Kosugi}, G., {Totani}, T., {Park}, H.-S.,
  {MacFadyen}, A., {Hurley}, K.~C., Oct. 2005. {The afterglow of GRB 050709 and
  the nature of the short-hard {$\gamma$}-ray bursts}. \nat 437, 845--850.

\bibitem[{{Fox} and {M{\'e}sz{\'a}ros}(2006)}]{foxmeszaros06}
{Fox}, D.~B., {M{\'e}sz{\'a}ros}, P., Sep. 2006. {GRB fireball physics: prompt
  and early emission}. New Journal of Physics 8, 199--+.

\bibitem[{{Fox} et~al.(2003){Fox}, {Price}, {Soderberg}, {Berger}, {Kulkarni},
  {Sari}, {Frail}, {Harrison}, {Yost}, {Matthews}, {Peterson}, {Tanaka},
  {Christiansen}, and {Moriarty-Schieven}}]{fox03b}
{Fox}, D.~W., {Price}, P.~A., {Soderberg}, A.~M., {Berger}, E., {Kulkarni},
  S.~R., {Sari}, R., {Frail}, D.~A., {Harrison}, F.~A., {Yost}, S.~A.,
  {Matthews}, K., {Peterson}, B.~A., {Tanaka}, I., {Christiansen}, J.,
  {Moriarty-Schieven}, G.~H., Mar. 2003. {Discovery of Early Optical Emission
  from GRB 021211}. \apjl 586, L5--L8.

\bibitem[{{Frail} et~al.(1997){Frail}, {Kulkarni}, {Nicastro}, {Feroci}, and
  {Taylor}}]{frail97}
{Frail}, D.~A., {Kulkarni}, S.~R., {Nicastro}, L., {Feroci}, M., {Taylor},
  G.~B., Sep. 1997. {The radio afterglow from the {$\gamma$}-ray burst of 8 May
  1997}. \nat 389, 261--263.

\bibitem[{{Frail} et~al.(2001){Frail}, {Kulkarni}, {Sari}, {Djorgovski},
  {Bloom}, {Galama}, {Reichart}, {Berger}, {Harrison}, {Price}, {Yost},
  {Diercks}, {Goodrich}, and {Chaffee}}]{frail01}
{Frail}, D.~A., {Kulkarni}, S.~R., {Sari}, R., {Djorgovski}, S.~G., {Bloom},
  J.~S., {Galama}, T.~J., {Reichart}, D.~E., {Berger}, E., {Harrison}, F.~A.,
  {Price}, P.~A., {Yost}, S.~A., {Diercks}, A., {Goodrich}, R.~W., {Chaffee},
  F., Nov. 2001. {Beaming in Gamma-Ray Bursts: Evidence for a Standard Energy
  Reservoir}. \apjl 562, L55--L58.

\bibitem[{{Frail} et~al.(2000){Frail}, {Waxman}, and {Kulkarni}}]{frail00}
{Frail}, D.~A., {Waxman}, E., {Kulkarni}, S.~R., Jul. 2000. {A 450 Day Light
  Curve of the Radio Afterglow of GRB 970508: Fireball Calorimetry}. \apj 537,
  191--204.

\bibitem[{{Freedman} and {Waxman}(2001)}]{freedman01}
{Freedman}, D.~L., {Waxman}, E., Feb. 2001. {On the Energy of Gamma-Ray
  Bursts}. \apj 547, 922--928.

\bibitem[{{Frontera} et~al.(2013){Frontera}, {Amati}, {Farinelli}, {Dichiara},
  {Guidorzi}, {Landi}, and {Titarchuk}}]{frontera13}
{Frontera}, F., {Amati}, L., {Farinelli}, R., {Dichiara}, S., {Guidorzi}, C.,
  {Landi}, R., {Titarchuk}, L., Nov. 2013. {Comptonization signatures in the
  prompt emission of Gamma Ray Bursts}. ArXiv e-prints.

\bibitem[{{Frontera} et~al.(2012){Frontera}, {Amati}, {Guidorzi}, {Landi}, and
  {in't Zand}}]{frontera12}
{Frontera}, F., {Amati}, L., {Guidorzi}, C., {Landi}, R., {in't Zand}, J., Aug.
  2012. {Broadband Time-resolved E $_{ p, i }$-L $_{iso}$ Correlation in
  Gamma-Ray Bursts}. \apj 754, 138.

\bibitem[{{Frontera} et~al.(1998){Frontera}, {Costa}, {Piro}, {Muller},
  {Amati}, {Feroci}, {Fiore}, {Pizzichini}, {Tavani}, {Castro-Tirado},
  {Cusumano}, {dal Fiume}, {Heise}, {Hurley}, {Nicastro}, {Orlandini}, {Owens},
  {Palazzi}, {Parmar}, {in 't Zand}, and {Zavattini}}]{frontera98}
{Frontera}, F., {Costa}, E., {Piro}, L., {Muller}, J.~M., {Amati}, L.,
  {Feroci}, M., {Fiore}, F., {Pizzichini}, G., {Tavani}, M., {Castro-Tirado},
  A., {Cusumano}, G., {dal Fiume}, D., {Heise}, J., {Hurley}, K., {Nicastro},
  L., {Orlandini}, M., {Owens}, A., {Palazzi}, E., {Parmar}, A.~N., {in 't
  Zand}, J., {Zavattini}, G., Feb. 1998. {Spectral Properties of the Prompt
  X-Ray Emission and Afterglow from the Gamma-Ray Burst of 1997 February 28}.
  \apjl 493, L67.

\bibitem[{{Fruchter} et~al.(2006){Fruchter}, {Levan}, {Strolger}, {Vreeswijk},
  {Thorsett}, {Bersier}, {Burud}, {Castro Cer{\'o}n}, {Castro-Tirado},
  {Conselice}, {Dahlen}, {Ferguson}, {Fynbo}, {Garnavich}, {Gibbons},
  {Gorosabel}, {Gull}, {Hjorth}, {Holland}, {Kouveliotou}, {Levay}, {Livio},
  {Metzger}, {Nugent}, {Petro}, {Pian}, {Rhoads}, {Riess}, {Sahu}, {Smette},
  {Tanvir}, {Wijers}, and {Woosley}}]{fruchter06}
{Fruchter}, A.~S., {Levan}, A.~J., {Strolger}, L., {Vreeswijk}, P.~M.,
  {Thorsett}, S.~E., {Bersier}, D., {Burud}, I., {Castro Cer{\'o}n}, J.~M.,
  {Castro-Tirado}, A.~J., {Conselice}, C., {Dahlen}, T., {Ferguson}, H.~C.,
  {Fynbo}, J.~P.~U., {Garnavich}, P.~M., {Gibbons}, R.~A., {Gorosabel}, J.,
  {Gull}, T.~R., {Hjorth}, J., {Holland}, S.~T., {Kouveliotou}, C., {Levay},
  Z., {Livio}, M., {Metzger}, M.~R., {Nugent}, P.~E., {Petro}, L., {Pian}, E.,
  {Rhoads}, J.~E., {Riess}, A.~G., {Sahu}, K.~C., {Smette}, A., {Tanvir},
  N.~R., {Wijers}, R.~A.~M.~J., {Woosley}, S.~E., May 2006. {Long
  {$\gamma$}-ray bursts and core-collapse supernovae have different
  environments}. \nat 441, 463--468.

\bibitem[{{Fryer} et~al.(1999){Fryer}, {Woosley}, and {Hartmann}}]{fryer99}
{Fryer}, C.~L., {Woosley}, S.~E., {Hartmann}, D.~H., Nov. 1999. {Formation
  Rates of Black Hole Accretion Disk Gamma-Ray Bursts}. \apj 526, 152--177.

\bibitem[{{Fynbo} et~al.(2003){Fynbo}, {Jakobsson}, {M{\"o}ller}, {Hjorth},
  {Thomsen}, {Andersen}, {Fruchter}, {Gorosabel}, {Holland}, {Ledoux},
  {Pedersen}, {Rhoads}, {Weidinger}, and {Wijers}}]{fynbo03}
{Fynbo}, J.~P.~U., {Jakobsson}, P., {M{\"o}ller}, P., {Hjorth}, J., {Thomsen},
  B., {Andersen}, M.~I., {Fruchter}, A.~S., {Gorosabel}, J., {Holland}, S.~T.,
  {Ledoux}, C., {Pedersen}, H., {Rhoads}, J., {Weidinger}, M., {Wijers},
  R.~A.~M.~J., Jul. 2003. {On the Lyalpha emission from gamma-ray burst host
  galaxies: Evidence for low metallicities}. \aap 406, L63--L66.

\bibitem[{{Fynbo} et~al.(2006){Fynbo}, {Watson}, {Th{\"o}ne}, {Sollerman},
  {Bloom}, {Davis}, {Hjorth}, {Jakobsson}, {J{\o}rgensen}, {Graham},
  {Fruchter}, {Bersier}, {Kewley}, {Cassan}, {Castro Cer{\'o}n}, {Foley},
  {Gorosabel}, {Hinse}, {Horne}, {Jensen}, {Klose}, {Kocevski}, {Marquette},
  {Perley}, {Ramirez-Ruiz}, {Stritzinger}, {Vreeswijk}, {Wijers}, {Woller},
  {Xu}, and {Zub}}]{fynbo06}
{Fynbo}, J.~P.~U., {Watson}, D., {Th{\"o}ne}, C.~C., {Sollerman}, J., {Bloom},
  J.~S., {Davis}, T.~M., {Hjorth}, J., {Jakobsson}, P., {J{\o}rgensen}, U.~G.,
  {Graham}, J.~F., {Fruchter}, A.~S., {Bersier}, D., {Kewley}, L., {Cassan},
  A., {Castro Cer{\'o}n}, J.~M., {Foley}, S., {Gorosabel}, J., {Hinse}, T.~C.,
  {Horne}, K.~D., {Jensen}, B.~L., {Klose}, S., {Kocevski}, D., {Marquette},
  J.-B., {Perley}, D., {Ramirez-Ruiz}, E., {Stritzinger}, M.~D., {Vreeswijk},
  P.~M., {Wijers}, R.~A.~M., {Woller}, K.~G., {Xu}, D., {Zub}, M., Dec. 2006.
  {No supernovae associated with two long-duration {$\gamma$}-ray bursts}. \nat
  444, 1047--1049.

\bibitem[{{Gal-Yam} et~al.(2006){Gal-Yam}, {Fox}, {Price}, {Ofek}, {Davis},
  {Leonard}, {Soderberg}, {Schmidt}, {Lewis}, {Peterson}, {Kulkarni}, {Berger},
  {Cenko}, {Sari}, {Sharon}, {Frail}, {Moon}, {Brown}, {Cucchiara}, {Harrison},
  {Piran}, {Persson}, {McCarthy}, {Penprase}, {Chevalier}, and
  {MacFadyen}}]{galyam06}
{Gal-Yam}, A., {Fox}, D.~B., {Price}, P.~A., {Ofek}, E.~O., {Davis}, M.~R.,
  {Leonard}, D.~C., {Soderberg}, A.~M., {Schmidt}, B.~P., {Lewis}, K.~M.,
  {Peterson}, B.~A., {Kulkarni}, S.~R., {Berger}, E., {Cenko}, S.~B., {Sari},
  R., {Sharon}, K., {Frail}, D., {Moon}, D.-S., {Brown}, P.~J., {Cucchiara},
  A., {Harrison}, F., {Piran}, T., {Persson}, S.~E., {McCarthy}, P.~J.,
  {Penprase}, B.~E., {Chevalier}, R.~A., {MacFadyen}, A.~I., Dec. 2006. {A
  novel explosive process is required for the {$\gamma$}-ray burst GRB 060614}.
  \nat 444, 1053--1055.

\bibitem[{{Galama} et~al.(1998){Galama}, {Vreeswijk}, {van Paradijs},
  {Kouveliotou}, {Augusteijn}, {B{\"o}hnhardt}, {Brewer}, {Doublier},
  {Gonzalez}, {Leibundgut}, {Lidman}, {Hainaut}, {Patat}, {Heise}, {in't Zand},
  {Hurley}, {Groot}, {Strom}, {Mazzali}, {Iwamoto}, {Nomoto}, {Umeda},
  {Nakamura}, {Young}, {Suzuki}, {Shigeyama}, {Koshut}, {Kippen}, {Robinson},
  {de Wildt}, {Wijers}, {Tanvir}, {Greiner}, {Pian}, {Palazzi}, {Frontera},
  {Masetti}, {Nicastro}, {Feroci}, {Costa}, {Piro}, {Peterson}, {Tinney},
  {Boyle}, {Cannon}, {Stathakis}, {Sadler}, {Begam}, and {Ianna}}]{galama98}
{Galama}, T.~J., {Vreeswijk}, P.~M., {van Paradijs}, J., {Kouveliotou}, C.,
  {Augusteijn}, T., {B{\"o}hnhardt}, H., {Brewer}, J.~P., {Doublier}, V.,
  {Gonzalez}, J.-F., {Leibundgut}, B., {Lidman}, C., {Hainaut}, O.~R., {Patat},
  F., {Heise}, J., {in't Zand}, J., {Hurley}, K., {Groot}, P.~J., {Strom},
  R.~G., {Mazzali}, P.~A., {Iwamoto}, K., {Nomoto}, K., {Umeda}, H.,
  {Nakamura}, T., {Young}, T.~R., {Suzuki}, T., {Shigeyama}, T., {Koshut}, T.,
  {Kippen}, M., {Robinson}, C., {de Wildt}, P., {Wijers}, R.~A.~M.~J.,
  {Tanvir}, N., {Greiner}, J., {Pian}, E., {Palazzi}, E., {Frontera}, F.,
  {Masetti}, N., {Nicastro}, L., {Feroci}, M., {Costa}, E., {Piro}, L.,
  {Peterson}, B.~A., {Tinney}, C., {Boyle}, B., {Cannon}, R., {Stathakis}, R.,
  {Sadler}, E., {Begam}, M.~C., {Ianna}, P., Oct. 1998. {An unusual supernova
  in the error box of the {$\gamma$}-ray burst of 25 April 1998}. \nat 395,
  670--672.

\bibitem[{{Gao} et~al.(2013{\natexlab{a}}){Gao}, {Ding}, {Wu}, {Zhang}, and
  {Dai}}]{gao13a}
{Gao}, H., {Ding}, X., {Wu}, X.-F., {Zhang}, B., {Dai}, Z.-G., Jul.
  2013{\natexlab{a}}. {Bright Broadband Afterglows of Gravitational Wave Bursts
  from Mergers of Binary Neutron Stars}. \apj 771, 86.

\bibitem[{{Gao} et~al.(2013{\natexlab{b}}){Gao}, {Lei}, {Zou}, {Wu}, and
  {Zhang}}]{gao13b}
{Gao}, H., {Lei}, W.-H., {Zou}, Y.-C., {Wu}, X.-F., {Zhang}, B., Dec.
  2013{\natexlab{b}}. {A complete reference of the analytical synchrotron
  external shock models of gamma-ray bursts}. \nar 57, 141--190.

\bibitem[{{Gao} et~al.(2012){Gao}, {Zhang}, and {Zhang}}]{gao12}
{Gao}, H., {Zhang}, B.-B., {Zhang}, B., Apr. 2012. {Stepwise Filter Correlation
  Method and Evidence of Superposed Variability Components in Gamma-Ray Burst
  Prompt Emission Light Curves}. \apj 748, 134.

\bibitem[{{Gao} et~al.(2013{\natexlab{c}}){Gao}, {Kashiyama}, and
  {M{\'e}sz{\'a}ros}}]{gaoshan13}
{Gao}, S., {Kashiyama}, K., {M{\'e}sz{\'a}ros}, P., Jul. 2013{\natexlab{c}}.
  {On the Neutrino Non-detection of GRB 130427A}. \apjl 772, L4.

\bibitem[{{Gao} et~al.(2009){Gao}, {Mao}, {Xu}, and {Fan}}]{gao09}
{Gao}, W., {Mao}, J., {Xu}, D., {Fan}, Y., Nov. 2009. {GRB 080916C and GRB
  090510: The High-Energy Emission and the Afterglow}. \apjl 706, L33--L36.

\bibitem[{{Gao}(2009)}]{gao09a}
{Gao}, W.-H., Jun. 2009. {Optical/Infrared Flares of GRB 080129 from Late
  Internal Shocks}. \apj 697, 1044--1047.

\bibitem[{{Gao} and {Fan}(2006)}]{gaofan06}
{Gao}, W.-H., {Fan}, Y.-Z., Oct. 2006. {Short-living Supermassive Magnetar
  Model for the Early X-ray Flares Following Short GRBs}. Chinese Journal of
  Astronomy and Astrophysics 6, 513--516.

\bibitem[{{Gehrels} et~al.(2008){Gehrels}, {Barthelmy}, {Burrows}, {Cannizzo},
  {Chincarini}, {Fenimore}, {Kouveliotou}, {O'Brien}, {Palmer}, {Racusin},
  {Roming}, {Sakamoto}, {Tueller}, {Wijers}, and {Zhang}}]{gehrels08}
{Gehrels}, N., {Barthelmy}, S.~D., {Burrows}, D.~N., {Cannizzo}, J.~K.,
  {Chincarini}, G., {Fenimore}, E., {Kouveliotou}, C., {O'Brien}, P., {Palmer},
  D.~M., {Racusin}, J., {Roming}, P.~W.~A., {Sakamoto}, T., {Tueller}, J.,
  {Wijers}, R.~A.~M.~J., {Zhang}, B., Dec. 2008. {Correlations of Prompt and
  Afterglow Emission in Swift Long and Short Gamma-Ray Bursts}. \apj 689,
  1161--1172.

\bibitem[{{Gehrels} et~al.(2004){Gehrels}, {Chincarini}, {Giommi}, {Mason},
  {Nousek}, {Wells}, {White}, {Barthelmy}, {Burrows}, {Cominsky}, {Hurley},
  {Marshall}, {M{\'e}sz{\'a}ros}, {Roming}, {Angelini}, {Barbier}, {Belloni},
  {Campana}, {Caraveo}, {Chester}, {Citterio}, {Cline}, {Cropper}, {Cummings},
  {Dean}, {Feigelson}, {Fenimore}, {Frail}, {Fruchter}, {Garmire}, {Gendreau},
  {Ghisellini}, {Greiner}, {Hill}, {Hunsberger}, {Krimm}, {Kulkarni}, {Kumar},
  {Lebrun}, {Lloyd-Ronning}, {Markwardt}, {Mattson}, {Mushotzky}, {Norris},
  {Osborne}, {Paczynski}, {Palmer}, {Park}, {Parsons}, {Paul}, {Rees},
  {Reynolds}, {Rhoads}, {Sasseen}, {Schaefer}, {Short}, {Smale}, {Smith},
  {Stella}, {Tagliaferri}, {Takahashi}, {Tashiro}, {Townsley}, {Tueller},
  {Turner}, {Vietri}, {Voges}, {Ward}, {Willingale}, {Zerbi}, and
  {Zhang}}]{gehrels04}
{Gehrels}, N., {Chincarini}, G., {Giommi}, P., {Mason}, K.~O., {Nousek}, J.~A.,
  {Wells}, A.~A., {White}, N.~E., {Barthelmy}, S.~D., {Burrows}, D.~N.,
  {Cominsky}, L.~R., {Hurley}, K.~C., {Marshall}, F.~E., {M{\'e}sz{\'a}ros},
  P., {Roming}, P.~W.~A., {Angelini}, L., {Barbier}, L.~M., {Belloni}, T.,
  {Campana}, S., {Caraveo}, P.~A., {Chester}, M.~M., {Citterio}, O., {Cline},
  T.~L., {Cropper}, M.~S., {Cummings}, J.~R., {Dean}, A.~J., {Feigelson},
  E.~D., {Fenimore}, E.~E., {Frail}, D.~A., {Fruchter}, A.~S., {Garmire},
  G.~P., {Gendreau}, K., {Ghisellini}, G., {Greiner}, J., {Hill}, J.~E.,
  {Hunsberger}, S.~D., {Krimm}, H.~A., {Kulkarni}, S.~R., {Kumar}, P.,
  {Lebrun}, F., {Lloyd-Ronning}, N.~M., {Markwardt}, C.~B., {Mattson}, B.~J.,
  {Mushotzky}, R.~F., {Norris}, J.~P., {Osborne}, J., {Paczynski}, B.,
  {Palmer}, D.~M., {Park}, H.-S., {Parsons}, A.~M., {Paul}, J., {Rees}, M.~J.,
  {Reynolds}, C.~S., {Rhoads}, J.~E., {Sasseen}, T.~P., {Schaefer}, B.~E.,
  {Short}, A.~T., {Smale}, A.~P., {Smith}, I.~A., {Stella}, L., {Tagliaferri},
  G., {Takahashi}, T., {Tashiro}, M., {Townsley}, L.~K., {Tueller}, J.,
  {Turner}, M.~J.~L., {Vietri}, M., {Voges}, W., {Ward}, M.~J., {Willingale},
  R., {Zerbi}, F.~M., {Zhang}, W.~W., Aug. 2004. {The Swift Gamma-Ray Burst
  Mission}. \apj 611, 1005--1020.

\bibitem[{{Gehrels} et~al.(2006){Gehrels}, {Norris}, {Barthelmy}, {Granot},
  {Kaneko}, {Kouveliotou}, {Markwardt}, {M{\'e}sz{\'a}ros}, {Nakar}, {Nousek},
  {O'Brien}, {Page}, {Palmer}, {Parsons}, {Roming}, {Sakamoto}, {Sarazin},
  {Schady}, {Stamatikos}, and {Woosley}}]{gehrels06}
{Gehrels}, N., {Norris}, J.~P., {Barthelmy}, S.~D., {Granot}, J., {Kaneko}, Y.,
  {Kouveliotou}, C., {Markwardt}, C.~B., {M{\'e}sz{\'a}ros}, P., {Nakar}, E.,
  {Nousek}, J.~A., {O'Brien}, P.~T., {Page}, M., {Palmer}, D.~M., {Parsons},
  A.~M., {Roming}, P.~W.~A., {Sakamoto}, T., {Sarazin}, C.~L., {Schady}, P.,
  {Stamatikos}, M., {Woosley}, S.~E., Dec. 2006. {A new {$\gamma$}-ray burst
  classification scheme from GRB060614}. \nat 444, 1044--1046.

\bibitem[{{Gehrels} et~al.(2009){Gehrels}, {Ramirez-Ruiz}, and
  {Fox}}]{gehrels09}
{Gehrels}, N., {Ramirez-Ruiz}, E., {Fox}, D.~B., Sep. 2009. {Gamma-Ray Bursts
  in the Swift Era}. \araa 47, 567--617.

\bibitem[{{Gehrels} et~al.(2005){Gehrels}, {Sarazin}, {O'Brien}, {Zhang},
  {Barbier}, {Barthelmy}, {Blustin}, {Burrows}, {Cannizzo}, {Cummings}, {Goad},
  {Holland}, {Hurkett}, {Kennea}, {Levan}, {Markwardt}, {Mason},
  {M\'esz\'aros}, {Page}, {Palmer}, {Rol}, {Sakamoto}, {Willingale},
  {Angelini}, {Beardmore}, {Boyd}, {Breeveld}, {Campana}, {Chester},
  {Chincarini}, {Cominsky}, {Cusumano}, {de Pasquale}, {Fenimore}, {Giommi},
  {Gronwall}, {Grupe}, {Hill}, {Hinshaw}, {Hjorth}, {Hullinger}, {Hurley},
  {Klose}, {Kobayashi}, {Kouveliotou}, {Krimm}, {Mangano}, {Marshall},
  {McGowan}, {Moretti}, {Mushotzky}, {Nakazawa}, {Norris}, {Nousek}, {Osborne},
  {Page}, {Parsons}, {Patel}, {Perri}, {Poole}, {Romano}, {Roming}, {Rosen},
  {Sato}, {Schady}, {Smale}, {Sollerman}, {Starling}, {Still}, {Suzuki},
  {Tagliaferri}, {Takahashi}, {Tashiro}, {Tueller}, {Wells}, {White}, and
  {Wijers}}]{gehrels05}
{Gehrels}, N., {Sarazin}, C.~L., {O'Brien}, P.~T., {Zhang}, B., {Barbier}, L.,
  {Barthelmy}, S.~D., {Blustin}, A., {Burrows}, D.~N., {Cannizzo}, J.,
  {Cummings}, J.~R., {Goad}, M., {Holland}, S.~T., {Hurkett}, C.~P., {Kennea},
  J.~A., {Levan}, A., {Markwardt}, C.~B., {Mason}, K.~O., {M\'esz\'aros}, P.,
  {Page}, M., {Palmer}, D.~M., {Rol}, E., {Sakamoto}, T., {Willingale}, R.,
  {Angelini}, L., {Beardmore}, A., {Boyd}, P.~T., {Breeveld}, A., {Campana},
  S., {Chester}, M.~M., {Chincarini}, G., {Cominsky}, L.~R., {Cusumano}, G.,
  {de Pasquale}, M., {Fenimore}, E.~E., {Giommi}, P., {Gronwall}, C., {Grupe},
  D., {Hill}, J.~E., {Hinshaw}, D., {Hjorth}, J., {Hullinger}, D., {Hurley},
  K.~C., {Klose}, S., {Kobayashi}, S., {Kouveliotou}, C., {Krimm}, H.~A.,
  {Mangano}, V., {Marshall}, F.~E., {McGowan}, K., {Moretti}, A., {Mushotzky},
  R.~F., {Nakazawa}, K., {Norris}, J.~P., {Nousek}, J.~A., {Osborne}, J.~P.,
  {Page}, K., {Parsons}, A.~M., {Patel}, S., {Perri}, M., {Poole}, T.,
  {Romano}, P., {Roming}, P.~W.~A., {Rosen}, S., {Sato}, G., {Schady}, P.,
  {Smale}, A.~P., {Sollerman}, J., {Starling}, R., {Still}, M., {Suzuki}, M.,
  {Tagliaferri}, G., {Takahashi}, T., {Tashiro}, M., {Tueller}, J., {Wells},
  A.~A., {White}, N.~E., {Wijers}, R.~A.~M.~J., Oct. 2005. {A short
  {$\gamma$}-ray burst apparently associated with an elliptical galaxy at
  redshift z = 0.225}. \nat 437, 851--854.

\bibitem[{{Gendre} et~al.(2012){Gendre}, {Atteia}, {Bo{\"e}r}, {Colas},
  {Klotz}, {Kugel}, {Laas-Bourez}, {Rinner}, {Strajnic}, {Stratta}, and
  {Vachier}}]{gendre12}
{Gendre}, B., {Atteia}, J.~L., {Bo{\"e}r}, M., {Colas}, F., {Klotz}, A.,
  {Kugel}, F., {Laas-Bourez}, M., {Rinner}, C., {Strajnic}, J., {Stratta}, G.,
  {Vachier}, F., Mar. 2012. {GRB 110205A: Anatomy of a Long Gamma-Ray Burst}.
  \apj 748, 59.

\bibitem[{{Gendre} et~al.(2013){Gendre}, {Stratta}, {Atteia}, {Basa},
  {Bo{\"e}r}, {Coward}, {Cutini}, {D'Elia}, {Howell}, {Klotz}, and
  {Piro}}]{gendre13}
{Gendre}, B., {Stratta}, G., {Atteia}, J.~L., {Basa}, S., {Bo{\"e}r}, M.,
  {Coward}, D.~M., {Cutini}, S., {D'Elia}, V., {Howell}, E.~J., {Klotz}, A.,
  {Piro}, L., Mar. 2013. {The Ultra-long Gamma-Ray Burst 111209A: The Collapse
  of a Blue Supergiant?} \apj 766, 30.

\bibitem[{{Genet} et~al.(2007){Genet}, {Daigne}, and {Mochkovitch}}]{genet07}
{Genet}, F., {Daigne}, F., {Mochkovitch}, R., Oct. 2007. {Can the early X-ray
  afterglow of gamma-ray bursts be explained by a contribution from the reverse
  shock?} \mnras 381, 732--740.

\bibitem[{{Genet} and {Granot}(2009)}]{genet09}
{Genet}, F., {Granot}, J., Nov. 2009. {Realistic analytic model for the prompt
  and high-latitude emission in GRBs}. \mnras 399, 1328--1346.

\bibitem[{{Geng} et~al.(2014){Geng}, {Wu}, {Li}, {Huang}, and {Dai}}]{geng14}
{Geng}, J.~J., {Wu}, X.~F., {Li}, L., {Huang}, Y.~F., {Dai}, Z.~G., Sep. 2014.
  {Revisiting the Emission from Relativistic Blast Waves in a Density-jump
  Medium}. \apj 792, 31.

\bibitem[{{Ghirlanda} et~al.(2007){Ghirlanda}, {Bosnjak}, {Ghisellini},
  {Tavecchio}, and {Firmani}}]{ghirlanda07}
{Ghirlanda}, G., {Bosnjak}, Z., {Ghisellini}, G., {Tavecchio}, F., {Firmani},
  C., Jul. 2007. {Blackbody components in gamma-ray bursts spectra?} \mnras
  379, 73--85.

\bibitem[{{Ghirlanda} et~al.(2006){Ghirlanda}, {Ghisellini}, {Firmani}, {Nava},
  {Tavecchio}, and {Lazzati}}]{ghirlanda06}
{Ghirlanda}, G., {Ghisellini}, G., {Firmani}, C., {Nava}, L., {Tavecchio}, F.,
  {Lazzati}, D., Jun. 2006. {Cosmological constraints with GRBs: homogeneous
  medium vs. wind density profile}. \aap 452, 839--844.

\bibitem[{{Ghirlanda} et~al.(2004{\natexlab{a}}){Ghirlanda}, {Ghisellini}, and
  {Lazzati}}]{ghirlanda04}
{Ghirlanda}, G., {Ghisellini}, G., {Lazzati}, D., Nov. 2004{\natexlab{a}}. {The
  Collimation-corrected Gamma-Ray Burst Energies Correlate with the Peak Energy
  of Their {$\nu F_\nu$} Spectrum}. \apj 616, 331--338.

\bibitem[{{Ghirlanda} et~al.(2004{\natexlab{b}}){Ghirlanda}, {Ghisellini},
  {Lazzati}, and {Firmani}}]{ghirlanda04b}
{Ghirlanda}, G., {Ghisellini}, G., {Lazzati}, D., {Firmani}, C., Sep.
  2004{\natexlab{b}}. {Gamma-Ray Bursts: New Rulers to Measure the Universe}.
  \apjl 613, L13--L16.

\bibitem[{{Ghirlanda} et~al.(2011){Ghirlanda}, {Ghisellini}, {Nava}, and
  {Burlon}}]{ghirlanda11}
{Ghirlanda}, G., {Ghisellini}, G., {Nava}, L., {Burlon}, D., Jan. 2011.
  {Spectral evolution of Fermi/GBM short gamma-ray bursts}. \mnras 410,
  L47--L51.

\bibitem[{{Ghirlanda} et~al.(2009){Ghirlanda}, {Nava}, {Ghisellini}, {Celotti},
  and {Firmani}}]{ghirlanda09}
{Ghirlanda}, G., {Nava}, L., {Ghisellini}, G., {Celotti}, A., {Firmani}, C.,
  Mar. 2009. {Short versus long gamma-ray bursts: spectra, energetics, and
  luminosities}. \aap 496, 585--595.

\bibitem[{{Ghirlanda} et~al.(2008){Ghirlanda}, {Nava}, {Ghisellini}, {Firmani},
  and {Cabrera}}]{ghirlanda08}
{Ghirlanda}, G., {Nava}, L., {Ghisellini}, G., {Firmani}, C., {Cabrera}, J.~I.,
  Jun. 2008. {The E$_{peak}$-E$_{iso}$ plane of long gamma-ray bursts and
  selection effects}. \mnras 387, 319--330.

\bibitem[{{Ghisellini}(2012)}]{ghisellini12}
{Ghisellini}, G., Feb. 2012. {Radiative Processes in High Energy Astrophysics}.
  ArXiv e-prints.

\bibitem[{{Ghisellini} and {Celotti}(1999)}]{ghisellini99}
{Ghisellini}, G., {Celotti}, A., Feb. 1999. {Quasi-thermal Comptonization and
  Gamma-Ray Bursts}. \apjl 511, L93--L96.

\bibitem[{{Ghisellini} et~al.(2000){Ghisellini}, {Celotti}, and
  {Lazzati}}]{ghisellini00}
{Ghisellini}, G., {Celotti}, A., {Lazzati}, D., Mar. 2000. {Constraints on the
  emission mechanisms of gamma-ray bursts}. \mnras 313, L1--L5.

\bibitem[{{Ghisellini} et~al.(2010){Ghisellini}, {Ghirlanda}, {Nava}, and
  {Celotti}}]{ghisellini10}
{Ghisellini}, G., {Ghirlanda}, G., {Nava}, L., {Celotti}, A., Apr. 2010. {GeV
  emission from gamma-ray bursts: a radiative fireball?} \mnras 403, 926--937.

\bibitem[{{Ghisellini} et~al.(2007){Ghisellini}, {Ghirlanda}, {Nava}, and
  {Firmani}}]{ghisellini07}
{Ghisellini}, G., {Ghirlanda}, G., {Nava}, L., {Firmani}, C., Apr. 2007.
  {``Late Prompt'' Emission in Gamma-Ray Bursts?} \apjl 658, L75--L78.

\bibitem[{{Ghisellini} et~al.(2009){Ghisellini}, {Nardini}, {Ghirlanda}, and
  {Celotti}}]{ghisellini09}
{Ghisellini}, G., {Nardini}, M., {Ghirlanda}, G., {Celotti}, A., Feb. 2009. {A
  unifying view of gamma-ray burst afterglows}. \mnras 393, 253--271.

\bibitem[{{Giacomazzo} and {Perna}(2013)}]{giacomazzo13}
{Giacomazzo}, B., {Perna}, R., Jul. 2013. {Formation of Stable Magnetars from
  Binary Neutron Star Mergers}. \apjl 771, L26.

\bibitem[{{Giacomazzo} et~al.(2011){Giacomazzo}, {Rezzolla}, and
  {Baiotti}}]{giacomazzo11}
{Giacomazzo}, B., {Rezzolla}, L., {Baiotti}, L., Feb. 2011. {Accurate
  evolutions of inspiralling and magnetized neutron stars: Equal-mass
  binaries}. \prd 83~(4), 044014.

\bibitem[{{Giannios}(2006)}]{giannios06}
{Giannios}, D., Aug. 2006. {Flares in GRB afterglows from delayed magnetic
  dissipation}. \aap 455, L5--L8.

\bibitem[{{Giannios}(2008)}]{giannios08}
{Giannios}, D., Mar. 2008. {Prompt GRB emission from gradual energy
  dissipation}. \aap 480, 305--312.

\bibitem[{{Giannios}(2010)}]{giannios10}
{Giannios}, D., Oct. 2010. {UHECRs from magnetic reconnection in relativistic
  jets}. \mnras 408, L46--L50.

\bibitem[{{Giannios} and {Spruit}(2007)}]{giannios07}
{Giannios}, D., {Spruit}, H.~C., Jul. 2007. {Spectral and timing properties of
  a dissipative {$\gamma$}-ray burst photosphere}. \aap 469, 1--9.

\bibitem[{{Goldreich} and {Julian}(1970)}]{goldreich70}
{Goldreich}, P., {Julian}, W.~H., Jun. 1970. {Stellar Winds}. \apj 160, 971.

\bibitem[{{Golenetskii} et~al.(1983){Golenetskii}, {Mazets}, {Aptekar}, and
  {Ilinskii}}]{golenetskii83}
{Golenetskii}, S.~V., {Mazets}, E.~P., {Aptekar}, R.~L., {Ilinskii}, V.~N.,
  Dec. 1983. {Correlation between luminosity and temperature in gamma-ray burst
  sources}. \nat 306, 451--453.

\bibitem[{{Gomboc} et~al.(2008){Gomboc}, {Kobayashi}, {Guidorzi}, {Melandri},
  {Mangano}, {Sbarufatti}, {Mundell}, {Schady}, {Smith}, {Updike}, {Kann},
  {Misra}, {Rol}, {Pozanenko}, {Castro-Tirado}, {Anupama}, {Bersier}, {Bode},
  {Carter}, {Curran}, {Fruchter}, {Graham}, {Hartmann}, {Ibrahimov}, {Levan},
  {Monfardini}, {Mottram}, {O'Brien}, {Prema}, {Sahu}, {Steele}, {Tanvir}, and
  {Wiersema}}]{gomboc08}
{Gomboc}, A., {Kobayashi}, S., {Guidorzi}, C., {Melandri}, A., {Mangano}, V.,
  {Sbarufatti}, B., {Mundell}, C.~G., {Schady}, P., {Smith}, R.~J., {Updike},
  A.~C., {Kann}, D.~A., {Misra}, K., {Rol}, E., {Pozanenko}, A.,
  {Castro-Tirado}, A.~J., {Anupama}, G.~C., {Bersier}, D., {Bode}, M.~F.,
  {Carter}, D., {Curran}, P., {Fruchter}, A., {Graham}, J., {Hartmann}, D.~H.,
  {Ibrahimov}, M., {Levan}, A., {Monfardini}, A., {Mottram}, C.~J., {O'Brien},
  P.~T., {Prema}, P., {Sahu}, D.~K., {Steele}, I.~A., {Tanvir}, N.~R.,
  {Wiersema}, K., Nov. 2008. {Multiwavelength Analysis of the Intriguing GRB
  061126: The Reverse Shock Scenario and Magnetization}. \apj 687, 443--455.

\bibitem[{{Gonz{\'a}lez} et~al.(2003){Gonz{\'a}lez}, {Dingus}, {Kaneko},
  {Preece}, {Dermer}, and {Briggs}}]{gonzalez03}
{Gonz{\'a}lez}, M.~M., {Dingus}, B.~L., {Kaneko}, Y., {Preece}, R.~D.,
  {Dermer}, C.~D., {Briggs}, M.~S., Aug. 2003. {A {$\gamma$}-ray burst with a
  high-energy spectral component inconsistent with the synchrotron shock
  model}. \nat 424, 749--751.

\bibitem[{{Goodman}(1986)}]{goodman86}
{Goodman}, J., Sep. 1986. {Are gamma-ray bursts optically thick?} \apjl 308,
  L47--L50.

\bibitem[{{Goodman}(1997)}]{goodman97}
{Goodman}, J., Nov. 1997. {Radio scintillation of gamma-ray-burst afterglows}.
  \na 2, 449--460.

\bibitem[{{Goodman} and {MacFadyen}(2008)}]{goodman08}
{Goodman}, J., {MacFadyen}, A., 2008. {Ultra-relativistic geometrical shock
  dynamics and vorticity}. Journal of Fluid Mechanics 604, 325--338.

\bibitem[{{Gou} et~al.(2004){Gou}, {M{\'e}sz{\'a}ros}, {Abel}, and
  {Zhang}}]{gou04}
{Gou}, L.~J., {M{\'e}sz{\'a}ros}, P., {Abel}, T., {Zhang}, B., Apr. 2004.
  {Detectability of Long Gamma-Ray Burst Afterglows from Very High Redshifts}.
  \apj 604, 508--520.

\bibitem[{{Graham} and {Fruchter}(2013)}]{graham13}
{Graham}, J.~F., {Fruchter}, A.~S., Sep. 2013. {The Metal Aversion of
  Long-duration Gamma-Ray Bursts}. \apj 774, 119.

\bibitem[{{Granot} et~al.(2008){Granot}, {Cohen-Tanugi}, and {do Couto e
  Silva}}]{granot08}
{Granot}, J., {Cohen-Tanugi}, J., {do Couto e Silva}, E., Apr. 2008. {Opacity
  Buildup in Impulsive Relativistic Sources}. \apj 677, 92--126.

\bibitem[{{Granot} and {Guetta}(2003)}]{granotguetta03}
{Granot}, J., {Guetta}, D., Nov. 2003. {Explaining the High-Energy Spectral
  Component in GRB 941017}. \apjl 598, L11--L14.

\bibitem[{{Granot} et~al.(2011){Granot}, {Komissarov}, and
  {Spitkovsky}}]{granot11}
{Granot}, J., {Komissarov}, S.~S., {Spitkovsky}, A., Feb. 2011. {Impulsive
  acceleration of strongly magnetized relativistic flows}. \mnras 411,
  1323--1353.

\bibitem[{{Granot} et~al.(2006){Granot}, {K{\"o}nigl}, and {Piran}}]{granot06b}
{Granot}, J., {K{\"o}nigl}, A., {Piran}, T., Aug. 2006. {Implications of the
  early X-ray afterglow light curves of Swift gamma-ray bursts}. \mnras 370,
  1946--1960.

\bibitem[{{Granot} and {Kumar}(2003)}]{granotkumar03}
{Granot}, J., {Kumar}, P., Jul. 2003. {Constraining the Structure of Gamma-Ray
  Burst Jets through the Afterglow Light Curves}. \apj 591, 1086--1096.

\bibitem[{{Granot} and {Kumar}(2006)}]{granot06}
{Granot}, J., {Kumar}, P., Feb. 2006. {Distribution of gamma-ray burst ejecta
  energy with Lorentz factor}. \mnras 366, L13--L16.

\bibitem[{{Granot} et~al.(2001){Granot}, {Miller}, {Piran}, {Suen}, and
  {Hughes}}]{granot01}
{Granot}, J., {Miller}, M., {Piran}, T., {Suen}, W.~M., {Hughes}, P.~A., 2001.
  {Light Curves from an Expanding Relativistic Jet}. In: {Costa}, E.,
  {Frontera}, F., {Hjorth}, J. (Eds.), Gamma-ray Bursts in the Afterglow Era.
  p. 312.

\bibitem[{{Granot} et~al.(2003){Granot}, {Nakar}, and {Piran}}]{granot03b}
{Granot}, J., {Nakar}, E., {Piran}, T., Nov. 2003. {Astrophysics: refreshed
  shocks from a {$\gamma$}-ray burst}. \nat 426, 138--139.

\bibitem[{{Granot} et~al.(2002){Granot}, {Panaitescu}, {Kumar}, and
  {Woosley}}]{granot02}
{Granot}, J., {Panaitescu}, A., {Kumar}, P., {Woosley}, S.~E., May 2002.
  {Off-Axis Afterglow Emission from Jetted Gamma-Ray Bursts}. \apjl 570,
  L61--L64.

\bibitem[{{Granot} and {Piran}(2012)}]{granotpiran12}
{Granot}, J., {Piran}, T., Mar. 2012. {On the lateral expansion of gamma-ray
  burst jets}. \mnras 421, 570--587.

\bibitem[{{Granot} and {Sari}(2002)}]{granotsari02}
{Granot}, J., {Sari}, R., Apr. 2002. {The Shape of Spectral Breaks in Gamma-Ray
  Burst Afterglows}. \apj 568, 820--829.

\bibitem[{{Greiner} et~al.(2009{\natexlab{a}}){Greiner}, {Clemens},
  {Kr{\"u}hler}, {von Kienlin}, {Rau}, {Sari}, {Fox}, {Kawai}, {Afonso},
  {Ajello}, {Berger}, {Cenko}, {Cucchiara}, {Filgas}, {Klose}, {K{\"u}pc{\"u}
  Yolda{\c s}}, {Lichti}, {L{\"o}w}, {McBreen}, {Nagayama}, {Rossi}, {Sato},
  {Szokoly}, {Yolda{\c s}}, and {Zhang}}]{greiner09b}
{Greiner}, J., {Clemens}, C., {Kr{\"u}hler}, T., {von Kienlin}, A., {Rau}, A.,
  {Sari}, R., {Fox}, D.~B., {Kawai}, N., {Afonso}, P., {Ajello}, M., {Berger},
  E., {Cenko}, S.~B., {Cucchiara}, A., {Filgas}, R., {Klose}, S.,
  {K{\"u}pc{\"u} Yolda{\c s}}, A., {Lichti}, G.~G., {L{\"o}w}, S., {McBreen},
  S., {Nagayama}, T., {Rossi}, A., {Sato}, S., {Szokoly}, G., {Yolda{\c s}},
  A., {Zhang}, X.-L., Apr. 2009{\natexlab{a}}. {The redshift and afterglow of
  the extremely energetic gamma-ray burst GRB 080916C}. \aap 498, 89--94.

\bibitem[{{Greiner} et~al.(2009{\natexlab{b}}){Greiner}, {Kr{\"u}hler},
  {Fynbo}, {Rossi}, {Schwarz}, {Klose}, {Savaglio}, {Tanvir}, {McBreen},
  {Totani}, {Zhang}, {Wu}, {Watson}, {Barthelmy}, {Beardmore}, {Ferrero},
  {Gehrels}, {Kann}, {Kawai}, {Yolda{\c s}}, {M{\'e}sz{\'a}ros},
  {Milvang-Jensen}, {Oates}, {Pierini}, {Schady}, {Toma}, {Vreeswijk},
  {Yolda{\c s}}, {Zhang}, {Afonso}, {Aoki}, {Burrows}, {Clemens}, {Filgas},
  {Haiman}, {Hartmann}, {Hasinger}, {Hjorth}, {Jehin}, {Levan}, {Liang},
  {Malesani}, {Pyo}, {Schulze}, {Szokoly}, {Terada}, and
  {Wiersema}}]{greiner09}
{Greiner}, J., {Kr{\"u}hler}, T., {Fynbo}, J.~P.~U., {Rossi}, A., {Schwarz},
  R., {Klose}, S., {Savaglio}, S., {Tanvir}, N.~R., {McBreen}, S., {Totani},
  T., {Zhang}, B.~B., {Wu}, X.~F., {Watson}, D., {Barthelmy}, S.~D.,
  {Beardmore}, A.~P., {Ferrero}, P., {Gehrels}, N., {Kann}, D.~A., {Kawai}, N.,
  {Yolda{\c s}}, A.~K., {M{\'e}sz{\'a}ros}, P., {Milvang-Jensen}, B., {Oates},
  S.~R., {Pierini}, D., {Schady}, P., {Toma}, K., {Vreeswijk}, P.~M., {Yolda{\c
  s}}, A., {Zhang}, B., {Afonso}, P., {Aoki}, K., {Burrows}, D.~N., {Clemens},
  C., {Filgas}, R., {Haiman}, Z., {Hartmann}, D.~H., {Hasinger}, G., {Hjorth},
  J., {Jehin}, E., {Levan}, A.~J., {Liang}, E.~W., {Malesani}, D., {Pyo}, T.,
  {Schulze}, S., {Szokoly}, G., {Terada}, K., {Wiersema}, K., Mar.
  2009{\natexlab{b}}. {GRB 080913 at Redshift 6.7}. \apj 693, 1610--1620.

\bibitem[{{Grossan} et~al.(2012){Grossan}, {Park}, {Ahmad}, {Ahn}, {Barrillon},
  {Brandt}, {Budtz-J{\o}rgensen}, {Castro-Tirado}, {Chen}, {Choi}, {Choi},
  {Connell}, {Dagoret-Campagne}, {De La Taille}, {Eyles}, {Hermann}, {Huang},
  {Jung}, {Jeong}, {Kim}, {Kim}, {Kim}, {Kim}, {Lee}, {Lim}, {Linder}, {Liu},
  {Lund}, {Min}, {Na}, {Nam}, {Panasyuk}, {Ripa}, {Reglero}, {Rodrigo},
  {Smoot}, {Suh}, {Svertilov}, {Vedenkin}, {Wang}, {Yashin}, and
  {Zhao}}]{grossan12}
{Grossan}, B., {Park}, I.~H., {Ahmad}, S., {Ahn}, K.~B., {Barrillon}, P.,
  {Brandt}, S., {Budtz-J{\o}rgensen}, C., {Castro-Tirado}, A.~J., {Chen}, P.,
  {Choi}, H.~S., {Choi}, Y.~J., {Connell}, P., {Dagoret-Campagne}, S., {De La
  Taille}, C., {Eyles}, C., {Hermann}, I., {Huang}, M.-H.~A., {Jung}, A.,
  {Jeong}, S., {Kim}, J.~E., {Kim}, M., {Kim}, S.-W., {Kim}, Y.~W., {Lee}, J.,
  {Lim}, H., {Linder}, E.~V., {Liu}, T.-C., {Lund}, N., {Min}, K.~W., {Na},
  G.~W., {Nam}, J.~W., {Panasyuk}, M.~I., {Ripa}, J., {Reglero}, V., {Rodrigo},
  J.~M., {Smoot}, G.~F., {Suh}, J.~E., {Svertilov}, S., {Vedenkin}, N., {Wang},
  M.-Z., {Yashin}, I., {Zhao}, M.~H., Sep. 2012. {A next generation Ultra-Fast
  Flash Observatory (UFFO-100) for IR/optical observations of the rise phase of
  gamma-ray bursts}. In: Society of Photo-Optical Instrumentation Engineers
  (SPIE) Conference Series. Vol. 8443 of Society of Photo-Optical
  Instrumentation Engineers (SPIE) Conference Series.

\bibitem[{{Grupe} et~al.(2007){Grupe}, {Gronwall}, {Wang}, {Roming},
  {Cummings}, {Zhang}, {M{\'e}sz{\'a}ros}, {Trigo}, {O'Brien}, {Page},
  {Beardmore}, {Godet}, {vanden Berk}, {Brown}, {Koch}, {Morris}, {Stroh},
  {Burrows}, {Nousek}, {McMath Chester}, {Immler}, {Mangano}, {Romano},
  {Chincarini}, {Osborne}, {Sakamoto}, and {Gehrels}}]{grupe07}
{Grupe}, D., {Gronwall}, C., {Wang}, X.-Y., {Roming}, P.~W.~A., {Cummings}, J.,
  {Zhang}, B., {M{\'e}sz{\'a}ros}, P., {Trigo}, M.~D., {O'Brien}, P.~T.,
  {Page}, K.~L., {Beardmore}, A., {Godet}, O., {vanden Berk}, D.~E., {Brown},
  P.~J., {Koch}, S., {Morris}, D., {Stroh}, M., {Burrows}, D.~N., {Nousek},
  J.~A., {McMath Chester}, M., {Immler}, S., {Mangano}, V., {Romano}, P.,
  {Chincarini}, G., {Osborne}, J., {Sakamoto}, T., {Gehrels}, N., Jun. 2007.
  {Swift and XMM-Newton Observations of the Extraordinary Gamma-Ray Burst
  060729: More than 125 Days of X-Ray Afterglow}. \apj 662, 443--458.

\bibitem[{{Guetta} et~al.(2004){Guetta}, {Perna}, {Stella}, and
  {Vietri}}]{guetta04}
{Guetta}, D., {Perna}, R., {Stella}, L., {Vietri}, M., Nov. 2004. {Are All
  Gamma-Ray Bursts Like GRB 980425, GRB 030329, and GRB 031203?} \apjl 615,
  L73--L76.

\bibitem[{{Guetta} and {Piran}(2005)}]{guetta05b}
{Guetta}, D., {Piran}, T., May 2005. {The luminosity and redshift distributions
  of short-duration GRBs}. \aap 435, 421--426.

\bibitem[{{Guetta} and {Piran}(2006)}]{guetta06}
{Guetta}, D., {Piran}, T., Jul. 2006. {The BATSE-Swift luminosity and redshift
  distributions of short-duration GRBs}. \aap 453, 823--828.

\bibitem[{{Guetta} and {Stella}(2009)}]{guetta09}
{Guetta}, D., {Stella}, L., May 2009. {Short {$\gamma$}-ray bursts and
  gravitational waves from dynamically formed merging binaries}. \aap 498,
  329--333.

\bibitem[{{Guidorzi} et~al.(2005){Guidorzi}, {Frontera}, {Montanari}, {Rossi},
  {Amati}, {Gomboc}, {Hurley}, and {Mundell}}]{guidorzi05}
{Guidorzi}, C., {Frontera}, F., {Montanari}, E., {Rossi}, F., {Amati}, L.,
  {Gomboc}, A., {Hurley}, K., {Mundell}, C.~G., Oct. 2005. {The gamma-ray burst
  variability-peak luminosity correlation: new results}. \mnras 363, 315--325.

\bibitem[{{Guiriec} et~al.(2011){Guiriec}, {Connaughton}, {Briggs}, {Burgess},
  {Ryde}, {Daigne}, {M{\'e}sz{\'a}ros}, {Goldstein}, {McEnery}, {Omodei},
  {Bhat}, {Bissaldi}, {Camero-Arranz}, {Chaplin}, {Diehl}, {Fishman}, {Foley},
  {Gibby}, {Giles}, {Greiner}, {Gruber}, {von Kienlin}, {Kippen},
  {Kouveliotou}, {McBreen}, {Meegan}, {Paciesas}, {Preece}, {Rau}, {Tierney},
  {van der Horst}, and {Wilson-Hodge}}]{guiriec11}
{Guiriec}, S., {Connaughton}, V., {Briggs}, M.~S., {Burgess}, M., {Ryde}, F.,
  {Daigne}, F., {M{\'e}sz{\'a}ros}, P., {Goldstein}, A., {McEnery}, J.,
  {Omodei}, N., {Bhat}, P.~N., {Bissaldi}, E., {Camero-Arranz}, A., {Chaplin},
  V., {Diehl}, R., {Fishman}, G., {Foley}, S., {Gibby}, M., {Giles}, M.~M.,
  {Greiner}, J., {Gruber}, D., {von Kienlin}, A., {Kippen}, M., {Kouveliotou},
  C., {McBreen}, S., {Meegan}, C.~A., {Paciesas}, W., {Preece}, R., {Rau}, A.,
  {Tierney}, D., {van der Horst}, A.~J., {Wilson-Hodge}, C., Feb. 2011.
  {Detection of a Thermal Spectral Component in the Prompt Emission of GRB
  100724B}. \apjl 727, L33.

\bibitem[{{Guiriec} et~al.(2013){Guiriec}, {Daigne}, {Hasco{\"e}t}, {Vianello},
  {Ryde}, {Mochkovitch}, {Kouveliotou}, {Xiong}, {Bhat}, {Foley}, {Gruber},
  {Burgess}, {McGlynn}, {McEnery}, and {Gehrels}}]{guiriec13}
{Guiriec}, S., {Daigne}, F., {Hasco{\"e}t}, R., {Vianello}, G., {Ryde}, F.,
  {Mochkovitch}, R., {Kouveliotou}, C., {Xiong}, S., {Bhat}, P.~N., {Foley},
  S., {Gruber}, D., {Burgess}, J.~M., {McGlynn}, S., {McEnery}, J., {Gehrels},
  N., Jun. 2013. {Evidence for a Photospheric Component in the Prompt Emission
  of the Short GRB 120323A and Its Effects on the GRB Hardness-Luminosity
  Relation}. \apj 770, 32.

\bibitem[{{Gupta} and {Zhang}(2007{\natexlab{a}})}]{gupta07a}
{Gupta}, N., {Zhang}, B., Jun. 2007{\natexlab{a}}. {Neutrino spectra from low
  and high luminosity populations of gamma ray bursts}. Astroparticle Physics
  27, 386--391.

\bibitem[{{Gupta} and {Zhang}(2007{\natexlab{b}})}]{gupta07b}
{Gupta}, N., {Zhang}, B., Sep. 2007{\natexlab{b}}. {Prompt emission of
  high-energy photons from gamma ray bursts}. \mnras 380, 78--92.

\bibitem[{{Gupta} and {Zhang}(2008)}]{gupta08}
{Gupta}, N., {Zhang}, B., Feb. 2008. {Diagnosing the site of gamma-ray burst
  prompt emission with spectral cut-off energy}. \mnras 384, L11--L15.

\bibitem[{{Hakkila} et~al.(2003){Hakkila}, {Giblin}, {Roiger}, {Haglin},
  {Paciesas}, and {Meegan}}]{hakkila03}
{Hakkila}, J., {Giblin}, T.~W., {Roiger}, R.~J., {Haglin}, D.~J., {Paciesas},
  W.~S., {Meegan}, C.~A., Jan. 2003. {How Sample Completeness Affects Gamma-Ray
  Burst Classification}. \apj 582, 320--329.

\bibitem[{{Hakkila} and {Preece}(2011)}]{hakkila11}
{Hakkila}, J., {Preece}, R.~D., Oct. 2011. {Unification of Pulses in Long and
  Short Gamma-Ray Bursts: Evidence from Pulse Properties and Their
  Correlations}. \apj 740, 104.

\bibitem[{{Harrison} et~al.(1999){Harrison}, {Bloom}, {Frail}, {Sari},
  {Kulkarni}, {Djorgovski}, {Axelrod}, {Mould}, {Schmidt}, {Wieringa}, {Wark},
  {Subrahmanyan}, {McConnell}, {McCarthy}, {Schaefer}, {McMahon}, {Markze},
  {Firth}, {Soffitta}, and {Amati}}]{harrison99}
{Harrison}, F.~A., {Bloom}, J.~S., {Frail}, D.~A., {Sari}, R., {Kulkarni},
  S.~R., {Djorgovski}, S.~G., {Axelrod}, T., {Mould}, J., {Schmidt}, B.~P.,
  {Wieringa}, M.~H., {Wark}, R.~M., {Subrahmanyan}, R., {McConnell}, D.,
  {McCarthy}, P.~J., {Schaefer}, B.~E., {McMahon}, R.~G., {Markze}, R.~O.,
  {Firth}, E., {Soffitta}, P., {Amati}, L., Oct. 1999. {Optical and Radio
  Observations of the Afterglow from GRB 990510: Evidence for a Jet}. \apjl
  523, L121--L124.

\bibitem[{{Hasco{\"e}t} et~al.(2012{\natexlab{a}}){Hasco{\"e}t}, {Daigne}, and
  {Mochkovitch}}]{hascoet12}
{Hasco{\"e}t}, R., {Daigne}, F., {Mochkovitch}, R., Jun. 2012{\natexlab{a}}.
  {Accounting for the XRT early steep decay in models of the prompt gamma-ray
  burst emission}. \aap 542, L29.

\bibitem[{{Hasco{\"e}t} et~al.(2012{\natexlab{b}}){Hasco{\"e}t}, {Daigne},
  {Mochkovitch}, and {Vennin}}]{hascoet12b}
{Hasco{\"e}t}, R., {Daigne}, F., {Mochkovitch}, R., {Vennin}, V., Mar.
  2012{\natexlab{b}}. {Do Fermi Large Area Telescope observations imply very
  large Lorentz factors in gamma-ray burst outflows?} \mnras 421, 525--545.

\bibitem[{{He} et~al.(2012){He}, {Liu}, {Wang}, {Nagataki}, {Murase}, and
  {Dai}}]{he12}
{He}, H.-N., {Liu}, R.-Y., {Wang}, X.-Y., {Nagataki}, S., {Murase}, K., {Dai},
  Z.-G., Jun. 2012. {Icecube Nondetection of Gamma-Ray Bursts: Constraints on
  the Fireball Properties}. \apj 752, 29.

\bibitem[{{He} et~al.(2011){He}, {Wu}, {Toma}, {Wang}, and
  {M{\'e}sz{\'a}ros}}]{he11}
{He}, H.-N., {Wu}, X.-F., {Toma}, K., {Wang}, X.-Y., {M{\'e}sz{\'a}ros}, P.,
  May 2011. {On the High-energy Emission of the Short GRB 090510}. \apj 733,
  22.

\bibitem[{{Heinz} and {Begelman}(2000)}]{heinz00}
{Heinz}, S., {Begelman}, M.~C., May 2000. {Jet Acceleration by Tangled Magnetic
  Fields}. \apj 535, 104--117.

\bibitem[{{Hesse} and {Zenitani}(2007)}]{hesse07}
{Hesse}, M., {Zenitani}, S., Nov. 2007. {Dissipation in relativistic
  pair-plasma reconnection}. Physics of Plasmas 14~(11), 112102.

\bibitem[{{Hjorth} and {Bloom}(2011)}]{hjorth11}
{Hjorth}, J., {Bloom}, J.~S., Apr. 2011. {The Gamma-Ray Burst - Supernova
  Connection}. ArXiv e-prints.

\bibitem[{{Hjorth} et~al.(2003){Hjorth}, {Sollerman}, {M{\o}ller}, {Fynbo},
  {Woosley}, {Kouveliotou}, {Tanvir}, {Greiner}, {Andersen}, {Castro-Tirado},
  {Castro Cer{\'o}n}, {Fruchter}, {Gorosabel}, {Jakobsson}, {Kaper}, {Klose},
  {Masetti}, {Pedersen}, {Pedersen}, {Pian}, {Palazzi}, {Rhoads}, {Rol}, {van
  den Heuvel}, {Vreeswijk}, {Watson}, and {Wijers}}]{hjorth03}
{Hjorth}, J., {Sollerman}, J., {M{\o}ller}, P., {Fynbo}, J.~P.~U., {Woosley},
  S.~E., {Kouveliotou}, C., {Tanvir}, N.~R., {Greiner}, J., {Andersen}, M.~I.,
  {Castro-Tirado}, A.~J., {Castro Cer{\'o}n}, J.~M., {Fruchter}, A.~S.,
  {Gorosabel}, J., {Jakobsson}, P., {Kaper}, L., {Klose}, S., {Masetti}, N.,
  {Pedersen}, H., {Pedersen}, K., {Pian}, E., {Palazzi}, E., {Rhoads}, J.~E.,
  {Rol}, E., {van den Heuvel}, E.~P.~J., {Vreeswijk}, P.~M., {Watson}, D.,
  {Wijers}, R.~A.~M.~J., Jun. 2003. {A very energetic supernova associated with
  the {$\gamma$}-ray burst of 29 March 2003}. \nat 423, 847--850.

\bibitem[{{Holland} et~al.(2010){Holland}, {Sbarufatti}, {Shen}, {Schady},
  {Cummings}, {Fonseca}, {Fynbo}, {Jakobsson}, {Leitet}, {Linn{\'e}}, {Roming},
  {Still}, and {Zhang}}]{holland10}
{Holland}, S.~T., {Sbarufatti}, B., {Shen}, R., {Schady}, P., {Cummings},
  J.~R., {Fonseca}, E., {Fynbo}, J.~P.~U., {Jakobsson}, P., {Leitet}, E.,
  {Linn{\'e}}, S., {Roming}, P.~W.~A., {Still}, M., {Zhang}, B., Jul. 2010.
  {GRB 090417B and its Host Galaxy: A Step Toward an Understanding of Optically
  Dark Gamma-ray Bursts}. \apj 717, 223--234.

\bibitem[{{Holland} et~al.(2003){Holland}, {Weidinger}, {Fynbo}, {Gorosabel},
  {Hjorth}, {Pedersen}, {M{\'e}ndez Alvarez}, {Augusteijn}, {Castro Cer{\'o}n},
  {Castro-Tirado}, {Dahle}, {Egholm}, {Jakobsson}, {Jensen}, {Levan},
  {M{\o}ller}, {Pedersen}, {Pursimo}, {Ruiz-Lapuente}, and
  {Thomsen}}]{holland03}
{Holland}, S.~T., {Weidinger}, M., {Fynbo}, J.~P.~U., {Gorosabel}, J.,
  {Hjorth}, J., {Pedersen}, K., {M{\'e}ndez Alvarez}, J., {Augusteijn}, T.,
  {Castro Cer{\'o}n}, J.~M., {Castro-Tirado}, A., {Dahle}, H., {Egholm}, M.~P.,
  {Jakobsson}, P., {Jensen}, B.~L., {Levan}, A., {M{\o}ller}, P., {Pedersen},
  H., {Pursimo}, T., {Ruiz-Lapuente}, P., {Thomsen}, B., May 2003. {Optical
  Photometry of GRB 021004: The First Month}. \aj 125, 2291--2298.

\bibitem[{{Horv{\'a}th}(1998)}]{horvath98}
{Horv{\'a}th}, I., Dec. 1998. {A Third Class of Gamma-Ray Bursts?} \apj 508,
  757--759.

\bibitem[{{Horv{\'a}th} et~al.(2010){Horv{\'a}th}, {Bagoly}, {Bal{\'a}zs}, {de
  Ugarte Postigo}, {Veres}, and {M{\'e}sz{\'a}ros}}]{horvath10}
{Horv{\'a}th}, I., {Bagoly}, Z., {Bal{\'a}zs}, L.~G., {de Ugarte Postigo}, A.,
  {Veres}, P., {M{\'e}sz{\'a}ros}, A., Apr. 2010. {Detailed Classification of
  Swift 's Gamma-ray Bursts}. \apj 713, 552--557.

\bibitem[{{Hu} et~al.(2014){Hu}, {Liang}, {Xi}, {Peng}, {Lu}, {L{\"u}}, and
  {Zhang}}]{hu14}
{Hu}, Y.-D., {Liang}, E.-W., {Xi}, S.-Q., {Peng}, F.-K., {Lu}, R.-J., {L{\"u}},
  L.-Z., {Zhang}, B., Jul. 2014. {Internal Energy Dissipation of Gamma-Ray
  Bursts Observed with Swift: Precursors, Prompt Gamma-Rays, Extended Emission,
  and Late X-Ray Flares}. \apj 789, 145.

\bibitem[{{Huang} et~al.(2007){Huang}, {Urata}, {Kuo}, {Ip}, {Ioka}, {Aoki},
  {Chen}, {Chen}, {Isogai}, {Lin}, {Makishima}, {Mito}, {Miyata}, {Nakada},
  {Nishiura}, {Onda}, {Qiu}, {Soyano}, {Tamagawa}, {Tarusawa}, {Tashiro}, and
  {Yoshioka}}]{huang07}
{Huang}, K.~Y., {Urata}, Y., {Kuo}, P.~H., {Ip}, W.~H., {Ioka}, K., {Aoki}, T.,
  {Chen}, C.~W., {Chen}, W.~P., {Isogai}, M., {Lin}, H.~C., {Makishima}, K.,
  {Mito}, H., {Miyata}, T., {Nakada}, Y., {Nishiura}, S., {Onda}, K., {Qiu},
  Y., {Soyano}, T., {Tamagawa}, T., {Tarusawa}, K., {Tashiro}, M., {Yoshioka},
  T., Jan. 2007. {Multicolor Shallow Decay and Chromatic Breaks in the GRB
  050319 Optical Afterglow}. \apjl 654, L25--L28.

\bibitem[{{Huang} and {Cheng}(2003)}]{huang03}
{Huang}, Y.~F., {Cheng}, K.~S., May 2003. {Gamma-ray bursts: optical afterglows
  in the deep Newtonian phase}. \mnras 341, 263--269.

\bibitem[{{Huang} et~al.(1999){Huang}, {Dai}, and {Lu}}]{huang99}
{Huang}, Y.~F., {Dai}, Z.~G., {Lu}, T., Oct. 1999. {A generic dynamical model
  of gamma-ray burst remnants}. \mnras 309, 513--516.

\bibitem[{{Huang} et~al.(2002){Huang}, {Dai}, and {Lu}}]{huang02}
{Huang}, Y.~F., {Dai}, Z.~G., {Lu}, T., May 2002. {Failed gamma-ray bursts and
  orphan afterglows}. \mnras 332, 735--740.

\bibitem[{{Huang} et~al.(2004){Huang}, {Wu}, {Dai}, {Ma}, and {Lu}}]{huang04}
{Huang}, Y.~F., {Wu}, X.~F., {Dai}, Z.~G., {Ma}, H.~T., {Lu}, T., Apr. 2004.
  {Rebrightening of XRF 030723: Further Evidence for a Two-Component Jet in a
  Gamma-Ray Burst}. \apj 605, 300--306.

\bibitem[{{H{\"u}mmer} et~al.(2012){H{\"u}mmer}, {Baerwald}, and
  {Winter}}]{hummer12}
{H{\"u}mmer}, S., {Baerwald}, P., {Winter}, W., Jun. 2012. {Neutrino Emission
  from Gamma-Ray Burst Fireballs, Revised}. Physical Review Letters 108~(23),
  231101.

\bibitem[{{Hurley} et~al.(1994){Hurley}, {Dingus}, {Mukherjee}, {Sreekumar},
  {Kouveliotou}, {Meegan}, {Fishman}, {Band}, {Ford}, {Bertsch}, {Cline},
  {Fichtel}, {Hartman}, {Hunter}, {Thompson}, {Kanbach}, {Mayer-Hasselwander},
  {von Montigny}, {Sommer}, {Lin}, {Nolan}, {Michelson}, {Kniffen}, {Mattox},
  {Schneid}, {Boer}, and {Niel}}]{hurley94}
{Hurley}, K., {Dingus}, B.~L., {Mukherjee}, R., {Sreekumar}, P., {Kouveliotou},
  C., {Meegan}, C., {Fishman}, G.~J., {Band}, D., {Ford}, L., {Bertsch}, D.,
  {Cline}, T., {Fichtel}, C., {Hartman}, R., {Hunter}, S., {Thompson}, D.~J.,
  {Kanbach}, G., {Mayer-Hasselwander}, H., {von Montigny}, C., {Sommer}, M.,
  {Lin}, Y., {Nolan}, P., {Michelson}, P., {Kniffen}, D., {Mattox}, J.,
  {Schneid}, E., {Boer}, M., {Niel}, M., Dec. 1994. {Detection of a Gamma-Ray
  Burst of Very Long Duration and Very High Energy}. \nat 372, 652--+.

\bibitem[{{Inoue} et~al.(2013){Inoue}, {Granot}, {O'Brien}, {Asano}, {Bouvier},
  {Carosi}, {Connaughton}, {Garczarczyk}, {Gilmore}, {Hinton}, {Inoue}, {Ioka},
  {Kakuwa}, {Markoff}, {Murase}, {Osborne}, {Otte}, {Starling}, {Tajima},
  {Teshima}, {Toma}, {Wagner}, {Wijers}, {Williams}, {Yamamoto}, {Yamazaki},
  and {CTA Consortium}}]{inoue13}
{Inoue}, S., {Granot}, J., {O'Brien}, P.~T., {Asano}, K., {Bouvier}, A.,
  {Carosi}, A., {Connaughton}, V., {Garczarczyk}, M., {Gilmore}, R., {Hinton},
  J., {Inoue}, Y., {Ioka}, K., {Kakuwa}, J., {Markoff}, S., {Murase}, K.,
  {Osborne}, J.~P., {Otte}, A.~N., {Starling}, R., {Tajima}, H., {Teshima}, M.,
  {Toma}, K., {Wagner}, S., {Wijers}, R.~A.~M.~J., {Williams}, D.~A.,
  {Yamamoto}, T., {Yamazaki}, R., {CTA Consortium}, Mar. 2013. {Gamma-ray burst
  science in the era of the Cherenkov Telescope Array}. Astroparticle Physics
  43, 252--275.

\bibitem[{{Inoue} et~al.(2011){Inoue}, {Asano}, and {Ioka}}]{inoue11}
{Inoue}, T., {Asano}, K., {Ioka}, K., Jun. 2011. {Three-dimensional Simulations
  of Magnetohydrodynamic Turbulence Behind Relativistic Shock Waves and Their
  Implications for Gamma-Ray Bursts}. \apj 734, 77.

\bibitem[{{Ioka}(2010)}]{ioka10}
{Ioka}, K., Oct. 2010. {Very High Lorentz Factor Fireballs and Gamma-Ray Burst
  Spectra}. Progress of Theoretical Physics 124, 667--710.

\bibitem[{{Ioka} et~al.(2005){Ioka}, {Kobayashi}, and {Zhang}}]{ioka05}
{Ioka}, K., {Kobayashi}, S., {Zhang}, B., Sep. 2005. {Variabilities of
  Gamma-Ray Burst Afterglows: Long-acting Engine, Anisotropic Jet, or Many
  Fluctuating Regions?} \apj 631, 429--434.

\bibitem[{{Ioka} and {M{\'e}sz{\'a}ros}(2005)}]{ioka05b}
{Ioka}, K., {M{\'e}sz{\'a}ros}, P., Feb. 2005. {Radio Afterglows of Gamma-Ray
  Bursts and Hypernovae at High Redshift and Their Potential for 21 Centimeter
  Absorption Studies}. \apj 619, 684--696.

\bibitem[{{Ioka} et~al.(2007){Ioka}, {Murase}, {Toma}, {Nagataki}, and
  {Nakamura}}]{ioka07}
{Ioka}, K., {Murase}, K., {Toma}, K., {Nagataki}, S., {Nakamura}, T., Dec.
  2007. {Unstable GRB Photospheres and $e^{+/-}$ Annihilation Lines}. \apjl
  670, L77--L80.

\bibitem[{{Ioka} et~al.(2011){Ioka}, {Ohira}, {Kawanaka}, and
  {Mizuta}}]{ioka11}
{Ioka}, K., {Ohira}, Y., {Kawanaka}, N., {Mizuta}, A., Sep. 2011. {Gamma-Ray
  Burst without Baryonic and Magnetic Load?} Progress of Theoretical Physics
  126, 555--564.

\bibitem[{{Ioka} et~al.(2006){Ioka}, {Toma}, {Yamazaki}, and
  {Nakamura}}]{ioka06}
{Ioka}, K., {Toma}, K., {Yamazaki}, R., {Nakamura}, T., Oct. 2006. {Efficiency
  crisis of swift gamma-ray bursts with shallow X-ray afterglows: prior
  activity or time-dependent microphysics?} \aap 458, 7--12.

\bibitem[{{Ito} et~al.(2013){Ito}, {Nagataki}, {Ono}, {Lee}, {Mao}, {Yamada},
  {Pe'er}, {Mizuta}, and {Harikae}}]{ito13}
{Ito}, H., {Nagataki}, S., {Ono}, M., {Lee}, S.-H., {Mao}, J., {Yamada}, S.,
  {Pe'er}, A., {Mizuta}, A., {Harikae}, S., Jun. 2013. {Photospheric emission
  from stratified jets}. ArXiv e-prints.

\bibitem[{{Jaroschek} et~al.(2004){Jaroschek}, {Treumann}, {Lesch}, and
  {Scholer}}]{jaroschek04}
{Jaroschek}, C.~H., {Treumann}, R.~A., {Lesch}, H., {Scholer}, M., Mar. 2004.
  {Fast reconnection in relativistic pair plasmas: Analysis of particle
  acceleration in self-consistent full particle simulations}. Physics of
  Plasmas 11, 1151--1163.

\bibitem[{{Jin} and {Fan}(2007)}]{jinfan07}
{Jin}, Z.~P., {Fan}, Y.~Z., Jul. 2007. {GRB 060418 and 060607A: the medium
  surrounding the progenitor and the weak reverse shock emission}. \mnras 378,
  1043--1048.

\bibitem[{{Kagan} et~al.(2013){Kagan}, {Milosavljevi{\'c}}, and
  {Spitkovsky}}]{kagan13}
{Kagan}, D., {Milosavljevi{\'c}}, M., {Spitkovsky}, A., Sep. 2013. {A Flux Rope
  Network and Particle Acceleration in Three-dimensional Relativistic Magnetic
  Reconnection}. \apj 774, 41.

\bibitem[{{Kakuwa} et~al.(2012){Kakuwa}, {Murase}, {Toma}, {Inoue}, {Yamazaki},
  and {Ioka}}]{kakuwa12}
{Kakuwa}, J., {Murase}, K., {Toma}, K., {Inoue}, S., {Yamazaki}, R., {Ioka},
  K., Sep. 2012. {Prospects for detecting gamma-ray bursts at very high
  energies with the Cherenkov Telescope Array}. \mnras 425, 514--526.

\bibitem[{{Kalemci} et~al.(2007){Kalemci}, {Boggs}, {Kouveliotou}, {Finger},
  and {Baring}}]{kalemci07}
{Kalemci}, E., {Boggs}, S.~E., {Kouveliotou}, C., {Finger}, M., {Baring},
  M.~G., Mar. 2007. {Search for Polarization from the Prompt Gamma-Ray Emission
  of GRB 041219a with SPI on INTEGRAL}. \apjs 169, 75--82.

\bibitem[{{Kann} et~al.(2011){Kann}, {Klose}, {Zhang}, {Covino}, {Butler},
  {Malesani}, {Nakar}, {Wilson}, {Antonelli}, {Chincarini}, {Cobb}, {D'Avanzo},
  {D'Elia}, {Della Valle}, {Ferrero}, {Fugazza}, {Gorosabel}, {Israel},
  {Mannucci}, {Piranomonte}, {Schulze}, {Stella}, {Tagliaferri}, and
  {Wiersema}}]{kann11}
{Kann}, D.~A., {Klose}, S., {Zhang}, B., {Covino}, S., {Butler}, N.~R.,
  {Malesani}, D., {Nakar}, E., {Wilson}, A.~C., {Antonelli}, L.~A.,
  {Chincarini}, G., {Cobb}, B.~E., {D'Avanzo}, P., {D'Elia}, V., {Della Valle},
  M., {Ferrero}, P., {Fugazza}, D., {Gorosabel}, J., {Israel}, G.~L.,
  {Mannucci}, F., {Piranomonte}, S., {Schulze}, S., {Stella}, L.,
  {Tagliaferri}, G., {Wiersema}, K., Jun. 2011. {The Afterglows of Swift-era
  Gamma-Ray Bursts. II. Type I GRB versus Type II GRB Optical Afterglows}. \apj
  734, 96.

\bibitem[{{Kann} et~al.(2010){Kann}, {Klose}, {Zhang}, {Malesani}, {Nakar},
  {Pozanenko}, {Wilson}, {Butler}, {Jakobsson}, {Schulze}, {Andreev},
  {Antonelli}, {Bikmaev}, {Biryukov}, {B{\"o}ttcher}, {Burenin}, {Castro
  Cer{\'o}n}, {Castro-Tirado}, {Chincarini}, {Cobb}, {Covino}, {D'Avanzo},
  {D'Elia}, {Della Valle}, {de Ugarte Postigo}, {Efimov}, {Ferrero}, {Fugazza},
  {Fynbo}, {G{\aa}lfalk}, {Grundahl}, {Gorosabel}, {Gupta}, {Guziy}, {Hafizov},
  {Hjorth}, {Holhjem}, {Ibrahimov}, {Im}, {Israel}, {Je{\'l}inek}, {Jensen},
  {Karimov}, {Khamitov}, {Kizilo{\v g}lu}, {Klunko}, {Kub{\'a}nek}, {Kutyrev},
  {Laursen}, {Levan}, {Mannucci}, {Martin}, {Mescheryakov}, {Mirabal},
  {Norris}, {Ovaldsen}, {Paraficz}, {Pavlenko}, {Piranomonte}, {Rossi},
  {Rumyantsev}, {Salinas}, {Sergeev}, {Sharapov}, {Sollerman}, {Stecklum},
  {Stella}, {Tagliaferri}, {Tanvir}, {Telting}, {Testa}, {Updike}, {Volnova},
  {Watson}, {Wiersema}, and {Xu}}]{kann10}
{Kann}, D.~A., {Klose}, S., {Zhang}, B., {Malesani}, D., {Nakar}, E.,
  {Pozanenko}, A., {Wilson}, A.~C., {Butler}, N.~R., {Jakobsson}, P.,
  {Schulze}, S., {Andreev}, M., {Antonelli}, L.~A., {Bikmaev}, I.~F.,
  {Biryukov}, V., {B{\"o}ttcher}, M., {Burenin}, R.~A., {Castro Cer{\'o}n},
  J.~M., {Castro-Tirado}, A.~J., {Chincarini}, G., {Cobb}, B.~E., {Covino}, S.,
  {D'Avanzo}, P., {D'Elia}, V., {Della Valle}, M., {de Ugarte Postigo}, A.,
  {Efimov}, Y., {Ferrero}, P., {Fugazza}, D., {Fynbo}, J.~P.~U., {G{\aa}lfalk},
  M., {Grundahl}, F., {Gorosabel}, J., {Gupta}, S., {Guziy}, S., {Hafizov}, B.,
  {Hjorth}, J., {Holhjem}, K., {Ibrahimov}, M., {Im}, M., {Israel}, G.~L.,
  {Je{\'l}inek}, M., {Jensen}, B.~L., {Karimov}, R., {Khamitov}, I.~M.,
  {Kizilo{\v g}lu}, {\"U}., {Klunko}, E., {Kub{\'a}nek}, P., {Kutyrev}, A.~S.,
  {Laursen}, P., {Levan}, A.~J., {Mannucci}, F., {Martin}, C.~M.,
  {Mescheryakov}, A., {Mirabal}, N., {Norris}, J.~P., {Ovaldsen}, J.-E.,
  {Paraficz}, D., {Pavlenko}, E., {Piranomonte}, S., {Rossi}, A., {Rumyantsev},
  V., {Salinas}, R., {Sergeev}, A., {Sharapov}, D., {Sollerman}, J.,
  {Stecklum}, B., {Stella}, L., {Tagliaferri}, G., {Tanvir}, N.~R., {Telting},
  J., {Testa}, V., {Updike}, A.~C., {Volnova}, A., {Watson}, D., {Wiersema},
  K., {Xu}, D., Sep. 2010. {The Afterglows of Swift-era Gamma-ray Bursts. I.
  Comparing pre-Swift and Swift-era Long/Soft (Type II) GRB Optical
  Afterglows}. \apj 720, 1513--1558.

\bibitem[{{Katz}(1976)}]{katz76}
{Katz}, J.~I., Jun. 1976. {Nonrelativistic Compton scattering and models of
  quasars}. \apj 206, 910--916.

\bibitem[{{Katz}(1994)}]{katz94}
{Katz}, J.~I., Feb. 1994. {Two populations and models of gamma-ray bursts}.
  \apj 422, 248--259.

\bibitem[{{Katz}(1997)}]{katz97a}
{Katz}, J.~I., Dec. 1997. {Yet Another Model of Gamma-Ray Bursts}. \apj 490,
  633--641.

\bibitem[{{Katz} and {Piran}(1997)}]{katz97}
{Katz}, J.~I., {Piran}, T., Dec. 1997. {Persistent Counterparts to Gamma-Ray
  Bursts}. \apj 490, 772.

\bibitem[{{Katz} and {Piran}(1998)}]{Katz98a}
{Katz}, J.~I., {Piran}, T., May 1998. {What have we learned from GRB
  afterglows?} In: {Meegan}, C.~A., {Preece}, R.~D., {Koshut}, T.~M. (Eds.),
  Gamma-Ray Bursts, 4th Hunstville Symposium. Vol. 428 of American Institute of
  Physics Conference Series. pp. 689--698.

\bibitem[{{Katz} et~al.(1998){Katz}, {Piran}, and {Sari}}]{katz98}
{Katz}, J.~I., {Piran}, T., {Sari}, R., Feb. 1998. {Implications of the Visible
  and X-Ray Counterparts to GRB 970228}. Physical Review Letters 80,
  1580--1581.

\bibitem[{{Kawai} et~al.(2006){Kawai}, {Kosugi}, {Aoki}, {Yamada}, {Totani},
  {Ohta}, {Iye}, {Hattori}, {Aoki}, {Furusawa}, {Hurley}, {Kawabata},
  {Kobayashi}, {Komiyama}, {Mizumoto}, {Nomoto}, {Noumaru}, {Ogasawara},
  {Sato}, {Sekiguchi}, {Shirasaki}, {Suzuki}, {Takata}, {Tamagawa}, {Terada},
  {Watanabe}, {Yatsu}, and {Yoshida}}]{kawai06}
{Kawai}, N., {Kosugi}, G., {Aoki}, K., {Yamada}, T., {Totani}, T., {Ohta}, K.,
  {Iye}, M., {Hattori}, T., {Aoki}, W., {Furusawa}, H., {Hurley}, K.,
  {Kawabata}, K.~S., {Kobayashi}, N., {Komiyama}, Y., {Mizumoto}, Y., {Nomoto},
  K., {Noumaru}, J., {Ogasawara}, R., {Sato}, R., {Sekiguchi}, K., {Shirasaki},
  Y., {Suzuki}, M., {Takata}, T., {Tamagawa}, T., {Terada}, H., {Watanabe}, J.,
  {Yatsu}, Y., {Yoshida}, A., Mar. 2006. {An optical spectrum of the afterglow
  of a {$\gamma$}-ray burst at a redshift of z = 6.295}. \nat 440, 184--186.

\bibitem[{{Kazanas} et~al.(2002){Kazanas}, {Georganopoulos}, and
  {Mastichiadis}}]{kazanas02}
{Kazanas}, D., {Georganopoulos}, M., {Mastichiadis}, A., Oct. 2002. {The
  ``Supercritical Pile'' Model for Gamma-Ray Bursts: Getting the {$\nu$}F$_{Ω}$
  Peak at 1 MeV}. \apjl 578, L15--L18.

\bibitem[{{Kennel} and {Coroniti}(1984{\natexlab{a}})}]{kennel84}
{Kennel}, C.~F., {Coroniti}, F.~V., Aug. 1984{\natexlab{a}}. {Confinement of
  the Crab pulsar's wind by its supernova remnant}. \apj 283, 694--709.

\bibitem[{{Kennel} and {Coroniti}(1984{\natexlab{b}})}]{kennel84b}
{Kennel}, C.~F., {Coroniti}, F.~V., Aug. 1984{\natexlab{b}}.
  {Magnetohydrodynamic model of Crab nebula radiation}. \apj 283, 710--730.

\bibitem[{{King} et~al.(2005){King}, {O'Brien}, {Goad}, {Osborne}, {Olsson},
  and {Page}}]{king05}
{King}, A., {O'Brien}, P.~T., {Goad}, M.~R., {Osborne}, J., {Olsson}, E.,
  {Page}, K., Sep. 2005. {Gamma-Ray Bursts: Restarting the Engine}. \apjl 630,
  L113--L115.

\bibitem[{{Kirk} et~al.(2000){Kirk}, {Guthmann}, {Gallant}, and
  {Achterberg}}]{kirk00}
{Kirk}, J.~G., {Guthmann}, A.~W., {Gallant}, Y.~A., {Achterberg}, A., Oct.
  2000. {Particle Acceleration at Ultrarelativistic Shocks: An Eigenfunction
  Method}. \apj 542, 235--242.

\bibitem[{{Kistler} et~al.(2008){Kistler}, {Y{\"u}ksel}, {Beacom}, and
  {Stanek}}]{kistler08}
{Kistler}, M.~D., {Y{\"u}ksel}, H., {Beacom}, J.~F., {Stanek}, K.~Z., Feb.
  2008. {An Unexpectedly Swift Rise in the Gamma-Ray Burst Rate}. \apjl 673,
  L119--L122.

\bibitem[{{Kiuchi} et~al.(2012){Kiuchi}, {Kyutoku}, and {Shibata}}]{kiuchi12}
{Kiuchi}, K., {Kyutoku}, K., {Shibata}, M., Sep. 2012. {Three-dimensional
  evolution of differentially rotating magnetized neutron stars}. \prd 86~(6),
  064008.

\bibitem[{{Klebesadel} et~al.(1973){Klebesadel}, {Strong}, and
  {Olson}}]{klebesadel73}
{Klebesadel}, R.~W., {Strong}, I.~B., {Olson}, R.~A., Jun. 1973. {Observations
  of Gamma-Ray Bursts of Cosmic Origin}. \apjl 182, L85.

\bibitem[{{Klu{\'z}niak} and {Ruderman}(1998)}]{kluzniak98}
{Klu{\'z}niak}, W., {Ruderman}, M., Oct. 1998. {The Central Engine of Gamma-Ray
  Bursters}. \apjl 505, L113--L117.

\bibitem[{{Kobayashi}(2000)}]{kobayashi00}
{Kobayashi}, S., Dec. 2000. {Light Curves of Gamma-Ray Burst Optical Flashes}.
  \apj 545, 807--812.

\bibitem[{{Kobayashi} and {M{\'e}sz{\'a}ros}(2003)}]{kobayashimeszaros03}
{Kobayashi}, S., {M{\'e}sz{\'a}ros}, P., Jun. 2003. {Gravitational Radiation
  from Gamma-Ray Burst Progenitors}. \apj 589, 861--870.

\bibitem[{{Kobayashi} et~al.(1997){Kobayashi}, {Piran}, and
  {Sari}}]{kobayashi97}
{Kobayashi}, S., {Piran}, T., {Sari}, R., Nov. 1997. {Can Internal Shocks
  Produce the Variability in Gamma-Ray Bursts?} \apj 490, 92--+.

\bibitem[{{Kobayashi} et~al.(1999){Kobayashi}, {Piran}, and
  {Sari}}]{kobayashi99}
{Kobayashi}, S., {Piran}, T., {Sari}, R., Mar. 1999. {Hydrodynamics of a
  Relativistic Fireball: The Complete Evolution}. \apj 513, 669--678.

\bibitem[{{Kobayashi} and {Sari}(2001)}]{kobayashisari01}
{Kobayashi}, S., {Sari}, R., Apr. 2001. {Ultraefficient Internal Shocks}. \apj
  551, 934--939.

\bibitem[{{Kobayashi} and {Zhang}(2003{\natexlab{a}})}]{kobayashizhang03b}
{Kobayashi}, S., {Zhang}, B., Nov. 2003{\natexlab{a}}. {Early Optical
  Afterglows from Wind-Type Gamma-Ray Bursts}. \apj 597, 455--458.

\bibitem[{{Kobayashi} and {Zhang}(2003{\natexlab{b}})}]{kobayashizhang03a}
{Kobayashi}, S., {Zhang}, B., Jan. 2003{\natexlab{b}}. {GRB 021004: Reverse
  Shock Emission}. \apjl 582, L75--L78.

\bibitem[{{Kocevski}(2012)}]{kocevski12b}
{Kocevski}, D., Mar. 2012. {On the Origin of High-energy Correlations in
  Gamma-Ray Bursts}. \apj 747, 146.

\bibitem[{{Kocevski} and {Butler}(2008)}]{kocevski08}
{Kocevski}, D., {Butler}, N., Jun. 2008. {Gamma-Ray Burst Energetics in the
  Swift Era}. \apj 680, 531--538.

\bibitem[{{Kocevski} and {Petrosian}(2013)}]{kocevski13}
{Kocevski}, D., {Petrosian}, V., Mar. 2013. {On the Lack of Time Dilation
  Signatures in Gamma-Ray Burst Light Curves}. \apj 765, 116.

\bibitem[{{Kochanek} and {Piran}(1993)}]{kochanek93}
{Kochanek}, C.~S., {Piran}, T., Nov. 1993. {Gravitational Waves and gamma -Ray
  Bursts}. \apjl 417, L17+.

\bibitem[{{Kodama} et~al.(2008){Kodama}, {Yonetoku}, {Murakami}, {Tanabe},
  {Tsutsui}, and {Nakamura}}]{kodama08}
{Kodama}, Y., {Yonetoku}, D., {Murakami}, T., {Tanabe}, S., {Tsutsui}, R.,
  {Nakamura}, T., Nov. 2008. {Gamma-ray bursts between z of 1.8 and 5.6 suggest
  that the time variation of the dark energy is small}. \mnras 391, L1--L4.

\bibitem[{{Kohri} and {Mineshige}(2002)}]{kohri02}
{Kohri}, K., {Mineshige}, S., Sep. 2002. {Can Neutrino-cooled Accretion Disks
  Be an Origin of Gamma-Ray Bursts?} \apj 577, 311--321.

\bibitem[{{Kohri} et~al.(2005){Kohri}, {Narayan}, and {Piran}}]{kohri05}
{Kohri}, K., {Narayan}, R., {Piran}, T., Aug. 2005. {Neutrino-dominated
  Accretion and Supernovae}. \apj 629, 341--361.

\bibitem[{{Komissarov}(2001)}]{komissarov01}
{Komissarov}, S.~S., Sep. 2001. {Direct numerical simulations of the
  Blandford-Znajek effect}. \mnras 326, L41--L44.

\bibitem[{{Komissarov}(2002)}]{komissarov02}
{Komissarov}, S.~S., Nov. 2002. {Time-dependent, force-free, degenerate
  electrodynamics}. \mnras 336, 759--766.

\bibitem[{{Komissarov}(2004)}]{komissarov04}
{Komissarov}, S.~S., May 2004. {Electrodynamics of black hole magnetospheres}.
  \mnras 350, 427--448.

\bibitem[{{Komissarov}(2007)}]{komissarov07a}
{Komissarov}, S.~S., Dec. 2007. {Multidimensional numerical scheme for
  resistive relativistic magnetohydrodynamics}. \mnras 382, 995--1004.

\bibitem[{{Komissarov} et~al.(2007){Komissarov}, {Barkov}, {Vlahakis}, and
  {K{\"o}nigl}}]{komissarov07b}
{Komissarov}, S.~S., {Barkov}, M.~V., {Vlahakis}, N., {K{\"o}nigl}, A., Sep.
  2007. {Magnetic acceleration of relativistic active galactic nucleus jets}.
  \mnras 380, 51--70.

\bibitem[{{Komissarov} et~al.(2010){Komissarov}, {Vlahakis}, and
  {K{\"o}nigl}}]{komissarov10a}
{Komissarov}, S.~S., {Vlahakis}, N., {K{\"o}nigl}, A., Sep. 2010. {Rarefaction
  acceleration of ultrarelativistic magnetized jets in gamma-ray burst
  sources}. \mnras 407, 17--28.

\bibitem[{{Komissarov} et~al.(2009){Komissarov}, {Vlahakis}, {K{\"o}nigl}, and
  {Barkov}}]{komissarov09}
{Komissarov}, S.~S., {Vlahakis}, N., {K{\"o}nigl}, A., {Barkov}, M.~V., Apr.
  2009. {Magnetic acceleration of ultrarelativistic jets in gamma-ray burst
  sources}. \mnras 394, 1182--1212.

\bibitem[{{Kopa{\v c}} et~al.(2013){Kopa{\v c}}, {Kobayashi}, {Gomboc},
  {Japelj}, {Mundell}, {Guidorzi}, {Melandri}, {Bersier}, {Cano}, {Smith},
  {Steele}, and {Virgili}}]{kopac13}
{Kopa{\v c}}, D., {Kobayashi}, S., {Gomboc}, A., {Japelj}, J., {Mundell},
  C.~G., {Guidorzi}, C., {Melandri}, A., {Bersier}, D., {Cano}, Z., {Smith},
  R.~J., {Steele}, I.~A., {Virgili}, F.~J., Jul. 2013. {GRB 090727 and
  Gamma-Ray Bursts with Early-time Optical Emission}. \apj 772, 73.

\bibitem[{{Koshut} et~al.(1995){Koshut}, {Kouveliotou}, {Paciesas}, {van
  Paradijs}, {Pendleton}, {Briggs}, {Fishman}, and {Meegan}}]{koshut95}
{Koshut}, T.~M., {Kouveliotou}, C., {Paciesas}, W.~S., {van Paradijs}, J.,
  {Pendleton}, G.~N., {Briggs}, M.~S., {Fishman}, G.~J., {Meegan}, C.~A., Oct.
  1995. {Gamma-Ray Burst Precursor Activity as Observed with BATSE}. \apj 452,
  145.

\bibitem[{{Kouveliotou} et~al.(1993){Kouveliotou}, {Meegan}, {Fishman}, {Bhat},
  {Briggs}, {Koshut}, {Paciesas}, and {Pendleton}}]{kouveliotou93}
{Kouveliotou}, C., {Meegan}, C.~A., {Fishman}, G.~J., {Bhat}, N.~P., {Briggs},
  M.~S., {Koshut}, T.~M., {Paciesas}, W.~S., {Pendleton}, G.~N., Aug. 1993.
  {Identification of two classes of gamma-ray bursts}. \apjl 413, L101--L104.

\bibitem[{{Kowal} et~al.(2009){Kowal}, {Lazarian}, {Vishniac}, and
  {Otmianowska-Mazur}}]{kowal09}
{Kowal}, G., {Lazarian}, A., {Vishniac}, E.~T., {Otmianowska-Mazur}, K., Jul.
  2009. {Numerical Tests of Fast Reconnection in Weakly Stochastic Magnetic
  Fields}. \apj 700, 63--85.

\bibitem[{{Krolik}(1999)}]{krolik99}
{Krolik}, J.~H., 1999. {Active galactic nuclei : from the central black hole to
  the galactic environment}. Princeton University Press, 1999.

\bibitem[{{Kulkarni}(2005)}]{kulkarni05}
{Kulkarni}, S.~R., Oct. 2005. {Modeling Supernova-like Explosions Associated
  with Gamma-ray Bursts with Short Durations}. ArXiv Astrophysics e-prints.

\bibitem[{{Kulkarni} et~al.(1999{\natexlab{a}}){Kulkarni}, {Djorgovski},
  {Odewahn}, {Bloom}, {Gal}, {Koresko}, {Harrison}, {Lubin}, {Armus}, {Sari},
  {Illingworth}, {Kelson}, {Magee}, {Dokkum}, {Frail}, {Mulchaey}, {Malkan},
  {McClean}, {Teplitz}, {Koerner}, {Kirkpatrick}, {Kobayashi}, {Yadigaroglu},
  {Halpern}, {Piran}, {Goodrich}, {Chaffee}, {Feroci}, and
  {Costa}}]{kulkarni99}
{Kulkarni}, S.~R., {Djorgovski}, S.~G., {Odewahn}, S.~C., {Bloom}, J.~S.,
  {Gal}, R.~R., {Koresko}, C.~D., {Harrison}, F.~A., {Lubin}, L.~M., {Armus},
  L., {Sari}, R., {Illingworth}, G.~D., {Kelson}, D.~D., {Magee}, D.~K.,
  {Dokkum}, P.~G.~V., {Frail}, D.~A., {Mulchaey}, J.~S., {Malkan}, M.~A.,
  {McClean}, I.~S., {Teplitz}, H.~I., {Koerner}, D., {Kirkpatrick}, D.,
  {Kobayashi}, N., {Yadigaroglu}, I.-A., {Halpern}, J., {Piran}, T.,
  {Goodrich}, R.~W., {Chaffee}, F.~H., {Feroci}, M., {Costa}, E., Apr.
  1999{\natexlab{a}}. {The afterglow, redshift and extreme energetics of the
  {$\gamma$}-ray burst of 23 January 1999}. \nat 398, 389--394.

\bibitem[{{Kulkarni} et~al.(1999{\natexlab{b}}){Kulkarni}, {Frail}, {Sari},
  {Moriarty-Schieven}, {Shepherd}, {Udomprasert}, {Readhead}, {Bloom},
  {Feroci}, and {Costa}}]{kulkarni99b}
{Kulkarni}, S.~R., {Frail}, D.~A., {Sari}, R., {Moriarty-Schieven}, G.~H.,
  {Shepherd}, D.~S., {Udomprasert}, P., {Readhead}, A.~C.~S., {Bloom}, J.~S.,
  {Feroci}, M., {Costa}, E., Sep. 1999{\natexlab{b}}. {Discovery of a Radio
  Flare from GRB 990123}. \apjl 522, L97--L100.

\bibitem[{{Kulkarni} et~al.(1998){Kulkarni}, {Frail}, {Wieringa}, {Ekers},
  {Sadler}, {Wark}, {Higdon}, {Phinney}, and {Bloom}}]{kulkarni98}
{Kulkarni}, S.~R., {Frail}, D.~A., {Wieringa}, M.~H., {Ekers}, R.~D., {Sadler},
  E.~M., {Wark}, R.~M., {Higdon}, J.~L., {Phinney}, E.~S., {Bloom}, J.~S., Oct.
  1998. {Radio emission from the unusual supernova 1998bw and its association
  with the {$\gamma$}-ray burst of 25 April 1998}. \nat 395, 663--669.

\bibitem[{{Kulsrud}(2005)}]{kulsrud05}
{Kulsrud}, R.~M., 2005. {Plasma physics for astrophysics}. Princeton U
  niversity Press, 2005.

\bibitem[{{Kumar}(1999)}]{kumar99}
{Kumar}, P., Oct. 1999. {Gamma-Ray Burst Energetics}. \apjl 523, L113--L116.

\bibitem[{{Kumar}(2000)}]{kumar00c}
{Kumar}, P., Oct. 2000. {The distribution of burst energy and shock parameters
  for GRBs}. \apjl 538, L125--L128.

\bibitem[{{Kumar} and {Barniol Duran}(2009)}]{kumar09}
{Kumar}, P., {Barniol Duran}, R., Nov. 2009. {On the generation of high-energy
  photons detected by the Fermi Satellite from gamma-ray bursts}. \mnras 400,
  L75--L79.

\bibitem[{{Kumar} and {Barniol Duran}(2010)}]{kumar10}
{Kumar}, P., {Barniol Duran}, R., Nov. 2010. {External forward shock origin of
  high-energy emission for three gamma-ray bursts detected by Fermi}. \mnras
  409, 226--236.

\bibitem[{{Kumar} and {Granot}(2003)}]{kumargranot03}
{Kumar}, P., {Granot}, J., Jul. 2003. {The Evolution of a Structured
  Relativistic Jet and Gamma-Ray Burst Afterglow Light Curves}. \apj 591,
  1075--1085.

\bibitem[{{Kumar} et~al.(2012){Kumar}, {Hern{\'a}ndez}, {Bo{\v s}njak}, and
  {Duran}}]{kumarhernandez12}
{Kumar}, P., {Hern{\'a}ndez}, R.~A., {Bo{\v s}njak}, {\v Z}., {Duran}, R.~B.,
  Nov. 2012. {Maximum synchrotron frequency for shock-accelerated particles}.
  \mnras 427, L40--L44.

\bibitem[{{Kumar} and {McMahon}(2008)}]{kumarmcmahon08}
{Kumar}, P., {McMahon}, E., Feb. 2008. {A general scheme for modelling
  {$\gamma$}-ray burst prompt emission}. \mnras 384, 33--63.

\bibitem[{{Kumar} et~al.(2007){Kumar}, {McMahon}, {Panaitescu}, {Willingale},
  {O'Brien}, {Burrows}, {Cummings}, {Gehrels}, {Holland}, {Pandey}, {vanden
  Berk}, and {Zane}}]{kumar07}
{Kumar}, P., {McMahon}, E., {Panaitescu}, A., {Willingale}, R., {O'Brien}, P.,
  {Burrows}, D., {Cummings}, J., {Gehrels}, N., {Holland}, S., {Pandey}, S.~B.,
  {vanden Berk}, D., {Zane}, S., Mar. 2007. {The nature of the outflow in
  gamma-ray bursts}. \mnras 376, L57--L61.

\bibitem[{{Kumar} and {Narayan}(2009)}]{kumarnarayan09}
{Kumar}, P., {Narayan}, R., May 2009. {GRB 080319B: evidence for relativistic
  turbulence, not internal shocks}. \mnras 395, 472--489.

\bibitem[{{Kumar} et~al.(2008{\natexlab{a}}){Kumar}, {Narayan}, and
  {Johnson}}]{kumar08b}
{Kumar}, P., {Narayan}, R., {Johnson}, J.~L., Jul. 2008{\natexlab{a}}. {Mass
  fall-back and accretion in the central engine of gamma-ray bursts}. \mnras,
  750--+.

\bibitem[{{Kumar} et~al.(2008{\natexlab{b}}){Kumar}, {Narayan}, and
  {Johnson}}]{kumar08a}
{Kumar}, P., {Narayan}, R., {Johnson}, J.~L., Jul. 2008{\natexlab{b}}.
  {Properties of Gamma-Ray Burst Progenitor Stars}. Science 321, 376--.

\bibitem[{{Kumar} and {Panaitescu}(2000{\natexlab{a}})}]{kumar00}
{Kumar}, P., {Panaitescu}, A., Oct. 2000{\natexlab{a}}. {Afterglow Emission
  from Naked Gamma-Ray Bursts}. \apjl 541, L51--L54.

\bibitem[{{Kumar} and {Panaitescu}(2000{\natexlab{b}})}]{kumar00b}
{Kumar}, P., {Panaitescu}, A., Sep. 2000{\natexlab{b}}. {Steepening of
  Afterglow Decay for Jets Interacting with Stratified Media}. \apjl 541,
  L9--L12.

\bibitem[{{Kumar} and {Panaitescu}(2003)}]{kumar03}
{Kumar}, P., {Panaitescu}, A., Dec. 2003. {A unified treatment of the gamma-ray
  burst 021211 and its afterglow}. \mnras 346, 905--914.

\bibitem[{{Kumar} and {Panaitescu}(2004)}]{kumarpanaitescu04}
{Kumar}, P., {Panaitescu}, A., Oct. 2004. {Creation of electron-positron wind
  in gamma-ray bursts and its effect on the early afterglow emission}. \mnras
  354, 252--258.

\bibitem[{{Kumar} and {Panaitescu}(2008)}]{kumarpanaitescu08}
{Kumar}, P., {Panaitescu}, A., Nov. 2008. {What did we learn from gamma-ray
  burst 080319B?} \mnras 391, L19--L23.

\bibitem[{{Kumar} and {Piran}(2000{\natexlab{a}})}]{kumarpiran00b}
{Kumar}, P., {Piran}, T., May 2000{\natexlab{a}}. {Energetics and Luminosity
  Function of Gamma-Ray Bursts}. \apj 535, 152--157.

\bibitem[{{Kumar} and {Piran}(2000{\natexlab{b}})}]{kumarpiran00a}
{Kumar}, P., {Piran}, T., Mar. 2000{\natexlab{b}}. {Some Observational
  Consequences of Gamma-Ray Burst Shock Models}. \apj 532, 286--293.

\bibitem[{{Kyutoku} et~al.(2012){Kyutoku}, {Ioka}, and {Shibata}}]{kyutoku13a}
{Kyutoku}, K., {Ioka}, K., {Shibata}, M., Sep. 2012. {Ultra-Relativistic
  Counterparts to Binary Neutron Star Mergers in Every Direction,
  X-ray-to-Radio Bands and Second-to-Day Timescales}. ArXiv e-prints.

\bibitem[{{Kyutoku} et~al.(2013){Kyutoku}, {Ioka}, and {Shibata}}]{kyutoku13b}
{Kyutoku}, K., {Ioka}, K., {Shibata}, M., May 2013. {Anisotropic mass ejection
  from black hole-neutron star binaries: Diversity of electromagnetic
  counterparts}. ArXiv e-prints.

\bibitem[{{Lamb} and {Reichart}(2000)}]{lamb00}
{Lamb}, D.~Q., {Reichart}, D.~E., Jun. 2000. {Gamma-Ray Bursts as a Probe of
  the Very High Redshift Universe}. \apj 536, 1--18.

\bibitem[{{Larrabee} et~al.(2003){Larrabee}, {Lovelace}, and
  {Romanova}}]{larrabee03}
{Larrabee}, D.~A., {Lovelace}, R.~V.~E., {Romanova}, M.~M., Mar. 2003. {Lepton
  Acceleration by Relativistic Collisionless Magnetic Reconnection}. \apj 586,
  72--78.

\bibitem[{{Lazar} et~al.(2009){Lazar}, {Nakar}, and {Piran}}]{lazar09}
{Lazar}, A., {Nakar}, E., {Piran}, T., Apr. 2009. {Gamma-Ray Burst Light Curves
  in the Relativistic Turbulence and Relativistic Subjet Models}. \apjl 695,
  L10--L14.

\bibitem[{{Lazarian} and {Vishniac}(1999)}]{lazarian99}
{Lazarian}, A., {Vishniac}, E.~T., Jun. 1999. {Reconnection in a Weakly
  Stochastic Field}. \apj 517, 700--718.

\bibitem[{{Lazzati}(2005)}]{lazzati05}
{Lazzati}, D., Feb. 2005. {Precursor activity in bright, long BATSE gamma-ray
  bursts}. \mnras 357, 722--731.

\bibitem[{{Lazzati} and {Begelman}(2006)}]{lazzati06}
{Lazzati}, D., {Begelman}, M.~C., Apr. 2006. {Thick Fireballs and the Steep
  Decay in the Early X-Ray Afterglow of Gamma-Ray Bursts}. \apj 641, 972--977.

\bibitem[{{Lazzati} and {Begelman}(2010)}]{lazzati10}
{Lazzati}, D., {Begelman}, M.~C., Dec. 2010. {Non-thermal Emission from the
  Photospheres of Gamma-ray Burst Outflows. I. High-Frequency Tails}. \apj 725,
  1137--1145.

\bibitem[{{Lazzati} et~al.(2000){Lazzati}, {Ghisellini}, {Celotti}, and
  {Rees}}]{lazzati00}
{Lazzati}, D., {Ghisellini}, G., {Celotti}, A., {Rees}, M.~J., Jan. 2000.
  {Compton-dragged Gamma-Ray Bursts Associated with Supernovae}. \apjl 529,
  L17--L20.

\bibitem[{{Lazzati} et~al.(2009){Lazzati}, {Morsony}, and
  {Begelman}}]{lazzati09}
{Lazzati}, D., {Morsony}, B.~J., {Begelman}, M.~C., Jul. 2009. {Very High
  Efficiency Photospheric Emission in Long-Duration {$\gamma$}-Ray Bursts}.
  \apjl 700, L47--L50.

\bibitem[{{Lazzati} et~al.(2013){Lazzati}, {Morsony}, {Margutti}, and
  {Begelman}}]{lazzati13}
{Lazzati}, D., {Morsony}, B.~J., {Margutti}, R., {Begelman}, M.~C., Mar. 2013.
  {Photospheric Emission as the Dominant Radiation Mechanism in Long-duration
  Gamma-Ray Bursts}. \apj 765, 103.

\bibitem[{{Lazzati} and {Perna}(2007)}]{lazzati07}
{Lazzati}, D., {Perna}, R., Feb. 2007. {X-ray flares and the duration of engine
  activity in gamma-ray bursts}. \mnras 375, L46--L50.

\bibitem[{{Lazzati} et~al.(2002){Lazzati}, {Rossi}, {Covino}, {Ghisellini}, and
  {Malesani}}]{lazzati02}
{Lazzati}, D., {Rossi}, E., {Covino}, S., {Ghisellini}, G., {Malesani}, D.,
  Dec. 2002. {The afterglow of GRB 021004: Surfing on density waves}. \aap 396,
  L5--L9.

\bibitem[{{Lee} et~al.(2000){Lee}, {Wijers}, and {Brown}}]{lee00}
{Lee}, H.~K., {Wijers}, R.~A.~M.~J., {Brown}, G.~E., 2000. {The
  Blandford-Znajek process as a central engine for a gamma-ray burst}. \physrep
  325, 83--114.

\bibitem[{{Lee} and {Ramirez-Ruiz}(2007)}]{lee07}
{Lee}, W.~H., {Ramirez-Ruiz}, E., Jan. 2007. {The progenitors of short
  gamma-ray bursts}. New Journal of Physics 9, 17--+.

\bibitem[{{Lee} et~al.(2009){Lee}, {Ramirez-Ruiz}, and
  {L{\'o}pez-C{\'a}mara}}]{lee09}
{Lee}, W.~H., {Ramirez-Ruiz}, E., {L{\'o}pez-C{\'a}mara}, D., Jul. 2009. {Phase
  Transitions and He-Synthesis-Driven Winds in Neutrino Cooled Accretion Disks:
  Prospects for Late Flares in Short Gamma-Ray Bursts}. \apjl 699, L93--L96.

\bibitem[{{Lei} et~al.(2007){Lei}, {Wang}, {Gong}, and {Huang}}]{lei07}
{Lei}, W.~H., {Wang}, D.~X., {Gong}, B.~P., {Huang}, C.~Y., Jun. 2007. {A model
  of the light curves of gamma-ray bursts}. \aap 468, 563--569.

\bibitem[{{Lei} et~al.(2009){Lei}, {Wang}, {Zhang}, {Gan}, {Zou}, and
  {Xie}}]{lei09}
{Lei}, W.~H., {Wang}, D.~X., {Zhang}, L., {Gan}, Z.~M., {Zou}, Y.~C., {Xie},
  Y., Aug. 2009. {Magnetically Torqued Neutrino-dominated Accretion Flows for
  Gamma-ray Bursts}. \apj 700, 1970--1976.

\bibitem[{{Lei} et~al.(2013){Lei}, {Zhang}, and {Liang}}]{lei13}
{Lei}, W.-H., {Zhang}, B., {Liang}, E.-W., Mar. 2013. {Hyperaccreting Black
  Hole as Gamma-Ray Burst Central Engine. I. Baryon Loading in Gamma-Ray Burst
  Jets}. \apj 765, 125.

\bibitem[{{Lemoine} et~al.(2013){Lemoine}, {Li}, and {Wang}}]{lemoine13}
{Lemoine}, M., {Li}, Z., {Wang}, X.-Y., May 2013. {On the magnetisation of
  gamma-ray burst blast waves}. ArXiv e-prints.

\bibitem[{{Lemoine} and {Pelletier}(2003)}]{lemoine03}
{Lemoine}, M., {Pelletier}, G., Jun. 2003. {Particle Transport in Tangled
  Magnetic Fields and Fermi Acceleration at Relativistic Shocks}. \apjl 589,
  L73--L76.

\bibitem[{{Levan} et~al.(2013){Levan}, {Tanvir}, {Fruchter}, {Hjorth}, {Pian},
  {Mazzali}, {Perley}, {Cano}, {Graham}, {Hounsell}, {Cenko}, {Fynbo},
  {Kouveliotou}, {Pe'er}, {Misra}, and {Wiersema}}]{levan13}
{Levan}, A.~J., {Tanvir}, N.~R., {Fruchter}, A.~S., {Hjorth}, J., {Pian}, E.,
  {Mazzali}, P., {Perley}, D.~A., {Cano}, Z., {Graham}, J., {Hounsell}, R.~A.,
  {Cenko}, S.~B., {Fynbo}, J.~P.~U., {Kouveliotou}, C., {Pe'er}, A., {Misra},
  K., {Wiersema}, K., Jul. 2013. {Hubble Space Telescope observations of the
  afterglow, supernova and host galaxy associated with the extremely bright GRB
  130427A}. ArXiv e-prints.

\bibitem[{{Levan} et~al.(2014){Levan}, {Tanvir}, {Starling}, {Wiersema},
  {Page}, {Perley}, {Schulze}, {Wynn}, {Chornock}, {Hjorth}, {Cenko},
  {Fruchter}, {O'Brien}, {Brown}, {Tunnicliffe}, {Malesani}, {Jakobsson},
  {Watson}, {Berger}, {Bersier}, {Cobb}, {Covino}, {Cucchiara}, {de Ugarte
  Postigo}, {Fox}, {Gal-Yam}, {Goldoni}, {Gorosabel}, {Kaper}, {Kr{\"u}hler},
  {Karjalainen}, {Osborne}, {Pian}, {S{\'a}nchez-Ram{\'{\i}}rez}, {Schmidt},
  {Skillen}, {Tagliaferri}, {Th{\"o}ne}, {Vaduvescu}, {Wijers}, and
  {Zauderer}}]{levan14}
{Levan}, A.~J., {Tanvir}, N.~R., {Starling}, R.~L.~C., {Wiersema}, K., {Page},
  K.~L., {Perley}, D.~A., {Schulze}, S., {Wynn}, G.~A., {Chornock}, R.,
  {Hjorth}, J., {Cenko}, S.~B., {Fruchter}, A.~S., {O'Brien}, P.~T., {Brown},
  G.~C., {Tunnicliffe}, R.~L., {Malesani}, D., {Jakobsson}, P., {Watson}, D.,
  {Berger}, E., {Bersier}, D., {Cobb}, B.~E., {Covino}, S., {Cucchiara}, A.,
  {de Ugarte Postigo}, A., {Fox}, D.~B., {Gal-Yam}, A., {Goldoni}, P.,
  {Gorosabel}, J., {Kaper}, L., {Kr{\"u}hler}, T., {Karjalainen}, R.,
  {Osborne}, J.~P., {Pian}, E., {S{\'a}nchez-Ram{\'{\i}}rez}, R., {Schmidt},
  B., {Skillen}, I., {Tagliaferri}, G., {Th{\"o}ne}, C., {Vaduvescu}, O.,
  {Wijers}, R.~A.~M.~J., {Zauderer}, B.~A., Jan. 2014. {A New Population of
  Ultra-long Duration Gamma-Ray Bursts}. \apj 781, 13.

\bibitem[{{Levesque} et~al.(2010){Levesque}, {Bloom}, {Butler}, {Perley},
  {Cenko}, {Prochaska}, {Kewley}, {Bunker}, {Chen}, {Chornock}, {Filippenko},
  {Glazebrook}, {Lopez}, {Masiero}, {Modjaz}, {Morgan}, and
  {Poznanski}}]{levesque10}
{Levesque}, E.~M., {Bloom}, J.~S., {Butler}, N.~R., {Perley}, D.~A., {Cenko},
  S.~B., {Prochaska}, J.~X., {Kewley}, L.~J., {Bunker}, A., {Chen}, H.,
  {Chornock}, R., {Filippenko}, A.~V., {Glazebrook}, K., {Lopez}, S.,
  {Masiero}, J., {Modjaz}, M., {Morgan}, A., {Poznanski}, D., Jan. 2010.
  {GRB090426: the environment of a rest-frame 0.35-s gamma-ray burst at a
  redshift of 2.609}. \mnras 401, 963--972.

\bibitem[{{Levesque} et~al.(2012){Levesque}, {Chornock}, {Soderberg}, {Berger},
  and {Lunnan}}]{levesque12}
{Levesque}, E.~M., {Chornock}, R., {Soderberg}, A.~M., {Berger}, E., {Lunnan},
  R., Oct. 2012. {Host Galaxy Properties of the Subluminous GRB 120422A/SN
  2012bz}. \apj 758, 92.

\bibitem[{{Levinson} and {Begelman}(2013)}]{levinson13}
{Levinson}, A., {Begelman}, M.~C., Feb. 2013. {Collimation and Confinement of
  Magnetic Jets by External Media}. \apj 764, 148.

\bibitem[{{Levinson} and {Bromberg}(2008)}]{levinson08}
{Levinson}, A., {Bromberg}, O., Apr. 2008. {Relativistic Photon Mediated
  Shocks}. Physical Review Letters 100~(13), 131101.

\bibitem[{{Levinson} and {Eichler}(2003)}]{levinson03}
{Levinson}, A., {Eichler}, D., Sep. 2003. {Baryon Loading of Gamma-Ray Burst by
  Neutron Pickup}. \apjl 594, L19--L22.

\bibitem[{{Levinson} et~al.(2002){Levinson}, {Ofek}, {Waxman}, and
  {Gal-Yam}}]{levinson02}
{Levinson}, A., {Ofek}, E.~O., {Waxman}, E., {Gal-Yam}, A., Sep. 2002. {Orphan
  Gamma-Ray Burst Radio Afterglows: Candidates and Constraints on Beaming}.
  \apj 576, 923--931.

\bibitem[{{Li} and {Fenimore}(1996)}]{lidermer96}
{Li}, H., {Fenimore}, E.~E., Oct. 1996. {Log-normal Distributions in Gamma-Ray
  Burst Time Histories}. \apjl 469, L115.

\bibitem[{{Li}(2000)}]{li00b}
{Li}, L.-X., Apr. 2000. {Toy model for the Blandford-Znajek mechanism}. \prd
  61~(8), 084016.

\bibitem[{{Li} and {Paczy{\'n}ski}(1998)}]{lipaczynski98}
{Li}, L.-X., {Paczy{\'n}ski}, B., Nov. 1998. {Transient Events from Neutron
  Star Mergers}. \apjl 507, L59--L62.

\bibitem[{{Li} et~al.(2003){Li}, {Filippenko}, {Chornock}, and {Jha}}]{li03}
{Li}, W., {Filippenko}, A.~V., {Chornock}, R., {Jha}, S., Mar. 2003. {The Early
  Light Curve of the Optical Afterglow of GRB 021211}. \apjl 586, L9--L12.

\bibitem[{{Li}(2010)}]{li10}
{Li}, Z., Jan. 2010. {Prompt GeV Emission from Residual Collisions in Gamma-Ray
  Burst Outflows: Evidence from Fermi Observations of Grb 080916c}. \apj 709,
  525--534.

\bibitem[{{Li}(2012)}]{lizhuo12}
{Li}, Z., Jan. 2012. {Note on the normalization of predicted gamma-ray burst
  neutrino flux}. \prd 85~(2), 027301.

\bibitem[{{Li} and {Waxman}(2008)}]{liwaxman08}
{Li}, Z., {Waxman}, E., Feb. 2008. {Prompt Optical Emission from Residual
  Collisions in Gamma-Ray Burst Outflows}. \apjl 674, L65--L68.

\bibitem[{{Li} et~al.(1992){Li}, {Chiueh}, and {Begelman}}]{li92}
{Li}, Z.-Y., {Chiueh}, T., {Begelman}, M.~C., Aug. 1992. {Electromagnetically
  driven relativistic jets - A class of self-similar solutions}. \apj 394,
  459--471.

\bibitem[{{Liang} and {Zhang}(2005)}]{liangzhang05}
{Liang}, E., {Zhang}, B., Nov. 2005. {Model-independent Multivariable Gamma-Ray
  Burst Luminosity Indicator and Its Possible Cosmological Implications}. \apj
  633, 611--623.

\bibitem[{{Liang} and {Zhang}(2006)}]{liangzhang06b}
{Liang}, E., {Zhang}, B., Jun. 2006. {Calibration of gamma-ray burst luminosity
  indicators}. \mnras 369, L37--L41.

\bibitem[{{Liang} et~al.(2007{\natexlab{a}}){Liang}, {Zhang}, {Virgili}, and
  {Dai}}]{liang07}
{Liang}, E., {Zhang}, B., {Virgili}, F., {Dai}, Z.~G., Jun. 2007{\natexlab{a}}.
  {Low-Luminosity Gamma-Ray Bursts as a Unique Population: Luminosity Function,
  Local Rate, and Beaming Factor}. \apj 662, 1111--1118.

\bibitem[{{Liang} et~al.(2004){Liang}, {Dai}, and {Wu}}]{liang04}
{Liang}, E.~W., {Dai}, Z.~G., {Wu}, X.~F., May 2004. {The Luminosity-$E_{p}$
  Relation within Gamma-Ray Bursts and the Implications for Fireball Models}.
  \apjl 606, L29--L32.

\bibitem[{{Liang} et~al.(2009){Liang}, {L{\"u}}, {Hou}, {Zhang}, and
  {Zhang}}]{liang09}
{Liang}, E.-W., {L{\"u}}, H.-J., {Hou}, S.-J., {Zhang}, B.-B., {Zhang}, B.,
  Dec. 2009. {A Comprehensive Analysis of Swift/X-Ray Telescope Data. IV.
  Single Power-Law Decaying Light Curves Versus Canonical Light Curves and
  Implications for a Unified Origin of X-Rays}. \apj 707, 328--342.

\bibitem[{{Liang} et~al.(2008{\natexlab{a}}){Liang}, {Racusin}, {Zhang},
  {Zhang}, and {Burrows}}]{liang08}
{Liang}, E.-W., {Racusin}, J.~L., {Zhang}, B., {Zhang}, B.-B., {Burrows},
  D.~N., Mar. 2008{\natexlab{a}}. {A Comprehensive Analysis of Swift XRT Data.
  III. Jet Break Candidates in X-Ray and Optical Afterglow Light Curves}. \apj
  675, 528--552.

\bibitem[{{Liang} et~al.(2010){Liang}, {Yi}, {Zhang}, {L{\"u}}, {Zhang}, and
  {Zhang}}]{liang10}
{Liang}, E.-W., {Yi}, S.-X., {Zhang}, J., {L{\"u}}, H.-J., {Zhang}, B.-B.,
  {Zhang}, B., Dec. 2010. {Constraining Gamma-ray Burst Initial Lorentz Factor
  with the Afterglow Onset Feature and Discovery of a Tight {$\Gamma$}$_{0}$-E
  $_{≥,iso}$ Correlation}. \apj 725, 2209--2224.

\bibitem[{{Liang} et~al.(2006{\natexlab{a}}){Liang}, {Zhang}, {O'Brien},
  {Willingale}, {Angelini}, {Burrows}, {Campana}, {Chincarini}, {Falcone},
  {Gehrels}, {Goad}, {Grupe}, {Kobayashi}, {M{\'e}sz{\'a}ros}, {Nousek},
  {Osborne}, {Page}, and {Tagliaferri}}]{liang06}
{Liang}, E.~W., {Zhang}, B., {O'Brien}, P.~T., {Willingale}, R., {Angelini},
  L., {Burrows}, D.~N., {Campana}, S., {Chincarini}, G., {Falcone}, A.,
  {Gehrels}, N., {Goad}, M.~R., {Grupe}, D., {Kobayashi}, S.,
  {M{\'e}sz{\'a}ros}, P., {Nousek}, J.~A., {Osborne}, J.~P., {Page}, K.~L.,
  {Tagliaferri}, G., Jul. 2006{\natexlab{a}}. {Testing the Curvature Effect and
  Internal Origin of Gamma-Ray Burst Prompt Emissions and X-Ray Flares with
  Swift Data}. \apj 646, 351--357.

\bibitem[{{Liang} et~al.(2006{\natexlab{b}}){Liang}, {Zhang}, {Stamatikos},
  {Zhang}, {Norris}, {Gehrels}, {Zhang}, and {Dai}}]{liang06b}
{Liang}, E.-W., {Zhang}, B.-B., {Stamatikos}, M., {Zhang}, B., {Norris}, J.,
  {Gehrels}, N., {Zhang}, J., {Dai}, Z.~G., Dec. 2006{\natexlab{b}}. {Temporal
  Profiles and Spectral Lags of XRF 060218}. \apjl 653, L81--L84.

\bibitem[{{Liang} et~al.(2007{\natexlab{b}}){Liang}, {Zhang}, and
  {Zhang}}]{liang07b}
{Liang}, E.-W., {Zhang}, B.-B., {Zhang}, B., Nov. 2007{\natexlab{b}}. {A
  Comprehensive Analysis of Swift XRT Data. II. Diverse Physical Origins of the
  Shallow Decay Segment}. \apj 670, 565--583.

\bibitem[{{Liang} et~al.(2008{\natexlab{b}}){Liang}, {Xiao}, {Liu}, and
  {Zhang}}]{liangn08}
{Liang}, N., {Xiao}, W.~K., {Liu}, Y., {Zhang}, S.~N., Sep. 2008{\natexlab{b}}.
  {A Cosmology-Independent Calibration of Gamma-Ray Burst Luminosity Relations
  and the Hubble Diagram}. \apj 685, 354--360.

\bibitem[{{Lindner} et~al.(2010){Lindner}, {Milosavljevi{\'c}}, {Couch}, and
  {Kumar}}]{lindner10}
{Lindner}, C.~C., {Milosavljevi{\'c}}, M., {Couch}, S.~M., {Kumar}, P., Apr.
  2010. {Collapsar Accretion and the Gamma-Ray Burst X-Ray Light Curve}. \apj
  713, 800--815.

\bibitem[{{Lipkin} et~al.(2004){Lipkin}, {Ofek}, {Gal-Yam}, {Leibowitz},
  {Poznanski}, {Kaspi}, {Polishook}, {Kulkarni}, {Fox}, {Berger}, {Mirabal},
  {Halpern}, {Bureau}, {Fathi}, {Price}, {Peterson}, {Frebel}, {Schmidt},
  {Orosz}, {Fitzgerald}, {Bloom}, {van Dokkum}, {Bailyn}, {Buxton}, and
  {Barsony}}]{lipkin04}
{Lipkin}, Y.~M., {Ofek}, E.~O., {Gal-Yam}, A., {Leibowitz}, E.~M., {Poznanski},
  D., {Kaspi}, S., {Polishook}, D., {Kulkarni}, S.~R., {Fox}, D.~W., {Berger},
  E., {Mirabal}, N., {Halpern}, J., {Bureau}, M., {Fathi}, K., {Price}, P.~A.,
  {Peterson}, B.~A., {Frebel}, A., {Schmidt}, B., {Orosz}, J.~A., {Fitzgerald},
  J.~B., {Bloom}, J.~S., {van Dokkum}, P.~G., {Bailyn}, C.~D., {Buxton}, M.~M.,
  {Barsony}, M., May 2004. {The Detailed Optical Light Curve of GRB 030329}.
  \apj 606, 381--394.

\bibitem[{{Lithwick} and {Sari}(2001)}]{lithwick01}
{Lithwick}, Y., {Sari}, R., Jul. 2001. {Lower Limits on Lorentz Factors in
  Gamma-Ray Bursts}. \apj 555, 540--545.

\bibitem[{{Littlejohns} et~al.(2013){Littlejohns}, {Tanvir}, {Willingale},
  {Evans}, {O'Brien}, and {Levan}}]{littlejohns13}
{Littlejohns}, O.~M., {Tanvir}, N.~R., {Willingale}, R., {Evans}, P.~A.,
  {O'Brien}, P.~T., {Levan}, A.~J., Dec. 2013. {Are gamma-ray bursts the same
  at high redshift and low redshift?} \mnras 436, 3640--3655.

\bibitem[{{Liu} and {Wang}(2011)}]{liu11}
{Liu}, R.-Y., {Wang}, X.-Y., Mar. 2011. {Modeling the Broadband Emission of GRB
  090902B}. \apj 730, 1.

\bibitem[{{Liu} et~al.(2013){Liu}, {Wang}, and {Wu}}]{liu13}
{Liu}, R.-Y., {Wang}, X.-Y., {Wu}, X.-F., Aug. 2013. {Interpretation of the
  Unprecedentedly Long-lived High-energy Emission of GRB 130427A}. \apjl 773,
  L20.

\bibitem[{{Liu} et~al.(2010){Liu}, {Liang}, {Gu}, {Zhao}, {Dai}, and
  {Lu}}]{liu10}
{Liu}, T., {Liang}, E.-W., {Gu}, W.-M., {Zhao}, X.-H., {Dai}, Z.-G., {Lu},
  J.-F., Jun. 2010. {Jet precession driven by neutrino-cooled disk for
  gamma-ray bursts}. \aap 516, A16.

\bibitem[{{Liu} et~al.(2011){Liu}, {Li}, {Yin}, {Albright}, {Bowers}, and
  {Liang}}]{liu11b}
{Liu}, W., {Li}, H., {Yin}, L., {Albright}, B.~J., {Bowers}, K.~J., {Liang},
  E.~P., May 2011. {Particle energization in 3D magnetic reconnection of
  relativistic pair plasmas}. Physics of Plasmas 18~(5), 052105.

\bibitem[{{Liu} et~al.(2008){Liu}, {Shapiro}, {Etienne}, and
  {Taniguchi}}]{liu08}
{Liu}, Y.~T., {Shapiro}, S.~L., {Etienne}, Z.~B., {Taniguchi}, K., Jul. 2008.
  {General relativistic simulations of magnetized binary neutron star mergers}.
  \prd 78~(2), 024012.

\bibitem[{{Livio} and {Waxman}(2000)}]{livio00}
{Livio}, M., {Waxman}, E., Jul. 2000. {Toward a Model for the Progenitors of
  Gamma-Ray Bursts}. \apj 538, 187--191.

\bibitem[{{Lloyd-Ronning} et~al.(2004){Lloyd-Ronning}, {Dai}, and
  {Zhang}}]{lloydronning04}
{Lloyd-Ronning}, N.~M., {Dai}, X., {Zhang}, B., Jan. 2004. {On the Structure of
  Quasi-universal Jets for Gamma-Ray Bursts}. \apj 601, 371--379.

\bibitem[{{Longair}(2010)}]{longair10}
{Longair}, M.~S., 2010. {High Energy Astrophysics}. Cambridge University Press,
  2010.

\bibitem[{{Lorimer}(2005)}]{lorimer05}
{Lorimer}, D.~R., Nov. 2005. {Binary and Millisecond Pulsars}. Living Reviews
  in Relativity 8, 7.

\bibitem[{{Loureiro} et~al.(2007){Loureiro}, {Schekochihin}, and
  {Cowley}}]{loureiro07}
{Loureiro}, N.~F., {Schekochihin}, A.~A., {Cowley}, S.~C., Oct. 2007.
  {Instability of current sheets and formation of plasmoid chains}. Physics of
  Plasmas 14~(10), 100703.

\bibitem[{{L{\"u}} et~al.(2010){L{\"u}}, {Liang}, {Zhang}, and {Zhang}}]{lv10}
{L{\"u}}, H.-J., {Liang}, E.-W., {Zhang}, B.-B., {Zhang}, B., Dec. 2010. {A New
  Classification Method for Gamma-ray Bursts}. \apj 725, 1965--1970.

\bibitem[{{L{\"u}} and {Zhang}(2014)}]{luzhang14}
{L{\"u}}, H.-J., {Zhang}, B., Apr. 2014. {A Test of the Millisecond Magnetar
  Central Engine Model of Gamma-Ray Bursts with Swift Data}. \apj 785, 74.

\bibitem[{{L{\"u}} et~al.(2014){L{\"u}}, {Zhang}, {Liang}, {Zhang}, and
  {Sakamoto}}]{lv14}
{L{\"u}}, H.-J., {Zhang}, B., {Liang}, E.-W., {Zhang}, B.-B., {Sakamoto}, T.,
  Aug. 2014. {The `amplitude' parameter of gamma-ray bursts and its
  implications for GRB classification}. \mnras 442, 1922--1929.

\bibitem[{{L{\"u}} et~al.(2012){L{\"u}}, {Zou}, {Lei}, {Zhang}, {Wu}, {Wang},
  {Liang}, and {L{\"u}}}]{lv12}
{L{\"u}}, J., {Zou}, Y.-C., {Lei}, W.-H., {Zhang}, B., {Wu}, Q., {Wang}, D.-X.,
  {Liang}, E.-W., {L{\"u}}, H.-J., May 2012.
  {Lorentz-factor-Isotropic-luminosity/Energy Correlations of Gamma-Ray Bursts
  and Their Interpretation}. \apj 751, 49.

\bibitem[{{Lu} et~al.(2010){Lu}, {Hou}, and {Liang}}]{lu10}
{Lu}, R.-J., {Hou}, S.-J., {Liang}, E.-W., Sep. 2010. {The E $_{p}$-flux
  Correlation in the Rising and Decaying Phases of gamma-ray Burst Pulses:
  Evidence for Viewing Angle Effect?} \apj 720, 1146--1154.

\bibitem[{{Lu} et~al.(2012){Lu}, {Wei}, {Liang}, {Zhang}, {L{\"u}}, {L{\"u}},
  {Lei}, and {Zhang}}]{lu12}
{Lu}, R.-J., {Wei}, J.-J., {Liang}, E.-W., {Zhang}, B.-B., {L{\"u}}, H.-J.,
  {L{\"u}}, L.-Z., {Lei}, W.-H., {Zhang}, B., Sep. 2012. {A Comprehensive
  Analysis of Fermi Gamma-Ray Burst Data. II. E $_{p}$ Evolution Patterns and
  Implications for the Observed Spectrum-Luminosity Relations}. \apj 756, 112.

\bibitem[{{Lundman} et~al.(2013){Lundman}, {Pe'er}, and {Ryde}}]{lundman13}
{Lundman}, C., {Pe'er}, A., {Ryde}, F., Jan. 2013. {A theory of photospheric
  emission from relativistic, collimated outflows}. \mnras 428, 2430--2442.

\bibitem[{{Lynden-Bell}(1996)}]{lynden-bell96}
{Lynden-Bell}, D., Mar. 1996. {Magnetic collimation by accretion discs of
  quasars and stars}. \mnras 279, 389--401.

\bibitem[{{Lyons} et~al.(2010){Lyons}, {O'Brien}, {Zhang}, {Willingale},
  {Troja}, and {Starling}}]{lyons10}
{Lyons}, N., {O'Brien}, P.~T., {Zhang}, B., {Willingale}, R., {Troja}, E.,
  {Starling}, R.~L.~C., Feb. 2010. {Can X-ray emission powered by a
  spinning-down magnetar explain some gamma-ray burst light-curve features?}
  \mnras 402, 705--712.

\bibitem[{{Lyubarsky}(2005)}]{lyubarsky05}
{Lyubarsky}, Y.~E., Mar. 2005. {On the relativistic magnetic reconnection}.
  \mnras 358, 113--119.

\bibitem[{{Lyubarsky}(2010)}]{lyubarsky10}
{Lyubarsky}, Y.~E., Feb. 2010. {Transformation of the Poynting flux into
  kinetic energy in relativistic jets}. \mnras 402, 353--361.

\bibitem[{{Lyutikov}(2006)}]{lyutikov06}
{Lyutikov}, M., Jun. 2006. {Did Swift measure gamma-ray burst prompt emission
  radii?} \mnras 369, L5--L8.

\bibitem[{{Lyutikov} and {Blandford}(2003)}]{lyutikov03}
{Lyutikov}, M., {Blandford}, R., Dec. 2003. {Gamma Ray Bursts as
  Electromagnetic Outflows}. ArXiv Astrophysics e-prints.

\bibitem[{{Lyutikov} and {Uzdensky}(2003)}]{lyutikov03c}
{Lyutikov}, M., {Uzdensky}, D., Jun. 2003. {Dynamics of Relativistic
  Reconnection}. \apj 589, 893--901.

\bibitem[{{MacFadyen} et~al.(2005){MacFadyen}, {Ramirez-Ruiz}, and
  {Zhang}}]{macfadyen05}
{MacFadyen}, A.~I., {Ramirez-Ruiz}, E., {Zhang}, W., Oct. 2005. {X-ray flares
  following short gamma-ray bursts from shock heating of binary stellar
  companions}. ArXiv Astrophysics e-prints.

\bibitem[{{MacFadyen} and {Woosley}(1999)}]{macfadyen99}
{MacFadyen}, A.~I., {Woosley}, S.~E., Oct. 1999. {Collapsars: Gamma-Ray Bursts
  and Explosions in ``Failed Supernovae''}. \apj 524, 262--289.

\bibitem[{{Madau} and {Thompson}(2000)}]{madau00}
{Madau}, P., {Thompson}, C., May 2000. {Relativistic Winds from Compact
  Gamma-Ray Sources. I. Radiative Acceleration in the Klein-Nishina Regime}.
  \apj 534, 239--247.

\bibitem[{{Malesani} et~al.(2004){Malesani}, {Tagliaferri}, {Chincarini},
  {Covino}, {Della Valle}, {Fugazza}, {Mazzali}, {Zerbi}, {D'Avanzo},
  {Kalogerakos}, {Simoncelli}, {Antonelli}, {Burderi}, {Campana}, {Cucchiara},
  {Fiore}, {Ghirlanda}, {Goldoni}, {G{\"o}tz}, {Mereghetti}, {Mirabel},
  {Romano}, {Stella}, {Minezaki}, {Yoshii}, and {Nomoto}}]{malesani04}
{Malesani}, D., {Tagliaferri}, G., {Chincarini}, G., {Covino}, S., {Della
  Valle}, M., {Fugazza}, D., {Mazzali}, P.~A., {Zerbi}, F.~M., {D'Avanzo}, P.,
  {Kalogerakos}, S., {Simoncelli}, A., {Antonelli}, L.~A., {Burderi}, L.,
  {Campana}, S., {Cucchiara}, A., {Fiore}, F., {Ghirlanda}, G., {Goldoni}, P.,
  {G{\"o}tz}, D., {Mereghetti}, S., {Mirabel}, I.~F., {Romano}, P., {Stella},
  L., {Minezaki}, T., {Yoshii}, Y., {Nomoto}, K., Jul. 2004. {SN 2003lw and GRB
  031203: A Bright Supernova for a Faint Gamma-Ray Burst}. \apjl 609, L5--L8.

\bibitem[{{Mangano} et~al.(2007){Mangano}, {La Parola}, {Cusumano}, {Mineo},
  {Malesani}, {Dyks}, {Campana}, {Capalbi}, {Chincarini}, {Giommi}, {Moretti},
  {Perri}, {Romano}, {Tagliaferri}, {Burrows}, {Gehrels}, {Godet}, {Holland},
  {Kennea}, {Page}, {Racusin}, {Roming}, and {Zhang}}]{mangano07}
{Mangano}, V., {La Parola}, V., {Cusumano}, G., {Mineo}, T., {Malesani}, D.,
  {Dyks}, J., {Campana}, S., {Capalbi}, M., {Chincarini}, G., {Giommi}, P.,
  {Moretti}, A., {Perri}, M., {Romano}, P., {Tagliaferri}, G., {Burrows},
  D.~N., {Gehrels}, N., {Godet}, O., {Holland}, S.~T., {Kennea}, J.~A., {Page},
  K.~L., {Racusin}, J.~L., {Roming}, P.~W.~A., {Zhang}, B., Jan. 2007. {Swift
  XRT Observations of the Afterglow of XRF 050416A}. \apj 654, 403--412.

\bibitem[{{Mangano} and {Sbarufatti}(2011)}]{mangano11}
{Mangano}, V., {Sbarufatti}, B., Apr. 2011. {Modeling the spectral evolution in
  the decaying tail of gamma-ray bursts observed by Swift}. Advances in Space
  Research 47, 1367--1373.

\bibitem[{{Mao} and {Paczynski}(1992)}]{mao92}
{Mao}, S., {Paczynski}, B., Apr. 1992. {On the cosmological origin of gamma-ray
  bursts}. \apjl 388, L45--L48.

\bibitem[{{Margutti} et~al.(2011){Margutti}, {Bernardini}, {Barniol Duran},
  {Guidorzi}, {Shen}, and {Chincarini}}]{margutti11}
{Margutti}, R., {Bernardini}, G., {Barniol Duran}, R., {Guidorzi}, C., {Shen},
  R.~F., {Chincarini}, G., Jan. 2011. {On the average gamma-ray burst X-ray
  flaring activity}. \mnras 410, 1064--1075.

\bibitem[{{Margutti} et~al.(2010){Margutti}, {Guidorzi}, {Chincarini},
  {Bernardini}, {Genet}, {Mao}, and {Pasotti}}]{margutti10}
{Margutti}, R., {Guidorzi}, C., {Chincarini}, G., {Bernardini}, M.~G., {Genet},
  F., {Mao}, J., {Pasotti}, F., Aug. 2010. {Lag-luminosity relation in
  {$\gamma$}-ray burst X-ray flares: a direct link to the prompt emission}.
  \mnras 406, 2149--2167.

\bibitem[{{Matzner}(2003)}]{matzner03}
{Matzner}, C.~D., Oct. 2003. {Supernova hosts for gamma-ray burst jets:
  dynamical constraints}. \mnras 345, 575--589.

\bibitem[{{Maxham} and {Zhang}(2009)}]{maxham09}
{Maxham}, A., {Zhang}, B., Dec. 2009. {Modeling Gamma-Ray Burst X-Ray Flares
  Within the Internal Shock Model}. \apj 707, 1623--1633.

\bibitem[{{Maxham} et~al.(2011){Maxham}, {Zhang}, and {Zhang}}]{maxham11}
{Maxham}, A., {Zhang}, B.-B., {Zhang}, B., Jul. 2011. {Is GeV emission from
  Gamma-Ray Bursts of external shock origin?} \mnras 415, 77--82.

\bibitem[{{McBreen} et~al.(1994){McBreen}, {Hurley}, {Long}, and
  {Metcalfe}}]{mcbreen94}
{McBreen}, B., {Hurley}, K.~J., {Long}, R., {Metcalfe}, L., Dec. 1994.
  {Lognormal Distributions in Gamma-Ray Bursts and Cosmic Lightning}. \mnras
  271, 662.

\bibitem[{{McBreen} et~al.(2008){McBreen}, {Foley}, {Watson}, {Hanlon},
  {Malesani}, {Fynbo}, {Kann}, {Gehrels}, {McGlynn}, and {Palmer}}]{mcbreen08}
{McBreen}, S., {Foley}, S., {Watson}, D., {Hanlon}, L., {Malesani}, D.,
  {Fynbo}, J.~P.~U., {Kann}, D.~A., {Gehrels}, N., {McGlynn}, S., {Palmer}, D.,
  Apr. 2008. {The Spectral Lag of GRB 060505: A Likely Member of the
  Long-Duration Class}. \apjl 677, L85--L88.

\bibitem[{{McGlynn} et~al.(2007){McGlynn}, {Clark}, {Dean}, {Hanlon},
  {McBreen}, {Willis}, {McBreen}, {Bird}, and {Foley}}]{mcglynn07}
{McGlynn}, S., {Clark}, D.~J., {Dean}, A.~J., {Hanlon}, L., {McBreen}, S.,
  {Willis}, D.~R., {McBreen}, B., {Bird}, A.~J., {Foley}, S., May 2007.
  {Polarisation studies of the prompt gamma-ray emission from GRB 041219a using
  the spectrometer aboard INTEGRAL}. \aap 466, 895--904.

\bibitem[{{McKinney}(2005)}]{mckinney05}
{McKinney}, J.~C., Sep. 2005. {Total and Jet Blandford-Znajek Power in the
  Presence of an Accretion Disk}. \apjl 630, L5--L8.

\bibitem[{{McKinney}(2006)}]{mckinney06}
{McKinney}, J.~C., Apr. 2006. {General relativistic force-free electrodynamics:
  a new code and applications to black hole magnetospheres}. \mnras 367,
  1797--1807.

\bibitem[{{McKinney} and {Blandford}(2009)}]{mckinney09}
{McKinney}, J.~C., {Blandford}, R.~D., Mar. 2009. {Stability of relativistic
  jets from rotating, accreting black holes via fully three-dimensional
  magnetohydrodynamic simulations}. \mnras 394, L126--L130.

\bibitem[{{McKinney} and {Narayan}(2007)}]{mckinney07}
{McKinney}, J.~C., {Narayan}, R., Feb. 2007. {Disc-jet coupling in black hole
  accretion systems - I. General relativistic magnetohydrodynamical models}.
  \mnras 375, 513--530.

\bibitem[{{McKinney} and {Uzdensky}(2012)}]{mckinney12}
{McKinney}, J.~C., {Uzdensky}, D.~A., Jan. 2012. {A reconnection switch to
  trigger gamma-ray burst jet dissipation}. \mnras 419, 573--607.

\bibitem[{{McMahon} et~al.(2006){McMahon}, {Kumar}, and {Piran}}]{mcmahon06}
{McMahon}, E., {Kumar}, P., {Piran}, T., Feb. 2006. {Reverse shock emission as
  a probe of gamma-ray burst ejecta}. \mnras 366, 575--585.

\bibitem[{{Medvedev} and {Loeb}(1999)}]{medvedev99}
{Medvedev}, M.~V., {Loeb}, A., Dec. 1999. {Generation of Magnetic Fields in the
  Relativistic Shock of Gamma-Ray Burst Sources}. \apj 526, 697--706.

\bibitem[{{Meegan} et~al.(2009){Meegan}, {Lichti}, {Bhat}, {Bissaldi},
  {Briggs}, {Connaughton}, {Diehl}, {Fishman}, {Greiner}, {Hoover}, {van der
  Horst}, {von Kienlin}, {Kippen}, {Kouveliotou}, {McBreen}, {Paciesas},
  {Preece}, {Steinle}, {Wallace}, {Wilson}, and {Wilson-Hodge}}]{meegan09}
{Meegan}, C., {Lichti}, G., {Bhat}, P.~N., {Bissaldi}, E., {Briggs}, M.~S.,
  {Connaughton}, V., {Diehl}, R., {Fishman}, G., {Greiner}, J., {Hoover},
  A.~S., {van der Horst}, A.~J., {von Kienlin}, A., {Kippen}, R.~M.,
  {Kouveliotou}, C., {McBreen}, S., {Paciesas}, W.~S., {Preece}, R., {Steinle},
  H., {Wallace}, M.~S., {Wilson}, R.~B., {Wilson-Hodge}, C., Sep. 2009. {The
  Fermi Gamma-ray Burst Monitor}. \apj 702, 791--804.

\bibitem[{{Meegan} et~al.(1992){Meegan}, {Fishman}, {Wilson}, {Horack},
  {Brock}, {Paciesas}, {Pendleton}, and {Kouveliotou}}]{meegan92}
{Meegan}, C.~A., {Fishman}, G.~J., {Wilson}, R.~B., {Horack}, J.~M., {Brock},
  M.~N., {Paciesas}, W.~S., {Pendleton}, G.~N., {Kouveliotou}, C., Jan. 1992.
  {Spatial distribution of gamma-ray bursts observed by BATSE}. \nat 355,
  143--145.

\bibitem[{{Melandri} et~al.(2012){Melandri}, {Pian}, {Ferrero}, {D'Elia},
  {Walker}, {Ghirlanda}, {Covino}, {Amati}, {D'Avanzo}, {Mazzali}, {Della
  Valle}, {Guidorzi}, {Antonelli}, {Bernardini}, {Bersier}, {Bufano},
  {Campana}, {Castro-Tirado}, {Chincarini}, {Deng}, {Filippenko}, {Fugazza},
  {Ghisellini}, {Kouveliotou}, {Maeda}, {Marconi}, {Masetti}, {Nomoto},
  {Palazzi}, {Patat}, {Piranomonte}, {Salvaterra}, {Saviane}, {Starling},
  {Tagliaferri}, {Tanaka}, and {Vergani}}]{melandri12}
{Melandri}, A., {Pian}, E., {Ferrero}, P., {D'Elia}, V., {Walker}, E.~S.,
  {Ghirlanda}, G., {Covino}, S., {Amati}, L., {D'Avanzo}, P., {Mazzali}, P.~A.,
  {Della Valle}, M., {Guidorzi}, C., {Antonelli}, L.~A., {Bernardini}, M.~G.,
  {Bersier}, D., {Bufano}, F., {Campana}, S., {Castro-Tirado}, A.~J.,
  {Chincarini}, G., {Deng}, J., {Filippenko}, A.~V., {Fugazza}, D.,
  {Ghisellini}, G., {Kouveliotou}, C., {Maeda}, K., {Marconi}, G., {Masetti},
  N., {Nomoto}, K., {Palazzi}, E., {Patat}, F., {Piranomonte}, S.,
  {Salvaterra}, R., {Saviane}, I., {Starling}, R.~L.~C., {Tagliaferri}, G.,
  {Tanaka}, M., {Vergani}, S.~D., Nov. 2012. {The optical SN 2012bz associated
  with the long GRB 120422A}. \aap 547, A82.

\bibitem[{{M{\'e}sz{\'a}ros}(2002)}]{meszaros02}
{M{\'e}sz{\'a}ros}, P., 2002. {Theories of Gamma-Ray Bursts}. \araa 40,
  137--169.

\bibitem[{{M\'esz\'aros} et~al.(1993){M\'esz\'aros}, {Laguna}, and
  {Rees}}]{meszaros93}
{M\'esz\'aros}, P., {Laguna}, P., {Rees}, M.~J., Sep. 1993. {Gasdynamics of
  relativistically expanding gamma-ray burst sources - Kinematics, energetics,
  magnetic fields, and efficiency}. \apj 415, 181--190.

\bibitem[{{M{\'e}sz{\'a}ros} et~al.(2001){M{\'e}sz{\'a}ros}, {Ramirez-Ruiz},
  and {Rees}}]{meszaros01}
{M{\'e}sz{\'a}ros}, P., {Ramirez-Ruiz}, E., {Rees}, M.~J., Jun. 2001.
  {e$^{+/-}$ Pair Cascades and Precursors in Gamma-Ray Bursts}. \apj 554,
  660--666.

\bibitem[{{M{\'e}sz{\'a}ros} et~al.(2002){M{\'e}sz{\'a}ros}, {Ramirez-Ruiz},
  {Rees}, and {Zhang}}]{meszaros02b}
{M{\'e}sz{\'a}ros}, P., {Ramirez-Ruiz}, E., {Rees}, M.~J., {Zhang}, B., Oct.
  2002. {X-Ray-rich Gamma-Ray Bursts, Photospheres, and Variability}. \apj 578,
  812--817.

\bibitem[{{Meszaros} and {Rees}(1993)}]{meszarosrees93b}
{Meszaros}, P., {Rees}, M.~J., Dec. 1993. {Gamma-Ray Bursts: Multiwaveband
  Spectral Predictions for Blast Wave Models}. \apjl 418, L59.

\bibitem[{{M\'esz\'aros} and {Rees}(1993)}]{meszarosrees93}
{M\'esz\'aros}, P., {Rees}, M.~J., Mar. 1993. {Relativistic fireballs and their
  impact on external matter - Models for cosmological gamma-ray bursts}. \apj
  405, 278--284.

\bibitem[{{M\'esz\'aros} and {Rees}(1994)}]{meszarosrees94}
{M\'esz\'aros}, P., {Rees}, M.~J., Jul. 1994. {Delayed GEV Emission from
  Cosmological Gamma-Ray Bursts - Impact of a Relativistic Wind on External
  Matter}. \mnras 269, L41+.

\bibitem[{{M\'esz\'aros} and {Rees}(1997{\natexlab{a}})}]{meszarosrees97}
{M\'esz\'aros}, P., {Rees}, M.~J., Feb. 1997{\natexlab{a}}. {Optical and
  Long-Wavelength Afterglow from Gamma-Ray Bursts}. \apj 476, 232.

\bibitem[{{M\'esz\'aros} and {Rees}(1997{\natexlab{b}})}]{meszarosrees97b}
{M\'esz\'aros}, P., {Rees}, M.~J., Jun. 1997{\natexlab{b}}. {Poynting Jets from
  Black Holes and Cosmological Gamma-Ray Bursts}. \apjl 482, L29.

\bibitem[{{M{\'e}sz{\'a}ros} and {Rees}(1999)}]{meszarosrees99}
{M{\'e}sz{\'a}ros}, P., {Rees}, M.~J., Jul. 1999. {GRB 990123: reverse and
  internal shock flashes and late afterglow behaviour}. \mnras 306, L39--L43.

\bibitem[{{M{\'e}sz{\'a}ros} and {Rees}(2000{\natexlab{a}})}]{meszarosrees00b}
{M{\'e}sz{\'a}ros}, P., {Rees}, M.~J., Sep. 2000{\natexlab{a}}. {Multi-GEV
  Neutrinos from Internal Dissipation in Gamma-Ray Burst Fireballs}. \apjl 541,
  L5--L8.

\bibitem[{{M{\'e}sz{\'a}ros} and {Rees}(2000{\natexlab{b}})}]{meszarosrees00}
{M{\'e}sz{\'a}ros}, P., {Rees}, M.~J., Feb. 2000{\natexlab{b}}. {Steep Slopes
  and Preferred Breaks in Gamma-Ray Burst Spectra: The Role of Photospheres and
  Comptonization}. \apj 530, 292--298.

\bibitem[{{M{\'e}sz{\'a}ros} and {Rees}(2001)}]{meszarosrees01}
{M{\'e}sz{\'a}ros}, P., {Rees}, M.~J., Jul. 2001. {Collapsar Jets, Bubbles, and
  Fe Lines}. \apjl 556, L37--L40.

\bibitem[{{M{\'e}sz{\'a}ros} and {Rees}(2011)}]{meszarosrees11}
{M{\'e}sz{\'a}ros}, P., {Rees}, M.~J., Jun. 2011. {GeV Emission from
  Collisional Magnetized Gamma-Ray Bursts}. \apjl 733, L40.

\bibitem[{{M\'esz\'aros} et~al.(1994){M\'esz\'aros}, {Rees}, and
  {Papathanassiou}}]{meszaros94}
{M\'esz\'aros}, P., {Rees}, M.~J., {Papathanassiou}, H., Sep. 1994. {Spectral
  properties of blast-wave models of gamma-ray burst sources}. \apj 432,
  181--193.

\bibitem[{{M\'esz\'aros} et~al.(1998){M\'esz\'aros}, {Rees}, and
  {Wijers}}]{meszaros98}
{M\'esz\'aros}, P., {Rees}, M.~J., {Wijers}, R.~A.~M.~J., May 1998. {Viewing
  Angle and Environment Effects in Gamma-Ray Bursts: Sources of Afterglow
  Diversity}. \apj 499, 301--+.

\bibitem[{{M{\'e}sz{\'a}ros} and {Waxman}(2001)}]{meszaroswaxman01}
{M{\'e}sz{\'a}ros}, P., {Waxman}, E., Oct. 2001. {TeV Neutrinos from Successful
  and Choked Gamma-Ray Bursts}. Physical Review Letters 87~(17), 171102--+.

\bibitem[{{Metzger} and {Berger}(2012)}]{metzger12}
{Metzger}, B.~D., {Berger}, E., Feb. 2012. {What is the Most Promising
  Electromagnetic Counterpart of a Neutron Star Binary Merger?} \apj 746, 48.

\bibitem[{{Metzger} et~al.(2011){Metzger}, {Giannios}, {Thompson},
  {Bucciantini}, and {Quataert}}]{metzger11}
{Metzger}, B.~D., {Giannios}, D., {Thompson}, T.~A., {Bucciantini}, N.,
  {Quataert}, E., May 2011. {The protomagnetar model for gamma-ray bursts}.
  \mnras 413, 2031--2056.

\bibitem[{{Metzger} et~al.(2010){Metzger}, {Mart{\'{\i}}nez-Pinedo}, {Darbha},
  {Quataert}, {Arcones}, {Kasen}, {Thomas}, {Nugent}, {Panov}, and
  {Zinner}}]{metzger10}
{Metzger}, B.~D., {Mart{\'{\i}}nez-Pinedo}, G., {Darbha}, S., {Quataert}, E.,
  {Arcones}, A., {Kasen}, D., {Thomas}, R., {Nugent}, P., {Panov}, I.~V.,
  {Zinner}, N.~T., Aug. 2010. {Electromagnetic counterparts of compact object
  mergers powered by the radioactive decay of r-process nuclei}. \mnras 406,
  2650--2662.

\bibitem[{{Metzger} and {Piro}(2014)}]{metzger14}
{Metzger}, B.~D., {Piro}, A.~L., Apr. 2014. {Optical and X-ray emission from
  stable millisecond magnetars formed from the merger of binary neutron stars}.
  \mnras 439, 3916--3930.

\bibitem[{{Metzger} et~al.(2008){Metzger}, {Quataert}, and
  {Thompson}}]{metzger08}
{Metzger}, B.~D., {Quataert}, E., {Thompson}, T.~A., Apr. 2008. {Short-duration
  gamma-ray bursts with extended emission from protomagnetar spin-down}. \mnras
  385, 1455--1460.

\bibitem[{{Milgrom} and {Usov}(1995)}]{milgrom95}
{Milgrom}, M., {Usov}, V., Aug. 1995. {Possible Association of
  Ultra--High-Energy Cosmic-Ray Events with Strong Gamma-Ray Bursts}. \apjl
  449, L37+.

\bibitem[{{Milosavljevi{\'c}} and {Nakar}(2006)}]{milosavljevic06a}
{Milosavljevi{\'c}}, M., {Nakar}, E., Nov. 2006. {The Cosmic-Ray Precursor of
  Relativistic Collisionless Shocks: A Missing Link in Gamma-Ray Burst
  Afterglows}. \apj 651, 979--984.

\bibitem[{{Mimica} et~al.(2009){Mimica}, {Giannios}, and {Aloy}}]{mimica09}
{Mimica}, P., {Giannios}, D., {Aloy}, M.~A., Feb. 2009. {Deceleration of
  arbitrarily magnetized GRB ejecta: the complete evolution}. \aap 494,
  879--890.

\bibitem[{{Mizuta} et~al.(2011){Mizuta}, {Nagataki}, and {Aoi}}]{mizuta11}
{Mizuta}, A., {Nagataki}, S., {Aoi}, J., May 2011. {Thermal Radiation from
  Gamma-ray Burst Jets}. \apj 732, 26.

\bibitem[{{Modjaz} et~al.(2008){Modjaz}, {Kewley}, {Kirshner}, {Stanek},
  {Challis}, {Garnavich}, {Greene}, {Kelly}, and {Prieto}}]{modjaz08}
{Modjaz}, M., {Kewley}, L., {Kirshner}, R.~P., {Stanek}, K.~Z., {Challis}, P.,
  {Garnavich}, P.~M., {Greene}, J.~E., {Kelly}, P.~L., {Prieto}, J.~L., Apr.
  2008. {Measured Metallicities at the Sites of Nearby Broad-Lined Type Ic
  Supernovae and Implications for the Supernovae Gamma-Ray Burst Connection}.
  \aj 135, 1136--1150.

\bibitem[{{Modjaz} et~al.(2006){Modjaz}, {Stanek}, {Garnavich}, {Berlind},
  {Blondin}, {Brown}, {Calkins}, {Challis}, {Diamond-Stanic}, {Hao}, {Hicken},
  {Kirshner}, and {Prieto}}]{modjaz06}
{Modjaz}, M., {Stanek}, K.~Z., {Garnavich}, P.~M., {Berlind}, P., {Blondin},
  S., {Brown}, W., {Calkins}, M., {Challis}, P., {Diamond-Stanic}, A.~M.,
  {Hao}, H., {Hicken}, M., {Kirshner}, R.~P., {Prieto}, J.~L., Jul. 2006.
  {Early-Time Photometry and Spectroscopy of the Fast Evolving SN 2006aj
  Associated with GRB 060218}. \apjl 645, L21--L24.

\bibitem[{{Molinari} et~al.(2007){Molinari}, {Vergani}, {Malesani}, {Covino},
  {D'Avanzo}, {Chincarini}, {Zerbi}, {Antonelli}, {Conconi}, {Testa}, {Tosti},
  {Vitali}, {D'Alessio}, {Malaspina}, {Nicastro}, {Palazzi}, {Guetta},
  {Campana}, {Goldoni}, {Masetti}, {Meurs}, {Monfardini}, {Norci}, {Pian},
  {Piranomonte}, {Rizzuto}, {Stefanon}, {Stella}, {Tagliaferri}, {Ward},
  {Ihle}, {Gonzalez}, {Pizarro}, {Sinclaire}, and {Valenzuela}}]{molinari07}
{Molinari}, E., {Vergani}, S.~D., {Malesani}, D., {Covino}, S., {D'Avanzo}, P.,
  {Chincarini}, G., {Zerbi}, F.~M., {Antonelli}, L.~A., {Conconi}, P., {Testa},
  V., {Tosti}, G., {Vitali}, F., {D'Alessio}, F., {Malaspina}, G., {Nicastro},
  L., {Palazzi}, E., {Guetta}, D., {Campana}, S., {Goldoni}, P., {Masetti}, N.,
  {Meurs}, E.~J.~A., {Monfardini}, A., {Norci}, L., {Pian}, E., {Piranomonte},
  S., {Rizzuto}, D., {Stefanon}, M., {Stella}, L., {Tagliaferri}, G., {Ward},
  P.~A., {Ihle}, G., {Gonzalez}, L., {Pizarro}, A., {Sinclaire}, P.,
  {Valenzuela}, J., Jul. 2007. {REM observations of GRB 060418 and GRB 060607A:
  the onset of the afterglow and the initial fireball Lorentz factor
  determination}. \aap 469, L13--L16.

\bibitem[{{Morsony} et~al.(2007){Morsony}, {Lazzati}, and
  {Begelman}}]{morsony07}
{Morsony}, B.~J., {Lazzati}, D., {Begelman}, M.~C., Aug. 2007. {Temporal and
  Angular Properties of Gamma-Ray Burst Jets Emerging from Massive Stars}. \apj
  665, 569--598.

\bibitem[{{M{\"u}cke} et~al.(2003){M{\"u}cke}, {Protheroe}, {Engel}, {Rachen},
  and {Stanev}}]{mucke03}
{M{\"u}cke}, A., {Protheroe}, R.~J., {Engel}, R., {Rachen}, J.~P., {Stanev},
  T., Mar. 2003. {BL Lac objects in the synchrotron proton blazar model}.
  Astroparticle Physics 18, 593--613.

\bibitem[{{Mukherjee} et~al.(1998){Mukherjee}, {Feigelson}, {Jogesh Babu},
  {Murtagh}, {Fraley}, and {Raftery}}]{mukherjee98}
{Mukherjee}, S., {Feigelson}, E.~D., {Jogesh Babu}, G., {Murtagh}, F.,
  {Fraley}, C., {Raftery}, A., Nov. 1998. {Three Types of Gamma-Ray Bursts}.
  \apj 508, 314--327.

\bibitem[{{Mundell} et~al.(2013){Mundell}, {Kopa{\v c}}, {Arnold}, {Steele},
  {Gomboc}, {Kobayashi}, {Harrison}, {Smith}, {Guidorzi}, {Virgili},
  {Melandri}, and {Japelj}}]{mundell13}
{Mundell}, C.~G., {Kopa{\v c}}, D., {Arnold}, D.~M., {Steele}, I.~A., {Gomboc},
  A., {Kobayashi}, S., {Harrison}, R.~M., {Smith}, R.~J., {Guidorzi}, C.,
  {Virgili}, F.~J., {Melandri}, A., {Japelj}, J., Dec. 2013. {Highly polarized
  light from stable ordered magnetic fields in GRB120308A}. \nat 504, 119--121.

\bibitem[{{Mundell} et~al.(2007){Mundell}, {Steele}, {Smith}, {Kobayashi},
  {Melandri}, {Guidorzi}, {Gomboc}, {Mottram}, {Clarke}, {Monfardini},
  {Carter}, and {Bersier}}]{mundell07}
{Mundell}, C.~G., {Steele}, I.~A., {Smith}, R.~J., {Kobayashi}, S., {Melandri},
  A., {Guidorzi}, C., {Gomboc}, A., {Mottram}, C.~J., {Clarke}, D.,
  {Monfardini}, A., {Carter}, D., {Bersier}, D., Mar. 2007. {Early Optical
  Polarization of a Gamma-Ray Burst Afterglow}. Science 315, 1822--.

\bibitem[{{Murase}(2008)}]{murase08b}
{Murase}, K., Nov. 2008. {Prompt high-energy neutrinos from gamma-ray bursts in
  photospheric and synchrotron self-Compton scenarios}. \prd 78~(10), 101302.

\bibitem[{{Murase} et~al.(2012){Murase}, {Asano}, {Terasawa}, and
  {M{\'e}sz{\'a}ros}}]{murase12}
{Murase}, K., {Asano}, K., {Terasawa}, T., {M{\'e}sz{\'a}ros}, P., Feb. 2012.
  {The Role of Stochastic Acceleration in the Prompt Emission of Gamma-Ray
  Bursts: Application to Hadronic Injection}. \apj 746, 164.

\bibitem[{{Murase} and {Ioka}(2008)}]{muraseioka08}
{Murase}, K., {Ioka}, K., Apr. 2008. {Closure Relations for e$^{+/-}$ Pair
  Signatures in Gamma-Ray Bursts}. \apj 676, 1123--1129.

\bibitem[{{Murase} and {Ioka}(2013)}]{murase13}
{Murase}, K., {Ioka}, K., Sep. 2013. {TeV-PeV Neutrinos from Low-Power
  Gamma-Ray Burst Jets inside Stars}. Physical Review Letters 111~(12), 121102.

\bibitem[{{Murase} et~al.(2006){Murase}, {Ioka}, {Nagataki}, and
  {Nakamura}}]{murase06}
{Murase}, K., {Ioka}, K., {Nagataki}, S., {Nakamura}, T., Nov. 2006.
  {High-Energy Neutrinos and Cosmic Rays from Low-Luminosity Gamma-Ray Bursts?}
  \apjl 651, L5--L8.

\bibitem[{{Murase} et~al.(2013){Murase}, {Kashiyama}, and
  {Meszaros}}]{murase13b}
{Murase}, K., {Kashiyama}, K., {Meszaros}, P., Jan. 2013. {Subphotospheric
  Neutrinos from Gamma-Ray Bursts: The Role of Neutrons}. ArXiv e-prints.

\bibitem[{{Murase} and {Nagataki}(2006)}]{murase06b}
{Murase}, K., {Nagataki}, S., Mar. 2006. {High energy neutrino emission and
  neutrino background from gamma-ray bursts in the internal shock model}. \prd
  73~(6), 063002.

\bibitem[{{Murase} et~al.(2009){Murase}, {Zhang}, {Takahashi}, and
  {Nagataki}}]{murase09}
{Murase}, K., {Zhang}, B., {Takahashi}, K., {Nagataki}, S., Jul. 2009.
  {Possible effects of pair echoes on gamma-ray burst afterglow emission}.
  \mnras 396, 1825--1832.

\bibitem[{{Nagakura} et~al.(2014){Nagakura}, {Hotokezaka}, {Sekiguchi},
  {Shibata}, and {Ioka}}]{nagakura14}
{Nagakura}, H., {Hotokezaka}, K., {Sekiguchi}, Y., {Shibata}, M., {Ioka}, K.,
  Apr. 2014. {Jet Collimation in the Ejecta of Double Neutron Star Mergers: A
  New Canonical Picture of Short Gamma-Ray Bursts}. \apjl 784, L28.

\bibitem[{{Nagakura} et~al.(2011){Nagakura}, {Ito}, {Kiuchi}, and
  {Yamada}}]{nagakura11}
{Nagakura}, H., {Ito}, H., {Kiuchi}, K., {Yamada}, S., Apr. 2011. {Jet
  Propagations, Breakouts, and Photospheric Emissions in Collapsing Massive
  Progenitors of Long-duration Gamma-ray Bursts}. \apj 731, 80.

\bibitem[{{Nagakura} et~al.(2012){Nagakura}, {Suwa}, and {Ioka}}]{nagakura12}
{Nagakura}, H., {Suwa}, Y., {Ioka}, K., Aug. 2012. {Population III Gamma-Ray
  Bursts and Breakout Criteria for Accretion-powered Jets}. \apj 754, 85.

\bibitem[{{Nagamine} et~al.(2008){Nagamine}, {Zhang}, and
  {Hernquist}}]{nagamine08}
{Nagamine}, K., {Zhang}, B., {Hernquist}, L., Oct. 2008. {Incidence Rate of
  GRB-Host DLAs at High Redshift}. \apjl 686, L57--L60.

\bibitem[{{Nagataki}(2009)}]{nagataki09}
{Nagataki}, S., Oct. 2009. {Development of a General Relativistic
  Magnetohydrodynamic Code and Its Application to the Central Engine of Long
  Gamma-Ray Bursts}. \apj 704, 937--950.

\bibitem[{{Nagataki}(2011)}]{nagataki11}
{Nagataki}, S., Dec. 2011. {Rotating Black Holes as Central Engines of Long
  Gamma-Ray Bursts: Faster is Better}. \pasj 63, 1243--1249.

\bibitem[{{Nakamura} and {Umemura}(2001)}]{nakamura01}
{Nakamura}, F., {Umemura}, M., Feb. 2001. {On the Initial Mass Function of
  Population III Stars}. \apj 548, 19--32.

\bibitem[{{Nakar}(2007)}]{nakar07}
{Nakar}, E., Apr. 2007. {Short-hard gamma-ray bursts}. \physrep 442, 166--236.

\bibitem[{{Nakar} et~al.(2009){Nakar}, {Ando}, and {Sari}}]{nakar09}
{Nakar}, E., {Ando}, S., {Sari}, R., Sep. 2009. {Klein-Nishina Effects on
  Optically Thin Synchrotron and Synchrotron Self-Compton Spectrum}. \apj 703,
  675--691.

\bibitem[{{Nakar} et~al.(2006{\natexlab{a}}){Nakar}, {Gal-Yam}, and
  {Fox}}]{nakar06}
{Nakar}, E., {Gal-Yam}, A., {Fox}, D.~B., Oct. 2006{\natexlab{a}}. {The Local
  Rate and the Progenitor Lifetimes of Short-Hard Gamma-Ray Bursts: Synthesis
  and Predictions for the Laser Interferometer Gravitational-Wave Observatory}.
  \apj 650, 281--290.

\bibitem[{{Nakar} et~al.(2006{\natexlab{b}}){Nakar}, {Gal-Yam}, {Piran}, and
  {Fox}}]{nakar06b}
{Nakar}, E., {Gal-Yam}, A., {Piran}, T., {Fox}, D.~B., Apr. 2006{\natexlab{b}}.
  {The Distances of Short-Hard Gamma-Ray Bursts and the Soft Gamma-Ray Repeater
  Connection}. \apj 640, 849--853.

\bibitem[{{Nakar} and {Granot}(2007)}]{nakargranot07}
{Nakar}, E., {Granot}, J., Oct. 2007. {Smooth light curves from a bumpy ride:
  relativistic blast wave encounters a density jump}. \mnras 380, 1744--1760.

\bibitem[{{Nakar} et~al.(2004){Nakar}, {Granot}, and {Guetta}}]{nakar04}
{Nakar}, E., {Granot}, J., {Guetta}, D., May 2004. {Testing the Predictions of
  the Universal Structured Gamma-Ray Burst Jet Model}. \apjl 606, L37--L40.

\bibitem[{{Nakar} and {Piran}(2002{\natexlab{a}})}]{nakarpiran02}
{Nakar}, E., {Piran}, T., Mar. 2002{\natexlab{a}}. {Temporal properties of
  short gamma-ray bursts}. \mnras 330, 920--926.

\bibitem[{{Nakar} and {Piran}(2002{\natexlab{b}})}]{nakarpiran02b}
{Nakar}, E., {Piran}, T., Mar. 2002{\natexlab{b}}. {Time-scales in long
  gamma-ray bursts}. \mnras 331, 40--44.

\bibitem[{{Nakar} and {Piran}(2004)}]{nakarpiran04}
{Nakar}, E., {Piran}, T., Sep. 2004. {Early afterglow emission from a reverse
  shock as a diagnostic tool for gamma-ray burst outflows}. \mnras 353,
  647--653.

\bibitem[{{Nakar} and {Piran}(2005)}]{nakarpiran05}
{Nakar}, E., {Piran}, T., Jun. 2005. {Outliers to the peak energy-isotropic
  energy relation in gamma-ray bursts}. \mnras 360, L73--L76.

\bibitem[{{Nakar} and {Piran}(2011)}]{nakar11}
{Nakar}, E., {Piran}, T., Oct. 2011. {Detectable radio flares following
  gravitational waves from mergers of binary neutron stars}. \nat 478, 82--84.

\bibitem[{{Nakar} et~al.(2002){Nakar}, {Piran}, and {Granot}}]{nakar02}
{Nakar}, E., {Piran}, T., {Granot}, J., Nov. 2002. {The Detectability of Orphan
  Afterglows}. \apj 579, 699--705.

\bibitem[{{Nakar} et~al.(2003){Nakar}, {Piran}, and {Granot}}]{nakar03}
{Nakar}, E., {Piran}, T., {Granot}, J., Jul. 2003. {Variability in GRB
  afterglows and GRB 021004}. New Astronomy 8, 495--505.

\bibitem[{{Nakar} et~al.(2005){Nakar}, {Piran}, and {Sari}}]{nakar05}
{Nakar}, E., {Piran}, T., {Sari}, R., Dec. 2005. {Pure and Loaded Fireballs in
  Soft Gamma-Ray Repeater Giant Flares}. \apj 635, 516--521.

\bibitem[{{Nakar} and {Sari}(2012)}]{nakar12}
{Nakar}, E., {Sari}, R., Mar. 2012. {Relativistic Shock Breakouts - A Variety
  of Gamma-Ray Flares: From Low-luminosity Gamma-Ray Bursts to Type Ia
  Supernovae}. \apj 747, 88.

\bibitem[{{Narayan} and {Kumar}(2009)}]{narayan09}
{Narayan}, R., {Kumar}, P., Mar. 2009. {A turbulent model of gamma-ray burst
  variability}. \mnras 394, L117--L120.

\bibitem[{{Narayan} et~al.(2011){Narayan}, {Kumar}, and
  {Tchekhovskoy}}]{narayan11}
{Narayan}, R., {Kumar}, P., {Tchekhovskoy}, A., Sep. 2011. {Constraints on cold
  magnetized shocks in gamma-ray bursts}. \mnras 416, 2193--2201.

\bibitem[{{Narayan} et~al.(1992){Narayan}, {Paczynski}, and
  {Piran}}]{narayan92}
{Narayan}, R., {Paczynski}, B., {Piran}, T., Aug. 1992. {Gamma-ray bursts as
  the death throes of massive binary stars}. \apjl 395, L83--L86.

\bibitem[{{Narayan} et~al.(2001){Narayan}, {Piran}, and {Kumar}}]{narayan01}
{Narayan}, R., {Piran}, T., {Kumar}, P., Aug. 2001. {Accretion Models of
  Gamma-Ray Bursts}. \apj 557, 949--957.

\bibitem[{{Nardini} et~al.(2011){Nardini}, {Greiner}, {Kr{\"u}hler}, {Filgas},
  {Klose}, {Afonso}, {Clemens}, {Guelbenzu}, {Olivares E.}, {Rau}, {Rossi},
  {Updike}, {K{\"u}pc{\"u} Yolda{\c s}}, {Yolda{\c s}}, {Burlon}, {Elliott},
  and {Kann}}]{nardini11}
{Nardini}, M., {Greiner}, J., {Kr{\"u}hler}, T., {Filgas}, R., {Klose}, S.,
  {Afonso}, P., {Clemens}, C., {Guelbenzu}, A.~N., {Olivares E.}, F., {Rau},
  A., {Rossi}, A., {Updike}, A., {K{\"u}pc{\"u} Yolda{\c s}}, A., {Yolda{\c
  s}}, A., {Burlon}, D., {Elliott}, J., {Kann}, D.~A., Jul. 2011. {On the
  nature of the extremely fast optical rebrightening of the afterglow of GRB
  081029}. \aap 531, A39.

\bibitem[{{Nava} et~al.(2011){Nava}, {Ghirlanda}, {Ghisellini}, and
  {Celotti}}]{nava11}
{Nava}, L., {Ghirlanda}, G., {Ghisellini}, G., {Celotti}, A., Jun. 2011.
  {Spectral properties of 438 GRBs detected by Fermi/GBM}. \aap 530, A21.

\bibitem[{{Nava} et~al.(2013){Nava}, {Sironi}, {Ghisellini}, {Celotti}, and
  {Ghirlanda}}]{nava13}
{Nava}, L., {Sironi}, L., {Ghisellini}, G., {Celotti}, A., {Ghirlanda}, G.,
  Aug. 2013. {Afterglow emission in gamma-ray bursts - I. Pair-enriched ambient
  medium and radiative blast waves}. \mnras 433, 2107--2121.

\bibitem[{{Niino} et~al.(2011){Niino}, {Choi}, {Kobayashi}, {Nagamine},
  {Totani}, and {Zhang}}]{niino11}
{Niino}, Y., {Choi}, J.-H., {Kobayashi}, M.~A.~R., {Nagamine}, K., {Totani},
  T., {Zhang}, B., Jan. 2011. {Luminosity Distribution of Gamma-ray Burst Host
  Galaxies at Redshift z = 1 in Cosmological Smoothed Particle Hydrodynamic
  Simulations: Implications for the Metallicity Dependence of GRBs}. \apj 726,
  88.

\bibitem[{{Nomoto} et~al.(2006){Nomoto}, {Tominaga}, {Tanaka}, {Maeda},
  {Suzuki}, {Deng}, and {Mazzali}}]{nomoto06}
{Nomoto}, K., {Tominaga}, N., {Tanaka}, M., {Maeda}, K., {Suzuki}, T., {Deng},
  J.~S., {Mazzali}, P.~A., Oct. 2006. {Diversity of the supernova--gamma-ray
  burst connection}. Nuovo Cimento B Serie 121, 1207--1222.

\bibitem[{{Norris}(2002)}]{norris02}
{Norris}, J.~P., Nov. 2002. {Implications of the Lag-Luminosity Relationship
  for Unified Gamma-Ray Burst Paradigms}. \apj 579, 386--403.

\bibitem[{{Norris} and {Bonnell}(2006)}]{norris06}
{Norris}, J.~P., {Bonnell}, J.~T., May 2006. {Short Gamma-Ray Bursts with
  Extended Emission}. \apj 643, 266--275.

\bibitem[{{Norris} et~al.(2005){Norris}, {Bonnell}, {Kazanas}, {Scargle},
  {Hakkila}, and {Giblin}}]{norris05}
{Norris}, J.~P., {Bonnell}, J.~T., {Kazanas}, D., {Scargle}, J.~D., {Hakkila},
  J., {Giblin}, T.~W., Jul. 2005. {Long-Lag, Wide-Pulse Gamma-Ray Bursts}. \apj
  627, 324--345.

\bibitem[{{Norris} et~al.(2000){Norris}, {Marani}, and {Bonnell}}]{norris00}
{Norris}, J.~P., {Marani}, G.~F., {Bonnell}, J.~T., May 2000. {Connection
  between Energy-dependent Lags and Peak Luminosity in Gamma-Ray Bursts}. \apj
  534, 248--257.

\bibitem[{{Norris} et~al.(1986){Norris}, {Share}, {Messina}, {Dennis}, {Desai},
  {Cline}, {Matz}, and {Chupp}}]{norris86}
{Norris}, J.~P., {Share}, G.~H., {Messina}, D.~C., {Dennis}, B.~R., {Desai},
  U.~D., {Cline}, T.~L., {Matz}, S.~M., {Chupp}, E.~L., Feb. 1986. {Spectral
  evolution of pulse structures in gamma-ray bursts}. \apj 301, 213--219.

\bibitem[{{Nousek} et~al.(2006){Nousek}, {Kouveliotou}, {Grupe}, {Page},
  {Granot}, {Ramirez-Ruiz}, {Patel}, {Burrows}, {Mangano}, {Barthelmy},
  {Beardmore}, {Campana}, {Capalbi}, {Chincarini}, {Cusumano}, {Falcone},
  {Gehrels}, {Giommi}, {Goad}, {Godet}, {Hurkett}, {Kennea}, {Moretti},
  {O'Brien}, {Osborne}, {Romano}, {Tagliaferri}, and {Wells}}]{nousek06}
{Nousek}, J.~A., {Kouveliotou}, C., {Grupe}, D., {Page}, K.~L., {Granot}, J.,
  {Ramirez-Ruiz}, E., {Patel}, S.~K., {Burrows}, D.~N., {Mangano}, V.,
  {Barthelmy}, S., {Beardmore}, A.~P., {Campana}, S., {Capalbi}, M.,
  {Chincarini}, G., {Cusumano}, G., {Falcone}, A.~D., {Gehrels}, N., {Giommi},
  P., {Goad}, M.~R., {Godet}, O., {Hurkett}, C.~P., {Kennea}, J.~A., {Moretti},
  A., {O'Brien}, P.~T., {Osborne}, J.~P., {Romano}, P., {Tagliaferri}, G.,
  {Wells}, A.~A., May 2006. {Evidence for a Canonical Gamma-Ray Burst Afterglow
  Light Curve in the Swift XRT Data}. \apj 642, 389--400.

\bibitem[{{Nysewander} et~al.(2009){Nysewander}, {Fruchter}, and
  {Pe'er}}]{nysewander09}
{Nysewander}, M., {Fruchter}, A.~S., {Pe'er}, A., Aug. 2009. {A Comparison of
  the Afterglows of Short- and Long-duration Gamma-ray Bursts}. \apj 701,
  824--836.

\bibitem[{{O'Brien} et~al.(2006){O'Brien}, {Willingale}, {Osborne}, {Goad},
  {Page}, {Vaughan}, {Rol}, {Beardmore}, {Godet}, {Hurkett}, {Wells}, {Zhang},
  {Kobayashi}, {Burrows}, {Nousek}, {Kennea}, {Falcone}, {Grupe}, {Gehrels},
  {Barthelmy}, {Cannizzo}, {Cummings}, {Hill}, {Krimm}, {Chincarini},
  {Tagliaferri}, {Campana}, {Moretti}, {Giommi}, {Perri}, {Mangano}, and
  {LaParola}}]{obrien06}
{O'Brien}, P.~T., {Willingale}, R., {Osborne}, J., {Goad}, M.~R., {Page},
  K.~L., {Vaughan}, S., {Rol}, E., {Beardmore}, A., {Godet}, O., {Hurkett},
  C.~P., {Wells}, A., {Zhang}, B., {Kobayashi}, S., {Burrows}, D.~N., {Nousek},
  J.~A., {Kennea}, J.~A., {Falcone}, A., {Grupe}, D., {Gehrels}, N.,
  {Barthelmy}, S., {Cannizzo}, J., {Cummings}, J., {Hill}, J.~E., {Krimm}, H.,
  {Chincarini}, G., {Tagliaferri}, G., {Campana}, S., {Moretti}, A., {Giommi},
  P., {Perri}, M., {Mangano}, V., {LaParola}, V., Aug. 2006. {The Early X-Ray
  Emission from GRBs}. \apj 647, 1213--1237.

\bibitem[{{Ott} et~al.(2012){Ott}, {Abdikamalov}, {O'Connor}, {Reisswig},
  {Haas}, {Kalmus}, {Drasco}, {Burrows}, and {Schnetter}}]{ott12}
{Ott}, C.~D., {Abdikamalov}, E., {O'Connor}, E., {Reisswig}, C., {Haas}, R.,
  {Kalmus}, P., {Drasco}, S., {Burrows}, A., {Schnetter}, E., Jul. 2012.
  {Correlated gravitational wave and neutrino signals from general-relativistic
  rapidly rotating iron core collapse}. \prd 86~(2), 024026.

\bibitem[{{Paciesas} et~al.(2012){Paciesas}, {Meegan}, {von Kienlin}, {Bhat},
  {Bissaldi}, {Briggs}, {Burgess}, {Chaplin}, {Connaughton}, {Diehl},
  {Fishman}, {Fitzpatrick}, {Foley}, {Gibby}, {Giles}, {Goldstein}, {Greiner},
  {Gruber}, {Guiriec}, {van der Horst}, {Kippen}, {Kouveliotou}, {Lichti},
  {Lin}, {McBreen}, {Preece}, {Rau}, {Tierney}, and
  {Wilson-Hodge}}]{paciesas12}
{Paciesas}, W.~S., {Meegan}, C.~A., {von Kienlin}, A., {Bhat}, P.~N.,
  {Bissaldi}, E., {Briggs}, M.~S., {Burgess}, J.~M., {Chaplin}, V.,
  {Connaughton}, V., {Diehl}, R., {Fishman}, G.~J., {Fitzpatrick}, G., {Foley},
  S., {Gibby}, M., {Giles}, M., {Goldstein}, A., {Greiner}, J., {Gruber}, D.,
  {Guiriec}, S., {van der Horst}, A.~J., {Kippen}, R.~M., {Kouveliotou}, C.,
  {Lichti}, G., {Lin}, L., {McBreen}, S., {Preece}, R.~D., {Rau}, A.,
  {Tierney}, D., {Wilson-Hodge}, C., Mar. 2012. {The Fermi GBM Gamma-Ray Burst
  Catalog: The First Two Years}. \apjs 199, 18.

\bibitem[{{Pacz\'ynski}(1986)}]{paczynski86}
{Pacz\'ynski}, B., Sep. 1986. {Gamma-ray bursters at cosmological distances}.
  \apjl 308, L43--L46.

\bibitem[{{Pacz\'ynski}(1991)}]{paczynski91}
{Pacz\'ynski}, B., 1991. {Cosmological gamma-ray bursts}. Acta Astronomica 41,
  257--267.

\bibitem[{{Pacz\'ynski}(1998)}]{paczynski98}
{Pacz\'ynski}, B., Feb. 1998. {Are Gamma-Ray Bursts in Star-Forming Regions?}
  \apjl 494, L45+.

\bibitem[{{Pacz\'ynski} and {Rhoads}(1993)}]{paczynski93}
{Pacz\'ynski}, B., {Rhoads}, J.~E., Nov. 1993. {Radio Transients from Gamma-Ray
  Bursters}. \apjl 418, L5+.

\bibitem[{{Page} et~al.(2007){Page}, {Willingale}, {Osborne}, {Zhang}, {Godet},
  {Marshall}, {Melandri}, {Norris}, {O'Brien}, {Pal'shin}, {Rol}, {Romano},
  {Starling}, {Schady}, {Yost}, {Barthelmy}, {Beardmore}, {Cusumano},
  {Burrows}, {De Pasquale}, {Ehle}, {Evans}, {Gehrels}, {Goad}, {Golenetskii},
  {Guidorzi}, {Mundell}, {Page}, {Ricker}, {Sakamoto}, {Schaefer},
  {Stamatikos}, {Troja}, {Ulanov}, {Yuan}, and {Ziaeepour}}]{page07}
{Page}, K.~L., {Willingale}, R., {Osborne}, J.~P., {Zhang}, B., {Godet}, O.,
  {Marshall}, F.~E., {Melandri}, A., {Norris}, J.~P., {O'Brien}, P.~T.,
  {Pal'shin}, V., {Rol}, E., {Romano}, P., {Starling}, R.~L.~C., {Schady}, P.,
  {Yost}, S.~A., {Barthelmy}, S.~D., {Beardmore}, A.~P., {Cusumano}, G.,
  {Burrows}, D.~N., {De Pasquale}, M., {Ehle}, M., {Evans}, P.~A., {Gehrels},
  N., {Goad}, M.~R., {Golenetskii}, S., {Guidorzi}, C., {Mundell}, C., {Page},
  M.~J., {Ricker}, G., {Sakamoto}, T., {Schaefer}, B.~E., {Stamatikos}, M.,
  {Troja}, E., {Ulanov}, M., {Yuan}, F., {Ziaeepour}, H., Jul. 2007. {GRB
  061121: Broadband Spectral Evolution through the Prompt and Afterglow Phases
  of a Bright Burst}. \apj 663, 1125--1138.

\bibitem[{{Palmer} et~al.(2005){Palmer}, {Barthelmy}, {Gehrels}, {Kippen},
  {Cayton}, {Kouveliotou}, {Eichler}, {Wijers}, {Woods}, {Granot}, {Lyubarsky},
  {Ramirez-Ruiz}, {Barbier}, {Chester}, {Cummings}, {Fenimore}, {Finger},
  {Gaensler}, {Hullinger}, {Krimm}, {Markwardt}, {Nousek}, {Parsons}, {Patel},
  {Sakamoto}, {Sato}, {Suzuki}, and {Tueller}}]{palmer05}
{Palmer}, D.~M., {Barthelmy}, S., {Gehrels}, N., {Kippen}, R.~M., {Cayton}, T.,
  {Kouveliotou}, C., {Eichler}, D., {Wijers}, R.~A.~M.~J., {Woods}, P.~M.,
  {Granot}, J., {Lyubarsky}, Y.~E., {Ramirez-Ruiz}, E., {Barbier}, L.,
  {Chester}, M., {Cummings}, J., {Fenimore}, E.~E., {Finger}, M.~H.,
  {Gaensler}, B.~M., {Hullinger}, D., {Krimm}, H., {Markwardt}, C.~B.,
  {Nousek}, J.~A., {Parsons}, A., {Patel}, S., {Sakamoto}, T., {Sato}, G.,
  {Suzuki}, M., {Tueller}, J., Apr. 2005. {A giant {$\gamma$}-ray flare from
  the magnetar SGR 1806 - 20}. \nat 434, 1107--1109.

\bibitem[{{Panaitescu}(2005)}]{panaitescu05}
{Panaitescu}, A., Nov. 2005. {Jets, structured outflows and energy injection in
  gamma-ray burst afterglows: numerical modelling}. \mnras 363, 1409--1423.

\bibitem[{{Panaitescu}(2006)}]{panaitescu06}
{Panaitescu}, A., Mar. 2006. {The energetics and environment of the short-GRB
  afterglows 050709 and 050724}. \mnras 367, L42--L46.

\bibitem[{{Panaitescu} and {Kumar}(2000)}]{panaitescu00}
{Panaitescu}, A., {Kumar}, P., Nov. 2000. {Analytic Light Curves of Gamma-Ray
  Burst Afterglows: Homogeneous versus Wind External Media}. \apj 543, 66--76.

\bibitem[{{Panaitescu} and {Kumar}(2001)}]{panaitescu01}
{Panaitescu}, A., {Kumar}, P., Oct. 2001. {Fundamental Physical Parameters of
  Collimated Gamma-Ray Burst Afterglows}. \apjl 560, L49--L53.

\bibitem[{{Panaitescu} and {Kumar}(2002)}]{panaitescu02}
{Panaitescu}, A., {Kumar}, P., Jun. 2002. {Properties of Relativistic Jets in
  Gamma-Ray Burst Afterglows}. \apj 571, 779--789.

\bibitem[{{Panaitescu} et~al.(2001){Panaitescu}, {Kumar}, and
  {Narayan}}]{panaitescu01b}
{Panaitescu}, A., {Kumar}, P., {Narayan}, R., Nov. 2001. {Observational
  Prospects for Afterglows of Short-Duration Gamma-Ray Bursts}. \apjl 561,
  L171--L174.

\bibitem[{{Panaitescu} and {M{\'e}sz{\'a}ros}(1999)}]{panaitescu99b}
{Panaitescu}, A., {M{\'e}sz{\'a}ros}, P., Dec. 1999. {Dynamical Evolution,
  Light Curves, and Spectra of Spherical and Collimated Gamma-Ray Burst
  Remnants}. \apj 526, 707--715.

\bibitem[{{Panaitescu} et~al.(2006{\natexlab{a}}){Panaitescu},
  {M{\'e}sz{\'a}ros}, {Burrows}, {Nousek}, {Gehrels}, {O'Brien}, and
  {Willingale}}]{panai06b}
{Panaitescu}, A., {M{\'e}sz{\'a}ros}, P., {Burrows}, D., {Nousek}, J.,
  {Gehrels}, N., {O'Brien}, P., {Willingale}, R., Jul. 2006{\natexlab{a}}.
  {Evidence for chromatic X-ray light-curve breaks in Swift gamma-ray burst
  afterglows and their theoretical implications}. \mnras 369, 2059--2064.

\bibitem[{{Panaitescu} et~al.(2006{\natexlab{b}}){Panaitescu},
  {M{\'e}sz{\'a}ros}, {Gehrels}, {Burrows}, and {Nousek}}]{panai06a}
{Panaitescu}, A., {M{\'e}sz{\'a}ros}, P., {Gehrels}, N., {Burrows}, D.,
  {Nousek}, J., Mar. 2006{\natexlab{b}}. {Analysis of the X-ray emission of
  nine Swift afterglows}. \mnras 366, 1357--1366.

\bibitem[{{Panaitescu} et~al.(1999){Panaitescu}, {Spada}, and
  {M{\'e}sz{\'a}ros}}]{panaitescu99}
{Panaitescu}, A., {Spada}, M., {M{\'e}sz{\'a}ros}, P., Sep. 1999. {Power
  Density Spectra of Gamma-Ray Bursts in the Internal Shock Model}. \apjl 522,
  L105--L108.

\bibitem[{{Park} et~al.(2013){Park}, {Brandt}, {Budtz-J{\o}rgensen},
  {Castro-Tirado}, {Chen}, {Connell}, {Eyles}, {Grossan}, {Huang}, {Jeong},
  {Jung}, {Kim}, {Kim}, {Lee}, {Lim}, {Linder}, {Liu}, {Min}, {Na}, {Nam},
  {Panasyuk}, {Reglero}, {Ripa}, {Rodrigo}, {Smoot}, {Svertilov}, {Vedenkin},
  and {Yashin}}]{park13}
{Park}, I.~H., {Brandt}, S., {Budtz-J{\o}rgensen}, C., {Castro-Tirado}, A.~J.,
  {Chen}, P., {Connell}, P., {Eyles}, C., {Grossan}, B., {Huang}, M.-H.~A.,
  {Jeong}, S., {Jung}, A., {Kim}, J.~E., {Kim}, S.-W., {Lee}, J., {Lim}, H.,
  {Linder}, E.~V., {Liu}, T.-C., {Min}, K.~W., {Na}, G.~W., {Nam}, J.~W.,
  {Panasyuk}, M.~I., {Reglero}, V., {Ripa}, J., {Rodrigo}, J.~M., {Smoot},
  G.~F., {Svertilov}, S., {Vedenkin}, N., {Yashin}, I., Feb. 2013. {Ultra-Fast
  Flash Observatory for the observation of early photons from gamma-ray
  bursts}. New Journal of Physics 15~(2), 023031.

\bibitem[{{Parker}(1957)}]{parker57}
{Parker}, E.~N., Dec. 1957. {Sweet's Mechanism for Merging Magnetic Fields in
  Conducting Fluids}. \jgr 62, 509--520.

\bibitem[{{Paul} et~al.(2011){Paul}, {Wei}, {Basa}, and {Zhang}}]{paul12}
{Paul}, J., {Wei}, J., {Basa}, S., {Zhang}, S.-N., Apr. 2011. {The
  Chinese-French SVOM mission for gamma-ray burst studies}. Comptes Rendus
  Physique 12, 298--308.

\bibitem[{{Pe'er}(2008)}]{peer08}
{Pe'er}, A., Jul. 2008. {Temporal Evolution of Thermal Emission from
  Relativistically Expanding Plasma}. \apj 682, 463--473.

\bibitem[{{Pe'er}(2012)}]{peer12}
{Pe'er}, A., Jun. 2012. {Dynamical Model of an Expanding Shell}. \apjl 752, L8.

\bibitem[{{Pe'er} et~al.(2006{\natexlab{a}}){Pe'er}, {M{\'e}sz{\'a}ros}, and
  {Rees}}]{peer06b}
{Pe'er}, A., {M{\'e}sz{\'a}ros}, P., {Rees}, M.~J., Nov. 2006{\natexlab{a}}.
  {Radiation from an Expanding Cocoon as an Explanation of the Steep Decay
  Observed in GRB Early Afterglow Light Curves}. \apj 652, 482--489.

\bibitem[{{Pe'er} et~al.(2006{\natexlab{b}}){Pe'er}, {M{\'e}sz{\'a}ros}, and
  {Rees}}]{peer06}
{Pe'er}, A., {M{\'e}sz{\'a}ros}, P., {Rees}, M.~J., May 2006{\natexlab{b}}.
  {The Observable Effects of a Photospheric Component on GRB and XRF Prompt
  Emission Spectrum}. \apj 642, 995--1003.

\bibitem[{{Pe'er} and {Waxman}(2004)}]{peerwaxman04b}
{Pe'er}, A., {Waxman}, E., Mar. 2004. {The High-Energy Tail of GRB 941017:
  Comptonization of Synchrotron Self-absorbed Photons}. \apjl 603, L1--L4.

\bibitem[{{Pe'er} and {Wijers}(2006)}]{peerwijers06}
{Pe'er}, A., {Wijers}, R.~A.~M.~J., Jun. 2006. {The Signature of a Wind Reverse
  Shock in Gamma-Ray Burst Afterglows}. \apj 643, 1036--1046.

\bibitem[{{Pe'er} and {Zhang}(2006)}]{peerzhang06}
{Pe'er}, A., {Zhang}, B., Dec. 2006. {Synchrotron Emission in Small-Scale
  Magnetic Fields as a Possible Explanation for Prompt Emission Spectra of
  Gamma-Ray Bursts}. \apj 653, 454--461.

\bibitem[{{Peng} et~al.(2005){Peng}, {K{\"o}nigl}, and {Granot}}]{peng05}
{Peng}, F., {K{\"o}nigl}, A., {Granot}, J., Jun. 2005. {Two-Component Jet
  Models of Gamma-Ray Burst Sources}. \apj 626, 966--977.

\bibitem[{{Perley} et~al.(2013{\natexlab{a}}){Perley}, {Cenko}, {Corsi},
  {Tanvir}, {Levan}, {Kann}, {Sonbas}, {Wiersema}, {Zheng}, {Zhao}, {Bai},
  {Chang}, {Clubb}, {Frail}, {Fruchter}, {G{\"o}{\u g}{\"u}{\c s}}, {Greiner},
  {G{\"u}ver}, {Horesh}, {Filippenko}, {Klose}, {Mao}, {Morgan}, {Schmidl},
  {Stecklum}, {Tanga}, {Wang}, and {Xin}}]{perley13}
{Perley}, D.~A., {Cenko}, S.~B., {Corsi}, A., {Tanvir}, N.~R., {Levan}, A.~J.,
  {Kann}, D.~A., {Sonbas}, E., {Wiersema}, K., {Zheng}, W., {Zhao}, X.-H.,
  {Bai}, J.-M., {Chang}, L., {Clubb}, K., {Frail}, D., {Fruchter}, A.,
  {G{\"o}{\u g}{\"u}{\c s}}, E., {Greiner}, J., {G{\"u}ver}, T., {Horesh}, A.,
  {Filippenko}, A.~V., {Klose}, S., {Mao}, J., {Morgan}, A.~N., {Schmidl}, S.,
  {Stecklum}, B., {Tanga}, M., {Wang}, J.-G., {Xin}, Y.-X., Jul.
  2013{\natexlab{a}}. {The Afterglow of GRB 130427A from 1 to 10\^{}16 GHz}.
  ArXiv e-prints.

\bibitem[{{Perley} et~al.(2013{\natexlab{b}}){Perley}, {Levan}, {Tanvir},
  {Cenko}, {Bloom}, {Hjorth}, {Kruehler}, {Filippenko}, {Fruchter}, {Fynbo},
  {Jakobsson}, {Kalirai}, {Milvang-Jensen}, {Morgan}, {Prochaska}, and
  {Silverman}}]{perley13b}
{Perley}, D.~A., {Levan}, A.~J., {Tanvir}, N.~R., {Cenko}, S.~B., {Bloom},
  J.~S., {Hjorth}, J., {Kruehler}, T., {Filippenko}, A.~V., {Fruchter}, A.,
  {Fynbo}, J.~P.~U., {Jakobsson}, P., {Kalirai}, J., {Milvang-Jensen}, B.,
  {Morgan}, A.~N., {Prochaska}, J.~X., {Silverman}, J.~M., Jan.
  2013{\natexlab{b}}. {A Population of Massive, Luminous Galaxies Hosting
  Heavily Dust-Obscured Gamma-Ray Bursts: Implications for the Use of GRBs as
  Tracers of Cosmic Star Formation}. ArXiv e-prints.

\bibitem[{{Perna} et~al.(2006){Perna}, {Armitage}, and {Zhang}}]{perna06}
{Perna}, R., {Armitage}, P.~J., {Zhang}, B., Jan. 2006. {Flares in Long and
  Short Gamma-Ray Bursts: A Common Origin in a Hyperaccreting Accretion Disk}.
  \apjl 636, L29--L32.

\bibitem[{{Perna} et~al.(2003){Perna}, {Sari}, and {Frail}}]{perna03}
{Perna}, R., {Sari}, R., {Frail}, D., Sep. 2003. {Jets in Gamma-Ray Bursts:
  Tests and Predictions for the Structured Jet Model}. \apj 594, 379--384.

\bibitem[{{P{\'e}tri} and {Lyubarsky}(2007)}]{petri07}
{P{\'e}tri}, J., {Lyubarsky}, Y., Oct. 2007. {Magnetic reconnection at the
  termination shock in a striped pulsar wind}. \aap 473, 683--700.

\bibitem[{{Petropoulou} et~al.(2011){Petropoulou}, {Mastichiadis}, and
  {Piran}}]{petropoulou11}
{Petropoulou}, M., {Mastichiadis}, A., {Piran}, T., Jul. 2011. {Effects of a
  low electron distribution cutoff on multiwavelength spectra and light curves
  of GRB afterglows}. \aap 531, A76.

\bibitem[{{Petschek}(1964)}]{petschek64}
{Petschek}, H.~E., 1964. {Magnetic Field Annihilation}. NASA Special
  Publication 50, 425--+.

\bibitem[{{Pian} et~al.(2006){Pian}, {Mazzali}, {Masetti}, {Ferrero}, {Klose},
  {Palazzi}, {Ramirez-Ruiz}, {Woosley}, {Kouveliotou}, {Deng}, {Filippenko},
  {Foley}, {Fynbo}, {Kann}, {Li}, {Hjorth}, {Nomoto}, {Patat}, {Sauer},
  {Sollerman}, {Vreeswijk}, {Guenther}, {Levan}, {O'Brien}, {Tanvir}, {Wijers},
  {Dumas}, {Hainaut}, {Wong}, {Baade}, {Wang}, {Amati}, {Cappellaro},
  {Castro-Tirado}, {Ellison}, {Frontera}, {Fruchter}, {Greiner}, {Kawabata},
  {Ledoux}, {Maeda}, {M{\o}ller}, {Nicastro}, {Rol}, and {Starling}}]{pian06}
{Pian}, E., {Mazzali}, P.~A., {Masetti}, N., {Ferrero}, P., {Klose}, S.,
  {Palazzi}, E., {Ramirez-Ruiz}, E., {Woosley}, S.~E., {Kouveliotou}, C.,
  {Deng}, J., {Filippenko}, A.~V., {Foley}, R.~J., {Fynbo}, J.~P.~U., {Kann},
  D.~A., {Li}, W., {Hjorth}, J., {Nomoto}, K., {Patat}, F., {Sauer}, D.~N.,
  {Sollerman}, J., {Vreeswijk}, P.~M., {Guenther}, E.~W., {Levan}, A.,
  {O'Brien}, P., {Tanvir}, N.~R., {Wijers}, R.~A.~M.~J., {Dumas}, C.,
  {Hainaut}, O., {Wong}, D.~S., {Baade}, D., {Wang}, L., {Amati}, L.,
  {Cappellaro}, E., {Castro-Tirado}, A.~J., {Ellison}, S., {Frontera}, F.,
  {Fruchter}, A.~S., {Greiner}, J., {Kawabata}, K., {Ledoux}, C., {Maeda}, K.,
  {M{\o}ller}, P., {Nicastro}, L., {Rol}, E., {Starling}, R., Aug. 2006. {An
  optical supernova associated with the X-ray flash XRF 060218}. \nat 442,
  1011--1013.

\bibitem[{{Piran}(1992)}]{piran92}
{Piran}, T., Apr. 1992. {The implications of the Compton (GRO) observations for
  cosmological gamma-ray bursts}. \apjl 389, L45--L48.

\bibitem[{{Piran}(1999)}]{piran99}
{Piran}, T., Jun. 1999. {Gamma-ray bursts and the fireball model}. \physrep
  314, 575--667.

\bibitem[{{Piran}(2000)}]{piran00}
{Piran}, T., Aug. 2000. {Gamma-ray bursts - a puzzle being resolved}. \physrep
  333, 529--553.

\bibitem[{{Piran}(2002)}]{piran02}
{Piran}, T., Sep. 2002. {Gamma-Ray Bursts - a Primer for Relativists}. In:
  {Bishop}, N.~T., {Maharaj}, S.~D. (Eds.), General Relativity and Gravitation.
  pp. 259--275.

\bibitem[{{Piran}(2004)}]{piran04}
{Piran}, T., Oct. 2004. {The physics of gamma-ray bursts}. Reviews of Modern
  Physics 76, 1143--1210.

\bibitem[{{Piran} and {Nakar}(2010)}]{piran10}
{Piran}, T., {Nakar}, E., Aug. 2010. {On the External Shock Synchrotron Model
  for Gamma-ray Bursts' GeV Emission}. \apjl 718, L63--L67.

\bibitem[{{Piran} et~al.(2013){Piran}, {Nakar}, and {Rosswog}}]{piran13}
{Piran}, T., {Nakar}, E., {Rosswog}, S., Apr. 2013. {The electromagnetic
  signals of compact binary mergers}. \mnras 430, 2121--2136.

\bibitem[{{Piran} et~al.(2009){Piran}, {Sari}, and {Zou}}]{piran09}
{Piran}, T., {Sari}, R., {Zou}, Y.-C., Mar. 2009. {Observational limits on
  inverse Compton processes in gamma-ray bursts}. \mnras 393, 1107--1113.

\bibitem[{{Piran} et~al.(1993){Piran}, {Shemi}, and {Narayan}}]{piran93}
{Piran}, T., {Shemi}, A., {Narayan}, R., Aug. 1993. {Hydrodynamics of
  Relativistic Fireballs}. \mnras 263, 861--+.

\bibitem[{{Piro} et~al.(2005){Piro}, {De Pasquale}, {Soffitta}, {Lazzati},
  {Amati}, {Costa}, {Feroci}, {Frontera}, {Guidorzi}, {in't Zand}, {Montanari},
  and {Nicastro}}]{piro05}
{Piro}, L., {De Pasquale}, M., {Soffitta}, P., {Lazzati}, D., {Amati}, L.,
  {Costa}, E., {Feroci}, M., {Frontera}, F., {Guidorzi}, C., {in't Zand},
  J.~M.~J., {Montanari}, E., {Nicastro}, L., Apr. 2005. {Probing the
  Environment in Gamma-Ray Bursts: The Case of an X-Ray Precursor, Afterglow
  Late Onset, and Wind Versus Constant Density Profile in GRB 011121 and GRB
  011211}. \apj 623, 314--324.

\bibitem[{{Plaga}(1995)}]{plaga95}
{Plaga}, R., Mar. 1995. {Detecting Intergalactic Magnetic Fields Using Time
  Delays in Pulses of Gamma-Rays}. \nat 374, 430--+.

\bibitem[{{Pontzen} et~al.(2010){Pontzen}, {Deason}, {Governato}, {Pettini},
  {Wadsley}, {Quinn}, {Brooks}, {Bellovary}, and {Fynbo}}]{pontzen10}
{Pontzen}, A., {Deason}, A., {Governato}, F., {Pettini}, M., {Wadsley}, J.,
  {Quinn}, T., {Brooks}, A., {Bellovary}, J., {Fynbo}, J.~P.~U., Mar. 2010.
  {The nature of HIabsorbers in gamma-ray burst afterglows: clues from
  hydrodynamic simulations}. \mnras 402, 1523--1535.

\bibitem[{{Popham} et~al.(1999){Popham}, {Woosley}, and {Fryer}}]{popham99}
{Popham}, R., {Woosley}, S.~E., {Fryer}, C., Jun. 1999. {Hyperaccreting Black
  Holes and Gamma-Ray Bursts}. \apj 518, 356--374.

\bibitem[{{Preece} et~al.(2014){Preece}, {Burgess}, {von Kienlin}, {Bhat},
  {Briggs}, {Byrne}, {Chaplin}, {Cleveland}, {Collazzi}, and
  {Connaughton}}]{preece14}
{Preece}, R., {Burgess}, J.~M., {von Kienlin}, A., {Bhat}, P.~N., {Briggs},
  M.~S., {Byrne}, D., {Chaplin}, V., {Cleveland}, W., {Collazzi}, A.~C.,
  {Connaughton}, V. e.~a., Jan. 2014. {The First Pulse of the Extremely Bright
  GRB 130427A: A Test Lab for Synchrotron Shocks}. Science 343, 51--54.

\bibitem[{{Preece} et~al.(2000){Preece}, {Briggs}, {Mallozzi}, {Pendleton},
  {Paciesas}, and {Band}}]{preece00}
{Preece}, R.~D., {Briggs}, M.~S., {Mallozzi}, R.~S., {Pendleton}, G.~N.,
  {Paciesas}, W.~S., {Band}, D.~L., Jan. 2000. {The BATSE Gamma-Ray Burst
  Spectral Catalog. I. High Time Resolution Spectroscopy of Bright Bursts Using
  High Energy Resolution Data}. \apjs 126, 19--36.

\bibitem[{{Prochaska} et~al.(2004){Prochaska}, {Bloom}, {Chen}, {Hurley},
  {Melbourne}, {Dressler}, {Graham}, {Osip}, and {Vacca}}]{prochaska04}
{Prochaska}, J.~X., {Bloom}, J.~S., {Chen}, H.-W., {Hurley}, K.~C.,
  {Melbourne}, J., {Dressler}, A., {Graham}, J.~R., {Osip}, D.~J., {Vacca},
  W.~D., Aug. 2004. {The Host Galaxy of GRB 031203: Implications of Its Low
  Metallicity, Low Redshift, and Starburst Nature}. \apj 611, 200--207.

\bibitem[{{Proga} and {Zhang}(2006)}]{proga06}
{Proga}, D., {Zhang}, B., Jul. 2006. {The late time evolution of gamma-ray
  bursts: ending hyperaccretion and producing flares}. \mnras 370, L61--L65.

\bibitem[{{Qian} and {Woosley}(1996)}]{qian96}
{Qian}, Y.-Z., {Woosley}, S.~E., Nov. 1996. {Nucleosynthesis in Neutrino-driven
  Winds. I. The Physical Conditions}. \apj 471, 331.

\bibitem[{{Qin} et~al.(1998){Qin}, {Wu}, {Chu}, {Fang}, and {Hu}}]{qin98}
{Qin}, B., {Wu}, X.-P., {Chu}, M.-C., {Fang}, L.-Z., {Hu}, J.-Y., Feb. 1998.
  {The Collapse of Neutron Stars in High-Mass Binaries as the Energy Source for
  the Gamma-Ray Bursts}. \apjl 494, L57.

\bibitem[{{Qin} et~al.(2010){Qin}, {Liang}, {Lu}, {Wei}, and {Zhang}}]{qin10}
{Qin}, S., {Liang}, E., {Lu}, R., {Wei}, J., {Zhang}, S., Jul. 2010.
  {Simulations on high-z long gamma-ray burst rate}. \mnras 406, 558--565.

\bibitem[{{Qin} et~al.(2013){Qin}, {Liang}, {Liang}, {Yi}, {Lin}, {Zhang},
  {Zhang}, {L{\"u}}, {Lu}, {L{\"u}}, and {Zhang}}]{qin13}
{Qin}, Y., {Liang}, E.-W., {Liang}, Y.-F., {Yi}, S.-X., {Lin}, L., {Zhang},
  B.-B., {Zhang}, J., {L{\"u}}, H.-J., {Lu}, R.-J., {L{\"u}}, L.-Z., {Zhang},
  B., Jan. 2013. {A Comprehensive Analysis of Fermi Gamma-Ray Burst Data. III.
  Energy-dependent T $_{90}$ Distributions of GBM GRBs and Instrumental
  Selection Effect on Duration Classification}. \apj 763, 15.

\bibitem[{{Rachen}(1996)}]{rachen96}
{Rachen}, J.~P., 1996. {Interaction Processes and Statistical Properties of the
  Propagation of Cosmic Rays in Photon Backgrounds}. PhD thesis, University of
  Bonn, 1996.~159 p.

\bibitem[{{Rachen} and {M{\'e}sz{\'a}ros}(1998)}]{rachen98}
{Rachen}, J.~P., {M{\'e}sz{\'a}ros}, P., Dec. 1998. {Photohadronic neutrinos
  from transients in astrophysical sources}. \prd 58~(12), 123005--+.

\bibitem[{{Racusin} et~al.(2008){Racusin}, {Karpov}, {Sokolowski}, {Granot},
  {Wu}, {Pal'Shin}, {Covino}, {van der Horst}, {Oates}, {Schady}, {Smith},
  {Cummings}, {Starling}, {Piotrowski}, {Zhang}, {Evans}, {Holland}, {Malek},
  {Page}, {Vetere}, {Margutti}, {Guidorzi}, {Kamble}, {Curran}, {Beardmore},
  {Kouveliotou}, {Mankiewicz}, {Melandri}, {O'Brien}, {Page}, {Piran},
  {Tanvir}, {Wrochna}, {Aptekar}, {Barthelmy}, {Bartolini}, {Beskin}, {Bondar},
  {Bremer}, {Campana}, {Castro-Tirado}, {Cucchiara}, {Cwiok}, {D'Avanzo},
  {D'Elia}, {Valle}, {de Ugarte Postigo}, {Dominik}, {Falcone}, {Fiore}, {Fox},
  {Frederiks}, {Fruchter}, {Fugazza}, {Garrett}, {Gehrels}, {Golenetskii},
  {Gomboc}, {Gorosabel}, {Greco}, {Guarnieri}, {Immler}, {Jelinek},
  {Kasprowicz}, {La Parola}, {Levan}, {Mangano}, {Mazets}, {Molinari},
  {Moretti}, {Nawrocki}, {Oleynik}, {Osborne}, {Pagani}, {Pandey}, {Paragi},
  {Perri}, {Piccioni}, {Ramirez-Ruiz}, {Roming}, {Steele}, {Strom}, {Testa},
  {Tosti}, {Ulanov}, {Wiersema}, {Wijers}, {Winters}, {Zarnecki}, {Zerbi},
  {M{\'e}sz{\'a}ros}, {Chincarini}, and {Burrows}}]{racusin08}
{Racusin}, J.~L., {Karpov}, S.~V., {Sokolowski}, M., {Granot}, J., {Wu}, X.~F.,
  {Pal'Shin}, V., {Covino}, S., {van der Horst}, A.~J., {Oates}, S.~R.,
  {Schady}, P., {Smith}, R.~J., {Cummings}, J., {Starling}, R.~L.~C.,
  {Piotrowski}, L.~W., {Zhang}, B., {Evans}, P.~A., {Holland}, S.~T., {Malek},
  K., {Page}, M.~T., {Vetere}, L., {Margutti}, R., {Guidorzi}, C., {Kamble},
  A.~P., {Curran}, P.~A., {Beardmore}, A., {Kouveliotou}, C., {Mankiewicz}, L.,
  {Melandri}, A., {O'Brien}, P.~T., {Page}, K.~L., {Piran}, T., {Tanvir},
  N.~R., {Wrochna}, G., {Aptekar}, R.~L., {Barthelmy}, S., {Bartolini}, C.,
  {Beskin}, G.~M., {Bondar}, S., {Bremer}, M., {Campana}, S., {Castro-Tirado},
  A., {Cucchiara}, A., {Cwiok}, M., {D'Avanzo}, P., {D'Elia}, V., {Valle},
  M.~D., {de Ugarte Postigo}, A., {Dominik}, W., {Falcone}, A., {Fiore}, F.,
  {Fox}, D.~B., {Frederiks}, D.~D., {Fruchter}, A.~S., {Fugazza}, D.,
  {Garrett}, M.~A., {Gehrels}, N., {Golenetskii}, S., {Gomboc}, A.,
  {Gorosabel}, J., {Greco}, G., {Guarnieri}, A., {Immler}, S., {Jelinek}, M.,
  {Kasprowicz}, G., {La Parola}, V., {Levan}, A.~J., {Mangano}, V., {Mazets},
  E.~P., {Molinari}, E., {Moretti}, A., {Nawrocki}, K., {Oleynik}, P.~P.,
  {Osborne}, J.~P., {Pagani}, C., {Pandey}, S.~B., {Paragi}, Z., {Perri}, M.,
  {Piccioni}, A., {Ramirez-Ruiz}, E., {Roming}, P.~W.~A., {Steele}, I.~A.,
  {Strom}, R.~G., {Testa}, V., {Tosti}, G., {Ulanov}, M.~V., {Wiersema}, K.,
  {Wijers}, R.~A.~M.~J., {Winters}, J.~M., {Zarnecki}, A.~F., {Zerbi}, F.,
  {M{\'e}sz{\'a}ros}, P., {Chincarini}, G., {Burrows}, D.~N., Sep. 2008.
  {Broadband observations of the naked-eye {$\gamma$}-ray burst GRB080319B}.
  \nat 455, 183--188.

\bibitem[{{Racusin} et~al.(2009){Racusin}, {Liang}, {Burrows}, {Falcone},
  {Sakamoto}, {Zhang}, {Zhang}, {Evans}, and {Osborne}}]{racusin09}
{Racusin}, J.~L., {Liang}, E.~W., {Burrows}, D.~N., {Falcone}, A., {Sakamoto},
  T., {Zhang}, B.~B., {Zhang}, B., {Evans}, P., {Osborne}, J., Jun. 2009. {Jet
  Breaks and Energetics of Swift Gamma-Ray Burst X-Ray Afterglows}. \apj 698,
  43--74.

\bibitem[{{Racusin} et~al.(2011){Racusin}, {Oates}, {Schady}, {Burrows}, {de
  Pasquale}, {Donato}, {Gehrels}, {Koch}, {McEnery}, {Piran}, {Roming},
  {Sakamoto}, {Swenson}, {Troja}, {Vasileiou}, {Virgili}, {Wanderman}, and
  {Zhang}}]{racusin11}
{Racusin}, J.~L., {Oates}, S.~R., {Schady}, P., {Burrows}, D.~N., {de
  Pasquale}, M., {Donato}, D., {Gehrels}, N., {Koch}, S., {McEnery}, J.,
  {Piran}, T., {Roming}, P., {Sakamoto}, T., {Swenson}, C., {Troja}, E.,
  {Vasileiou}, V., {Virgili}, F., {Wanderman}, D., {Zhang}, B., Sep. 2011.
  {Fermi and Swift Gamma-ray Burst Afterglow Population Studies}. \apj 738,
  138.

\bibitem[{{Ramirez-Ruiz} et~al.(2002){Ramirez-Ruiz}, {Celotti}, and
  {Rees}}]{ramirezruiz02}
{Ramirez-Ruiz}, E., {Celotti}, A., {Rees}, M.~J., Dec. 2002. {Events in the
  life of a cocoon surrounding a light, collapsar jet}. \mnras 337, 1349--1356.

\bibitem[{{Ramirez-Ruiz} et~al.(2001){Ramirez-Ruiz}, {Dray}, {Madau}, and
  {Tout}}]{ramirezruiz01}
{Ramirez-Ruiz}, E., {Dray}, L.~M., {Madau}, P., {Tout}, C.~A., Nov. 2001.
  {Winds from massive stars: implications for the afterglows of {$\gamma$}-ray
  bursts}. \mnras 327, 829--840.

\bibitem[{{Razzaque} et~al.(2010){Razzaque}, {Dermer}, and
  {Finke}}]{razzaque10}
{Razzaque}, S., {Dermer}, C.~D., {Finke}, J.~D., Aug. 2010. {Synchrotron
  Radiation from Ultra-High Energy Protons and the Fermi Observations of GRB
  080916C}. The Open Astronomy Journal 3, 150--155.

\bibitem[{{Razzaque} et~al.(2003{\natexlab{a}}){Razzaque}, {M{\'e}sz{\'a}ros},
  and {Waxman}}]{razzaque03}
{Razzaque}, S., {M{\'e}sz{\'a}ros}, P., {Waxman}, E., Jun. 2003{\natexlab{a}}.
  {High Energy Neutrinos from Gamma-Ray Bursts with Precursor Supernovae}.
  Physical Review Letters 90~(24), 241103--+.

\bibitem[{{Razzaque} et~al.(2003{\natexlab{b}}){Razzaque}, {M{\'e}sz{\'a}ros},
  and {Waxman}}]{razzaque03b}
{Razzaque}, S., {M{\'e}sz{\'a}ros}, P., {Waxman}, E., Oct. 2003{\natexlab{b}}.
  {Neutrino tomography of gamma ray bursts and massive stellar collapses}. \prd
  68~(8), 083001.

\bibitem[{{Rees} and {M\'esz\'aros}(1992)}]{rees92}
{Rees}, M.~J., {M\'esz\'aros}, P., Sep. 1992. {Relativistic fireballs - Energy
  conversion and time-scales}. \mnras 258, 41P--43P.

\bibitem[{{Rees} and {M\'esz\'aros}(1994)}]{rees94}
{Rees}, M.~J., {M\'esz\'aros}, P., Aug. 1994. {Unsteady outflow models for
  cosmological gamma-ray bursts}. \apjl 430, L93--L96.

\bibitem[{{Rees} and {M\'esz\'aros}(1998)}]{rees98}
{Rees}, M.~J., {M\'esz\'aros}, P., Mar. 1998. {Refreshed Shocks and Afterglow
  Longevity in Gamma-Ray Bursts}. \apjl 496, L1+.

\bibitem[{{Rees} and {M{\'e}sz{\'a}ros}(2005)}]{rees05}
{Rees}, M.~J., {M{\'e}sz{\'a}ros}, P., Aug. 2005. {Dissipative Photosphere
  Models of Gamma-Ray Bursts and X-Ray Flashes}. \apj 628, 847--852.

\bibitem[{{Reichart} et~al.(2001){Reichart}, {Lamb}, {Fenimore},
  {Ramirez-Ruiz}, {Cline}, and {Hurley}}]{reichart01}
{Reichart}, D.~E., {Lamb}, D.~Q., {Fenimore}, E.~E., {Ramirez-Ruiz}, E.,
  {Cline}, T.~L., {Hurley}, K., May 2001. {A Possible Cepheid-like Luminosity
  Estimator for the Long Gamma-Ray Bursts}. \apj 552, 57--71.

\bibitem[{{Reimer} et~al.(2004){Reimer}, {Protheroe}, and {Donea}}]{reimer04}
{Reimer}, A., {Protheroe}, R.~J., {Donea}, A.-C., Apr. 2004. {M87 - a
  misaligned synchrotron-proton blazar?} \nar 48, 411--413.

\bibitem[{{Resmi} and {Bhattacharya}(2008)}]{resmi08}
{Resmi}, L., {Bhattacharya}, D., Jul. 2008. {Hard electron energy distribution
  in the relativistic shocks of gamma-ray burst afterglows}. \mnras 388,
  144--158.

\bibitem[{{Resmi} and {Zhang}(2012)}]{resmi12}
{Resmi}, L., {Zhang}, B., Oct. 2012. {Gamma-ray burst prompt emission
  variability in synchrotron and synchrotron self-Compton light curves}. \mnras
  426, 1385--1395.

\bibitem[{{Rezzolla} et~al.(2010){Rezzolla}, {Baiotti}, {Giacomazzo}, {Link},
  and {Font}}]{rezzolla10}
{Rezzolla}, L., {Baiotti}, L., {Giacomazzo}, B., {Link}, D., {Font}, J.~A.,
  Jun. 2010. {Accurate evolutions of unequal-mass neutron-star binaries:
  properties of the torus and short GRB engines}. Classical and Quantum Gravity
  27~(11), 114105.

\bibitem[{{Rezzolla} et~al.(2011){Rezzolla}, {Giacomazzo}, {Baiotti}, {Granot},
  {Kouveliotou}, and {Aloy}}]{rezzolla11}
{Rezzolla}, L., {Giacomazzo}, B., {Baiotti}, L., {Granot}, J., {Kouveliotou},
  C., {Aloy}, M.~A., May 2011. {The Missing Link: Merging Neutron Stars
  Naturally Produce Jet-like Structures and Can Power Short Gamma-ray Bursts}.
  \apjl 732, L6.

\bibitem[{{Rezzolla} and {Zanotti}(2013)}]{rezzolla13}
{Rezzolla}, L., {Zanotti}, O., Sep. 2013. {Relativistic Hydrodynamics}.

\bibitem[{{Rhoads}(1997)}]{rhoads97}
{Rhoads}, J.~E., Sep. 1997. {How to Tell a Jet from a Balloon: A Proposed Test
  for Beaming in Gamma-Ray Bursts}. \apjl 487, L1+.

\bibitem[{{Rhoads}(1999)}]{rhoads99}
{Rhoads}, J.~E., Nov. 1999. {The Dynamics and Light Curves of Beamed Gamma-Ray
  Burst Afterglows}. \apj 525, 737--749.

\bibitem[{{Robertson} and {Ellis}(2012)}]{robertson12}
{Robertson}, B.~E., {Ellis}, R.~S., Jan. 2012. {Connecting the Gamma Ray Burst
  Rate and the Cosmic Star Formation History: Implications for Reionization and
  Galaxy Evolution}. \apj 744, 95.

\bibitem[{{Romano} et~al.(2006){Romano}, {Moretti}, {Banat}, {Burrows},
  {Campana}, {Chincarini}, {Covino}, {Malesani}, {Tagliaferri}, {Kobayashi},
  {Zhang}, {Falcone}, {Angelini}, {Barthelmy}, {Beardmore}, {Capalbi},
  {Cusumano}, {Giommi}, {Goad}, {Godet}, {Grupe}, {Hill}, {Kennea}, {La
  Parola}, {Mangano}, {M{\'e}sz{\'a}ros}, {Morris}, {Nousek}, {O'Brien},
  {Osborne}, {Parsons}, {Perri}, {Pagani}, {Page}, {Wells}, and
  {Gehrels}}]{romano06}
{Romano}, P., {Moretti}, A., {Banat}, P.~L., {Burrows}, D.~N., {Campana}, S.,
  {Chincarini}, G., {Covino}, S., {Malesani}, D., {Tagliaferri}, G.,
  {Kobayashi}, S., {Zhang}, B., {Falcone}, A.~D., {Angelini}, L., {Barthelmy},
  S., {Beardmore}, A.~P., {Capalbi}, M., {Cusumano}, G., {Giommi}, P., {Goad},
  M.~R., {Godet}, O., {Grupe}, D., {Hill}, J.~E., {Kennea}, J.~A., {La Parola},
  V., {Mangano}, V., {M{\'e}sz{\'a}ros}, P., {Morris}, D.~C., {Nousek}, J.~A.,
  {O'Brien}, P.~T., {Osborne}, J.~P., {Parsons}, A., {Perri}, M., {Pagani}, C.,
  {Page}, K.~L., {Wells}, A.~A., {Gehrels}, N., Apr. 2006. {X-ray flare in XRF
  050406: evidence for prolonged engine activity}. \aap 450, 59--68.

\bibitem[{{Romanova} and {Lovelace}(1992)}]{romanova92}
{Romanova}, M.~M., {Lovelace}, R.~V.~E., Aug. 1992. {Magnetic field,
  reconnection, and particle acceleration in extragalactic jets}. \aap 262,
  26--36.

\bibitem[{{Romero} et~al.(2010){Romero}, {Reynoso}, and
  {Christiansen}}]{romero10}
{Romero}, G.~E., {Reynoso}, M.~M., {Christiansen}, H.~R., Dec. 2010.
  {Gravitational radiation from precessing accretion disks in gamma-ray
  bursts}. \aap 524, A4.

\bibitem[{{Roming} et~al.(2005){Roming}, {Kennedy}, {Mason}, {Nousek}, {Ahr},
  {Bingham}, {Broos}, {Carter}, {Hancock}, {Huckle}, {Hunsberger}, {Kawakami},
  {Killough}, {Koch}, {McLelland}, {Smith}, {Smith}, {Soto}, {Boyd},
  {Breeveld}, {Holland}, {Ivanushkina}, {Pryzby}, {Still}, and
  {Stock}}]{roming05}
{Roming}, P.~W.~A., {Kennedy}, T.~E., {Mason}, K.~O., {Nousek}, J.~A., {Ahr},
  L., {Bingham}, R.~E., {Broos}, P.~S., {Carter}, M.~J., {Hancock}, B.~K.,
  {Huckle}, H.~E., {Hunsberger}, S.~D., {Kawakami}, H., {Killough}, R., {Koch},
  T.~S., {McLelland}, M.~K., {Smith}, K., {Smith}, P.~J., {Soto}, J.~C.,
  {Boyd}, P.~T., {Breeveld}, A.~A., {Holland}, S.~T., {Ivanushkina}, M.,
  {Pryzby}, M.~S., {Still}, M.~D., {Stock}, J., Oct. 2005. {The Swift
  Ultra-Violet/Optical Telescope}. Space Science Reviews 120, 95--142.

\bibitem[{{Rossi} et~al.(2011){Rossi}, {Schulze}, {Klose}, {Kann}, {Rau},
  {Krimm}, {J{\'o}hannesson}, {Panaitescu}, {Yuan}, {Ferrero}, {Kr{\"u}hler},
  {Greiner}, {Schady}, {Pandey}, {Amati}, {Afonso}, {Akerlof}, {Arnold},
  {Clemens}, {Filgas}, {Hartmann}, {K{\"u}pc{\"u} Yolda{\c s}}, {McBreen},
  {McKay}, {Nicuesa Guelbenzu}, {Olivares}, {Paciesas}, {Rykoff}, {Szokoly},
  {Updike}, and {Yolda{\c s}}}]{rossi11}
{Rossi}, A., {Schulze}, S., {Klose}, S., {Kann}, D.~A., {Rau}, A., {Krimm},
  H.~A., {J{\'o}hannesson}, G., {Panaitescu}, A., {Yuan}, F., {Ferrero}, P.,
  {Kr{\"u}hler}, T., {Greiner}, J., {Schady}, P., {Pandey}, S.~B., {Amati}, L.,
  {Afonso}, P.~M.~J., {Akerlof}, C.~W., {Arnold}, L.~A., {Clemens}, C.,
  {Filgas}, R., {Hartmann}, D.~H., {K{\"u}pc{\"u} Yolda{\c s}}, A., {McBreen},
  S., {McKay}, T.~A., {Nicuesa Guelbenzu}, A., {Olivares}, F.~E., {Paciesas},
  B., {Rykoff}, E.~S., {Szokoly}, G., {Updike}, A.~C., {Yolda{\c s}}, A., May
  2011. {The Swift/Fermi GRB 080928 from 1 eV to 150 keV}. \aap 529, A142.

\bibitem[{{Rossi} et~al.(2002){Rossi}, {Lazzati}, and {Rees}}]{rossi02}
{Rossi}, E., {Lazzati}, D., {Rees}, M.~J., Jun. 2002. {Afterglow light curves,
  viewing angle and the jet structure of {$\gamma$}-ray bursts}. \mnras 332,
  945--950.

\bibitem[{{Rossi} and {Rees}(2003)}]{rossi03}
{Rossi}, E., {Rees}, M.~J., Mar. 2003. {Gamma-ray burst afterglow emission with
  a decaying magnetic field}. \mnras 339, 881--886.

\bibitem[{{Rossi} et~al.(2006){Rossi}, {Beloborodov}, and {Rees}}]{rossi06}
{Rossi}, E.~M., {Beloborodov}, A.~M., {Rees}, M.~J., Jul. 2006. {Neutron-loaded
  outflows in gamma-ray bursts}. \mnras 369, 1797--1807.

\bibitem[{{Rosswog} et~al.(2013){Rosswog}, {Piran}, and {Nakar}}]{rosswog13}
{Rosswog}, S., {Piran}, T., {Nakar}, E., Apr. 2013. {The multimessenger picture
  of compact object encounters: binary mergers versus dynamical collisions}.
  \mnras 430, 2585--2604.

\bibitem[{{Rosswog} et~al.(2003){Rosswog}, {Ramirez-Ruiz}, and
  {Davies}}]{rosswog03}
{Rosswog}, S., {Ramirez-Ruiz}, E., {Davies}, M.~B., Nov. 2003. {High-resolution
  calculations of merging neutron stars - III. Gamma-ray bursts}. \mnras 345,
  1077--1090.

\bibitem[{{Rowlinson} et~al.(2013){Rowlinson}, {O'Brien}, {Metzger}, {Tanvir},
  and {Levan}}]{rowlinson13}
{Rowlinson}, A., {O'Brien}, P.~T., {Metzger}, B.~D., {Tanvir}, N.~R., {Levan},
  A.~J., Apr. 2013. {Signatures of magnetar central engines in short GRB light
  curves}. \mnras 430, 1061--1087.

\bibitem[{{Rowlinson} et~al.(2010){Rowlinson}, {O'Brien}, {Tanvir}, {Zhang},
  {Evans}, {Lyons}, {Levan}, {Willingale}, {Page}, {Onal}, {Burrows},
  {Beardmore}, {Ukwatta}, {Berger}, {Hjorth}, {Fruchter}, {Tunnicliffe}, {Fox},
  and {Cucchiara}}]{rowlinson10}
{Rowlinson}, A., {O'Brien}, P.~T., {Tanvir}, N.~R., {Zhang}, B., {Evans},
  P.~A., {Lyons}, N., {Levan}, A.~J., {Willingale}, R., {Page}, K.~L., {Onal},
  O., {Burrows}, D.~N., {Beardmore}, A.~P., {Ukwatta}, T.~N., {Berger}, E.,
  {Hjorth}, J., {Fruchter}, A.~S., {Tunnicliffe}, R.~L., {Fox}, D.~B.,
  {Cucchiara}, A., Dec. 2010. {The unusual X-ray emission of the short Swift
  GRB 090515: evidence for the formation of a magnetar?} \mnras 409, 531--540.

\bibitem[{{Ruderman} et~al.(2000){Ruderman}, {Tao}, and
  {Klu{\'z}niak}}]{ruderman00}
{Ruderman}, M.~A., {Tao}, L., {Klu{\'z}niak}, W., Oct. 2000. {A Central Engine
  for Cosmic Gamma-Ray Burst Sources}. \apj 542, 243--250.

\bibitem[{{Ruffert} and {Janka}(1999)}]{ruffert99}
{Ruffert}, M., {Janka}, H.-T., Apr. 1999. {Gamma-ray bursts from accreting
  black holes in neutron star mergers}. \aap 344, 573--606.

\bibitem[{{Ruffini} et~al.(2001{\natexlab{a}}){Ruffini}, {Bianco},
  {Fraschetti}, {Xue}, and {Chardonnet}}]{ruffini01b}
{Ruffini}, R., {Bianco}, C.~L., {Fraschetti}, F., {Xue}, S.-S., {Chardonnet},
  P., Jul. 2001{\natexlab{a}}. {On the Interpretation of the Burst Structure of
  Gamma-Ray Bursts}. \apjl 555, L113--L116.

\bibitem[{{Ruffini} et~al.(2001{\natexlab{b}}){Ruffini}, {Bianco},
  {Fraschetti}, {Xue}, and {Chardonnet}}]{ruffini01a}
{Ruffini}, R., {Bianco}, C.~L., {Fraschetti}, F., {Xue}, S.-S., {Chardonnet},
  P., Jul. 2001{\natexlab{b}}. {Relative Spacetime Transformations in Gamma-Ray
  Bursts}. \apjl 555, L107--L111.

\bibitem[{{Ruffini} et~al.(2010){Ruffini}, {Vereshchagin}, and
  {Xue}}]{ruffini10}
{Ruffini}, R., {Vereshchagin}, G., {Xue}, S.-S., Feb. 2010. {Electron-positron
  pairs in physics and astrophysics: From heavy nuclei to black holes}.
  \physrep 487, 1--140.

\bibitem[{{Rutledge} and {Fox}(2004)}]{rutledge04}
{Rutledge}, R.~E., {Fox}, D.~B., Jun. 2004. {Re-analysis of polarization in the
  {$\gamma$}-ray flux of GRB 021206}. \mnras 350, 1288--1300.

\bibitem[{Ryan} et~al.(2014)]{ryan14} Ryan, G., van Eerten, H.,  
MacFadyen, A., \& Zhang, B.-B.\ 2014, arXiv:1405.5516  

\bibitem[{{Rybicki} and {Lightman}(1979)}]{rybicki79}
{Rybicki}, G.~B., {Lightman}, A.~P., 1979. {Radiative processes in
  astrophysics}. New York, Wiley-Interscience, 1979.~393 p.

\bibitem[{{Ryde}(2004)}]{ryde04}
{Ryde}, F., Oct. 2004. {The Cooling Behavior of Thermal Pulses in Gamma-Ray
  Bursts}. \apj 614, 827--846.

\bibitem[{{Ryde} et~al.(2010){Ryde}, {Axelsson}, {Zhang}, {McGlynn}, {Pe'er},
  {Lundman}, {Larsson}, {Battelino}, {Zhang}, {Bissaldi}, {Bregeon}, {Briggs},
  {Chiang}, {de Palma}, {Guiriec}, {Larsson}, {Longo}, {McBreen}, {Omodei},
  {Petrosian}, {Preece}, and {van der Horst}}]{ryde10}
{Ryde}, F., {Axelsson}, M., {Zhang}, B.~B., {McGlynn}, S., {Pe'er}, A.,
  {Lundman}, C., {Larsson}, S., {Battelino}, M., {Zhang}, B., {Bissaldi}, E.,
  {Bregeon}, J., {Briggs}, M.~S., {Chiang}, J., {de Palma}, F., {Guiriec}, S.,
  {Larsson}, J., {Longo}, F., {McBreen}, S., {Omodei}, N., {Petrosian}, V.,
  {Preece}, R., {van der Horst}, A.~J., Feb. 2010. {Identification and
  Properties of the Photospheric Emission in GRB090902B}. \apjl 709,
  L172--L177.

\bibitem[{{Ryde} and {Pe'er}(2009)}]{ryde09}
{Ryde}, F., {Pe'er}, A., Sep. 2009. {Quasi-blackbody Component and Radiative
  Efficiency of the Prompt Emission of Gamma-ray Bursts}. \apj 702, 1211--1229.

\bibitem[{{Rykoff} et~al.(2009){Rykoff}, {Aharonian}, {Akerlof}, {Ashley},
  {Barthelmy}, {Flewelling}, {Gehrels}, {G{\"o}{\v g}{\"u}{\c s}}, {G{\"u}ver},
  {Kizilo{\v g}lu}, {Krimm}, {McKay}, {{\"O}zel}, {Phillips}, {Quimby},
  {Rowell}, {Rujopakarn}, {Schaefer}, {Smith}, {Vestrand}, {Wheeler}, {Wren},
  {Yuan}, and {Yost}}]{rykoff09}
{Rykoff}, E.~S., {Aharonian}, F., {Akerlof}, C.~W., {Ashley}, M.~C.~B.,
  {Barthelmy}, S.~D., {Flewelling}, H.~A., {Gehrels}, N., {G{\"o}{\v g}{\"u}{\c
  s}}, E., {G{\"u}ver}, T., {Kizilo{\v g}lu}, {\"U}., {Krimm}, H.~A., {McKay},
  T.~A., {{\"O}zel}, M., {Phillips}, A., {Quimby}, R.~M., {Rowell}, G.,
  {Rujopakarn}, W., {Schaefer}, B.~E., {Smith}, D.~A., {Vestrand}, W.~T.,
  {Wheeler}, J.~C., {Wren}, J., {Yuan}, F., {Yost}, S.~A., Sep. 2009. {Looking
  Into the Fireball: ROTSE-III and Swift Observations of Early Gamma-ray Burst
  Afterglows}. \apj 702, 489--505.

\bibitem[{{Sakamoto} et~al.(2006){Sakamoto}, {Barbier}, {Barthelmy},
  {Cummings}, {Fenimore}, {Gehrels}, {Hullinger}, {Krimm}, {Markwardt},
  {Palmer}, {Parsons}, {Sato}, and {Tueller}}]{sakamoto06}
{Sakamoto}, T., {Barbier}, L., {Barthelmy}, S.~D., {Cummings}, J.~R.,
  {Fenimore}, E.~E., {Gehrels}, N., {Hullinger}, D., {Krimm}, H.~A.,
  {Markwardt}, C.~B., {Palmer}, D.~M., {Parsons}, A.~M., {Sato}, G., {Tueller},
  J., Jan. 2006. {Confirmation of the $E^{src}_{peak}-E_{iso}$ (Amati) Relation
  from the X-Ray Flash XRF 050416AObserved by the Swift Burst Alert Telescope}.
  \apjl 636, L73--L76.

\bibitem[{{Sakamoto} et~al.(2008{\natexlab{a}}){Sakamoto}, {Barthelmy},
  {Barbier}, {Cummings}, {Fenimore}, {Gehrels}, {Hullinger}, {Krimm},
  {Markwardt}, {Palmer}, {Parsons}, {Sato}, {Stamatikos}, {Tueller}, {Ukwatta},
  and {Zhang}}]{sakamoto08}
{Sakamoto}, T., {Barthelmy}, S.~D., {Barbier}, L., {Cummings}, J.~R.,
  {Fenimore}, E.~E., {Gehrels}, N., {Hullinger}, D., {Krimm}, H.~A.,
  {Markwardt}, C.~B., {Palmer}, D.~M., {Parsons}, A.~M., {Sato}, G.,
  {Stamatikos}, M., {Tueller}, J., {Ukwatta}, T.~N., {Zhang}, B., Mar.
  2008{\natexlab{a}}. {The First Swift BAT Gamma-Ray Burst Catalog}. \apjs 175,
  179--190.

\bibitem[{{Sakamoto} et~al.(2011){Sakamoto}, {Barthelmy}, {Baumgartner},
  {Cummings}, {Fenimore}, {Gehrels}, {Krimm}, {Markwardt}, {Palmer}, {Parsons},
  {Sato}, {Stamatikos}, {Tueller}, {Ukwatta}, and {Zhang}}]{sakamoto11}
{Sakamoto}, T., {Barthelmy}, S.~D., {Baumgartner}, W.~H., {Cummings}, J.~R.,
  {Fenimore}, E.~E., {Gehrels}, N., {Krimm}, H.~A., {Markwardt}, C.~B.,
  {Palmer}, D.~M., {Parsons}, A.~M., {Sato}, G., {Stamatikos}, M., {Tueller},
  J., {Ukwatta}, T.~N., {Zhang}, B., Jul. 2011. {The Second Swift Burst Alert
  Telescope Gamma-Ray Burst Catalog}. \apjs 195, 2.

\bibitem[{{Sakamoto} et~al.(2008{\natexlab{b}}){Sakamoto}, {Hullinger}, {Sato},
  {Yamazaki}, {Barbier}, {Barthelmy}, {Cummings}, {Fenimore}, {Gehrels},
  {Krimm}, {Lamb}, {Markwardt}, {Osborne}, {Palmer}, {Parsons}, {Stamatikos},
  and {Tueller}}]{sakamoto08a}
{Sakamoto}, T., {Hullinger}, D., {Sato}, G., {Yamazaki}, R., {Barbier}, L.,
  {Barthelmy}, S.~D., {Cummings}, J.~R., {Fenimore}, E.~E., {Gehrels}, N.,
  {Krimm}, H.~A., {Lamb}, D.~Q., {Markwardt}, C.~B., {Osborne}, J.~P.,
  {Palmer}, D.~M., {Parsons}, A.~M., {Stamatikos}, M., {Tueller}, J., May
  2008{\natexlab{b}}. {Global Properties of X-Ray Flashes and X-Ray-Rich
  Gamma-Ray Bursts Observed by Swift}. \apj 679, 570--586.

\bibitem[{{Sakamoto} et~al.(2005){Sakamoto}, {Lamb}, {Kawai}, {Yoshida},
  {Graziani}, {Fenimore}, {Donaghy}, {Matsuoka}, {Suzuki}, {Ricker}, {Atteia},
  {Shirasaki}, {Tamagawa}, {Torii}, {Galassi}, {Doty}, {Vanderspek}, {Crew},
  {Villasenor}, {Butler}, {Prigozhin}, {Jernigan}, {Barraud}, {Boer},
  {Dezalay}, {Olive}, {Hurley}, {Levine}, {Monnelly}, {Martel}, {Morgan},
  {Woosley}, {Cline}, {Braga}, {Manchanda}, {Pizzichini}, {Takagishi}, and
  {Yamauchi}}]{sakamoto05}
{Sakamoto}, T., {Lamb}, D.~Q., {Kawai}, N., {Yoshida}, A., {Graziani}, C.,
  {Fenimore}, E.~E., {Donaghy}, T.~Q., {Matsuoka}, M., {Suzuki}, M., {Ricker},
  G., {Atteia}, J.-L., {Shirasaki}, Y., {Tamagawa}, T., {Torii}, K., {Galassi},
  M., {Doty}, J., {Vanderspek}, R., {Crew}, G.~B., {Villasenor}, J., {Butler},
  N., {Prigozhin}, G., {Jernigan}, J.~G., {Barraud}, C., {Boer}, M., {Dezalay},
  J.-P., {Olive}, J.-F., {Hurley}, K., {Levine}, A., {Monnelly}, G., {Martel},
  F., {Morgan}, E., {Woosley}, S.~E., {Cline}, T., {Braga}, J., {Manchanda},
  R., {Pizzichini}, G., {Takagishi}, K., {Yamauchi}, M., Aug. 2005. {Global
  Characteristics of X-Ray Flashes and X-Ray-Rich Gamma-Ray Bursts Observed by
  HETE-2}. \apj 629, 311--327.

\bibitem[{{Salvaterra} et~al.(2009){Salvaterra}, {Della Valle}, {Campana},
  {Chincarini}, {Covino}, {D'Avanzo}, {Fern{\'a}ndez-Soto}, {Guidorzi},
  {Mannucci}, {Margutti}, {Th{\"o}ne}, {Antonelli}, {Barthelmy}, {de Pasquale},
  {D'Elia}, {Fiore}, {Fugazza}, {Hunt}, {Maiorano}, {Marinoni}, {Marshall},
  {Molinari}, {Nousek}, {Pian}, {Racusin}, {Stella}, {Amati}, {Andreuzzi},
  {Cusumano}, {Fenimore}, {Ferrero}, {Giommi}, {Guetta}, {Holland}, {Hurley},
  {Israel}, {Mao}, {Markwardt}, {Masetti}, {Pagani}, {Palazzi}, {Palmer},
  {Piranomonte}, {Tagliaferri}, and {Testa}}]{salvaterra09}
{Salvaterra}, R., {Della Valle}, M., {Campana}, S., {Chincarini}, G., {Covino},
  S., {D'Avanzo}, P., {Fern{\'a}ndez-Soto}, A., {Guidorzi}, C., {Mannucci}, F.,
  {Margutti}, R., {Th{\"o}ne}, C.~C., {Antonelli}, L.~A., {Barthelmy}, S.~D.,
  {de Pasquale}, M., {D'Elia}, V., {Fiore}, F., {Fugazza}, D., {Hunt}, L.~K.,
  {Maiorano}, E., {Marinoni}, S., {Marshall}, F.~E., {Molinari}, E., {Nousek},
  J., {Pian}, E., {Racusin}, J.~L., {Stella}, L., {Amati}, L., {Andreuzzi}, G.,
  {Cusumano}, G., {Fenimore}, E.~E., {Ferrero}, P., {Giommi}, P., {Guetta}, D.,
  {Holland}, S.~T., {Hurley}, K., {Israel}, G.~L., {Mao}, J., {Markwardt},
  C.~B., {Masetti}, N., {Pagani}, C., {Palazzi}, E., {Palmer}, D.~M.,
  {Piranomonte}, S., {Tagliaferri}, G., {Testa}, V., Oct. 2009. {GRB090423 at a
  redshift of z\~{}8.1}. \nat 461, 1258--1260.

\bibitem[{{Samtaney} et~al.(2009){Samtaney}, {Loureiro}, {Uzdensky},
  {Schekochihin}, and {Cowley}}]{samtaney09}
{Samtaney}, R., {Loureiro}, N.~F., {Uzdensky}, D.~A., {Schekochihin}, A.~A.,
  {Cowley}, S.~C., Sep. 2009. {Formation of Plasmoid Chains in Magnetic
  Reconnection}. Physical Review Letters 103~(10), 105004.

\bibitem[{{Santana} et~al.(2014){Santana}, {Barniol Duran}, and
  {Kumar}}]{santana13}
{Santana}, R., {Barniol Duran}, R., {Kumar}, P., Apr. 2014. {Magnetic Fields in
  Relativistic Collisionless Shocks}. \apj 785, 29.

\bibitem[{{Sari} and {Esin}(2001)}]{sari01}
{Sari}, R., {Esin}, A.~A., Feb. 2001. {On the Synchrotron Self-Compton Emission
  from Relativistic Shocks and Its Implications for Gamma-Ray Burst
  Afterglows}. \apj 548, 787--799.

\bibitem[{{Sari} and {M{\'e}sz{\'a}ros}(2000)}]{sarimeszaros00}
{Sari}, R., {M{\'e}sz{\'a}ros}, P., May 2000. {Impulsive and Varying Injection
  in Gamma-Ray Burst Afterglows}. \apjl 535, L33--L37.

\bibitem[{{Sari} et~al.(1996){Sari}, {Narayan}, and {Piran}}]{sari96}
{Sari}, R., {Narayan}, R., {Piran}, T., Dec. 1996. {Cooling Timescales and
  Temporal Structure of Gamma-Ray Bursts}. \apj 473, 204.

\bibitem[{{Sari} and {Piran}(1995)}]{sari95}
{Sari}, R., {Piran}, T., Dec. 1995. {Hydrodynamic Timescales and Temporal
  Structure of Gamma-Ray Bursts}. \apjl 455, L143+.

\bibitem[{{Sari} and {Piran}(1997)}]{sari97}
{Sari}, R., {Piran}, T., Aug. 1997. {Variability in Gamma-Ray Bursts: A Clue}.
  \apj 485, 270--+.

\bibitem[{{Sari} and {Piran}(1999{\natexlab{a}})}]{saripiran99}
{Sari}, R., {Piran}, T., Jun. 1999{\natexlab{a}}. {GRB 990123: The Optical
  Flash and the Fireball Model}. \apjl 517, L109--L112.

\bibitem[{{Sari} and {Piran}(1999{\natexlab{b}})}]{saripiran99b}
{Sari}, R., {Piran}, T., Aug. 1999{\natexlab{b}}. {Predictions for the Very
  Early Afterglow and the Optical Flash}. \apj 520, 641--649.

\bibitem[{{Sari} et~al.(1999){Sari}, {Piran}, and {Halpern}}]{sari99}
{Sari}, R., {Piran}, T., {Halpern}, J.~P., Jul. 1999. {Jets in Gamma-Ray
  Bursts}. \apjl 519, L17--L20.

\bibitem[{{Sari} et~al.(1998){Sari}, {Piran}, and {Narayan}}]{sari98}
{Sari}, R., {Piran}, T., {Narayan}, R., Apr. 1998. {Spectra and Light Curves of
  Gamma-Ray Burst Afterglows}. \apjl 497, L17+.

\bibitem[{{Sato} et~al.(2007){Sato}, {Yamazaki}, {Ioka}, {Sakamoto},
  {Takahashi}, {Nakazawa}, {Nakamura}, {Toma}, {Hullinger}, {Tashiro},
  {Parsons}, {Krimm}, {Barthelmy}, {Gehrels}, {Burrows}, {O'Brien}, {Osborne},
  {Chincarini}, and {Lamb}}]{sato07}
{Sato}, G., {Yamazaki}, R., {Ioka}, K., {Sakamoto}, T., {Takahashi}, T.,
  {Nakazawa}, K., {Nakamura}, T., {Toma}, K., {Hullinger}, D., {Tashiro}, M.,
  {Parsons}, A.~M., {Krimm}, H.~A., {Barthelmy}, S.~D., {Gehrels}, N.,
  {Burrows}, D.~N., {O'Brien}, P.~T., {Osborne}, J.~P., {Chincarini}, G.,
  {Lamb}, D.~Q., Mar. 2007. {Swift Discovery of Gamma-Ray Bursts without a Jet
  Break Feature in Their X-Ray Afterglows}. \apj 657, 359--366.

\bibitem[{{Savaglio} et~al.(2009){Savaglio}, {Glazebrook}, and {Le
  Borgne}}]{savaglio09}
{Savaglio}, S., {Glazebrook}, K., {Le Borgne}, D., Jan. 2009. {The Galaxy
  Population Hosting Gamma-Ray Bursts}. \apj 691, 182--211.

\bibitem[{{Schaefer}(2003)}]{schaefer03}
{Schaefer}, B.~E., Feb. 2003. {Gamma-Ray Burst Hubble Diagram to z=4.5}. \apjl
  583, L67--L70.

\bibitem[{{Schaefer}(2007)}]{schaefer07}
{Schaefer}, B.~E., May 2007. {The Hubble Diagram to Redshift $>$6 from 69
  Gamma-Ray Bursts}. \apj 660, 16--46.

\bibitem[{{Schaerer}(2002)}]{schaerer02}
{Schaerer}, D., Jan. 2002. {On the properties of massive Population III stars
  and metal-free stellar populations}. \aap 382, 28--42.

\bibitem[{{Shao} and {Dai}(2005)}]{shao05}
{Shao}, L., {Dai}, Z.~G., Nov. 2005. {A Reverse-Shock Model for the Early
  Afterglow of GRB 050525A}. \apj 633, 1027--1030.

\bibitem[{{Shao} and {Dai}(2007)}]{shao07}
{Shao}, L., {Dai}, Z.~G., May 2007. {Behavior of X-Ray Dust Scattering and
  Implications for X-Ray Afterglows of Gamma-Ray Bursts}. \apj 660, 1319--1325.

\bibitem[{{Shapiro} et~al.(1976){Shapiro}, {Lightman}, and
  {Eardley}}]{shapiro76}
{Shapiro}, S.~L., {Lightman}, A.~P., {Eardley}, D.~M., Feb. 1976. {A
  two-temperature accretion disk model for Cygnus X-1 - Structure and
  spectrum}. \apj 204, 187--199.

\bibitem[{{Shemi} and {Piran}(1990)}]{shemi90}
{Shemi}, A., {Piran}, T., Dec. 1990. {The appearance of cosmic fireballs}.
  \apjl 365, L55--L58.

\bibitem[{{Shen} et~al.(2006){Shen}, {Kumar}, and {Robinson}}]{shen06}
{Shen}, R., {Kumar}, P., {Robinson}, E.~L., Sep. 2006. {No universality for the
  electron power-law index (p) in gamma-ray bursts and other relativistic
  sources}. \mnras 371, 1441--1447.

\bibitem[{{Shen} and {Matzner}(2012)}]{shen12}
{Shen}, R., {Matzner}, C.~D., Jan. 2012. {Coasting External Shock in Wind
  Medium: An Origin for the X-Ray Plateau Decay Component in Swift Gamma-Ray
  Burst Afterglows}. \apj 744, 36.

\bibitem[{{Shen} et~al.(2009){Shen}, {Willingale}, {Kumar}, {O'Brien}, and
  {Evans}}]{shen09b}
{Shen}, R., {Willingale}, R., {Kumar}, P., {O'Brien}, P.~T., {Evans}, P.~A.,
  Feb. 2009. {The dust scattering model cannot explain the shallow X-ray decay
  in GRB afterglows}. \mnras 393, 598--606.

\bibitem[{{Shen} and {Zhang}(2009)}]{shen09}
{Shen}, R., {Zhang}, B., Oct. 2009. {Prompt optical emission and synchrotron
  self-absorption constraints on emission site of GRBs}. \mnras 398,
  1936--1950.

\bibitem[{{Shibata} et~al.(2011){Shibata}, {Suwa}, {Kiuchi}, and
  {Ioka}}]{shibata11}
{Shibata}, M., {Suwa}, Y., {Kiuchi}, K., {Ioka}, K., Jun. 2011. {Afterglow of a
  Binary Neutron Star Merger}. \apjl 734, L36.

\bibitem[{{Siegel} et~al.(2014){Siegel}, {Ciolfi}, and {Rezzolla}}]{siegel14}
{Siegel}, D.~M., {Ciolfi}, R., {Rezzolla}, L., Apr. 2014. {Magnetically Driven
  Winds from Differentially Rotating Neutron Stars and X-Ray Afterglows of
  Short Gamma-Ray Bursts}. \apjl 785, L6.

\bibitem[{{Silva} et~al.(2003){Silva}, {Fonseca}, {Tonge}, {Dawson}, {Mori},
  and {Medvedev}}]{silva03}
{Silva}, L.~O., {Fonseca}, R.~A., {Tonge}, J.~W., {Dawson}, J.~M., {Mori},
  W.~B., {Medvedev}, M.~V., Oct. 2003. {Interpenetrating Plasma Shells:
  Near-equipartition Magnetic Field Generation and Nonthermal Particle
  Acceleration}. \apjl 596, L121--L124.

\bibitem[{{Sironi} and {Goodman}(2007)}]{sironi07}
{Sironi}, L., {Goodman}, J., Dec. 2007. {Production of Magnetic Energy by
  Macroscopic Turbulence in GRB Afterglows}. \apj 671, 1858--1867.

\bibitem[{{Sironi} and {Spitkovsky}(2011)}]{sironi11a}
{Sironi}, L., {Spitkovsky}, A., Jan. 2011. {Particle Acceleration in
  Relativistic Magnetized Collisionless Electron-Ion Shocks}. \apj 726, 75.

\bibitem[{{Sironi} and {Spitkovsky}(2012)}]{sironi12}
{Sironi}, L., {Spitkovsky}, A., 2012. {Particle Acceleration at the Termination
  Shock of Striped Pulsar Winds}. International Journal of Modern Physics
  Conference Series 8, 144.

\bibitem[{{Soderberg}(2007)}]{soderberg07}
{Soderberg}, A.~M., Oct. 2007. {The Radio Properties of Type Ibc Supernovae}.
  In: {Immler}, S., {Weiler}, K., {McCray}, R. (Eds.), Supernova 1987A: 20
  Years After: Supernovae and Gamma-Ray Bursters. Vol. 937 of American
  Institute of Physics Conference Series. pp. 492--499.

\bibitem[{{Soderberg} et~al.(2006{\natexlab{a}}){Soderberg}, {Berger},
  {Kasliwal}, {Frail}, {Price}, {Schmidt}, {Kulkarni}, {Fox}, {Cenko},
  {Gal-Yam}, {Nakar}, and {Roth}}]{soderberg06b}
{Soderberg}, A.~M., {Berger}, E., {Kasliwal}, M., {Frail}, D.~A., {Price},
  P.~A., {Schmidt}, B.~P., {Kulkarni}, S.~R., {Fox}, D.~B., {Cenko}, S.~B.,
  {Gal-Yam}, A., {Nakar}, E., {Roth}, K.~C., Oct. 2006{\natexlab{a}}. {The
  Afterglow, Energetics, and Host Galaxy of the Short-Hard Gamma-Ray Burst
  051221a}. \apj 650, 261--271.

\bibitem[{{Soderberg} et~al.(2006{\natexlab{b}}){Soderberg}, {Kulkarni},
  {Nakar}, {Berger}, {Cameron}, {Fox}, {Frail}, {Gal-Yam}, {Sari}, {Cenko},
  {Kasliwal}, {Chevalier}, {Piran}, {Price}, {Schmidt}, {Pooley}, {Moon},
  {Penprase}, {Ofek}, {Rau}, {Gehrels}, {Nousek}, {Burrows}, {Persson}, and
  {McCarthy}}]{soderberg06}
{Soderberg}, A.~M., {Kulkarni}, S.~R., {Nakar}, E., {Berger}, E., {Cameron},
  P.~B., {Fox}, D.~B., {Frail}, D., {Gal-Yam}, A., {Sari}, R., {Cenko}, S.~B.,
  {Kasliwal}, M., {Chevalier}, R.~A., {Piran}, T., {Price}, P.~A., {Schmidt},
  B.~P., {Pooley}, G., {Moon}, D.-S., {Penprase}, B.~E., {Ofek}, E., {Rau}, A.,
  {Gehrels}, N., {Nousek}, J.~A., {Burrows}, D.~N., {Persson}, S.~E.,
  {McCarthy}, P.~J., Aug. 2006{\natexlab{b}}. {Relativistic ejecta from X-ray
  flash XRF 060218 and the rate of cosmic explosions}. \nat 442, 1014--1017.

\bibitem[{{Sparre} et~al.(2011){Sparre}, {Sollerman}, {Fynbo}, {Malesani},
  {Goldoni}, {de Ugarte Postigo}, {Covino}, {D'Elia}, {Flores}, {Hammer},
  {Hjorth}, {Jakobsson}, {Kaper}, {Leloudas}, {Levan}, {Milvang-Jensen},
  {Schulze}, {Tagliaferri}, {Tanvir}, {Watson}, {Wiersema}, and
  {Wijers}}]{sparre11}
{Sparre}, M., {Sollerman}, J., {Fynbo}, J.~P.~U., {Malesani}, D., {Goldoni},
  P., {de Ugarte Postigo}, A., {Covino}, S., {D'Elia}, V., {Flores}, H.,
  {Hammer}, F., {Hjorth}, J., {Jakobsson}, P., {Kaper}, L., {Leloudas}, G.,
  {Levan}, A.~J., {Milvang-Jensen}, B., {Schulze}, S., {Tagliaferri}, G.,
  {Tanvir}, N.~R., {Watson}, D.~J., {Wiersema}, K., {Wijers}, R.~A.~M.~J., Jul.
  2011. {Spectroscopic Evidence for SN 2010ma Associated with GRB 101219B}.
  \apjl 735, L24.

\bibitem[{{Stacy} et~al.(2010){Stacy}, {Greif}, and {Bromm}}]{stacy10}
{Stacy}, A., {Greif}, T.~H., {Bromm}, V., Mar. 2010. {The first stars:
  formation of binaries and small multiple systems}. \mnras 403, 45--60.

\bibitem[{{Stanek} et~al.(2003){Stanek}, {Matheson}, {Garnavich}, {Martini},
  {Berlind}, {Caldwell}, {Challis}, {Brown}, {Schild}, {Krisciunas}, {Calkins},
  {Lee}, {Hathi}, {Jansen}, {Windhorst}, {Echevarria}, {Eisenstein}, {Pindor},
  {Olszewski}, {Harding}, {Holland}, and {Bersier}}]{stanek03}
{Stanek}, K.~Z., {Matheson}, T., {Garnavich}, P.~M., {Martini}, P., {Berlind},
  P., {Caldwell}, N., {Challis}, P., {Brown}, W.~R., {Schild}, R.,
  {Krisciunas}, K., {Calkins}, M.~L., {Lee}, J.~C., {Hathi}, N., {Jansen},
  R.~A., {Windhorst}, R., {Echevarria}, L., {Eisenstein}, D.~J., {Pindor}, B.,
  {Olszewski}, E.~W., {Harding}, P., {Holland}, S.~T., {Bersier}, D., Jul.
  2003. {Spectroscopic Discovery of the Supernova 2003dh Associated with GRB
  030329}. \apjl 591, L17--L20.

\bibitem[{{Starling} et~al.(2011){Starling}, {Wiersema}, {Levan}, {Sakamoto},
  {Bersier}, {Goldoni}, {Oates}, {Rowlinson}, {Campana}, {Sollerman}, {Tanvir},
  {Malesani}, {Fynbo}, {Covino}, {D'Avanzo}, {O'Brien}, {Page}, {Osborne},
  {Vergani}, {Barthelmy}, {Burrows}, {Cano}, {Curran}, {de Pasquale}, {D'Elia},
  {Evans}, {Flores}, {Fruchter}, {Garnavich}, {Gehrels}, {Gorosabel}, {Hjorth},
  {Holland}, {van der Horst}, {Hurkett}, {Jakobsson}, {Kamble}, {Kouveliotou},
  {Kuin}, {Kaper}, {Mazzali}, {Nugent}, {Pian}, {Stamatikos}, {Th{\"o}ne}, and
  {Woosley}}]{starling11}
{Starling}, R.~L.~C., {Wiersema}, K., {Levan}, A.~J., {Sakamoto}, T.,
  {Bersier}, D., {Goldoni}, P., {Oates}, S.~R., {Rowlinson}, A., {Campana}, S.,
  {Sollerman}, J., {Tanvir}, N.~R., {Malesani}, D., {Fynbo}, J.~P.~U.,
  {Covino}, S., {D'Avanzo}, P., {O'Brien}, P.~T., {Page}, K.~L., {Osborne},
  J.~P., {Vergani}, S.~D., {Barthelmy}, S., {Burrows}, D.~N., {Cano}, Z.,
  {Curran}, P.~A., {de Pasquale}, M., {D'Elia}, V., {Evans}, P.~A., {Flores},
  H., {Fruchter}, A.~S., {Garnavich}, P., {Gehrels}, N., {Gorosabel}, J.,
  {Hjorth}, J., {Holland}, S.~T., {van der Horst}, A.~J., {Hurkett}, C.~P.,
  {Jakobsson}, P., {Kamble}, A.~P., {Kouveliotou}, C., {Kuin}, N.~P.~M.,
  {Kaper}, L., {Mazzali}, P.~A., {Nugent}, P.~E., {Pian}, E., {Stamatikos}, M.,
  {Th{\"o}ne}, C.~C., {Woosley}, S.~E., Mar. 2011. {Discovery of the nearby
  long, soft GRB 100316D with an associated supernova}. \mnras 411, 2792--2803.

\bibitem[{{Steele} et~al.(2009){Steele}, {Mundell}, {Smith}, {Kobayashi}, and
  {Guidorzi}}]{steele09}
{Steele}, I.~A., {Mundell}, C.~G., {Smith}, R.~J., {Kobayashi}, S., {Guidorzi},
  C., 2009. {Ten per cent polarized optical emission from GRB?090102.} \nat
  462, 767--+.

\bibitem[{{Sunyaev} and {Titarchuk}(1980)}]{sunyaev80}
{Sunyaev}, R.~A., {Titarchuk}, L.~G., Jun. 1980. {Comptonization of X-rays in
  plasma clouds - Typical radiation spectra}. \aap 86, 121--138.

\bibitem[{{Suwa} and {Ioka}(2011)}]{suwa11}
{Suwa}, Y., {Ioka}, K., Jan. 2011. {Can Gamma-ray Burst Jets Break Out the
  First Stars?} \apj 726, 107.

\bibitem[{{Sweet}(1958)}]{sweet58}
{Sweet}, P.~A., 1958. {The Neutral Point Theory of Solar Flares}. In:
  {B.~Lehnert} (Ed.), Electromagnetic Phenomena in Cosmical Physics. Vol.~6 of
  IAU Symposium. pp. 123--+.

\bibitem[{{Tagliaferri} et~al.(2005){Tagliaferri}, {Goad}, {Chincarini},
  {Moretti}, {Campana}, {Burrows}, {Perri}, {Barthelmy}, {Gehrels}, {Krimm},
  {Sakamoto}, {Kumar}, {M{\'e}sz{\'a}ros}, {Kobayashi}, {Zhang}, {Angelini},
  {Banat}, {Beardmore}, {Capalbi}, {Covino}, {Cusumano}, {Giommi}, {Godet},
  {Hill}, {Kennea}, {Mangano}, {Morris}, {Nousek}, {O'Brien}, {Osborne},
  {Pagani}, {Page}, {Romano}, {Stella}, and {Wells}}]{tagliaferri05}
{Tagliaferri}, G., {Goad}, M., {Chincarini}, G., {Moretti}, A., {Campana}, S.,
  {Burrows}, D.~N., {Perri}, M., {Barthelmy}, S.~D., {Gehrels}, N., {Krimm},
  H., {Sakamoto}, T., {Kumar}, P., {M{\'e}sz{\'a}ros}, P.~I., {Kobayashi}, S.,
  {Zhang}, B., {Angelini}, L., {Banat}, P., {Beardmore}, A.~P., {Capalbi}, M.,
  {Covino}, S., {Cusumano}, G., {Giommi}, P., {Godet}, O., {Hill}, J.~E.,
  {Kennea}, J.~A., {Mangano}, V., {Morris}, D.~C., {Nousek}, J.~A., {O'Brien},
  P.~T., {Osborne}, J.~P., {Pagani}, C., {Page}, K.~L., {Romano}, P., {Stella},
  L., {Wells}, A., Aug. 2005. {An unexpectedly rapid decline in the X-ray
  afterglow emission of long {$\gamma$}-ray bursts}. \nat 436, 985--988.

\bibitem[{{Takahashi} et~al.(2011){Takahashi}, {Kudoh}, {Masada}, and
  {Matsumoto}}]{takahashi11}
{Takahashi}, H.~R., {Kudoh}, T., {Masada}, Y., {Matsumoto}, J., Oct. 2011.
  {Scaling Law of Relativistic Sweet-Parker-type Magnetic Reconnection}. \apjl
  739, L53.

\bibitem[{{Tanvir} et~al.(2005){Tanvir}, {Chapman}, {Levan}, and
  {Priddey}}]{tanvir05}
{Tanvir}, N.~R., {Chapman}, R., {Levan}, A.~J., {Priddey}, R.~S., Dec. 2005.
  {An origin in the local Universe for some short {$\gamma$}-ray bursts}. \nat
  438, 991--993.

\bibitem[{{Tanvir} et~al.(2009){Tanvir}, {Fox}, {Levan}, {Berger}, {Wiersema},
  {Fynbo}, {Cucchiara}, {Kr{\"u}hler}, {Gehrels}, {Bloom}, {Greiner}, {Evans},
  {Rol}, {Olivares}, {Hjorth}, {Jakobsson}, {Farihi}, {Willingale}, {Starling},
  {Cenko}, {Perley}, {Maund}, {Duke}, {Wijers}, {Adamson}, {Allan}, {Bremer},
  {Burrows}, {Castro-Tirado}, {Cavanagh}, {de Ugarte Postigo}, {Dopita},
  {Fatkhullin}, {Fruchter}, {Foley}, {Gorosabel}, {Kennea}, {Kerr}, {Klose},
  {Krimm}, {Komarova}, {Kulkarni}, {Moskvitin}, {Mundell}, {Naylor}, {Page},
  {Penprase}, {Perri}, {Podsiadlowski}, {Roth}, {Rutledge}, {Sakamoto},
  {Schady}, {Schmidt}, {Soderberg}, {Sollerman}, {Stephens}, {Stratta},
  {Ukwatta}, {Watson}, {Westra}, {Wold}, and {Wolf}}]{tanvir09}
{Tanvir}, N.~R., {Fox}, D.~B., {Levan}, A.~J., {Berger}, E., {Wiersema}, K.,
  {Fynbo}, J.~P.~U., {Cucchiara}, A., {Kr{\"u}hler}, T., {Gehrels}, N.,
  {Bloom}, J.~S., {Greiner}, J., {Evans}, P.~A., {Rol}, E., {Olivares}, F.,
  {Hjorth}, J., {Jakobsson}, P., {Farihi}, J., {Willingale}, R., {Starling},
  R.~L.~C., {Cenko}, S.~B., {Perley}, D., {Maund}, J.~R., {Duke}, J., {Wijers},
  R.~A.~M.~J., {Adamson}, A.~J., {Allan}, A., {Bremer}, M.~N., {Burrows},
  D.~N., {Castro-Tirado}, A.~J., {Cavanagh}, B., {de Ugarte Postigo}, A.,
  {Dopita}, M.~A., {Fatkhullin}, T.~A., {Fruchter}, A.~S., {Foley}, R.~J.,
  {Gorosabel}, J., {Kennea}, J., {Kerr}, T., {Klose}, S., {Krimm}, H.~A.,
  {Komarova}, V.~N., {Kulkarni}, S.~R., {Moskvitin}, A.~S., {Mundell}, C.~G.,
  {Naylor}, T., {Page}, K., {Penprase}, B.~E., {Perri}, M., {Podsiadlowski},
  P., {Roth}, K., {Rutledge}, R.~E., {Sakamoto}, T., {Schady}, P., {Schmidt},
  B.~P., {Soderberg}, A.~M., {Sollerman}, J., {Stephens}, A.~W., {Stratta}, G.,
  {Ukwatta}, T.~N., {Watson}, D., {Westra}, E., {Wold}, T., {Wolf}, C., Oct.
  2009. {A {$\gamma$}-ray burst at a redshift of z\~{}8.2}. \nat 461,
  1254--1257.

\bibitem[{{Tanvir} et~al.(2013){Tanvir}, {Levan}, {Fruchter}, {Hjorth},
  {Hounsell}, {Wiersema}, and {Tunnicliffe}}]{tanvir13}
{Tanvir}, N.~R., {Levan}, A.~J., {Fruchter}, A.~S., {Hjorth}, J., {Hounsell},
  R.~A., {Wiersema}, K., {Tunnicliffe}, R.~L., Aug. 2013. {A `kilonova'
  associated with the short-duration {$\gamma$}-ray burst GRB130603B}. \nat
  500, 547--549.

\bibitem[{{Taylor} et~al.(2004){Taylor}, {Frail}, {Berger}, and
  {Kulkarni}}]{taylor04}
{Taylor}, G.~B., {Frail}, D.~A., {Berger}, E., {Kulkarni}, S.~R., Jul. 2004.
  {The Angular Size and Proper Motion of the Afterglow of GRB 030329}. \apjl
  609, L1--L4.

\bibitem[{{Tchekhovskoy} and {McKinney}(2012)}]{tchekhovskoy12}
{Tchekhovskoy}, A., {McKinney}, J.~C., Jun. 2012. {Prograde and retrograde
  black holes: whose jet is more powerful?} \mnras 423, L55--L59.

\bibitem[{{Tchekhovskoy} et~al.(2008){Tchekhovskoy}, {McKinney}, and
  {Narayan}}]{tchekhovskoy08}
{Tchekhovskoy}, A., {McKinney}, J.~C., {Narayan}, R., Aug. 2008. {Simulations
  of ultrarelativistic magnetodynamic jets from gamma-ray burst engines}.
  \mnras 388, 551--572.

\bibitem[{{Tchekhovskoy} et~al.(2009){Tchekhovskoy}, {McKinney}, and
  {Narayan}}]{tchekhovskoy09}
{Tchekhovskoy}, A., {McKinney}, J.~C., {Narayan}, R., Jul. 2009. {Efficiency of
  Magnetic to Kinetic Energy Conversion in a Monopole Magnetosphere}. \apj 699,
  1789--1808.

\bibitem[{{Tchekhovskoy} et~al.(2010){Tchekhovskoy}, {Narayan}, and
  {McKinney}}]{tchekhovskoy10}
{Tchekhovskoy}, A., {Narayan}, R., {McKinney}, J.~C., Nov. 2010.
  {Magnetohydrodynamic simulations of gamma-ray burst jets: Beyond the
  progenitor star}. \na 15, 749--754.

\bibitem[{{Tegmark} et~al.(1997){Tegmark}, {Silk}, {Rees}, {Blanchard}, {Abel},
  and {Palla}}]{tegmark97}
{Tegmark}, M., {Silk}, J., {Rees}, M.~J., {Blanchard}, A., {Abel}, T., {Palla},
  F., Jan. 1997. {How Small Were the First Cosmological Objects?} \apj 474, 1.

\bibitem[{{Thompson}(1994)}]{thompson94}
{Thompson}, C., Oct. 1994. {A Model of Gamma-Ray Bursts}. \mnras 270, 480--+.

\bibitem[{{Thompson} and {Madau}(2000)}]{thompson00}
{Thompson}, C., {Madau}, P., Jul. 2000. {Relativistic Winds from Compact
  Gamma-Ray Sources. II. Pair Loading and Radiative Acceleration in Gamma-Ray
  Bursts}. \apj 538, 105--114.

\bibitem[{{Thompson} et~al.(2007){Thompson}, {M{\'e}sz{\'a}ros}, and
  {Rees}}]{thompson07}
{Thompson}, C., {M{\'e}sz{\'a}ros}, P., {Rees}, M.~J., Sep. 2007.
  {Thermalization in Relativistic Outflows and the Correlation between Spectral
  Hardness and Apparent Luminosity in Gamma-Ray Bursts}. \apj 666, 1012--1023.

\bibitem[{{Th{\"o}ne} et~al.(2011){Th{\"o}ne}, {Campana}, {Lazzati}, {de Ugarte
  Postigo}, {Fynbo}, {Christensen}, {Levan}, {Aloy}, {Hjorth}, {Jakobsson},
  {Levesque}, {Malesani}, {Milvang-Jensen}, {Roming}, {Tanvir}, {Wiersema},
  {Gladders}, {Wuyts}, and {Dahle}}]{thone11}
{Th{\"o}ne}, C.~C., {Campana}, S., {Lazzati}, D., {de Ugarte Postigo}, A.,
  {Fynbo}, J.~P.~U., {Christensen}, L., {Levan}, A.~J., {Aloy}, M.~A.,
  {Hjorth}, J., {Jakobsson}, P., {Levesque}, E.~M., {Malesani}, D.,
  {Milvang-Jensen}, B., {Roming}, P.~W.~A., {Tanvir}, N.~R., {Wiersema}, K.,
  {Gladders}, M., {Wuyts}, E., {Dahle}, H., Jun. 2011. {Variable Ly{$\alpha$}
  sheds light on the environment surrounding GRB 090426}. \mnras 414, 479--488.

\bibitem[{{Titarchuk} et~al.(2012){Titarchuk}, {Farinelli}, {Frontera}, and
  {Amati}}]{titarchuk12}
{Titarchuk}, L., {Farinelli}, R., {Frontera}, F., {Amati}, L., Jun. 2012. {An
  Upscattering Spectral Formation Model for the Prompt Emission of Gamma-Ray
  Bursts}. \apj 752, 116.

\bibitem[{{Toma} et~al.(2011{\natexlab{a}}){Toma}, {Sakamoto}, and
  {M{\'e}sz{\'a}ros}}]{toma11b}
{Toma}, K., {Sakamoto}, T., {M{\'e}sz{\'a}ros}, P., Apr. 2011{\natexlab{a}}.
  {Population III Gamma-ray Burst Afterglows: Constraints on Stellar Masses and
  External Medium Densities}. \apj 731, 127.

\bibitem[{{Toma} et~al.(2009){Toma}, {Wu}, and {M{\'e}sz{\'a}ros}}]{toma09}
{Toma}, K., {Wu}, X., {M{\'e}sz{\'a}ros}, P., Dec. 2009. {An Up-Scattered
  Cocoon Emission Model of Gamma-Ray Burst High-Energy Lags}. \apj 707,
  1404--1416.

\bibitem[{{Toma} et~al.(2011{\natexlab{b}}){Toma}, {Wu}, and
  {M{\'e}sz{\'a}ros}}]{toma11}
{Toma}, K., {Wu}, X.-F., {M{\'e}sz{\'a}ros}, P., Aug. 2011{\natexlab{b}}.
  {Photosphere-internal shock model of gamma-ray bursts: case studies of
  Fermi/LAT bursts}. \mnras 415, 1663--1680.

\bibitem[{{Totani}(1998)}]{totani98}
{Totani}, T., Dec. 1998. {TEV Burst of Gamma-Ray Bursts and Ultra-High-Energy
  Cosmic Rays}. \apjl 509, L81--L84.

\bibitem[{{Totani} et~al.(2006){Totani}, {Kawai}, {Kosugi}, {Aoki}, {Yamada},
  {Iye}, {Ohta}, and {Hattori}}]{totani06}
{Totani}, T., {Kawai}, N., {Kosugi}, G., {Aoki}, K., {Yamada}, T., {Iye}, M.,
  {Ohta}, K., {Hattori}, T., Jun. 2006. {Implications for Cosmic Reionization
  from the Optical Afterglow Spectrum of the Gamma-Ray Burst 050904 at z =
  6.3}. \pasj 58, 485--498.

\bibitem[{{Totani} and {Panaitescu}(2002)}]{totani02}
{Totani}, T., {Panaitescu}, A., Sep. 2002. {Orphan Afterglows of Collimated
  Gamma-Ray Bursts: Rate Predictions and Prospects for Detection}. \apj 576,
  120--134.

\bibitem[{{Trenti} et~al.(2013){Trenti}, {Perna}, and {Tacchella}}]{trenti13}
{Trenti}, M., {Perna}, R., {Tacchella}, S., Aug. 2013. {Gamma-Ray Burst and
  Star Formation Rates: The Physical Origin for the Redshift Evolution of Their
  Ratio}. \apjl 773, L22.

\bibitem[{{Troja} et~al.(2007){Troja}, {Cusumano}, {O'Brien}, {Zhang},
  {Sbarufatti}, {Mangano}, {Willingale}, {Chincarini}, {Osborne}, {Marshall},
  {Burrows}, {Campana}, {Gehrels}, {Guidorzi}, {Krimm}, {La Parola}, {Liang},
  {Mineo}, {Moretti}, {Page}, {Romano}, {Tagliaferri}, {Zhang}, {Page}, and
  {Schady}}]{troja07}
{Troja}, E., {Cusumano}, G., {O'Brien}, P.~T., {Zhang}, B., {Sbarufatti}, B.,
  {Mangano}, V., {Willingale}, R., {Chincarini}, G., {Osborne}, J.~P.,
  {Marshall}, F.~E., {Burrows}, D.~N., {Campana}, S., {Gehrels}, N.,
  {Guidorzi}, C., {Krimm}, H.~A., {La Parola}, V., {Liang}, E.~W., {Mineo}, T.,
  {Moretti}, A., {Page}, K.~L., {Romano}, P., {Tagliaferri}, G., {Zhang},
  B.~B., {Page}, M.~J., {Schady}, P., Aug. 2007. {Swift Observations of GRB
  070110: An Extraordinary X-Ray Afterglow Powered by the Central Engine}. \apj
  665, 599--607.

\bibitem[{{Tsutsui} et~al.(2013){Tsutsui}, {Yonetoku}, {Nakamura}, {Takahashi},
  and {Morihara}}]{tsutsui13}
{Tsutsui}, R., {Yonetoku}, D., {Nakamura}, T., {Takahashi}, K., {Morihara}, Y.,
  May 2013. {Possible existence of the E$_{p}$-L$_{p}$ and E$_{p}$-E$_{iso}$
  correlations for short gamma-ray bursts with a factor 5-100 dimmer than those
  for long gamma-ray bursts}. \mnras 431, 1398--1404.

\bibitem[{{Tumlinson} and {Shull}(2000)}]{tumlinson00}
{Tumlinson}, J., {Shull}, J.~M., Jan. 2000. {Zero-Metallicity Stars and the
  Effects of the First Stars on Reionization}. \apjl 528, L65--L68.

\bibitem[{{Uehara} et~al.(2012){Uehara}, {Toma}, {Kawabata}, {Chiyonobu},
  {Fukazawa}, {Ikejiri}, {Inoue}, {Itoh}, {Komatsu}, {Miyamoto}, {Mizuno},
  {Nagae}, {Nakaya}, {Ohsugi}, {Sakimoto}, {Sasada}, {Tanaka}, {Uemura},
  {Yamanaka}, {Yamashita}, {Yamazaki}, and {Yoshida}}]{uehara12}
{Uehara}, T., {Toma}, K., {Kawabata}, K.~S., {Chiyonobu}, S., {Fukazawa}, Y.,
  {Ikejiri}, Y., {Inoue}, T., {Itoh}, R., {Komatsu}, T., {Miyamoto}, H.,
  {Mizuno}, T., {Nagae}, O., {Nakaya}, H., {Ohsugi}, T., {Sakimoto}, K.,
  {Sasada}, M., {Tanaka}, H., {Uemura}, M., {Yamanaka}, M., {Yamashita}, T.,
  {Yamazaki}, R., {Yoshida}, M., Jun. 2012. {GRB 091208B: First Detection of
  the Optical Polarization in Early Forward Shock Emission of a Gamma-Ray Burst
  Afterglow}. \apjl 752, L6.

\bibitem[{{Uhm} and {Beloborodov}(2007)}]{uhm07}
{Uhm}, Z.~L., {Beloborodov}, A.~M., Aug. 2007. {On the Mechanism of Gamma-Ray
  Burst Afterglows}. \apjl 665, L93--L96.

\bibitem[{{Uhm} and {Zhang}(2014{\natexlab{a}})}]{uhm14c}
{Uhm}, Z.~L., {Zhang}, B., Jul. 2014{\natexlab{a}}. {Dynamics and Afterglow
  Light Curves of Gamma-Ray Burst Blast Waves Encountering a Density Bump or
  Void}. \apj 789, 39.

\bibitem[{{Uhm} and {Zhang}(2014{\natexlab{b}})}]{uhm14}
{Uhm}, Z.~L., {Zhang}, B., May 2014{\natexlab{b}}. {Fast-cooling synchrotron
  radiation in a decaying magnetic field and {$\gamma$}-ray burst emission
  mechanism}. Nature Physics 10, 351--356.

\bibitem[{{Uhm} et~al.(2012){Uhm}, {Zhang}, {Hasco{\"e}t}, {Daigne},
  {Mochkovitch}, and {Park}}]{uhm12}
{Uhm}, Z.~L., {Zhang}, B., {Hasco{\"e}t}, R., {Daigne}, F., {Mochkovitch}, R.,
  {Park}, I.~H., Dec. 2012. {Dynamics and Afterglow Light Curves of Gamma-Ray
  Burst Blast Waves with a Long-lived Reverse Shock}. \apj 761, 147.

\bibitem[{{Ukwatta} et~al.(2012){Ukwatta}, {Dhuga}, {Stamatikos}, {Dermer},
  {Sakamoto}, {Sonbas}, {Parke}, {Maximon}, {Linnemann}, {Bhat}, {Eskandarian},
  {Gehrels}, {Abeysekara}, {Tollefson}, and {Norris}}]{ukwatta12}
{Ukwatta}, T.~N., {Dhuga}, K.~S., {Stamatikos}, M., {Dermer}, C.~D.,
  {Sakamoto}, T., {Sonbas}, E., {Parke}, W.~C., {Maximon}, L.~C., {Linnemann},
  J.~T., {Bhat}, P.~N., {Eskandarian}, A., {Gehrels}, N., {Abeysekara}, A.~U.,
  {Tollefson}, K., {Norris}, J.~P., Jan. 2012. {The lag-luminosity relation in
  the GRB source frame: an investigation with Swift BAT bursts}. \mnras 419,
  614--623.

\bibitem[{{Usov}(1992)}]{usov92}
{Usov}, V.~V., Jun. 1992. {Millisecond pulsars with extremely strong magnetic
  fields as a cosmological source of gamma-ray bursts}. \nat 357, 472--474.

\bibitem[{{Usov}(1994)}]{usov94}
{Usov}, V.~V., Apr. 1994. {On the Nature of Nonthermal Radiation from
  Cosmological Gamma-Ray Bursters}. \mnras 267, 1035--+.

\bibitem[{{Uzdensky} et~al.(2011){Uzdensky}, {Cerutti}, and
  {Begelman}}]{uzdensky11}
{Uzdensky}, D.~A., {Cerutti}, B., {Begelman}, M.~C., Aug. 2011.
  {Reconnection-powered Linear Accelerator and Gamma-Ray Flares in the Crab
  Nebula}. \apjl 737, L40.

\bibitem[{{Uzdensky} and {Kulsrud}(2000)}]{uzdensky00}
{Uzdensky}, D.~A., {Kulsrud}, R.~M., Oct. 2000. {Two-dimensional numerical
  simulation of the resistive reconnection layer}. Physics of Plasmas 7,
  4018--4030.

\bibitem[{{Uzdensky} and {MacFadyen}(2006)}]{uzdensky06}
{Uzdensky}, D.~A., {MacFadyen}, A.~I., Aug. 2006. {Stellar Explosions by
  Magnetic Towers}. \apj 647, 1192--1212.

\bibitem[{{van der Horst} et~al.(2008){van der Horst}, {Kamble}, {Resmi},
  {Wijers}, {Bhattacharya}, {Scheers}, {Rol}, {Strom}, {Kouveliotou},
  {Oosterloo}, and {Ishwara-Chandra}}]{vanderhorst08}
{van der Horst}, A.~J., {Kamble}, A., {Resmi}, L., {Wijers}, R.~A.~M.~J.,
  {Bhattacharya}, D., {Scheers}, B., {Rol}, E., {Strom}, R., {Kouveliotou}, C.,
  {Oosterloo}, T., {Ishwara-Chandra}, C.~H., Mar. 2008. {Detailed study of the
  GRB 030329 radio afterglow deep into the non-relativistic phase}. \aap 480,
  35--43.

\bibitem[{{van Eerten} et~al.(2012){van Eerten}, {van der Horst}, and
  {MacFadyen}}]{vaneerten12b}
{van Eerten}, H., {van der Horst}, A., {MacFadyen}, A., Apr. 2012. {Gamma-Ray
  Burst Afterglow Broadband Fitting Based Directly on Hydrodynamics
  Simulations}. \apj 749, 44.

\bibitem[{{van Eerten} and {MacFadyen}(2012)}]{vaneerten12a}
{van Eerten}, H.~J., {MacFadyen}, A.~I., Mar. 2012. {Gamma-Ray Burst Afterglow
  Scaling Relations for the Full Blast Wave Evolution}. \apjl 747, L30.

\bibitem[{{van Paradijs} et~al.(1997){van Paradijs}, {Groot}, {Galama},
  {Kouveliotou}, {Strom}, {Telting}, {Rutten}, {Fishman}, {Meegan}, {Pettini},
  {Tanvir}, {Bloom}, {Pedersen}, {N{\o}rdgaard-Nielsen}, {Linden-V{\o}rnle},
  {Melnick}, {van der Steene}, {Bremer}, {Naber}, {Heise}, {in't Zand},
  {Costa}, {Feroci}, {Piro}, {Frontera}, {Zavattini}, {Nicastro}, {Palazzi},
  {Bennet}, {Hanlon}, and {Parmar}}]{vanpara97}
{van Paradijs}, J., {Groot}, P.~J., {Galama}, T., {Kouveliotou}, C., {Strom},
  R.~G., {Telting}, J., {Rutten}, R.~G.~M., {Fishman}, G.~J., {Meegan}, C.~A.,
  {Pettini}, M., {Tanvir}, N., {Bloom}, J., {Pedersen}, H.,
  {N{\o}rdgaard-Nielsen}, H.~U., {Linden-V{\o}rnle}, M., {Melnick}, J., {van
  der Steene}, G., {Bremer}, M., {Naber}, R., {Heise}, J., {in't Zand}, J.,
  {Costa}, E., {Feroci}, M., {Piro}, L., {Frontera}, F., {Zavattini}, G.,
  {Nicastro}, L., {Palazzi}, E., {Bennet}, K., {Hanlon}, L., {Parmar}, A., Apr.
  1997. {Transient optical emission from the error box of the {$\gamma$}-ray
  burst of 28 February 1997}. \nat 386, 686--689.

\bibitem[{{Venkatesan} and {Truran}(2003)}]{venkatesan03}
{Venkatesan}, A., {Truran}, J.~W., Sep. 2003. {The Ionizing Efficiency of the
  First Stars}. \apjl 594, L1--L4.

\bibitem[{{Veres} et~al.(2010){Veres}, {Bagoly}, {Horv{\'a}th},
  {M{\'e}sz{\'a}ros}, and {Bal{\'a}zs}}]{veres10}
{Veres}, P., {Bagoly}, Z., {Horv{\'a}th}, I., {M{\'e}sz{\'a}ros}, A.,
  {Bal{\'a}zs}, L.~G., Dec. 2010. {A Distinct Peak-flux Distribution of the
  Third Class of Gamma-ray Bursts: A Possible Signature of X-ray Flashes?} \apj
  725, 1955--1964.

\bibitem[{{Veres} et~al.(2012){Veres}, {Zhang}, and
  {M{\'e}sz{\'a}ros}}]{veres12b}
{Veres}, P., {Zhang}, B.-B., {M{\'e}sz{\'a}ros}, P., Dec. 2012. {The Extremely
  High Peak Energy of GRB 110721A in the Context of a Dissipative Photosphere
  Synchrotron Emission Model}. \apjl 761, L18.

\bibitem[{{Vestrand} et~al.(2005){Vestrand}, {Wozniak}, {Wren}, {Fenimore},
  {Sakamoto}, {White}, {Casperson}, {Davis}, {Evans}, {Galassi}, {McGowan},
  {Schier}, {Asa}, {Barthelmy}, {Cummings}, {Gehrels}, {Hullinger}, {Krimm},
  {Markwardt}, {McLean}, {Palmer}, {Parsons}, and {Tueller}}]{vestrand05}
{Vestrand}, W.~T., {Wozniak}, P.~R., {Wren}, J.~A., {Fenimore}, E.~E.,
  {Sakamoto}, T., {White}, R.~R., {Casperson}, D., {Davis}, H., {Evans}, S.,
  {Galassi}, M., {McGowan}, K.~E., {Schier}, J.~A., {Asa}, J.~W., {Barthelmy},
  S.~D., {Cummings}, J.~R., {Gehrels}, N., {Hullinger}, D., {Krimm}, H.~A.,
  {Markwardt}, C.~B., {McLean}, K., {Palmer}, D., {Parsons}, A., {Tueller}, J.,
  May 2005. {A link between prompt optical and prompt {$\gamma$}-ray emission
  in {$\gamma$}-ray bursts}. \nat 435, 178--180.

\bibitem[{{Vestrand} et~al.(2006){Vestrand}, {Wren}, {Wozniak}, {Aptekar},
  {Golentskii}, {Pal'Shin}, {Sakamoto}, {White}, {Evans}, {Casperson}, and
  {Fenimore}}]{vestrand06}
{Vestrand}, W.~T., {Wren}, J.~A., {Wozniak}, P.~R., {Aptekar}, R.,
  {Golentskii}, S., {Pal'Shin}, V., {Sakamoto}, T., {White}, R.~R., {Evans},
  S., {Casperson}, D., {Fenimore}, E., Jul. 2006. {Energy input and response
  from prompt and early optical afterglow emission in {$\gamma$}-ray bursts}.
  \nat 442, 172--175.

\bibitem[{{Vetere} et~al.(2006){Vetere}, {Massaro}, {Costa}, {Soffitta}, and
  {Ventura}}]{vetere06}
{Vetere}, L., {Massaro}, E., {Costa}, E., {Soffitta}, P., {Ventura}, G., Feb.
  2006. {Slow and fast components in the X-ray light curves of gamma-ray
  bursts}. \aap 447, 499--513.

\bibitem[{{Vietri}(1995)}]{vietri95}
{Vietri}, M., Nov. 1995. {The Acceleration of Ultra--High-Energy Cosmic Rays in
  Gamma-Ray Bursts}. \apj 453, 883--+.

\bibitem[{{Virgili} et~al.(2009){Virgili}, {Liang}, and {Zhang}}]{virgili09}
{Virgili}, F.~J., {Liang}, E.-W., {Zhang}, B., Jan. 2009. {Low-luminosity
  gamma-ray bursts as a distinct GRB population: a firmer case from multiple
  criteria constraints}. \mnras 392, 91--103.

\bibitem[{{Virgili} et~al.(2013){Virgili}, {Mundell}, {Pal'shin}, {Guidorzi},
  {Margutti}, {Melandri}, {Harrison}, {Kobayashi}, {Chornock}, {Henden},
  {Updike}, {Cenko}, {Tanvir}, {Steele}, {Cucchiara}, {Gomboc}, {Levan},
  {Cano}, {Mottram}, {Clay}, {Bersier}, {Kopa{\v c}}, {Japelj}, {Filippenko},
  {Li}, {Svinkin}, {Golenetskii}, {Hartmann}, {Milne}, {Williams}, {O'Brien},
  {Fox}, and {Berger}}]{virgili13}
{Virgili}, F.~J., {Mundell}, C.~G., {Pal'shin}, V., {Guidorzi}, C., {Margutti},
  R., {Melandri}, A., {Harrison}, R., {Kobayashi}, S., {Chornock}, R.,
  {Henden}, A., {Updike}, A.~C., {Cenko}, S.~B., {Tanvir}, N.~R., {Steele},
  I.~A., {Cucchiara}, A., {Gomboc}, A., {Levan}, A., {Cano}, Z., {Mottram},
  C.~J., {Clay}, N.~R., {Bersier}, D., {Kopa{\v c}}, D., {Japelj}, J.,
  {Filippenko}, A.~V., {Li}, W., {Svinkin}, D., {Golenetskii}, S., {Hartmann},
  D.~H., {Milne}, P.~A., {Williams}, G., {O'Brien}, P.~T., {Fox}, D.~B.,
  {Berger}, E., Nov. 2013. {GRB 091024A and the Nature of Ultra-long Gamma-Ray
  Bursts}. \apj 778, 54.

\bibitem[{{Virgili} et~al.(2011){Virgili}, {Zhang}, {O'Brien}, and
  {Troja}}]{virgili11}
{Virgili}, F.~J., {Zhang}, B., {O'Brien}, P., {Troja}, E., Feb. 2011. {Are All
  Short-hard Gamma-ray Bursts Produced from Mergers of Compact Stellar
  Objects?} \apj 727, 109.

\bibitem[{{Vlahakis} and {K{\"o}nigl}(2003)}]{vlahakis03}
{Vlahakis}, N., {K{\"o}nigl}, A., Oct. 2003. {Relativistic Magnetohydrodynamics
  with Application to Gamma-Ray Burst Outflows. I. Theory and Semianalytic
  Trans-Alfv{\'e}nic Solutions}. \apj 596, 1080--1103.

\bibitem[{{Vlahakis} et~al.(2003){Vlahakis}, {Peng}, and
  {K{\"o}nigl}}]{vlahakis03b}
{Vlahakis}, N., {Peng}, F., {K{\"o}nigl}, A., Sep. 2003. {Neutron-rich
  Hydromagnetic Outflows in Gamma-Ray Burst Sources}. \apjl 594, L23--L26.

\bibitem[{{Vurm} et~al.(2011){Vurm}, {Beloborodov}, and {Poutanen}}]{vurm11}
{Vurm}, I., {Beloborodov}, A.~M., {Poutanen}, J., Sep. 2011. {Gamma-Ray Bursts
  from Magnetized Collisionally Heated Jets}. \apj 738, 77.

\bibitem[{{Vurm} et~al.(2013){Vurm}, {Lyubarsky}, and {Piran}}]{vurm13}
{Vurm}, I., {Lyubarsky}, Y., {Piran}, T., Feb. 2013. {On Thermalization in
  Gamma-Ray Burst Jets and the Peak Energies of Photospheric Spectra}. \apj
  764, 143.

\bibitem[{{Wanderman} and {Piran}(2010)}]{wanderman10}
{Wanderman}, D., {Piran}, T., Aug. 2010. {The luminosity function and the rate
  of Swift's gamma-ray bursts}. \mnras 406, 1944--1958.

\bibitem[{{Wanderman} and {Piran}(2014)}]{wanderman14}
{Wanderman}, D., {Piran}, T., May 2014. {The rate, luminosity function and time
  delay of non-Collapsar short GRBs}. ArXiv e-prints.

\bibitem[{{Wang} et~al.(2002){Wang}, {Xiao}, and {Lei}}]{wangdx02}
{Wang}, D.~X., {Xiao}, K., {Lei}, W.~H., Sep. 2002. {Evolution characteristics
  of the central black hole of a magnetized accretion disc}. \mnras 335,
  655--664.

\bibitem[{{Wang} et~al.(2004){Wang}, {Cheng}, {Dai}, and {Lu}}]{wang04}
{Wang}, X.~Y., {Cheng}, K.~S., {Dai}, Z.~G., {Lu}, T., Mar. 2004. {Constraining
  the Origin of TeV Photons from Gamma-Ray Bursts with Delayed MeV-GeV Emission
  Formed by Interaction with Cosmic Infrared/Microwave Background Photons}.
  \apj 604, 306--311.

\bibitem[{{Wang} and {Dai}(2009)}]{wangdai09}
{Wang}, X.-Y., {Dai}, Z.-G., Feb. 2009. {Prompt TeV Neutrinos from the
  Dissipative Photospheres of Gamma-ray Bursts}. \apjl 691, L67--L71.

\bibitem[{{Wang} et~al.(2001{\natexlab{a}}){Wang}, {Dai}, and {Lu}}]{wang01}
{Wang}, X.~Y., {Dai}, Z.~G., {Lu}, T., Jan. 2001{\natexlab{a}}. {Prompt
  High-Energy Gamma-Ray Emission from the Synchrotron Self-Compton Process in
  the Reverse Shocks of Gamma-Ray Bursts}. \apjl 546, L33--L37.

\bibitem[{{Wang} et~al.(2001{\natexlab{b}}){Wang}, {Dai}, and {Lu}}]{wang01b}
{Wang}, X.~Y., {Dai}, Z.~G., {Lu}, T., Aug. 2001{\natexlab{b}}. {The Inverse
  Compton Emission Spectra in the Very Early Afterglows of Gamma-Ray Bursts}.
  \apj 556, 1010--1016.

\bibitem[{{Wang} et~al.(2010){Wang}, {He}, {Li}, {Wu}, and {Dai}}]{wang10}
{Wang}, X.-Y., {He}, H.-N., {Li}, Z., {Wu}, X.-F., {Dai}, Z.-G., Apr. 2010.
  {Klein-Nishina Effects on the High-energy Afterglow Emission of Gamma-ray
  Bursts}. \apj 712, 1232--1240.

\bibitem[{{Wang} et~al.(2006){Wang}, {Li}, and {M{\'e}sz{\'a}ros}}]{wang06}
{Wang}, X.-Y., {Li}, Z., {M{\'e}sz{\'a}ros}, P., Apr. 2006. {GeV-TeV and X-Ray
  Flares from Gamma-Ray Bursts}. \apjl 641, L89--L92.

\bibitem[{{Wang} et~al.(2007){Wang}, {Li}, {Waxman}, and
  {M{\'e}sz{\'a}ros}}]{wang07}
{Wang}, X.-Y., {Li}, Z., {Waxman}, E., {M{\'e}sz{\'a}ros}, P., Aug. 2007.
  {Nonthermal Gamma-Ray/X-Ray Flashes from Shock Breakout in Gamma-Ray
  Burst-Associated Supernovae}. \apj 664, 1026--1032.

\bibitem[{{Waxman}(1995)}]{waxman95}
{Waxman}, E., Jul. 1995. {Cosmological Gamma-Ray Bursts and the Highest Energy
  Cosmic Rays}. Physical Review Letters 75, 386--389.

\bibitem[{{Waxman} and {Bahcall}(1997)}]{waxman97}
{Waxman}, E., {Bahcall}, J., Mar. 1997. {High Energy Neutrinos from
  Cosmological Gamma-Ray Burst Fireballs}. Physical Review Letters 78,
  2292--2295.

\bibitem[{{Waxman} and {Bahcall}(2000)}]{waxman00}
{Waxman}, E., {Bahcall}, J.~N., Oct. 2000. {Neutrino Afterglow from Gamma-Ray
  Bursts: $10^{18}$ EV}. \apj 541, 707--711.

\bibitem[{{Wei} and {Gao}(2003)}]{weigao03}
{Wei}, D.~M., {Gao}, W.~H., Nov. 2003. {Are there cosmological evolution trends
  on gamma-ray burst features?} \mnras 345, 743--746.

\bibitem[{{Wei} and {Lu}(1998)}]{weilu98}
{Wei}, D.~M., {Lu}, T., Sep. 1998. {Diverse Temporal Properties of Gamma-Ray
  Burst Afterglows}. \apj 505, 252--254.

\bibitem[{{Weibel}(1959)}]{weibel59}
{Weibel}, E.~S., Feb. 1959. {Spontaneously Growing Transverse Waves in a Plasma
  Due to an Anisotropic Velocity Distribution}. Physical Review Letters 2,
  83--84.

\bibitem[{{Wheeler} et~al.(2000){Wheeler}, {Yi}, {H{\"o}flich}, and
  {Wang}}]{wheeler00}
{Wheeler}, J.~C., {Yi}, I., {H{\"o}flich}, P., {Wang}, L., Jul. 2000.
  {Asymmetric Supernovae, Pulsars, Magnetars, and Gamma-Ray Bursts}. \apj 537,
  810--823.

\bibitem[{{Wiersema} et~al.(2014){Wiersema}, {Covino}, {Toma}, {van der Horst},
  {Varela}, {Min}, {Greiner}, {Starling}, {Tanvir}, {Wijers}, {Campana},
  {Curran}, {Fan}, {Fynbo}, {Gorosabel}, {Gomboc}, {Gotz}, {Hjorth}, {Jin},
  {Kobayashi}, {Kouveliotou}, {Mundell}, {O'Brien}, {Pian}, {Rowlinson},
  {Russell}, {Salvaterra}, {di Serego Alighieri}, {Tagliaferri}, {Vergani},
  {Elliott}, {Farina}, {Hartoog}, {Karjalainen}, {Klose}, {Knust}, {Levan},
  {Schady}, {Sudilovsky}, and {Willingale}}]{wiersema14}
{Wiersema}, K., {Covino}, S., {Toma}, K., {van der Horst}, A.~J., {Varela}, K.,
  {Min}, M., {Greiner}, J., {Starling}, R.~L.~C., {Tanvir}, N.~R., {Wijers},
  R.~A.~M.~J., {Campana}, S., {Curran}, P.~A., {Fan}, Y., {Fynbo}, J.~P.~U.,
  {Gorosabel}, J., {Gomboc}, A., {Gotz}, D., {Hjorth}, J., {Jin}, Z.~P.,
  {Kobayashi}, S., {Kouveliotou}, C., {Mundell}, C., {O'Brien}, P.~T., {Pian},
  E., {Rowlinson}, A., {Russell}, D.~M., {Salvaterra}, R., {di Serego
  Alighieri}, S., {Tagliaferri}, G., {Vergani}, S.~D., {Elliott}, J., {Farina},
  C., {Hartoog}, O.~E., {Karjalainen}, R., {Klose}, S., {Knust}, F., {Levan},
  A.~J., {Schady}, P., {Sudilovsky}, V., {Willingale}, R., May 2014. {Circular
  polarization in the optical afterglow of GRB 121024A.} \nat 509, 201--204.

\bibitem[{{Wijers} and {Galama}(1999)}]{wijers99}
{Wijers}, R.~A.~M.~J., {Galama}, T.~J., Sep. 1999. {Physical Parameters of GRB
  970508 and GRB 971214 from Their Afterglow Synchrotron Emission}. \apj 523,
  177--186.

\bibitem[{{Wijers} et~al.(1997){Wijers}, {Rees}, and {M\'esz\'aros}}]{wijers97}
{Wijers}, R.~A.~M.~J., {Rees}, M.~J., {M\'esz\'aros}, P., Jul. 1997. {Shocked
  by GRB 970228: the afterglow of a cosmological fireball}. \mnras 288,
  L51--L56.

\bibitem[{{Willingale} et~al.(2010){Willingale}, {Genet}, {Granot}, and
  {O'Brien}}]{willingale10}
{Willingale}, R., {Genet}, F., {Granot}, J., {O'Brien}, P.~T., Apr. 2010. {The
  spectral-temporal properties of the prompt pulses and rapid decay phase of
  gamma-ray bursts}. \mnras 403, 1296--1316.

\bibitem[{{Willingale} et~al.(2007){Willingale}, {O'Brien}, {Osborne}, {Godet},
  {Page}, {Goad}, {Burrows}, {Zhang}, {Rol}, {Gehrels}, and
  {Chincarini}}]{willingale07}
{Willingale}, R., {O'Brien}, P.~T., {Osborne}, J.~P., {Godet}, O., {Page},
  K.~L., {Goad}, M.~R., {Burrows}, D.~N., {Zhang}, B., {Rol}, E., {Gehrels},
  N., {Chincarini}, G., Jun. 2007. {Testing the Standard Fireball Model of
  Gamma-Ray Bursts Using Late X-Ray Afterglows Measured by Swift}. \apj 662,
  1093--1110.

\bibitem[{{Willis} et~al.(2005){Willis}, {Barlow}, {Bird}, {Clark}, {Dean},
  {McConnell}, {Moran}, {Shaw}, and {Sguera}}]{willis05}
{Willis}, D.~R., {Barlow}, E.~J., {Bird}, A.~J., {Clark}, D.~J., {Dean}, A.~J.,
  {McConnell}, M.~L., {Moran}, L., {Shaw}, S.~E., {Sguera}, V., Aug. 2005.
  {Evidence of polarisation in the prompt gamma-ray emission from GRB 930131
  and GRB 960924}. \aap 439, 245--253.

\bibitem[{{Woosley}(1993)}]{woosley93}
{Woosley}, S.~E., Mar. 1993. {Gamma-ray bursts from stellar mass accretion
  disks around black holes}. \apj 405, 273--277.

\bibitem[{{Woosley}(2011)}]{woosley11}
{Woosley}, S.~E., May 2011. {Models for Gamma-Ray Burst Progenitors and Central
  Engines}. ArXiv e-prints.

\bibitem[{{Woosley} and {Bloom}(2006)}]{woosley06}
{Woosley}, S.~E., {Bloom}, J.~S., Sep. 2006. {The Supernova Gamma-Ray Burst
  Connection}. \araa 44, 507--556.

\bibitem[{{Wu} et~al.(2003){Wu}, {Dai}, {Huang}, and {Lu}}]{wu03}
{Wu}, X.~F., {Dai}, Z.~G., {Huang}, Y.~F., {Lu}, T., Jul. 2003. {Optical
  flashes and very early afterglows in wind environments}. \mnras 342,
  1131--1138.

\bibitem[{{Wu} et~al.(2005){Wu}, {Dai}, {Huang}, and {Lu}}]{wu05b}
{Wu}, X.~F., {Dai}, Z.~G., {Huang}, Y.~F., {Lu}, T., Mar. 2005. {Gamma-ray
  bursts: polarization of afterglows from two-component jets}. \mnras 357,
  1197--1204.

\bibitem[{{Wu} et~al.(2006){Wu}, {Dai}, {Wang}, {Huang}, {Feng}, and
  {Lu}}]{wu06}
{Wu}, X.~F., {Dai}, Z.~G., {Wang}, X.~Y., {Huang}, Y.~F., {Feng}, L.~L., {Lu},
  T., 2006. {X-ray flares from late internal and late external shocks}. In:
  36th COSPAR Scientific Assembly. Vol.~36 of COSPAR, Plenary Meeting. pp.
  731--+.

\bibitem[{{Xin} et~al.(2011){Xin}, {Liang}, {Wei}, {Zhang}, {Lv}, {Zheng},
  {Urata}, {Im}, {Wang}, {Qiu}, {Deng}, {Huang}, {Hu}, {Jeon}, {Li}, and
  {Han}}]{xin11}
{Xin}, L.-P., {Liang}, E.-W., {Wei}, J.-Y., {Zhang}, B., {Lv}, H.-J., {Zheng},
  W.-K., {Urata}, Y., {Im}, M., {Wang}, J., {Qiu}, Y.-L., {Deng}, J.-S.,
  {Huang}, K.-Y., {Hu}, J.-Y., {Jeon}, Y., {Li}, H.-L., {Han}, X.-H., Jan.
  2011. {Probing the nature of high-z short GRB 090426 with its early optical
  and X-ray afterglows}. \mnras 410, 27--32.

\bibitem[{{Xu} et~al.(2005){Xu}, {Dai}, and {Liang}}]{xu05}
{Xu}, D., {Dai}, Z.~G., {Liang}, E.~W., Nov. 2005. {Can Gamma-Ray Bursts Be
  Used to Measure Cosmology? A Further Analysis}. \apj 633, 603--610.

\bibitem[{{Xu} et~al.(2013){Xu}, {de Ugarte Postigo}, {Leloudas}, {Kruhler},
  {Cano}, {Hjorth}, {Malesani}, {Fynbo}, {Thoene}, {Sanchez-Ramirez},
  {Schulze}, {Jakobsson}, {Kaper}, {Sollerman}, {Watson}, {Cabrera-Lavers},
  {Cao}, {Covino}, {Flores}, {Geier}, {Gorosabel}, {Hu}, {Milvang-Jensen},
  {Sparre}, {Xin}, {Zhang}, {Zheng}, and {Zou}}]{xu13}
{Xu}, D., {de Ugarte Postigo}, A., {Leloudas}, G., {Kruhler}, T., {Cano}, Z.,
  {Hjorth}, J., {Malesani}, D., {Fynbo}, J.~P.~U., {Thoene}, C.~C.,
  {Sanchez-Ramirez}, R., {Schulze}, S., {Jakobsson}, P., {Kaper}, L.,
  {Sollerman}, J., {Watson}, D.~J., {Cabrera-Lavers}, A., {Cao}, C., {Covino},
  S., {Flores}, H., {Geier}, S., {Gorosabel}, J., {Hu}, S.~M.,
  {Milvang-Jensen}, B., {Sparre}, M., {Xin}, L.~P., {Zhang}, T.~M., {Zheng},
  W.~K., {Zou}, Y.~C., May 2013. {Discovery of the broad-lined Type Ic SN
  2013cq associated with the very energetic GRB 130427A}. ArXiv e-prints.

\bibitem[{{Xu} et~al.(2009){Xu}, {Starling}, {Fynbo}, {Sollerman}, {Yost},
  {Watson}, {Foley}, {O'Brien}, and {Hjorth}}]{xu09}
{Xu}, D., {Starling}, R.~L.~C., {Fynbo}, J.~P.~U., {Sollerman}, J., {Yost}, S.,
  {Watson}, D., {Foley}, S., {O'Brien}, P.~T., {Hjorth}, J., May 2009. {In
  Search of Progenitors for Supernovaless Gamma-Ray Bursts 060505 and 060614:
  Re-examination of Their Afterglows}. \apj 696, 971--979.

\bibitem[{{Xu} and {Huang}(2012)}]{xuhuang12}
{Xu}, M., {Huang}, Y.~F., Feb. 2012. {New three-parameter correlation for
  gamma-ray bursts with a plateau phase in the afterglow}. \aap 538, A134.

\bibitem[{{Yamazaki}(2009)}]{yamazaki09}
{Yamazaki}, R., Jan. 2009. {Prior Emission Model for X-ray Plateau Phase of
  Gamma-Ray Burst Afterglows}. \apjl 690, L118--L121.

\bibitem[{{Yamazaki} et~al.(2004{\natexlab{a}}){Yamazaki}, {Ioka}, and
  {Nakamura}}]{yamazaki04b}
{Yamazaki}, R., {Ioka}, K., {Nakamura}, T., Jun. 2004{\natexlab{a}}. {A Unified
  Model of Short and Long Gamma-Ray Bursts, X-Ray-rich Gamma-Ray Bursts, and
  X-Ray Flashes}. \apjl 607, L103--L106.

\bibitem[{{Yamazaki} et~al.(2004{\natexlab{b}}){Yamazaki}, {Ioka}, and
  {Nakamura}}]{yamazaki04}
{Yamazaki}, R., {Ioka}, K., {Nakamura}, T., May 2004{\natexlab{b}}. {Peak
  Energy-Isotropic Energy Relation in the Off-Axis Gamma-Ray Burst Model}.
  \apjl 606, L33--L36.

\bibitem[{{Yi} et~al.(2006){Yi}, {Liang}, {Qin}, and {Lu}}]{yi06}
{Yi}, T., {Liang}, E., {Qin}, Y., {Lu}, R., Apr. 2006. {On the spectral lags of
  the short gamma-ray bursts}. \mnras 367, 1751--1756.

\bibitem[{{Yonetoku} et~al.(2012){Yonetoku}, {Murakami}, {Gunji}, {Mihara},
  {Toma}, {Morihara}, {Takahashi}, {Wakashima}, {Yonemochi}, {Sakashita},
  {Toukairin}, {Fujimoto}, and {Kodama}}]{yonetoku12}
{Yonetoku}, D., {Murakami}, T., {Gunji}, S., {Mihara}, T., {Toma}, K.,
  {Morihara}, Y., {Takahashi}, T., {Wakashima}, Y., {Yonemochi}, H.,
  {Sakashita}, T., {Toukairin}, N., {Fujimoto}, H., {Kodama}, Y., Oct. 2012.
  {Magnetic Structures in Gamma-Ray Burst Jets Probed by Gamma-Ray
  Polarization}. \apjl 758, L1.

\bibitem[{{Yonetoku} et~al.(2011){Yonetoku}, {Murakami}, {Gunji}, {Mihara},
  {Toma}, {Sakashita}, {Morihara}, {Takahashi}, {Toukairin}, {Fujimoto},
  {Kodama}, {Kubo}, and {IKAROS Demonstration Team}}]{yonetoku11}
{Yonetoku}, D., {Murakami}, T., {Gunji}, S., {Mihara}, T., {Toma}, K.,
  {Sakashita}, T., {Morihara}, Y., {Takahashi}, T., {Toukairin}, N.,
  {Fujimoto}, H., {Kodama}, Y., {Kubo}, S., {IKAROS Demonstration Team}, Dec.
  2011. {Detection of Gamma-Ray Polarization in Prompt Emission of GRB
  100826A}. \apjl 743, L30.

\bibitem[{{Yonetoku} et~al.(2004){Yonetoku}, {Murakami}, {Nakamura},
  {Yamazaki}, {Inoue}, and {Ioka}}]{yonetoku04}
{Yonetoku}, D., {Murakami}, T., {Nakamura}, T., {Yamazaki}, R., {Inoue}, A.~K.,
  {Ioka}, K., Jul. 2004. {Gamma-Ray Burst Formation Rate Inferred from the
  Spectral Peak Energy-Peak Luminosity Relation}. \apj 609, 935--951.

\bibitem[{{Yoshida} et~al.(2003){Yoshida}, {Abel}, {Hernquist}, and
  {Sugiyama}}]{yoshida03}
{Yoshida}, N., {Abel}, T., {Hernquist}, L., {Sugiyama}, N., Aug. 2003.
  {Simulations of Early Structure Formation: Primordial Gas Clouds}. \apj 592,
  645--663.

\bibitem[{{Yost} et~al.(2003){Yost}, {Harrison}, {Sari}, and {Frail}}]{yost03}
{Yost}, S.~A., {Harrison}, F.~A., {Sari}, R., {Frail}, D.~A., Nov. 2003. {A
  Study of the Afterglows of Four Gamma-Ray Bursts: Constraining the Explosion
  and Fireball Model}. \apj 597, 459--473.

\bibitem[{{Yost} et~al.(2007){Yost}, {Swan}, {Rykoff}, {Aharonian}, {Akerlof},
  {Alday}, {Ashley}, {Barthelmy}, {Burrows}, and {Depoy}}]{yost07}
{Yost}, S.~A., {Swan}, H.~F., {Rykoff}, E.~S., {Aharonian}, F., {Akerlof},
  C.~W., {Alday}, A., {Ashley}, M.~C.~B., {Barthelmy}, S., {Burrows}, D.,
  {Depoy}, e.~a., Mar. 2007. {Exploring Broadband GRB Behavior during
  {$\gamma$}-Ray Emission}. \apj 657, 925--941.

\bibitem[{{Yu} et~al.(2010){Yu}, {Cheng}, and {Cao}}]{yu10}
{Yu}, Y.-W., {Cheng}, K.~S., {Cao}, X.-F., May 2010. {The Role of Newly Born
  Magnetars in Gamma-ray Burst X-ray Afterglow Emission: Energy Injection and
  Internal Emission}. \apj 715, 477--484.

\bibitem[{{Yu} et~al.(2009){Yu}, {Wang}, and {Dai}}]{yu09}
{Yu}, Y.~W., {Wang}, X.~Y., {Dai}, Z.~G., Feb. 2009. {Optical and
  {$\gamma$}-ray Emissions from Internal Forward-Reverse Shocks: Application to
  GRB 080319B?} \apj 692, 1662--1668.

\bibitem[{{Yu} et~al.(2013){Yu}, {Zhang}, and {Gao}}]{yu13}
{Yu}, Y.-W., {Zhang}, B., {Gao}, H., Aug. 2013. {Bright ''merger-nova'' from
  the remnant of a neutron star binary merger: A signature of a newly born,
  massive, millisecond magnetar}. ArXiv e-prints.

\bibitem[{{Yuan} and {Zhang}(2012)}]{yuan12}
{Yuan}, F., {Zhang}, B., Sep. 2012. {Episodic Jets as the Central Engine of
  Gamma-Ray Bursts}. \apj 757, 56.

\bibitem[{{Y{\"u}ksel} et~al.(2008){Y{\"u}ksel}, {Kistler}, {Beacom}, and
  {Hopkins}}]{yuksel08}
{Y{\"u}ksel}, H., {Kistler}, M.~D., {Beacom}, J.~F., {Hopkins}, A.~M., Aug.
  2008. {Revealing the High-Redshift Star Formation Rate with Gamma-Ray
  Bursts}. \apjl 683, L5--L8.

\bibitem[{{Zalamea} and {Beloborodov}(2011)}]{zalamea11}
{Zalamea}, I., {Beloborodov}, A.~M., Feb. 2011. {Neutrino heating near
  hyper-accreting black holes}. \mnras 410, 2302--2308.

\bibitem[{{Zeh} et~al.(2004){Zeh}, {Klose}, and {Hartmann}}]{zeh04}
{Zeh}, A., {Klose}, S., {Hartmann}, D.~H., Jul. 2004. {A Systematic Analysis of
  Supernova Light in Gamma-Ray Burst Afterglows}. \apj 609, 952--961.

\bibitem[{{Zenitani} and {Hoshino}(2001)}]{zenitani01}
{Zenitani}, S., {Hoshino}, M., Nov. 2001. {The Generation of Nonthermal
  Particles in the Relativistic Magnetic Reconnection of Pair Plasmas}. \apjl
  562, L63--L66.

\bibitem[{{Zenitani} and {Hoshino}(2007)}]{zenitani07}
{Zenitani}, S., {Hoshino}, M., Nov. 2007. {Particle Acceleration and Magnetic
  Dissipation in Relativistic Current Sheet of Pair Plasmas}. \apj 670,
  702--726.

\bibitem[{{Zenitani} and {Hoshino}(2008)}]{zenitani08}
{Zenitani}, S., {Hoshino}, M., Apr. 2008. {The Role of the Guide Field in
  Relativistic Pair Plasma Reconnection}. \apj 677, 530--544.

\bibitem[{{Zhang}(2006)}]{zhangnature06}
{Zhang}, B., Dec. 2006. {Astrophysics: A burst of new ideas}. \nat 444,
  1010--1011.

\bibitem[{{Zhang}(2007)}]{zhangcjaa07}
{Zhang}, B., Feb. 2007. {Gamma-Ray Bursts in the Swift Era}. Chinese Journal of
  Astronomy and Astrophysics 7, 1--50.

\bibitem[{{Zhang}(2013)}]{zhang13}
{Zhang}, B., Jan. 2013. {Early X-Ray and Optical Afterglow of Gravitational
  Wave Bursts from Mergers of Binary Neutron Stars}. \apjl 763, L22.

\bibitem[{{Zhang}(2014)}]{zhang14}
{Zhang}, B., Jan. 2014. {A Possible Connection between Fast Radio Bursts and
  Gamma-Ray Bursts}. \apjl 780, L21.

\bibitem[{{Zhang} et~al.(2004{\natexlab{a}}){Zhang}, {Dai}, {Lloyd-Ronning},
  and {M{\'e}sz{\'a}ros}}]{zhang04}
{Zhang}, B., {Dai}, X., {Lloyd-Ronning}, N.~M., {M{\'e}sz{\'a}ros}, P., Feb.
  2004{\natexlab{a}}. {Quasi-universal Gaussian Jets: A Unified Picture for
  Gamma-Ray Bursts and X-Ray Flashes}. \apjl 601, L119--L122.

\bibitem[{{Zhang} et~al.(2006){Zhang}, {Fan}, {Dyks}, {Kobayashi},
  {M{\'e}sz{\'a}ros}, {Burrows}, {Nousek}, and {Gehrels}}]{zhang06}
{Zhang}, B., {Fan}, Y.~Z., {Dyks}, J., {Kobayashi}, S., {M{\'e}sz{\'a}ros}, P.,
  {Burrows}, D.~N., {Nousek}, J.~A., {Gehrels}, N., May 2006. {Physical
  Processes Shaping Gamma-Ray Burst X-Ray Afterglow Light Curves: Theoretical
  Implications from the Swift X-Ray Telescope Observations}. \apj 642,
  354--370.

\bibitem[{{Zhang} and {Kobayashi}(2005)}]{zhangkobayashi05}
{Zhang}, B., {Kobayashi}, S., Jul. 2005. {Gamma-Ray Burst Early Afterglows:
  Reverse Shock Emission from an Arbitrarily Magnetized Ejecta}. \apj 628,
  315--334.

\bibitem[{{Zhang} et~al.(2003{\natexlab{a}}){Zhang}, {Kobayashi}, and
  {M{\'e}sz{\'a}ros}}]{zhang03}
{Zhang}, B., {Kobayashi}, S., {M{\'e}sz{\'a}ros}, P., Oct. 2003{\natexlab{a}}.
  {Gamma-Ray Burst Early Optical Afterglows: Implications for the Initial
  Lorentz Factor and the Central Engine}. \apj 595, 950--954.

\bibitem[{{Zhang} and {Kumar}(2013)}]{zhangkumar13}
{Zhang}, B., {Kumar}, P., Mar. 2013. {Model-Dependent High-Energy Neutrino Flux
  from Gamma-Ray Bursts}. Physical Review Letters 110~(12), 121101.

\bibitem[{{Zhang} et~al.(2007{\natexlab{a}}){Zhang}, {Liang}, {Page}, {Grupe},
  {Zhang}, {Barthelmy}, {Burrows}, {Campana}, {Chincarini}, {Gehrels},
  {Kobayashi}, {M{\'e}sz{\'a}ros}, {Moretti}, {Nousek}, {O'Brien}, {Osborne},
  {Roming}, {Sakamoto}, {Schady}, and {Willingale}}]{zhang07a}
{Zhang}, B., {Liang}, E., {Page}, K.~L., {Grupe}, D., {Zhang}, B.-B.,
  {Barthelmy}, S.~D., {Burrows}, D.~N., {Campana}, S., {Chincarini}, G.,
  {Gehrels}, N., {Kobayashi}, S., {M{\'e}sz{\'a}ros}, P., {Moretti}, A.,
  {Nousek}, J.~A., {O'Brien}, P.~T., {Osborne}, J.~P., {Roming}, P.~W.~A.,
  {Sakamoto}, T., {Schady}, P., {Willingale}, R., Feb. 2007{\natexlab{a}}. {GRB
  Radiative Efficiencies Derived from the Swift Data: GRBs versus XRFs, Long
  versus Sh ort}. \apj 655, 989--1001.

\bibitem[{{Zhang} et~al.(2012{\natexlab{a}}){Zhang}, {Lu}, {Liang}, and
  {Wu}}]{zhang12}
{Zhang}, B., {Lu}, R.-J., {Liang}, E.-W., {Wu}, X.-F., Oct. 2012{\natexlab{a}}.
  {GRB 110721A: Photosphere ''Death Line'' and the Physical Origin of the GRB
  Band Function}. \apjl 758, L34.

\bibitem[{{Zhang} and
  {M{\'e}sz{\'a}ros}(2001{\natexlab{a}})}]{zhangmeszaros01a}
{Zhang}, B., {M{\'e}sz{\'a}ros}, P., May 2001{\natexlab{a}}. {Gamma-Ray Burst
  Afterglow with Continuous Energy Injection: Signature of a Highly Magnetized
  Millisecond Pulsar}. \apjl 552, L35--L38.

\bibitem[{{Zhang} and
  {M{\'e}sz{\'a}ros}(2001{\natexlab{b}})}]{zhangmeszaros01b}
{Zhang}, B., {M{\'e}sz{\'a}ros}, P., Sep. 2001{\natexlab{b}}. {High-Energy
  Spectral Components in Gamma-Ray Burst Afterglows}. \apj 559, 110--122.

\bibitem[{{Zhang} and
  {M{\'e}sz{\'a}ros}(2002{\natexlab{a}})}]{zhangmeszaros02c}
{Zhang}, B., {M{\'e}sz{\'a}ros}, P., Dec. 2002{\natexlab{a}}. {An Analysis of
  Gamma-Ray Burst Spectral Break Models}. \apj 581, 1236--1247.

\bibitem[{{Zhang} and
  {M{\'e}sz{\'a}ros}(2002{\natexlab{b}})}]{zhangmeszaros02b}
{Zhang}, B., {M{\'e}sz{\'a}ros}, P., Jun. 2002{\natexlab{b}}. {Gamma-Ray Burst
  Beaming: A Universal Configuration with a Standard Energy Reservoir?} \apj
  571, 876--879.

\bibitem[{{Zhang} and
  {M{\'e}sz{\'a}ros}(2002{\natexlab{c}})}]{zhangmeszaros02a}
{Zhang}, B., {M{\'e}sz{\'a}ros}, P., Feb. 2002{\natexlab{c}}. {Gamma-Ray Bursts
  with Continuous Energy Injection and Their Afterglow Signature}. \apj 566,
  712--722.

\bibitem[{{Zhang} and {M{\'e}sz{\'a}ros}(2004)}]{zhangmeszaros04}
{Zhang}, B., {M{\'e}sz{\'a}ros}, P., 2004. {Gamma-Ray Bursts: progress,
  problems \& prospects}. International Journal of Modern Physics A 19,
  2385--2472.

\bibitem[{{Zhang} and {Pe'er}(2009)}]{zhangpeer09}
{Zhang}, B., {Pe'er}, A., Aug. 2009. {Evidence of an Initially Magnetically
  Dominated Outflow in GRB 080916C}. \apjl 700, L65--L68.

\bibitem[{{Zhang} and {Yan}(2011)}]{zhangyan11}
{Zhang}, B., {Yan}, H., Jan. 2011. {The Internal-collision-induced Magnetic
  Reconnection and Turbulence (ICMART) Model of Gamma-ray Bursts}. \apj 726,
  90.

\bibitem[{{Zhang} and {Zhang}(2014)}]{zhangzhang14}
{Zhang}, B., {Zhang}, B., Feb. 2014. {Gamma-Ray Burst Prompt Emission Light
  Curves and Power Density Spectra in the ICMART Model}. \apj 782, 92.

\bibitem[{{Zhang} et~al.(2009{\natexlab{a}}){Zhang}, {Zhang}, {Virgili},
  {Liang}, {Kann}, {Wu}, {Proga}, {Lv}, {Toma}, {M{\'e}sz{\'a}ros}, {Burrows},
  {Roming}, and {Gehrels}}]{zhang09}
{Zhang}, B., {Zhang}, B., {Virgili}, F.~J., {Liang}, E., {Kann}, D.~A., {Wu},
  X., {Proga}, D., {Lv}, H., {Toma}, K., {M{\'e}sz{\'a}ros}, P., {Burrows},
  D.~N., {Roming}, P.~W.~A., {Gehrels}, N., Oct. 2009{\natexlab{a}}.
  {Discerning the Physical Origins of Cosmological Gamma-ray Bursts Based on
  Multiple Observational Criteria: The Cases of z = 6.7 GRB 080913, z = 8.2 GRB
  090423, and Some Short/Hard GRBs}. \apj 703, 1696--1724.

\bibitem[{{Zhang} et~al.(2007{\natexlab{b}}){Zhang}, {Zhang}, {Liang},
  {Gehrels}, {Burrows}, and {M{\'e}sz{\'a}ros}}]{zhang07b}
{Zhang}, B., {Zhang}, B.-B., {Liang}, E.-W., {Gehrels}, N., {Burrows}, D.~N.,
  {M{\'e}sz{\'a}ros}, P., Jan. 2007{\natexlab{b}}. {Making a Short Gamma-Ray
  Burst from a Long One: Implications for the Nature of GRB 060614}. \apjl 655,
  L25--L28.

\bibitem[{{Zhang} et~al.(2012{\natexlab{b}}){Zhang}, {Fan}, {Shen}, {Xu},
  {Zhang}, {Wei}, {Burrows}, {Zhang}, and {Gehrels}}]{zhangbb12}
{Zhang}, B.-B., {Fan}, Y.-Z., {Shen}, R.-F., {Xu}, D., {Zhang}, F.-W., {Wei},
  D.-M., {Burrows}, D.~N., {Zhang}, B., {Gehrels}, N., Sep. 2012{\natexlab{b}}.
  {GRB 120422A: A Low-luminosity Gamma-Ray Burst Driven by a Central Engine}.
  \apj 756, 190.

\bibitem[{{Zhang} et~al.(2007{\natexlab{c}}){Zhang}, {Liang}, and
  {Zhang}}]{zhangbb07}
{Zhang}, B.-B., {Liang}, E.-W., {Zhang}, B., Sep. 2007{\natexlab{c}}. {A
  Comprehensive Analysis of Swift XRT Data. I. Apparent Spectral Evolution of
  Gamma-Ray Burst X-Ray Tails}. \apj 666, 1002--1011.

\bibitem[{{Zhang} et~al.(2011){Zhang}, {Zhang}, {Liang}, {Fan}, {Wu}, {Pe'er},
  {Maxham}, {Gao}, and {Dong}}]{zhang11}
{Zhang}, B.-B., {Zhang}, B., {Liang}, E.-W., {Fan}, Y.-Z., {Wu}, X.-F.,
  {Pe'er}, A., {Maxham}, A., {Gao}, H., {Dong}, Y.-M., Apr. 2011. {A
  Comprehensive Analysis of Fermi Gamma-ray Burst Data. I. Spectral Components
  and the Possible Physical Origins of LAT/GBM GRBs}. \apj 730, 141.

\bibitem[{{Zhang} et~al.(2009{\natexlab{b}}){Zhang}, {Zhang}, {Liang}, and
  {Wang}}]{zhangbb09}
{Zhang}, B.-B., {Zhang}, B., {Liang}, E.-W., {Wang}, X.-Y., Jan.
  2009{\natexlab{b}}. {Curvature Effect of a Non-Power Spectrum and Spectral
  Evolution of GRB X-Ray Tails}. \apjl 690, L10--L13.

\bibitem[{{Zhang} et~al.(2014{\natexlab{a}}){Zhang}, {Zhang}, {Murase}, {Connaughton}, and
  {Briggs}}]{zhangbb14a}
{Zhang}, B.-B., {Zhang}, B., {Murase}, K., {Connaughton}, V., {Briggs}, M.~S.,
  May 2014a. {How Long does a Burst Burst?} \apj 787, 66.

\bibitem[Zhang et al.(2014{\natexlab{b}})]{zhangbb14b} Zhang, B.-B., van Eerten,  
  H., Burrows, D.~N., et al.\ 2014b, arXiv:1405.4867  

\bibitem[{{Zhang} et~al.(2012{\natexlab{c}}){Zhang}, {Shao}, {Yan}, and
  {Wei}}]{zhangfw12}
{Zhang}, F.-W., {Shao}, L., {Yan}, J.-Z., {Wei}, D.-M., May 2012{\natexlab{c}}.
  {Revisiting the Long/Soft-Short/Hard Classification of Gamma-Ray Bursts in
  the Fermi Era}. \apj 750, 88.

\bibitem[{{Zhang} and {MacFadyen}(2009)}]{zhangmacfadyen09}
{Zhang}, W., {MacFadyen}, A., Jun. 2009. {The Dynamics and Afterglow Radiation
  of Gamma-Ray Bursts. I. Constant Density Medium}. \apj 698, 1261--1272.

\bibitem[{{Zhang} et~al.(2004{\natexlab{b}}){Zhang}, {Woosley}, and
  {Heger}}]{zhangw04}
{Zhang}, W., {Woosley}, S.~E., {Heger}, A., Jun. 2004{\natexlab{b}}. {The
  Propagation and Eruption of Relativistic Jets from the Stellar Progenitors of
  Gamma-Ray Bursts}. \apj 608, 365--377.

\bibitem[{{Zhang} et~al.(2003{\natexlab{b}}){Zhang}, {Woosley}, and
  {MacFadyen}}]{zhangw03}
{Zhang}, W., {Woosley}, S.~E., {MacFadyen}, A.~I., Mar. 2003{\natexlab{b}}.
  {Relativistic Jets in Collapsars}. \apj 586, 356--371.

\bibitem[{{Zhao} et~al.(2011){Zhao}, {Li}, and {Bai}}]{zhao11}
{Zhao}, X.-H., {Li}, Z., {Bai}, J.-M., Jan. 2011. {The Bulk Lorentz Factors of
  Fermi-LAT Gamma Ray Bursts}. \apj 726, 89.

\bibitem[{{Zheng} et~al.(2012){Zheng}, {Shen}, {Sakamoto}, {Beardmore}, {De
  Pasquale}, {Wu}, {Gorosabel}, {Urata}, {Sugita}, and {Zhang}}]{zheng12}
{Zheng}, W., {Shen}, R.~F., {Sakamoto}, T., {Beardmore}, A.~P., {De Pasquale},
  M., {Wu}, X.~F., {Gorosabel}, J., {Urata}, Y., {Sugita}, S., {Zhang}, B.
  e.~a., Jun. 2012. {Panchromatic Observations of the Textbook GRB 110205A:
  Constraining Physical Mechanisms of Prompt Emission and Afterglow}. \apj 751,
  90.

\bibitem[{{Zou} et~al.(2009{\natexlab{a}}){Zou}, {Piran}, and {Sari}}]{zou09b}
{Zou}, Y., {Piran}, T., {Sari}, R., Feb. 2009{\natexlab{a}}. {Clues from the
  Prompt Emission of GRB 080319B}. \apjl 692, L92--L95.

\bibitem[{{Zou} et~al.(2009{\natexlab{b}}){Zou}, {Fan}, and {Piran}}]{zou09}
{Zou}, Y.-C., {Fan}, Y.-Z., {Piran}, T., Jun. 2009{\natexlab{b}}. {The possible
  high-energy emission from GRB 080319B and origins of the GeV emission of GRBs
  080514B, 080916C and 081024B}. \mnras 396, 1163--1170.

\bibitem[{{Zou} et~al.(2011){Zou}, {Fan}, and {Piran}}]{zou11}
{Zou}, Y.-C., {Fan}, Y.-Z., {Piran}, T., Jan. 2011. {A Revised Limit of the
  Lorentz Factors of Gamma-ray Bursts with Two Emitting Regions}. \apjl 726,
  L2.

\bibitem[{{Zou} et~al.(2007){Zou}, {Wu}, and {Dai}}]{zou07}
{Zou}, Y.~C., {Wu}, X.~F., {Dai}, Z.~G., Jan. 2007. {Estimation of the
  detectability of optical orphan afterglows}. \aap 461, 115--119.

\bibitem[{{Zrake} and {MacFadyen}(2012)}]{zrake12}
{Zrake}, J., {MacFadyen}, A.~I., Jan. 2012. {Numerical Simulations of Driven
  Relativistic Magnetohydrodynamic Turbulence}. \apj 744, 32.

\bibitem[{{Zweibel} and {Yamada}(2009)}]{zweibel09}
{Zweibel}, E.~G., {Yamada}, M., Sep. 2009. {Magnetic Reconnection in
  Astrophysical and Laboratory Plasmas}. \araa 47, 291--332.

\end{thebibliography}
\end{document}